\documentclass[fleqn]{mnras}

\usepackage{newtxtext,newtxmath}

\usepackage[T1]{fontenc}
\usepackage{ae,aecompl}

\usepackage{graphicx}	
\usepackage{amsmath}	
\usepackage{amssymb}	
\usepackage{multicol}
\usepackage{ragged2e}
\usepackage{booktabs}
\usepackage{float}
\usepackage{color}
\usepackage[table]{xcolor}
\usepackage{colortbl}
\usepackage{tabularx}
\usepackage{wrapfig}
\usepackage{xtab}
\usepackage{ltablex}
\usepackage{lscape}
\graphicspath{{images/}}
\newcolumntype{Y}{>{\centering\arraybackslash}X}

\def\etal{{\it et al.}\thinspace}

\def\eg{{\it e.g.}\thinspace}
\def\aj{{\it A.J.}\thinspace}
\def\apj{{\it Ap.J.}\thinspace}
\def\apjl{{\it Ap.J. Lett.}\thinspace}
\def\apjs{{\it Ap.J. Suppl.}\thinspace}
\def\aap{{\it A\&A}\thinspace}
\def\aaps{{\it A\&A Suppl.}\thinspace}
\def\jaa{{\it J. Ap. Ast.}\thinspace}
\def\mnras{{\it MNRAS}\thinspace}

\title[Arecibo 4.5/1.4/0.3-GHz Single-Pulse Survey]{Arecibo 4.5/1.4/0.33-GHz Polarimetric Single-Pulse Emission Survey}

\author[Olszanski, Mitra \& Rankin ]{
Timothy E. E. Olszanski,$^{1,2}$
Dipanjan Mitra,$^{1,3,4}$
and Joanna M. Rankin$^{1}$
\\
$^1$Department of Physics, University of Vermont, Burlington, VT, VT 05401, USA\\
$^2$Physics and Astronomy Department, PO Box 6315, West Virginia University, Morgantown, WV 26506 \\
$^3$National Centre for Radio Astrophysics, Tata Institute of Fundamental Research, Pune 411007, India\\
$^4$Janusz Gil Institute of Astronomy, University of Zielona G\'{o}ra, ul. Szafrana 2, PL-65-516 Zielona G\'{o}ra, Poland\\
}

\date{July 26, 2019}

\pubyear{2019}

\begin{document}
\label{firstpage}
\pagerange{\pageref{firstpage}--\pageref{lastpage}}
\maketitle

\begin{abstract}
We report on an Arecibo 4.5-GHz polarimetric single-pulse survey of the brightest pulsars at high frequency within its sky.  The high frequency profiles are accompanied by a collection of both previously published and unpublished high quality 1.4- and 0.33-GHz observations. Here our analyses and discussion primarily involve the average and statistical properties of the 46 pulsar's polarimetric pulse sequences, profile classification and frequency evolution, and polarimetric profiles and peak-occurrence histograms.  In most cases both the fractional linear polarization and profile widths decrease with frequency as expected, but there are some exceptions.  Similarly, we were able to review and/or extend the profile classifications for this population of pulsars and work out their beaming characteristics quantitatively showing that almost all show properties compatible with the core/double-cone emission beam model. The entirety of these observation's average profiles are accessible for download. 
\end{abstract}




\section{Introduction}
Pulsar high frequency radio emission has long been an interest of observational astronomers. Mainly, this is because of two factors; changes in component structure and changes in the polarization behaviour. In slow rotation powered pulsars, changes in component structure remain remarkably consistent and can be in most cases described by a core/double cone model. One such example of changes in component structure is the spectral evolution of components. In the context of the core/double cone model, it would suggest the varying frequencies of radio emission as originating from a continuum of emission heights (radius to frequency mapping, RFM; see Rankin 1983a). Thus, a study of high frequency radiation is expected to illuminate the physical conditions deeper within the pulsar magnetosphere.  Unfortunately, most pulsars have low flux densities at high frequencies, making single-pulse surveys above 2 GHz unattainable for all but the largest radio telescopes. As pulsar emission typically undergoes a frequency-dependent polarization, this makes single-pulse polarimetry at high frequencies even more difficult. 

The first high frequency observations of pulsar emission were carried out by astronomers at the Effelsberg Observatory, who pioneered early efforts before they were pursued more generally.  The classic survey of Morris, Graham, Sieber, Bartel \& Thomasson (1981; MGSBT) included a number of polarimetric profile measurements at 2.7 GHz, and other efforts at higher frequencies followed.  In particular, the 5-GHz survey of von Hoensbroech (1999) and von Hoensbroech \& Xilouris (1997; vHX) demonstrated the fascination and importance of pulsar studies in this regime of the radio spectrum.  

The Arecibo Observatory's 310-m primary and Gregorian reflector systems provide unmatched sensitivity to high frequency radio emission. Using recent 4.5-GHz as well as mostly existing 1400- and 327-MHz observations, we have herein assembled sets of multifrequency single-pulse, polarimetric observations on most of the brightest pulsars in the Arecibo sky.

\begin{table*}
\begin{tabular}{cccc|ccc|ccc|ccc}
\toprule
    Pulsar & P & DM & RM  & MJD & $N_{pulses}$ & Bins & MJD & $N_{pulses}$ & Bins & MJD & $N_{pulses}$ & Bins  \\
    (B1950) & (s) & pc/$cm^{3}$ & $rad m^{2}$ &   & & & & & & & &\\
    \midrule
    & & & & & \mbox{\textbf{(P-band)}} & & & \mbox{\textbf{(L-band)}} & & & \mbox{\textbf{(C-band)}} \\
    \midrule
    B0301+19	&	1.39	&	15.7	&	-8.3	&	57844	&	889	&	1024	&	57991	&	726	&	1024	&	57271	&	1157	&	1024	 \\
B0523+11	&	0.35	&	79.4	&	37.0	&	57956	&	1000	&	1024	&	57006	&	1681	&	1107	&	57271	&	1160	&	1024	 \\
B0525+21	&	3.75	&	50.9	&	-39.6	&	56164	&	267	&	1024	&	55522	&	630	&	1024	&	57271	&	1029	&	1024	 \\
B0540+23	&	0.25	&	77.7	&	2.7	&	57845	&	38813	&	635	&	57956	&	39210	&	1024	&	57271	&	1327	&	1024	 \\
B0609+37	&	0.30	&	27.2	&	23.0	&	56572	&	8784	&	1024	&	57113	&	4015	&	1164	&	57271	&	1177	&	1024	 \\
B0611+22	&	0.33	&	96.9	&	66.0	&	57846	&	24783	&	1082	&	57867	&	20872	&	1308	&	57292	&	2724	&	1024	 \\
B0626+24	&	0.48	&	84.2	&	69.5	&	54015	&	1878	&	1024	&	57113	&	1247	&	1036	&	57292	&	2113	&	1024	 \\
B0656+14	&	0.38	&	13.9	&	23.0	&	58106	&	24149	&	1020	&	56423	&	6069	&	1024	&	57292	&	2598	&	1024	 \\
B0751+32	&	1.44	&	40.0	&	-7.0	&	55988	&	4316	&	1024	&	55522	&	1928	&	1024	&	57292	&	905	&	1024	 \\
B0823+26	&	0.53	&	19.5	&	5.4	&	57902	&	15719	&	1054	&	58071	&	16012	&	1024	&	57292	&	2247	&	1024	 \\
B0834+06	&	1.27	&	12.9	&	25.3	&	53861*	&	1910	&	1024	&	56423	&	5126	&	1024	&	57292	&	590	&	1024	 \\
B0919+06	&	0.43	&	27.3	&	29.2	&	56417	&	14401	&	1024	&	56577	&	17638	&	1024	&	57292	&	2106	&	1024	 \\
B0950+08	&	0.25	&	3.0	&	-0.7	&	58206	&	14213	&	1046	&	55522	&	1929	&	1024	&	57292	&	2601	&	1024	 \\
B1133+16	&	1.19	&	4.8	&	4.0	&	53703*	&	2685	&	2048	&	55522	&	2094	&	1024	&	58206	&	6049	&	1024	 \\
B1237+25	&	1.38	&	9.3	&	-0.1	&	53378*	&	5199	&	1024	&	56550	&	863	&	1024	&	57307	&	1720	&	1024	 \\
B1530+27	&	1.12	&	14.7	&	1.0	&	55973	&	1855	&	1097	&	56249	&	6318	&	1024	&	57283	&	1110	&	1024	 \\
B1541+09	&	0.75	&	35.0	&	21.0	&	57954	&	1991	&	1024	&	55637	&	2047	&	1024	&	57307	&	1641	&	1024	 \\
B1612+07	&	1.21	&	21.4	&	40.0	&	56379	&	2714	&	1024	&	56415	&	1016	&	1024	&	57285	&	1463	&	1024	 \\
B1633+24	&	0.49	&	24.3	&	31.0	&	56157	&	4284	&	889	&	55982	&	4108	&	1033	&	57285	&	1700	&	1024	 \\
B1737+13	&	0.80	&	48.7	&	64.4	&	58125	&	4845	&	1037	&	55632	&	737	&	1024	&	57283	&	1447	&	1024	 \\
B1821+05	&	0.75	&	66.8	&	145.0	&	58125	&	979	&	1023	&	55633	&	1019	&	1024	&	57283	&	997	&	1024	 \\
B1839+09	&	0.38	&	49.2	&	53.0	&	57942	&	1568	&	1024	&	57114	&	1562	&	1059	&	57283	&	1243	&	1024	 \\
B1842+14	&	0.38	&	41.5	&	109.0	&	57942	&	2019	&	1024	&	57113	&	1586	&	1043	&	57285	&	1648	&	1024	 \\
B1848+12	&	1.21	&	70.6	&	158.0	&	58214	&	1138	&	1024	&	57942	&	318	&	1024	&	57285	&	732	&	1024	 \\
B1848+13	&	0.35	&	60.1	&	152.7	&	57878	&	2774	&	1275	&	57982	&	1201	&	1080	&	57283	&	1189	&	1024	 \\
B1853+01	&	0.27	&	96.7	&	-140.0	&	--	&	--	&	--	&	57941	&	1034	&	1024	&	57285	&	2212	&	1024	 \\
B1855+09	&	0.01	&	13.3	&	22.2	&	--	&	--	&	--	&	55637	&	111896	&	90	&	57285	&	51633	&	105	 \\
B1859+03	&	0.66	&	402.1	&	-237.4	&	--	&	--	&	--	&	56768	&	1011	&	1643	&	57285	&	1283	&	1024	 \\
B1900+01	&	0.73	&	245.2	&	72.3	&	--	&	--	&	--	&	57982	&	803	&	1024	&	57285	&	1160	&	1024	 \\
B1900+06	&	0.67	&	502.9	&	552.6	&	--	&	--	&	--	&	56415	&	1027	&	1024	&	57285	&	1485	&	1024	 \\
B1910+20	&	2.23	&	88.6	&	148.0	&	57878	&	1029	&	1088	&	56563	&	1014	&	1293	&	57285	&	525	&	1024	 \\
B1915+13	&	0.19	&	94.5	&	233.0	&	57940	&	14091	&	314	&	57941	&	12567	&	1024	&	57283	&	1186	&	1024	 \\
B1916+14	&	1.18	&	27.2	&	-41.7	&	57942	&	1062	&	1024	&	56415	&	1021	&	1024	&	57283	&	1071	&	1024	 \\
B1919+21	&	1.34	&	12.4	&	-16.99	&	58638	&	7073	&	1043	&	58639	&	7074	&	1036	&	58640	&	792	&	1024	 \\
B1924+16	&	0.58	&	176.9	&	320.0	&	57942	&	1109	&	1024	&	55982	&	1024	&	1004	&	57283	&	1224	&	1024	 \\
B1929+10	&	0.23	&	3.2	&	-6.9	&	55320	&	4629	&	1024	&	57995	&	39354	&	1024	&	57283	&	2151	&	1024	 \\
B1933+16	&	0.36	&	158.5	&	-10.2	&	--	&	--	&	--	&	57119	&	1046	&	1026	&	57286	&	4164	&	1024	 \\
B1935+25	&	0.20	&	53.2	&	26.0	&	57981	&	1027	&	1024	&	56419	&	2973	&	1024	&	57284	&	2630	&	1024	 \\
B1946+35	&	0.72	&	129.4	&	116.0	&	--	&	--	&	--	&	57638	&	6666	&	1024	&	57284	&	1227	&	1024	 \\
B1952+29	&	0.43	&	7.9	&	-18.0	&	56353	&	1343	&	1075	&	55632	&	2038	&	1024	&	57286	&	2387	&	1024	 \\
B2002+31	&	2.11	&	234.8	&	30.0	&	--	&	--	&	--	&	56564	&	1044	&	1223	&	57981	&	656	&	1024	 \\
B2016+28	&	0.56	&	14.2	&	-34.6	&	55320	&	942	&	1024	&	57114	&	1082	&	1089	&	57286	&	1427	&	1024	 \\
B2020+28	&	0.34	&	24.6	&	-74.7	&	55321	&	4200	&	1024	&	57114	&	2459	&	1041	&	57284	&	4125	&	1024	 \\
J0538+2817	&	0.14	&	39.6	&	-7.0	&	58390	&	16015	&	739	&	57004	&	4179	&	1118	&	57271	&	1198	&	1024	 \\
J0627+0649	&	0.35	&	86.6	&	180.0	&	57123	&	1719	&	1106	&	56932	&	1721	&	1353	&	57271	&	1570	&	1024	 \\
J1740+1000	&	0.15	&	23.9	&	23.8	&	57955	&	12151	&	995	&	56160	&	11610	&	1204	&	57283	&	2027	&	1024	 \\

    \bottomrule
   \end{tabular}
   \caption{Observation Information. MJDs marked by an asterisk identify WAPP observations.}
   \label{tab1}
\end{table*}

\begin{figure}
    \mbox{\includegraphics[height=95mm,,angle=0.]{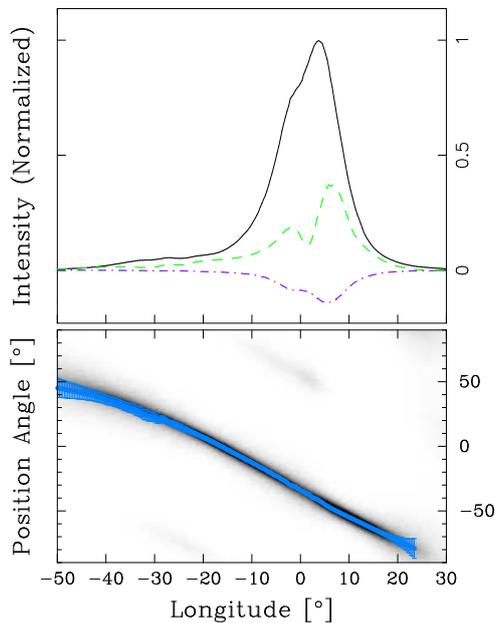}} 
        \caption{Sample average profile for pulsar B0950+08. The upper panel displays intensity $I$ (solid black), total linear polarization $L$ (dashed green), and circular $V$ (dash-dotted purple). The lower panel plots the single pulse PPAs in black, and the average PPA traverse for values greater than $3\sigma_{L}$ in blue.}
\label{fig1}
\end{figure}
\justifying

\section{Observations}
All the recent observations (2013 -- 2019) were carried out using Arecibo's Mock spectrometers\footnote{http://www.naic.edu/\textasciitilde phil/hardware/pdev/pdev.html} in single pixel mode\footnote{http://www.naic.edu/\textasciitilde phil/hardware/pdev/singlePixelSpecs.html}, but for a few cases where we lacked useable Mock observations, we instead included older (2005) WAPP\footnote{Wideband Arecibo Pulsar Processor; http://www.naic.edu/\textasciitilde wapp} observations. Each Mock spectrometer was configured for individual bandwidths of 170, 85 or 12.5 MHz at three frequencies; 4.6 GHz (C-band), 1.4 GHz (L-band), and 327 MHz (P-band).  Seven Mocks were used at C band for a total 1-GHz bandwidth; whereas, at L and P band four Mocks were used to provide a total nominal bandwidth of 350 or 50 MHz, respectively.  The WAPPs were used prior to early 2011 (MJD 55600), and their configurations were similar, apart from 100-MHz bandwidths at L band.  All three receiver systems have orthogonal linear feeds, but a circular hybrid was intermittently present in the P-band signal path for VLBI observations.

When taking data, the Mock spectrometers generate four uncalibrated (``search mode'')  data streams related to the Stokes parameters, which are subsequently written to disk for offline processing.  Each pulsar run was preceded by a short recording of a correlated 50-Hz calibration signal injected into the two signal paths; its use being both to determine the amplitude scaling and distinguish the real and imaginary parts of the voltage correlation.  Sufficient Fourier Transform lengths were used within the Mocks to provide nominal milliperiod dispersion resolution.  The output signals were then resampled modulo the pulsar rotation period using ephemerides either from the ATNF Pular Database (Manchester \etal\ 2005) or as updated by Ben Stappers (Private Communications), and combined to form the summed channel pulse sequence. Again, the sampling reflected the available resolution and was usually a milliperiod or less.  The band, Modified Julian Dates (MJDs), lengths, and modes of the observations are given in Table~\ref{tab1}.

\begin{figure}
    \mbox{\includegraphics[height=95mm,,angle=0.]{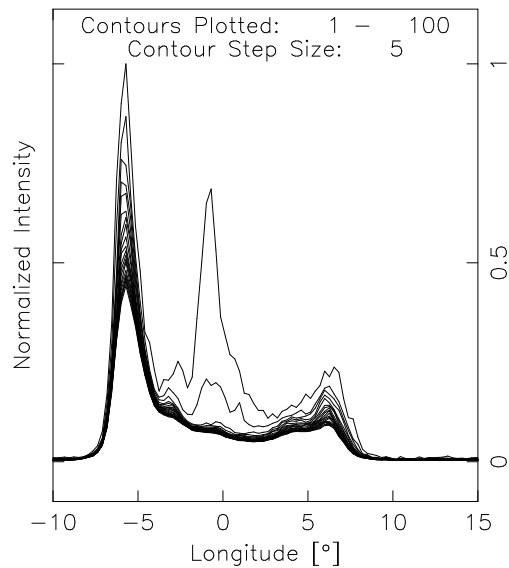}} 
        \caption{An example of a peak histogram plot for pulsar B1910+20. Each line represents a contour of the conjoined intensity maxima at each longitude. Indicated above are the plotted ordering of contours, from the maximum contour being 1 to the lowest ordering, and the ordering step size between each curve listed below.}
\label{fig2}
\end{figure}

\begin{table*}
\begin{center}
\begin{tabular}{cccccccc}
    \toprule
    Pulsar &  P & $\dot{P}$ & $\dot{E}$ & $\tau$ & $B_{surf}$ & $B_{12}/P^2$ & 1/Q  \\
    (B1950) & (s) & (s/s) & (ergs/s) & (yrs) & (G) &   &    \\
    \midrule
    \midrule
    B0301+19	&	1.3876	&	1.30E-15	&	1.9E+31	&	1.7E+07	&	1.4E+12	&	0.7	&	0.4	\\
B0523+11	&	0.3544	&	7.40E-17	&	6.5E+31	&	7.6E+07	&	1.6E+11	&	1.3	&	0.6	\\
B0525+21	&	3.7455	&	4.00E-14	&	3.0E+31	&	1.5E+06	&	1.2E+13	&	0.9	&	0.5	\\
B0540+23	&	0.2460	&	1.50E-14	&	4.1E+34	&	2.5E+05	&	2.0E+12	&	32.6	&	7.0	\\
B0609+37	&	0.2980	&	5.90E-17	&	8.9E+31	&	7.9E+07	&	1.3E+11	&	1.5	&	0.6	\\
B0611+22	&	0.3350	&	5.90E-14	&	6.2E+34	&	8.9E+04	&	4.5E+12	&	40.3	&	8.5	\\
B0626+24	&	0.4766	&	2.00E-15	&	7.3E+32	&	3.8E+06	&	9.9E+11	&	4.3	&	1.5	\\
B0656+14	&	0.3849	&	5.50E-14	&	3.8E+34	&	1.1E+05	&	4.7E+12	&	31.4	&	7.1	\\
B0751+32	&	1.4423	&	1.10E-15	&	1.4E+31	&	2.1E+07	&	1.3E+12	&	0.6	&	0.3	\\
B0823+26	&	0.5307	&	1.70E-15	&	4.5E+32	&	4.9E+06	&	9.6E+11	&	3.4	&	1.2	\\
B0834+06	&	1.2738	&	6.80E-15	&	1.3E+32	&	3.0E+06	&	3.0E+12	&	1.8	&	0.8	\\
B0919+06	&	0.4306	&	1.40E-14	&	6.8E+33	&	5.0E+05	&	2.5E+12	&	13.3	&	3.6	\\
B0950+08	&	0.2531	&	2.30E-16	&	5.6E+32	&	1.8E+07	&	2.4E+11	&	3.8	&	1.3	\\
B1133+16	&	1.1879	&	3.70E-15	&	8.8E+31	&	5.0E+06	&	2.1E+12	&	1.5	&	0.7	\\
B1237+25	&	1.3824	&	9.60E-16	&	1.4E+31	&	2.3E+07	&	1.2E+12	&	0.6	&	0.3	\\
B1530+27	&	1.1248	&	7.80E-16	&	2.2E+31	&	2.3E+07	&	9.5E+11	&	0.7	&	0.4	\\
B1541+09	&	0.7484	&	4.30E-16	&	4.1E+31	&	2.7E+07	&	5.8E+11	&	1.0	&	0.5	\\
B1612+07	&	1.2068	&	2.40E-15	&	5.3E+31	&	8.1E+06	&	1.7E+12	&	1.2	&	0.6	\\
B1633+24	&	0.4905	&	1.20E-16	&	4.0E+31	&	6.5E+07	&	2.4E+11	&	1.0	&	0.5	\\
B1737+13	&	0.8031	&	1.50E-15	&	1.1E+32	&	8.8E+06	&	1.1E+12	&	1.7	&	0.7	\\
B1821+05	&	0.7529	&	2.30E-16	&	2.1E+31	&	5.3E+07	&	4.2E+11	&	0.7	&	0.4	\\
B1839+09	&	0.3813	&	1.10E-15	&	7.8E+32	&	5.5E+06	&	6.5E+11	&	4.5	&	1.5	\\
B1842+14	&	0.3755	&	1.90E-15	&	1.4E+33	&	3.2E+06	&	8.5E+11	&	6.0	&	1.9	\\
B1848+12	&	1.2053	&	1.20E-14	&	2.6E+32	&	1.7E+06	&	3.8E+12	&	2.6	&	1.1	\\
B1848+13	&	0.3456	&	1.50E-15	&	1.4E+33	&	3.7E+06	&	7.3E+11	&	6.1	&	1.9	\\
B1853+01	&	0.2674	&	2.10E-13	&	4.3E+35	&	2.0E+04	&	7.6E+12	&	105.6	&	18.1	\\
B1855+09	&	0.0054	&	1.80E-20	&	4.6E+33	&	4.8E+09	&	3.1E+08	&	10.9	&	2.0	\\
B1859+03	&	0.6555	&	7.50E-15	&	1.0E+33	&	1.4E+06	&	2.2E+12	&	5.2	&	1.8	\\
B1900+01	&	0.7293	&	4.00E-15	&	4.1E+32	&	2.9E+06	&	1.7E+12	&	3.3	&	1.2	\\
B1900+06	&	0.6735	&	7.70E-15	&	1.0E+33	&	1.4E+06	&	2.3E+12	&	5.1	&	1.7	\\
B1910+20	&	2.2330	&	1.00E-14	&	3.6E+31	&	3.5E+06	&	4.8E+12	&	1.0	&	0.5	\\
B1915+13	&	0.1946	&	7.20E-15	&	3.9E+34	&	4.3E+05	&	1.2E+12	&	31.7	&	6.7	\\
B1916+14	&	1.1810	&	2.10E-13	&	5.1E+33	&	8.8E+04	&	1.6E+13	&	11.5	&	3.6	\\
B1919+21	&	1.3373	&	1.35E-15	&	2.2E+31	&	1.57E+07	&	1.4E+12	&	0.8	&	0.4 \\
B1924+16	&	0.5798	&	1.80E-14	&	3.6E+33	&	5.1E+05	&	3.3E+12	&	9.7	&	2.9	\\
B1929+10	&	0.2265	&	1.20E-15	&	3.9E+33	&	3.1E+06	&	5.2E+11	&	10.1	&	2.7	\\
B1933+16	&	0.3587	&	6.00E-15	&	5.1E+33	&	9.5E+05	&	1.5E+12	&	11.5	&	3.2	\\
B1935+25	&	0.2010	&	6.40E-16	&	3.1E+33	&	5.0E+06	&	3.6E+11	&	9.0	&	2.4	\\
B1946+35	&	0.7173	&	7.10E-15	&	7.6E+32	&	1.6E+06	&	2.3E+12	&	4.4	&	1.6	\\
B1952+29	&	0.4267	&	1.70E-18	&	8.7E+29	&	4.0E+09	&	2.7E+10	&	0.1	&	0.1	\\
B2002+31	&	2.1113	&	7.46E-14	&	3.1E+32	&	4.49E+05	&	1.3E+13	&	2.8 & 1.2 \\
B2016+28	&	0.5580	&	1.50E-16	&	3.4E+31	&	6.0E+07	&	2.9E+11	&	0.9	&	0.4	\\
B2020+28	&	0.3434	&	1.90E-15	&	1.8E+33	&	2.9E+06	&	8.2E+11	&	6.9	&	2.1	\\
J0538+2817	&	0.1432	&	3.70E-15	&	4.9E+34	&	6.2E+05	&	7.3E+11	&	35.8	&	7.1	\\
J0627+0649	&	0.3465	&	1.70E-15	&	1.6E+33	&	3.2E+06	&	7.8E+11	&	6.5	&	2.0	\\
J1740+1000	&	0.1541	&	2.10E-14	&	2.3E+35	&	1.1E+05	&	1.8E+12	&	77.5	&	13.3	\\

    \bottomrule
   \end{tabular}
   \raggedright{}\caption{Pulsar Parameters}\label{tab4}
\end{center}
\end{table*}

\subsection{Polarimetry}
Electromagnetic radiation is usually described and measured experimentally as a set of four Stokes Parameters. These are comprised of the intensity ($I$), and the two types of polarization a wave can take, (Linear -- $Q$, $U$ and Circular -- $V$).  The total linear polarization $L$ = $\sqrt{Q^2+U^2}$, and the polarization position angle (PPA) $\chi$ = $\frac{1}{2}\tan^{-1}(U/Q)$.  The four Stokes parameters were formed off-line in software, corrected for dispersion delays, Faraday rotation, and instrumental polarization effects.  

The polarimetry is absolute as follows:  values for the ionospheric Faraday rotation were estimated using GPS data and applied to model its time variation through the observation. Interstellar rotation measures (RM) were then measured for each L- and P-band observation by maximizing the aggregate linear polarization across the band [see Curtin \etal\ (2019) for the details of both procedures]---or in some cases the values from Sobey \etal\ (2019) were used when more accurate.  Nominal RM values or a nearby measurement were used at C band.  The measured PPAs were then derotated to infinite frequency, so as to have an absolute reference counterclockwise from north on the sky (see Morris \etal\ 1979; Johnston \etal\ 2005; Rankin 2015).  This leaves RM errors at C and L bands that are often as small as  0.25 rad-m$^2$, so that the PPAs can be directly compared at C and L band.  The measurement errors at P-band are often even smaller apart from errors in accommodating ionospheric RM variations, so the P-band PPAs often track the higher frequencies well.  A sample profile is given in Figure~\ref{fig1}.  Note the primary PPA track as well as the ``shadow'' corresponding to the secondary orthogonal polarization mode (OPM).

\subsection{RFI Mitigation}
Radio Frequency Interference (RFI) can degrade the observations, especially at the high frequencies we are using.  In general, RFI can be mitigated in two different ways. For RFI that is concentrated over a narrow range of frequencies, affected frequency channels can be excised before creating the final channel summed pulse sequence. With more sporadic broadband RFI, it is better mitigated by blanking single pulses from the summed channel pulse sequence as we did in the case of our observations. 

In at least one case (B1821+05/C-band), the off-pulse RFI was removed by blanking the affected profile section and injecting artificial noise (as the error of Stokes I follows a Gaussian distribution, the noise from the remaining part of the off-pulse can be used to generate the distribution).

\subsection{Average Profiles}

 \begin{figure}
    \mbox{\includegraphics[height=55mm,,angle=0.]{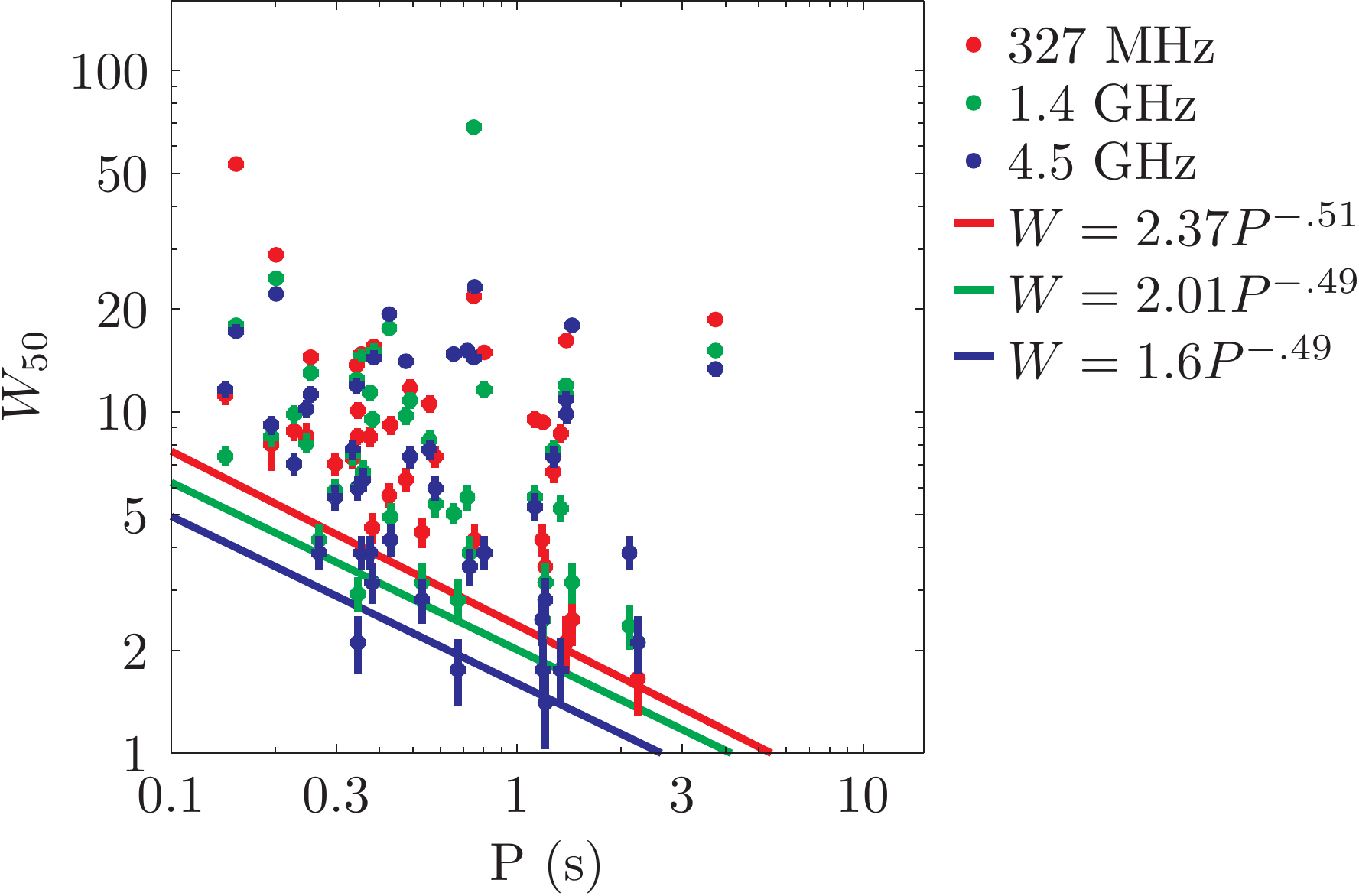}} 
    \mbox{\includegraphics[height=55mm,,angle=0.]{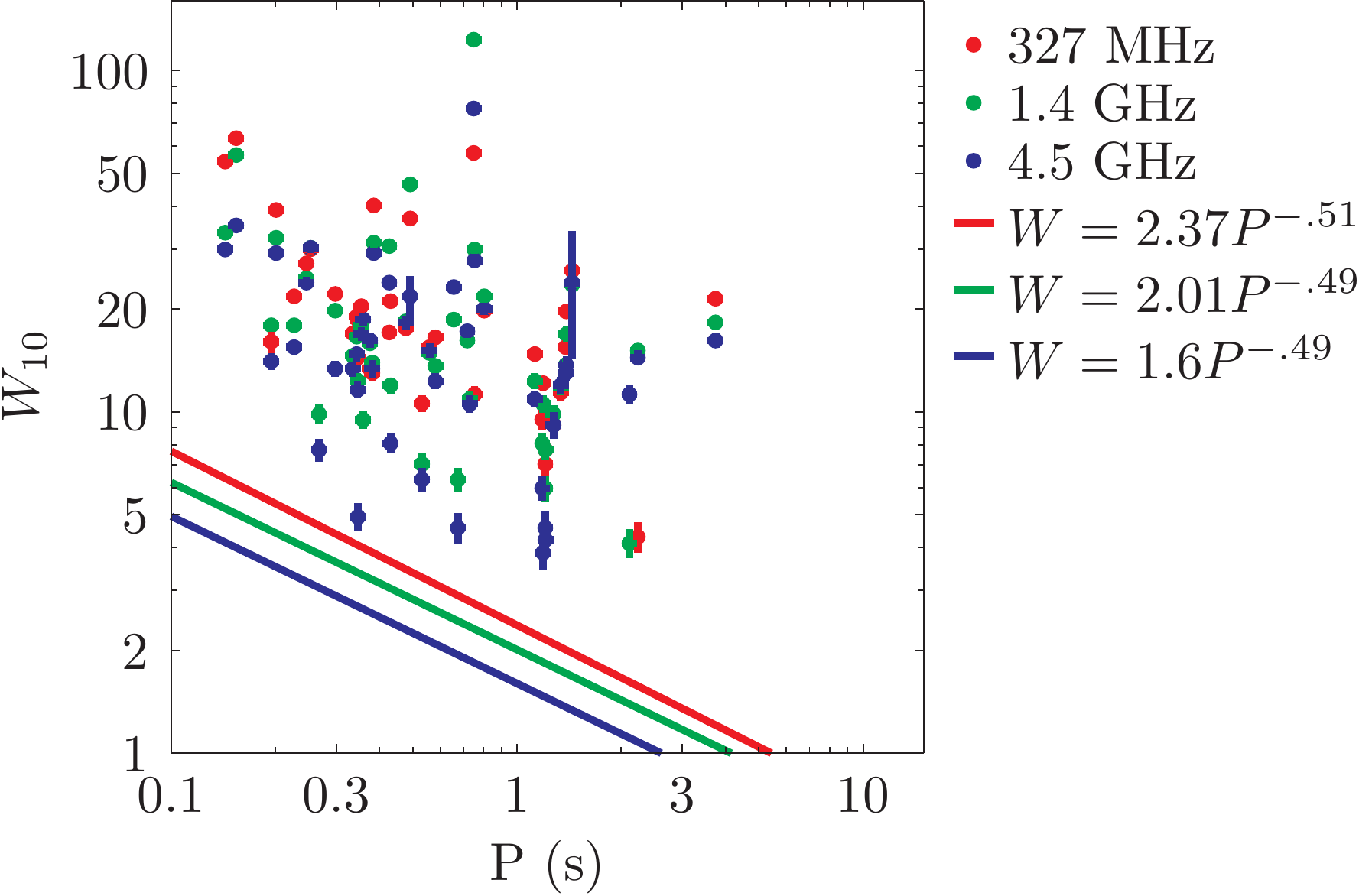}} 
    \mbox{\includegraphics[height=55mm,,angle=0.]{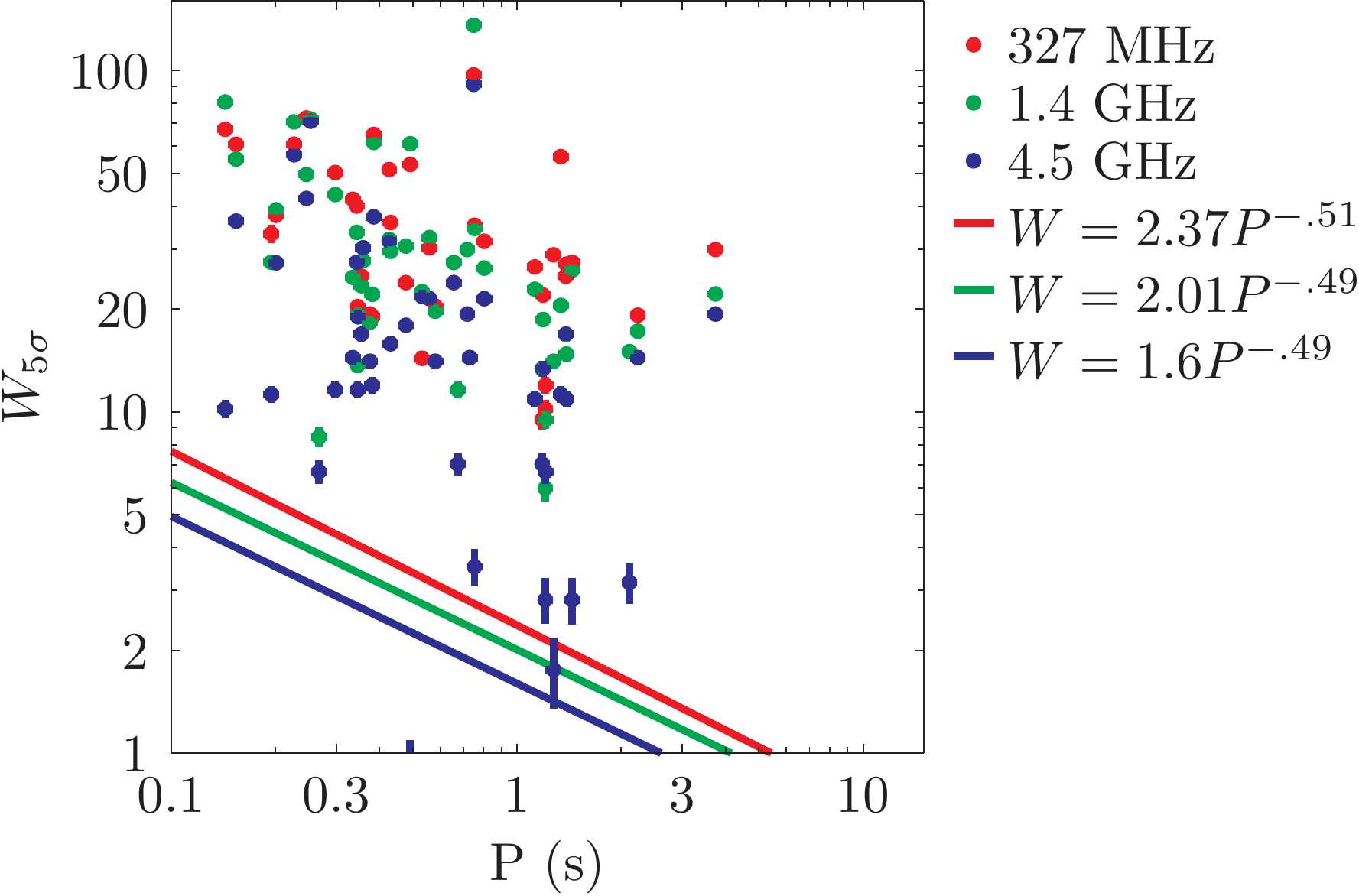}} 
\caption{$W_{50}$, $W_{10}$, and $W_{5\sigma}$ vs P. Lower boundary lines for the pulse half-width distribution have been taken from Skrzypczak \etal\ (2018).} 
\label{fig3}
\end{figure}

The total linear polarization at a particular bin/phase $k$ is $L_{k} = \sqrt{Q_{k}^{2}+U_{k}^{2}}$, while the fractional linear polarization is defined as $L/I = \frac{\sum_{k}^{} L_{k}}{\sum_{k}^{} I_{k}}$ where we have summed values of $I_{k}$ and $L_{k}$ for values three times their respective off-pulse noise value.

Assuming the errors in $Q$ and $U$ are Gaussian, the derived linear polarization error is not, and thus its uncertainties are better modeled by Monte Carlo methods (Mitra, Rankin \& Arjunwadkar 2016). Table~\ref{tab2} lists each quantity and its 2$\sigma$ error as determined from the sampled distribution. 

For each observation, we have plotted the average $I$, $L$, and $V$ along with the qualifying ($\geq2\sigma_{L}$) normalized single pulse polarization position angles (PPAs) for direct comparison between frequencies. These PPAs are plotted with error bars for values greater than $3\sigma_{L}$. Figure~\ref{fig1} provides an example display and the remainder appear in Appendix A. In each profile, the peak was rotated such that the central longitude corresponds with the pulse's center (and not the point of maximum intensity).

\subsection{Profile Width Estimation}
Uncertainties in the total profile widths were computed via
\begin{equation}
\sigma = \text{res x }\sqrt{1+\bigg(\frac{\text{rms}}{I}\bigg)^{2}}
\end{equation}
where $res$ is the sampling time per pulse bin (time resolution), $rms$ is the off-pulse-noise standard deviation for the summed channel profile, and Stokes $I$ is the measured signal intensity at the bin of interest, as described by Kijak \& Gil (1997). 

Profile widths were measured with respect to the profile's peak intensity while conal/core widths were measured with respect to the outermost/component half-power points. Table~\ref{tab3} lists 50\%, 10\% and 5$\sigma$ values for each observation together with its error.

\subsection{Peak Histogram Analyses}

\begin{table*}

\begin{tabular}{c|ccc|ccc|ccc}
    \toprule
    Pulsar &  $W_{50}$ & $W_{10}$ & $W_{5\sigma}$ & $W_{50}$ & $W_{10}$ & $W_{5\sigma}$ & $W_{50}$ & $W_{10}$ & $W_{5\sigma}$  \\
    (B1950) & (\degr)  & (\degr) & (\degr) & (\degr) & (\degr) & (\degr) & (\degr) & (\degr) & (\degr) \\
    \midrule
    & & \mbox{\textbf{(P-band)}} & & & \mbox{\textbf{(L-band)}} & & & \mbox{\textbf{(C-band)}} \\
    \midrule
    B0301+19	&	16.2	$ \pm $	0.4	&	19.7	$ \pm $	0.4	&	27.1	$ \pm $	0.4	&	11.2	$ \pm $	0.4	&	16.9	$ \pm $	0.5	&	14.8	$ \pm $	0.4	&	9.8	$ \pm $	0.4	&	13.7	$ \pm $	0.5	&	10.9	$ \pm $	0.4	 \\
B0523+11	&	14.8	$ \pm $	0.4	&	20.4	$ \pm $	0.4	&	25.0	$ \pm $	0.4	&	14.6	$ \pm $	0.3	&	17.9	$ \pm $	0.3	&	23.4	$ \pm $	0.3	&	3.9	$ \pm $	0.4	&	16.9	$ \pm $	0.4	&	16.9	$ \pm $	0.4	 \\
B0525+21	&	18.6	$ \pm $	0.4	&	21.4	$ \pm $	0.4	&	29.9	$ \pm $	0.4	&	15.1	$ \pm $	0.4	&	18.3	$ \pm $	0.4	&	22.1	$ \pm $	0.4	&	13.4	$ \pm $	0.4	&	16.2	$ \pm $	0.4	&	19.3	$ \pm $	0.4	 \\
B0540+23	&	8.5	$ \pm $	0.6	&	27.2	$ \pm $	0.6	&	72.6	$ \pm $	0.6	&	8.1	$ \pm $	0.4	&	24.6	$ \pm $	0.4	&	49.6	$ \pm $	0.4	&	10.2	$ \pm $	0.4	&	23.9	$ \pm $	0.4	&	42.2	$ \pm $	0.4	 \\
B0609+37	&	7.0	$ \pm $	0.4	&	22.1	$ \pm $	0.4	&	50.2	$ \pm $	0.4	&	5.9	$ \pm $	0.3	&	19.8	$ \pm $	0.3	&	43.3	$ \pm $	0.3	&	5.6	$ \pm $	0.4	&	13.4	$ \pm $	0.4	&	11.6	$ \pm $	0.4	 \\
B0611+22	&	7.3	$ \pm $	0.3	&	17.0	$ \pm $	0.3	&	41.9	$ \pm $	0.3	&	7.4	$ \pm $	0.3	&	14.6	$ \pm $	0.3	&	24.8	$ \pm $	0.3	&	7.7	$ \pm $	0.4	&	13.4	$ \pm $	0.4	&	14.4	$ \pm $	0.4	 \\
B0626+24	&	6.3	$ \pm $	0.4	&	17.6	$ \pm $	0.4	&	23.9	$ \pm $	0.4	&	9.7	$ \pm $	0.3	&	18.4	$ \pm $	0.3	&	30.6	$ \pm $	0.4	&	14.1	$ \pm $	0.4	&	18.3	$ \pm $	0.4	&	17.9	$ \pm $	0.4	 \\
B0656+14	&	15.5	$ \pm $	0.4	&	40.2	$ \pm $	0.4	&	64.9	$ \pm $	0.4	&	15.1	$ \pm $	0.4	&	31.3	$ \pm $	0.4	&	61.5	$ \pm $	0.4	&	14.4	$ \pm $	0.4	&	29.2	$ \pm $	0.4	&	37.3	$ \pm $	0.4	 \\
B0751+32	&	2.5	$ \pm $	0.4	&	25.9	$ \pm $	0.4	&	27.4	$ \pm $	0.4	&	3.2	$ \pm $	0.4	&	23.6	$ \pm $	0.4	&	26.0	$ \pm $	0.4	&	17.9	$ \pm $	0.4	&	23.9	$ \pm $	9.2	&	2.8	$ \pm $	0.4	 \\
B0823+26	&	4.4	$ \pm $	0.3	&	10.6	$ \pm $	0.3	&	14.3	$ \pm $	0.3	&	3.2	$ \pm $	0.4	&	7.0	$ \pm $	0.4	&	22.5	$ \pm $	0.4	&	2.8	$ \pm $	0.4	&	6.3	$ \pm $	0.4	&	21.8	$ \pm $	0.4	 \\
B0834+06	&	6.7	$ \pm $	0.4	&	9.1	$ \pm $	0.4	&	28.8	$ \pm $	0.4	&	7.7	$ \pm $	0.4	&	9.8	$ \pm $	0.4	&	14.1	$ \pm $	0.4	&	7.4	$ \pm $	0.4	&	9.1	$ \pm $	0.6	&	1.8	$ \pm $	0.4	 \\
B0919+06	&	9.1	$ \pm $	0.4	&	21.1	$ \pm $	0.4	&	35.9	$ \pm $	0.4	&	4.9	$ \pm $	0.4	&	12.0	$ \pm $	0.4	&	29.5	$ \pm $	0.4	&	4.2	$ \pm $	0.4	&	8.1	$ \pm $	0.4	&	15.8	$ \pm $	0.4	 \\
B0950+08	&	14.5	$ \pm $	0.3	&	29.9	$ \pm $	0.3	&	71.2	$ \pm $	0.4	&	13.0	$ \pm $	0.4	&	30.2	$ \pm $	0.4	&	72.1	$ \pm $	0.4	&	11.2	$ \pm $	0.4	&	30.2	$ \pm $	0.4	&	71.0	$ \pm $	0.4	 \\
B1133+16	&	9.3	$ \pm $	0.2	&	12.1	$ \pm $	0.2	&	22.0	$ \pm $	0.2	&	2.5	$ \pm $	0.4	&	10.5	$ \pm $	0.4	&	18.6	$ \pm $	0.4	&	1.8	$ \pm $	0.4	&	3.9	$ \pm $	0.4	&	13.4	$ \pm $	0.4	 \\
B1237+25	&	2.1	$ \pm $	0.4	&	15.5	$ \pm $	0.4	&	25.0	$ \pm $	0.4	&	12.0	$ \pm $	0.4	&	13.7	$ \pm $	0.4	&	16.9	$ \pm $	0.4	&	10.9	$ \pm $	0.4	&	13.0	$ \pm $	0.4	&	16.9	$ \pm $	0.4	 \\
B1530+27	&	9.5	$ \pm $	0.3	&	14.8	$ \pm $	0.3	&	26.6	$ \pm $	0.3	&	5.6	$ \pm $	0.4	&	12.3	$ \pm $	0.4	&	22.9	$ \pm $	0.4	&	5.3	$ \pm $	0.4	&	10.9	$ \pm $	0.4	&	10.9	$ \pm $	0.4	 \\
B1541+09	&	21.8	$ \pm $	0.4	&	57.3	$ \pm $	0.4	&	97.0	$ \pm $	0.4	&	68.2	$ \pm $	0.4	&	123.0	$ \pm $	0.4	&	135.7	$ \pm $	0.4	&	14.4	$ \pm $	0.4	&	77.3	$ \pm $	0.4	&	91.1	$ \pm $	0.4	 \\
B1612+07	&	3.2	$ \pm $	0.4	&	7.7	$ \pm $	0.4	&	12.0	$ \pm $	0.4	&	3.2	$ \pm $	0.4	&	7.7	$ \pm $	0.4	&	9.5	$ \pm $	0.4	&	1.4	$ \pm $	0.4	&	4.2	$ \pm $	0.4	&	6.7	$ \pm $	0.4	 \\
B1633+24	&	11.7	$ \pm $	0.4	&	36.9	$ \pm $	0.4	&	53.0	$ \pm $	0.4	&	10.8	$ \pm $	0.3	&	46.4	$ \pm $	0.3	&	61.0	$ \pm $	0.4	&	7.4	$ \pm $	0.4	&	21.8	$ \pm $	2.6	&	0.7	$ \pm $	0.4	 \\
B1737+13	&	14.9	$ \pm $	0.3	&	19.8	$ \pm $	0.3	&	31.6	$ \pm $	0.4	&	11.6	$ \pm $	0.4	&	21.8	$ \pm $	0.4	&	26.4	$ \pm $	0.4	&	3.9	$ \pm $	0.4	&	20.0	$ \pm $	0.4	&	21.4	$ \pm $	0.4	 \\
B1821+05	&	4.2	$ \pm $	0.4	&	11.3	$ \pm $	0.4	&	35.2	$ \pm $	0.4	&	23.2	$ \pm $	0.4	&	29.9	$ \pm $	0.4	&	34.5	$ \pm $	0.4	&	23.2	$ \pm $	0.4	&	27.8	$ \pm $	0.6	&	3.5	$ \pm $	0.4	 \\
B1839+09	&	4.6	$ \pm $	0.4	&	13.0	$ \pm $	0.4	&	19.0	$ \pm $	0.4	&	9.5	$ \pm $	0.3	&	13.9	$ \pm $	0.3	&	22.1	$ \pm $	0.3	&	3.2	$ \pm $	0.4	&	13.4	$ \pm $	0.4	&	12.0	$ \pm $	0.4	 \\
B1842+14	&	8.4	$ \pm $	0.4	&	13.4	$ \pm $	0.4	&	19.3	$ \pm $	0.4	&	11.4	$ \pm $	0.3	&	15.9	$ \pm $	0.3	&	18.3	$ \pm $	0.4	&	3.9	$ \pm $	0.4	&	16.2	$ \pm $	0.4	&	14.1	$ \pm $	0.4	 \\
B1848+12	&	3.5	$ \pm $	0.4	&	7.0	$ \pm $	0.4	&	10.2	$ \pm $	0.4	&	3.2	$ \pm $	0.4	&	6.0	$ \pm $	0.4	&	6.0	$ \pm $	0.4	&	2.8	$ \pm $	0.4	&	4.6	$ \pm $	0.4	&	2.8	$ \pm $	0.4	 \\
B1848+13	&	8.5	$ \pm $	0.3	&	14.4	$ \pm $	0.3	&	20.3	$ \pm $	0.3	&	6.0	$ \pm $	0.3	&	12.3	$ \pm $	0.3	&	13.7	$ \pm $	0.3	&	6.0	$ \pm $	0.4	&	11.6	$ \pm $	0.4	&	11.6	$ \pm $	0.4	 \\
B1853+01	&	--		&      --    &	      --	&	4.2	$ \pm $	0.4	&	9.8	$ \pm $	0.4	&	8.4	$ \pm $	0.4	&	3.9	$ \pm $	0.4	&	7.7	$ \pm $	0.4	&	6.7	$ \pm $	0.4	 \\
B1855+09	&	--		&      --    &	      --	&	44.0	$ \pm $	4.2	&	54.0	$ \pm $	6.9	&	44.0	$ \pm $	4.2	&	33	$ \pm $	 4	&	44	$ \pm $	5	&	??	$ \pm $		?? \\
B1859+03	&	--		&      --    &	      --	&	5.0	$ \pm $	0.2	&	18.6	$ \pm $	0.2	&	27.4	$ \pm $	0.2	&	14.8	$ \pm $	0.4	&	23.2	$ \pm $	0.4	&	23.9	$ \pm $	0.4	 \\
B1900+01	&	--		&      --    &	      --	&	3.9	$ \pm $	0.4	&	10.9	$ \pm $	0.4	&	14.4	$ \pm $	0.4	&	3.5	$ \pm $	0.4	&	10.5	$ \pm $	0.4	&	14.4	$ \pm $	0.4	 \\
B1900+06	&	--		&      --    &	      --	&	2.8	$ \pm $	0.4	&	6.3	$ \pm $	0.4	&	11.6	$ \pm $	0.4	&	1.8	$ \pm $	0.4	&	4.6	$ \pm $	0.4	&	7.0	$ \pm $	0.4	 \\
B1910+20	&	1.7	$ \pm $	0.3	&	4.3	$ \pm $	0.3	&	19.2	$ \pm $	0.3	&	2.1	$ \pm $	0.4	&	15.1	$ \pm $	0.4	&	17.2	$ \pm $	0.4	&	2.1	$ \pm $	0.4	&	14.4	$ \pm $	0.4	&	14.4	$ \pm $	0.4	 \\
B1915+13	&	8.0	$ \pm $	1.1	&	16.1	$ \pm $	1.1	&	33.2	$ \pm $	1.2	&	8.4	$ \pm $	0.4	&	17.9	$ \pm $	0.4	&	27.4	$ \pm $	0.4	&	9.1	$ \pm $	0.4	&	14.1	$ \pm $	0.4	&	11.2	$ \pm $	0.4	 \\
B1916+14	&	4.2	$ \pm $	0.4	&	9.5	$ \pm $	0.4	&	9.5	$ \pm $	0.4	&	2.5	$ \pm $	0.4	&	8.1	$ \pm $	0.4	&	13.4	$ \pm $	0.4	&	2.5	$ \pm $	0.4	&	6.0	$ \pm $	0.4	&	7.0	$ \pm $	0.4	 \\
B1919+21	&	8.6	$ \pm $	0.3	&	11.4	$ \pm $	0.3	&	55.9	$ \pm $	0.4	&	5.2	$ \pm $	0.3	&	11.8	$ \pm $	0.3	&	20.5	$ \pm $	0.4	&	1.8	$ \pm $	0.4	&	12.0	$ \pm $	0.4	&	11.2	$ \pm $	0.4	 \\
B1924+16	&	7.4	$ \pm $	0.4	&	16.5	$ \pm $	0.4	&	20.4	$ \pm $	0.4	&	5.4	$ \pm $	0.4	&	13.6	$ \pm $	0.4	&	19.7	$ \pm $	0.4	&	6.0	$ \pm $	0.4	&	12.3	$ \pm $	0.4	&	14.1	$ \pm $	0.4	 \\
B1929+10	&	8.8	$ \pm $	0.4	&	21.8	$ \pm $	0.4	&	60.8	$ \pm $	0.4	&	9.8	$ \pm $	0.4	&	17.9	$ \pm $	0.4	&	70.7	$ \pm $	0.4	&	7.0	$ \pm $	0.4	&	15.5	$ \pm $	0.4	&	56.6	$ \pm $	0.4	 \\
B1933+16	&	--		&      --    &	      --	&	6.7	$ \pm $	0.4	&	9.5	$ \pm $	0.4	&	27.7	$ \pm $	0.4	&	6.3	$ \pm $	0.4	&	18.6	$ \pm $	0.4	&	30.2	$ \pm $	0.4	 \\
B1935+25	&	28.8	$ \pm $	0.4	&	39.0	$ \pm $	0.4	&	37.7	$ \pm $	0.4	&	24.6	$ \pm $	0.4	&	32.3	$ \pm $	0.4	&	39.0	$ \pm $	0.4	&	22.1	$ \pm $	0.4	&	29.2	$ \pm $	0.4	&	27.3	$ \pm $	0.4	 \\
B1946+35	&	--		&      --    &	      --	&	5.6	$ \pm $	0.4	&	16.2	$ \pm $	0.4	&	29.9	$ \pm $	0.4	&	15.1	$ \pm $	0.4	&	17.2	$ \pm $	0.4	&	19.3	$ \pm $	0.4	 \\
B1952+29	&	5.7	$ \pm $	0.3	&	17.1	$ \pm $	0.3	&	51.2	$ \pm $	0.3	&	17.6	$ \pm $	0.4	&	30.6	$ \pm $	0.4	&	32.0	$ \pm $	0.4	&	19.3	$ \pm $	0.4	&	23.9	$ \pm $	0.4	&	31.6	$ \pm $	0.4	 \\
B2002+31	&	--	&	--	&	-- &	2.4	$ \pm $	0.3	&	4.1	$ \pm $	0.3	&	15.0	$ \pm $	0.3	&	3.9	$ \pm $	0.4	&	11.2	$ \pm $	0.4	&	3.2	$ \pm $	0.4	 \\
B2016+28	&	10.5	$ \pm $	0.4	&	15.5	$ \pm $	0.4	&	30.2	$ \pm $	0.4	&	8.3	$ \pm $	0.3	&	14.9	$ \pm $	0.3	&	32.4	$ \pm $	0.3	&	7.7	$ \pm $	0.4	&	15.1	$ \pm $	0.4	&	21.4	$ \pm $	0.4	 \\
B2020+28	&	13.7	$ \pm $	0.4	&	19.0	$ \pm $	0.4	&	40.1	$ \pm $	0.4	&	12.4	$ \pm $	0.3	&	16.6	$ \pm $	0.3	&	33.5	$ \pm $	0.4	&	12.0	$ \pm $	0.4	&	14.8	$ \pm $	0.4	&	27.4	$ \pm $	0.4	 \\
J0538+2817	&	11.2	$ \pm $	0.5	&	54.1	$ \pm $	0.5	&	67.2	$ \pm $	0.5	&	7.4	$ \pm $	0.3	&	33.5	$ \pm $	0.3	&	80.8	$ \pm $	0.3	&	11.6	$ \pm $	0.4	&	29.9	$ \pm $	0.5	&	10.2	$ \pm $	0.4	 \\
J0627+0649	&	10.1	$ \pm $	0.3	&	18.6	$ \pm $	0.3	&	19.9	$ \pm $	0.3	&	2.9	$ \pm $	0.3	&	16.8	$ \pm $	0.3	&	19.4	$ \pm $	0.3	&	2.1	$ \pm $	0.4	&	4.9	$ \pm $	0.4	&	19.0	$ \pm $	0.4	 \\
J1740+1000	&	53.2	$ \pm $	0.4	&	63.3	$ \pm $	0.4	&	60.8	$ \pm $	0.4	&	17.9	$ \pm $	0.3	&	56.5	$ \pm $	0.3	&	55.0	$ \pm $	0.3	&	17.2	$ \pm $	0.4	&	35.2	$ \pm $	0.4	&	36.2	$ \pm $	0.4	 \\

    \bottomrule
   \end{tabular}
   \caption{Profile Width Information}
   \label{tab3}
\end{table*}
\justifying

Pulsar emission on a single-pulse basis is well known to fluctuate broadly and on different time scales (Mitra \& Rankin 2011). Irregular peaks in single pulses can then be used to identify low-level components which are otherwise not detectable in average pulse profiles. To search for such sporadically emitting components, we have utilized a histogram based approach. For a 2-D pulse stack, we can treat each binned sample of the pulse as a 1-D intensity time-series. Flaring components will exhibit a wider range of intensities over steady state emission, thus showing up in these series as a higher intensity emission (this includes its maximum emission assuming a lack of RFI). Ordering from lowest to highest intensity, one would see that such a distribution is extremely sensitive to high intensity emission. Plotting each bin's greatest intensity then forms a contour of the brightest emission over the pulse region. Further contours then correspondingly relate to the lower numbered maxima. Essentially, we are plotting each bin's maximum intensity as a function of maxima. For the purposes of our work, we have defined a flaring component as one whose independent structure remains distinguishable over a wide range of intensity contours. This can also be thought of as forming contour outlines of the varying subpulse structure (hereafter Peak Histogram Plots PHP). An example for B1910+20 is shown in Figure~\ref{fig2} and the remainder appear in Appendix B.

\section{Survey Pulsars}
The 46 pulsars in this C-band single-pulse polarimetric survey are necessarily a population of some of the brightest pulsars in the Arecibo sky.  Some are bright across most of their observable band and some are bright because of relatively flat spectra.  As can be seen in Table~\ref{tab5}, they include all of the pulsar profile classes defined in Rankin (1993a,b).  Most of these stars have been studied extensively at lower frequencies, and we will try to include references to some of this work, but a full review of the published analyses is beyond the scope of this paper.  

Many survey pulsars were initially studied using the polarized multifrequency profiles then available---and in particular the early Effelsberg surveys of MGSBT and vHX, and these are referenced in Rankin (1993b).  The Rankin, Stinebring \& Weisberg (1989) 1.4-GHz survey included most of the pulsars below and demonstrates how to  classify them.  Weisberg \etal's (1999, 2004) Arecibo polarimetric surveys at 1.4 GHz and 430 MHz also include most of the current set and their classes.  Hankins \& Rankin (2010) published a single-pulse polarimetric survey of bright Arecibo pulsars at 430 and 1400 MHz.  Mitra \etal\ (2015) studied many of these pulsars at high time resolution.  Rankin (2015) explored the proper-motion alignments of many pulsars and showed that the physical (O,X) propagation modes can be identified in many cases.  Wahl \etal\ (2019) explore the polarization and profile geometry of many pulsars in our set at lower frequencies using the LOFAR High Band Survey of Sobey \etal\ (2015).  

The rotational periods ($P$), dispersion ($DM$) and rotation ($RM$) measures, along with the MJDs, lengths and resolutions of the observations are given in Table~\ref{tab1}.  In a few cases scattering prevented us from making meaningful P-band observations.  Fractional polarizations are given in Table~\ref{tab2} and the measured profile widths in Table~\ref{tab3}. The physical parameters derivable from a pulsar's period ($P$) and spindown ($\dot P$) are given in Table~\ref{tab4}---that is, the spindown energy ($\dot E$), spindown age ($\tau$), magnetic field ($B_{surf}$), acceleration parameter ($B_{12}/P^2$) and Beskin, Gurevich \& Istomin's (1993) (1/Q) parameter.  Finally, Table~\ref{tab5} gives the profile geometry and classifications, about which we will say more below.  However, one can see that the survey includes all of the pulsar profile classes defined in Rankin (1993a). 

Several studies in the literature have demonstrated that pulse width decreases  with increasing pulsar period (Lyne \& Manchester 1988, Rankin 1990, 1993, Gould \& Lyne 1998, Maciesiak \& Gil 2011, Maciesiak, Gil \& Melikidze 2012, Pilia \etal\ 2015, Mitra \etal\ 2016, Skrzypczak \etal\ 2018, Zhao \etal\ 2019, Johnston \& Karastergiou 2019), which is also seen in this survey as shown in Figure~\ref{fig4}. The importance of the existence of the lower boundary line (LBL) for both core and conal component $P^{-0.5}$ half-width scaling has also been demonstrated in recent works by Skrzypczak \etal\ (2018) and Zhao \etal\ (2019).

We find this scaling holds to at least 4.5 GHz, with at least three pulsars falling near the expected high frequency LBL as shown in Figures~\ref{fig3}. Interestingly, the pulsars that fall on this LBL (PSRs B1612+07, B1900+06, and B1910+20) exhibit similar PPA tracks and profiles.  The $P^{-0.5}$ dependence of conal profile widths has an obvious explanation in terms of the dipolar magnetic field geometry in the emission region (Rankin 1990), but the component width dependencies are more difficult to understand.  As Skrzypczak \etal\ (2018) point out, it is likely that an aspect of the radio emission mechanism is responsible. 

Figures~\ref{fig4} to \ref{fig6} give statistical population analyses of the tabulated values.  In Figure~\ref{fig4} the fractional linear polarization in the three bands is plotted against spindown age $\tau$, and energy $\dot E$, $P$ and $\dot P$.  Figure~\ref{fig5} plots the fractional circular polarization $V$ against the same quantities and Figure~\ref{fig6} against the absolute fractional $V$.  

A number of studies pertaining to specific pulsars are as follows---

{\bf B0301+19, B0525+21, B1133+16:} The three pulsars are classical examples of outer conal double profiles.  However, single pulse observations reveal both core and inner conal emission (Young \& Rankin 2012).  

{\bf J0538+2817 \& J1740+1000:} Full radio analyses of these two fast, high-energy pulsars are overdue given their interest and pertinence to key questions about pulsar physics and evolution. 

{\bf B0540+23, B0919+06, B1530+27, B1612+07, B1842+14, B1910+20, B1915+13 \& B1924+16} were found by Lyne \& Manchester (1988) to have ``partial cone'' profiles.  Mitra \& Rankin (2011) reobserved and reanalyzed this population and found that most have profiles of the types defined in Rankin (1993a).  

{\bf B0540+23, B0611+22 \& B0656+14:}  The three pulsars have long proven difficult to classify in part because of their large fractional linear polarization and large apparent aberration/retardation.  Lyne \& Manchester (1988) explored whether they could be examples of ``partial cones'', but Mitra \& Rankin (2011) argued for a different interpretation.  Seymour \etal's study identified the quasi-periodic shifting in the emission of B0611+22.  Our study of the three pulsars drawing on survey and other observations appears as Olszanski, Mitra \& Rankin (2019).  

{\bf B0823+26:} A multifrequency, polarimetric single pulse study was first conducted by Rankin \& Rathnasree (1995), and a further comprehensive analysis of both the pulsar's bright and weak modes by Young \etal\ (2012) and Sobey \etal\ 2012). Drawing on both the survey and other observations, our new study appears as Rankin, Olszanski \& Wright (2019). Single pulse observations suggest the presence of both an inner and outer emission cone.

{\bf B0834+06} exhibits interactions between its nulls and emission in a manner that may further illuminate the characteristics of subbeam carousels (Rankin \& Wright 2007).  

{\bf B0823+26, B0950+08, B1530+27 \& B1929+10:} The off-pulse structures in these pulsars manifesting as pre- and postcursor\footnote{An isolated emission region that lies between the main pulse and interpulse where a precursor precedes the main pulse and a postcursor follows the main pulse.} features are studied in an effort to understand their commonalities and emission geometries (Basu, Mitra \& Rankin 2015).  

{\bf B0919+06:} This pulsar exhibits one of the best examples of the ``swoosh'' phenomenon\footnote{The effect entails the emission window changing longitude gradually over a few pulses, remaining at the different longitude for a significant time, and then gradually returning to the usual longitude.} first identified by Rankin, Rodriguez \& Wright (2006).  A more detailed further study appears as Wahl \etal\ (2015). 

{\bf B1237+25:} This pulsar has been an exemplar of the five-component profile and was the object in which Don Backer identified bi-moding and null pulses as outlined in Srostlik \& Rankin (2005).  A more recent polarimetric analysis of the pulsar's normal and abnormal modes appear as Smith \etal\ (2013).  

{\bf B1541+09:} Nowakowski (1991) studied the single pulses of this very broad triple (T) profile pulsar.  

{\bf B1633+24:} Hankins \& Wolszczan (1987) showed the basic structure of this conal triple (cT) pulsar, but their work has never been adequately followed up.  

{\bf B1737+13} provides another rare example of a five-component (M) pulsar (Force \& Rankin 2010).

{\bf B1919+21:} This very first pulsar discovered by Jocelyn Bell was studied in detail in the 1970s, but has received little attention since.  Both its profile evolution and subpulse drift effects are interesting and will reward further investigation.  

{\bf B1929+10:} This fascinating pulsar with its main pulse, interpulse, and polarized emission over most of its rotation cycle remains one of the best polarization calibrators in the northern sky.  However, it has not been adequately studied in terms of its emission physics.  A beginning was made by Rankin \& Rathnasree (1997).  

{\bf B1933+16:} The pulsar is perhaps the brightest core-dominated pulsar in the Arecibo sky, and its characteristic core-single behavior in polarized single pulses is analyzed in Mitra \etal\ (2016).  

{\bf B1946+35:} A similar study was also carried out on this prominent core-single pulsar and appears as Mitra \& Rankin (2017).  

{\bf B2016+28} is one of the two Arecibo pulsars wherein the ``drifting subpulse'' phenomenon was discovered, but its irregular drift bands and non-orthogonal polarization modes (Ramachandran \etal\ 2004) have frustrated attempts to understand its individual pulse modulation (McKinnon 2003).  

{\bf B2020+28:} This fascinating pulsar, well studied in the past, may exhibit mode-switching as recently reported by Wen \etal\ (2018).  

\begin{table*}

\begin{tabular}{c|c|cc|ccc|ccc|ccc|cc}
    \toprule
    Pulsar & Class &  $W_{c}$ & $W_{cap}$ & $\alpha$ & $R$ & $\beta$ & $W_{cone1}$  & $\rho_1$ & $\beta/\rho_1$ & $W_{cone2}$ & $\rho_2$ & $\beta/\rho_2$ & $h_{cone1}$ & $h_{cone2}$  \\
    (B1950) &      & (\degr) & (\degr) & (\degr)  &(\degr/\degr) & (\degr) & (\degr) & (\degr) &(\degr) & (\degr) &(\degr) & (\degr) & (km) & (km)  \\
    \midrule
    B0301+19	&	D	&	---	&	2.1	&	30	& --17  &	+1.7 &	13.0 &	3.7	&  0.45	&	---	&	---	&	---	&	129	&	---	\\
B0523+11	&T?	&$\sim$4.2&	4.1	&	78	& --9.5 &  --5.9 &	---	 &	---	&	---	&  14.9	&	9.3	& --0.63&	---	&	206	\\
B0525+21	&	D	&	---	&	1.3	&	21	& +36	&	+0.6 &	---	 &	---	&	---	&  15.9	&	2.9	&  0.19 &	---	&	216	\\
B0540+23	&	S$_{t}$	&  7.7  &  4.9  &   40	& --3.5	&--10.6 &	 ---	& ---	& ---	&$\sim$20&  12.0&--0.88	& ---	& 235	\\
B0609+37	& S$_{t}$/T	&	5.1	&	4.5	&	62	& +24	&	+2.1 &	17.2 &	7.8	&  0.27	&	---	&	---	&	---	&	122	&	---	\\
B0611+22	&	S$_{t}$	&  4.4	&   4.2	&	74	& +11.1	&   +5.0 &  11.1 &	7.3	&  0.68	&   ---	&   ---	 &  ---	&	120	& --- \\
B0626+24	&cT/cQ?	& ---	&	3.5	&	63	& --12?	&	+4.3 &$\sim$10&	6.2	&  0.68	&$\sim$16&  8.4 & 0.51	&	123	&  225	\\
B0656+14	&	T	&$\sim$13&	3.9	&	18	& --3.8	&	+4.6 &	31? &	7.0	&  0.66	&  ---	&  ---	&	---	&	125	&	---	\\
B0751+32	&	D	&	4.8	&	2.0	&	26	& +25	&	+1.0 &	---	 &	---	&  ---	&  21.1	&	4.8	&  0.21	&	---	&	222	\\
B0823+26	&	S$_{t}$	&	3.6	&	3.4	&	84	& +30	&	+3.3 &$\sim$9&	5.5	&  0.59	&$\sim$14&	7.7	&  0.42	&	109	&	210	\\

B0834+06	&	D	&	3?	&	2.2	&	50	& +17	&	+2.6 &	7.7	 &	3.9	&  0.66	&	---	&	---	&	---	&	129	&	---	\\
B0919+06	& S$_{t}$/T	&$\sim$4.7&	3.7	&	53	& +9	&	+5.1 &	10.1 &	6.5	&  0.78	&	---	&	---	&	---	&	122	&	---	\\
B0950+08	&	S$_{d}$?	&  24.4	&	4.9	&	12	& --1.4 &	+8.5 &	75.4 &	8.7	&  0.98	&   ---	&   ---	&	---	&	128	&	---	\\
B1133+16	&	D	&	3?	&	2.2	&	46	& +10	&	+4.1 &	---	 &	---	&  ---	&   9.0	&	5.3	&  0.78	&	---	&	223	\\
B1237+25	&	M	&$\sim$2.6&	2.1	&	53	& --150	&  --0.3 &	9.4	 &	3.8	& --0.08&  12.0	&	4.8	& --0.06	&	131	&	213	\\

B1530+27	&	S$_{d}$	&	---	&	2.3	&	30	& +5.8	&	4.9	 &	---	 &	---	&	---	&	8.3	&	5.4	&  0.91	&	---	&	220	\\
B1541+09	&	T	& $\sim$29&	2.8	&	 6	& --2.2	&  --2.6 &	125	 &	5.1	&--0.51	&   ---	&	---	&   ---	&	128	&	---	\\
B1612+07	&	S$_{d}$/D	&	---	&	2.2	&	24	& --4.6	&	+5.1 &	---	 &	---	&	--- &	6.1	&	5.3	&  0.97	&	---	&	230	\\
B1633+24	&	cT/cQ	&	---	&	3.5	&	19	&	--4	&  --5.3 &$\sim$20&	6.0	&--0.89	&	47	&	8.4	&--0.64	&	118	&	230	\\
B1737+13	&	M	&$\sim$4.6&	2.7	&	36	& --12	&	+2.8 &	13.3 &	5.0	&  0.57	&  18.7	&	6.4	&  0.44	&	132	&	219	\\

B1821+05	&	T	&$\sim$6&	2.8	&	28	& --10	&	+2.7 &	--- &	---	&	---	&  24.5	&	6.6	&  0.41	&	---	&	217	\\
B1839+09	&	S$_{t}$	&$\sim$4?&	4.0	&	83	& --19	&	+3.0 &	10.2 &	6.3	&  0.47 &	---	&	---	&	---	&	102	&	---	\\
B1842+14	& S$_{t}$/T	&$\sim$4.6&	4.0	&	60	&	+12	&	4.2	 &	11.4 &	6.5	&  0.64 &	---	&	---	&	---	&	107	&	---	\\
B1848+12	&	S$_{d}$/D	& --- &	2.2	&	63	&	+36	&	1.4	 &	 8	 &	3.9	&  0.37 &	---	&	---	&	---	&	120	&	---	\\
B1848+13	&	S$_{t}$& $\sim$6&	4.2	&	44	&	---	&	---	 &	---	 & 	---	&	---	&	---	&	---	&	---	&	---	&	---	\\

B1853+01	&	S$_{d}$	&	---	&	4.7	&	---	& ---	&	---	 &	---	 &	---	&	---	&	---	&	---	&	---	&	---	&	---	\\
B1855+09	& S$_{t}$	&	35	&  33.5	&	73	& +1.3	&	47   &$\sim$60& 55.5& 0.85	&	---	&	---	&	---	&	110	&	---	\\
B1859+03	& S$_{t}$/T	&	5	&	3.0	&	39	& --10	&  --3.6 &	---	 &	---	&	---	&$\sim$20&	7.0	&--0.51	&	---	&	217	\\
B1900+01	&	S$_{t}$	&	3.6	&	2.9	&	53	&	+30	&	+1.5 &$\sim$12&	5.1	& 0.30	&	---	&	---	&	---	&	125	&	---	\\
B1900+06	&	T	&	3	&	3.0	&	84	& --16	&	3.6	 &	7.7	 &	5.2	& 0.68	&	---	&   ---	&	---	&	123	&	---	\\

B1910+20	&	M	&$\sim$3.4?&1.6	&	29	&	+18	&	1.5	 &$\sim$9.5&2.8	& 0.55	&   14	&	3.8	&  0.41	&	117	&	213	\\
B1915+13	& S$_{t}$		&	6	&	5.6	&	68	&	--8	&	+6.6 &	15	 &	9.7	& 0.68	&	---	&	---	&	---	&	122	&	---	\\
B1916+14	&	cT	&$\sim$2.3?&2.3	&	79	&   +45	&	+1.2 &	7.5	 &	3.9	& 0.32	&	---	&	---	&	---	&	119	&	---	\\
B1919+21	&	cQ?	&$\sim$3 & 2.1 &	45	&	--27	& 3.7	&$\sim$4 & 3.9 & 0.94	&  9.2	&	4.8	&  0.76	&	137	&	209	\\
B1924+16	& S$_{t}$		&	6.4	&	3.2	&	30	&	+6  &	+4.8 &	12	 &	5.8	& 0.83	&   ---	&	---	&  ---	&	129	&	---	\\
B1929+10m	& T/M?	& 5.15	&	5.1	&	88	&   --1.5	&	+6.4 &	13.2  &	42.2	& 0.99	&	---	&  ---	&  ---	&	2691	&	---	\\
B1929+10i	& T/M?	& 5.15	&	5.1	&	88	&	---	&	---	 &	---	 &	---	& ---	&	---	&	---	&	---	&	---	&	---	\\

B1933+16	& S$_{t}$		&$\sim$4.3&	4.1	&	72	&  --50	&	1.1	 &	---	 &	---	&	---	&$\sim$20&	9.6	&  0.11	&	---	&	220	\\
B1935+25	&	D	&	---	&	5.5	&	65	&	--9	&	5.8	 &	---	 &	---	&	---	&	25	&  12.9	&  0.45	&	---	&	224	\\
B1946+35	& S$_{t}$/T	&	5.7	&	2.9	&	32	&	+16	&   1.9  &	16.6 &	5.1	&  0.39 &	---	&	---	&	---	&	117	&	---	\\
B1952+29	& M/cQ?	&$\sim$6.5?&3.8	&	35	&	--7	&   4.7  &$\sim$13&	4.7	&  0.77 &$\sim$24&	8.7	&  0.54	&	108	&	215	\\
B2002+31	&	T	& 2.24	&	1.7	&	49	& $\infty$	& 0.0 &	---	 &	---	&	---	&	11.5	& 4.2	&  0.00	&	---	&	290	\\
B2016+28	&	S$_{d}$	&	---	&	3.3	&	39	&	+5	&	+7.2 &	---	 &	---	&	---	&	8.2	&	7.7	&  0.93	&	---	&	223	\\
B2020+28	& D/T?	&	6.2	&	4.2	&	88	&	+8	&	+7.2 &	---  &	---	&  ---	&  12.8	&   9.6	&  0.75	&	---	&	211	\\
J0538+2817	&	M	&$\sim$9&	6.5	&	46	&	+5	&	8.3	 &	19.7 & 11.2 &  0.74	&$\sim$34& 15.4 &  0.54	&	120	&	226	\\
J0627+0649	&	T?	&$\sim$6&	4.2	&	44	& --2.2	&  18.4  &	---  & ---	&	---	&	---	&	---	&  ---  &	---	&	---	\\
J1740+1000	&	T &	20.3 &	6.2	&	18	&	1.7	&  10.4 &	38.5	 &	11.1	& -.94	& --- &  --- & --- &	127	&	---	\\

    \bottomrule
   \end{tabular}
   \caption{Pulsar Geometry Information. Where $W_{cone1}$, $W_{cone2}$, and $W_{c}$ represent the inner cone, outer cone, and core half power width. Errors in width are similar to those listed in Table 3.}
   \label{tab5}
\end{table*}
\justifying

\section{Classification and Profile Geometry}
Most of the 46 pulsars in our survey have long been observed and studied and thus classified on the basis of the then available information in Rankin (1993a,b).  Subsequent studies have reexamined these classifications and in a few cases found evidence for changes (\eg Weisberg \etal\ 1999, 2004; Mitra \& Rankin 2011).  Our work here provides a further opportunity to examine these received conclusions and to confirm or better establish the quantitative geometry of this group of pulsars.  

The pulsar profile classification system in Rankin (1993a,b) envisions a core/double cone emission beam configuration, wherein two concentric annular conal beams surround a central core (``pencil'') beam (depicted in Figure 1 of both Rankin 1993a and 2015).  It assumes dipolarity of the magnetic field in the radio emission region such that the rotating-vector model (RVM) is valid (Radhakrishnan \& Cooke 1969; Komesaroff 1970) and the  profiles scale as the $P^{-0.5}$ angular size of the polar flux tube.  Its classes reflect different sightline trajectories as well as different profile evolution with frequency, but the descriptors pertain to the profile form at 1 GHz.  

It begins by distinguishing between single profiles of the core type ({\bf S}$_t$) and conal type ({\bf S}$_d$).  {\bf S}$_t$ profiles reflect a central sightline trajectory that encounters only the core beam; whereas {\bf S}$_d$ profiles entail an oblique sightline that only grazes one of the conal beams.  {\bf S}$_t$ profiles remain single at low frequency but generally develop a pair of conal ``outriding'' components at high frequency; whereas, {\bf S}$_d$ profiles remain single at high frequency but often bifurcate at low frequency due to enlargement of the conal beam (RTF).  

In addition, conal double profiles ({\bf D}) reflect more central sightlines through conal beams; triple profiles ({\bf T}) encounter the core beam and central parts of a conal beam; and five-component ({\bf M}) profiles encounter the core and both conal beams.  Conal beams thus are of two types, inner and outer.  Conal triple profiles (c{\bf T}) reflect a traverse that passes well inside an outer cone while grazing the inner cone; similarly rare conal quadruple profiles (c{\bf Q}) represent traverses that cut both cones more centrally but do not encounter a core beam.  

Most radio pulsar profiles can be understood in terms of the core/double-cone beam system where one, two or all three beams are encountered in different configurations and somewhat different RF spectra.  

The three beams are found to have specific angular dimensions at 1 GHz in terms of a pulsar's polar cap, {$\Delta_{\rm PC}$} = $2.45\degr P^{-1/2}$.  The outside half-power radii of the inner and outer cones, {$\rho_{i}$} and {$\rho_{o}$} are $4.33\degr P^{-1/2}$ and $5.75\degr P^{-1/2}$.

In practice. the magnetic colatitude $\alpha$ can be estimated from the width of the core component when present, as its expected half-power width at 1 GHz, $W_{\rm core}$ is {$\Delta_{PC}/\sin\alpha$}.  The sightline impact angle $\beta$ can then in turn be estimated from the steepest gradient (SG) of the polarization angle (PPA) traverse (at the inflection point) using $R$=$|d\chi/d\varphi|$, where $R$ is the ratio $\sin\alpha/\sin\beta$.  Conal beam radii can in turn be estimated from the outside half-power width of a conal component or conal component pair at 1 GHz $W_{\rm cone}$ together with $\alpha$ and $\beta$ using eq.(4) in Rankin (1993a), and/or squared with the characteristic conal dimensions in order to estimate $\alpha$.  The emission characteristic heights can then be computed assuming dipolarity using its eq.(6).

These 1-GHz heights are then typically about 120 and 220 km.  However, it is important to recall that these are {\em characteristic} emission heights, not physical ones, estimated using the convenient but perhaps problematic assumption that the emission occurs adjacent to the ``last open'' field lines at the polar fluxtube edge.  More physical emission heights can be estimated using aberration/retardation (see Blaskiewicz, Cordes \& Wasserman 1991, as corrected by Dyks, Rudak \& Harding 2004), and these are typically 2 -- 3 times larger than the characteristic emission heights.  

Indeed, physical emission heights measured using aberration/retardation are available on a number of the survey pulsars: B0823+26, B1237+25, B1737+13, B1933+16, and B1946+35; see the references in \S 3.  

We will now discuss the most interesting finds in our survey.

\subsection{B0540+23}
The profiles (Figure A2) and PHP plots (Figure B2) show that this pulsar has a more complex emission pattern than the core-single type that it has sometimes been assumed to be.  A comprehensive analysis of its characteristics will be given in Olszanski \etal\ (2019).

\subsection{B0609+37}
While we classify B0609+37 as a ${\bf S}_{t}$, emission apparent on the profile's (Figure A2) trailing edge is difficult to explain, possibly a postcursor. It is frequency dependent in our observations, and appears to be just perceptible in some of the Lyne \& Gould (1998) profiles. We see it at both P and L band but at different longitudes. The PHP plots in Figure~B2 suggest a triple structure, the steep PPA traverse a central sightline cut, and Weltevrede \etal\ (2007) find evidence of what may be drift modulation.  All in all a {\bf T} classification prompts a reasonable core/inner cone geometry in Table~\ref{tab5}.

\subsection{B0611+22}
This pulsar shows quite complex emission patterns at the different frequencies, and a full analysis of its properties will appear in Olszanski \etal\ (2019).

\subsection{B0626+24}
The frequency development of this pulsar (Figure A3) is surprising in showing a full double structure only at C band. What immediately catches the eye is how the PPA track evolves. Over the three frequencies, the leading PPA displays a consistent change in angle from the rest of the track, going from nearly orthogonal at P-band to nearly aligned at C-band. This cannot be simply an OPM effect, as its location is frequency dependent.  The asymmetric lower two profiles are suggestive of a conal triple or quadruple structure like that in pulsar B2034+19 (Rankin 2018), and the rough quantitative geometry in Table~\ref{tab5} is compatible with this classification.

\subsection{B0656+14}
Many words have been written about the strangeness of this pulsar's single pulses, but OPM-dominance ``jumps'' on the edges of the C-band profile (Figure A3) provide some new clarification.  We are studying the pulsar in detail in a separate effort (Olszanski \etal\ 2019).

\subsection{B0823+26}
Again we base our analysis on a current analysis of this pulsar (Rankin \etal\ 2019) along with the work of Everett \& Weisberg (2001) where both inner and outer conal features can be identified in its main pulse together with the usually dominant core emission.  

\subsection{B0919+06}
The pulsar's ``swooshes'' are known to distort its average profile at P and L band.  However, a partial profile analysis (Rankin \etal\ 2006) permits the discernment of its underlying triple {\bf T} structure which is reported in Table~\ref{tab5}.

\subsection{B0950+08}
The strongest evidence for a {\bf S}$_d$ designation has been the component bifurcation apparent in low frequency profiles. On close-inspection of single pulse observations at 1.5 GHz and 4.5 GHz using our PHPs (Figure B5), we identify the outline of a component structure that shares nearly the same location and width as the low frequency profile, suggesting the components as inner cones.  This would appear at odds with the well studied geometry of this pulsar (Basu \etal\ 2015; Everett \& Weisberg 2001). It points the question whether this pulsar truly has a {\bf S}$_d$ profile.

\subsection{B1133+16}
As previously mentioned, core emission is a known facet of this pulsar's single pulses. From our PHP's (Figure B5), we identify a peaked feature at 1.5 GHz similar in location to Young \& Rankin's  noted core component. The derived geometry suggests the exterior components as outer cones, and this is in agreement with their frequency evolution. 

We also note that the core component appears to undergo a frequency dependent shift with respect to its surrounding cones. This is not unique to our observations, as Figure 7 of Young \& Rankin (2012) shows a similar such occurrence at L-band. The shift is seen to be most pronounced at 4.5 GHz, with the conal components appearing to centre around the SG. 

\subsection{B1633+24}
We can have little doubt that this pulsar shows an oblique double cone profile per the Hankins \& Wolszczan (1987) study.  However, a new analysis raising all the modern questions about such a configuration is needed.

\subsection{B1855+09}
A millisecond pulsar well known for its binary white-dwarf system, we find strong evidence for a core-single structure in this pulsar's profile (Figure A9). High resolution observations of this pulsar were conducted by Dai \etal\ (2015), and agree with our suggestion of a core-cone structure. Indeed, it is also interesting to point out that the high frequency RVM curve of Dai \etal\ appears more akin to that of a strongly abberated ``normal'' pulsar (\eg B0329+54). Only four other MSPs are known to show any semblance of a core-cone structure.

\subsection{B1919+21}
Our analyses basically confirm the c{\bf Q} geometry of this pulsar, although the PHP (Figure B12) diagrams suggest what appears to be vestige central core emission---something seen before in otherwise ``pure'' conal profiles.  This does nothing to explain why the putative outer cone fails to grow with wavelength or the character of the weak leading bridge of emission that seems to separate out into a weak component at C band (Figure A12).  The very first Cambridge pulsar needs a deep contemporary analysis.

\subsection{B1929+10}
As is well known, this pulsar gives contradictory evidence regarding its geometry:  It seems to be an orthogonal interpulsar, and the (putative core) widths of both the interpulse and central component of the main pulse would seem to support this.  However, the very shallow PPA traverse (--1.5\degr/\degr) (Figure A12) is totally at odds with this interpretation (see Rankin \& Rathnasree 1997).

\subsection{B1933+16}
The geometry in Table~\ref{tab5} incorporates the detailed single-pulse analysis of Mitra \etal\ (2016).

\subsection{B1946+35}
Our model in Table~\ref{tab5} is based on the pulse-sequence analysis of Mitra \& Rankin (2017).

\subsection{B1952+29}
Detailed aspects of this pulsar's profiles are very difficult to understand.  The model in Table~\ref{tab5} is therefore only a first approximation.

\subsection{B2002+31}
We confirm the nearly classic core/cone triple structure of this pulsar's profile (Figure A14), but while the conal outriders are weak at 21 cms, they are as strong as the central core at C band as expected given the usual relatively steeper core spectrum.

\subsection{J0538+2817}
PHPs (Figure B15) show clear evidence of a core and outer/inner cone structure and this is predominantly clear from the average profiles (Figure A15); we thus assign an {\bf M} classification. We were fortunate to observe this pulsar in both of its known emitting modes, and have concluded that the mode change primarily correlates with intensity changes in the core and trailing conal components. While for the survey's purpose we have only included a single profile of each pulsar, profiles of J0538+2817's two modes can be found in Anderson \etal\ (1996).

\subsection{J0627+0649}
No geometrical model is possible for this pulsar.  The profiles (Figure A15) seem to have a regular structure across the three bands, but the shallow linear PPA traverse seems incompatible with any model.  

\subsection{J1740+1000}
The 3-pronged structure at P-band (Figure A16) is suggestive of a core/cone structure and indeed the geometry agrees with a T classification. The central component's apparent bifurcation at higher frequencies is not unheard of in core components.

\section{Summary \& Conclusions}
We have carried out single-pulse polarimetric observations at 4.5 GHz on 46 pulsars benefiting from the unique sensitivity of the Arecibo Gregorian feed system and Mock spectrometers.  These have then been combined with complementary observations at nominal frequencies of 327-MHz (P band) and 1.4 GHz (L band) that were observed in a similar manner and thus have comparable sensitivity.  In a few cases, P-band observations are missing because interstellar scattering rendered them useless.  The observational properties of the survey pulsars are given in Table~\ref{tab1}.  

\begin{figure*}
   \begin{tabularx}{\textwidth}{YY}
   \toprule
    \multicolumn{2}{c}{} \\
     \includegraphics[page=1,width=\linewidth]{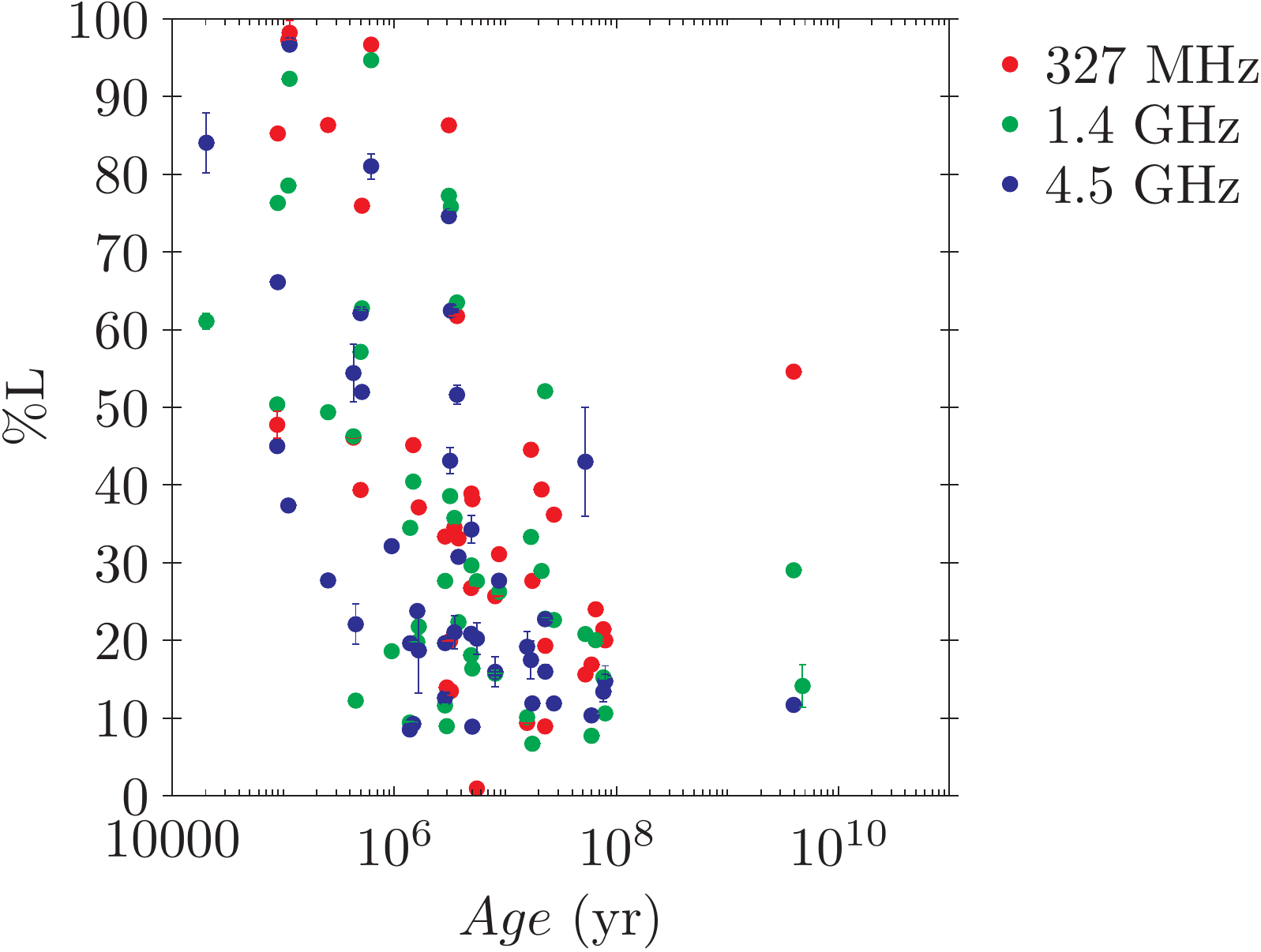}  &
     \includegraphics[page=1,width=\linewidth]{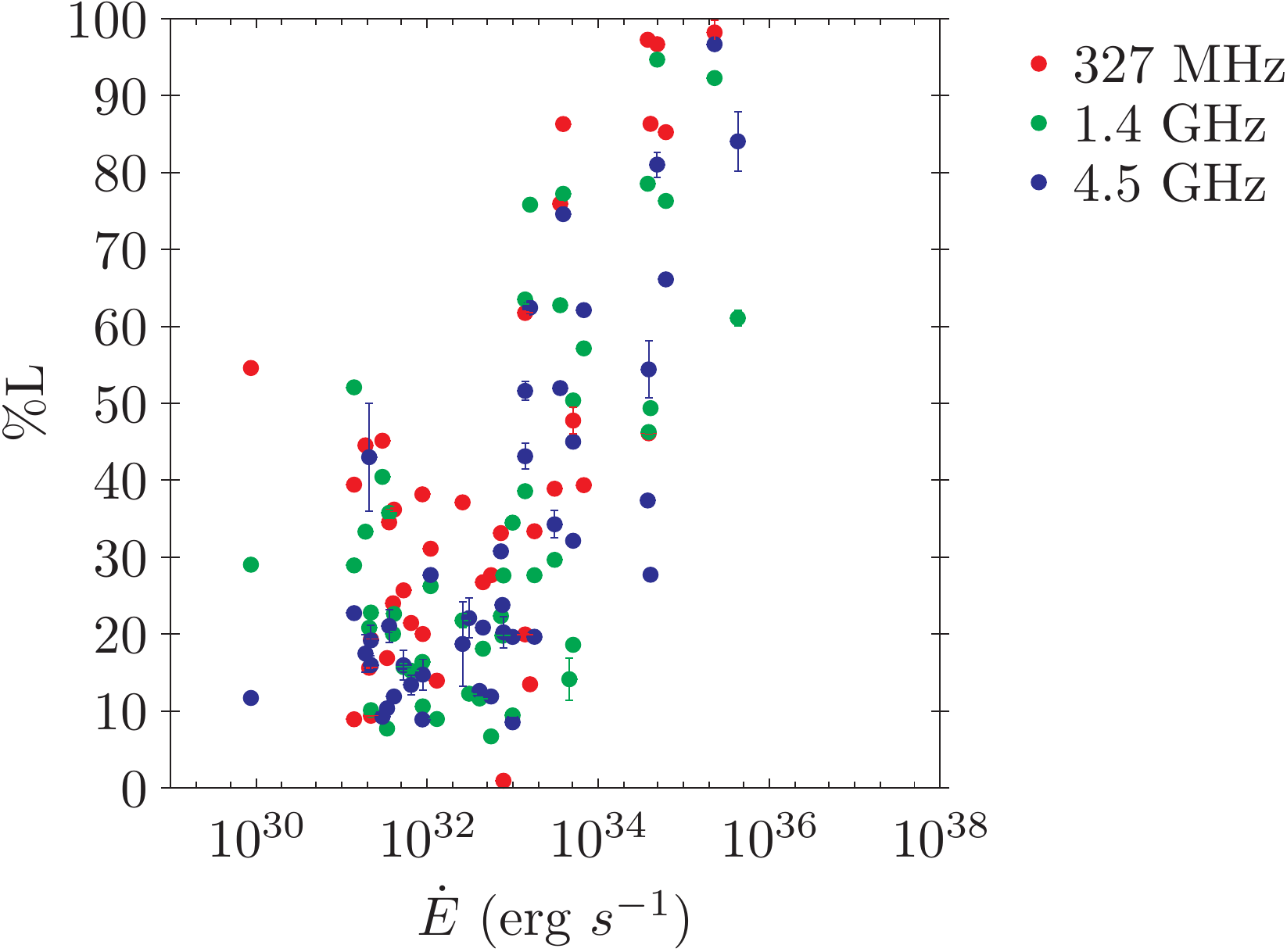}  \\
     \includegraphics[page=1,width=\linewidth]{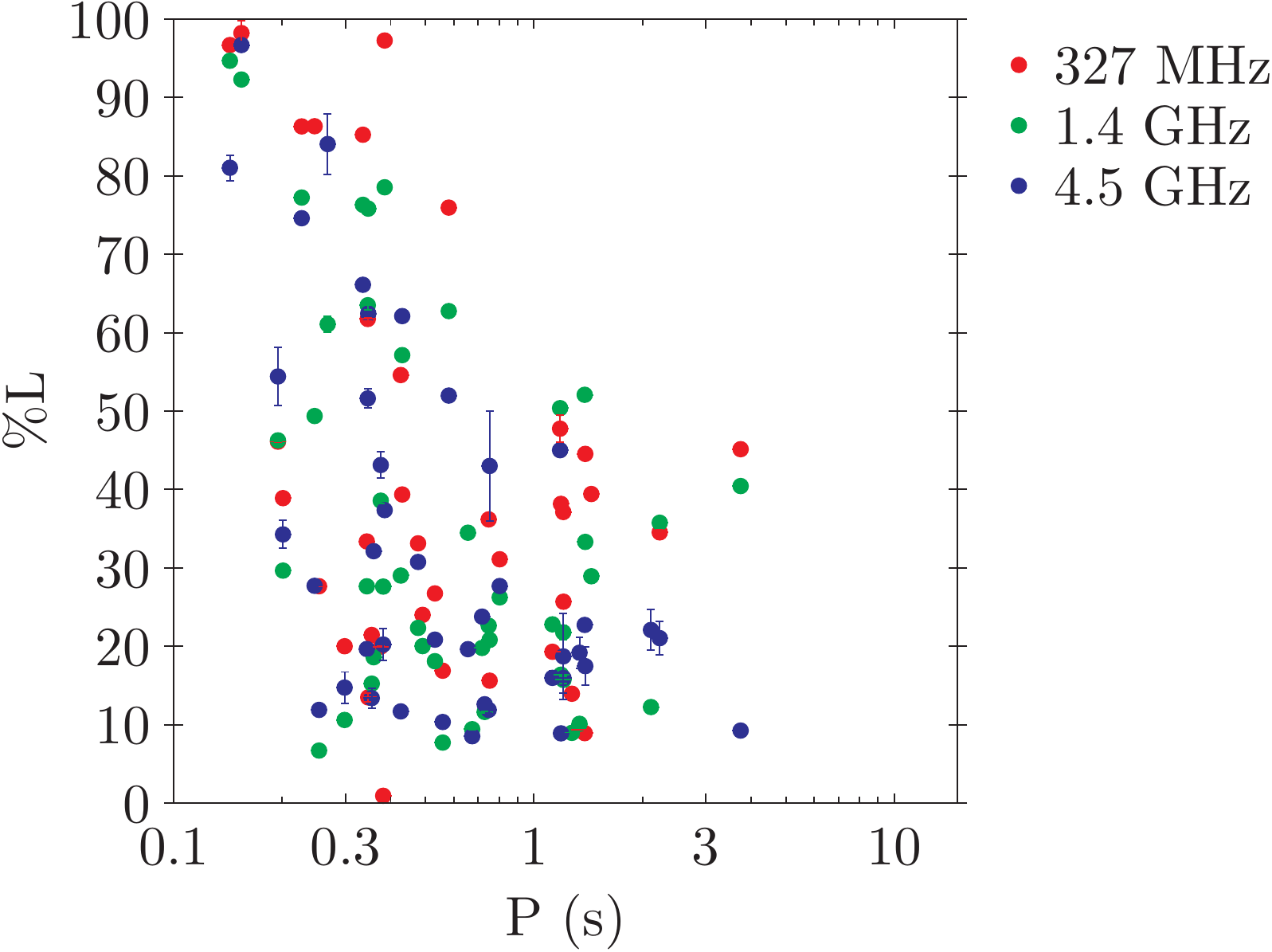}  &
     \includegraphics[page=1,width=\linewidth]{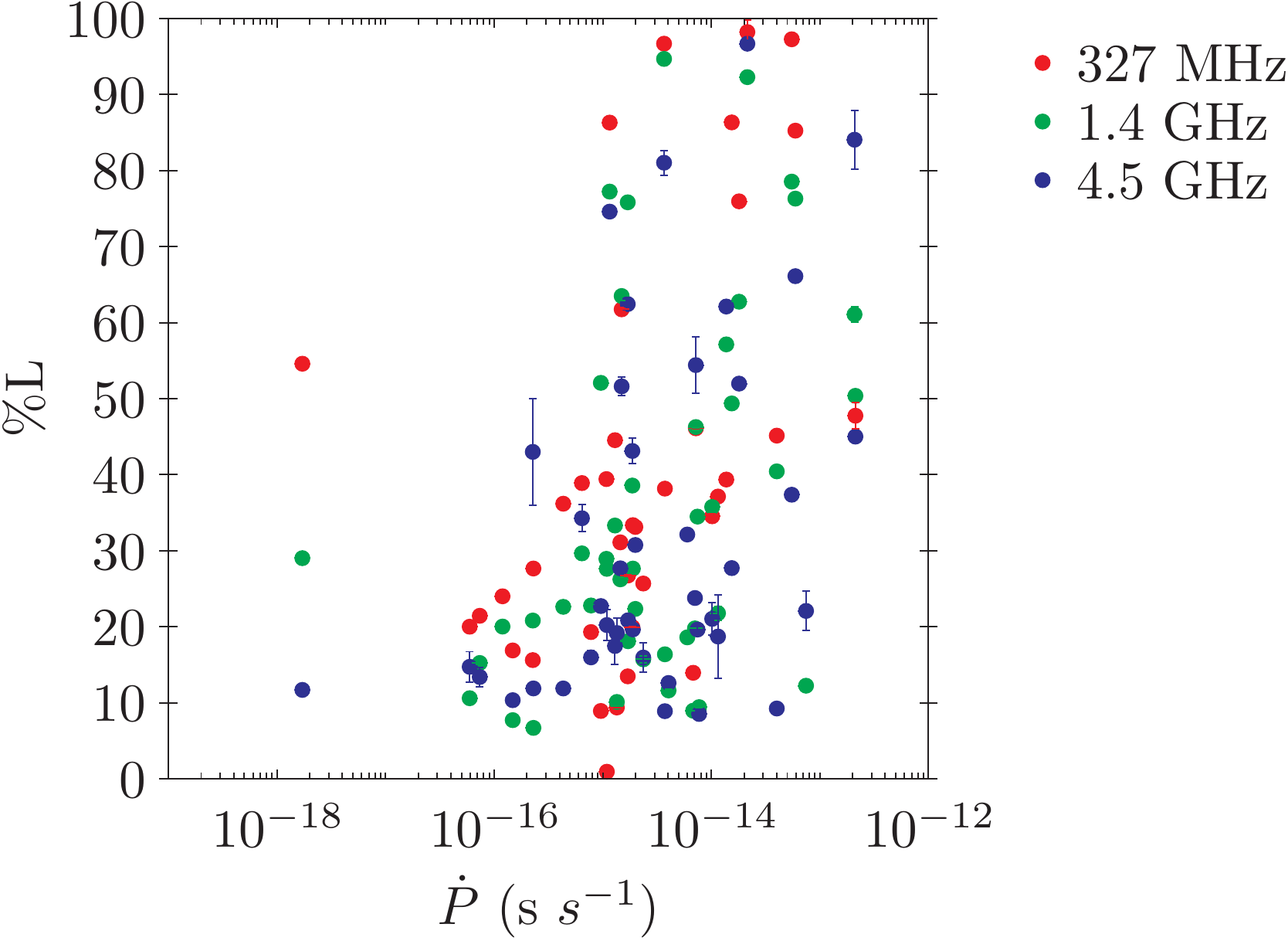}  \\
     \bottomrule

   \end{tabularx}
\caption{Population plots of \%L vs Age, $\dot{E}$, P, and $\dot{P}$ for the three frequencies along with error bars in polarization (327 MHz --red, 1.5 GHz -- green, 4.5 GHz -- blue.)} 
\label{fig4}
\end{figure*}

\begin{figure*}
   \begin{tabularx}{\textwidth}{YY}
   \toprule
    \multicolumn{2}{c}{} \\
     \includegraphics[page=1,width=\linewidth]{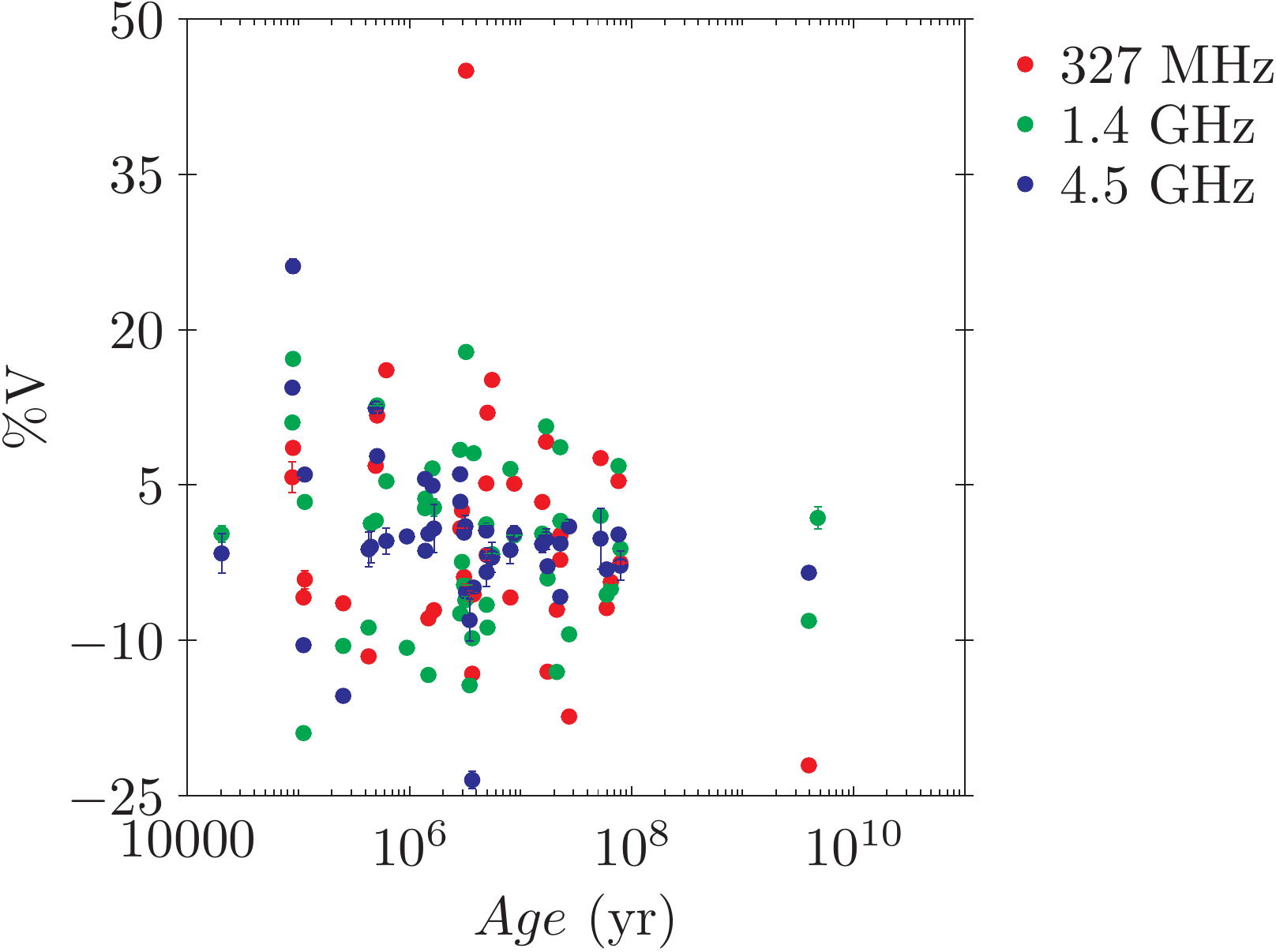}  &
     \includegraphics[page=1,width=\linewidth]{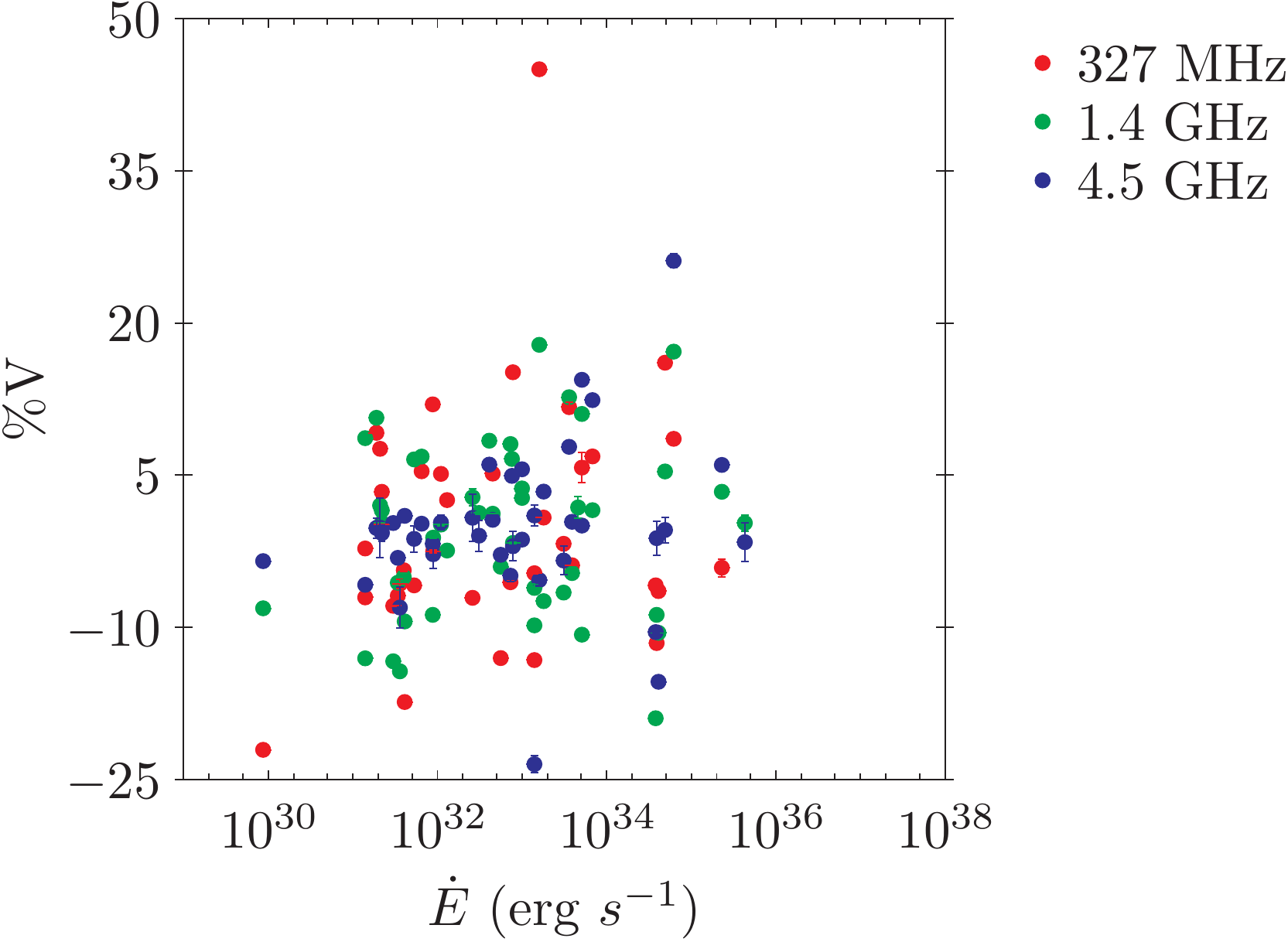}  \\
     \includegraphics[page=1,width=\linewidth]{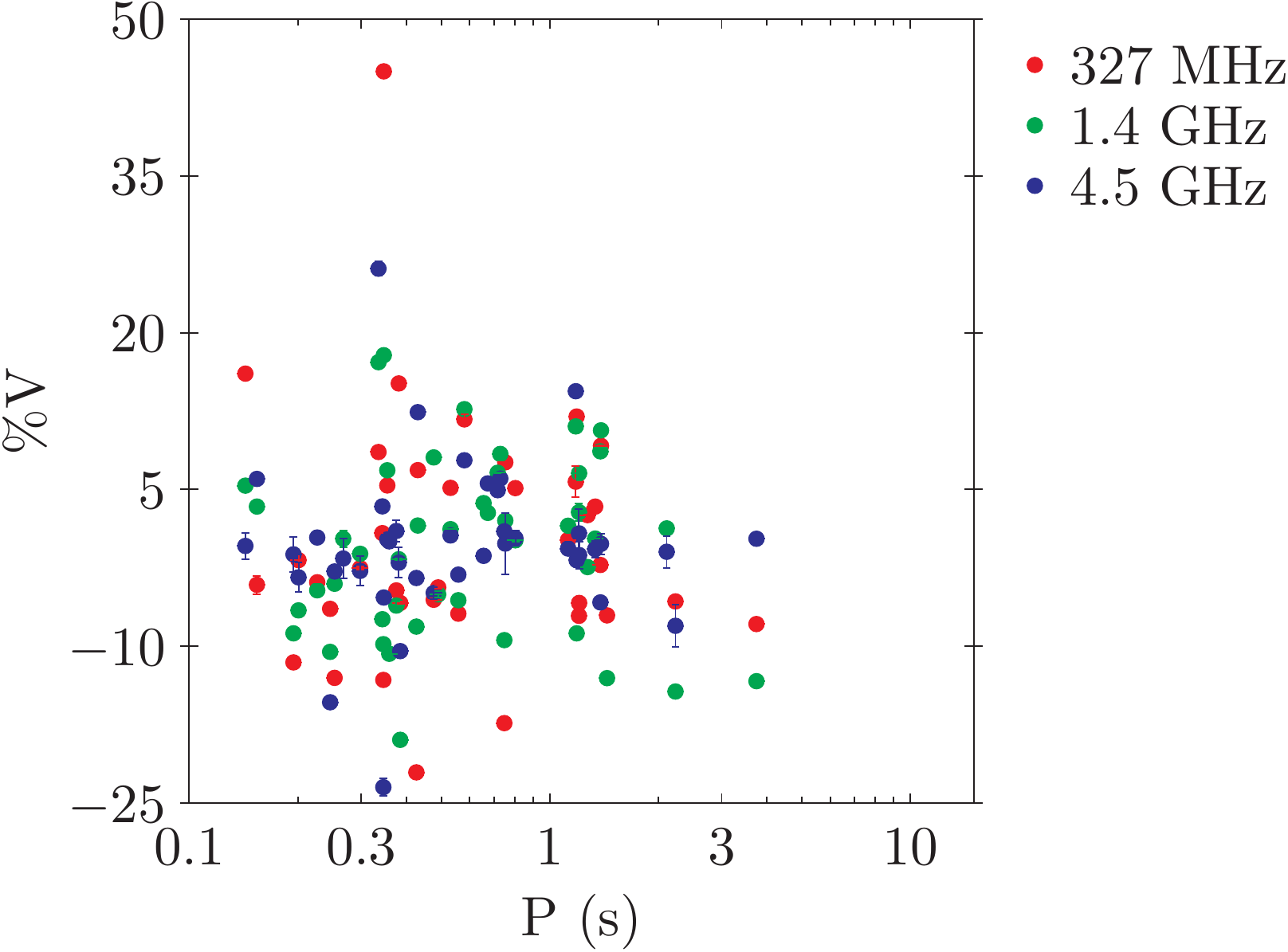}  &
     \includegraphics[page=1,width=\linewidth]{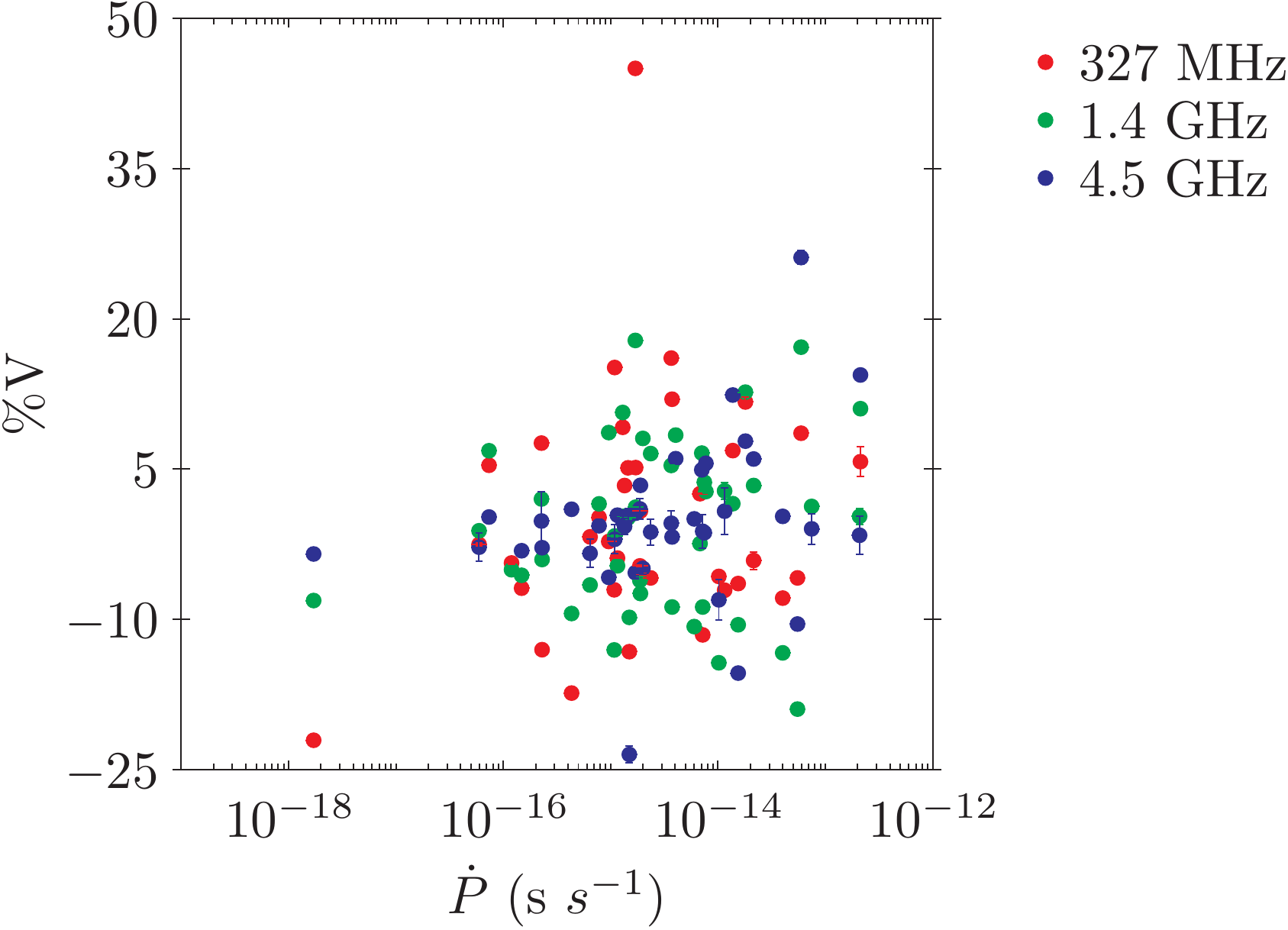}  \\
     \bottomrule

   \end{tabularx}
\caption{\%V vs Age, $\dot{E}$, P, and $\dot{P}$} 
\label{fig5}
\end{figure*}

\begin{figure*}
   \begin{tabularx}{\textwidth}{YY}
   \toprule
    \multicolumn{2}{c}{} \\
     \includegraphics[page=1,width=\linewidth]{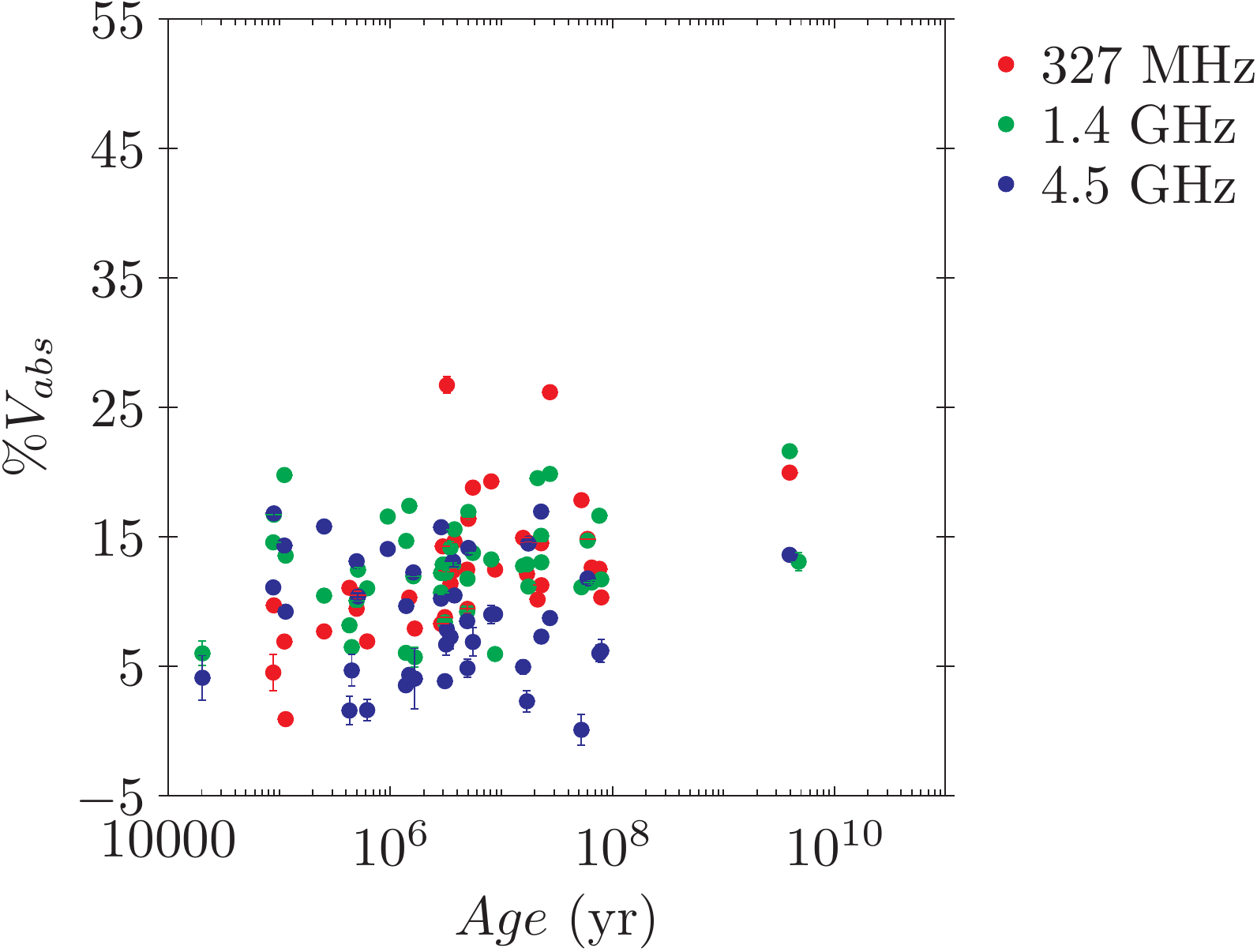}  &
     \includegraphics[page=1,width=\linewidth]{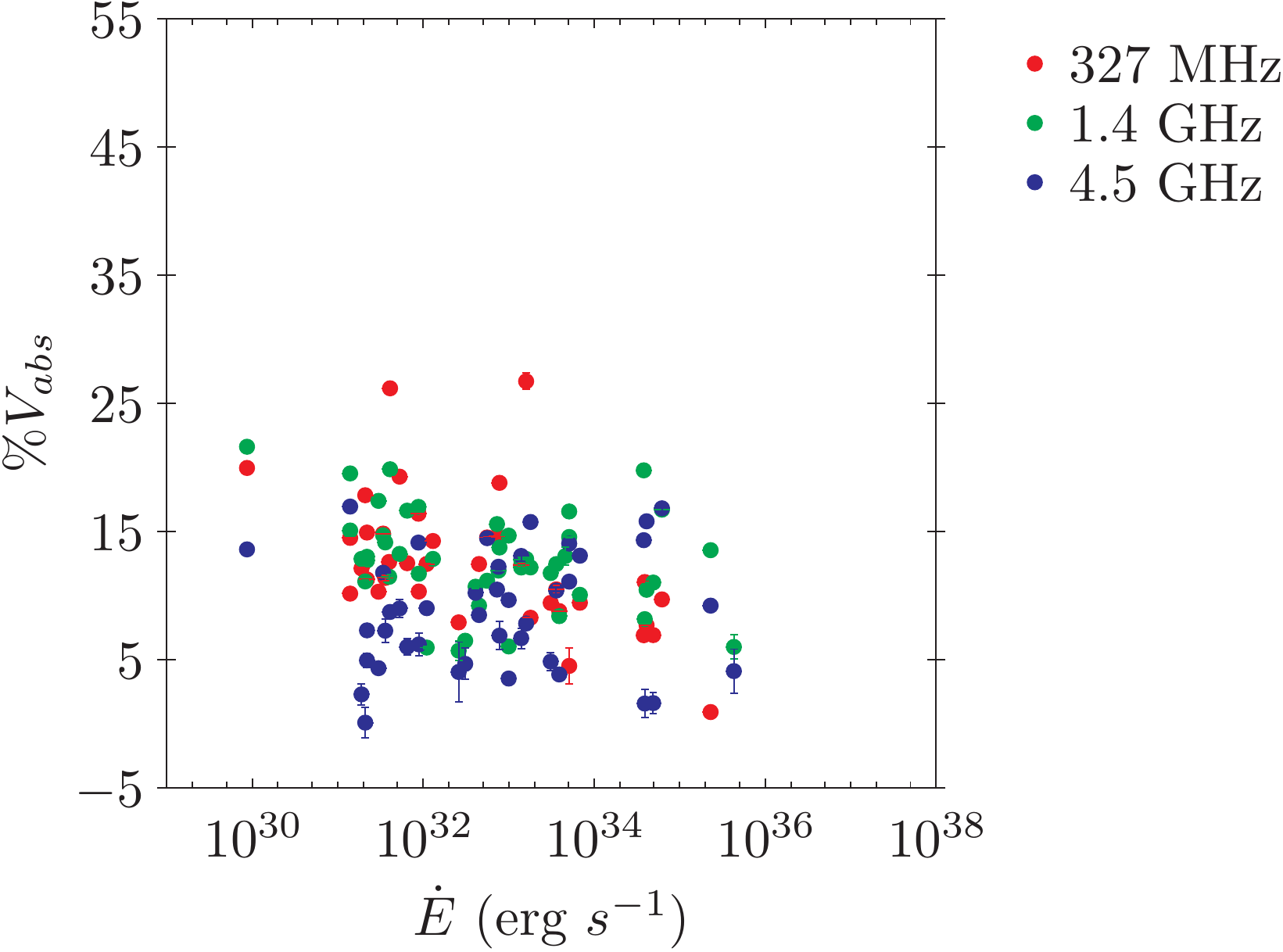}  \\
     \includegraphics[page=1,width=\linewidth]{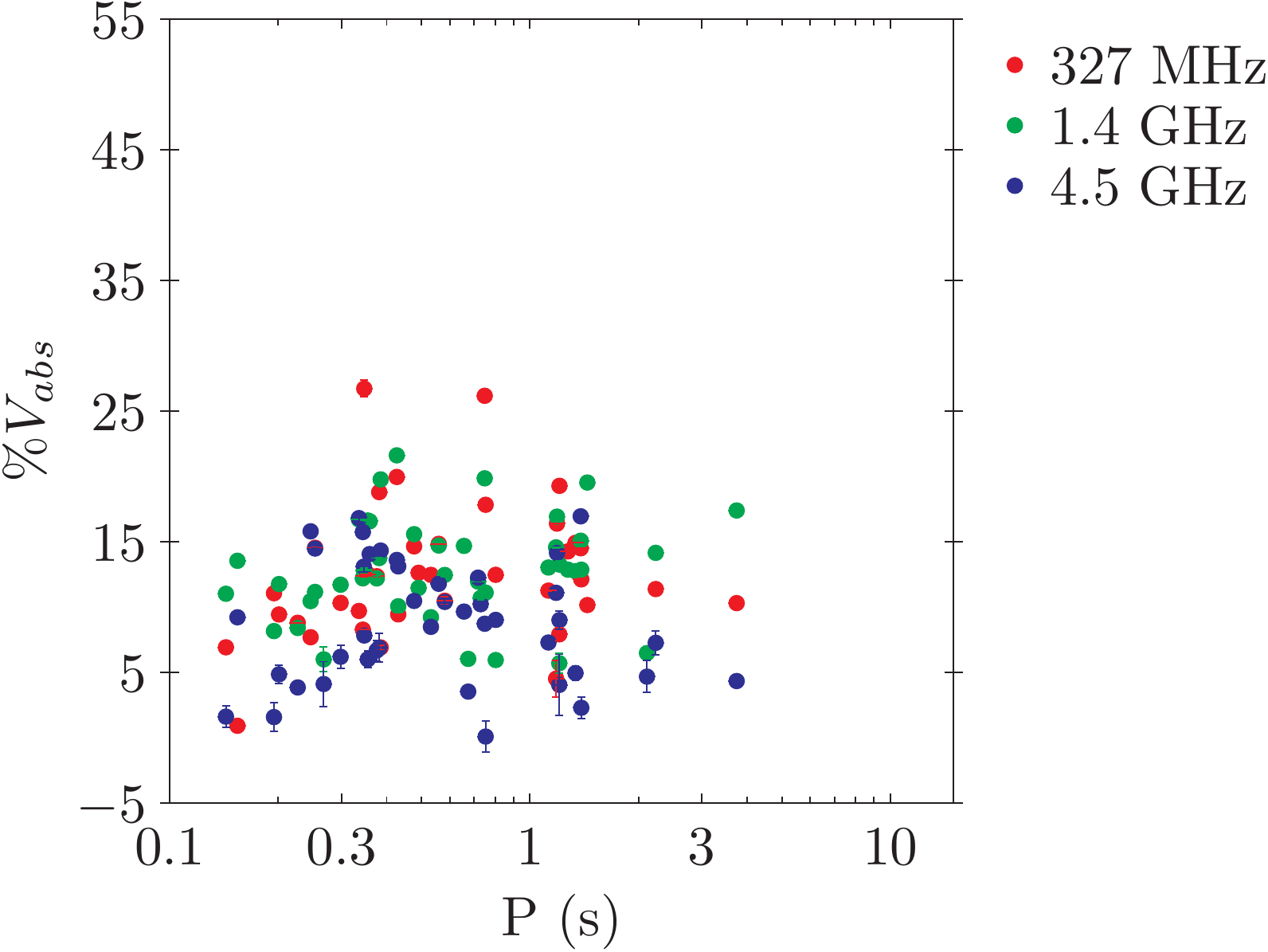}  &
     \includegraphics[page=1,width=\linewidth]{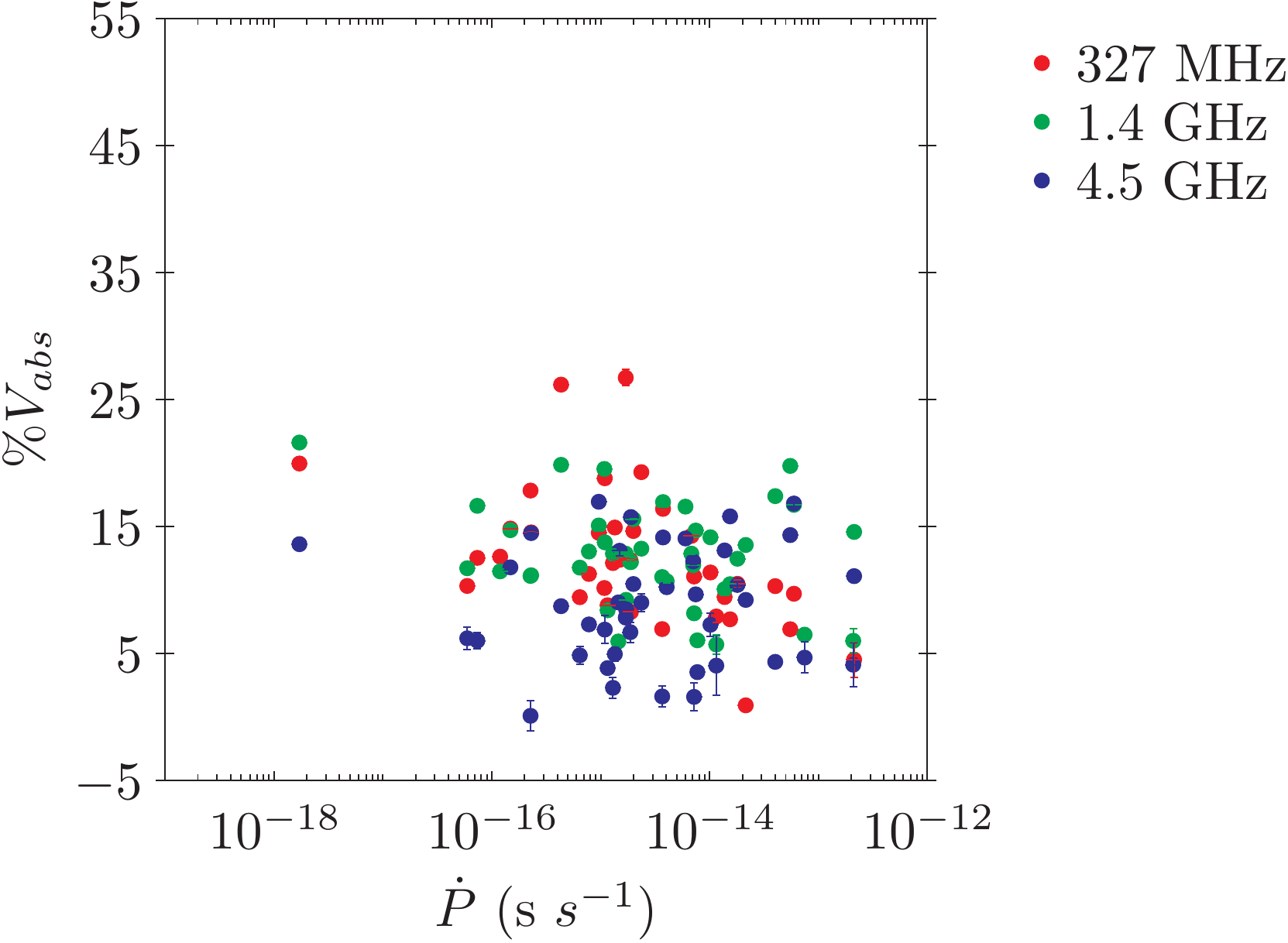}  \\
     \bottomrule

   \end{tabularx}
\caption{\%$V_{abs}$ vs Age, $\dot{E}$, P, and $\dot{P}$} 
\label{fig6}
\end{figure*}

The aggregate linear and circular polarization of all the profiles has been measured and tabulated in Table~\ref{tab2}.  These values are then plotted against age $\tau$ (=$P/\dot P$), spindown energy $\dot E$ (=$4\pi^2I\dot P/P^3$), $P$ and $\dot P$ in Figures~\ref{fig1} -- \ref{fig3} to explore their systematics.  As expected we see, for instance, that the aggregate fractional linear polarization declines with frequency in virtually every case.  Similarly, the profile width information from Table~\ref{tab3} is plotted as a function of $P$ in Figure~\ref{fig4}. 

The physical parameters for the survey pulsars are given in Table~\ref{tab2}.  B names are largely used because these are the discovery names for most of the survey objects.  The $P$, $\dot P$, $\dot E$, $\tau$ and $B$ from the ATNF Pulsar Catalog\footnote{http://www.atnf.csiro.au/research/pulsar/psrcat/}.  The acceleration parameter $B_{12}/P^2$ and Beskin \etal's (1993) 1/Q parameter are also given.  

Geometrical core/double-cone beam models are then given in Table~\ref{tab5}.  Each pulsar's class is given together with the estimated or measured magnetic colatitude $\alpha$, sightline impact angle $\beta$, together with the profile dimensions, beam dimensions and characteristic emission height of each cone.  

The Arecibo C-band Survey's observations will be available for public download in the form of Stokes-parameter integrated profiles at http://www.uvm.edu/\textasciitilde pulsar.

\onecolumn

\twocolumn

\onecolumn

\setcounter{table}{4}

\begin{table*}

\begin{tabular}{c|ccc|ccc|ccc}
    \toprule
    Pulsar &  \%L & \%V & \%$V_{abs}$ & \%L & \%V & \%$V_{abs}$ & \%L & \%V & \%$V_{abs}$  \\
    (B1950) &   &  & & & & & & & \\
    \midrule
    & & \mbox{\textbf{(P-band)}} & & & \mbox{\textbf{(L-band)}} & & & \mbox{\textbf{(C-band)}} \\
    \midrule
    B0301+19	&	44.5	$ \pm $	0.1	&	9.2	$ \pm $	0.1	&	12.1	$ \pm $	0.1	      &	33.3	$ \pm $	0.6	&	10.7	$ \pm $	0.4	&	12.9	$ \pm $	0.4	      &	17.5	$ \pm $	2.4	&	-0.2	$ \pm $	1.0	&	2.3	$ \pm $	0.8	 \\
B0523+11	&	21.4	$ \pm $	0.2	&	5.4	$ \pm $	0.2	&	12.5	$ \pm $	0.1	      &	15.2	$ \pm $	0.2	&	6.8	$ \pm $	0.2	&	16.6	$ \pm $	0.1	      &	13.4	$ \pm $	1.3	&	0.2	$ \pm $	0.6	&	6.0	$ \pm $	0.6	 \\
B0525+21	&	45.1	$ \pm $	0.0	&	-7.9	$ \pm $	0.0	&	10.3	$ \pm $	0.0	      &	40.4	$ \pm $	0.0	&	-13.3	$ \pm $	0.0	&	17.4	$ \pm $	0.0	      &	9.3	$ \pm $	0.1	&	0.3	$ \pm $	0.1	&	4.3	$ \pm $	0.0	 \\
B0540+23	&	86.3	$ \pm $	0.0	&	-6.4	$ \pm $	0.0	&	7.7	$ \pm $	0.0	      &	49.4	$ \pm $	0.0	&	-10.5	$ \pm $	0.0	&	10.5	$ \pm $	0.0	      &	27.7	$ \pm $	0.1	&	-15.4	$ \pm $	0.1	&	15.8	$ \pm $	0.0	 \\
B0609+37	&	20.0	$ \pm $	0.1	&	-2.5	$ \pm $	0.1	&	10.3	$ \pm $	0.0	      &	10.6	$ \pm $	0.1	&	-1.2	$ \pm $	0.1	&	11.7	$ \pm $	0.0	      &	14.7	$ \pm $	2.0	&	-2.8	$ \pm $	1.4	&	6.2	$ \pm $	0.9	 \\
B0611+22	&	85.2	$ \pm $	0.0	&	8.6	$ \pm $	0.0	&	9.7	$ \pm $	0.0	      &	76.3	$ \pm $	0.0	&	17.2	$ \pm $	0.0	&	16.7	$ \pm $	0.0	      &	66.1	$ \pm $	0.5	&	26.1	$ \pm $	0.7	&	16.8	$ \pm $	0.3	 \\
B0626+24	&	33.1	$ \pm $	0.1	&	-5.6	$ \pm $	0.1	&	14.6	$ \pm $	0.1	      &	22.3	$ \pm $	0.1	&	8.1	$ \pm $	0.1	&	15.6	$ \pm $	0.0	      &	30.8	$ \pm $	0.5	&	-4.9	$ \pm $	0.6	&	10.5	$ \pm $	0.3	 \\
B0656+14	&	97.3	$ \pm $	0.1	&	-5.9	$ \pm $	0.1	&	6.9	$ \pm $	0.2	      &	78.5	$ \pm $	0.1	&	-18.9	$ \pm $	0.0	&	19.8	$ \pm $	0.0	      &	37.4	$ \pm $	0.2	&	-10.5	$ \pm $	0.2	&	14.3	$ \pm $	0.1	 \\
B0751+32	&	39.4	$ \pm $	0.3	&	-7.0	$ \pm $	0.2	&	10.2	$ \pm $	0.5	      &	28.9	$ \pm $	0.2	&	-13.0	$ \pm $	0.3	&	19.5	$ \pm $	0.1	      &		--		&		--		& --	 \\
B0823+26	&	26.7	$ \pm $	0.1	&	5.2	$ \pm $	0.0	&	12.5	$ \pm $	0.0	      &	18.1	$ \pm $	0.0	&	1.2	$ \pm $	0.0	&	9.2	$ \pm $	0.0	      &	20.8	$ \pm $	0.0	&	0.6	$ \pm $	0.0	&	8.5	$ \pm $	0.0	 \\
B0834+06	&	13.9	$ \pm $	0.0	&	2.5	$ \pm $	0.0	&	14.3	$ \pm $	0.0	      &	9.0	$ \pm $	0.0	&	-2.4	$ \pm $	0.0	&	12.9	$ \pm $	0.0	      &		--		&		--		&	-- \\
B0919+06	&	39.4	$ \pm $	0.0	&	6.8	$ \pm $	0.0	&	9.4	$ \pm $	0.0	      &	57.1	$ \pm $	0.0	&	1.6	$ \pm $	0.0	&	10.1	$ \pm $	0.0	      &	62.1	$ \pm $	0.2	&	12.4	$ \pm $	0.1	&	13.1	$ \pm $	0.0	 \\
B0950+08	&	27.7	$ \pm $	0.0	&	-13.0	$ \pm $	0.0	&	14.5	$ \pm $	0.0	      &	6.7	$ \pm $	0.0	&	-4.0	$ \pm $	0.0	&	11.2	$ \pm $	0.0	      &	11.9	$ \pm $	0.0	&	-2.8	$ \pm $	0.0	&	14.5	$ \pm $	0.0	 \\
B1133+16	&	38.2	$ \pm $	0.0	&	12.0	$ \pm $	0.0	&	16.4	$ \pm $	0.0	      &	16.4	$ \pm $	0.0	&	-8.8	$ \pm $	0.0	&	16.9	$ \pm $	0.0	      &	8.9	$ \pm $	0.0	&	-1.8	$ \pm $	0.0	&	14.1	$ \pm $	0.0	 \\
B1237+25	&	8.9	$ \pm $	0.0	&	-2.2	$ \pm $	0.0	&	14.5	$ \pm $	0.0	      &	52.1	$ \pm $	0.1	&	8.6	$ \pm $	0.0	&	15.1	$ \pm $	0.0	      &	22.7	$ \pm $	0.1	&	-5.8	$ \pm $	0.1	&	16.9	$ \pm $	0.0	 \\
B1530+27	&	19.3	$ \pm $	0.0	&	0.2	$ \pm $	0.0	&	11.3	$ \pm $	0.0	      &	22.8	$ \pm $	0.0	&	1.5	$ \pm $	0.0	&	13.0	$ \pm $	0.0	      &	16.0	$ \pm $	0.9	&	-0.7	$ \pm $	0.6	&	7.3	$ \pm $	0.4	 \\
B1541+09	&	36.2	$ \pm $	0.2	&	-17.3	$ \pm $	0.1	&	26.2	$ \pm $	0.0	      &	22.6	$ \pm $	0.0	&	-9.4	$ \pm $	0.1	&	19.9	$ \pm $	0.0	      &	11.9	$ \pm $	0.4	&	1.0	$ \pm $	0.4	&	8.7	$ \pm $	0.2	 \\
B1612+07	&	25.7	$ \pm $	0.1	&	-5.9	$ \pm $	0.1	&	19.3	$ \pm $	0.0	      &	15.7	$ \pm $	0.5	&	6.6	$ \pm $	0.5	&	13.3	$ \pm $	0.3	      &	16.0	$ \pm $	1.9	&	-1.3	$ \pm $	1.3	&	9.0	$ \pm $	0.7	 \\
B1633+24	&	24.0	$ \pm $	0.1	&	-4.4	$ \pm $	0.1	&	12.6	$ \pm $	0.0	      &	20.0	$ \pm $	0.3	&	-5.0	$ \pm $	0.3	&	11.5	$ \pm $	0.2	      &		--		&		--		&   --	 \\
B1737+13	&	31.1	$ \pm $	0.1	&	5.1	$ \pm $	0.1	&	12.5	$ \pm $	0.0	      &	26.2	$ \pm $	0.1	&	0.1	$ \pm $	0.1	&	5.9	$ \pm $	0.1	      &	27.7	$ \pm $	0.7	&	0.3	$ \pm $	0.7	&	9.0	$ \pm $	0.4	 \\
B1821+05	&	15.6	$ \pm $	0.1	&	7.6	$ \pm $	0.1	&	17.8	$ \pm $	0.0	      &	20.8	$ \pm $	0.1	&	2.0	$ \pm $	0.1	&	11.1	$ \pm $	0.1	      &	43.0	$ \pm $	7.0	&	-0.2	$ \pm $	2.9	&	0.1	$ \pm $	1.2	 \\
B1839+09	&	0.9	$ \pm $	0.1	&	15.1	$ \pm $	0.1	&	18.8	$ \pm $	0.0	      &	27.6	$ \pm $	0.1	&	-1.7	$ \pm $	0.1	&	13.8	$ \pm $	0.0	      &	20.2	$ \pm $	2.0	&	-2.0	$ \pm $	1.4	&	6.9	$ \pm $	1.1	 \\
B1842+14	&	20.0	$ \pm $	0.1	&	-4.7	$ \pm $	0.1	&	12.4	$ \pm $	0.0	      &	38.6	$ \pm $	0.4	&	-6.1	$ \pm $	0.4	&	12.2	$ \pm $	0.2	      &	43.1	$ \pm $	1.7	&	1.0	$ \pm $	1.0	&	6.7	$ \pm $	0.8	 \\
B1848+12	&	37.1	$ \pm $	0.3	&	-7.1	$ \pm $	0.3	&	7.9	$ \pm $	0.5	      &	21.8	$ \pm $	0.9	&	2.8	$ \pm $	0.8	&	5.7	$ \pm $	0.7	      &	18.7	$ \pm $	5.5	&	0.8	$ \pm $	2.3	&	4.0	$ \pm $	2.3	 \\
B1848+13	&	61.7	$ \pm $	0.2	&	-13.2	$ \pm $	0.4	&	12.4	$ \pm $	0.1	      &	63.5	$ \pm $	0.6	&	-9.8	$ \pm $	0.3	&	12.8	$ \pm $	0.2	      &	51.6	$ \pm $	1.2	&	-23.5	$ \pm $	0.8	&	13.1	$ \pm $	0.4	 \\
B1853+01	&		--		&		--		&	 --     &	61.1	$ \pm $	1.1	&	0.3	$ \pm $	0.8	&	6.0	$ \pm $	0.9	      &	84.0	$ \pm $	3.9	&	-1.6	$ \pm $	1.9	&	4.1	$ \pm $	1.7	 \\
B1855+09	&		--		&		--		&	 --     &	14.1	$ \pm $	2.8	&	1.8	$ \pm $	1.1	&	13.1	$ \pm $	0.7	      &		--		&		--		&   --	 \\
B1859+03	&		--		&		--		&	 --     &	34.5	$ \pm $	0.1	&	3.7	$ \pm $	0.1	&	14.7	$ \pm $	0.0	      &	19.6	$ \pm $	0.6	&	-1.3	$ \pm $	0.6	&	9.7	$ \pm $	0.3	 \\
B1900+01	&		--		&		--		&	 --     &	11.6	$ \pm $	0.2	&	8.4	$ \pm $	0.2	&	10.7	$ \pm $	0.1	      &	12.6	$ \pm $	0.2	&	6.0	$ \pm $	0.2	&	10.2	$ \pm $	0.1	 \\
B1900+06	&		--		&		--		&	 --     &	9.4	$ \pm $	0.1	&	2.8	$ \pm $	0.1	&	6.0	$ \pm $	0.1	      &	8.5	$ \pm $	0.4	&	5.6	$ \pm $	0.6	&	3.5	$ \pm $	0.2	 \\
B1910+20	&	34.5	$ \pm $	0.4	&	-5.7	$ \pm $	0.4	&	11.4	$ \pm $	0.2	      &	35.8	$ \pm $	0.3	&	-14.3	$ \pm $	0.2	&	14.1	$ \pm $	0.5	      &	21.0	$ \pm $	2.1	&	-8.0	$ \pm $	2.0	&	7.3	$ \pm $	0.9	 \\
B1915+13	&	46.1	$ \pm $	0.1	&	-11.5	$ \pm $	0.1	&	11.1	$ \pm $	0.0	      &	46.3	$ \pm $	0.1	&	-8.8	$ \pm $	0.1	&	8.2	$ \pm $	0.0	      &	54.4	$ \pm $	3.7	&	-1.2	$ \pm $	1.7	&	1.6	$ \pm $	1.1	 \\
B1916+14	&	60.0	$ \pm $	1.9	&	6.0	$ \pm $	1.4	&	1.6	$ \pm $	2.6	      &	50.4	$ \pm $	0.1	&	11.0	$ \pm $	0.0	&	14.6	$ \pm $	0.0	      &	45.0	$ \pm $	0.5	&	14.4	$ \pm $	0.4	&	11.1	$ \pm $	0.1	 \\
B1919+21	&	9.4	$ \pm $	0.0	&	3.4	$ \pm $	0.0	&	14.9	$ \pm $	0.0	      &	10.1	$ \pm $	0.0	&	0.3	$ \pm $	0.0	&	12.7	$ \pm $	0.3	      &	19.2	$ \pm $	2.0	&	-0.7	$ \pm $	0.8	&	4.9	$ \pm $	0.5	 \\
B1924+16	&	75.9	$ \pm $	0.2	&	11.7	$ \pm $	0.5	&	10.5	$ \pm $	0.3	      &	62.7	$ \pm $	0.3	&	12.7	$ \pm $	0.2	&	12.5	$ \pm $	0.1	      &	52.0	$ \pm $	0.7	&	7.8	$ \pm $	0.5	&	10.4	$ \pm $	0.3	 \\
B1929+10	&	86.3	$ \pm $	0.0	&	-3.9	$ \pm $	0.0	&	8.8	$ \pm $	0.0	      &	77.2	$ \pm $	0.0	&	-4.6	$ \pm $	0.0	&	8.4	$ \pm $	0.0	      &	74.6	$ \pm $	0.0	&	0.4	$ \pm $	0.0	&	3.8	$ \pm $	0.0	 \\
B1933+16	&		--		&		--		&	  --    &	18.6	$ \pm $	0.0	&	-10.7	$ \pm $	0.0	&	16.6	$ \pm $	0.0	      &	32.1	$ \pm $	0.0	&	0.0	$ \pm $	0.0	&	14.1	$ \pm $	0.0	 \\
B1935+25	&	38.9	$ \pm $	0.8	&	-1.8	$ \pm $	0.6	&	9.4	$ \pm $	0.3	      &	29.7	$ \pm $	0.1	&	-6.6	$ \pm $	0.2	&	11.8	$ \pm $	0.1	      &	34.3	$ \pm $	1.8	&	-3.4	$ \pm $	1.4	&	4.9	$ \pm $	0.7	 \\
B1946+35	&		--		&		--		&	  --    &	19.8	$ \pm $	0.0	&	6.6	$ \pm $	0.0	&	12.0	$ \pm $	0.0	      &	23.8	$ \pm $	0.5	&	4.9	$ \pm $	0.4	&	12.2	$ \pm $	0.2	 \\
B1952+29	&	54.6	$ \pm $	0.1	&	-22.1	$ \pm $	0.2	&	20.0	$ \pm $	0.1	      &	29.0	$ \pm $	0.3	&	-8.1	$ \pm $	0.4	&	21.6	$ \pm $	0.4	      &	11.7	$ \pm $	0.2	&	-3.5	$ \pm $	0.2	&	13.6	$ \pm $	0.1	 \\
B2002+31	&	    --      &       --      &      --   & 12.2	$ \pm $	0.1	&	1.3	$ \pm $	0.1	&	6.5	$ \pm $	0.1	      &	    22.1	$ \pm $	2.6	&	-1.0	$ \pm $	1.5	&	4.7	$ \pm $	1.2	 \\
B2016+28	&	16.9	$ \pm $	0.0	&	-6.9	$ \pm $	0.0	&	14.8	$ \pm $	0.0	      &	7.7	$ \pm $	0.0	&	-5.6	$ \pm $	0.0	&	14.7	$ \pm $	0.0	      &	10.3	$ \pm $	0.1	&	-3.1	$ \pm $	0.1	&	11.8	$ \pm $	0.1	 \\
B2020+28	&	33.4	$ \pm $	0.0	&	0.8	$ \pm $	0.0	&	8.3	$ \pm $	0.0	      &	27.7	$ \pm $	0.0	&	-7.4	$ \pm $	0.0	&	12.2	$ \pm $	0.0	      &	19.7	$ \pm $	0.0	&	3.4	$ \pm $	0.0	&	15.7	$ \pm $	0.0	 \\
J0538+2817	&	96.7	$ \pm $	0.4	&	16.1	$ \pm $	0.3	&	6.9	$ \pm $	0.2		&	94.7	$ \pm $	0.0	&	5.4	$ \pm $	0.0	&	11.0	$ \pm $	0.0		&	81.0	$ \pm $	1.6	&	-0.4	$ \pm $	1.2	&	1.6	$ \pm $	0.8	 \\
J0627+0649	&	92.3	$ \pm $	0.8	&	-1.8	$ \pm $	0.6	&	??	$ \pm $	0.6		&	75.8	$ \pm $	0.5	&	17.8	$ \pm $	0.4	&	12.9	$ \pm $	0.2		&	62.4	$ \pm $	0.3	&	-5.3	$ \pm $	0.3	&	7.8	$ \pm $	0.1	 \\
J1740+1000	&	98.2	$ \pm $	1.5	&	-4.1	$ \pm $	0.9	&	0.9	$ \pm $	0.3		&	92.3	$ \pm $	0.1	&	3.4	$ \pm $	0.2	&	13.5	$ \pm $	0.1		&	96.7	$ \pm $	0.4	&	6.0	$ \pm $	0.6	&	9.2	$ \pm $	0.3	 \\

    \bottomrule
   \end{tabular}
   \caption{Polarization Information}
   \label{tab2}
\end{table*}

\twocolumn

\section*{Acknowledgements}
TO gratefully acknowledges funding from the Vermont Space Grant Consortium.  The authors wish to acknowledge Ben Stappers for providing timing solutions for some pulsars and Joel Weisberg for providing ionospheric rotation-measure estimates. Portions of this work were carried out with support from US National Science Foundation grants AST 0968296 and 1814397.  DM acknowledges funding from the grant ``Indo-French Centre for the Promotion of Advanced Research - CEFIPRA".  The National Astronomy and Ionosphere Center (aka Arecibo Observatory) is operated by the University of Central Florida under a cooperative agreement with the US National Science Foundation, and in alliance with Yang Enterprises and the Ana G. M\'endez-Universidad Metropolitana.  This work made use of the NASA ADS astronomical data system.  





\begin{thebibliography}{99}
\bibitem[\protect\citeauthoryear{Others}{2013}] {A++96} Anderson S., Cadwell B. J., Jacoby B. A., Wolszczan A., Foster R. S.,Kramer M., 1996, ApJ, 468, L55
\bibitem[\protect\citeauthoryear{Others}{2013}] {B++15} Basu R., Mitra D.,          Rankin J. M., 2015, \apj, 798, 105
\bibitem[Beskin \etal\ (1993)]{bgiii} Beskin V. S., Gurevich A. V., Istomin Ya. N, 1993, {\it Physics of the Pulsar Magnetosphere} Cambridge University Press
\bibitem[\protect\citeauthoryear{Others}{2013}] {BCW91} Blaskiewicz M., Cordes J.~M., Wasserman I., 1991, \apj, 370, 643
\bibitem[Curtin, \etal\ (2019)]{crw19} Curtin, A., Weisberg, J. M., Rankin, J. M., \& Venkataraman., A. 2019, \apj, submitted
\bibitem[Dai \etal\ (2015)]{D++15} Dai S. \etal\ , 2015, \mnras, 449, 3223
\bibitem[\protect\citeauthoryear{Others}{2013}] {DR01} Deshpande A.~A., Rankin J.~M, 2001, \mnras, 322, 438
\bibitem[\protect\citeauthoryear{Others}{2013}] {DKP11} Durant M., Kargaltsev O., Pavlov G.~G., 2011, \apj, 743, 38
\bibitem[Dyks \etal\ (2004)]{drh04} Dyks J., Rudak B., Harding A. K., 2004, \apj, 607, 939
\bibitem[\protect\citeauthoryear{Others}{2013}] {EW01} Everett J.~E., Weisberg J.~M., 2001, \apj, 553, 341
\bibitem[\protect\citeauthoryear{Others}{2013}] {FR10} Force M. M., Rankin J. M., 2010, \mnras, 406, 237
\bibitem[\protect\citeauthoryear{Others}{2013}] {GL98} Gould D.~M., Lyne A.~G., 1998, \mnras, 234, 477
\bibitem[Hankins \& Wolszczan (1987)]{HW87} Hankins T.H., Wolszcan A., 1987, \apj, 318, 410
\bibitem[Hankins \& Rankin (2010)]{HR10} Hankins T.H., Rankin J.M., 2010, \aj, 139, 168
\bibitem[\protect\citeauthoryear{von Hoensbroech} {1999}]{vh} von Hoensbroech, A., 1999, Ph.D. thesis, Univ. of Bonn
\bibitem[\protect\citeauthoryear{Others}{2013}] {HX97}von Hoensbroech A., Xilouris K.~M., 1997, \aaps, 126, 121
\bibitem[Johnston \etal\ (2005)]{JHVKWL} Johnston, S., Hobbs, G., Vigeland, S., Kramer, M., Weisberg, J. M., \& Lyne, A. G.  2005, \mnras, 364, 1397
\bibitem[\protect\citeauthoryear{Others} {2013}]{KJ06} Karastergiou A., Johnston S., 2006, \mnras, 365, 353
\bibitem[\protect\citeauthoryear{Others} {2013}]{JK19} Johnston S., Karastergiou A.,  2019, \mnras, 485, 640
\bibitem[\protect\citeauthoryear{Author}{2012}] {KG97} Kijak J., Gil J., 1997, \mnras, 288, 631
\bibitem[\protect\citeauthoryear{Author}{2012}] {K70} Komesaroff M.~M., 1970, Nature, 225, 612
\bibitem[\protect\citeauthoryear{Author}{2012}] {LM88} Lyne A.G.; Manchester R. N.,  1988, \mnras, 234, 477
\bibitem[\protect\citeauthoryear{Author}{2012}] {MG11} Maciesiak K., Gil J., 2011, \mnras, 417, 1444
\bibitem[\protect\citeauthoryear{Author}{2012}] {MGM12} Maciesiak K., Gil J., Melikidze G., 2012, \mnras, 424,1762
\bibitem[\protect\citrauthoryear{Author}{2012}] {MHTH06} Manchester R. N., Hobbs G.B., Teoh A., Hobbs M., 2005, \aj, 129, 1993
\bibitem[\protect\citeauthoryear{Author}{2012}] {M03} McKinnon M. M., 2003, \apj, 590, 1026
\bibitem[\protect\citeauthoryear{Author}{2012}] {MAR15} Mitra D., Arjunwadkar M., Rankin J. M., 2015, \apj, 806, 236
\bibitem[\protect\citeauthoryear{Author}{2012}] {M++16} Mitra D., Basu R., Maciesiak, K., Skrzypczak A., Melikidze G., Szary A., Krzeszowski K., 2016, \apj, 833, 28
\bibitem[\protect\citeauthoryear{Author}{2012}] {MR11} Mitra D., Rankin J.~M., 2011, \apj, 727, 92
\bibitem[\protect\citeauthoryear{Author}{2012}] {MRA16} Mitra, D., Rankin, J. M., Arjunwadkar, M., 2016, \mnras, 460, 3063
\bibitem[\protect\citeauthoryear{Author}{2012}] {MR17} Mitra D., Rankin J.~M., 2017, \mnras, 468, 4601
\bibitem[\protect\citeauthoryear{Morris \etal\ }{1979}]{n49} Morris, D., Graham, D. A., Sieber, W., Jones, B. B., Seiradakis, J. H.,  \& Thomasson, P. 1979, \aap, 73, 46
\bibitem[Morris \etal\ (1981)]{MGSBT} Morris, D., Graham D. A., Sieber W., Bartel N.,  Thomasson P., 1981, \aaps, 46, 421 
\bibitem[\protect\citeauthoryear{Author}{2012}] {N91} Nowakowski L.~A., 1991,      \apj, 377, 581
\bibitem[\protect\citeauthoryear{Others}{2013}] {OMR19} Olszanski T.E.E., Mitra D., Rankin J. M., 2019, \mnras, in preparation
\bibitem[\protect\citeauthoryear{Others}{2013}] {P++16} Pilia M. \etal\ , 2016, \aap, 586, A92
\bibitem[Popov \& Sieber (1990)]{PS90} Popov M.V., \& Sieber W., 1990, Soviet Astronomy, 34, 382
\bibitem[\protect\citeauthoryear{Others}{2013}] {RC69} Radhakrishnan V., Cooke D.~J., 1969, Astrophys. Lett., 3, 225
\bibitem[\protect\citeauthoryear{Author}{2012}] {RSLK++16} Rajwade K., Seymour A., Lorimer D.~R., Karastergiou A., Serylak M., McLaughlin M.~A., Griessmeier J.-M., 2016,  \mnras, 462, 2518
\bibitem[Ramachandran (2004)]{RBRWD} Ramachandran R., Backer D. C., Rankin J. M., Weisberg J. M., Devine K. E., 2004, \apj, 606, 1167
\bibitem[Rankin (1983a)]{R83a} Rankin J. M., 1983a, \apj, 274, 359
\bibitem[Rankin (1990)]{R90} Rankin J. M., 1990, \apj, 352, 247
\bibitem[Rankin (1993a)]{R93a} Rankin J. M., 1993a, \apj, 405, 285 
\bibitem[Rankin (1993b)]{R93b} Rankin J. M., 1993b, \apjs, 85, 145 
\bibitem[Rankin (2007)]{2007ApJ...664..443R} Rankin J. M., 2007, \apj, 664, 443
\bibitem[Rankin(2015)]{R15} Rankin J. M., 2015, \apj, 804, 112
\bibitem[Rankin(2017)]{R17} Rankin J. M., 2017, \jaa, 38, 53
\bibitem[\protect\citeauthoryear{Others}{2013}]     {ROW19} Rankin J. M., Olszanski T.E.E., Wright G. A. E., 2019, \apj, in preparation.
\bibitem[Rankin \& Rathnasree (1995)]{RR95} Rankin J. M., Rathnasree N., 1995, \jaa, 16, 327
\bibitem[Rankin \& Rathnasree (1997)]{RR97} Rankin J. M., Rathnasree N., 1997, \jaa, 18, 91
\bibitem[Rankin \etal\  (2006)]{RRW} Rankin, J. M., Rodriguez C., Wright G. A. E., 2006, \mnras, 370, 673
\bibitem[\protect\citeauthoryear{Author}{2012}] {RSW89}Rankin J. M., Stinebring D.  R., Weisberg J. M., 1989, \apj, 346, 869
\bibitem[Rankin \etal\  (2007)]{RW07} Rankin J. M., Wright G. A. E., 2007,      \mnras, 379, 507
\bibitem[\protect\citeauthoryear{Author}{2012}] {SLR14} Seymour A.~D., Lorimer D.~R.,     Ridley J.~P., 2014, \mnras, 439, 3951
\bibitem[Skrzypczak \etal\ (2018)] {S++18} Skrzypczak A., Basu R., Mitra D., Melikidze G. I., Maciesiak K., Koralewska O., Filothodoros A., 2018, \apj, 854, 162
\bibitem[Smith, Rankin, \& Mitra (2013)]{SRM13} Smith Z., Rankin J. M., Mitra D., 2013, \mnras 435, 1984, 1121
\bibitem[Sobey \etal\ (2015)]{S++15} Sobey C., Young N. J., Hessels W. T., Weltevrede P., Noutsos A., Stappers B. J., Kramer M., Bassa C., Lyne A. G., Kondratiev V. I. ++, 2012, \mnras, 451, 2493
\bibitem[Sobey\etal\ (2019)]{} Sobey, C., Bilous, A. V., Grie{\ss}meier, J.-M., Hessels, J.W.T., Karastergiou, A., Keane, E. F., Kondratiev, V. I., Kramer, M., Michilli, D., Noutsos, A., Pilia, M., Polzin, E. J., Stappers, B. W., Tan, C. M., van Leeuwen, J., Verbiest, J.P.W., Weltevrede, P., Heald, G., Alves, M.I.R., Carretti, E. En{\ss}lin, T., Haverkorn, M., Iacobelli, M., Reich, W., \& Van Eck, C. 2019, \mnras, 484, 3646
\bibitem[Srostlik \& Rankin (2005)]{SR05} Srostlik Z., Rankin  J. M., 2005, \mnras 362, 1121
\bibitem[Wahl \etal\ (2016)]{WORW16} Wahl H. M., Orfeo D. J., Rankin J. M., Weisberg J. M., 2016, \mnras, 461, 3740
\bibitem[Wahl \etal\ (2019)]{WORW19} Wahl H. M., Rankin J. M., Olszanski T. E. E., Venkataraman A., 2019, in preparation
\bibitem[\protect\citeauthoryear{Weisberg, \etal\ }{1999}]{w99}  Weisberg J. M.,         Cordes J. M., Lundgren S. C., Dawson, B.     R., Despotes J. T., Morgan J. J., Weitz     K. A., Zink E. C., Backer D. C., 1999,     \apjs, 121, 171
\bibitem[Weisberg \etal\ (2004)]{W++04} Weisberg J. M., Cordes J. M., Kuan B., Devine K. E., Green, J. T.,  Backer, D. C., 2004, \apjs, 150, 317
\bibitem[Weltevrede \etal\ 2006]{WSE06} Weltevrede P., Edwards R. T., Stappers B. W., 2006, \aap, 445, 243 
\bibitem[Weltevrede \etal\ 2007]{WES07} Weltevrede P., Stappers B. W., Edwards R. T., 2007, \aap, 469, 607
\bibitem[\protect\citeauthoryear{Others}{2013}] {WSRW06} Weltevrede P., Stappers B.~W., Rankin J.~M., Wright G.~A.~E., 2006, \apj, 645, L149
\bibitem[\protect\citeauthoryear{Others}{2013}] {WWSR06} Weltevrede P., Wright, G.~A.~E., Stappers B.~W., Rankin J.~M., 2006, A\&A, 458, 269
\bibitem[Young \& Rankin 2012]{YR12} Young S.A.E., Rankin J. M., 2012, \mnras, 424, 2477
\bibitem[Young \etal\ 2012]{YSWLK12} Young N. J., Stappers B. J., Weltevrede P., Lyne A. G., Kramer M., 2012, \mnras, 427, 114
\bibitem[Wen \etal\ 2018]{WWYYL18} Wen Z. G., Wang N., Yan W. M., Yuan J. P., Liu Z. Y., Chen M. Z., Chen J. L., 2018, \textit{Ap\&SS}, 361, 261
\bibitem[Wen \etal\ 2018]{ZY19} Zhao R. S., Yan Z., Wu X. J., Shen Z. Q., Manchester R. N., Liu J., Qiao, G. J., Xu, R. X., Lee K. J., 2019, preprint (arXiv:1901.10205 [astro-ph.HE])
\end{thebibliography}




\appendix
\clearpage
\section{Profile Polarization Plots}
\onecolumn
\begin{figure*} 
 \begin{tabularx}{\textwidth}{YYY}
    \multicolumn{3}{c}{} \\ \toprule
\includegraphics[page=1,width=\linewidth]{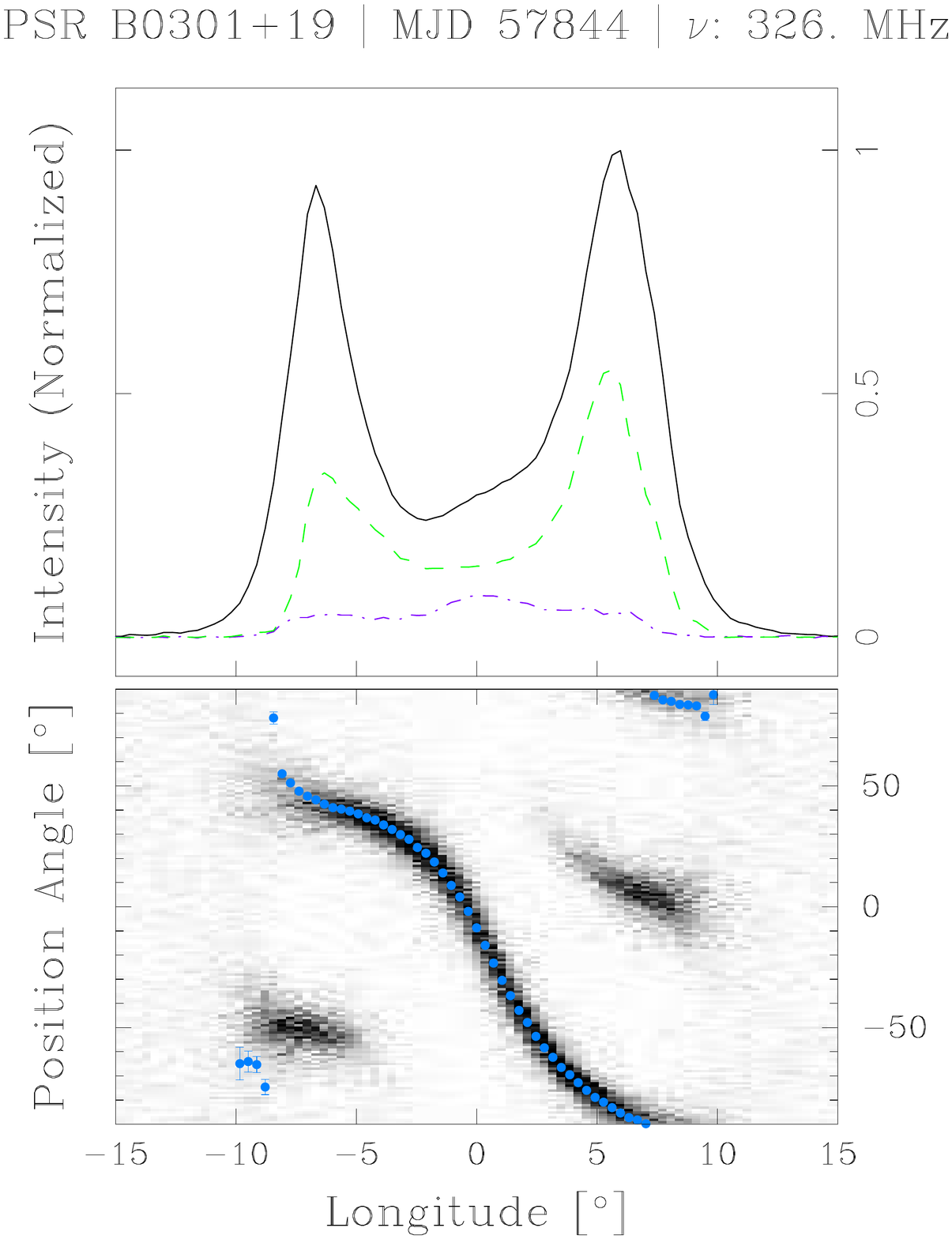} &
\includegraphics[page=1,width=\linewidth]{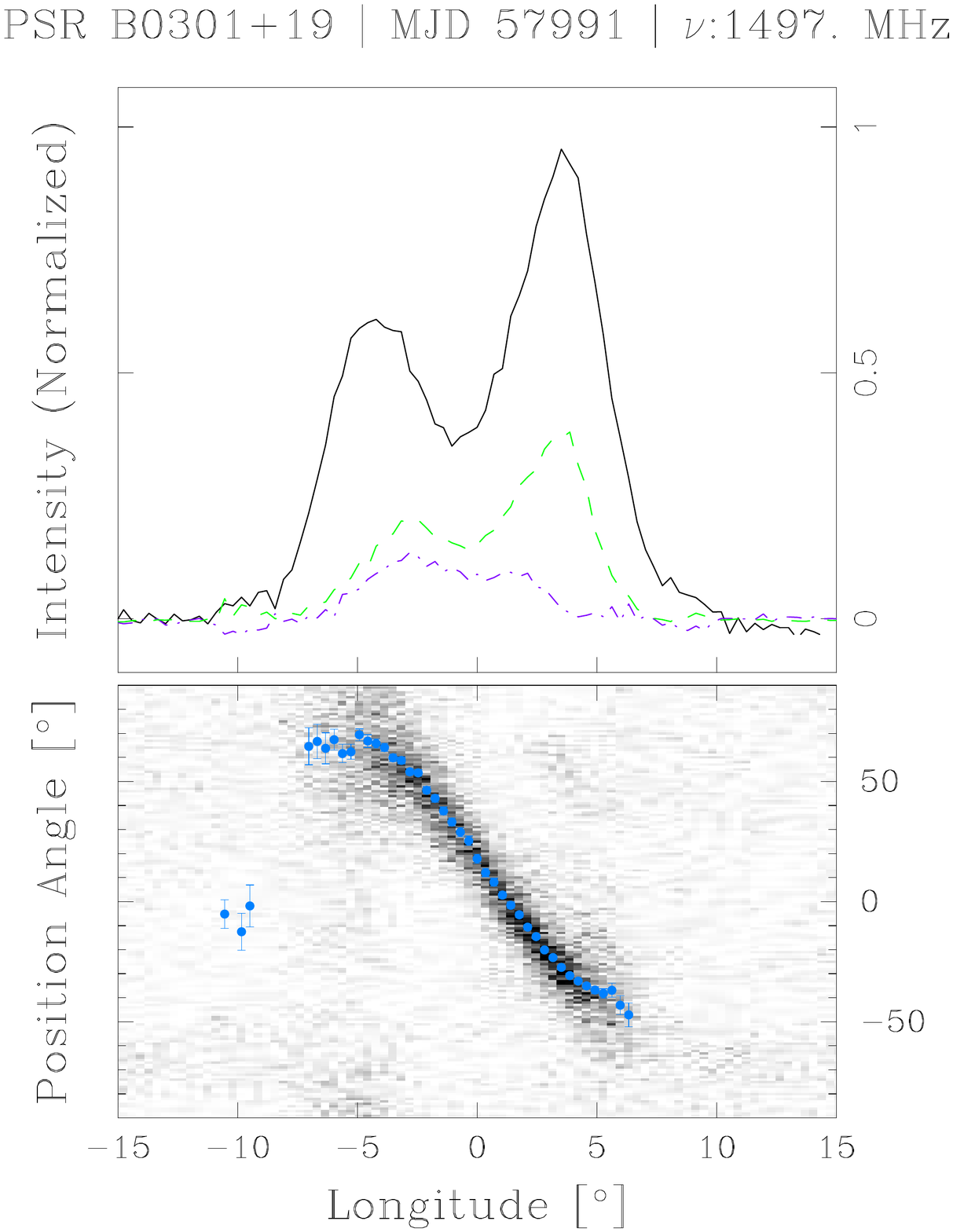} &
\includegraphics[page=1,width=\linewidth]{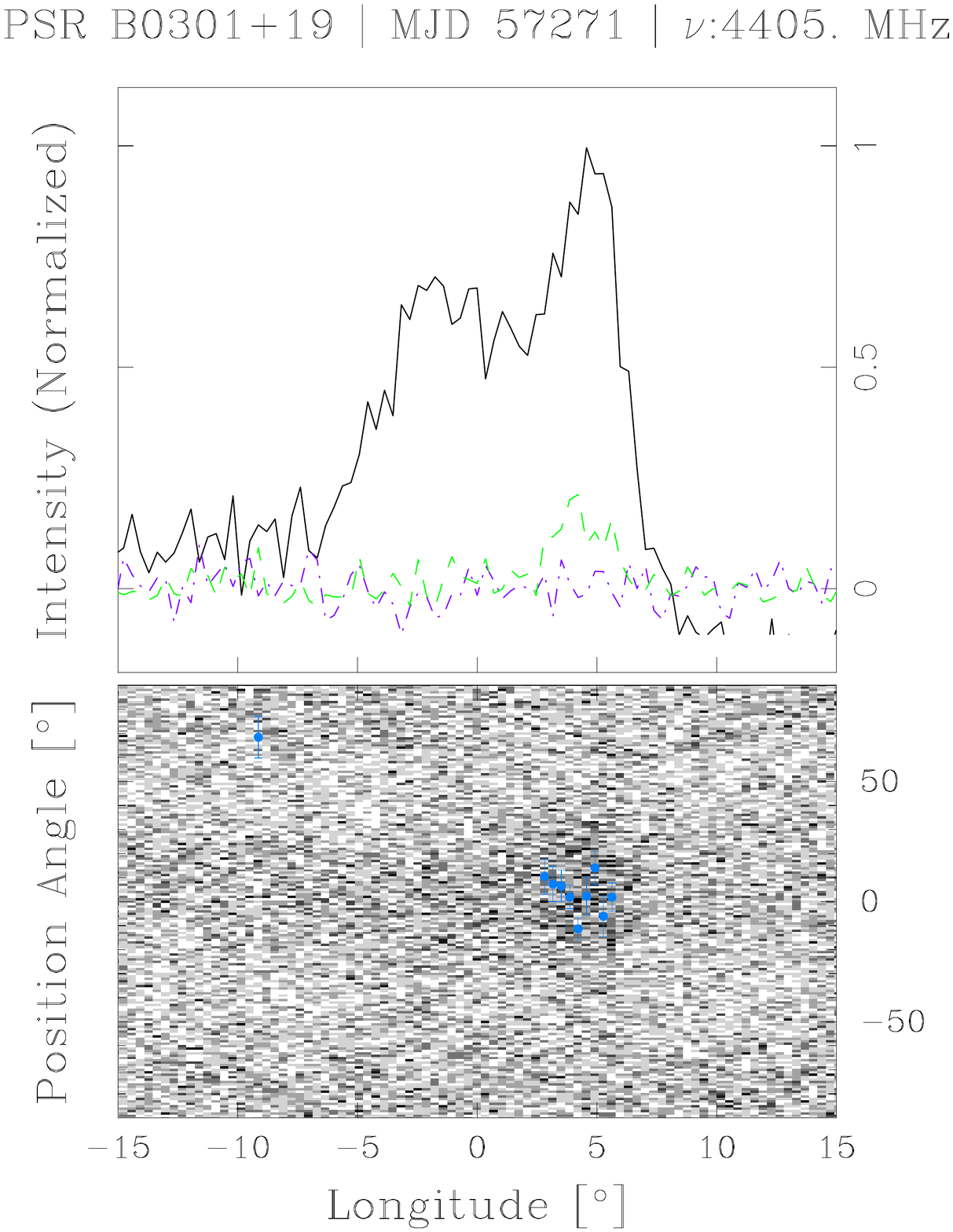} \\ \toprule
\includegraphics[page=1,width=\linewidth]{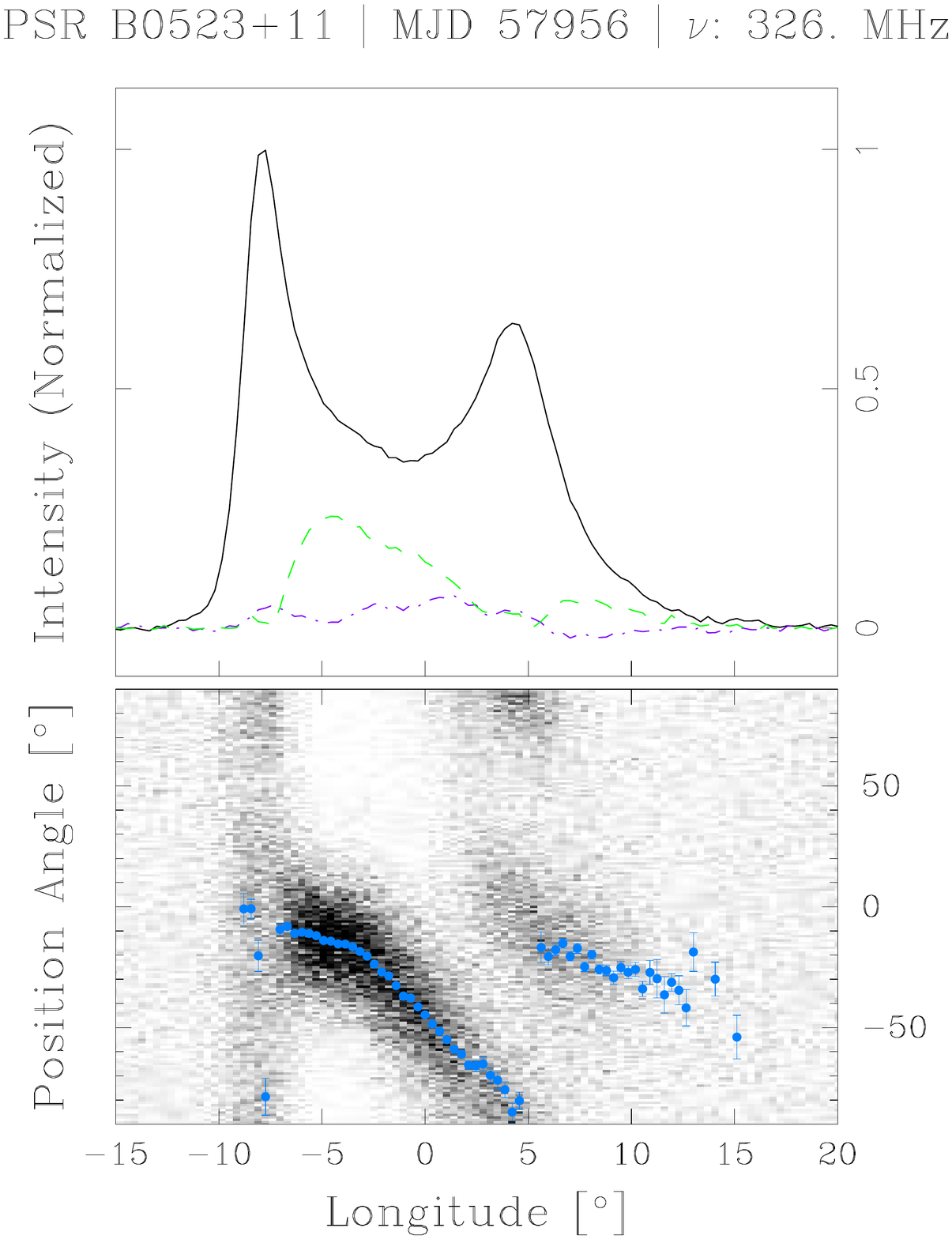} &
\includegraphics[page=1,width=\linewidth]{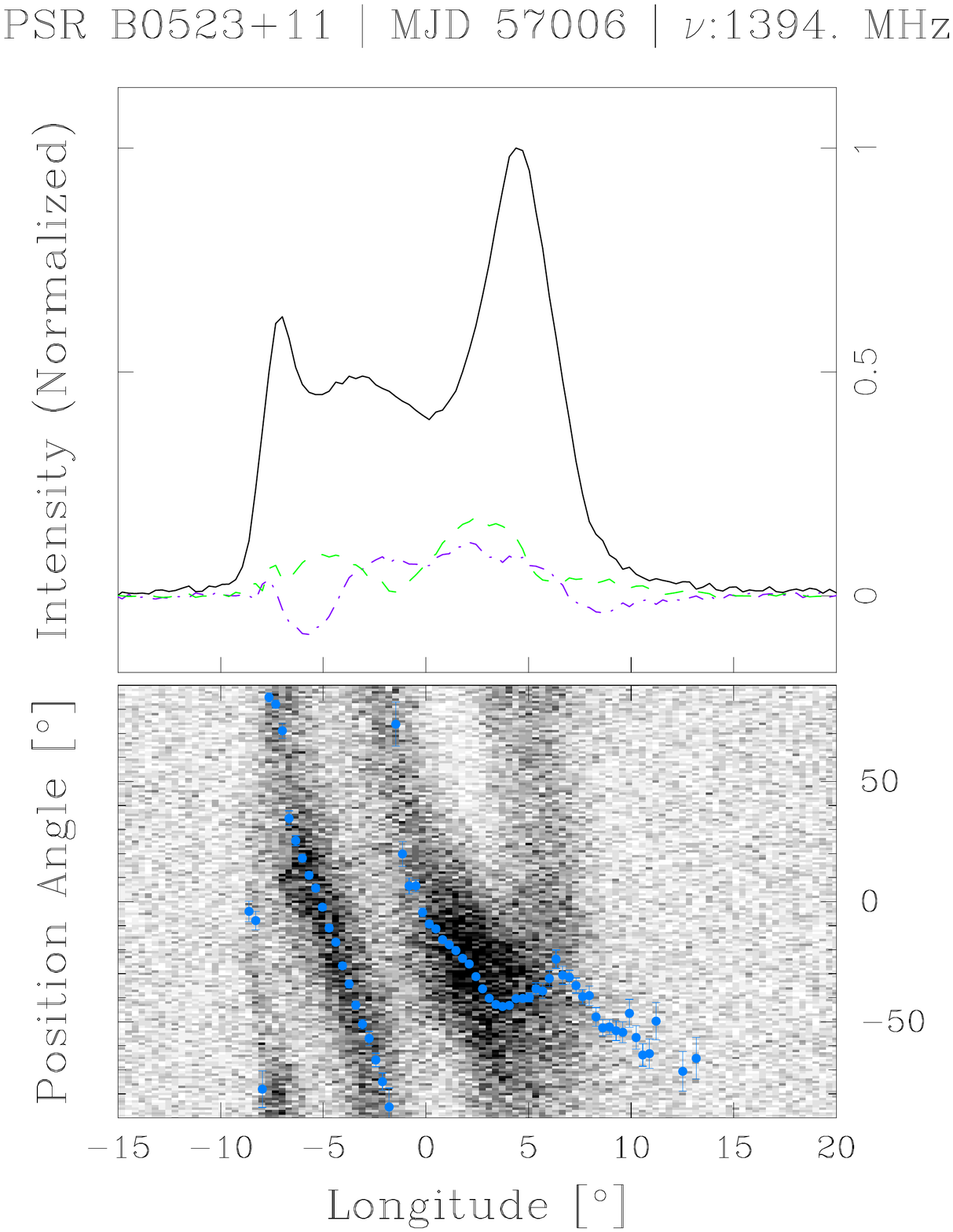} &
\includegraphics[page=1,width=\linewidth]{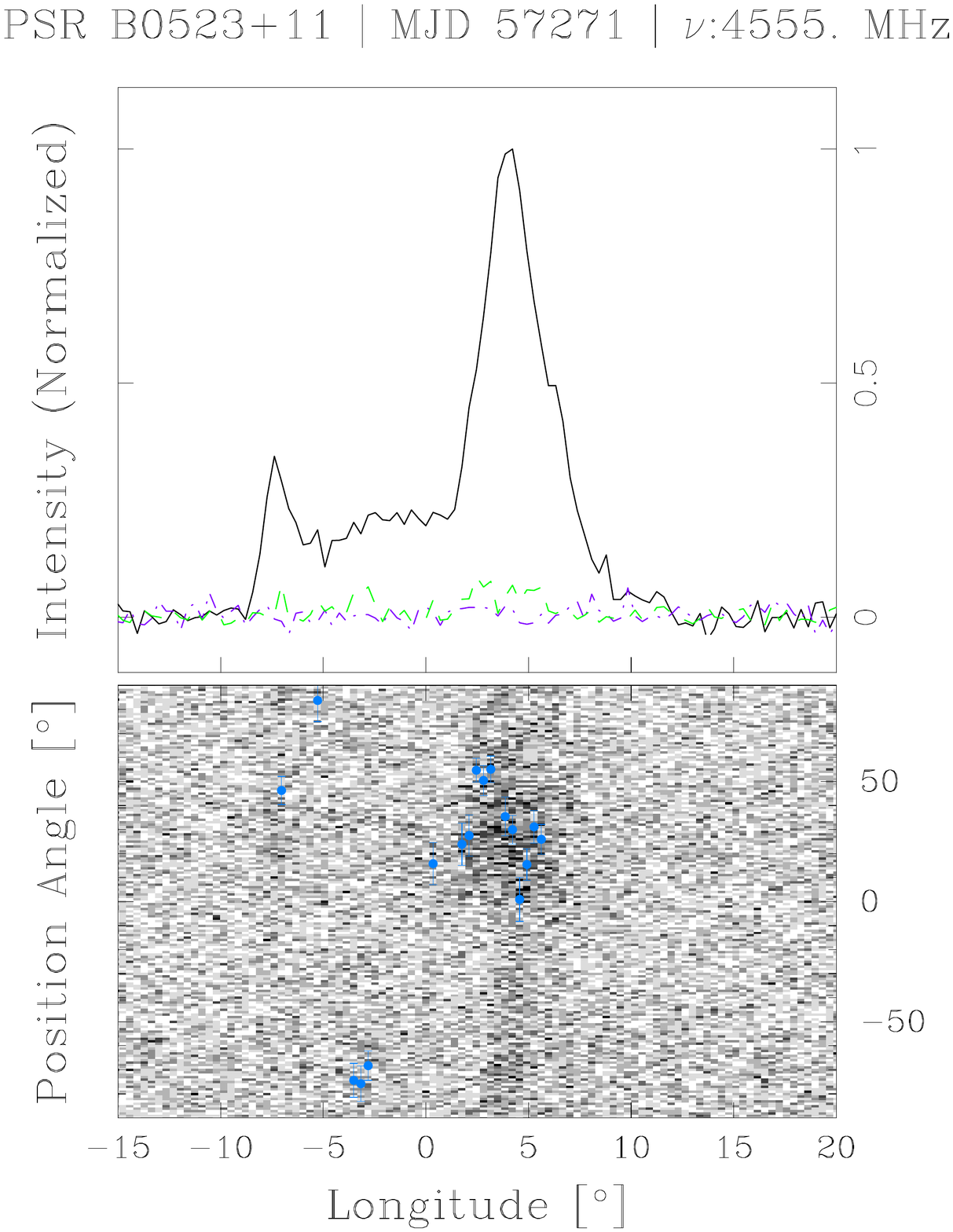} \\ \toprule
\includegraphics[page=1,width=\linewidth]{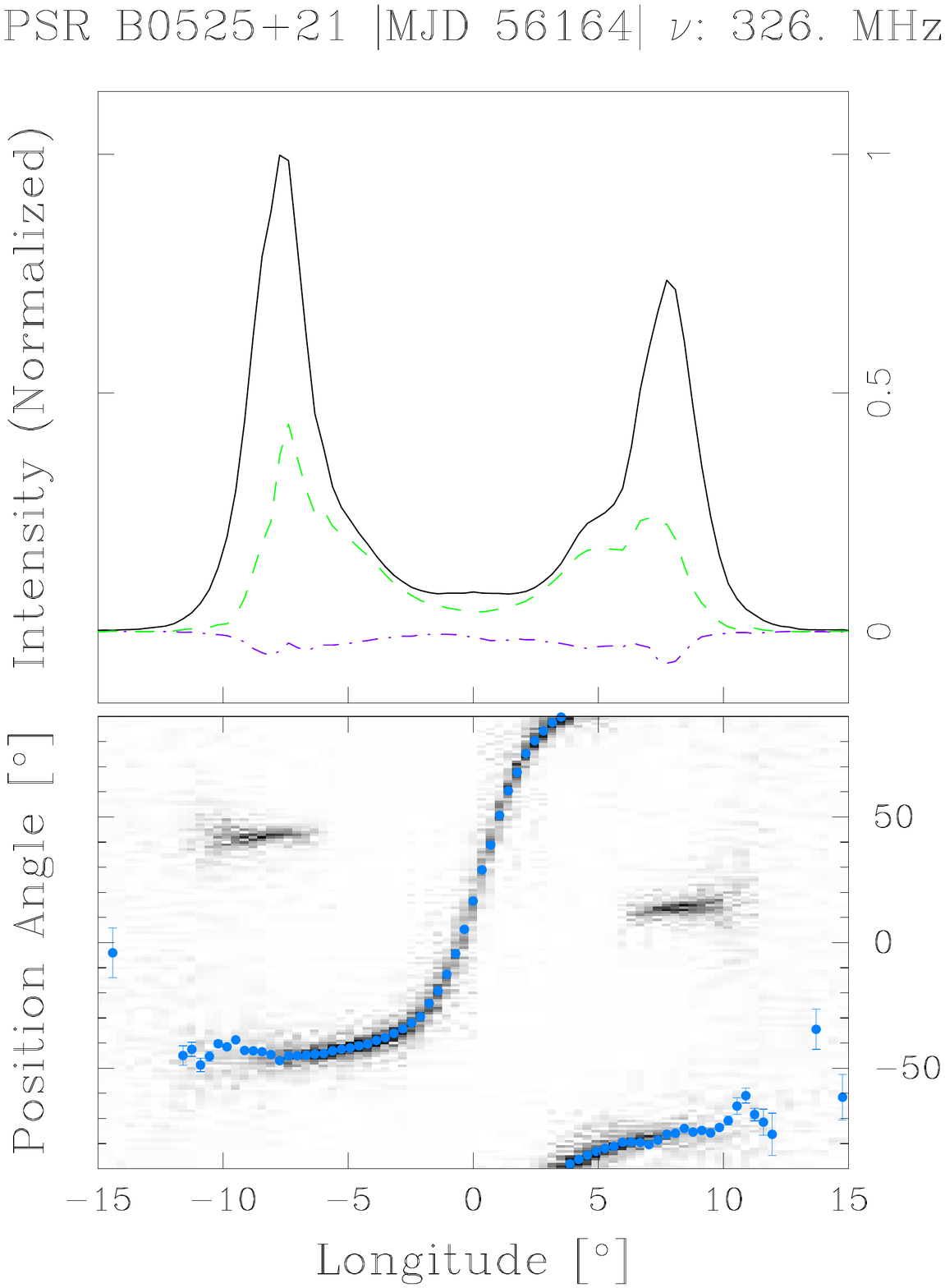} &
\includegraphics[page=1,width=\linewidth]{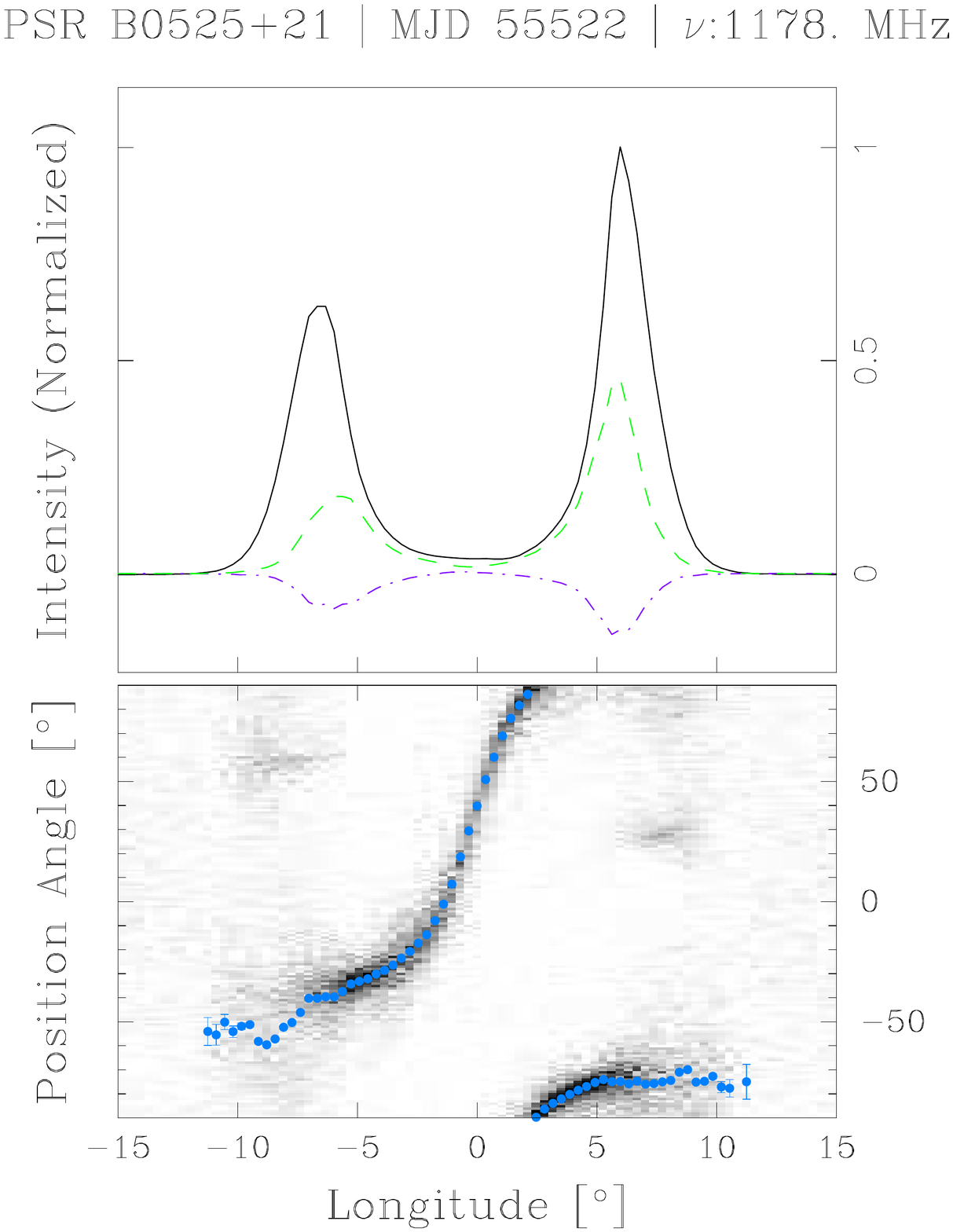} &
\includegraphics[page=1,width=\linewidth]{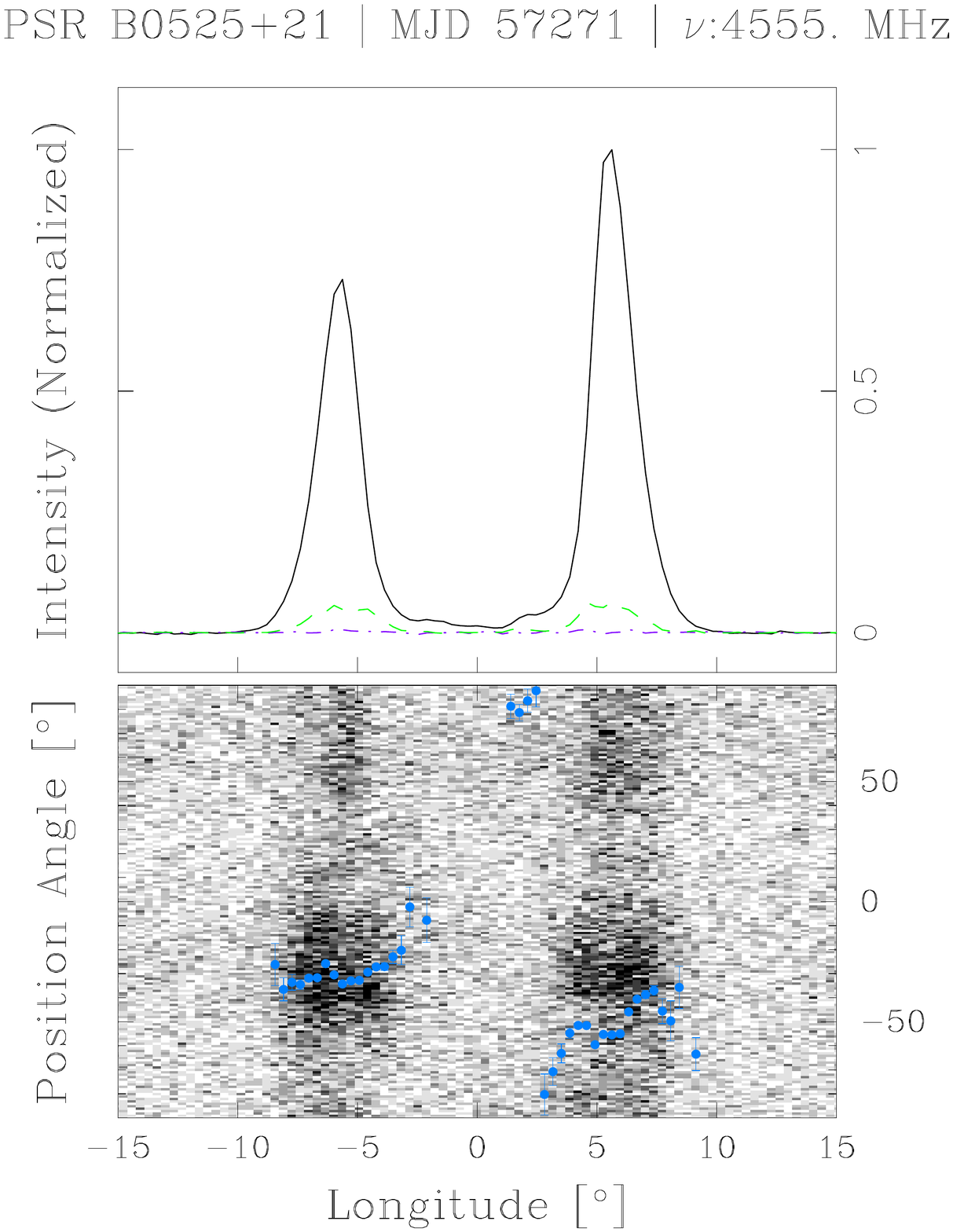} \\ 
     \bottomrule
   \end{tabularx} 
\caption{Average profiles of PSRs B0301+19, B0523+11, and B0525+21.}
 \end{figure*}
\vspace{1cm}

\begin{figure*} 
 \begin{tabularx}{\textwidth}{YYY}
 \multicolumn{3}{c}{} \\ \toprule
\includegraphics[page=1,width=\linewidth]{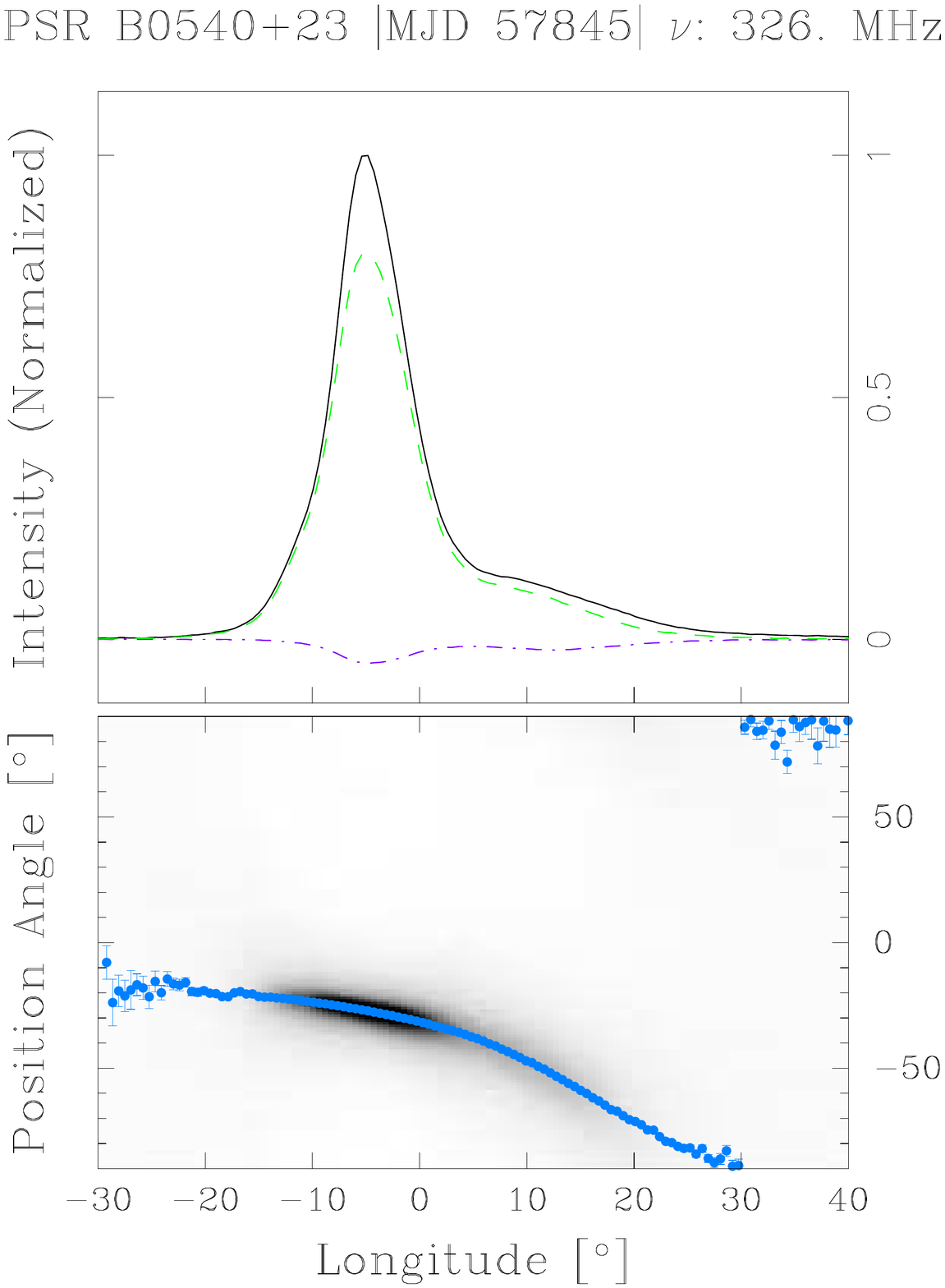} &
\includegraphics[page=1,width=\linewidth]{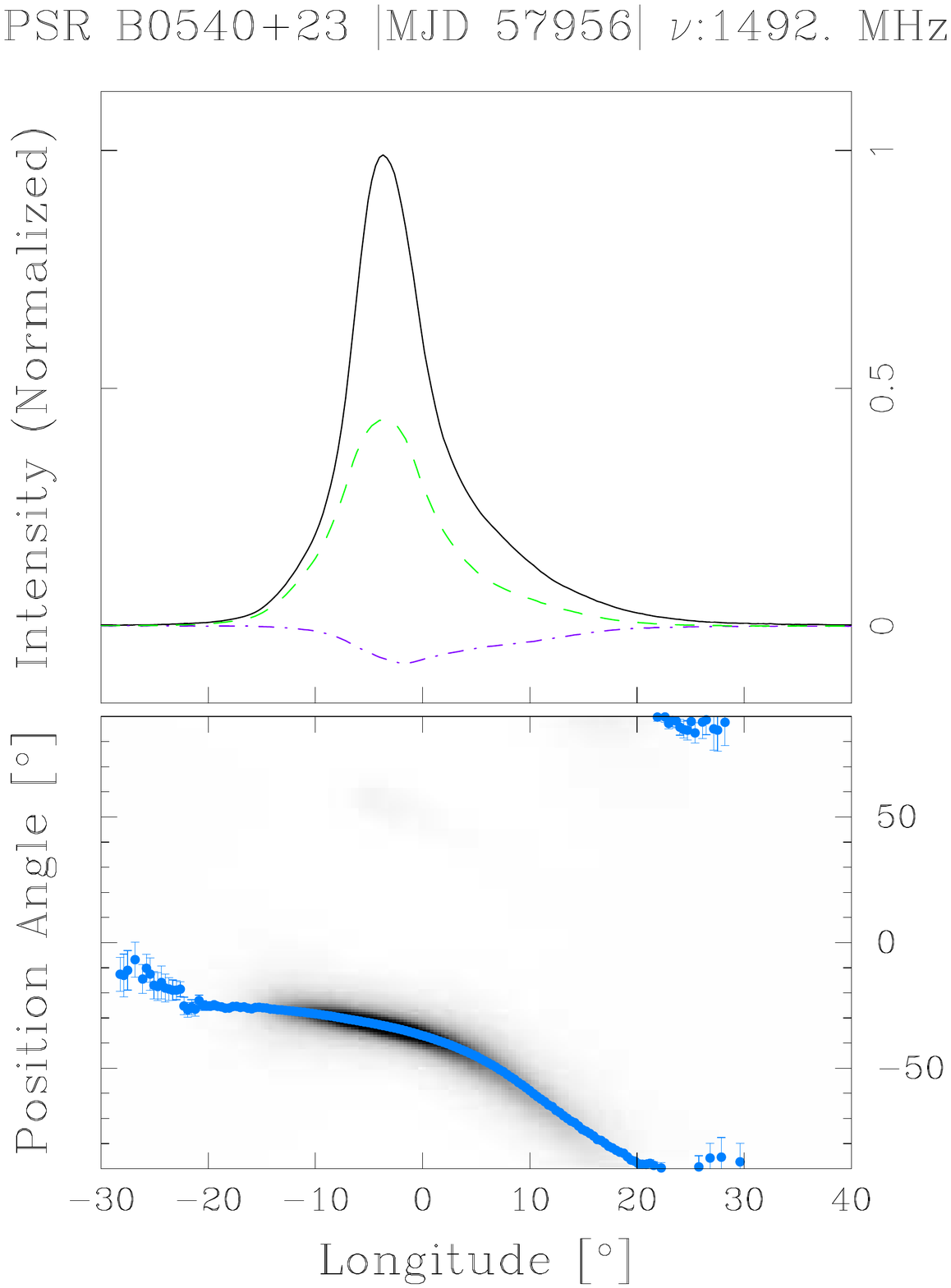} &
\includegraphics[page=1,width=\linewidth]{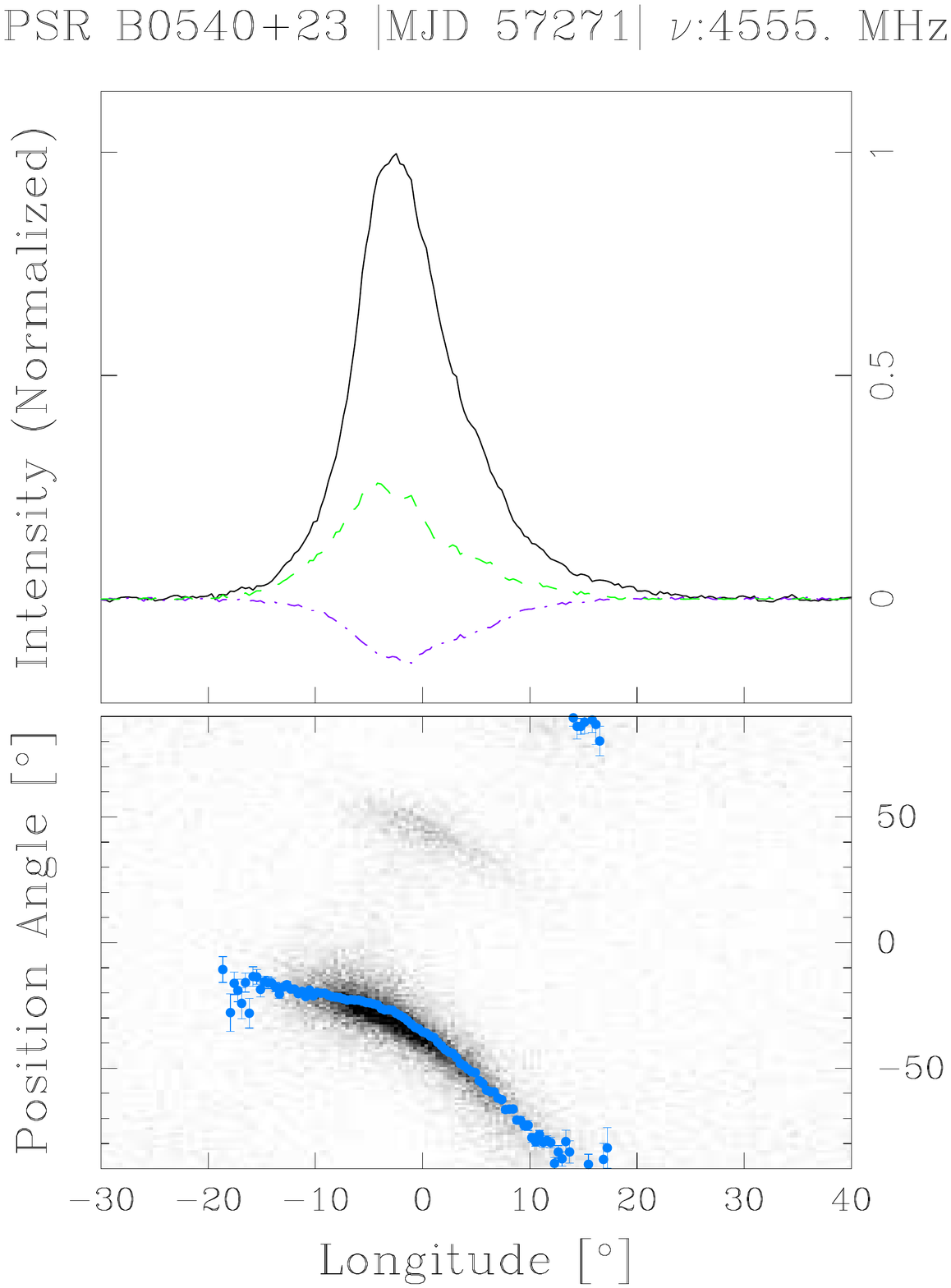} \\ \toprule
\includegraphics[page=1,width=\linewidth]{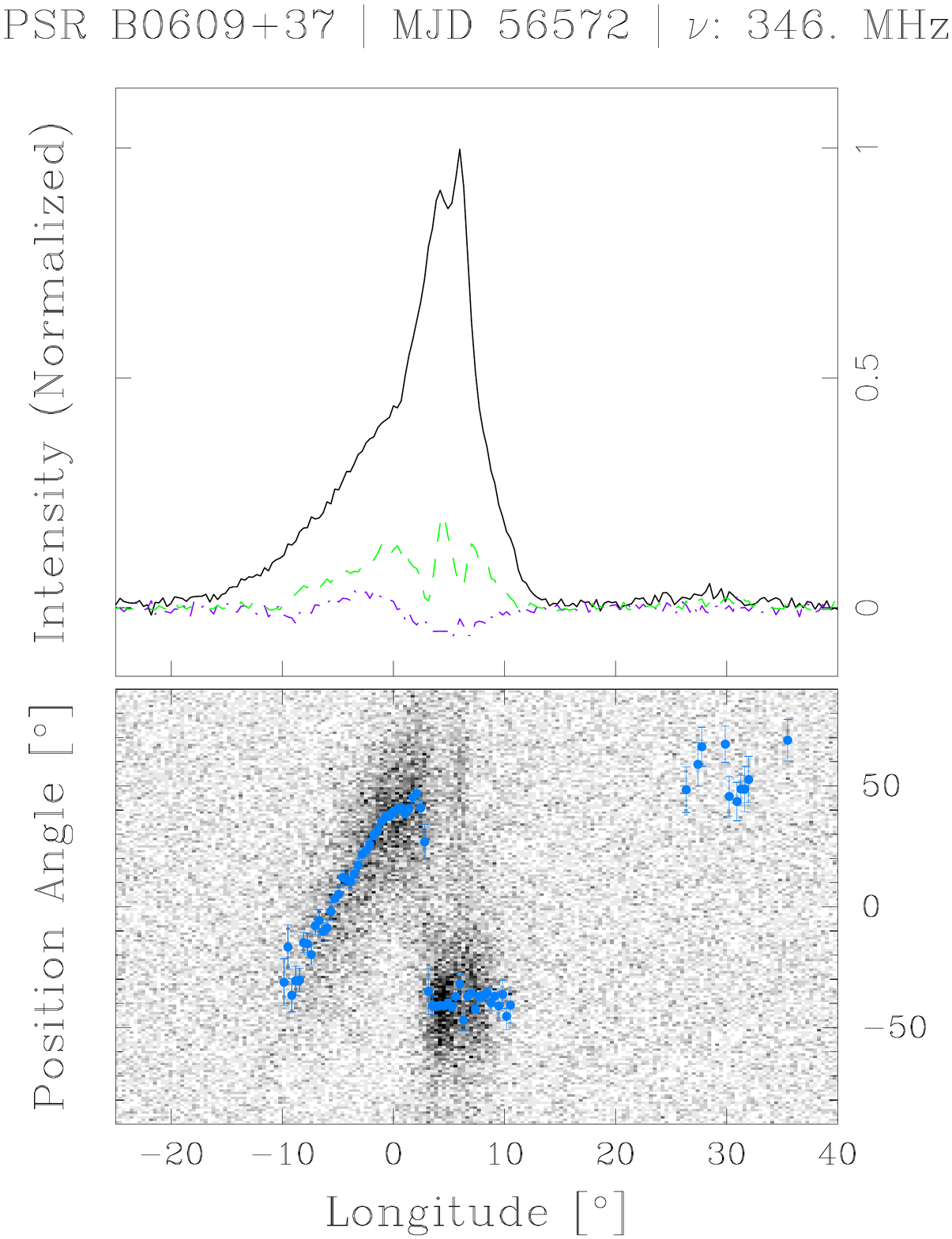} &
\includegraphics[page=1,width=\linewidth]{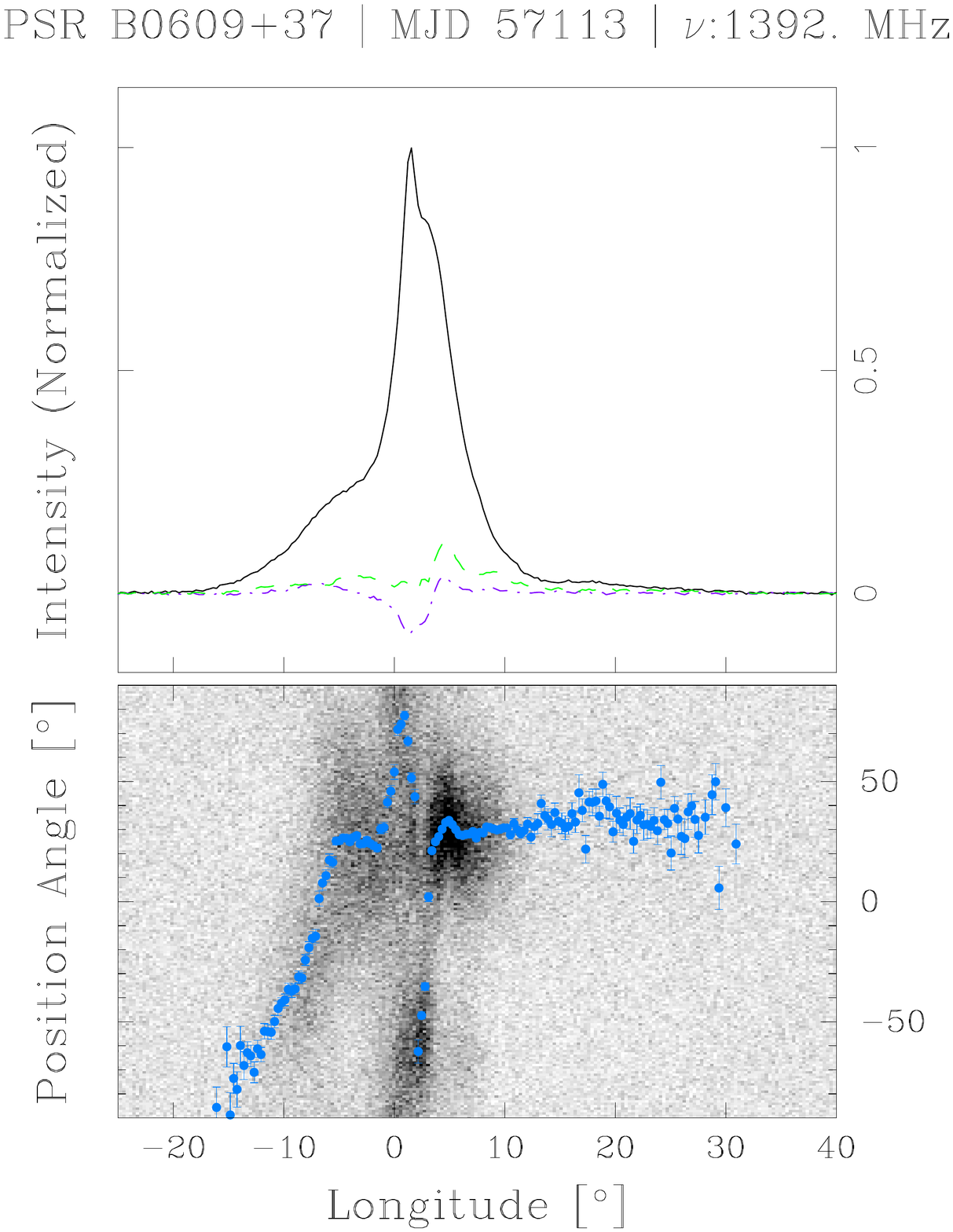} &
\includegraphics[page=1,width=\linewidth]{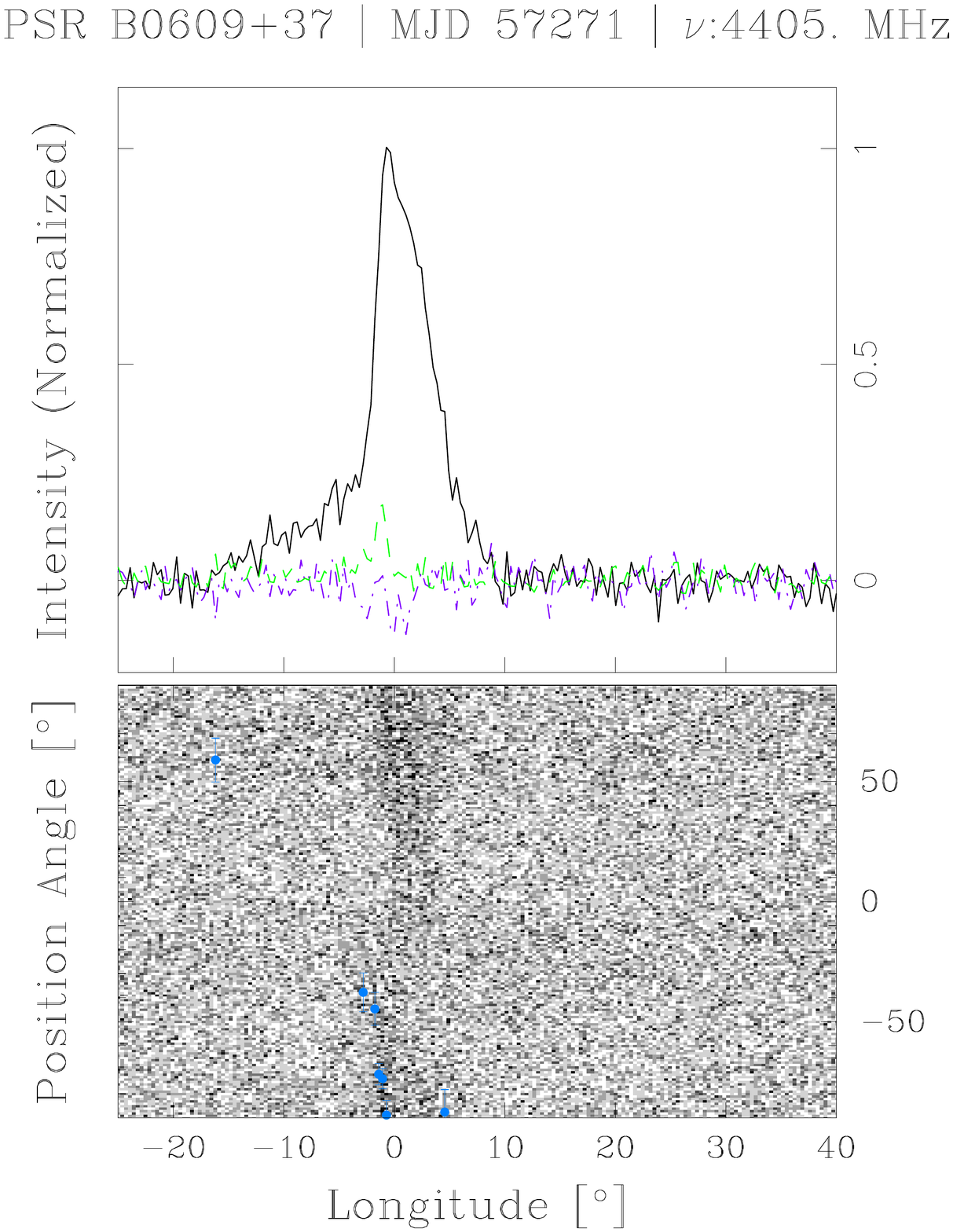} \\ \toprule
\includegraphics[page=1,width=\linewidth]{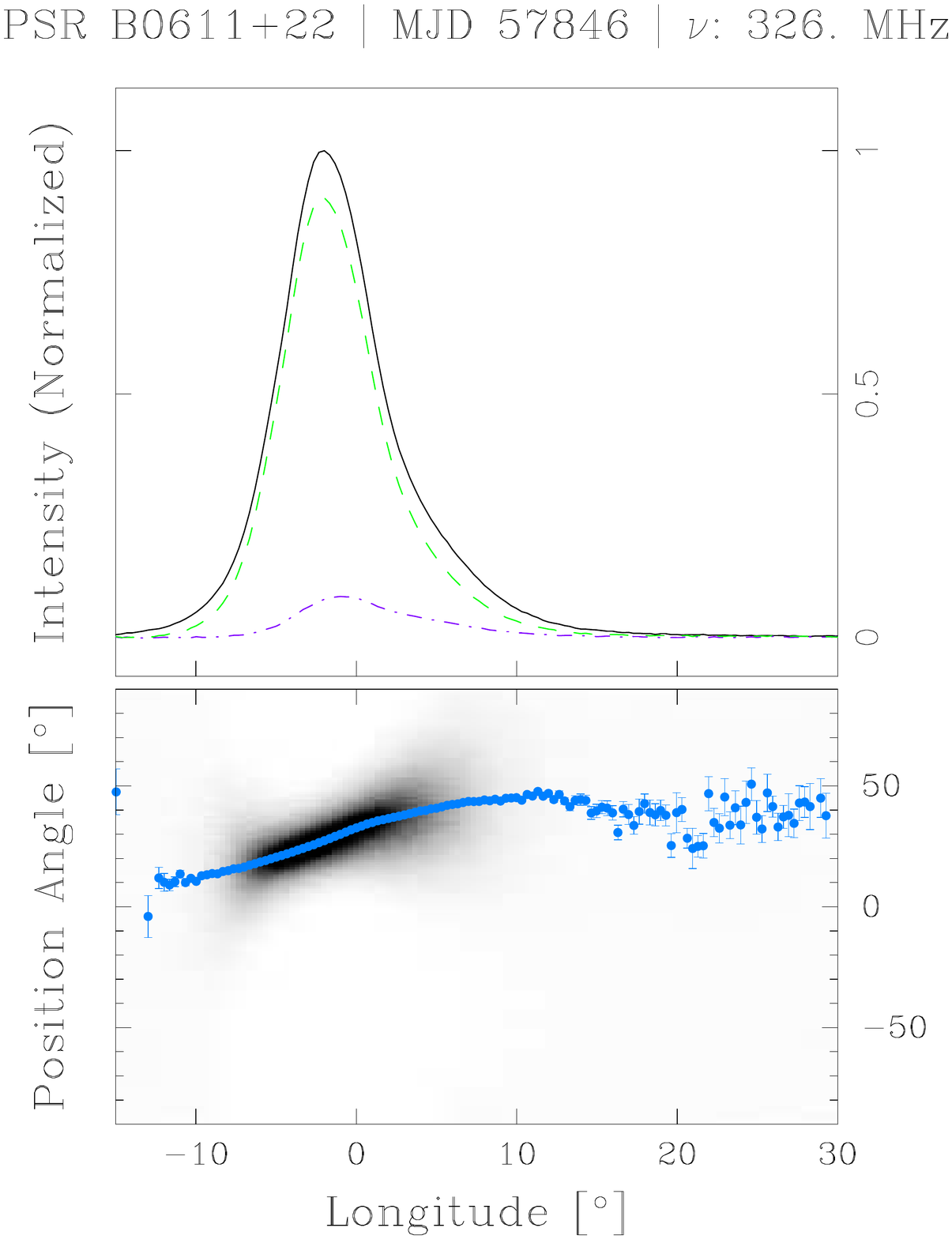} &
\includegraphics[page=1,width=\linewidth]{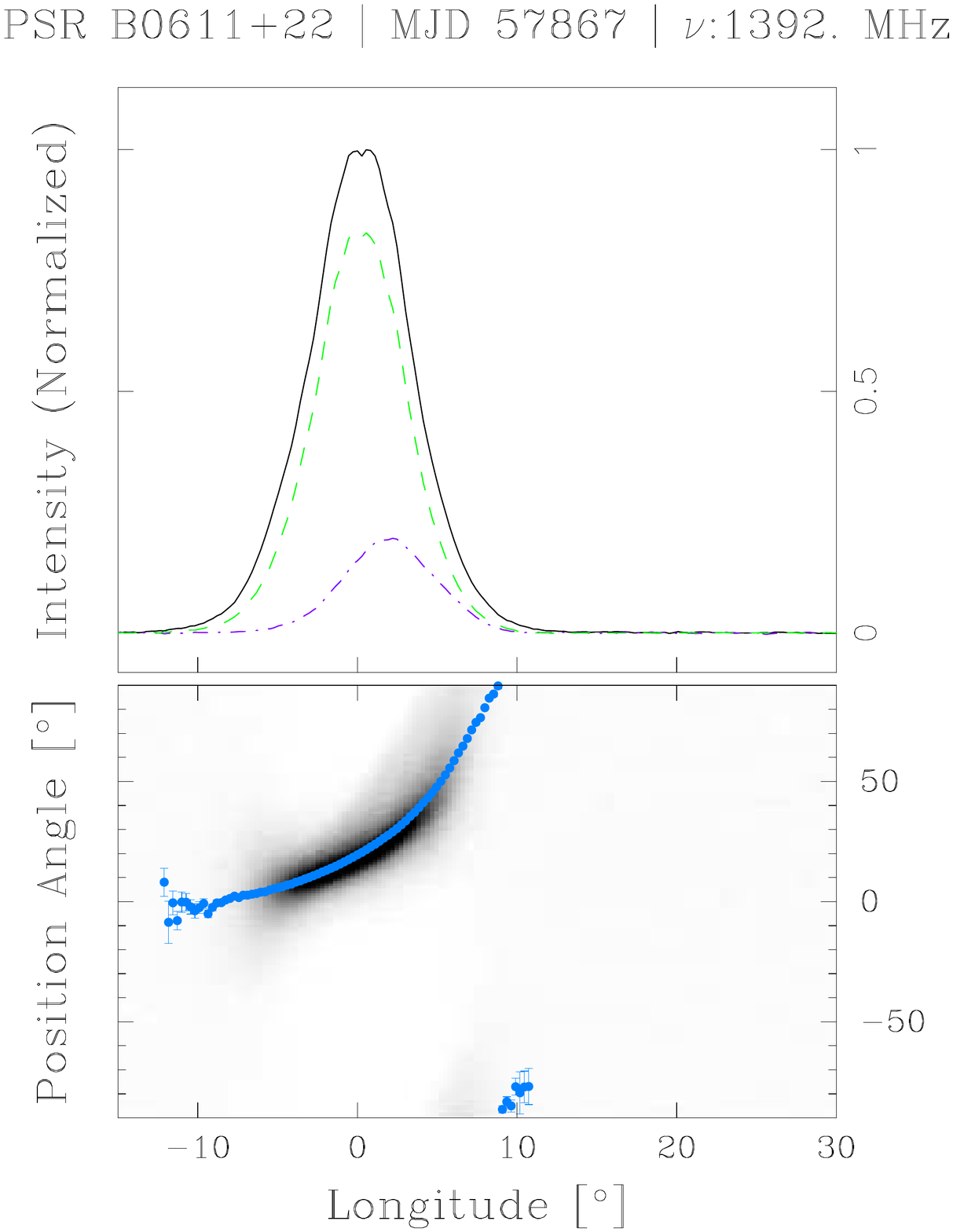} &
\includegraphics[page=1,width=\linewidth]{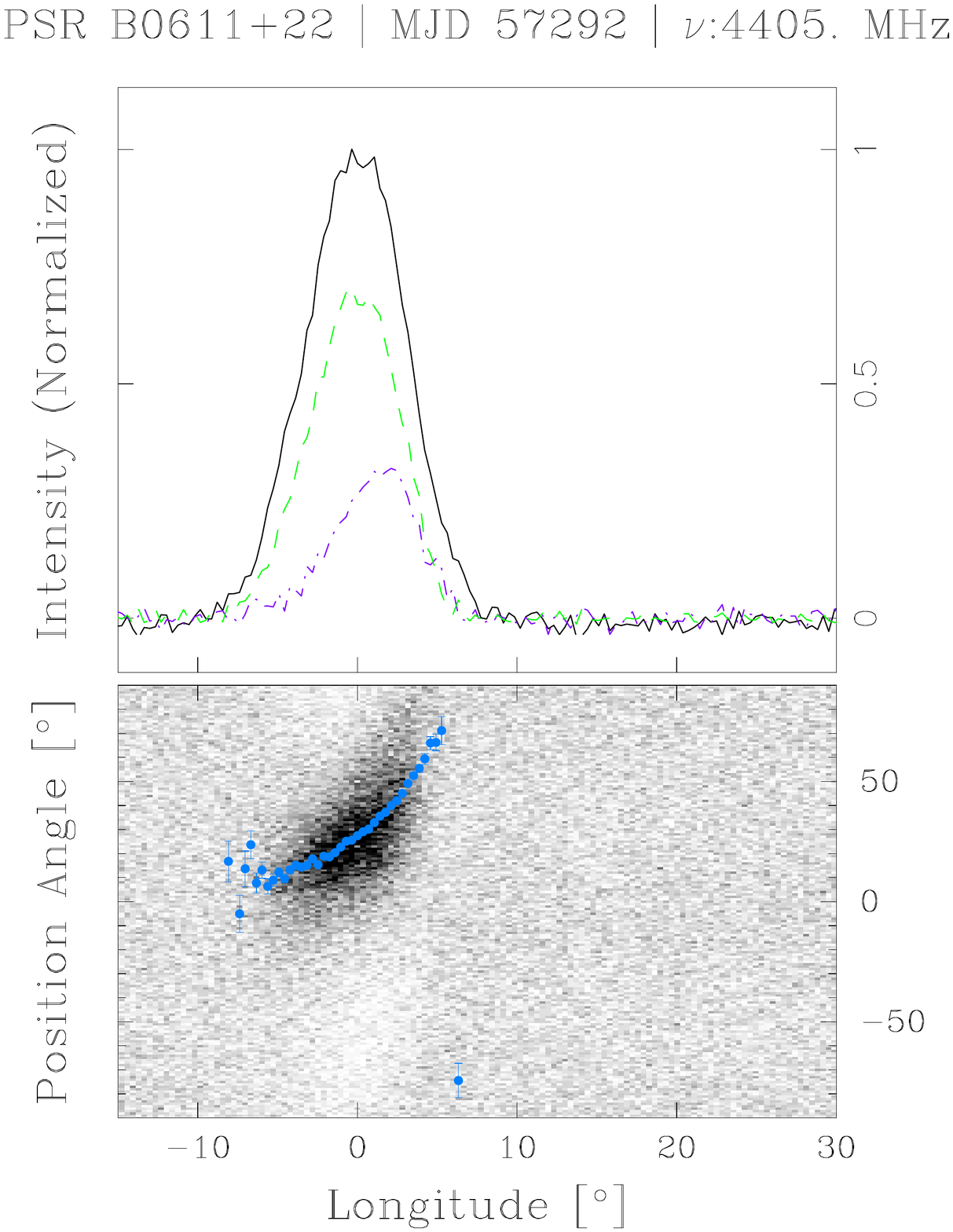} \\ 
     \bottomrule
   \end{tabularx} 
\caption{Average profiles of PSRs B0540+23, B0609+37, and B0611+22.}
\end{figure*}
\vspace{1cm}

\begin{figure*} 
 \begin{tabularx}{\textwidth}{YYY}
 \multicolumn{3}{c}{} \\ \toprule
\includegraphics[page=1,width=\linewidth]{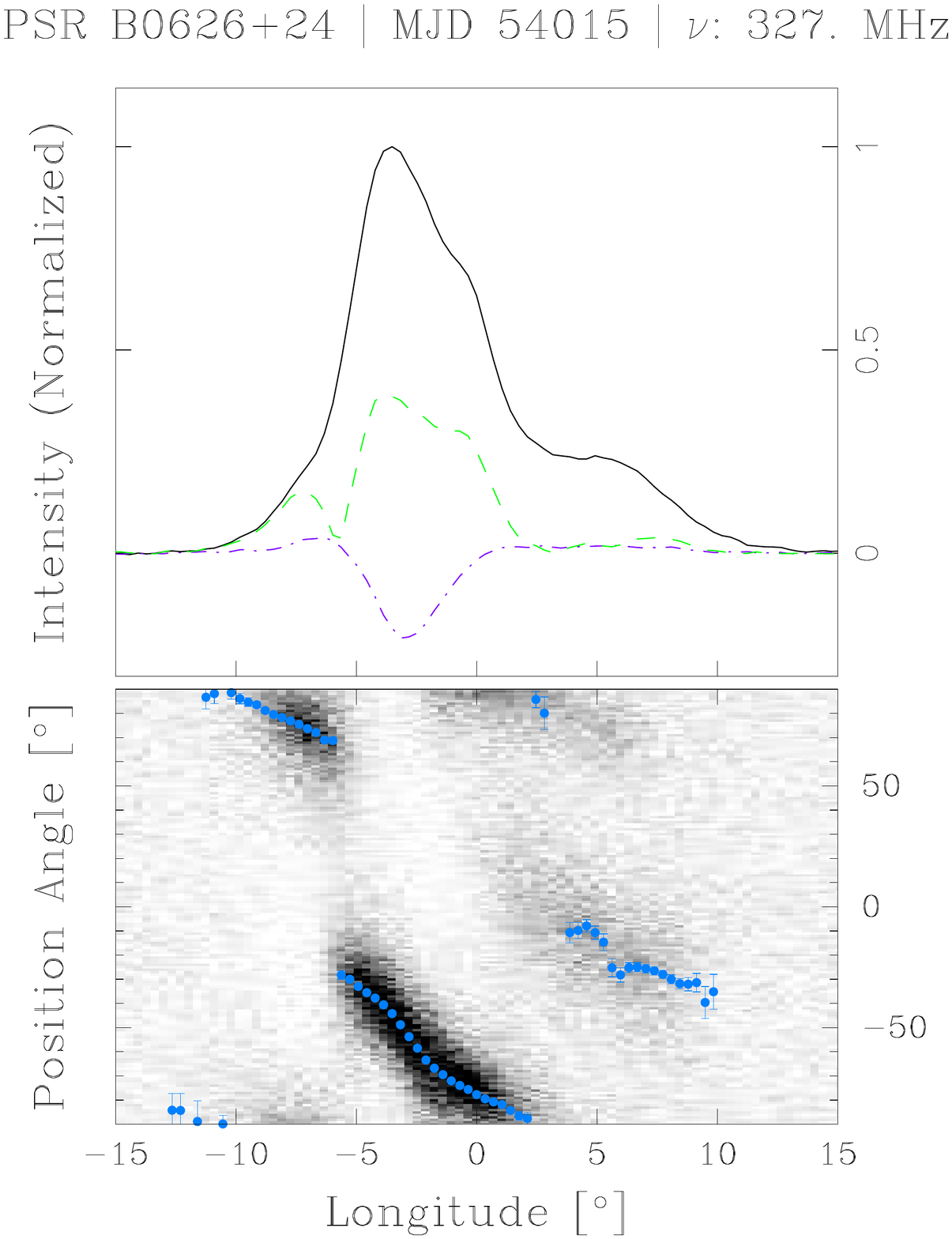} &
\includegraphics[page=1,width=\linewidth]{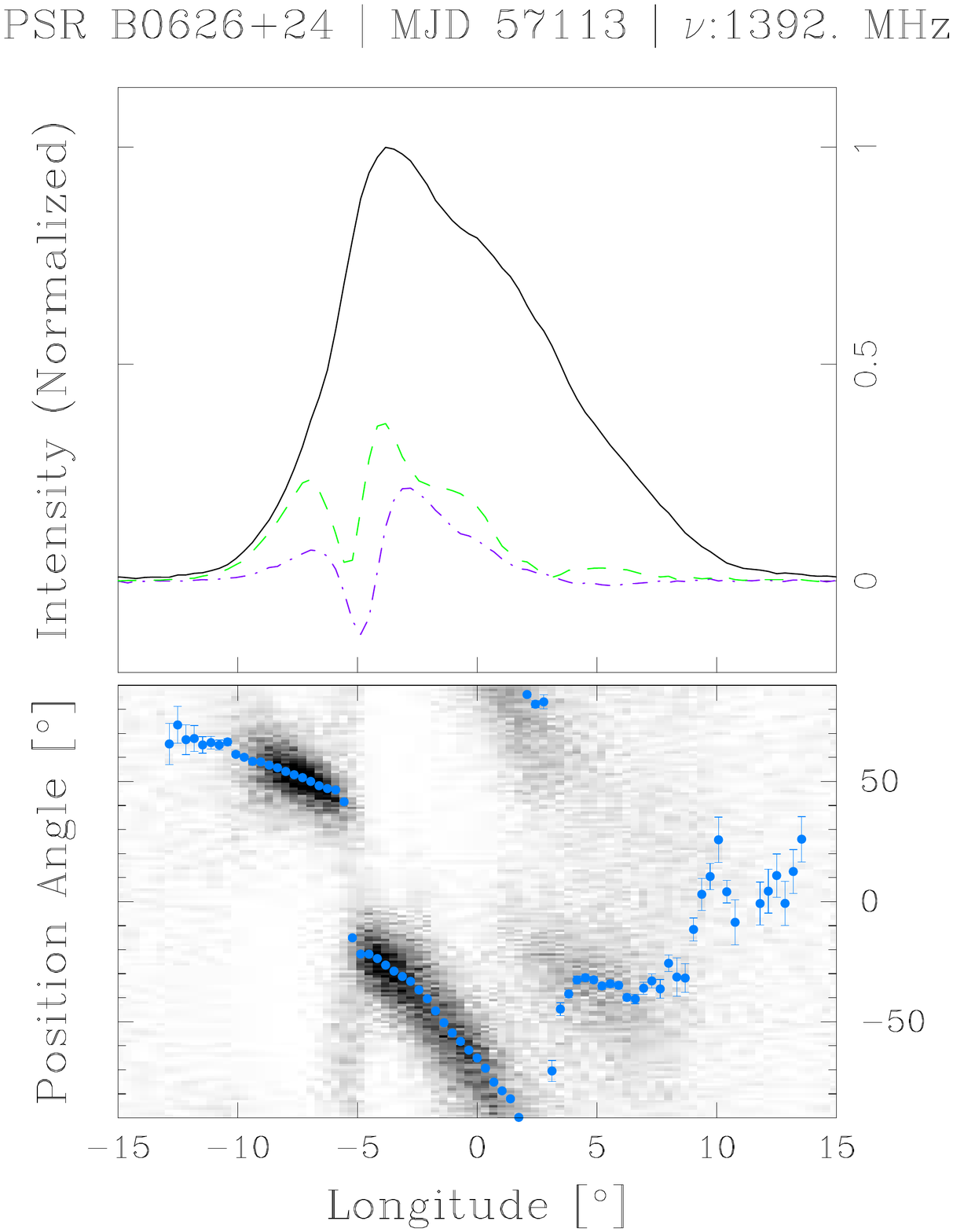} &
\includegraphics[page=1,width=\linewidth]{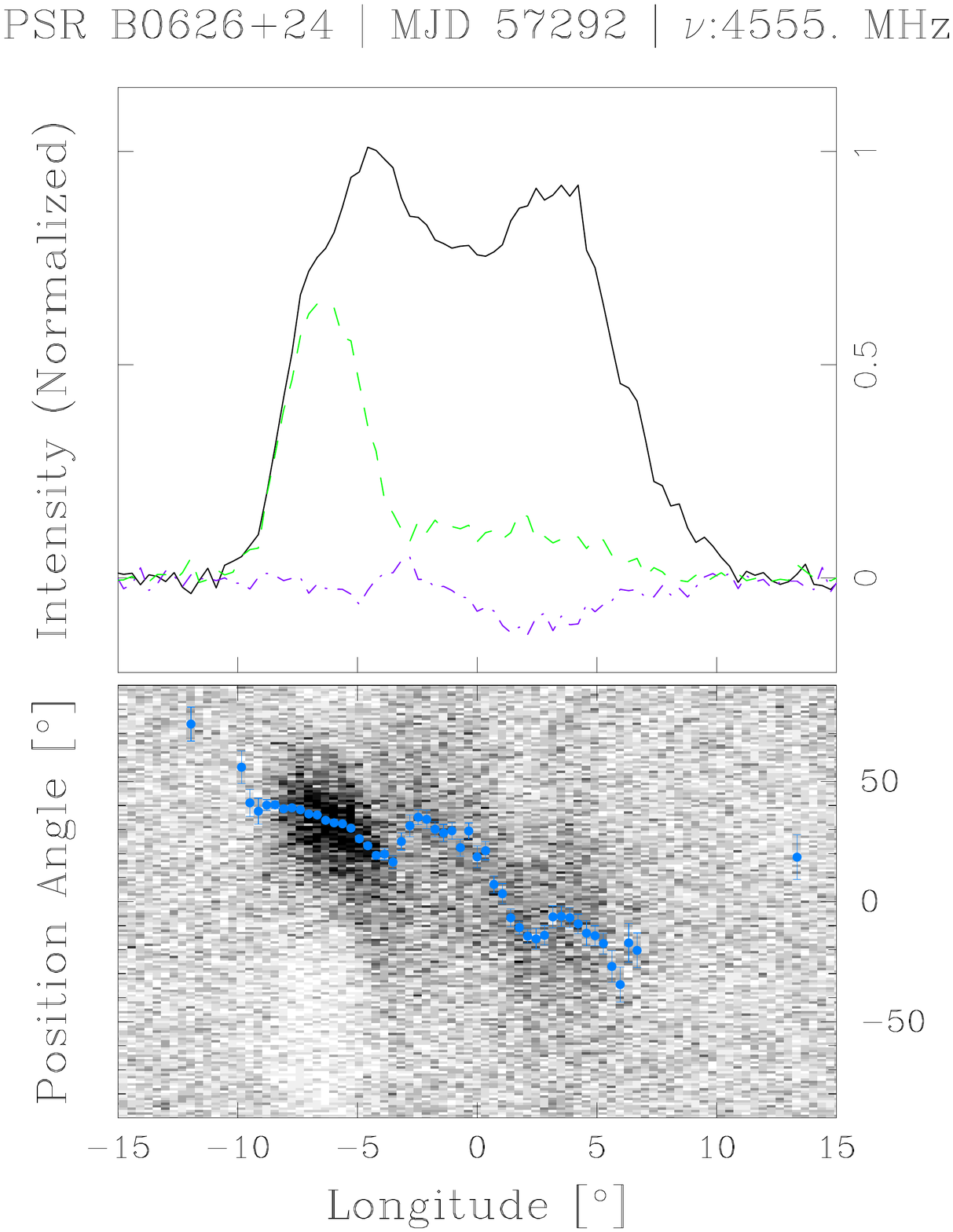} \\ \toprule
\includegraphics[page=1,width=\linewidth]{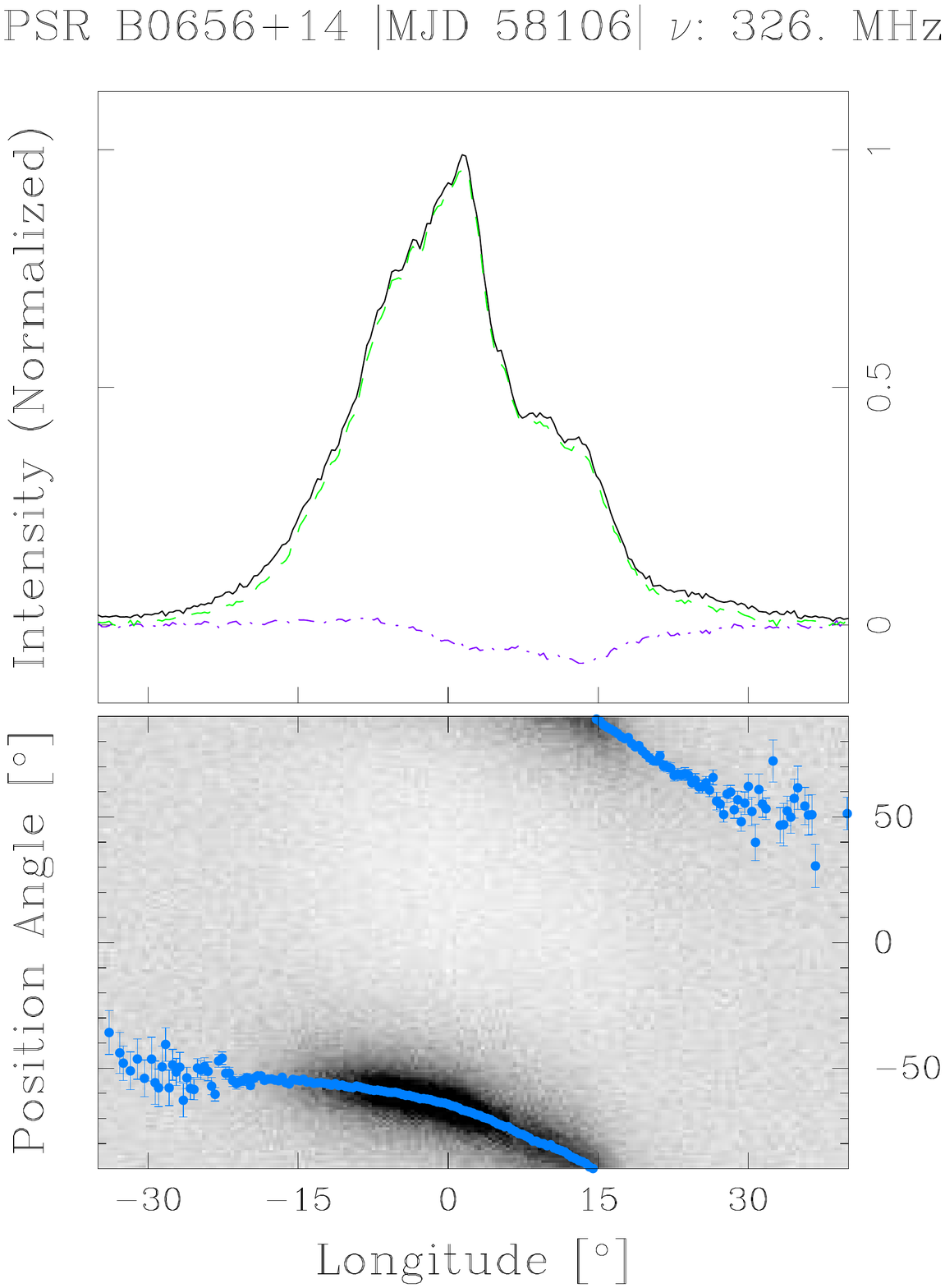} &
\includegraphics[page=1,width=\linewidth]{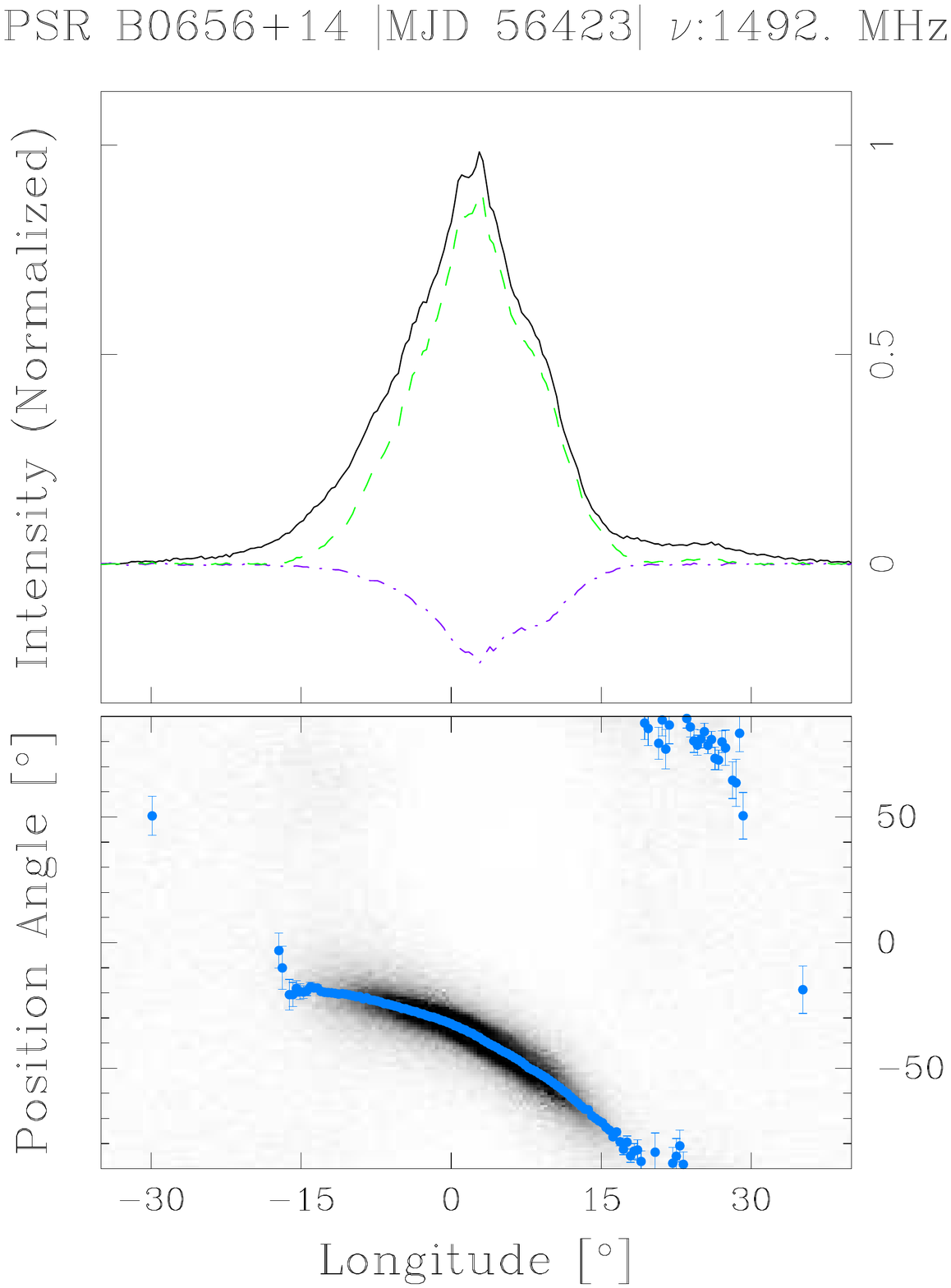} &
\includegraphics[page=1,width=\linewidth]{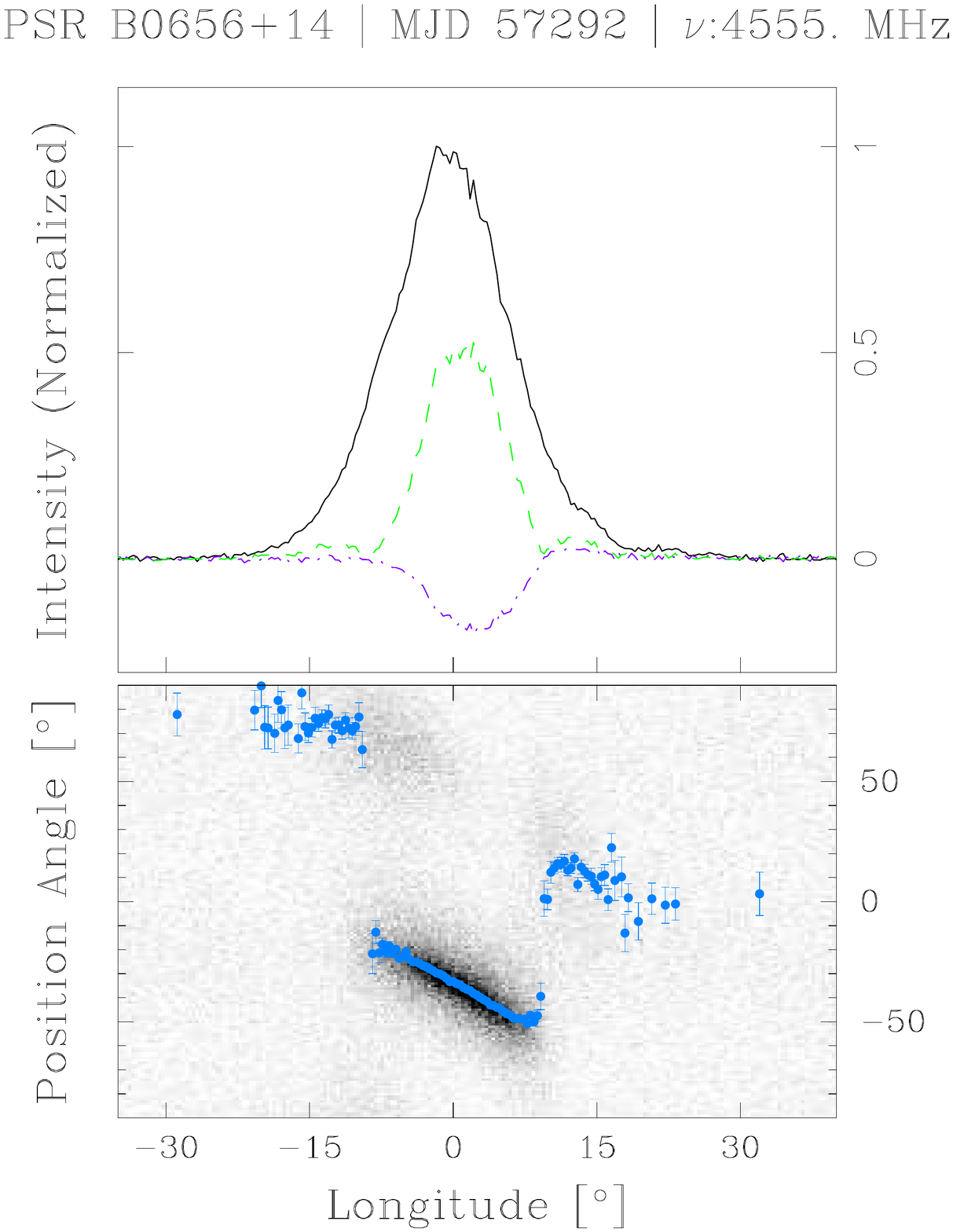} \\ \toprule
\includegraphics[page=1,width=\linewidth]{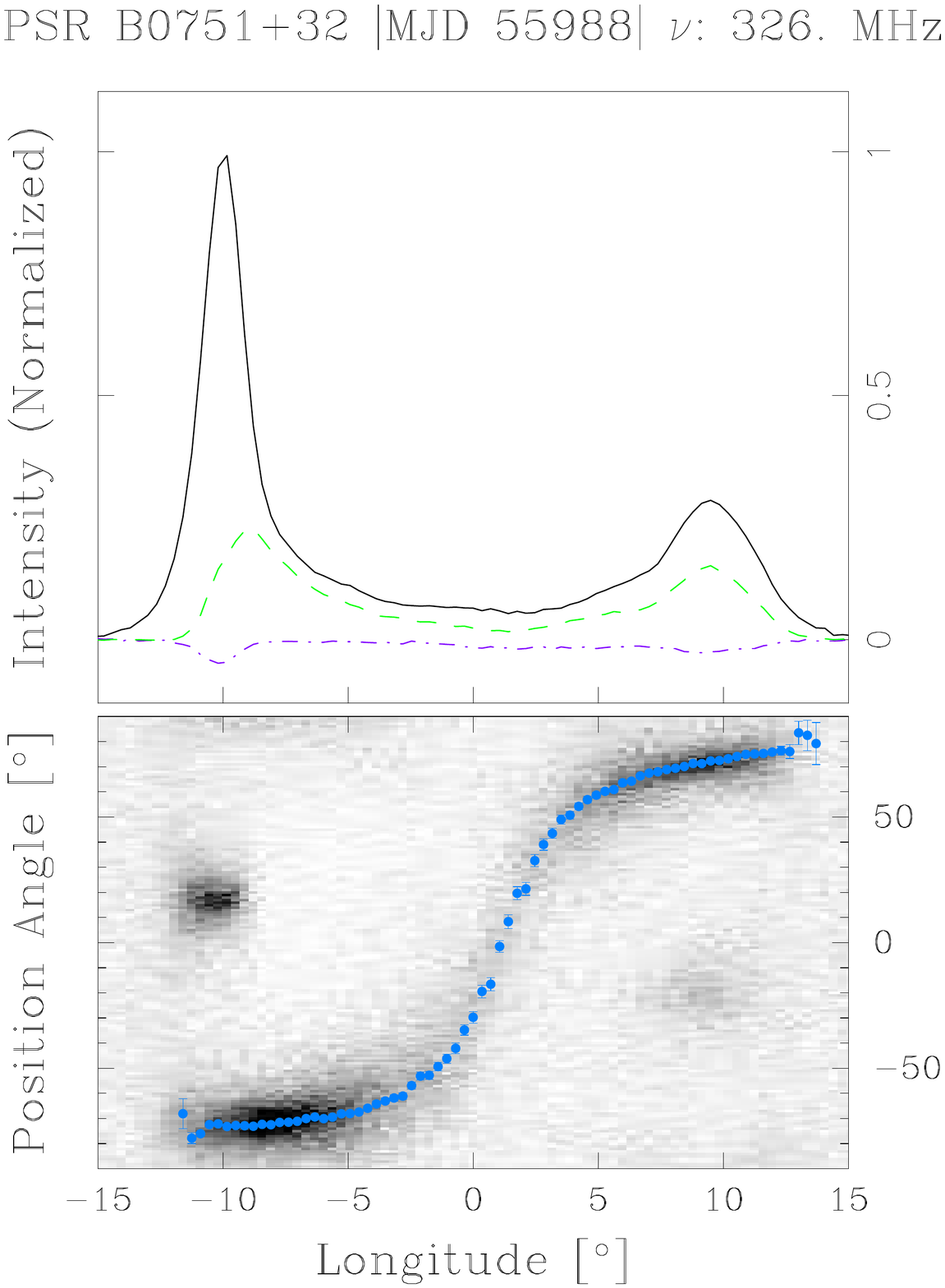} &
\includegraphics[page=1,width=\linewidth]{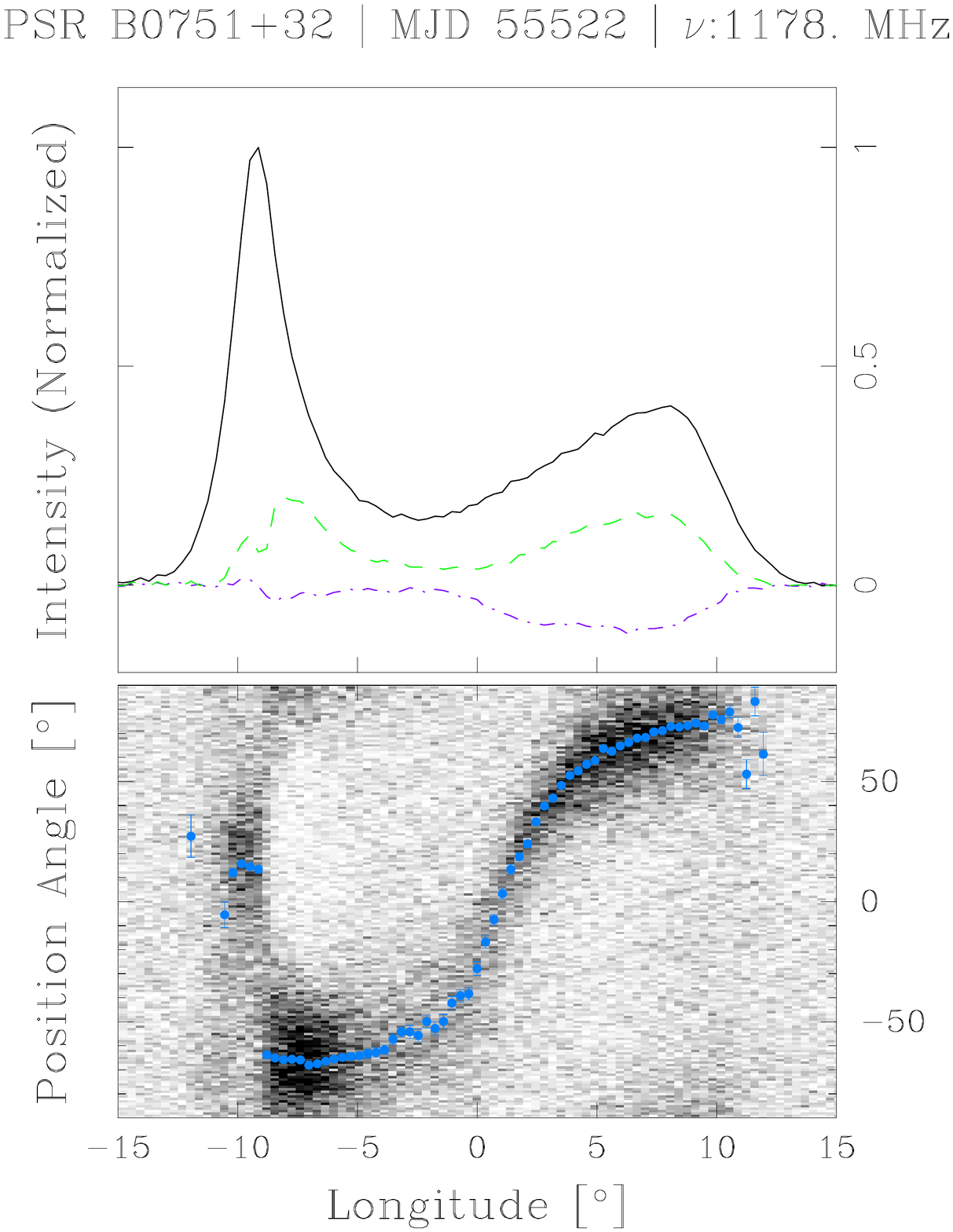} &
\includegraphics[page=1,width=\linewidth]{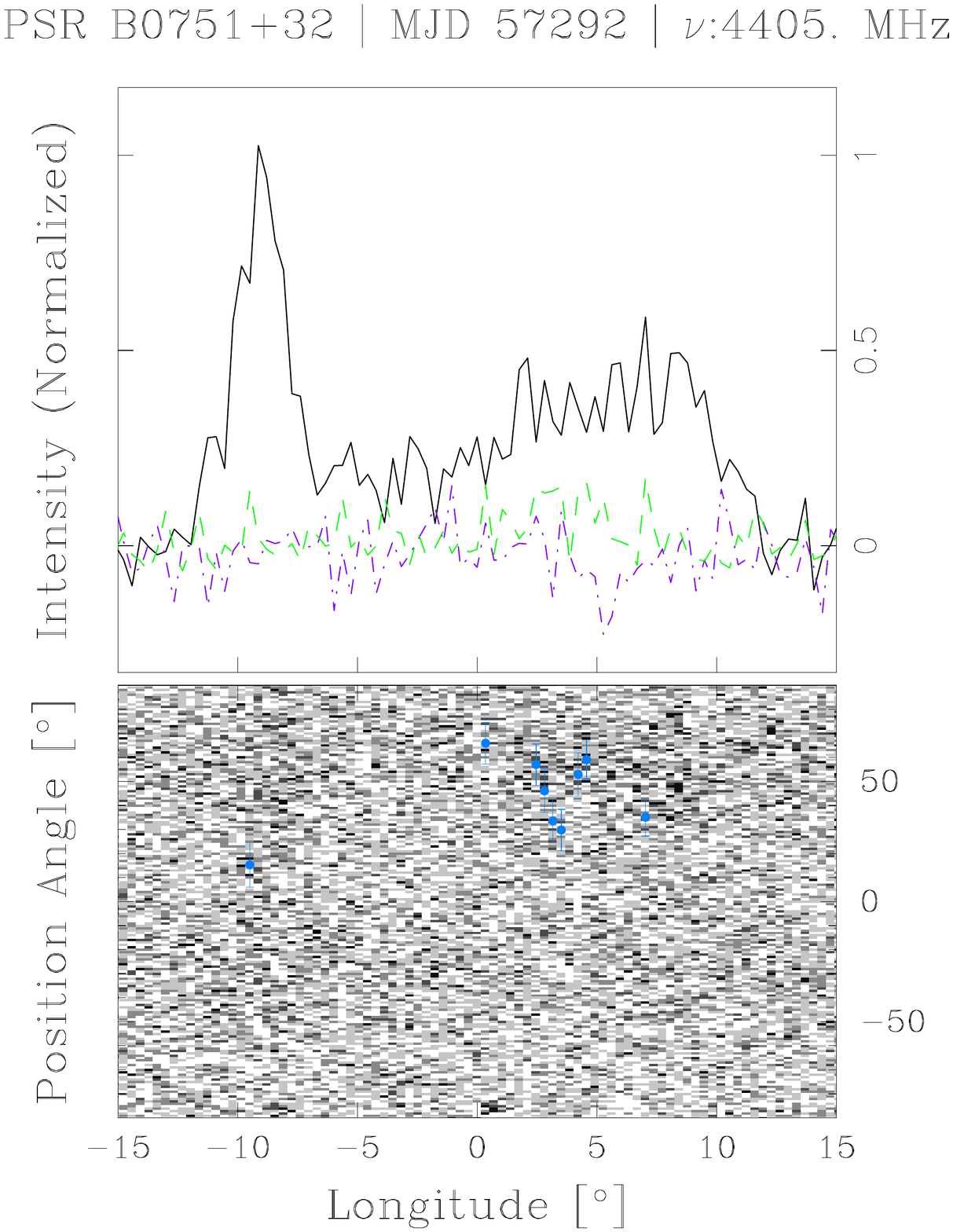} \\ 
     \bottomrule
   \end{tabularx} 
\caption{Average profiles of PSRs B0626+24, B0656+14, and B0751+32.}
 \end{figure*}
\vspace{1cm}

\begin{figure*} 
 \begin{tabularx}{\textwidth}{YYY}
 \multicolumn{3}{c}{} \\ \toprule
\includegraphics[page=1,width=\linewidth]{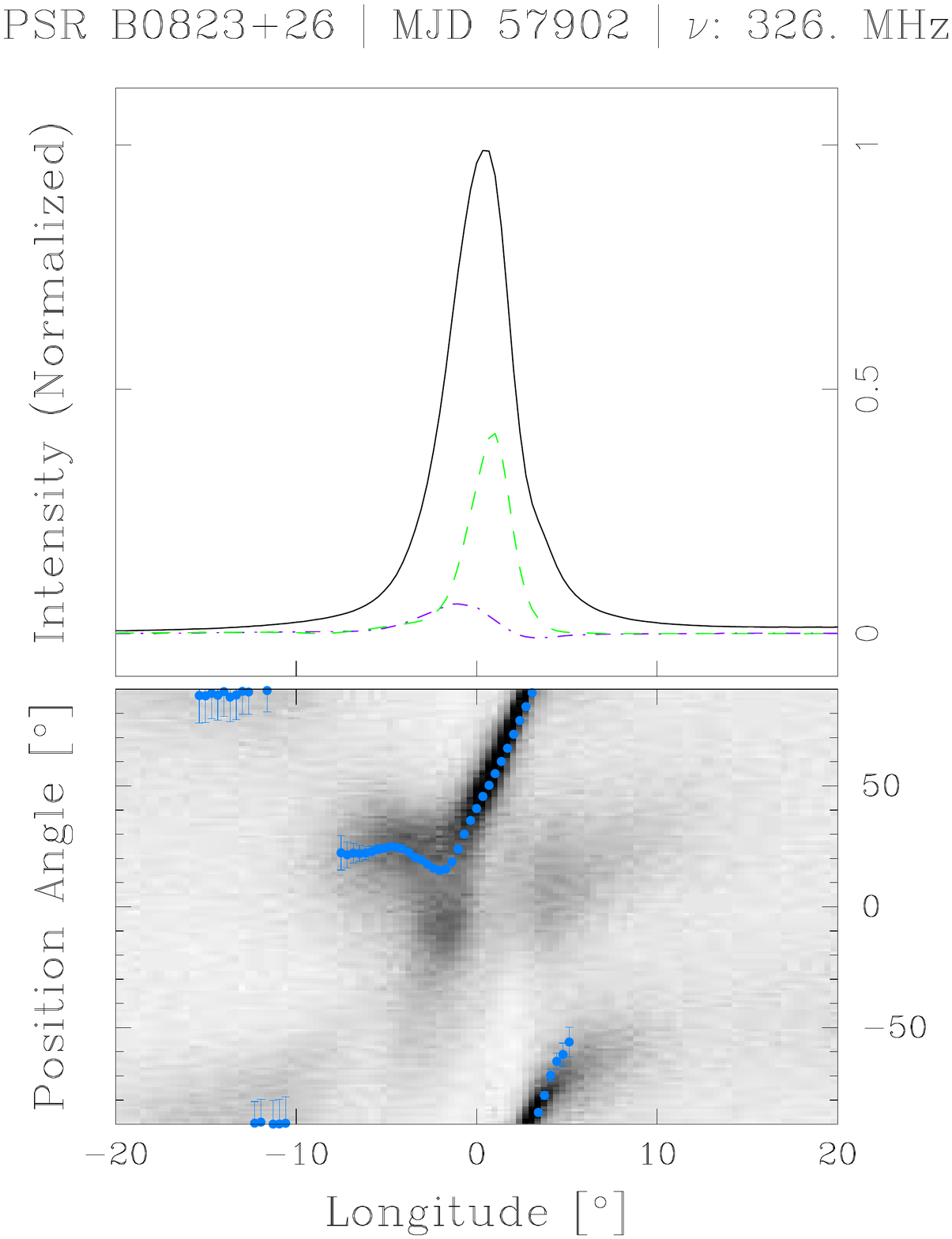} &
\includegraphics[page=1,width=\linewidth]{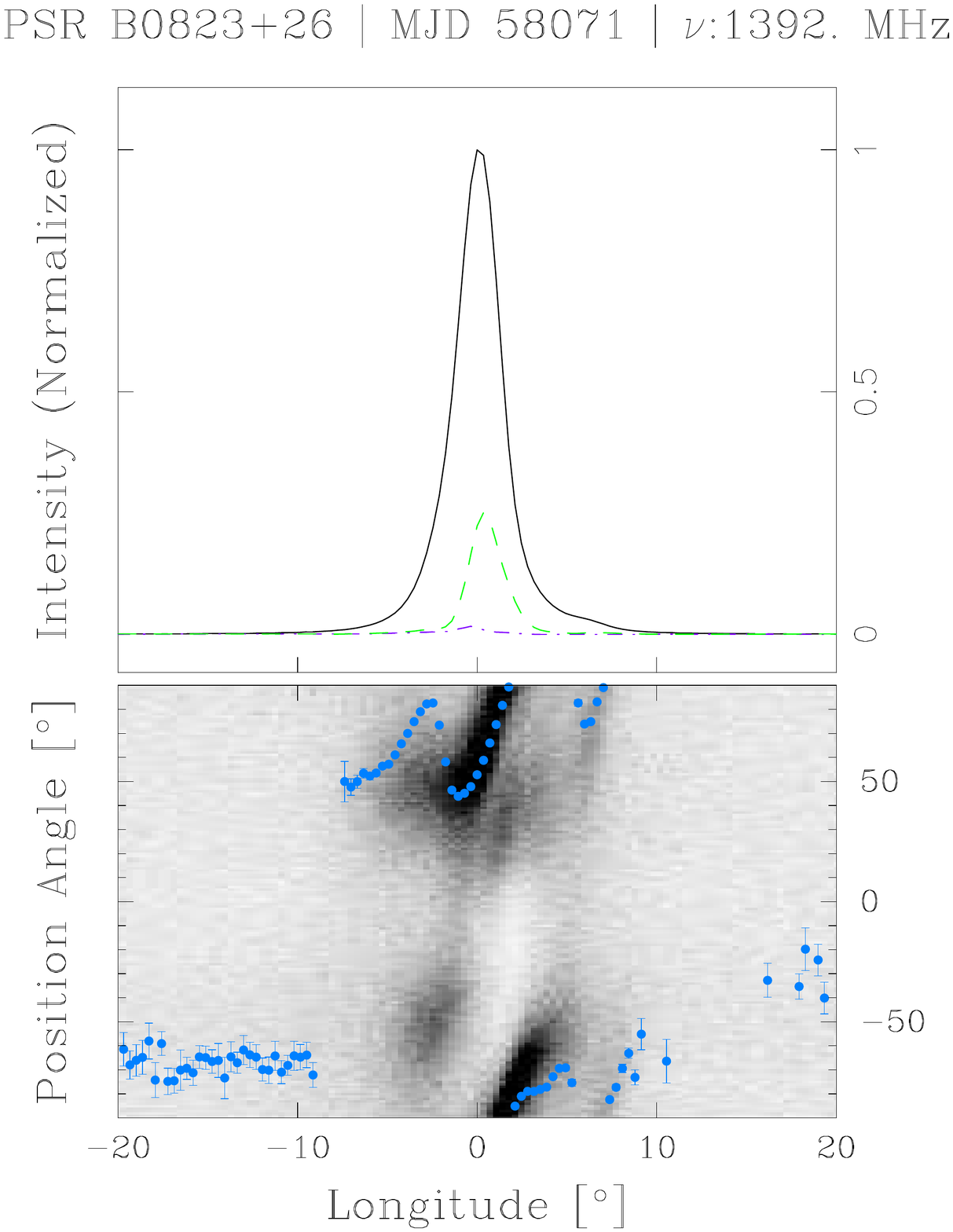} &
\includegraphics[page=1,width=\linewidth]{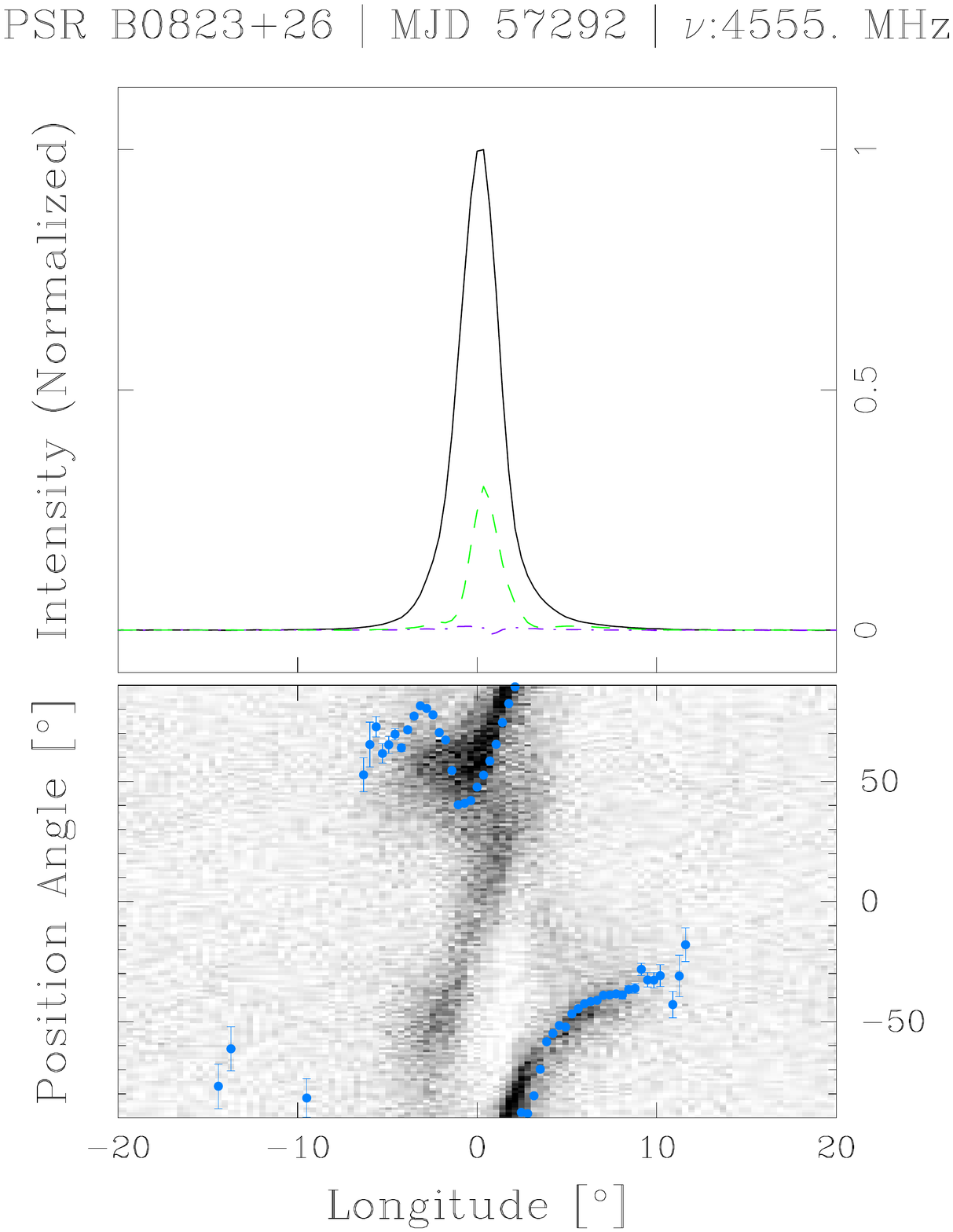} \\ \toprule

\includegraphics[page=1,width=\linewidth]{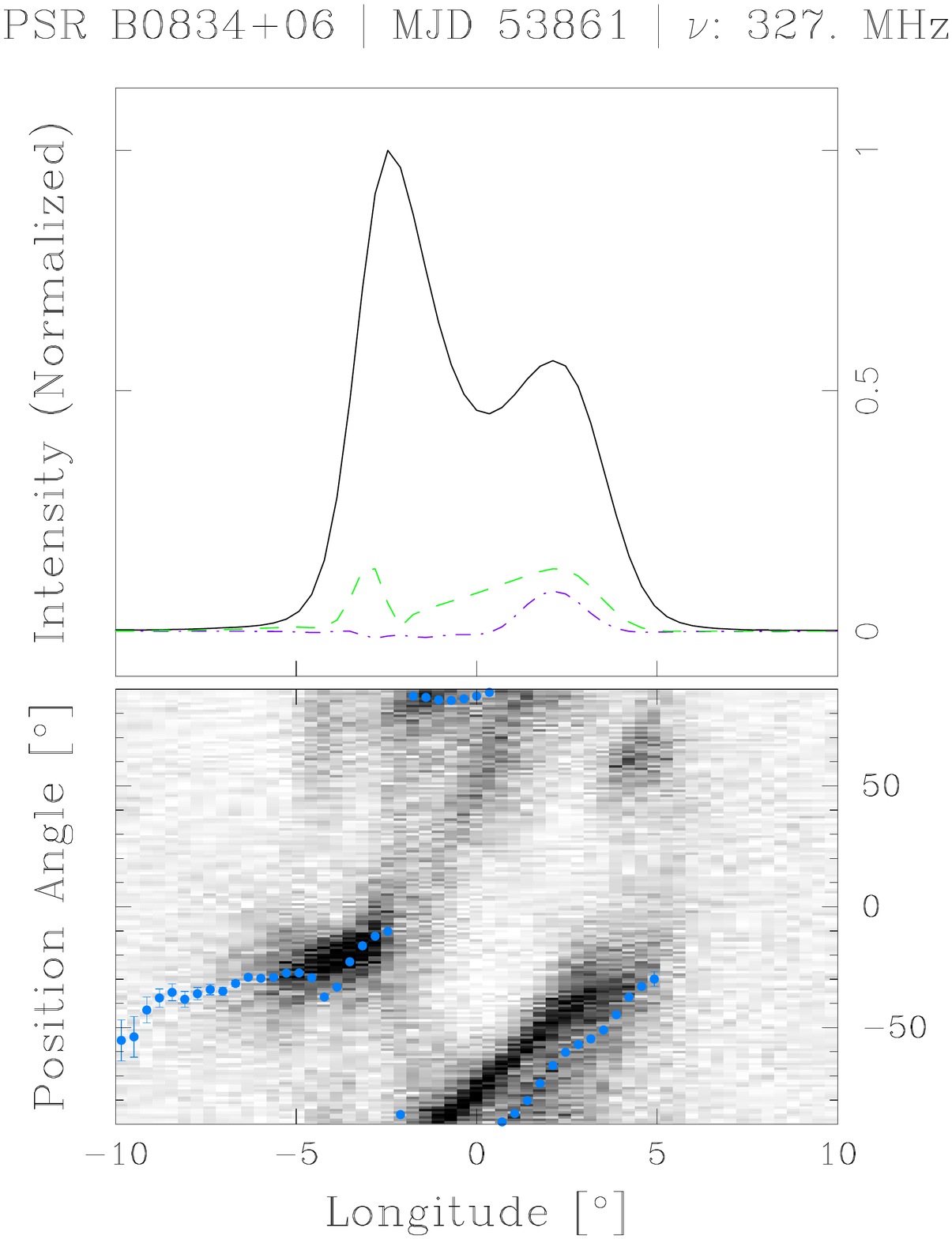} &
\includegraphics[page=1,width=\linewidth]{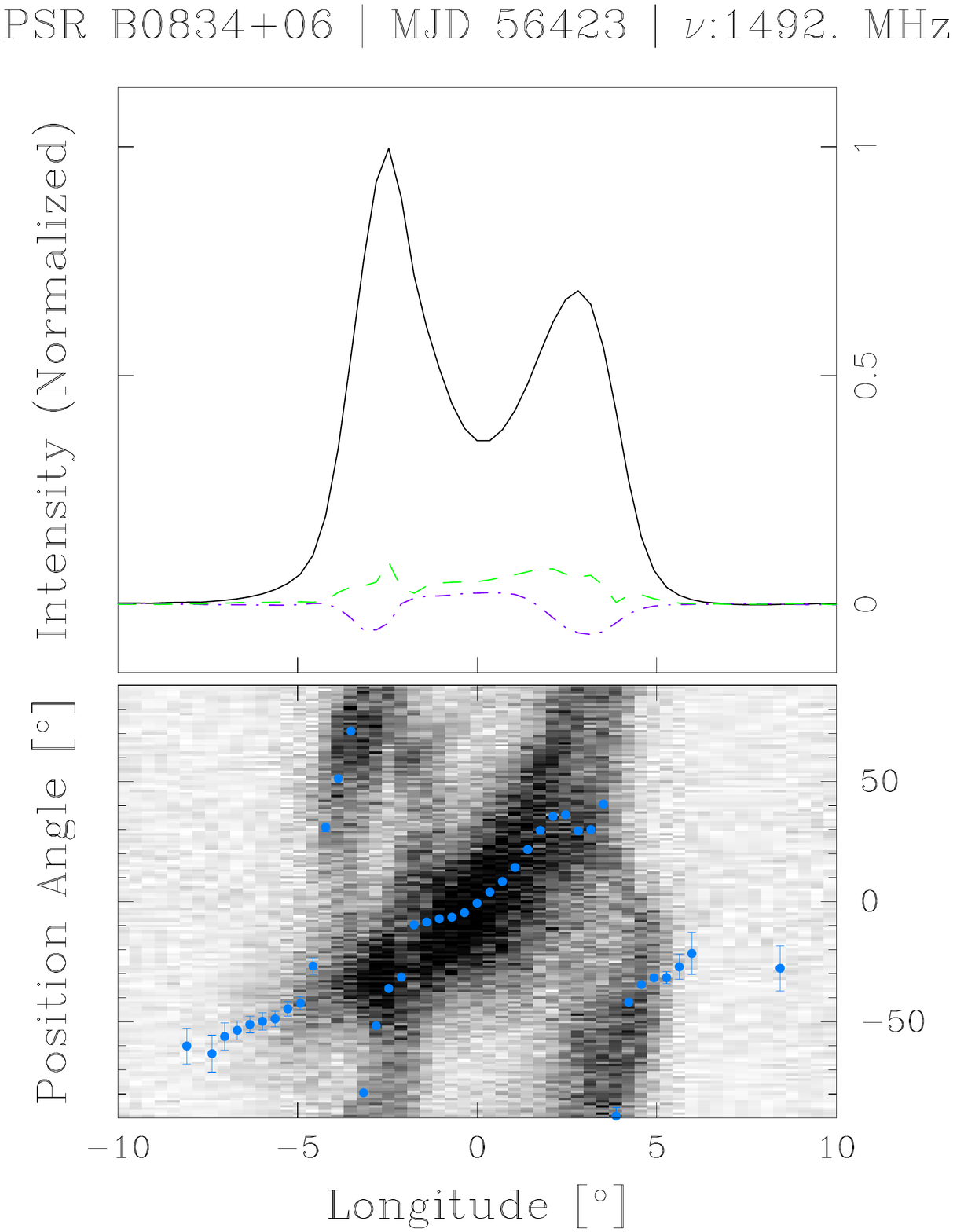} &
\includegraphics[page=1,width=\linewidth]{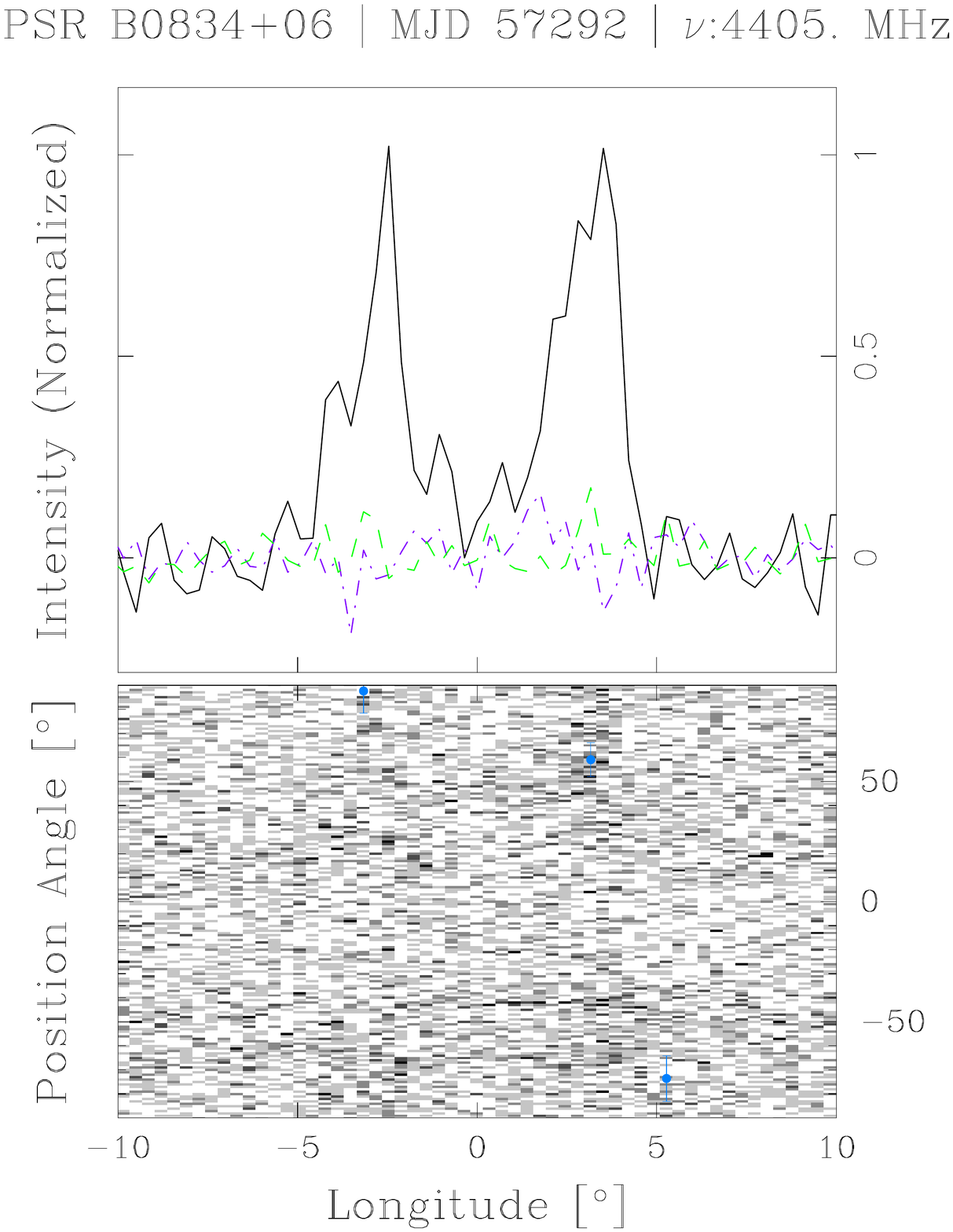} \\ \toprule
\includegraphics[page=1,width=\linewidth]{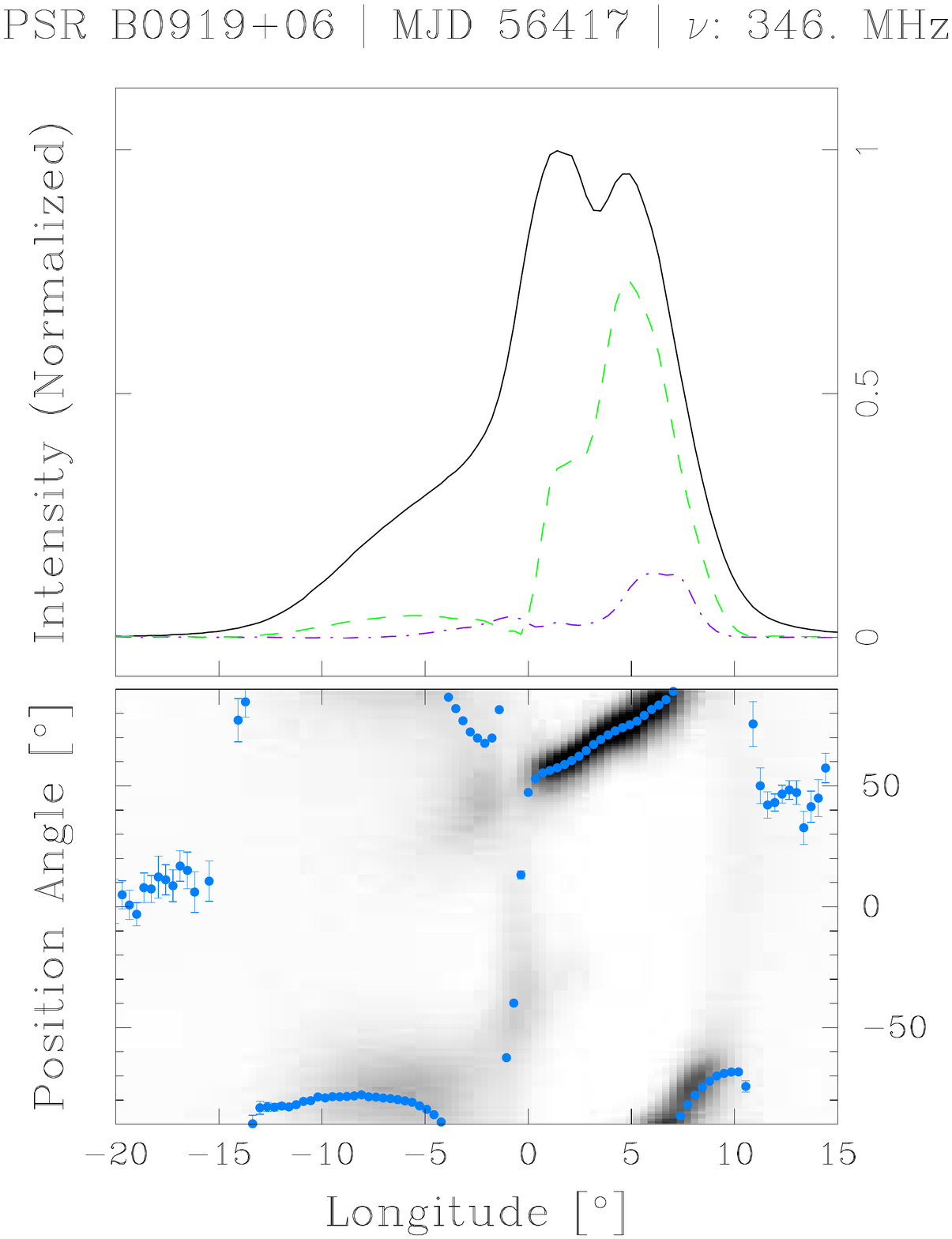} &
\includegraphics[page=1,width=\linewidth]{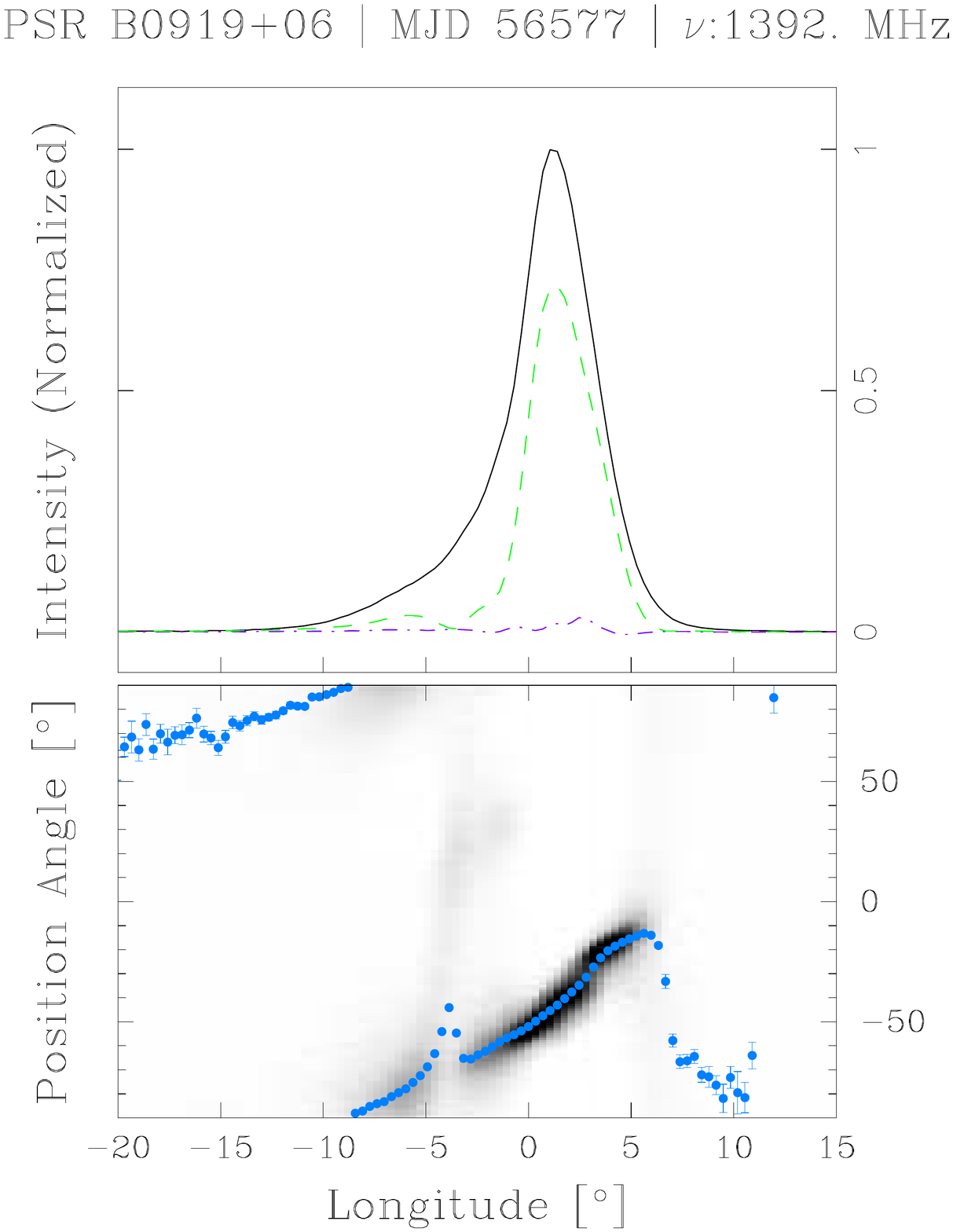} &
\includegraphics[page=1,width=\linewidth]{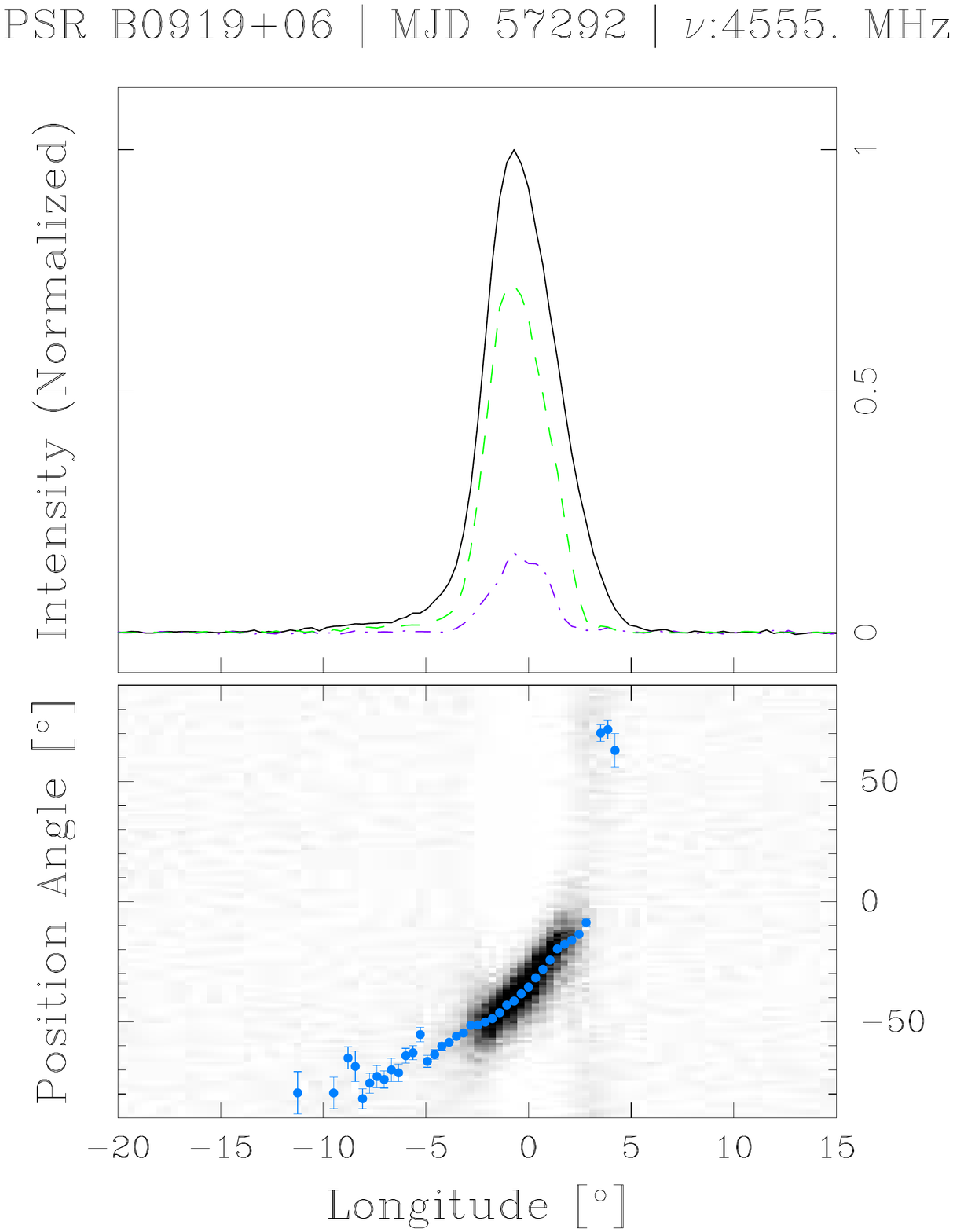} \\ 
     \bottomrule
   \end{tabularx} 
\caption{Average profiles of PSRs B0823+26, B0834+06, and B0919+06.}
 \end{figure*}
\vspace{1cm}

\begin{figure*} 
 \begin{tabularx}{\textwidth}{YYY}
    \multicolumn{3}{c}{} \\ \toprule
\includegraphics[page=1,width=\linewidth]{B0950+08pa_each_freq1_58206} &
\includegraphics[page=1,width=\linewidth]{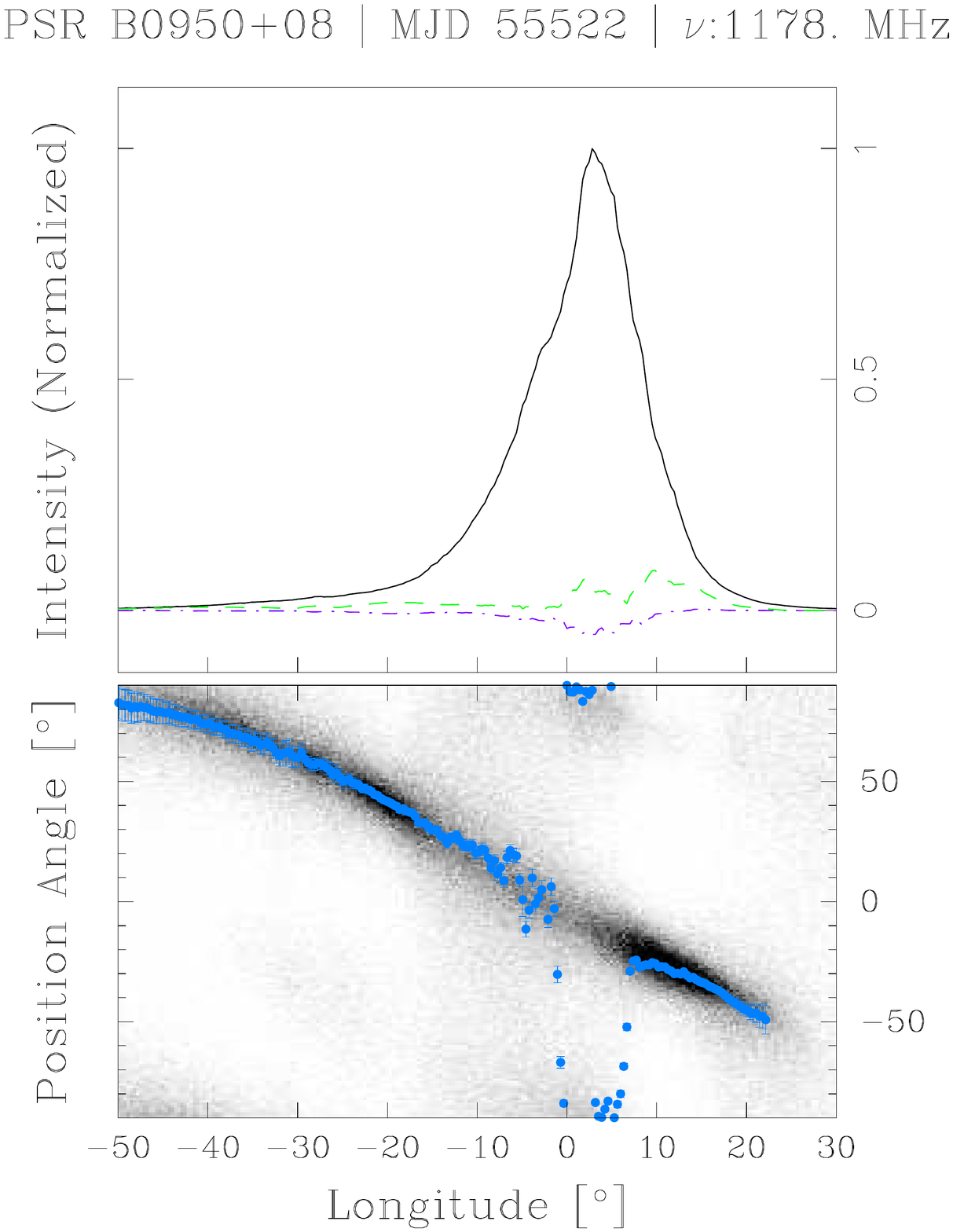} &
\includegraphics[page=1,width=\linewidth]{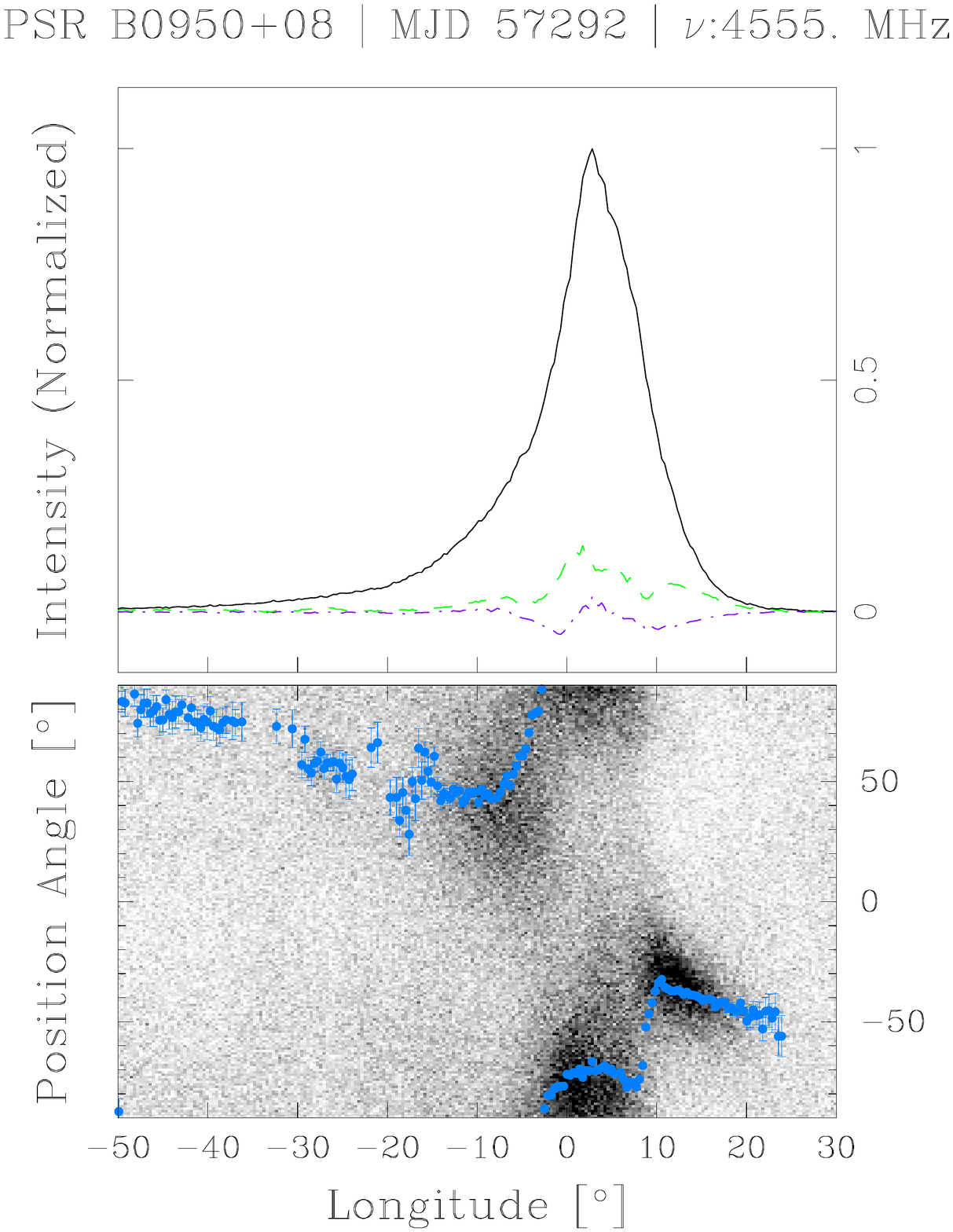} \\ \toprule

\includegraphics[page=1,width=\linewidth]{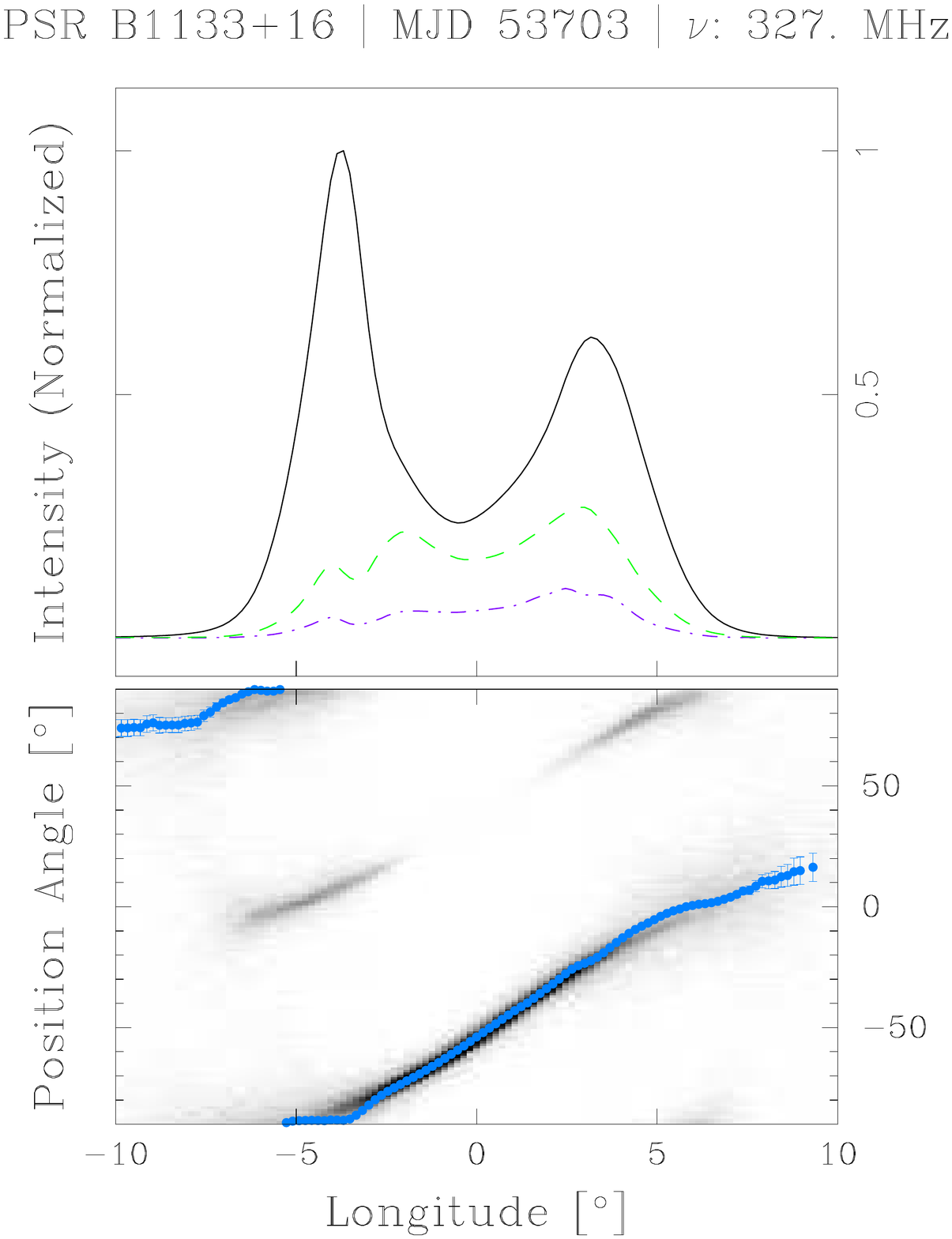} &
\includegraphics[page=1,width=\linewidth]{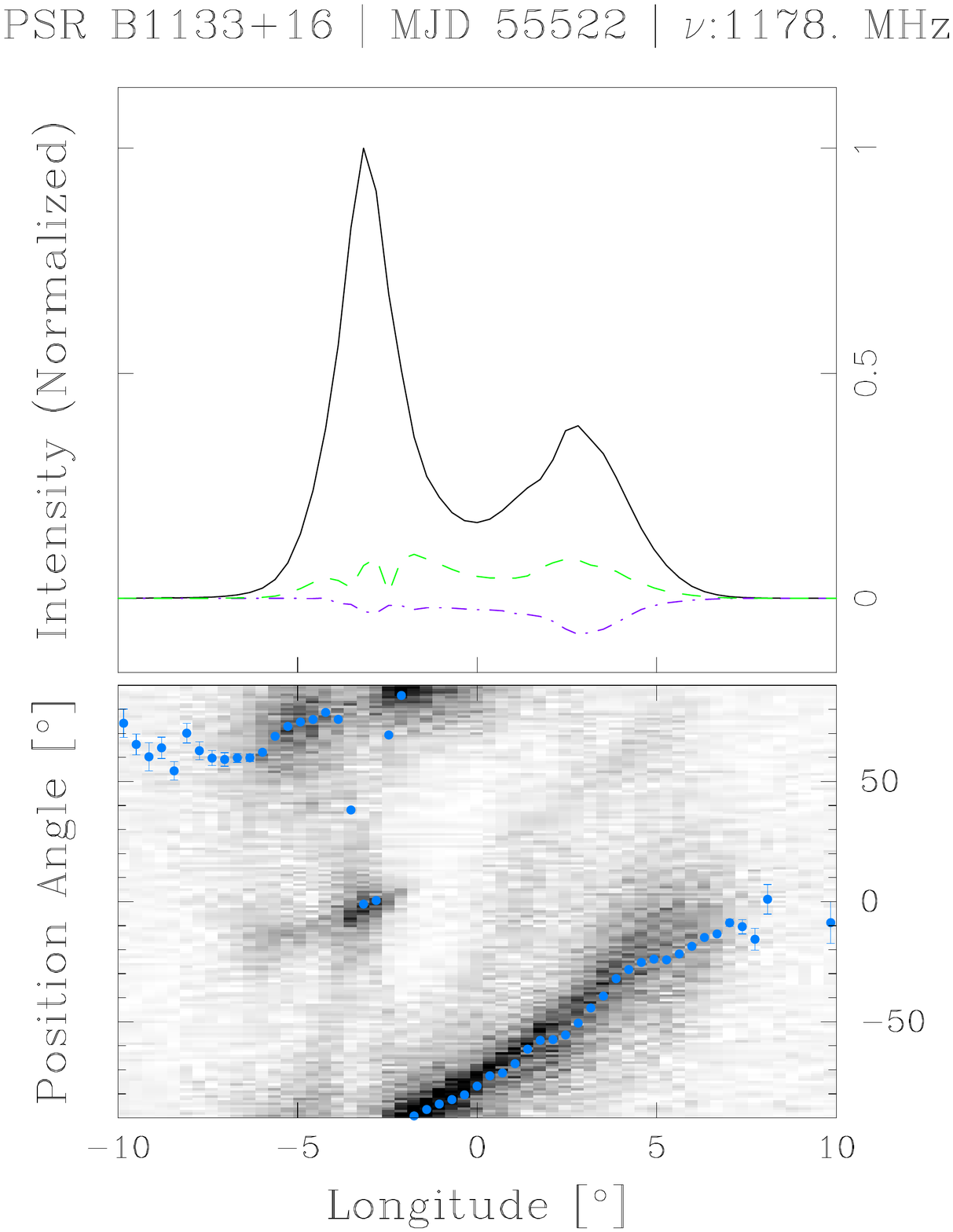} &
\includegraphics[page=1,width=\linewidth]{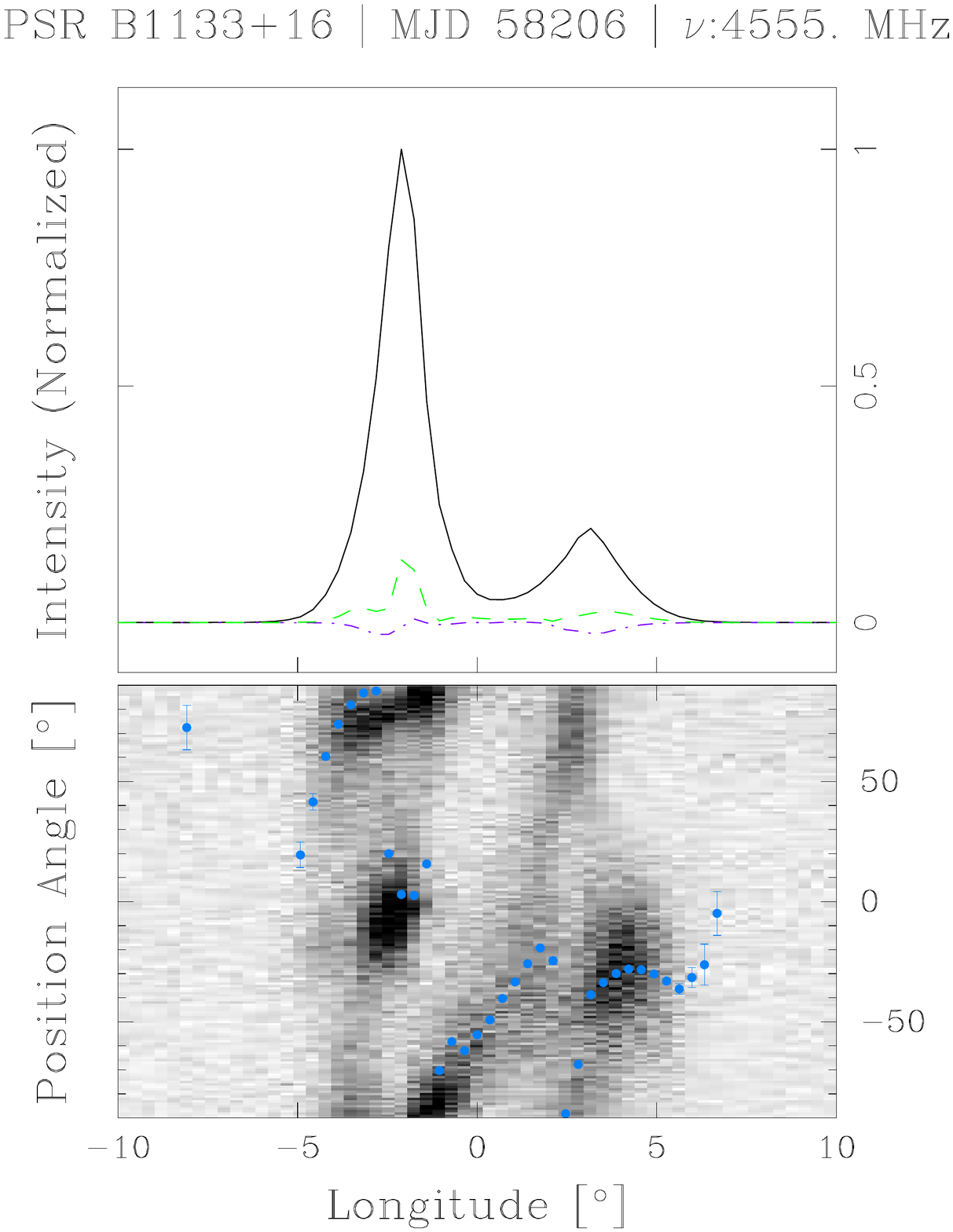} \\ \toprule
\includegraphics[page=1,width=\linewidth]{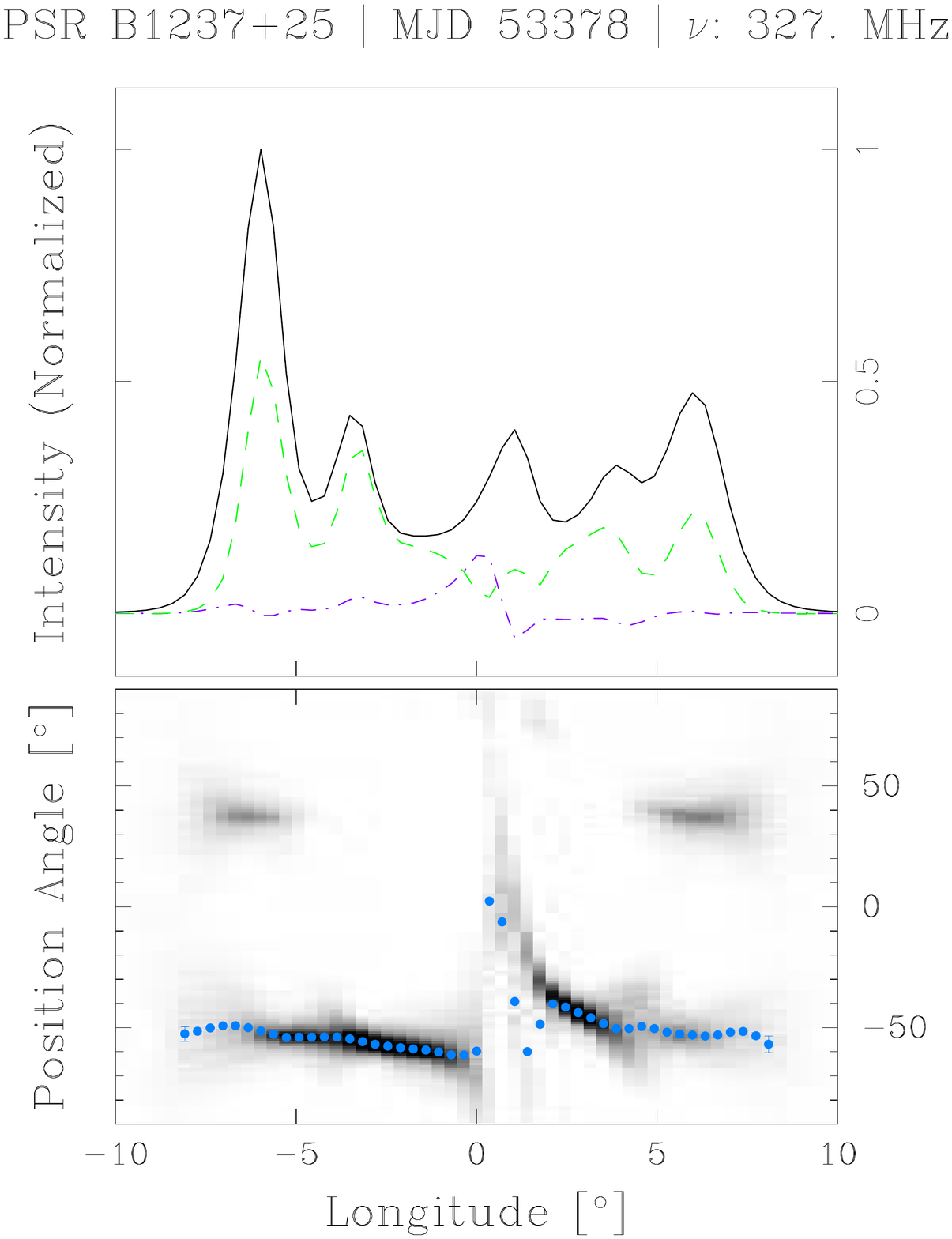} &
\includegraphics[page=1,width=\linewidth]{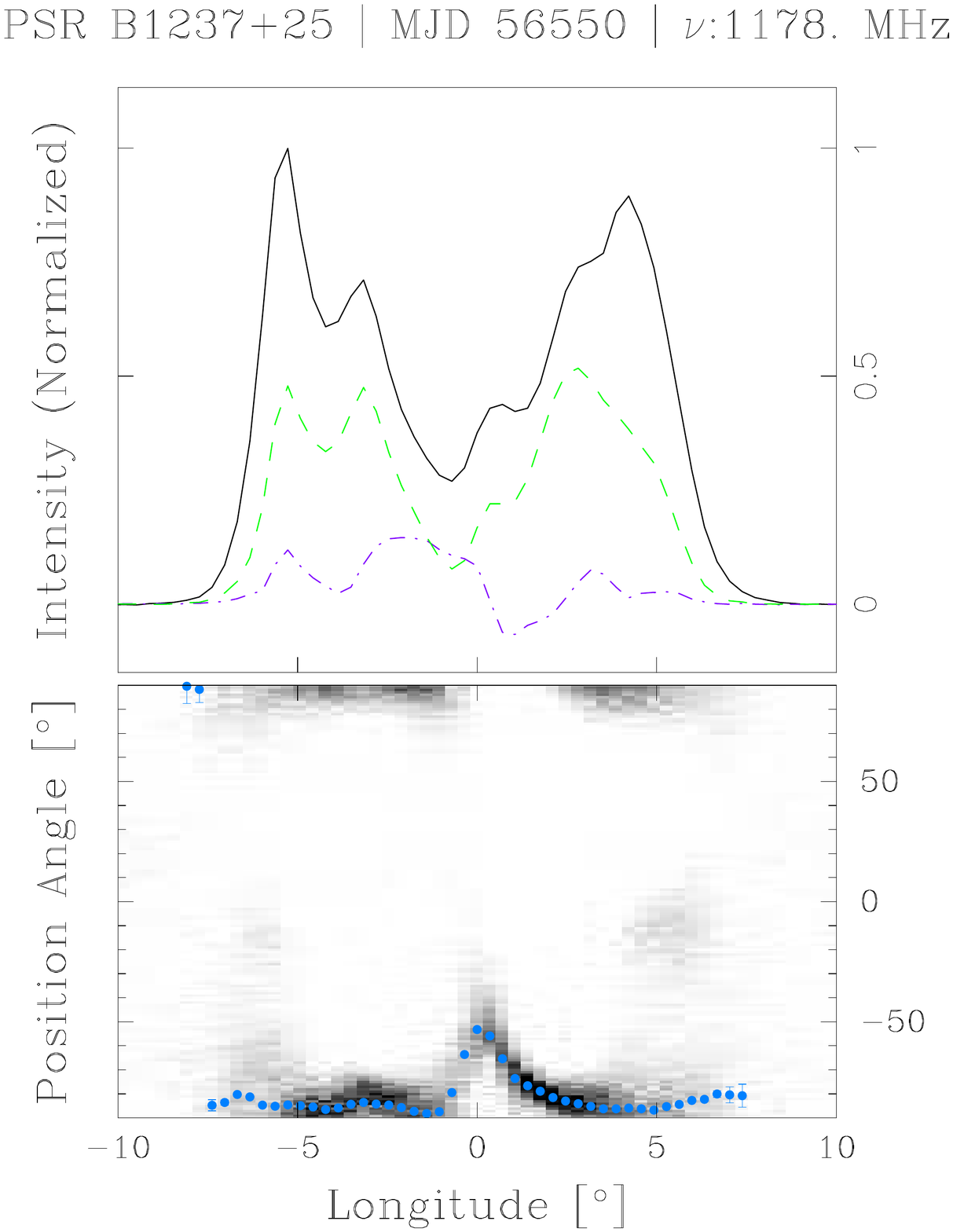} &
\includegraphics[page=1,width=\linewidth]{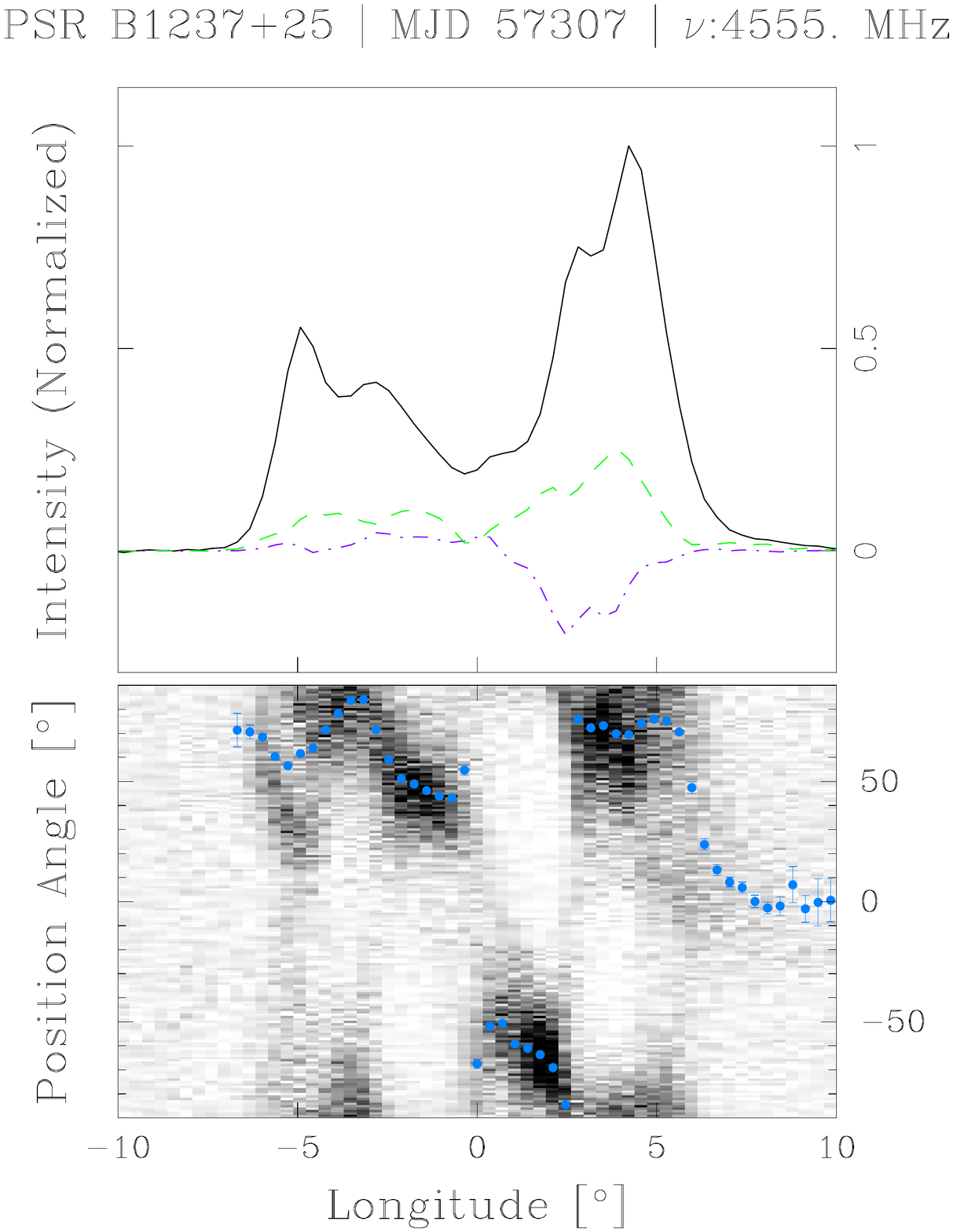} \\ 

     \bottomrule
   \end{tabularx} 
\caption{Average profiles of PSRs B0950+08, B1133+16, and B1237+25.}
 \end{figure*}
\vspace{1cm}

   \begin{figure*} 
 \begin{tabularx}{\textwidth}{YYY}
    \multicolumn{3}{c}{} \\ \toprule
\includegraphics[page=1,width=\linewidth]{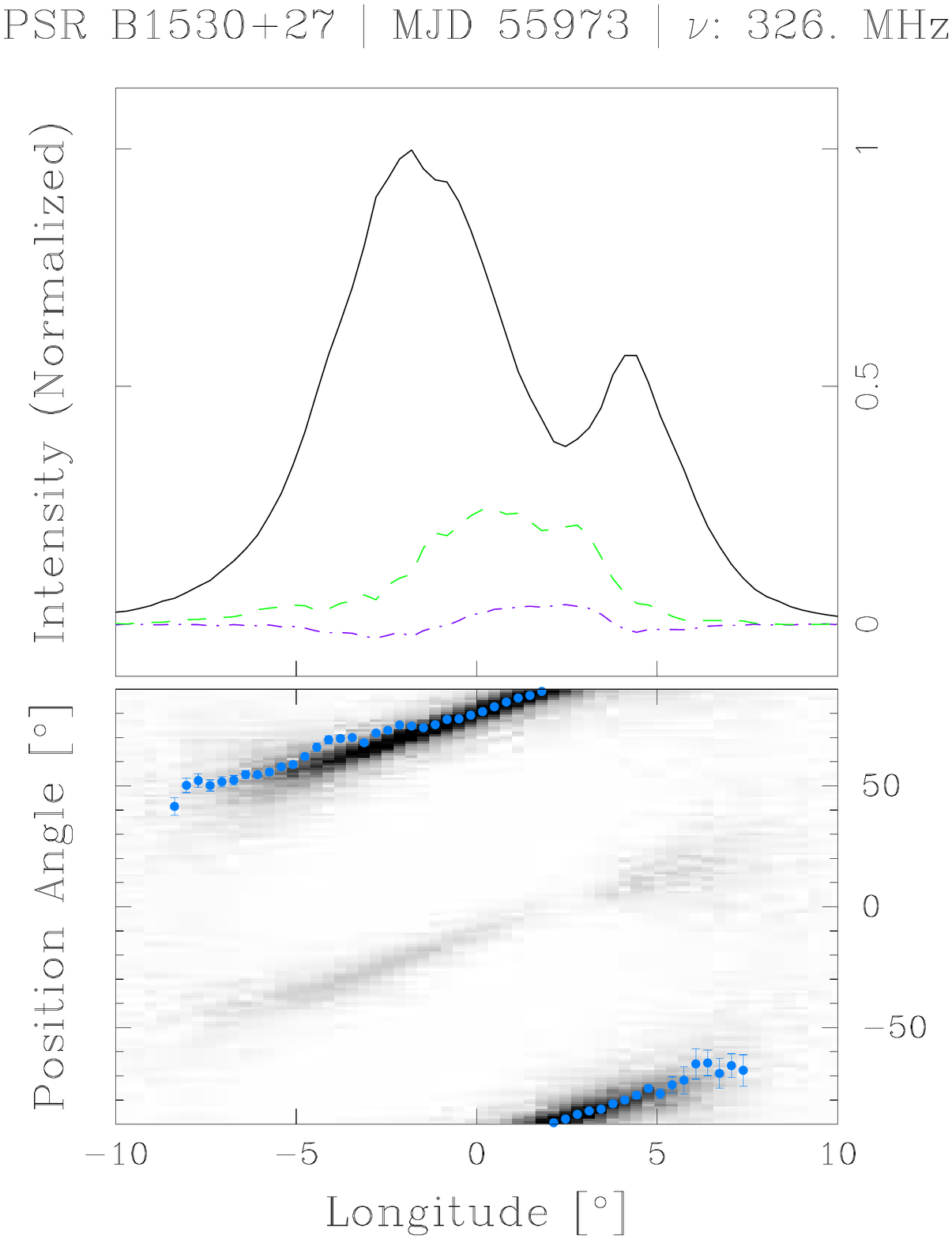} &
\includegraphics[page=1,width=\linewidth]{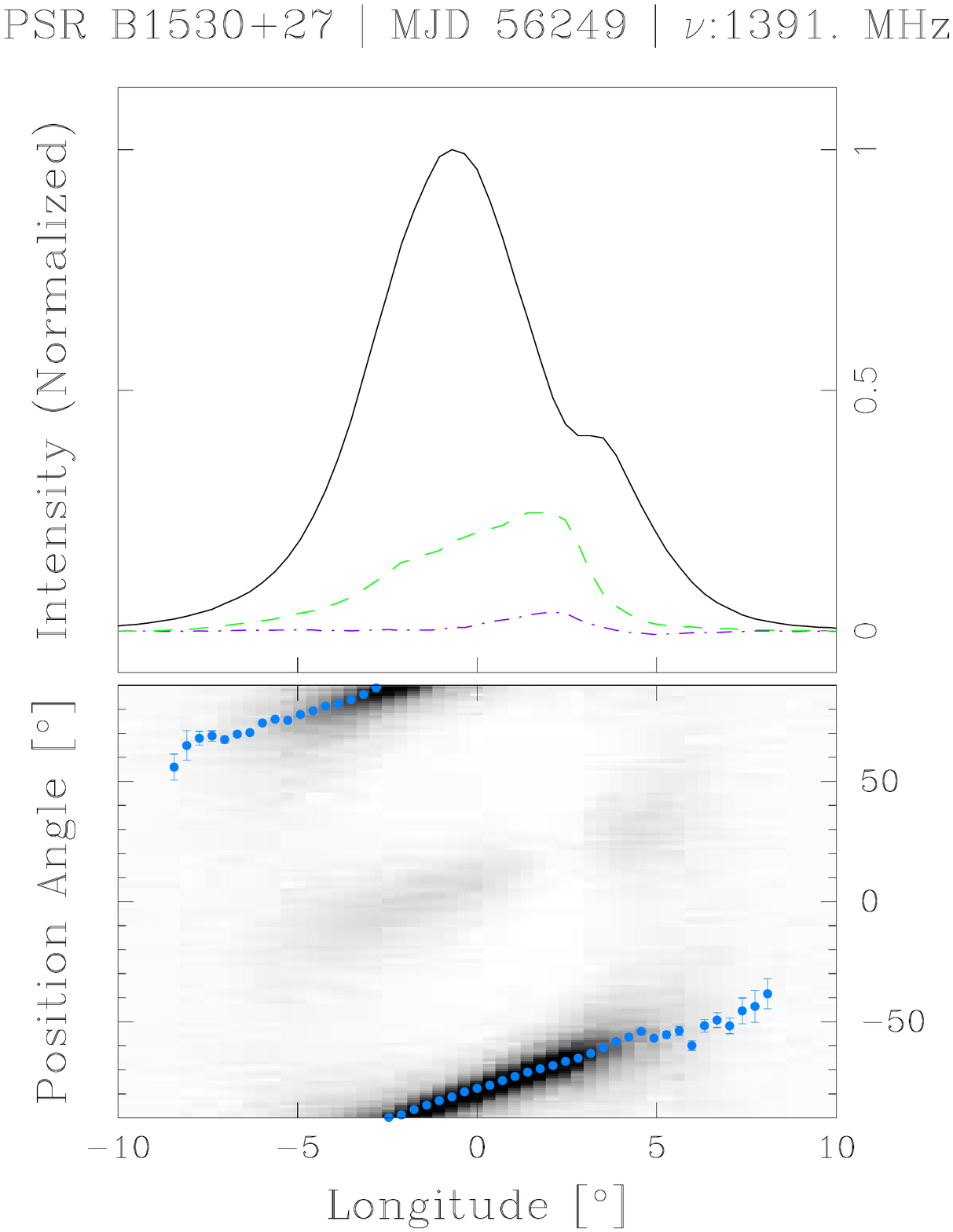} &
\includegraphics[page=1,width=\linewidth]{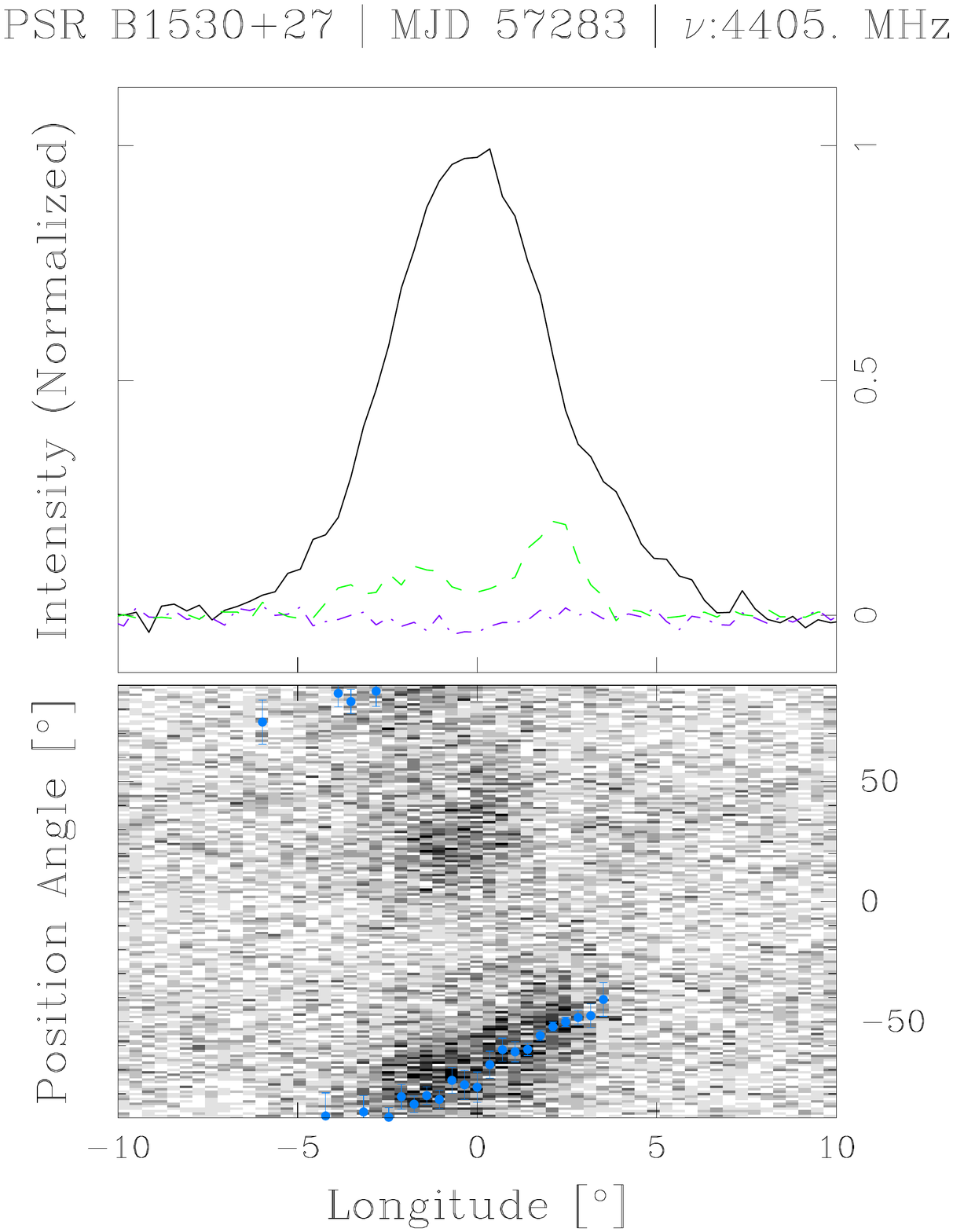} \\ \toprule
\includegraphics[page=1,width=\linewidth]{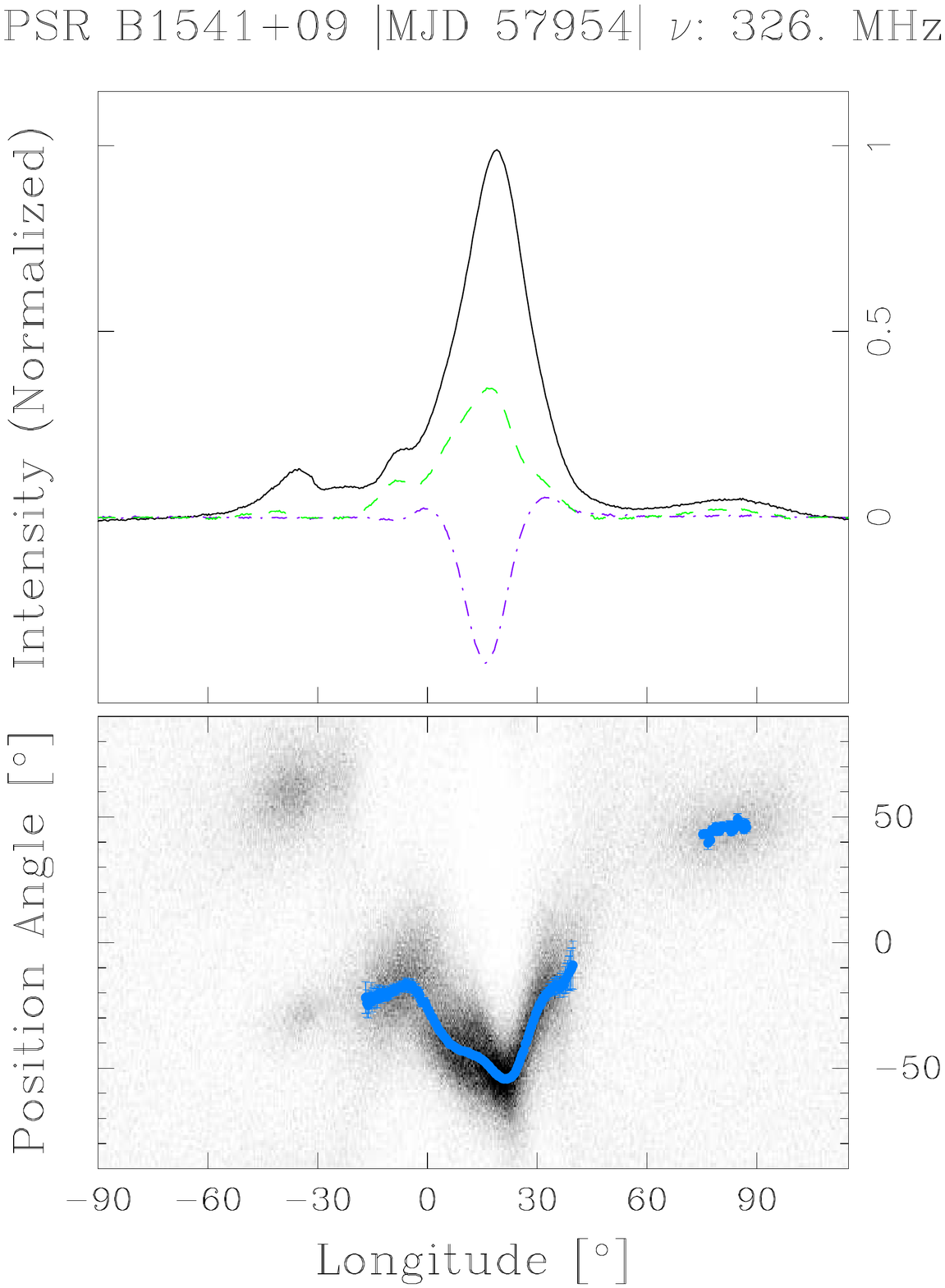} &
\includegraphics[page=1,width=\linewidth]{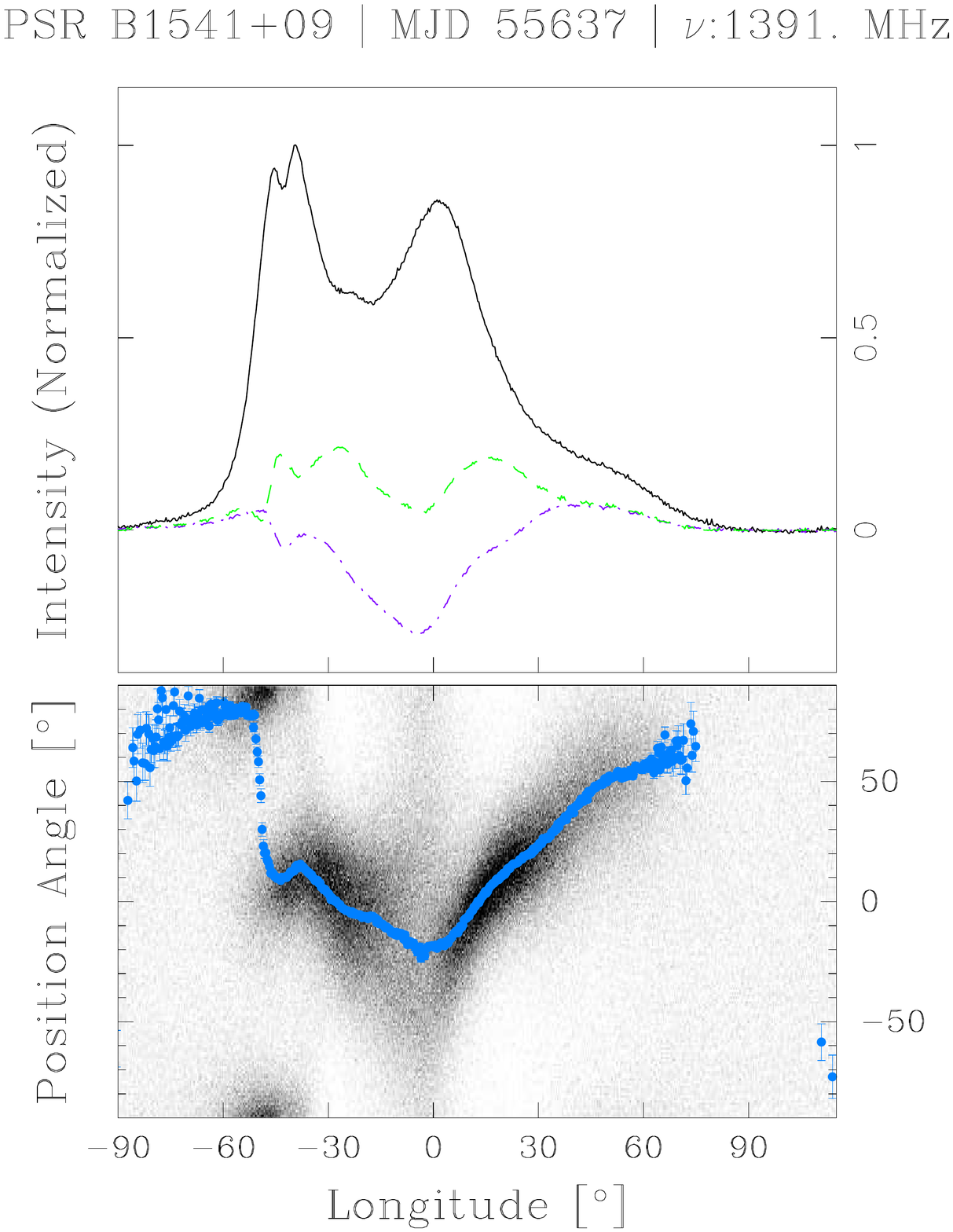} &
\includegraphics[page=1,width=\linewidth]{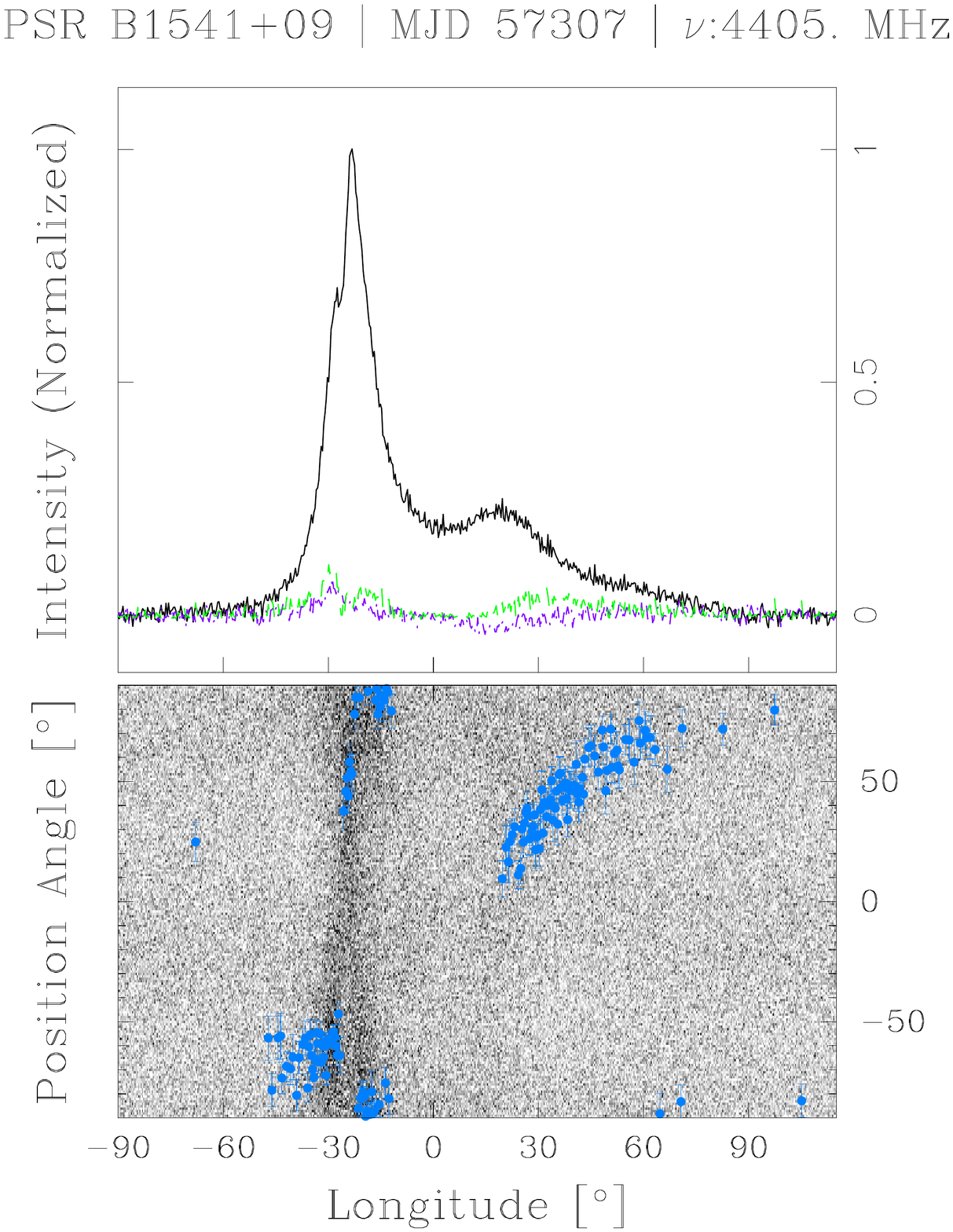} \\ \toprule
\includegraphics[page=1,width=\linewidth]{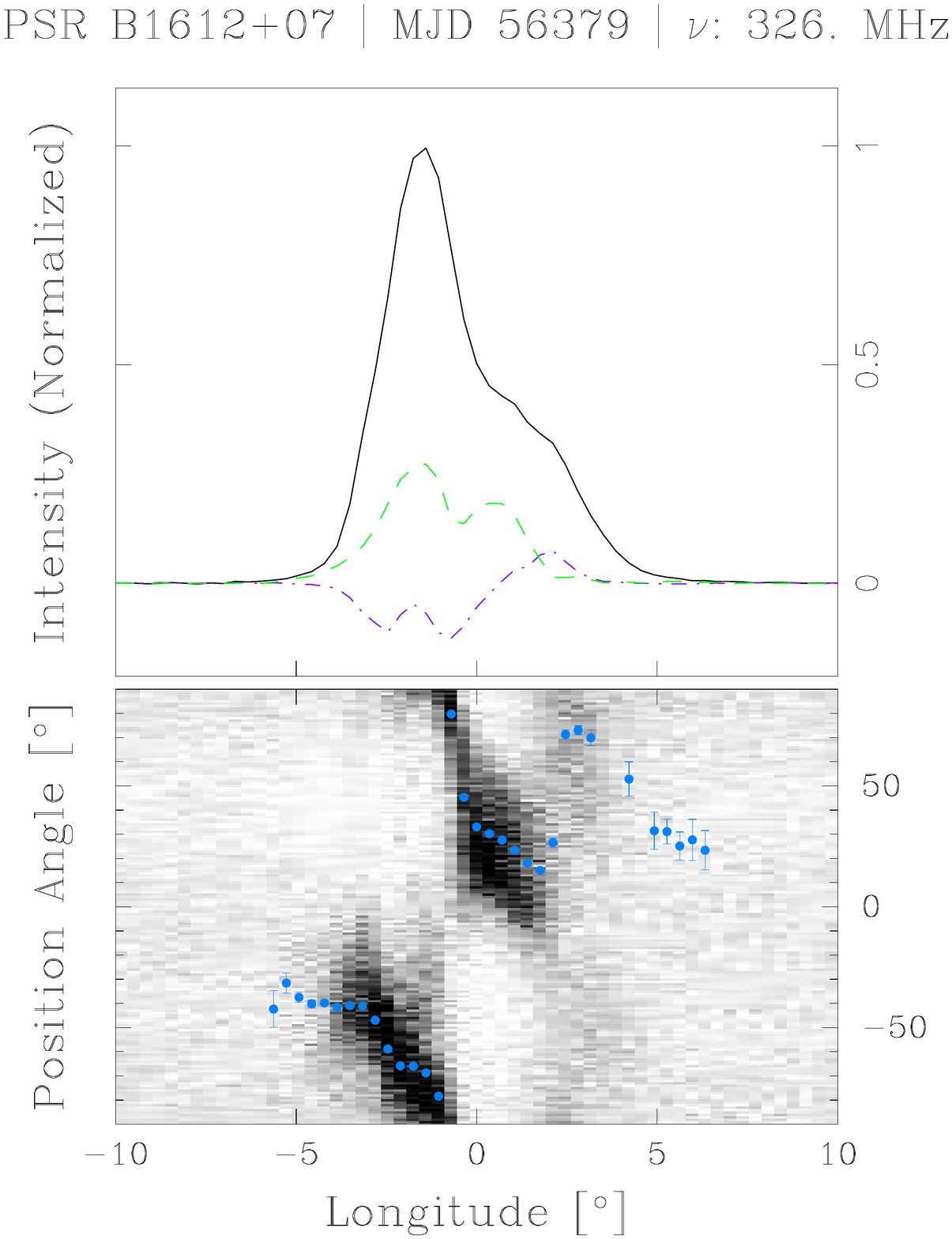} &
\includegraphics[page=1,width=\linewidth]{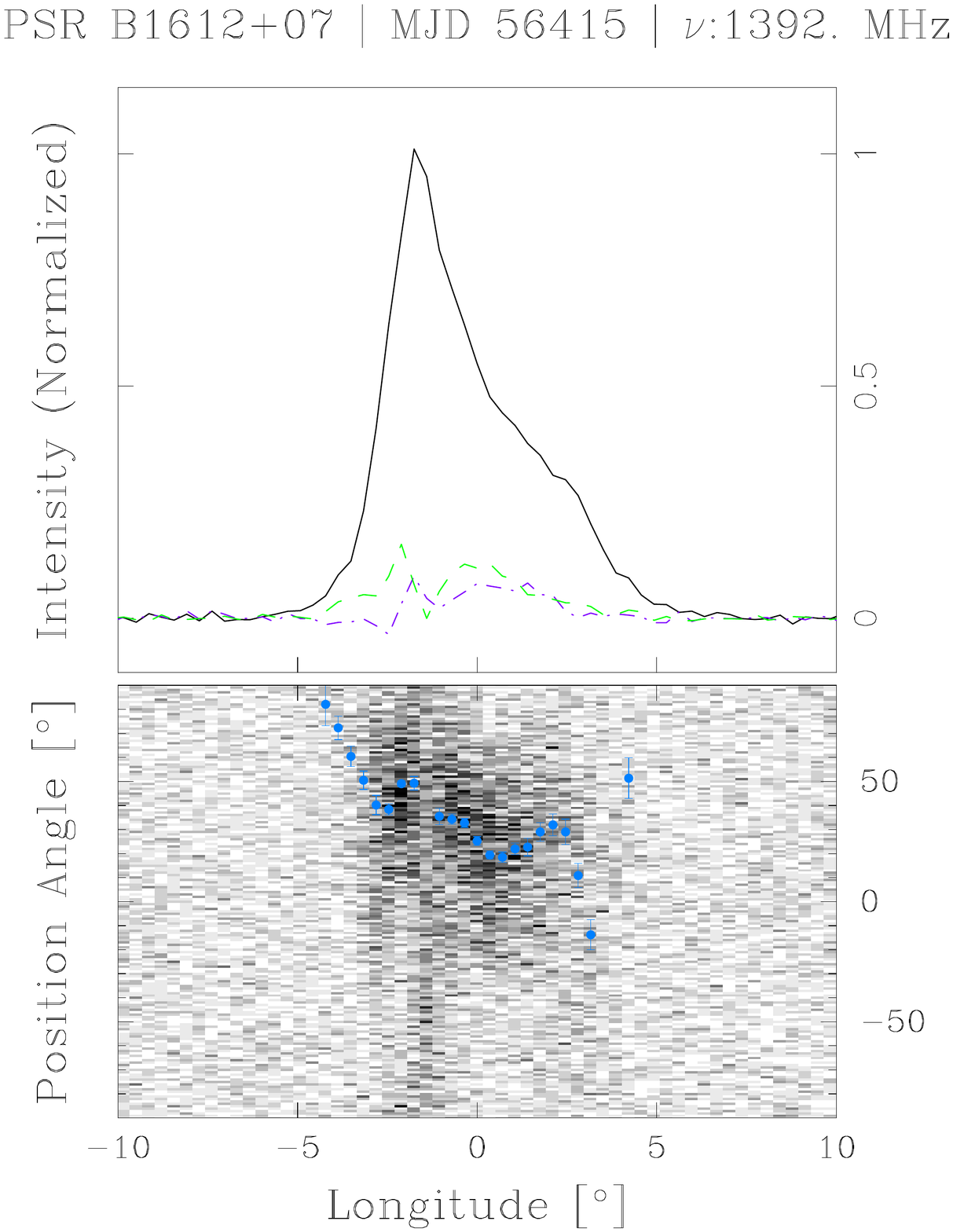} &
\includegraphics[page=1,width=\linewidth]{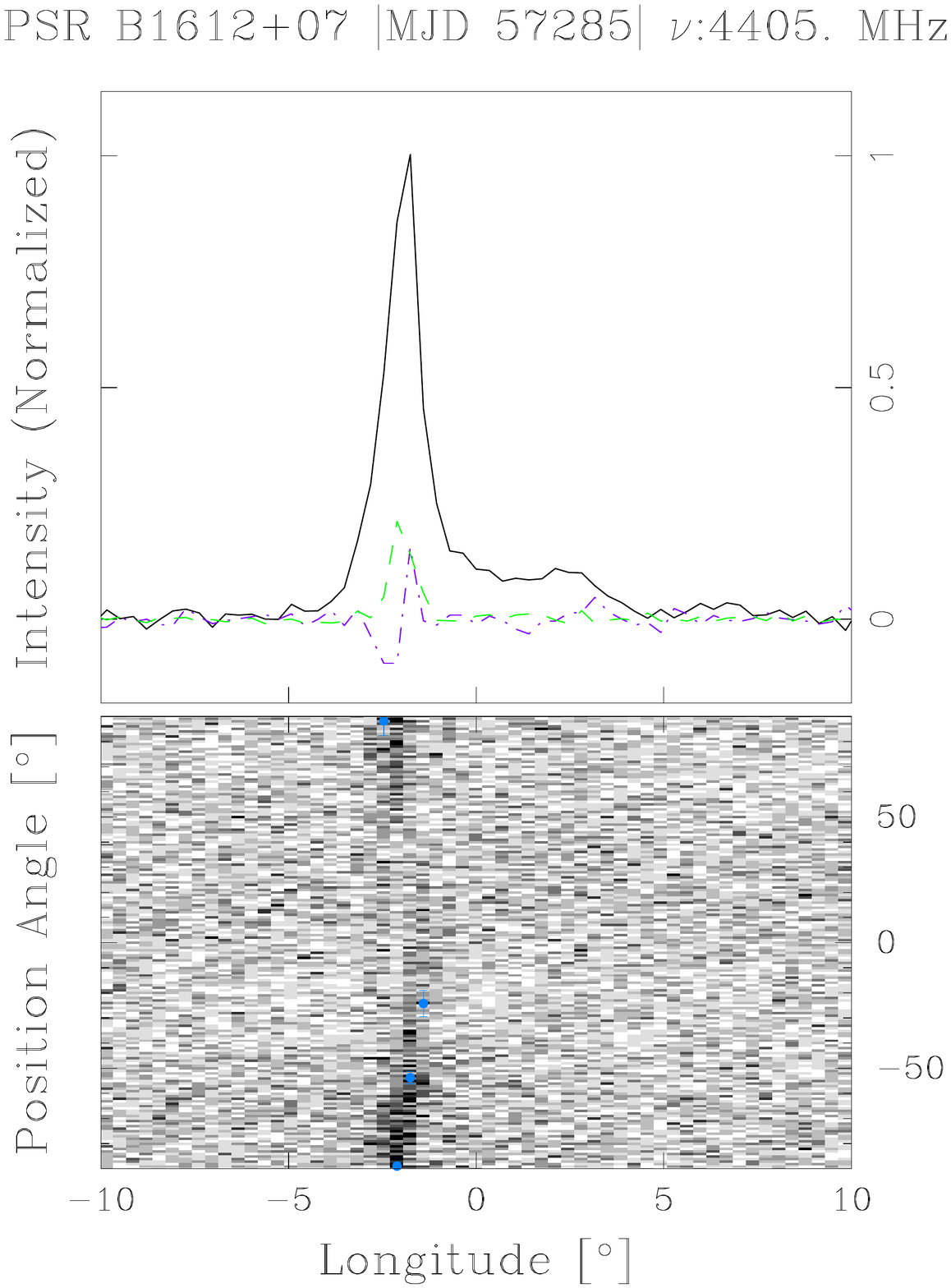} \\ 
     \bottomrule
   \end{tabularx} 
\caption{Average profiles of PSRs B1530+27, B1541+09, and B1612+07.}
 \end{figure*}
\vspace{1cm}

   \begin{figure*} 
 \begin{tabularx}{\textwidth}{YYY}
    \multicolumn{3}{c}{} \\ \toprule
\includegraphics[page=1,width=\linewidth]{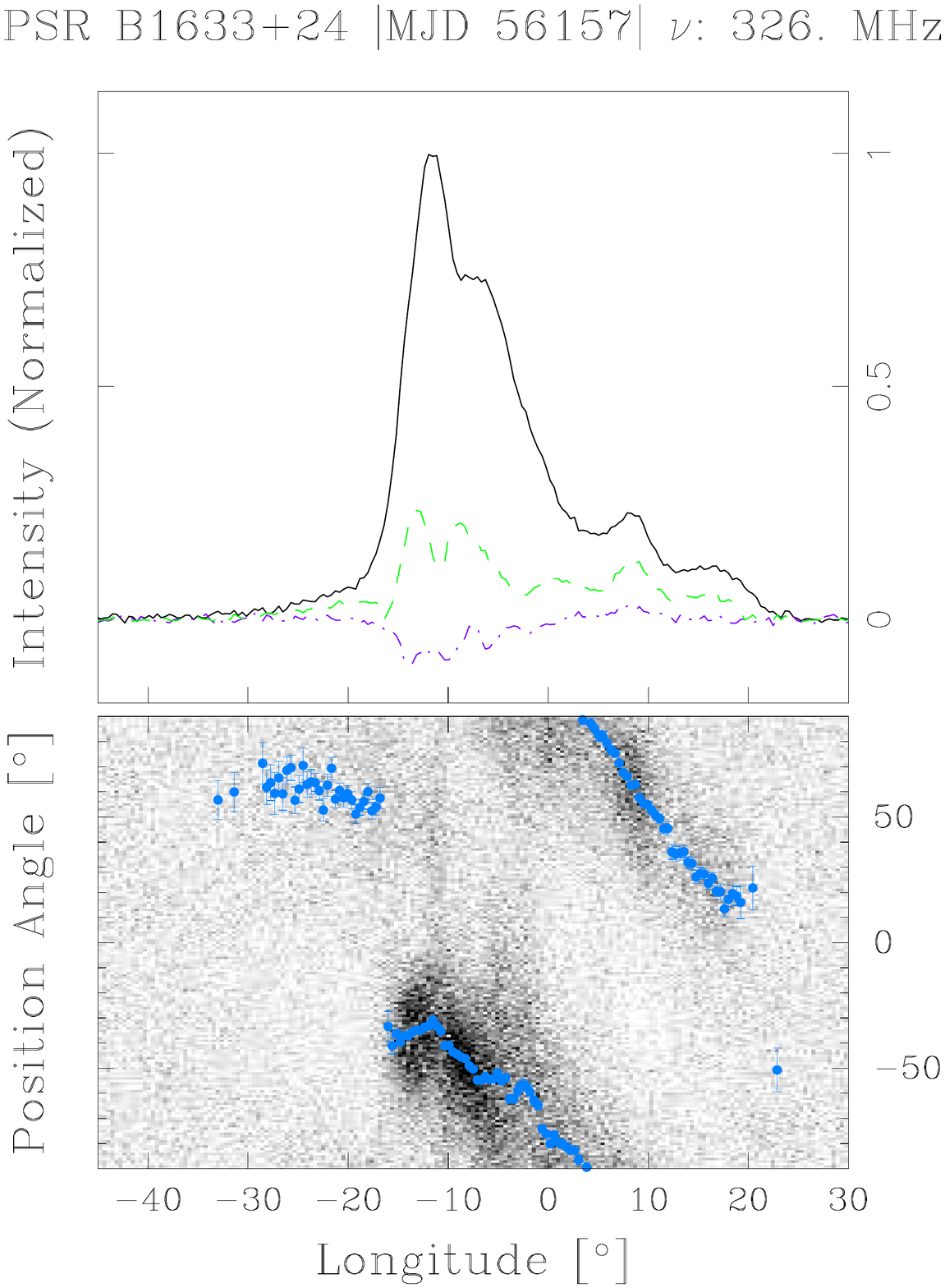} &
\includegraphics[page=1,width=\linewidth]{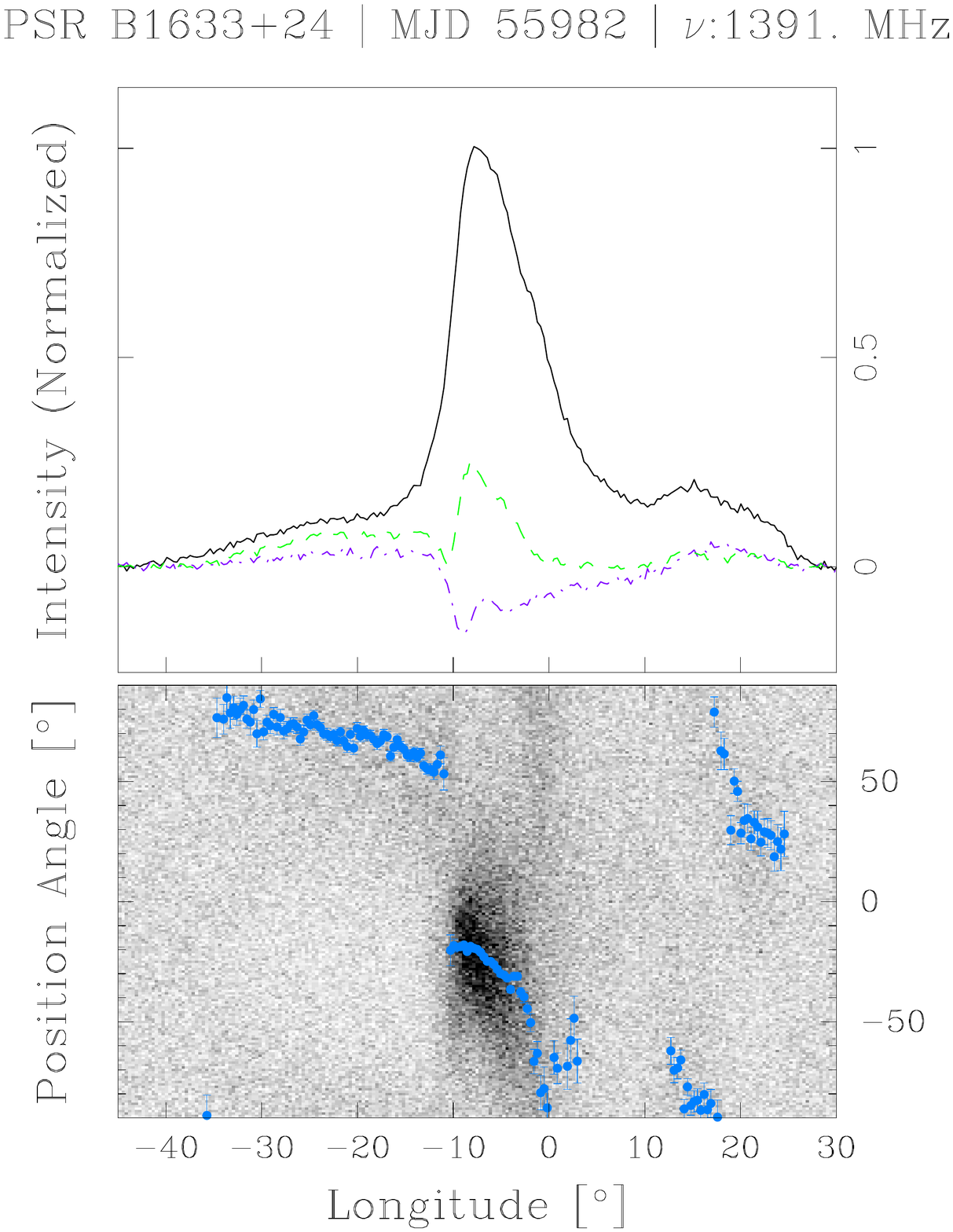} &
\includegraphics[page=1,width=\linewidth]{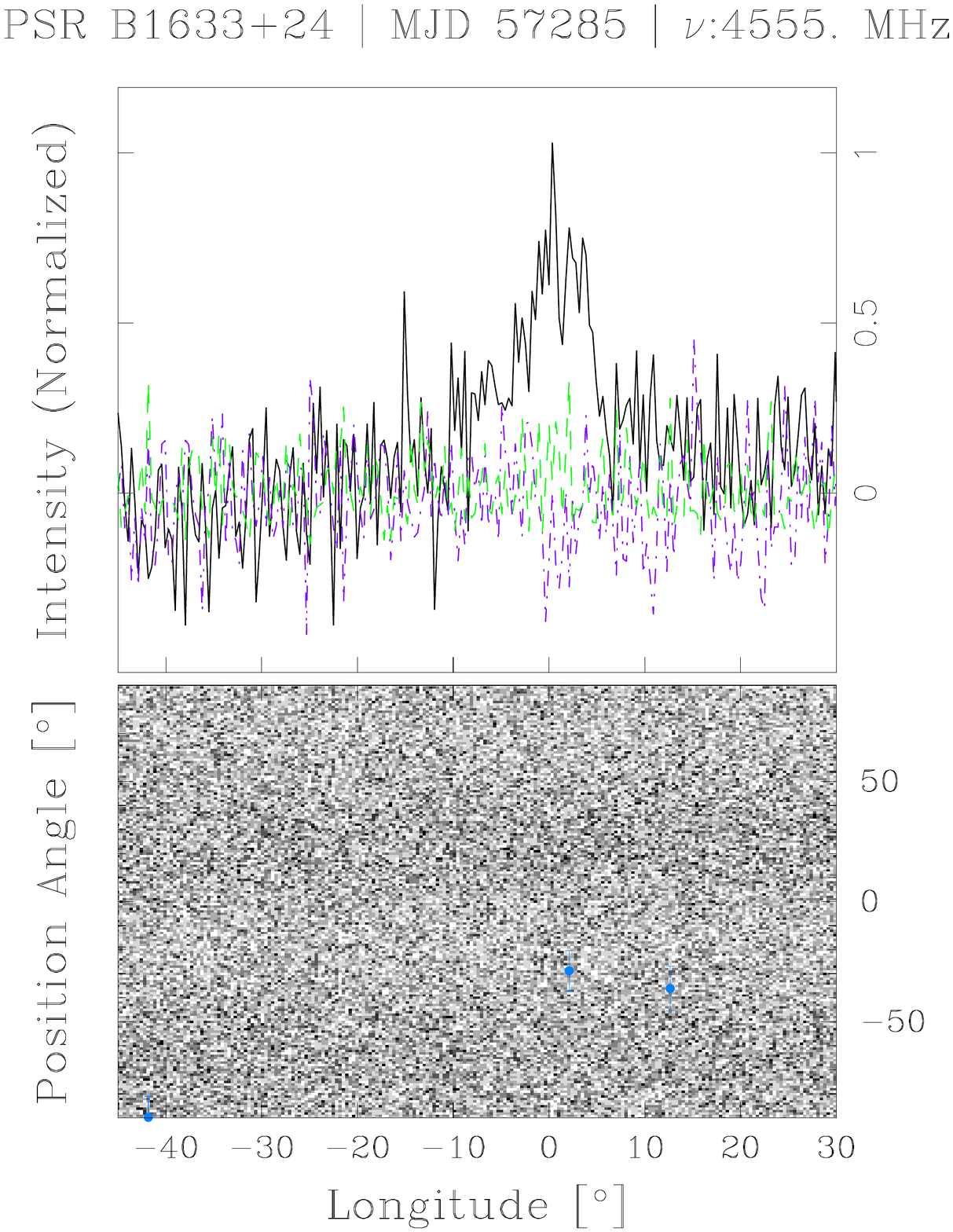} \\ \toprule

\includegraphics[page=1,width=\linewidth]{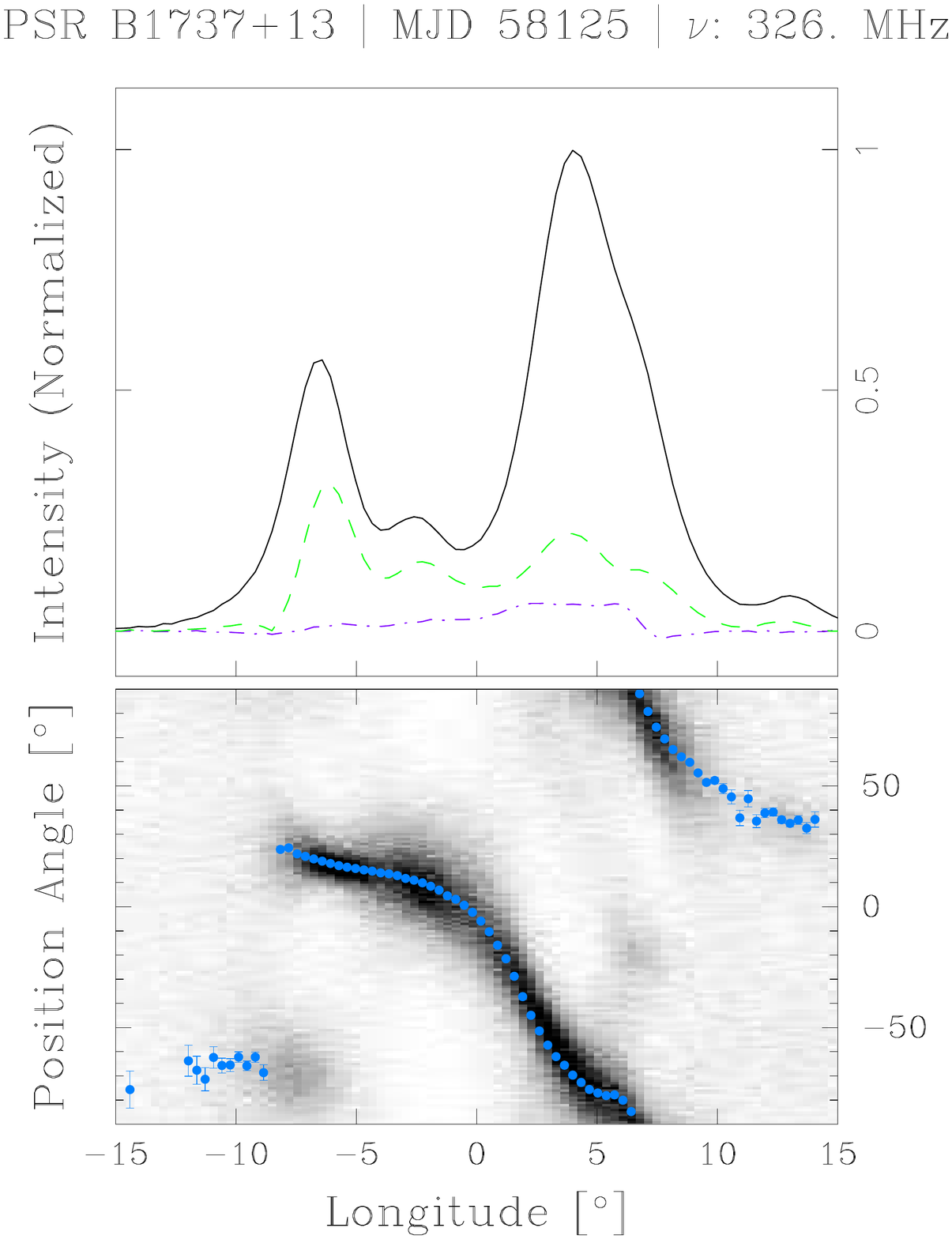} &
\includegraphics[page=1,width=\linewidth]{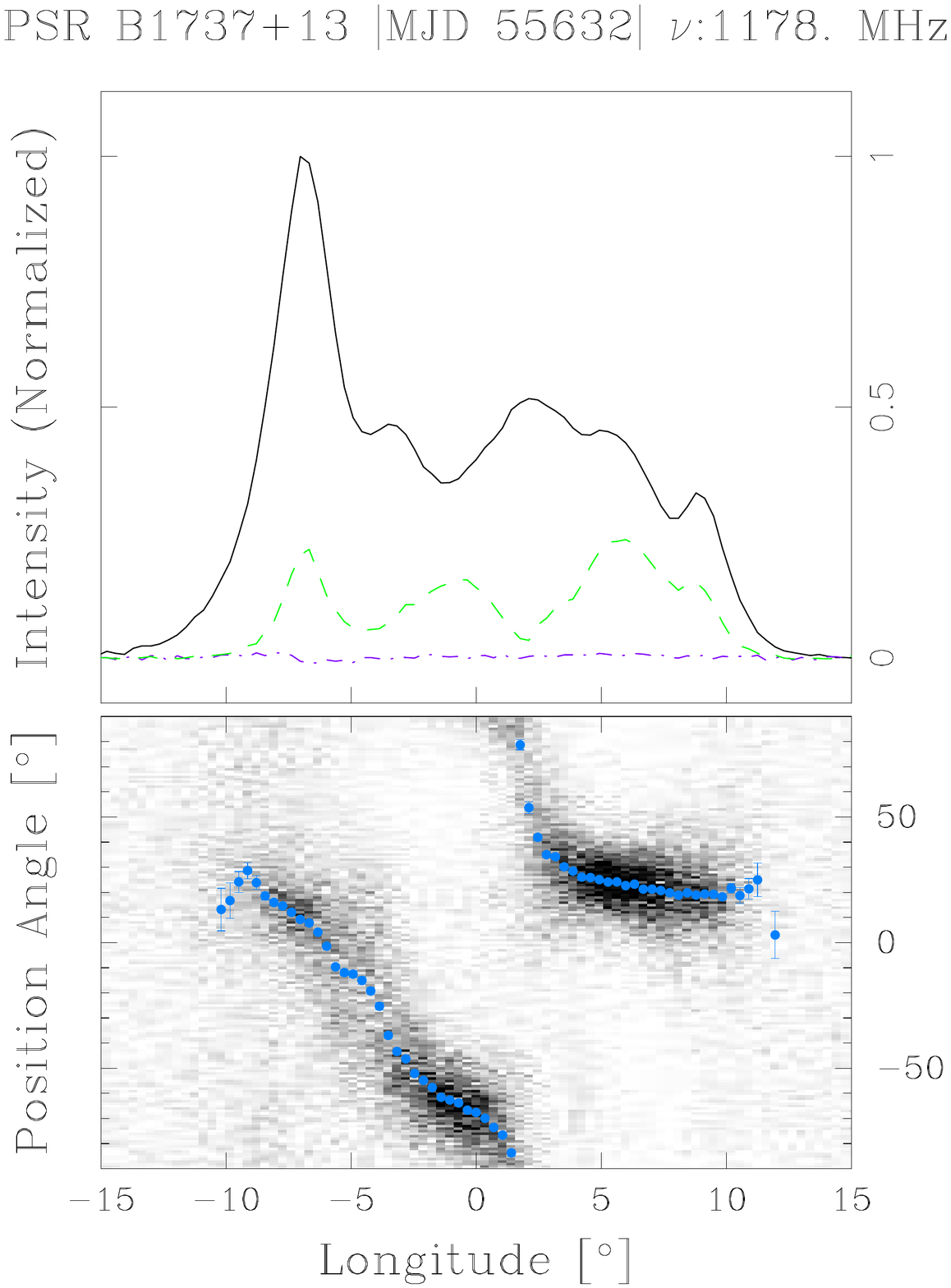} &
\includegraphics[page=1,width=\linewidth]{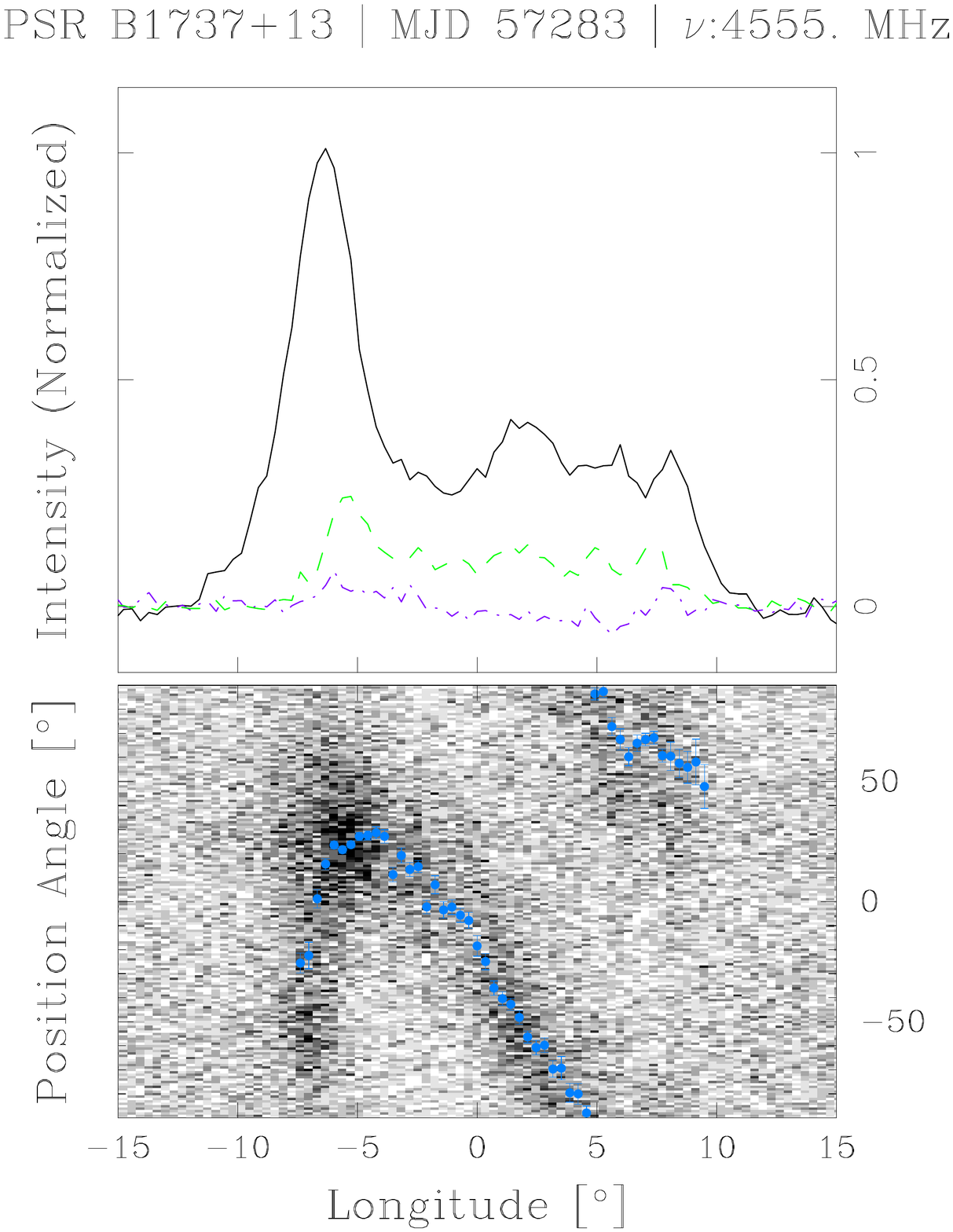} \\ \toprule

\includegraphics[page=1,width=\linewidth]{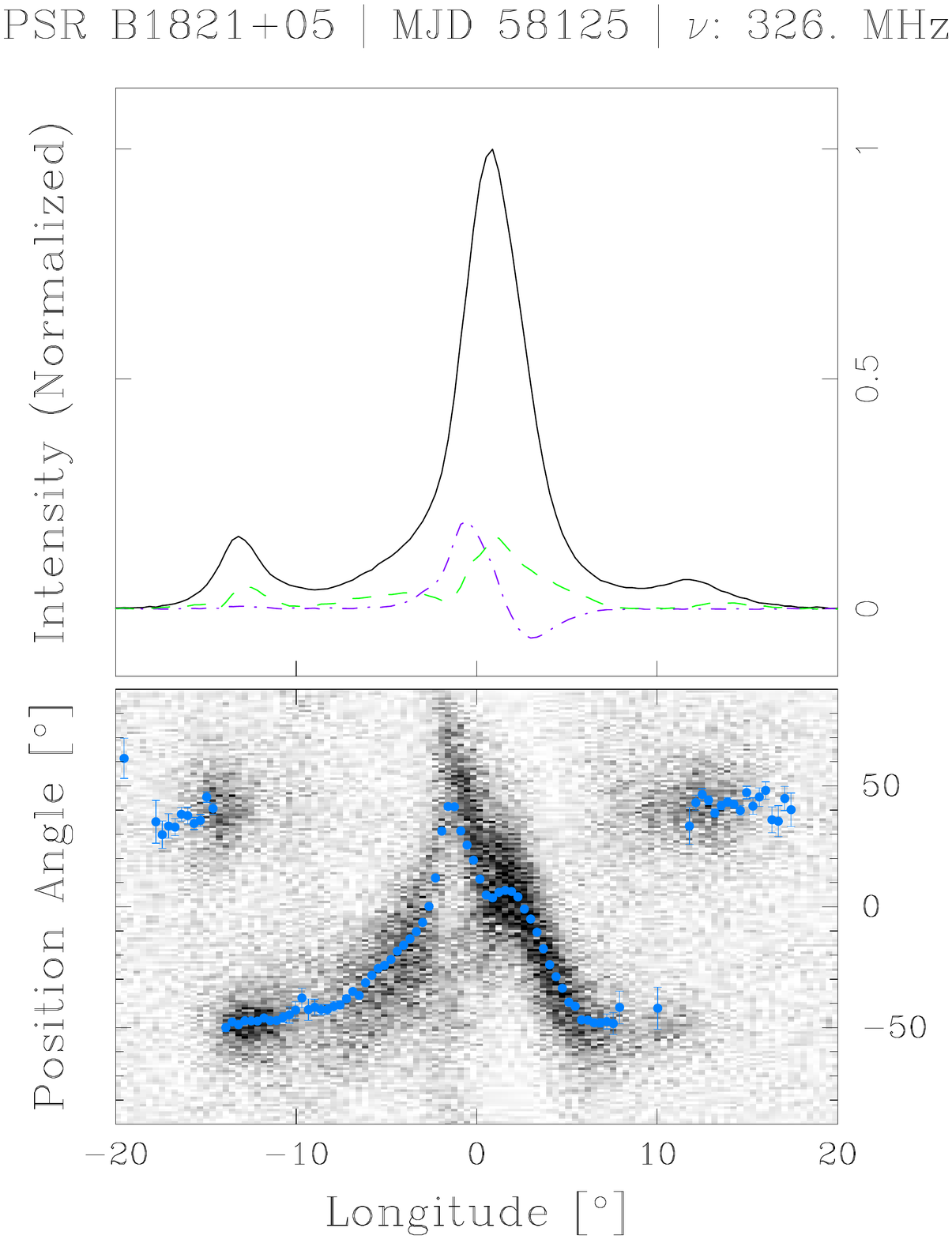} &
\includegraphics[page=1,width=\linewidth]{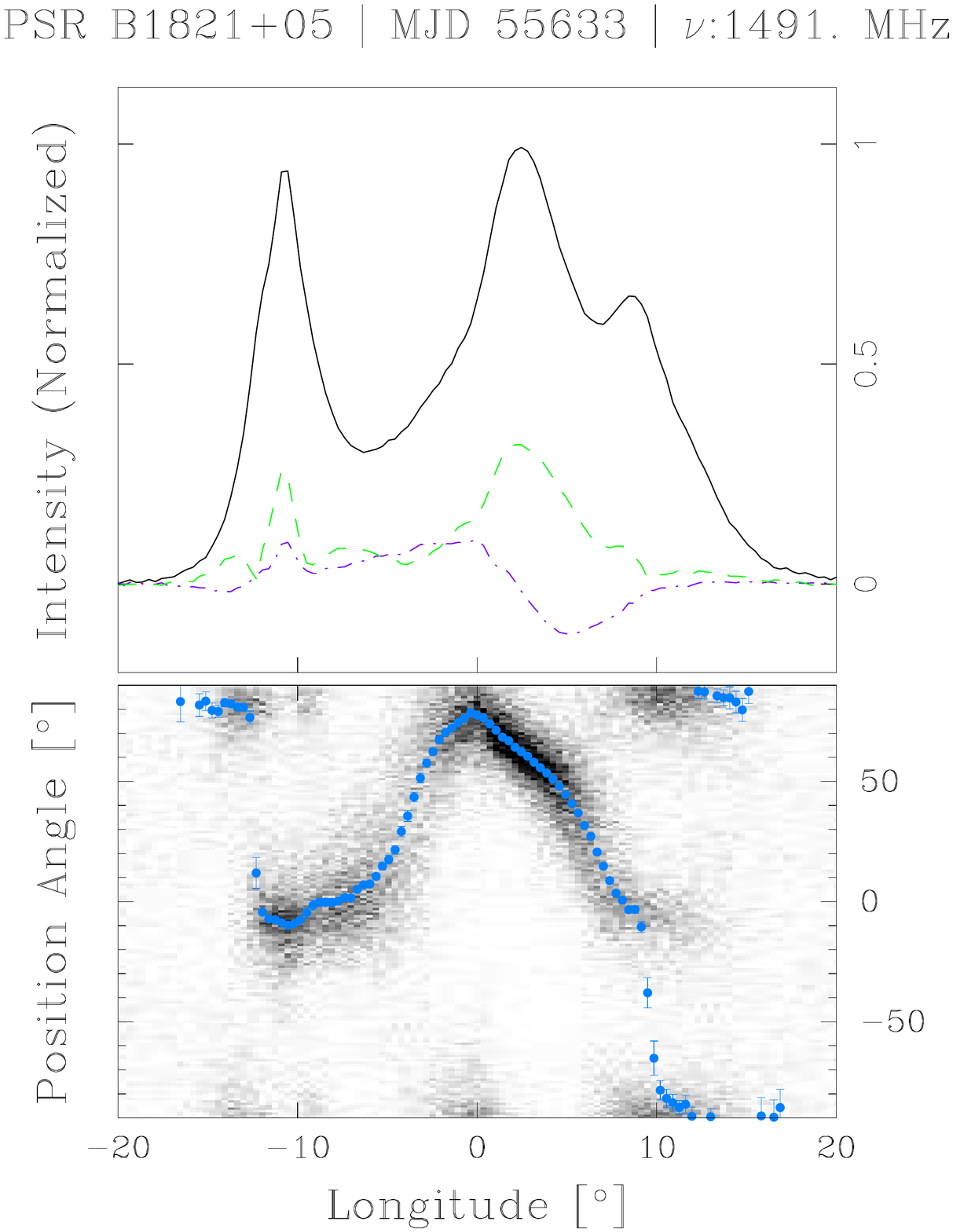} &
\includegraphics[page=1,width=\linewidth]{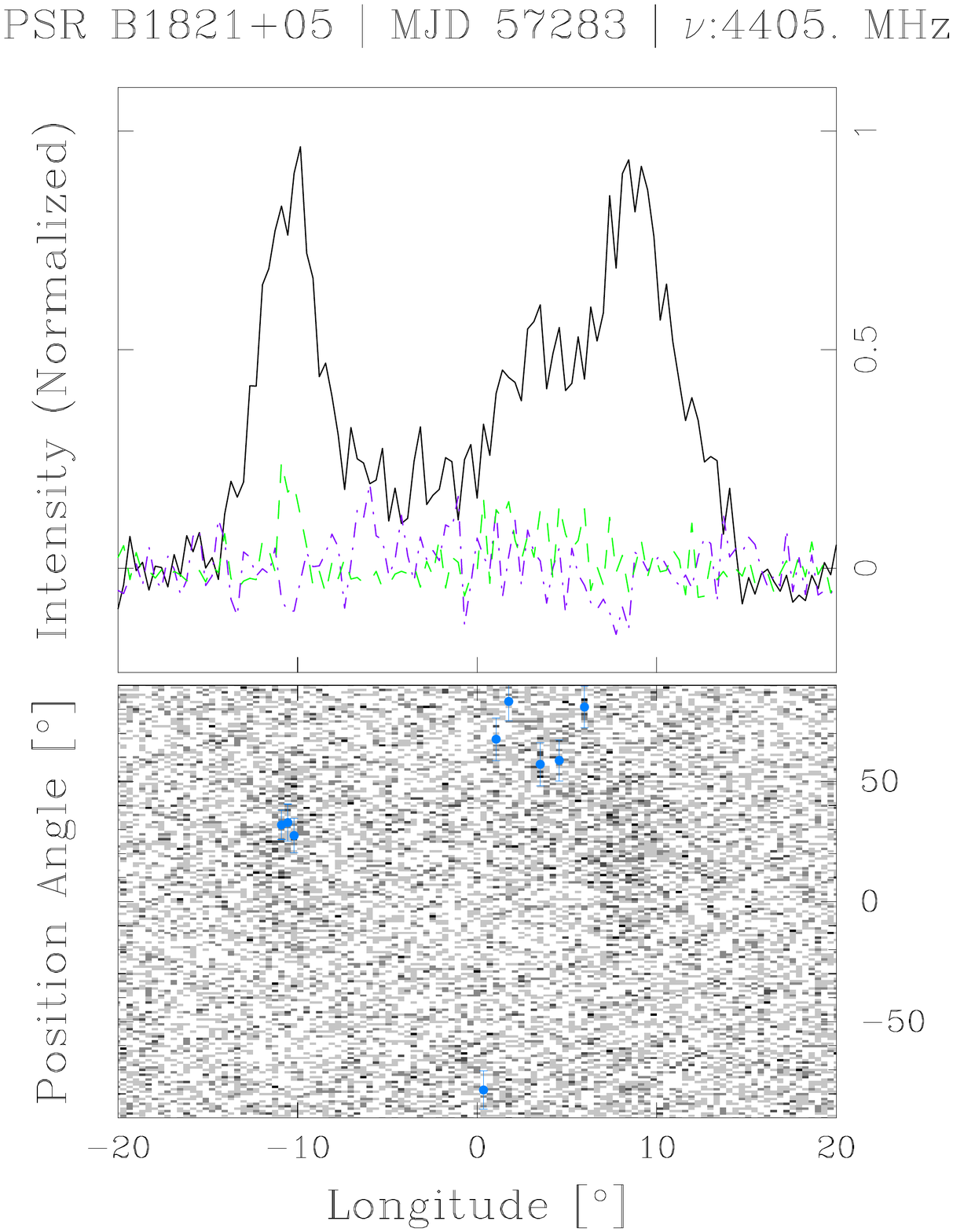} \\ 

     \bottomrule
   \end{tabularx} 
\caption{Average profiles of PSRs B1633+24, B1737+13, and B1821+05.}
 \end{figure*}
\vspace{1cm}

   \begin{figure*} 
 \begin{tabularx}{\textwidth}{YYY}
    \multicolumn{3}{c}{} \\ \toprule
\includegraphics[page=1,width=\linewidth]{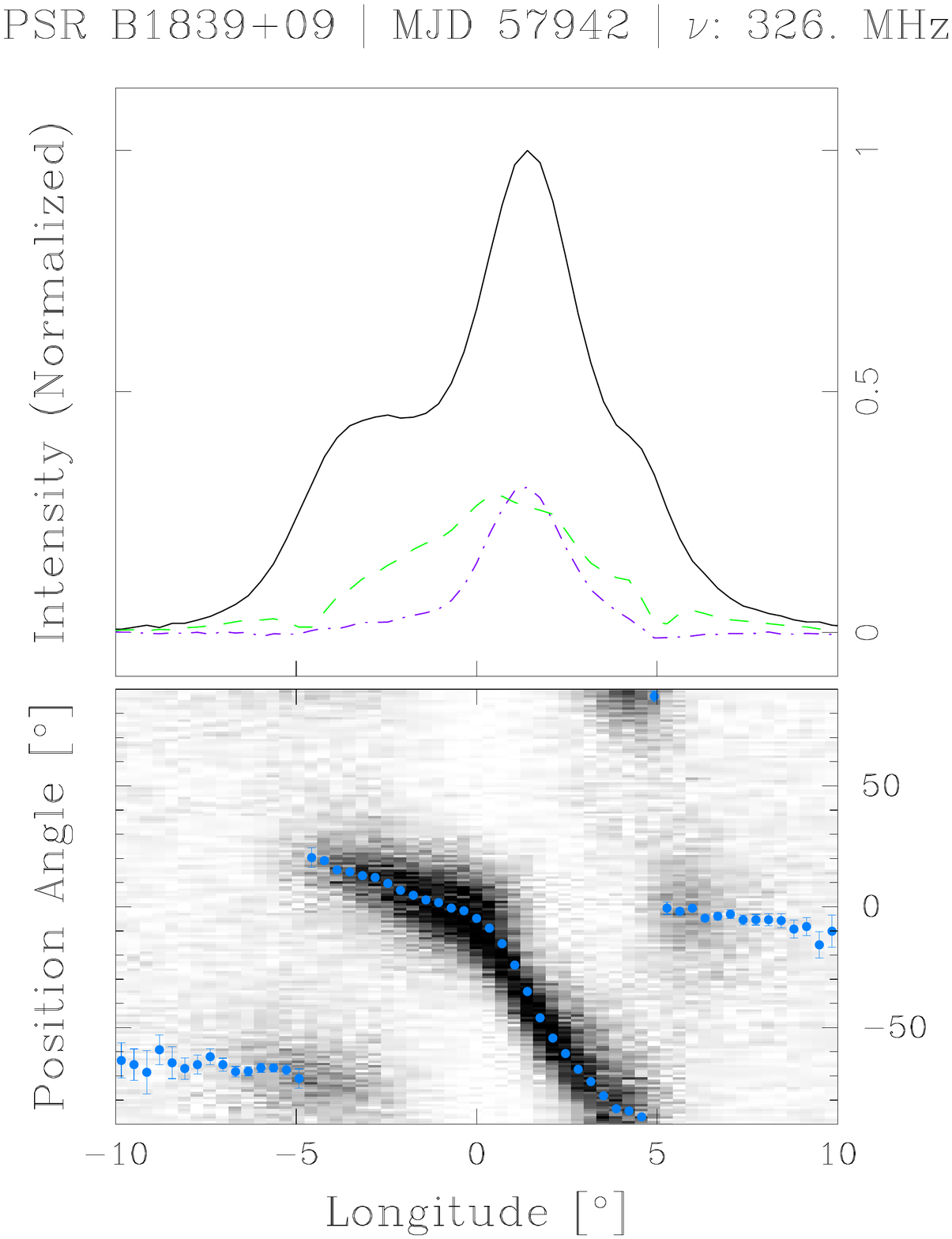} &
\includegraphics[page=1,width=\linewidth]{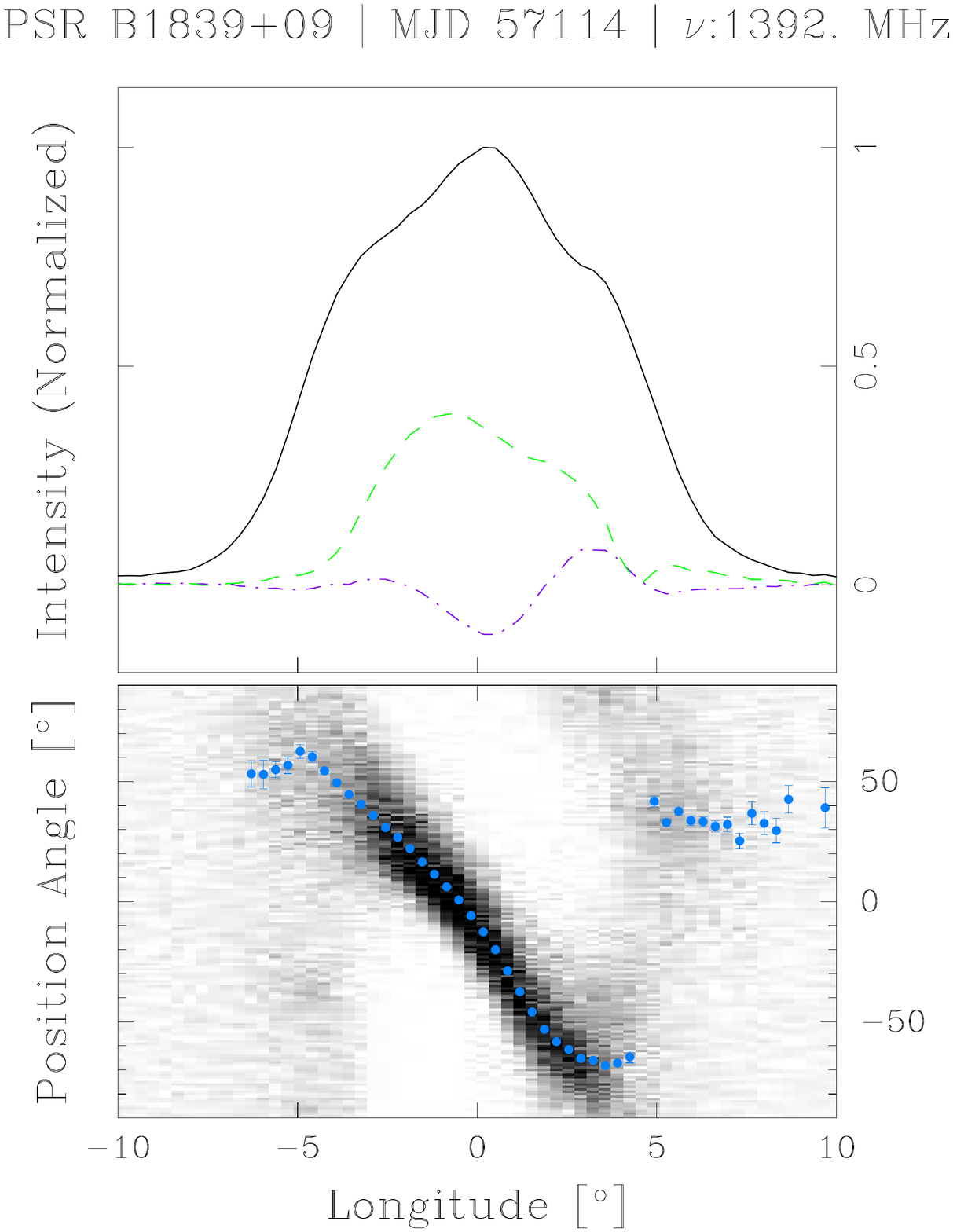} &
\includegraphics[page=1,width=\linewidth]{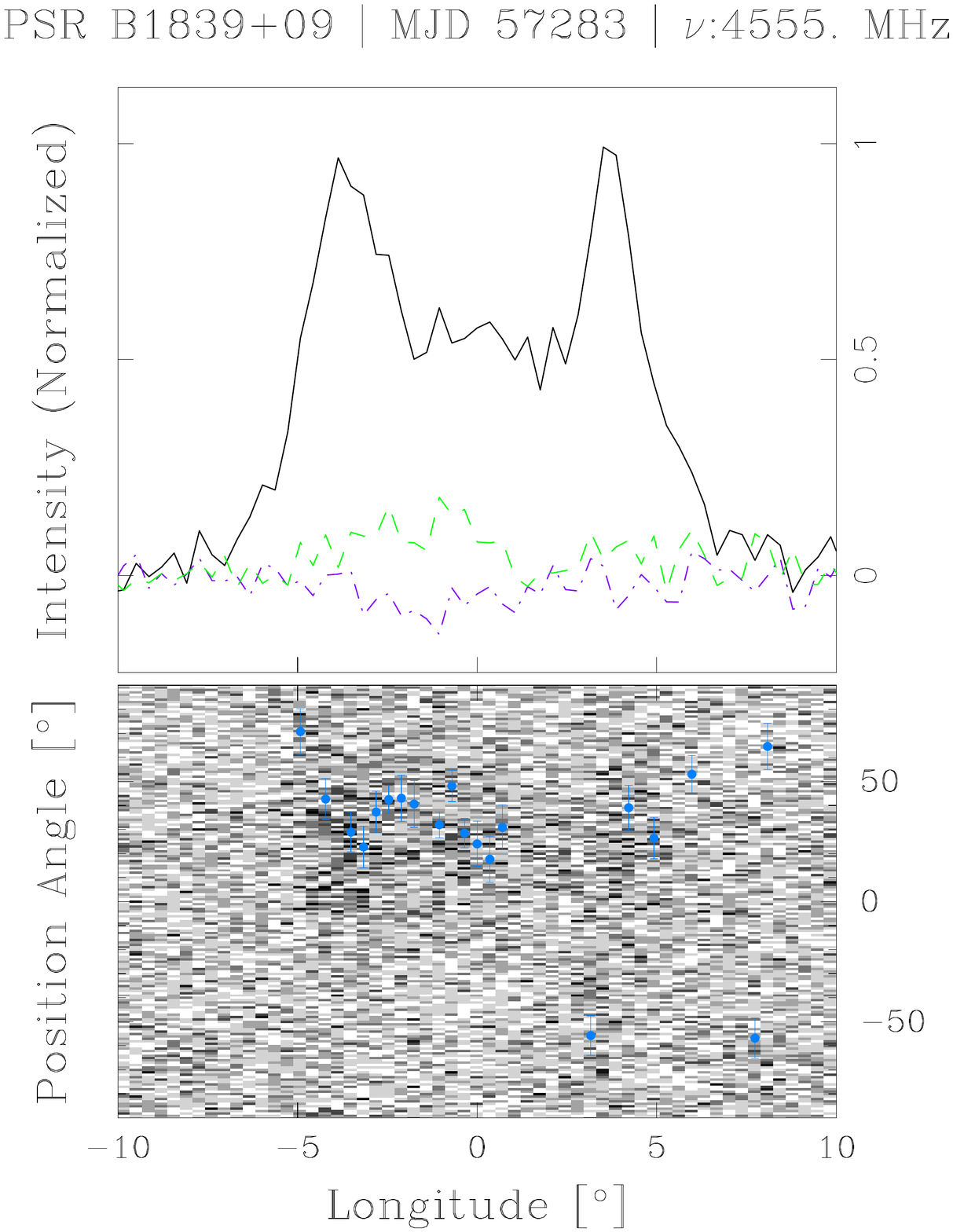} \\ \toprule
\includegraphics[page=1,width=\linewidth]{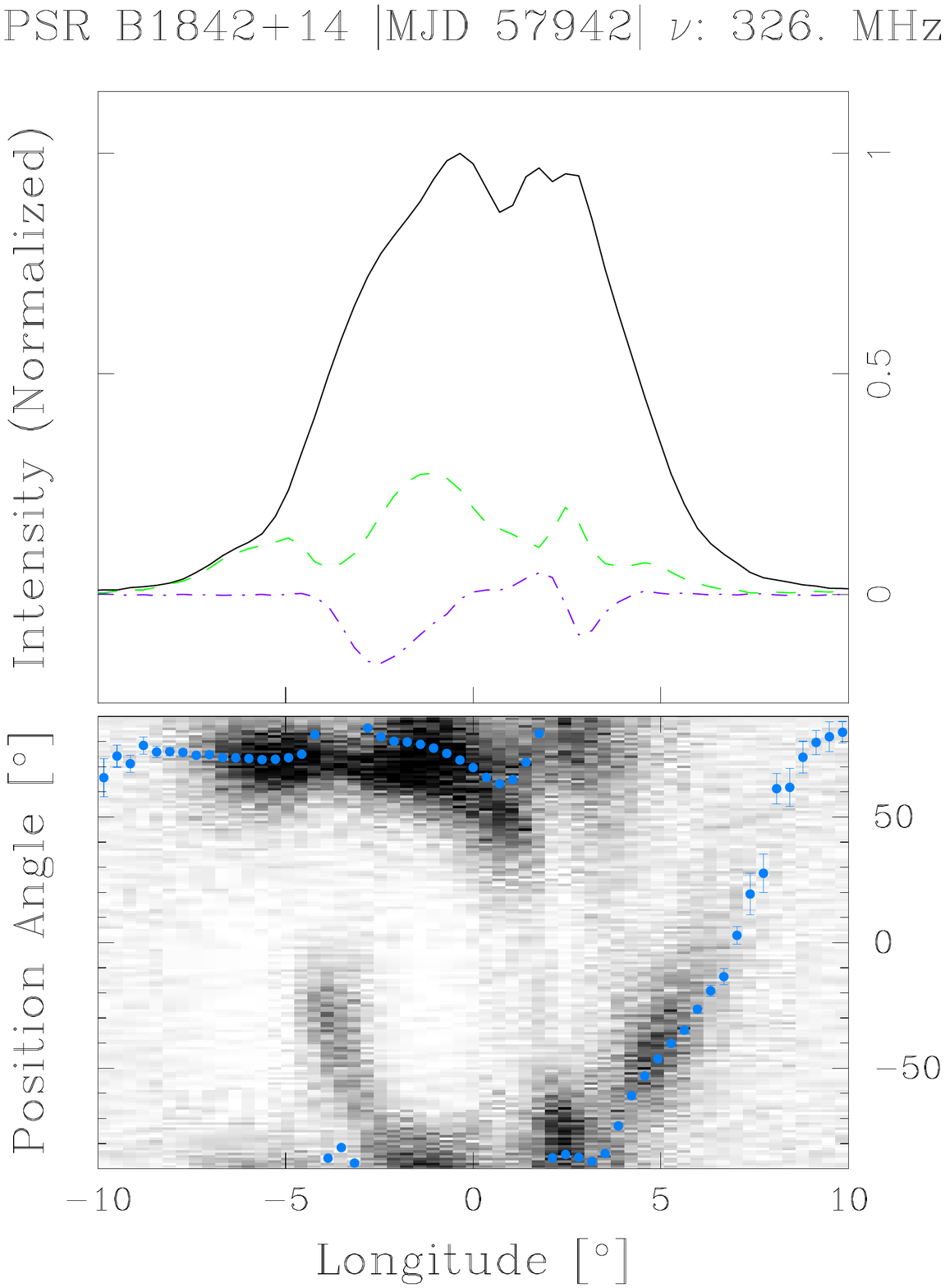} &
\includegraphics[page=1,width=\linewidth]{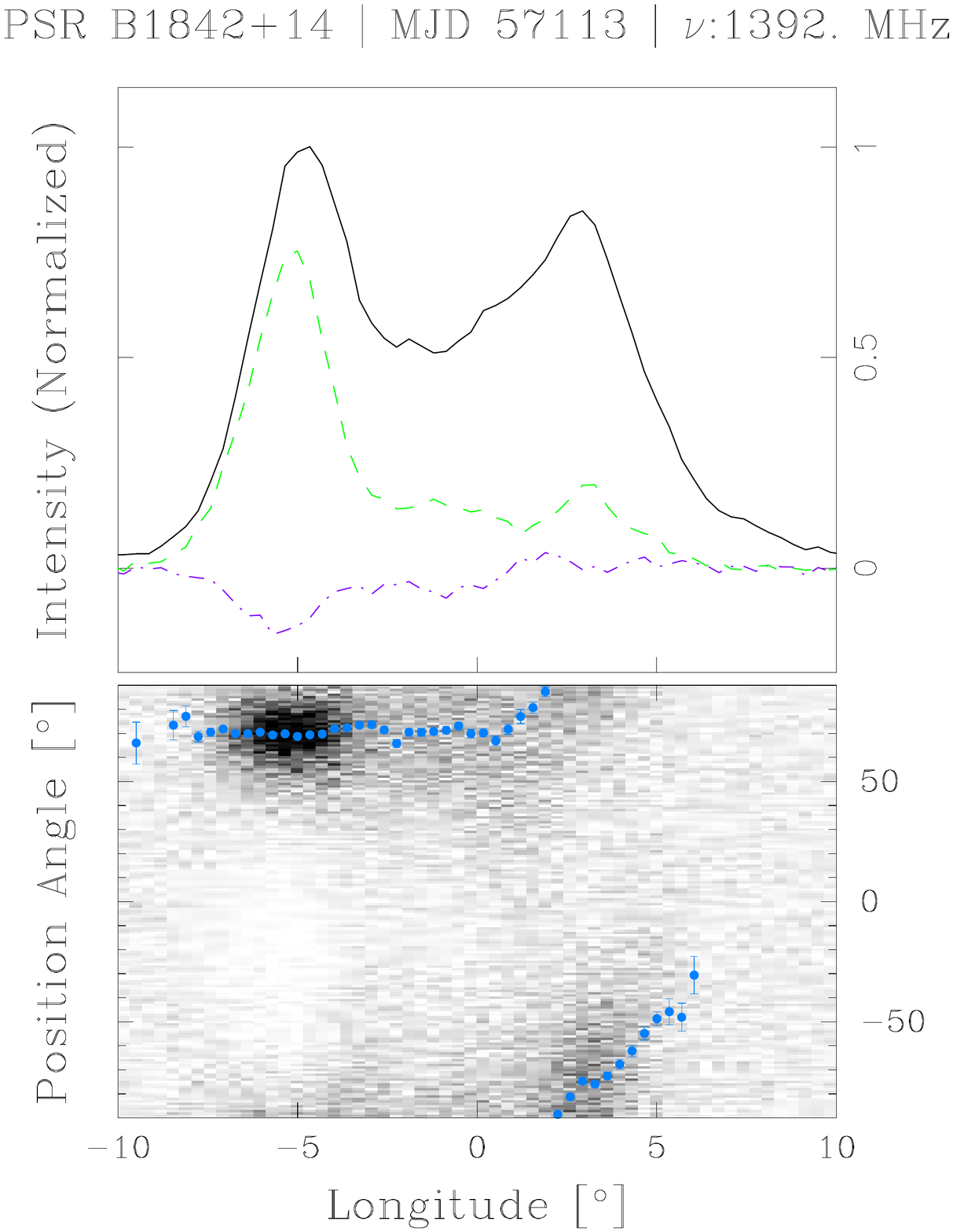} &
\includegraphics[page=1,width=\linewidth]{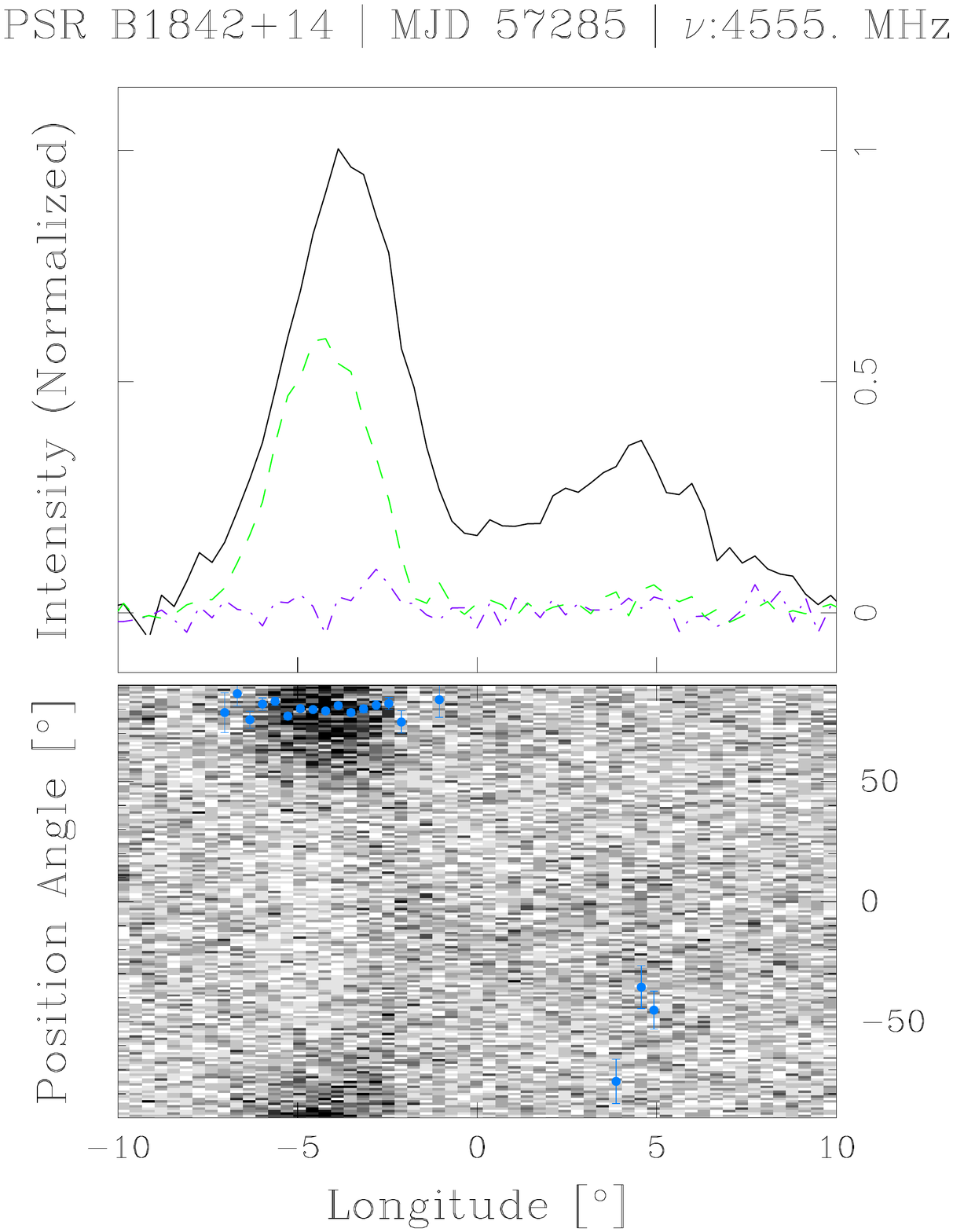} \\ \toprule
\includegraphics[page=1,width=\linewidth]{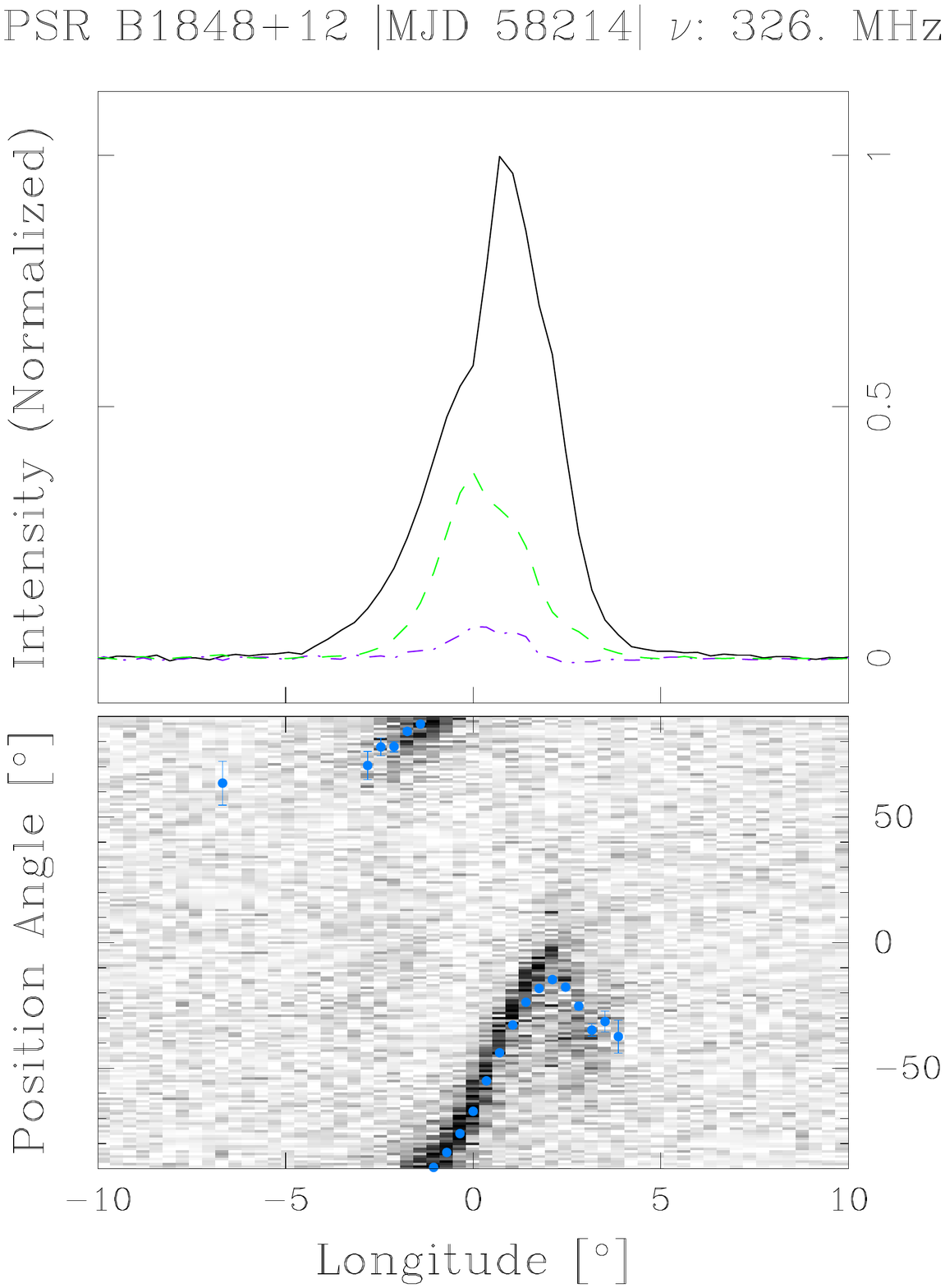} &
\includegraphics[page=1,width=\linewidth]{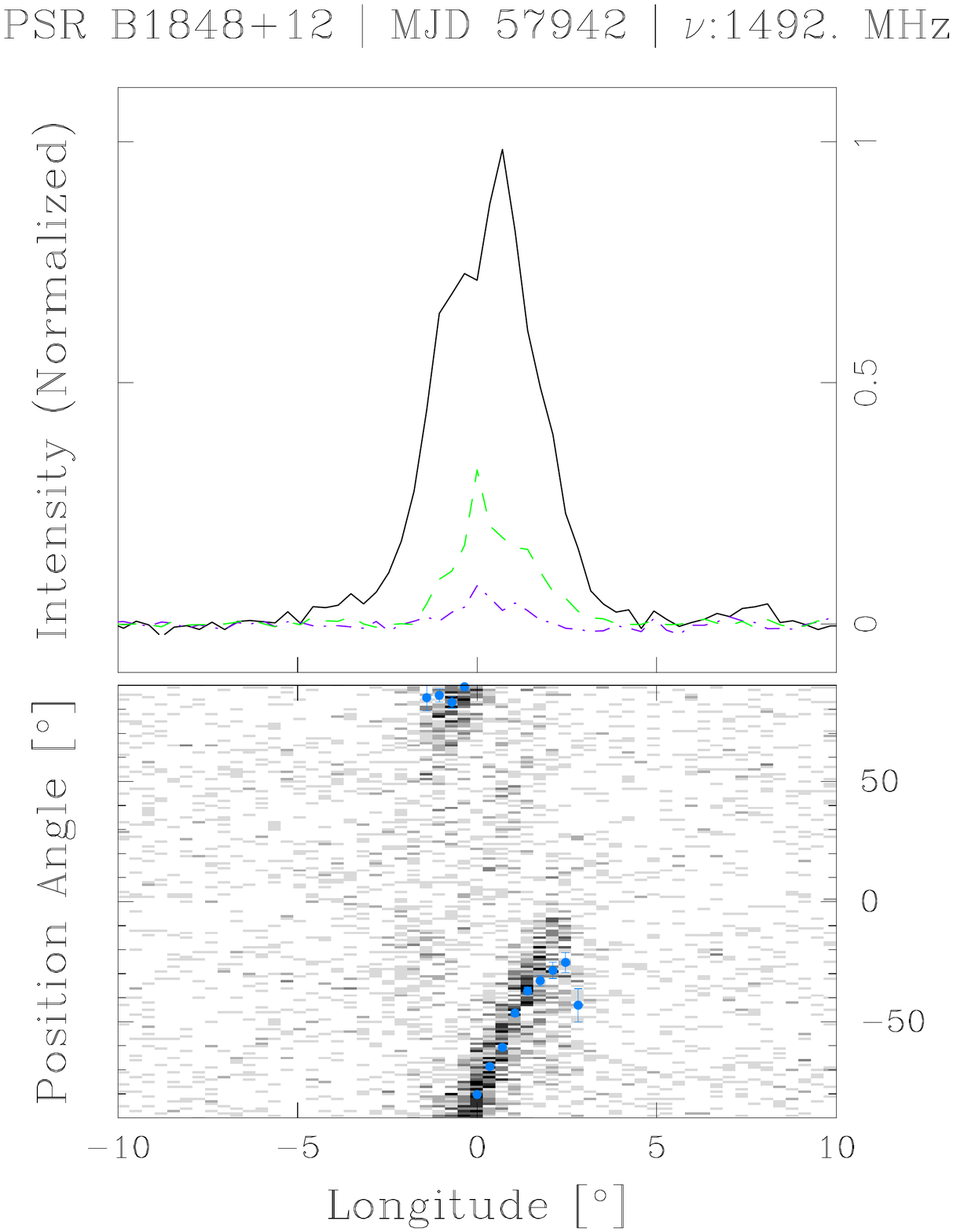} &
\includegraphics[page=1,width=\linewidth]{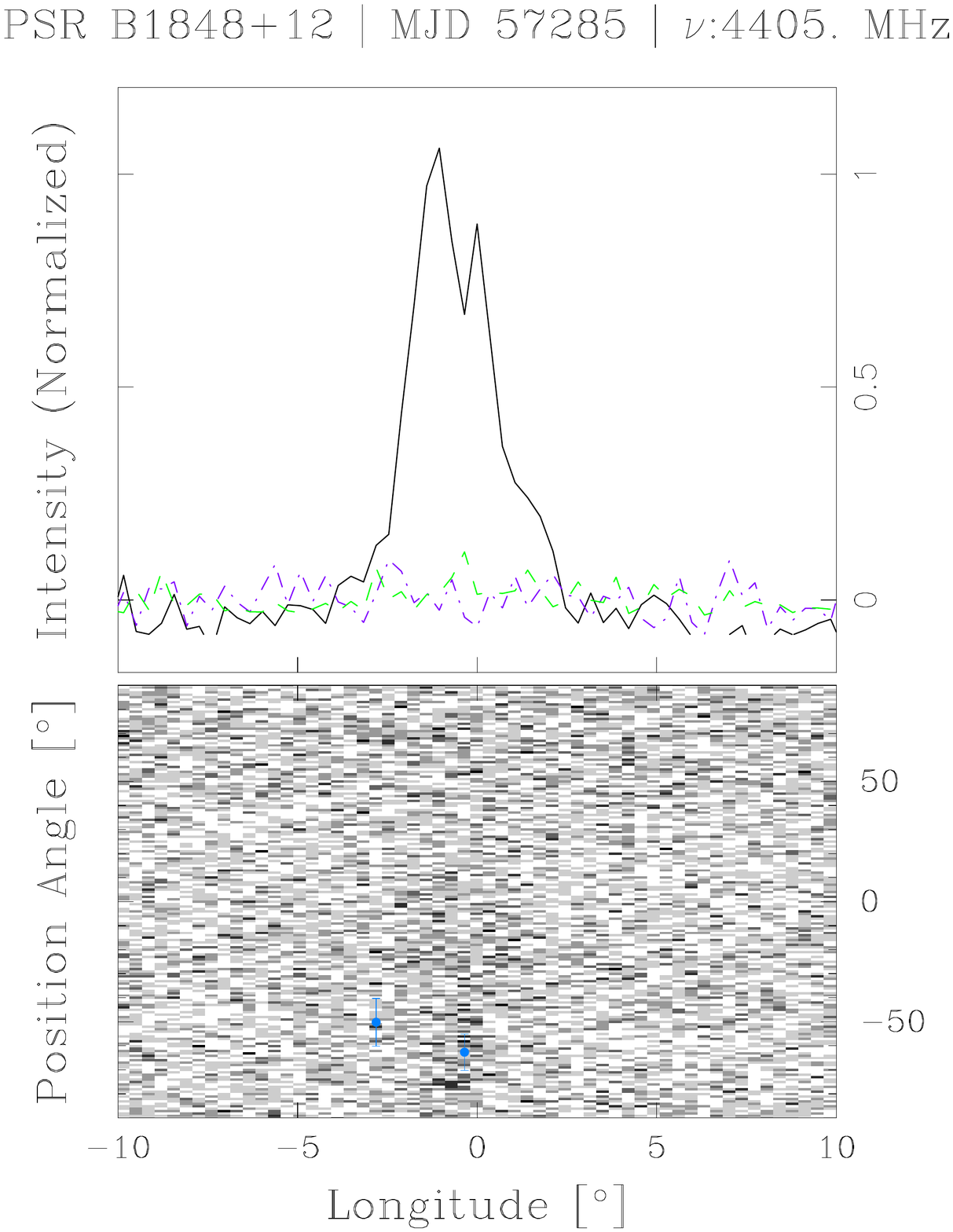} \\ 
     \bottomrule
   \end{tabularx} 
\caption{Average profiles of PSRs B1839+09, B1842+14, and B1848+12.}
 \end{figure*}
\vspace{1cm}

   \begin{figure*} 
 \begin{tabularx}{\textwidth}{YYY}
    \multicolumn{3}{c}{} \\ \toprule
\includegraphics[page=1,width=\linewidth]{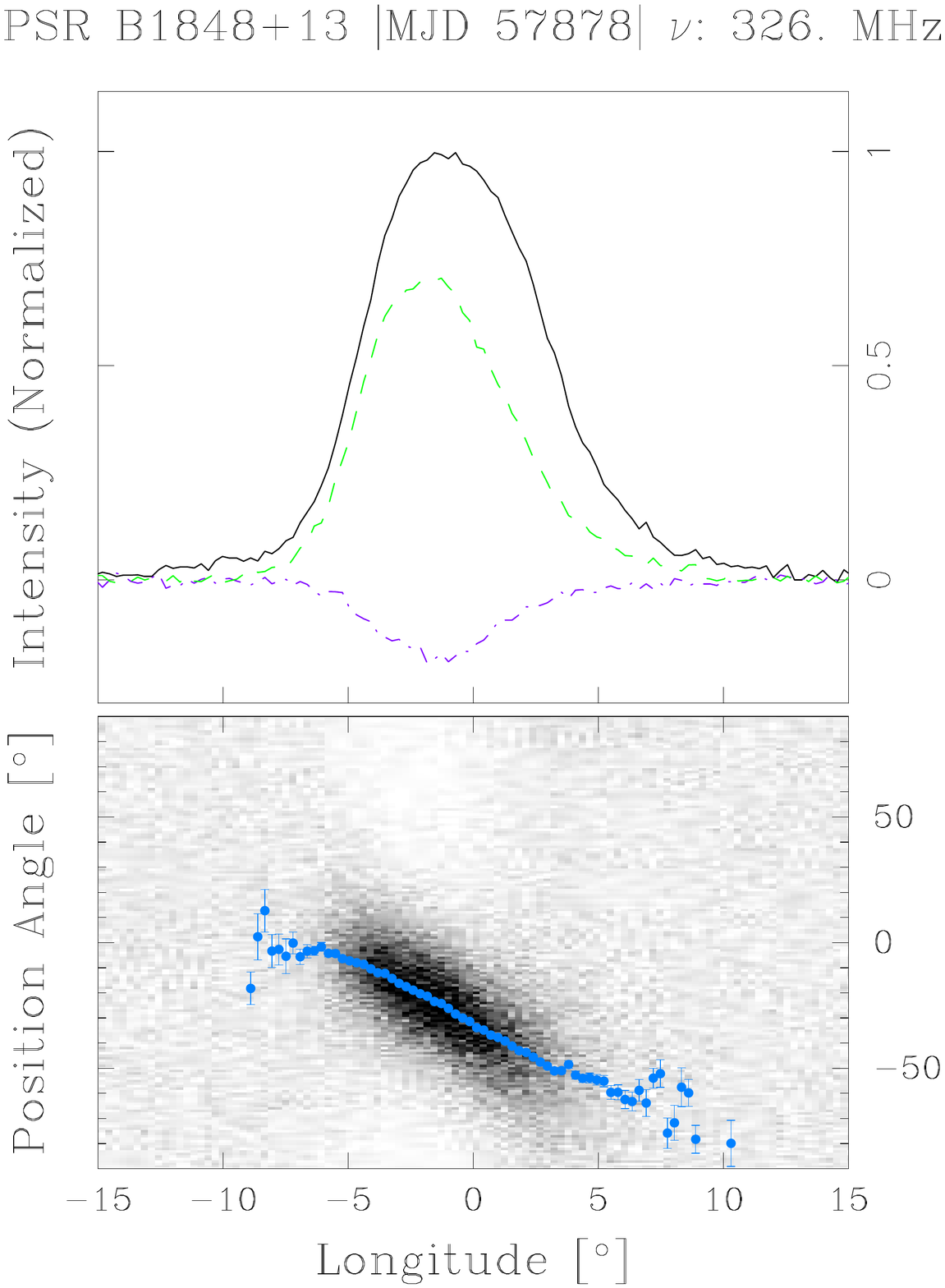} &
\includegraphics[page=1,width=\linewidth]{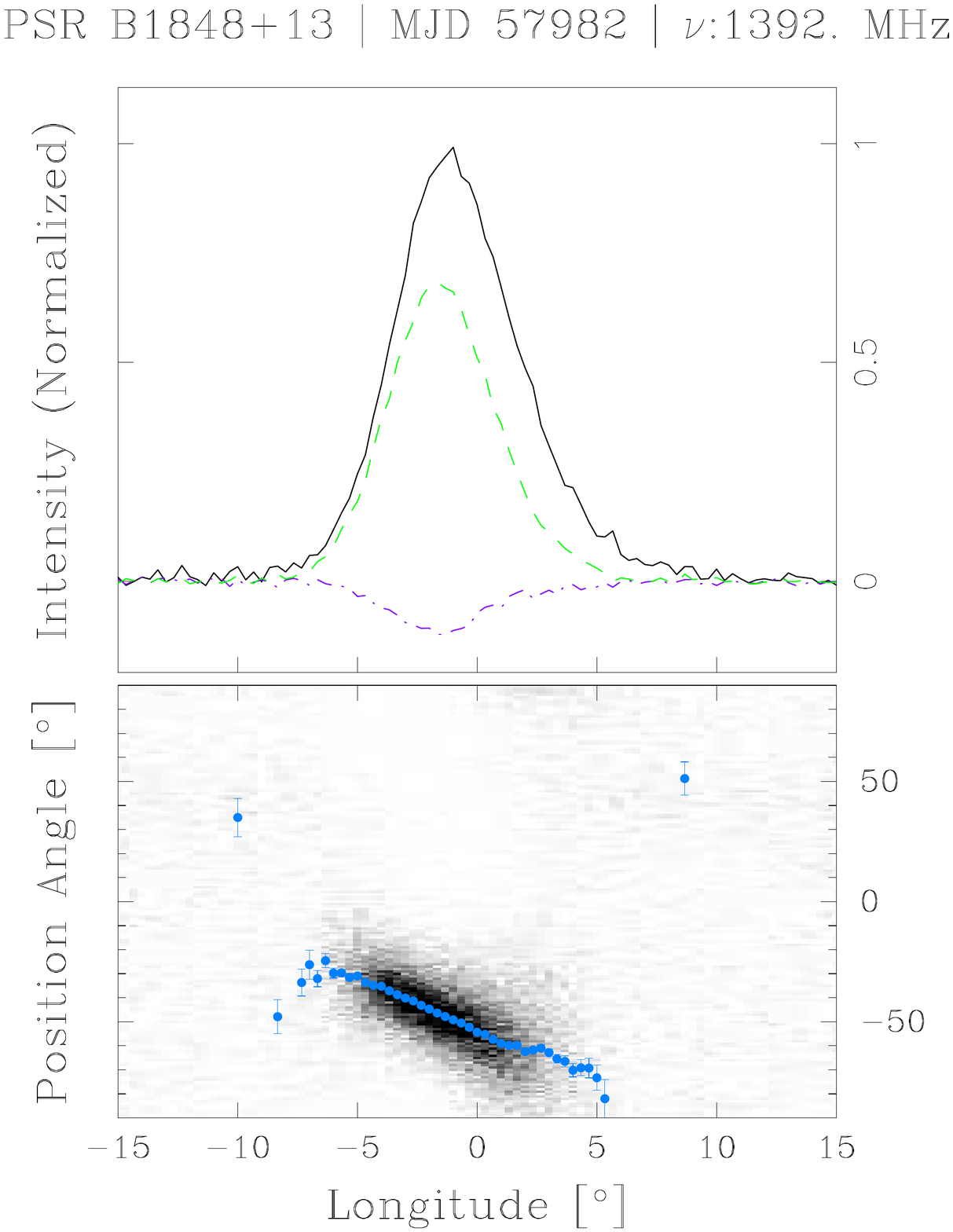} &
\includegraphics[page=1,width=\linewidth]{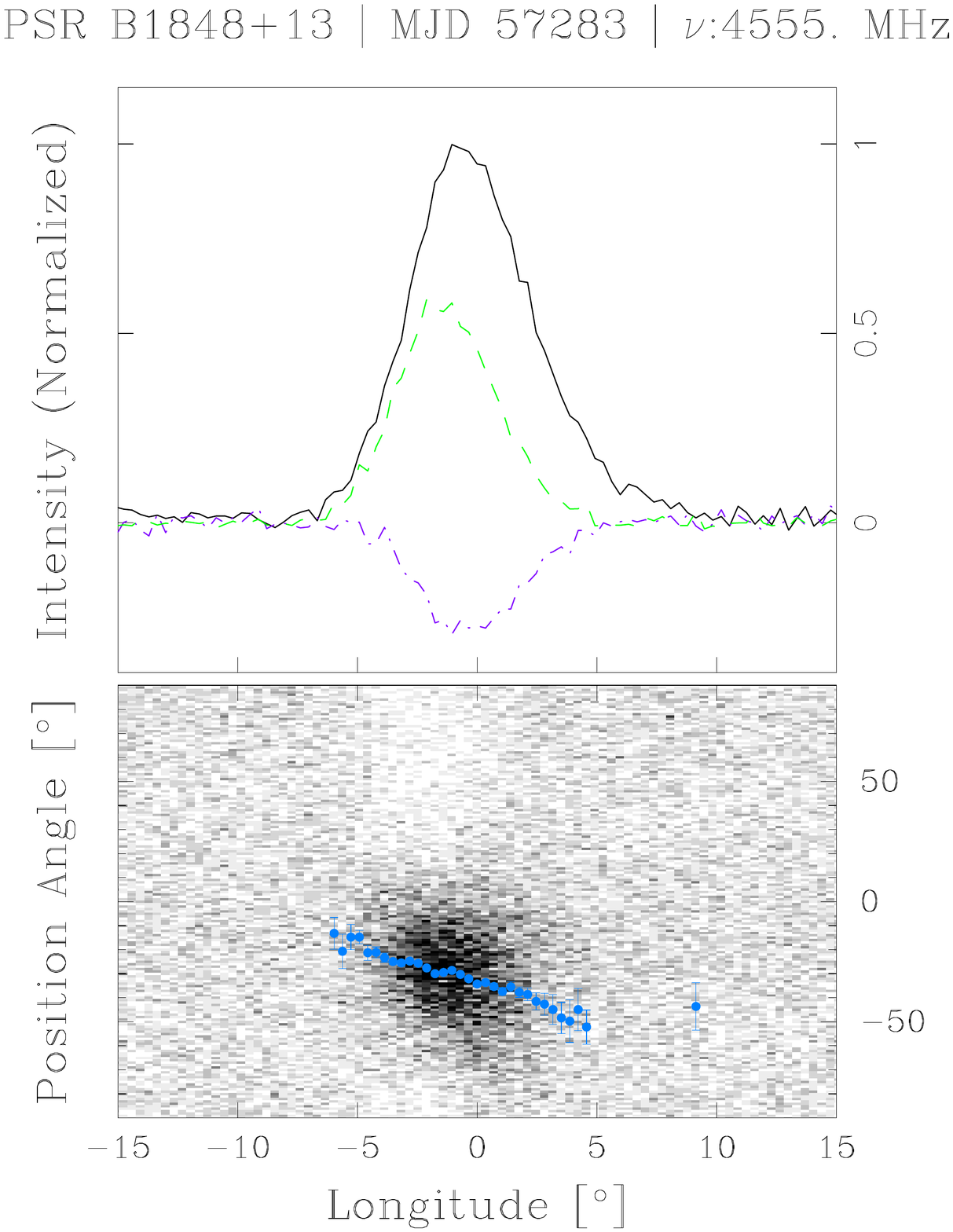} \\ \toprule
                &
\includegraphics[page=1,width=\linewidth]{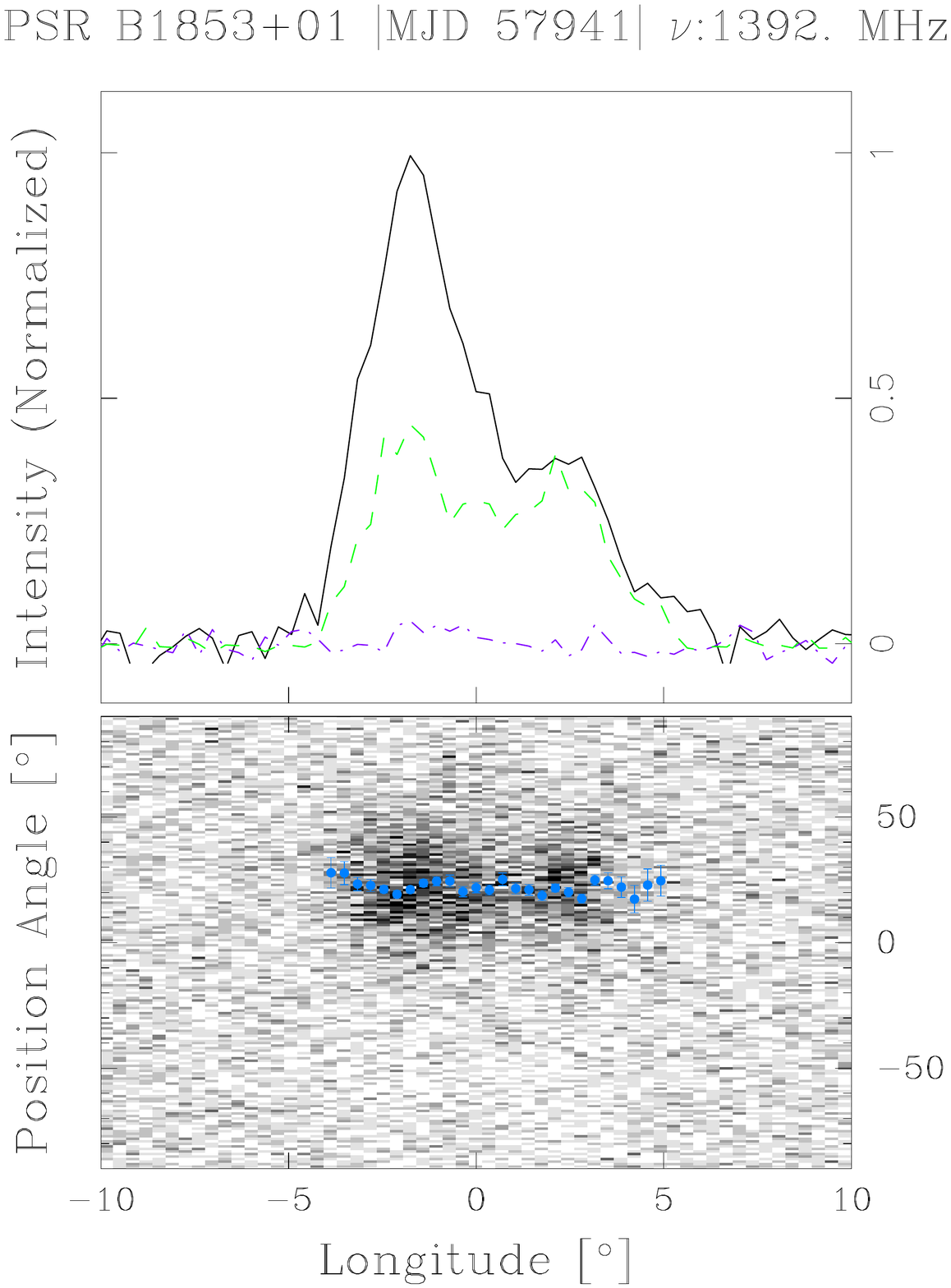} &
\includegraphics[page=1,width=\linewidth]{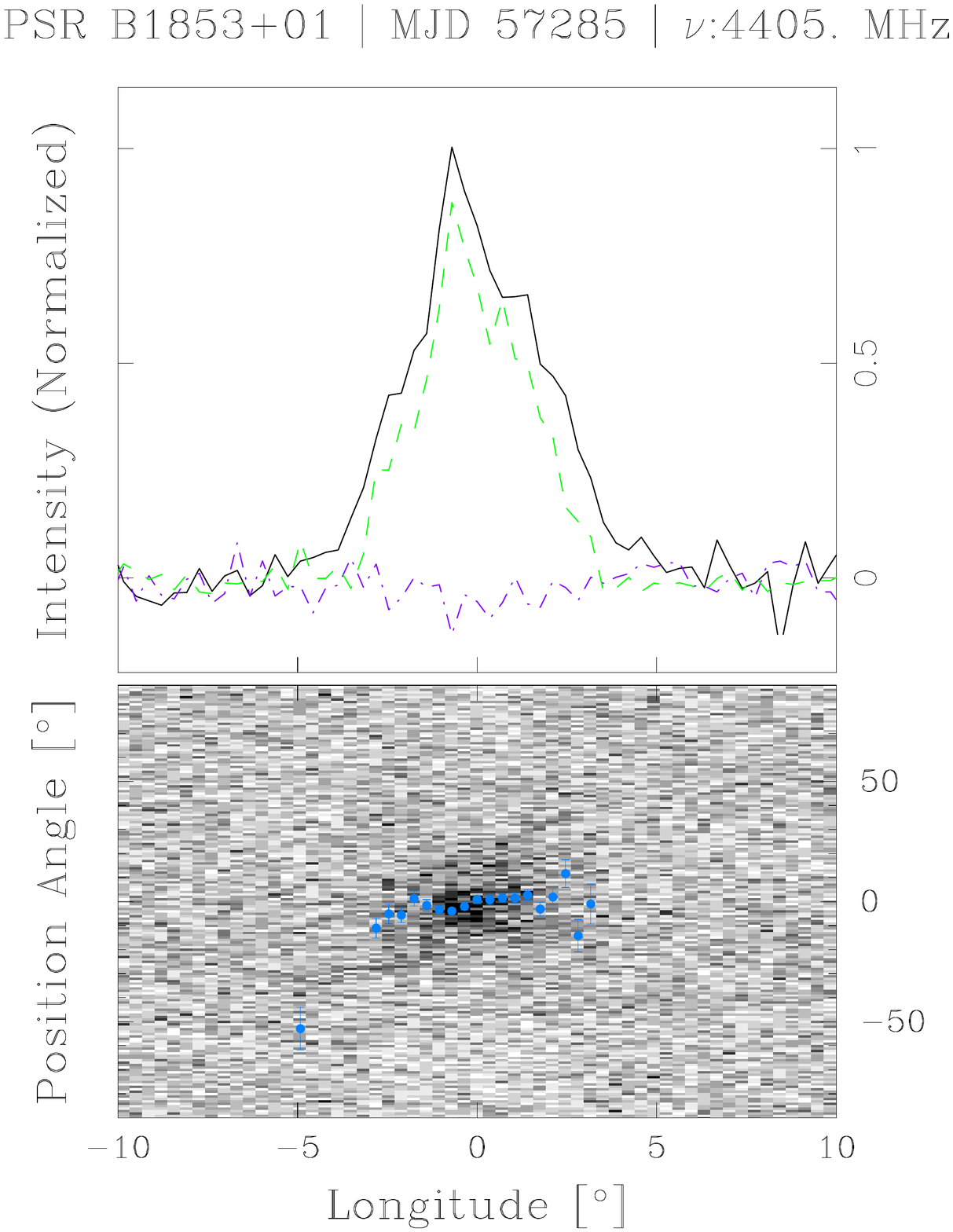} \\ \toprule
                &
\includegraphics[page=1,width=\linewidth]{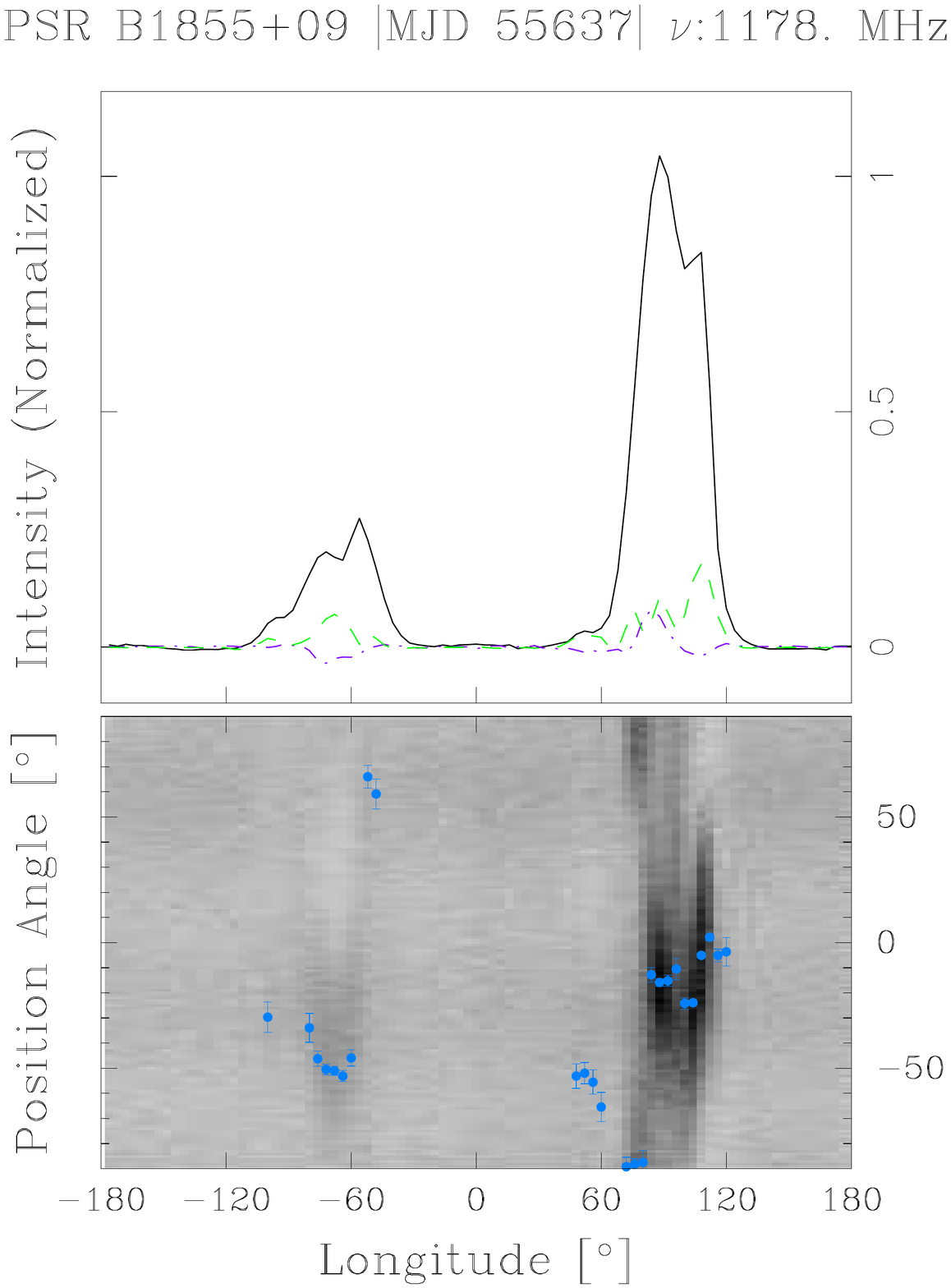} &
\includegraphics[page=1,width=\linewidth]{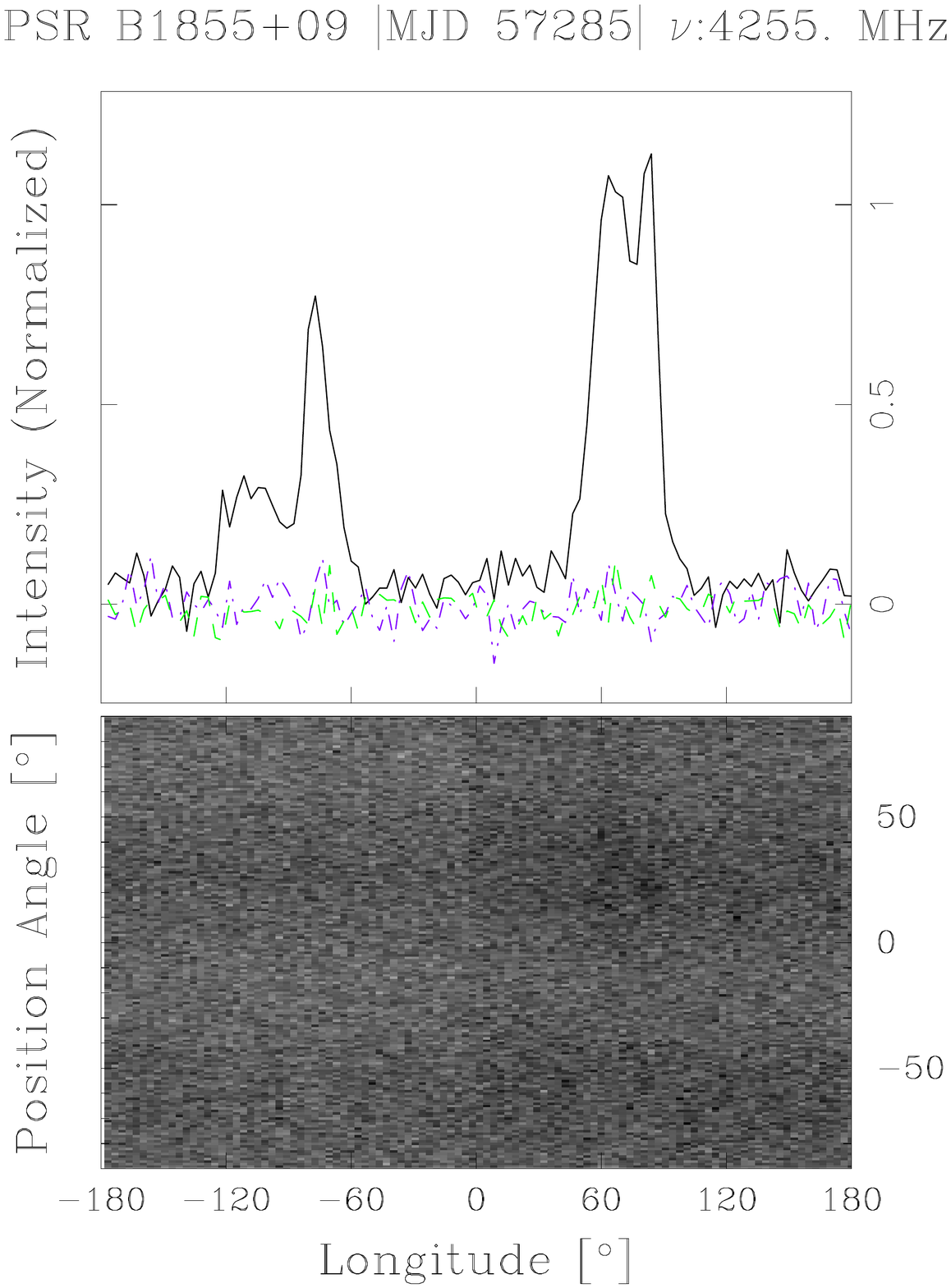} \\ 

     \bottomrule
   \end{tabularx} 
\caption{Average profiles of PSRs B1848+13, B1853+01, and B1855+09.}
 \end{figure*}
\vspace{1cm}
   \begin{figure*} 
 \begin{tabularx}{\textwidth}{YYY}
    \multicolumn{3}{c}{} \\ \toprule
                &
\includegraphics[page=1,width=\linewidth]{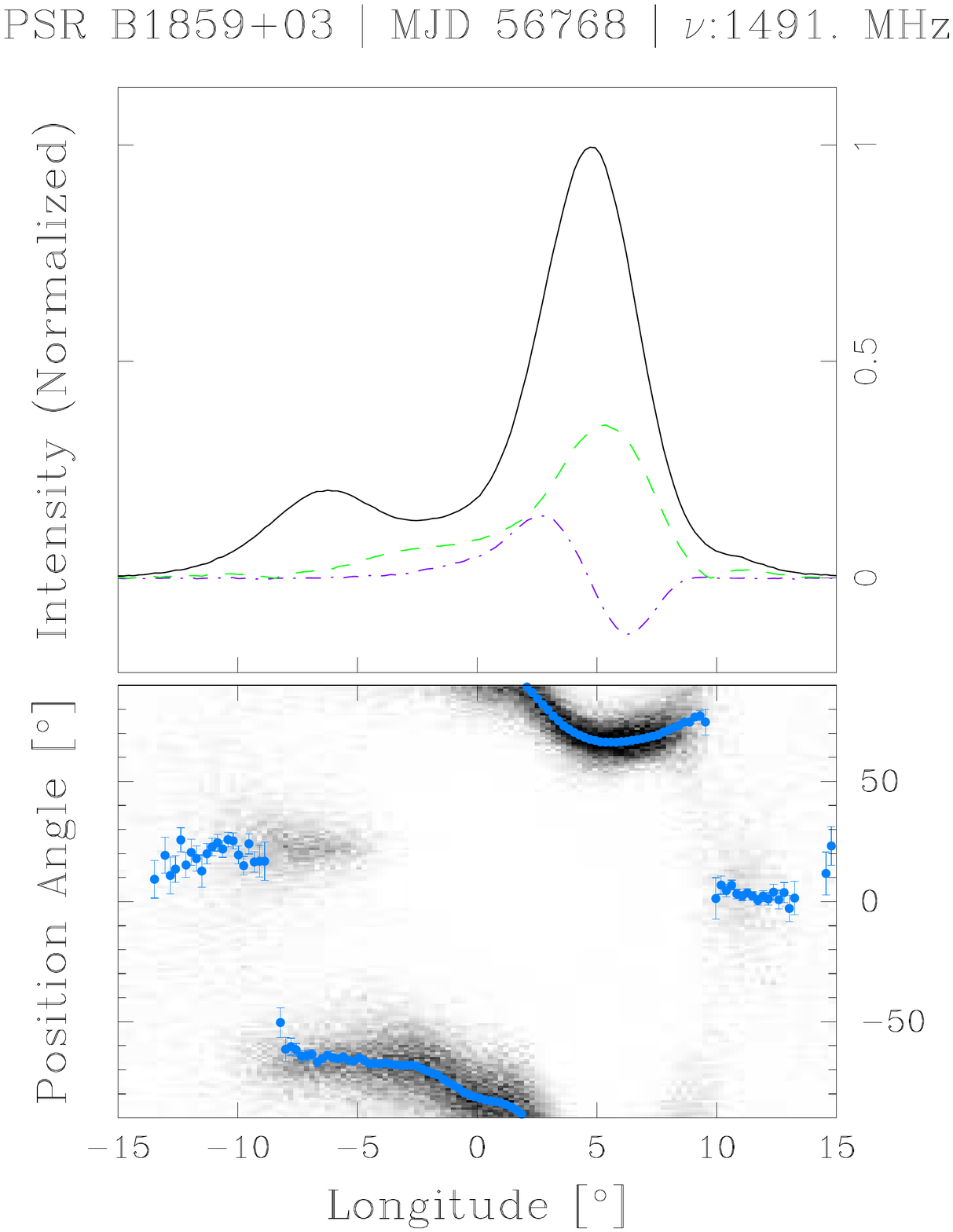} &
\includegraphics[page=1,width=\linewidth]{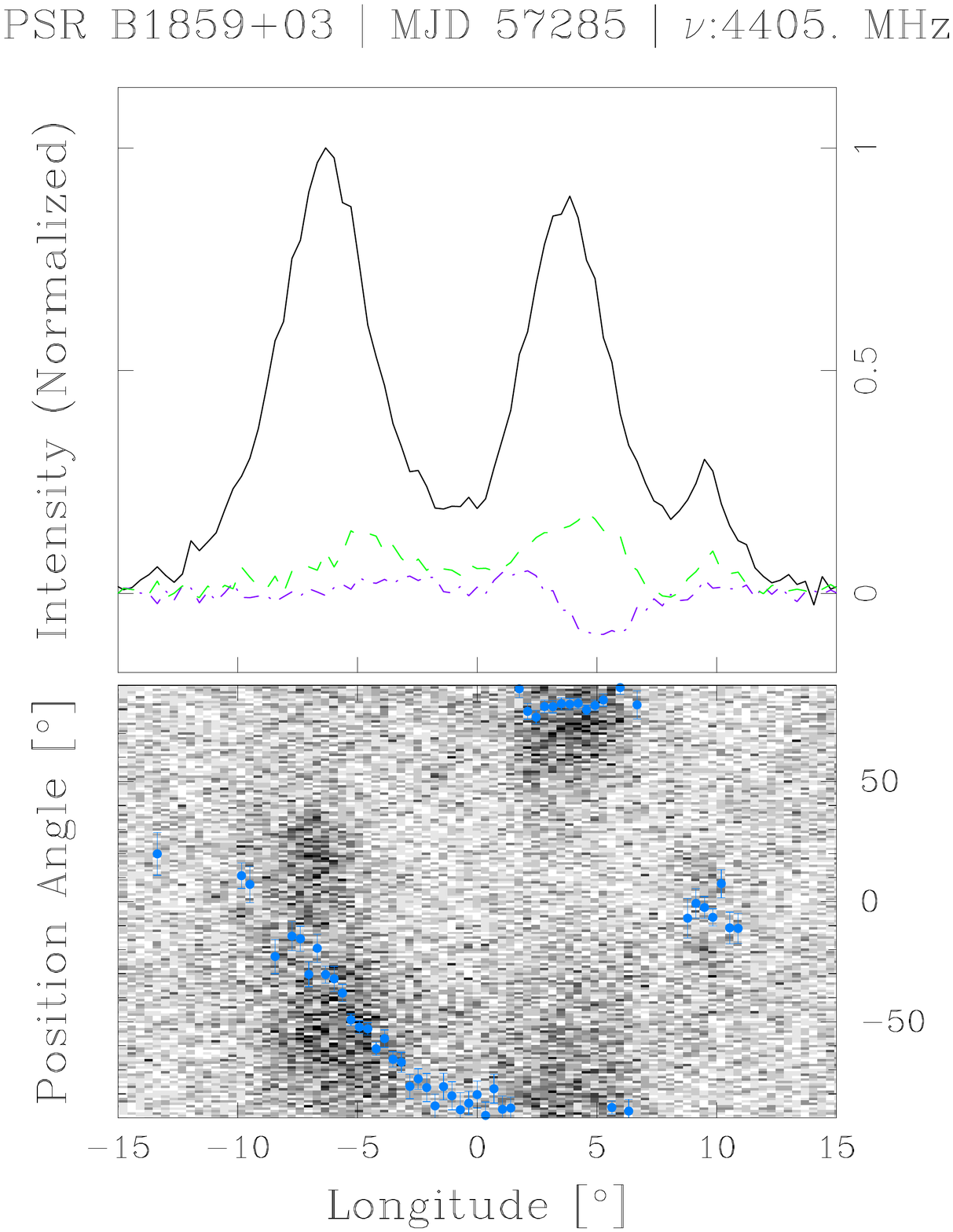} \\ \toprule
&
\includegraphics[page=1,width=\linewidth]{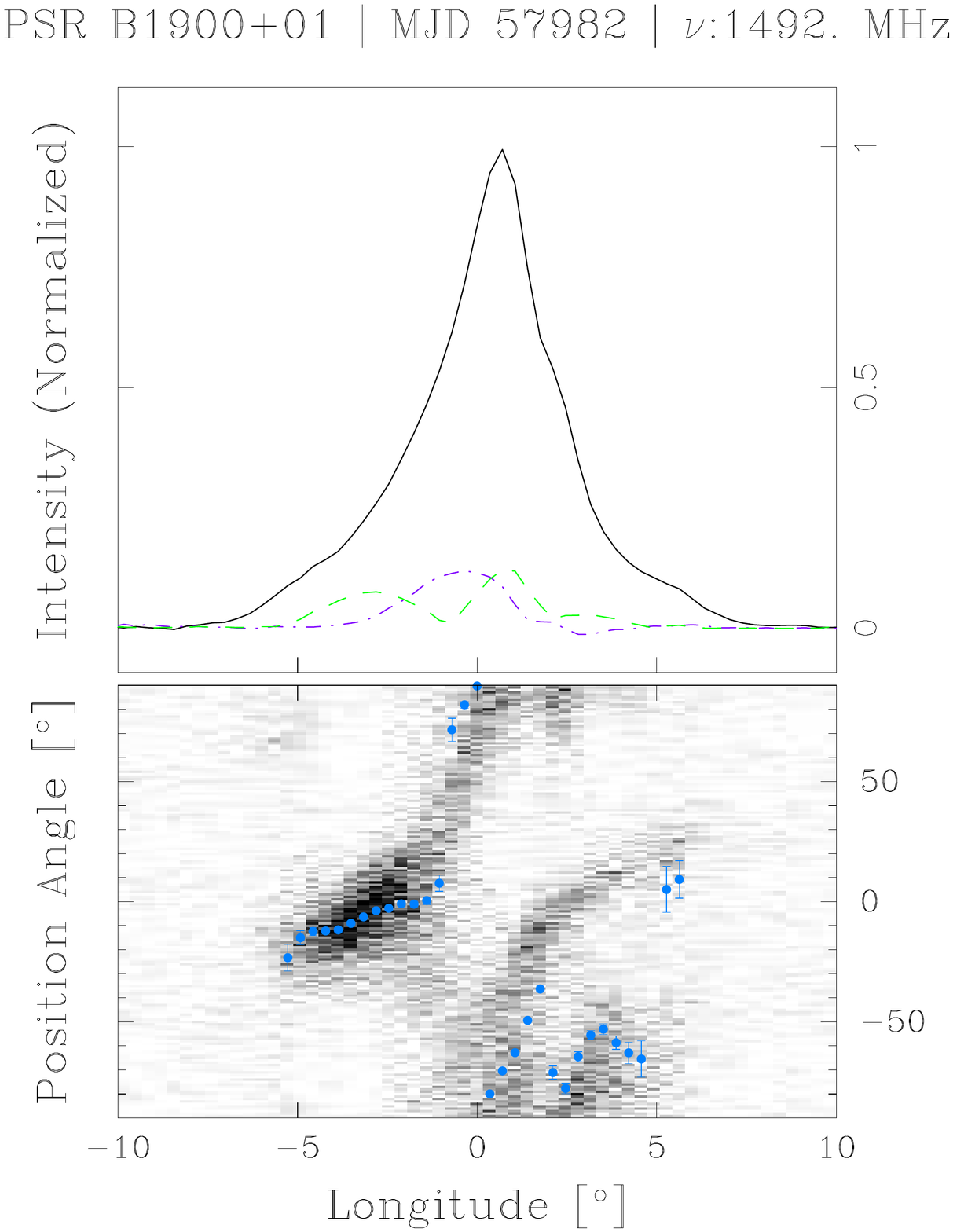}                &
\includegraphics[page=1,width=\linewidth]{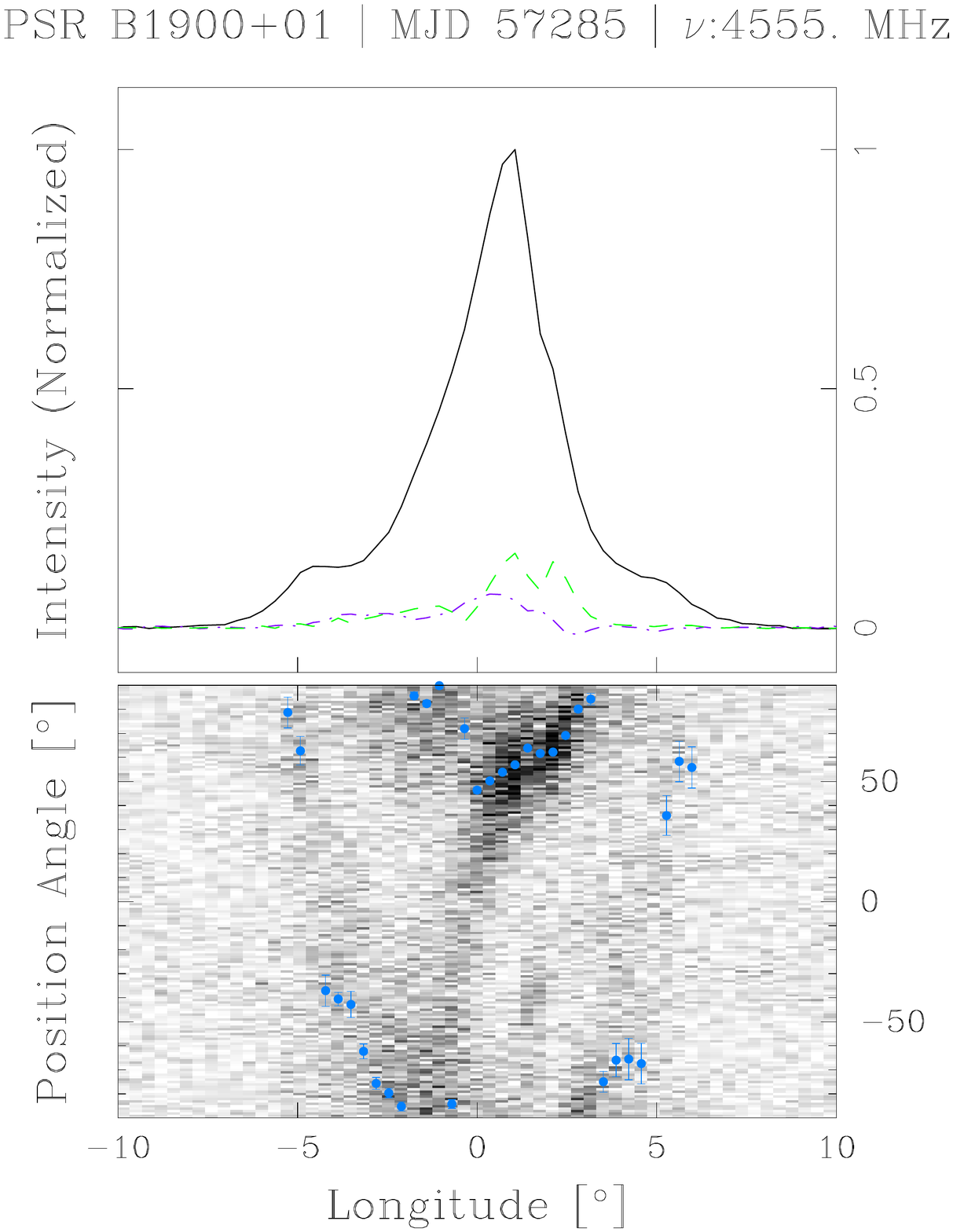} \\ \toprule
                &
\includegraphics[page=1,width=\linewidth]{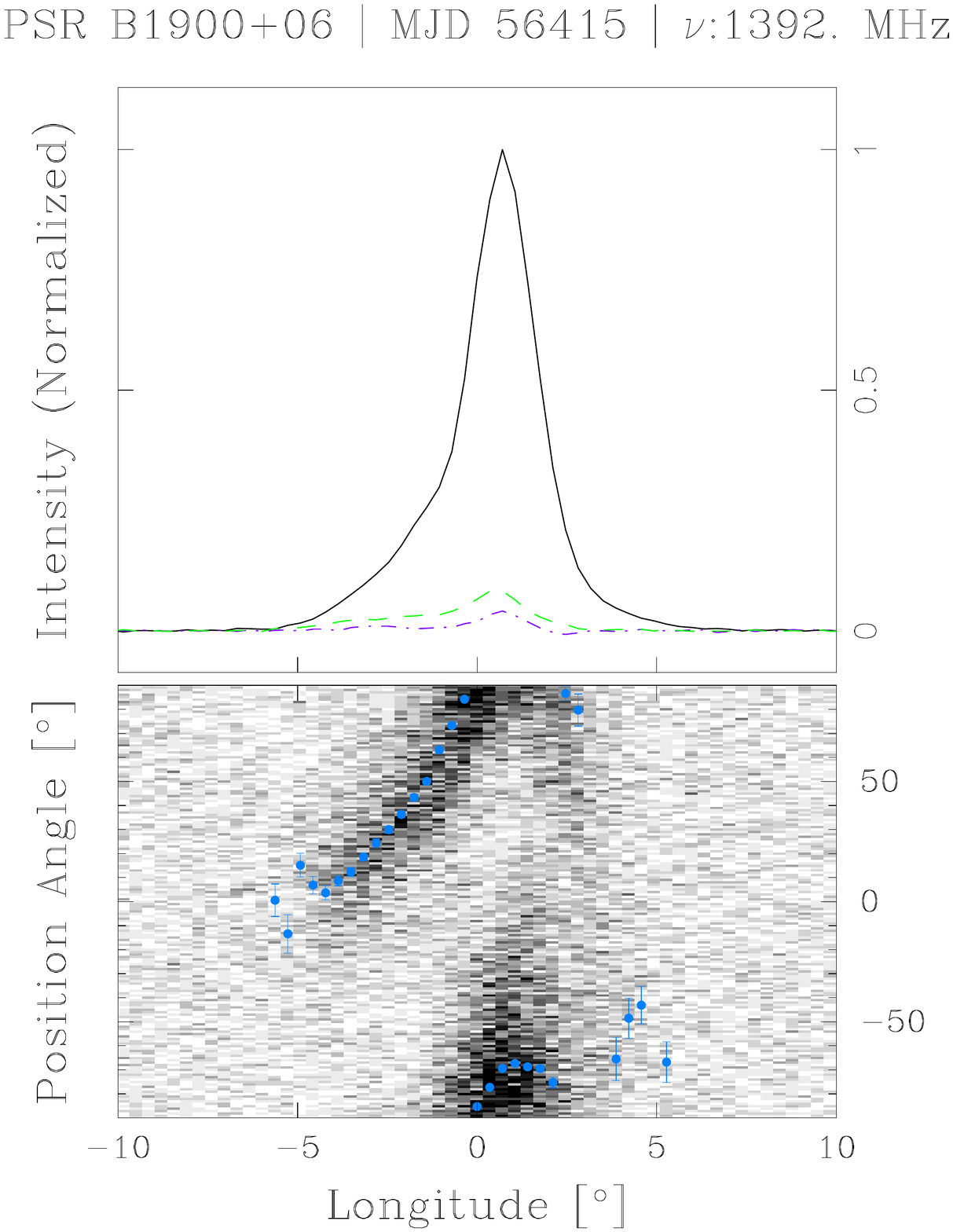} &
\includegraphics[page=1,width=\linewidth]{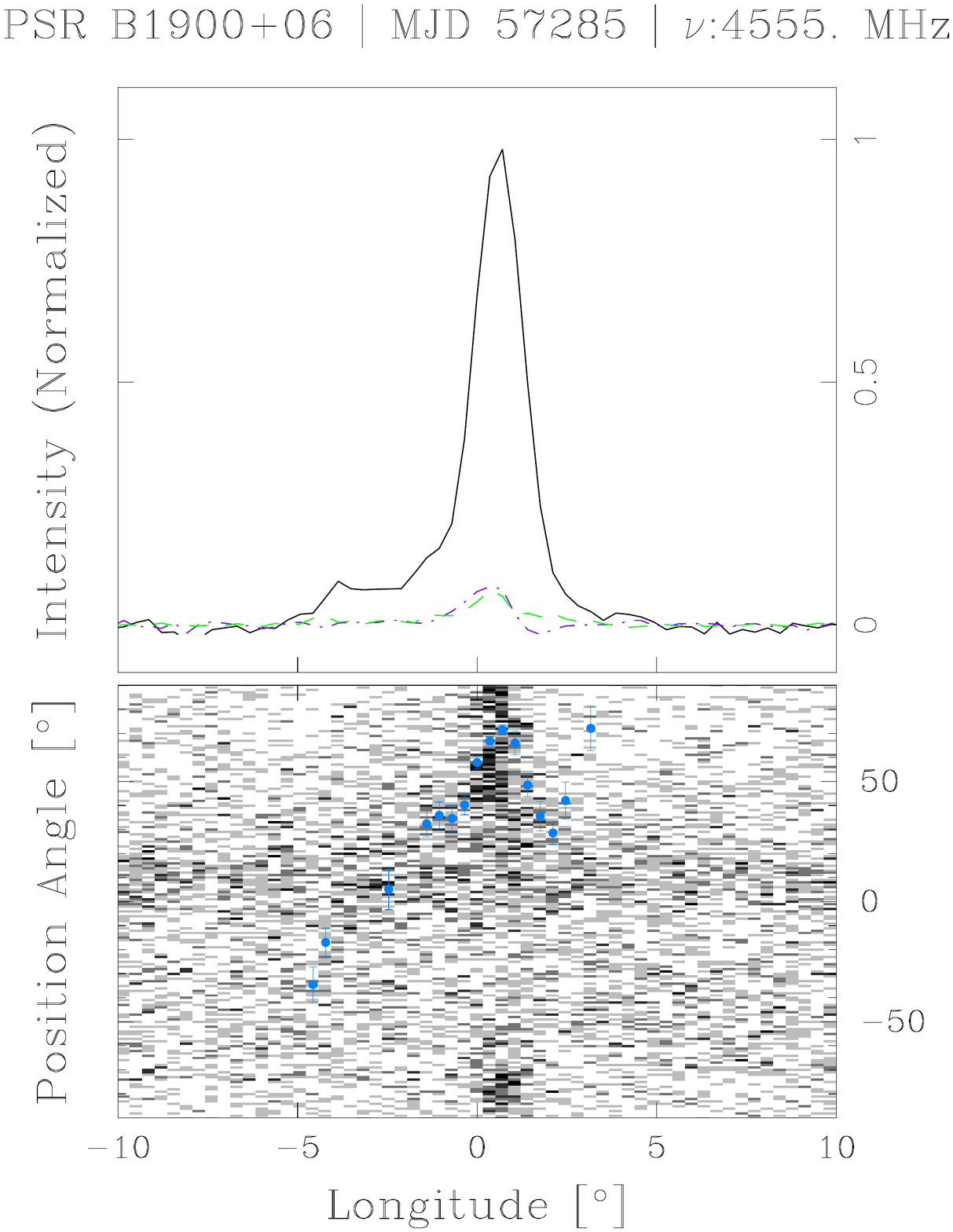} \\ 

     \bottomrule
   \end{tabularx} 
\caption{Average profiles of PSRs B1859+03, B1900+01, and B1900+06.}
 \end{figure*}
\vspace{1cm}
   \begin{figure*} 
 \begin{tabularx}{\textwidth}{YYY}
    \multicolumn{3}{c}{} \\ \toprule
\includegraphics[page=1,width=\linewidth]{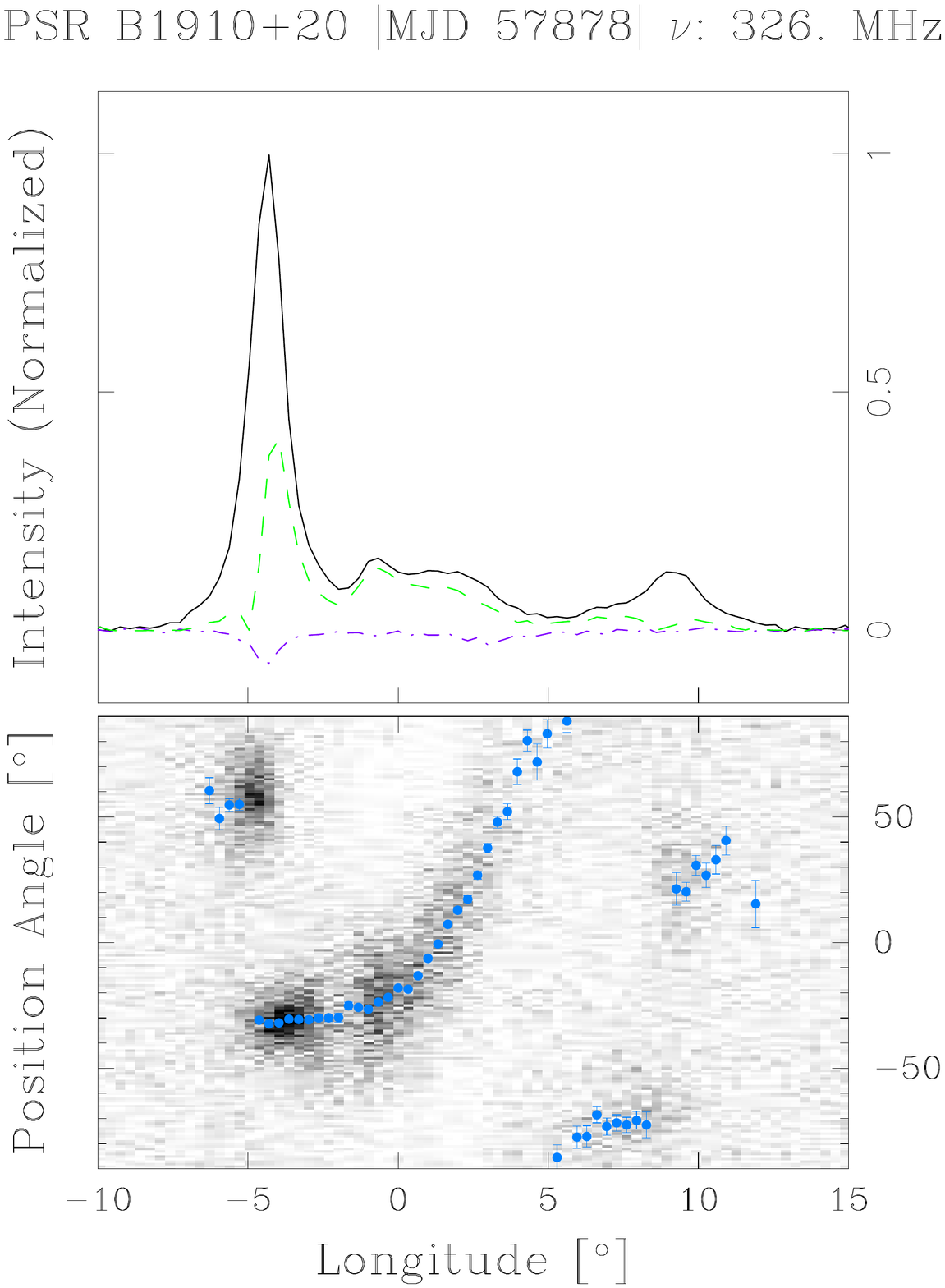} &
\includegraphics[page=1,width=\linewidth]{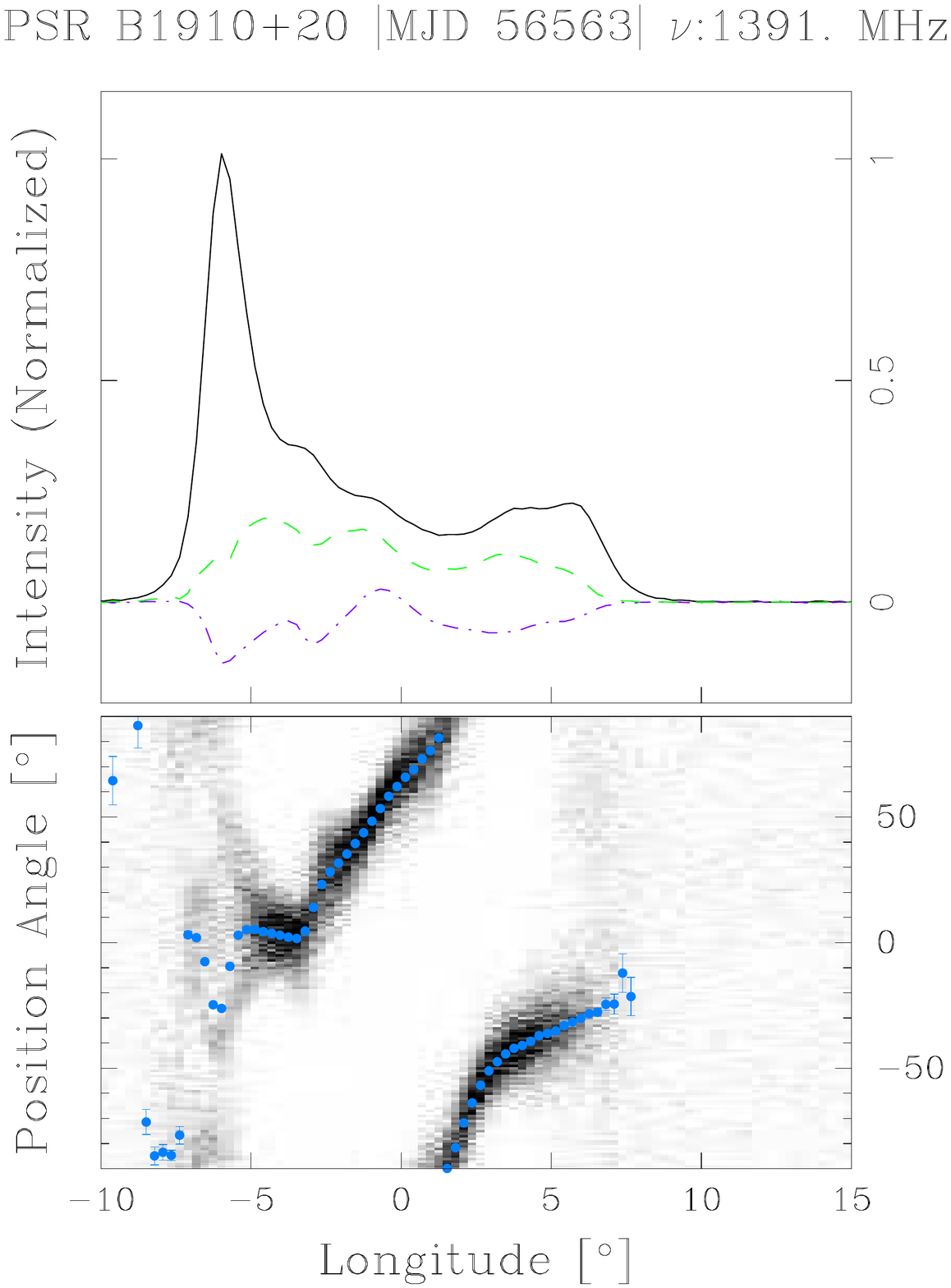} &
\includegraphics[page=1,width=\linewidth]{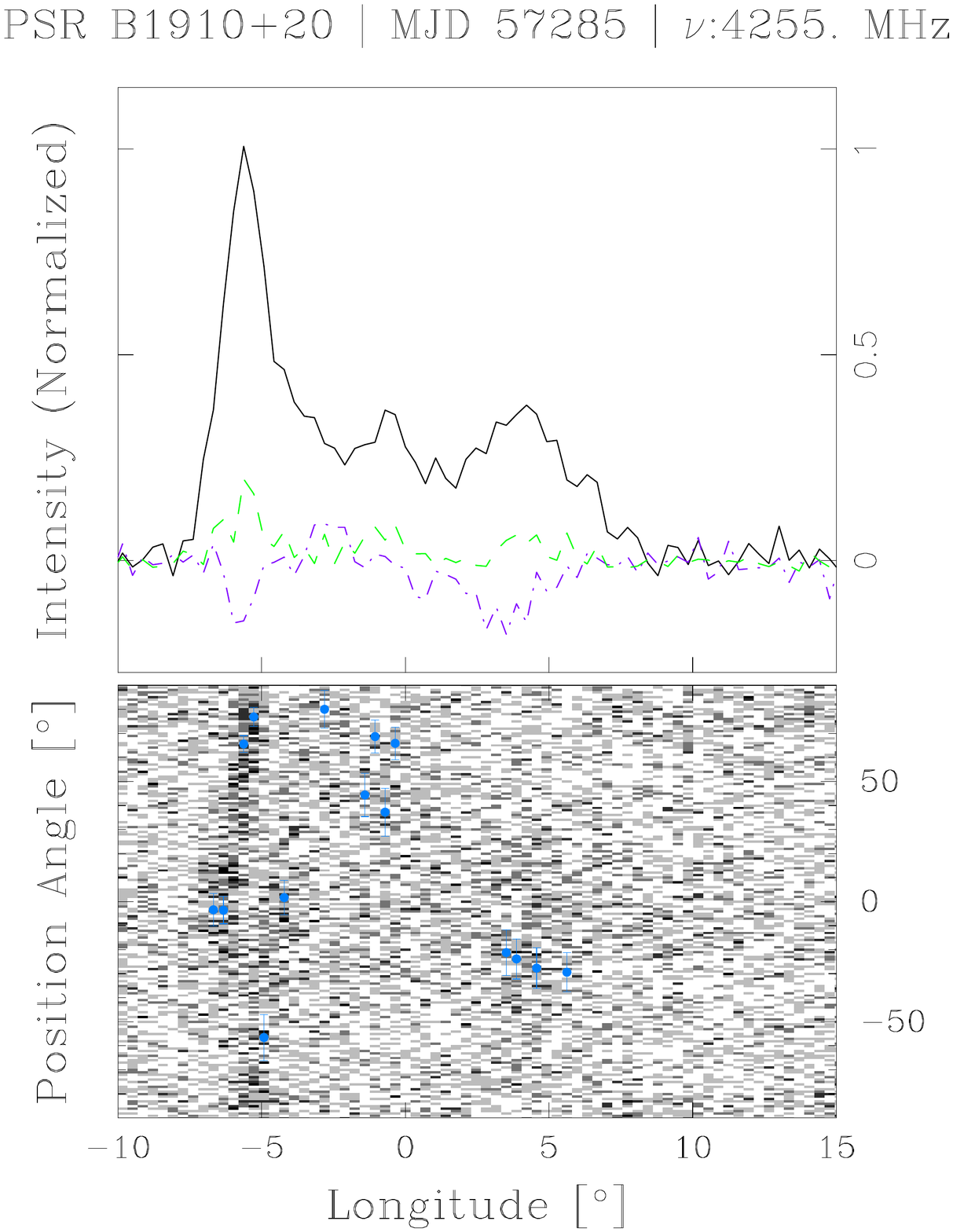} \\ \toprule
\includegraphics[page=1,width=\linewidth]{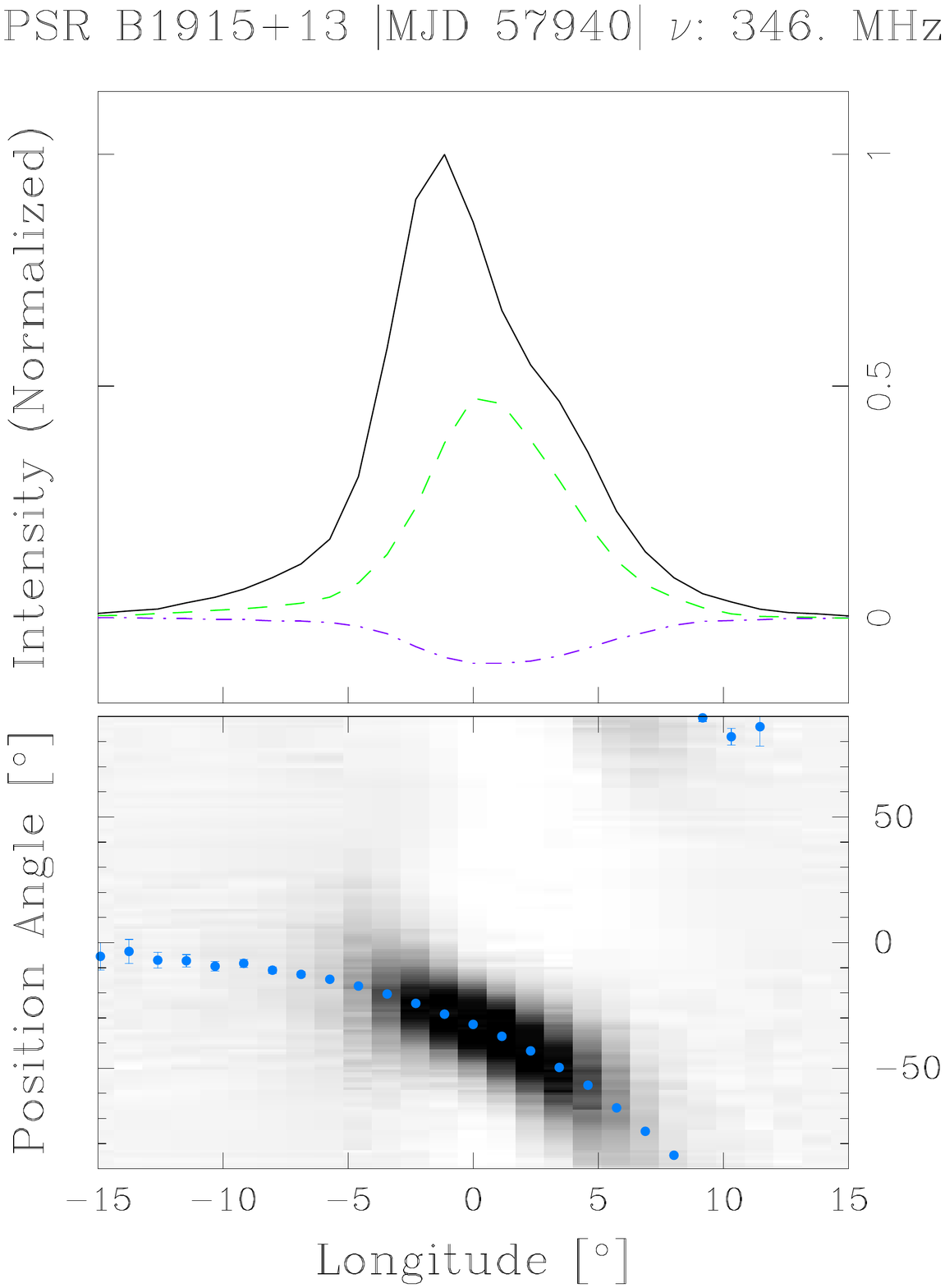} &
\includegraphics[page=1,width=\linewidth]{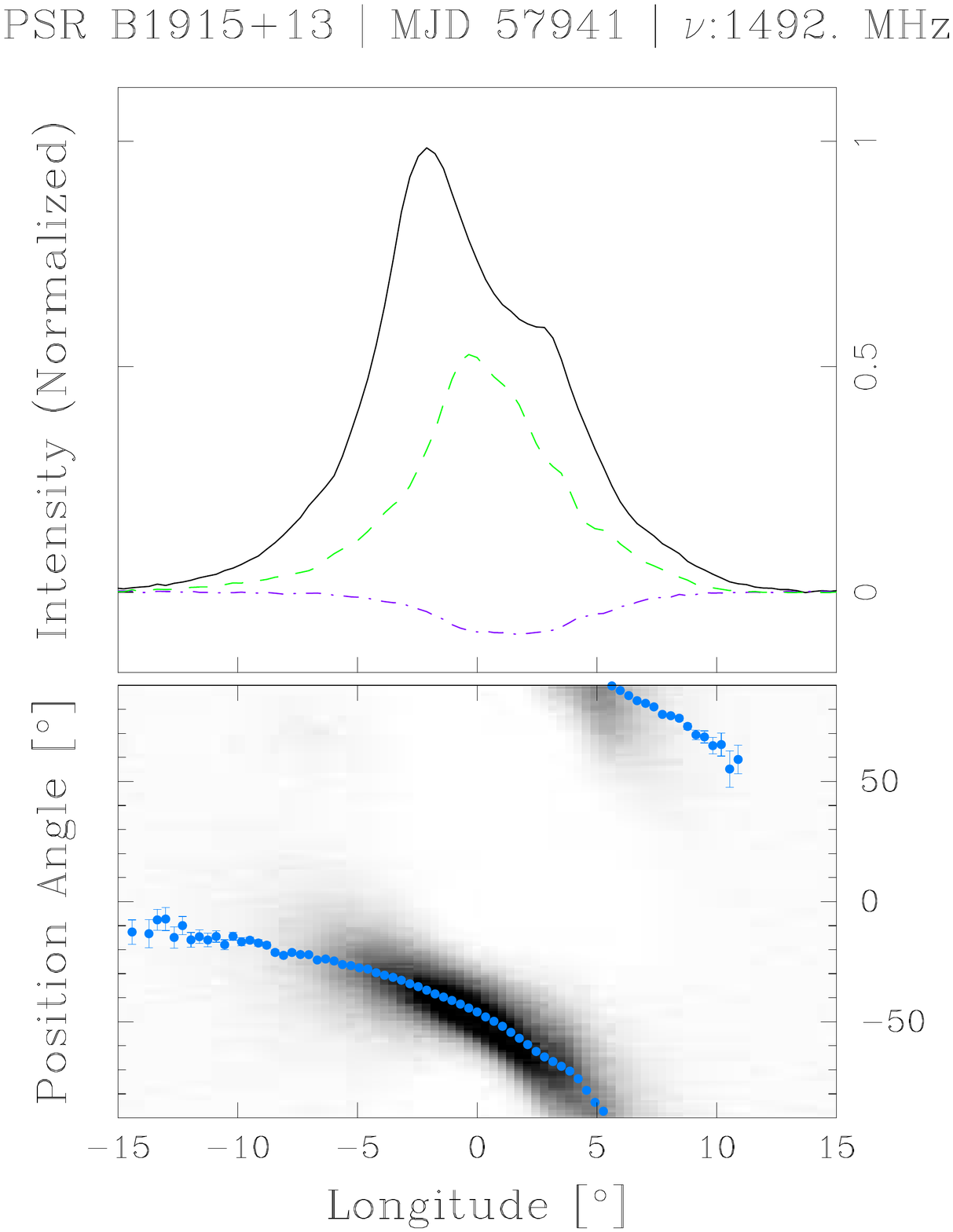} &
\includegraphics[page=1,width=\linewidth]{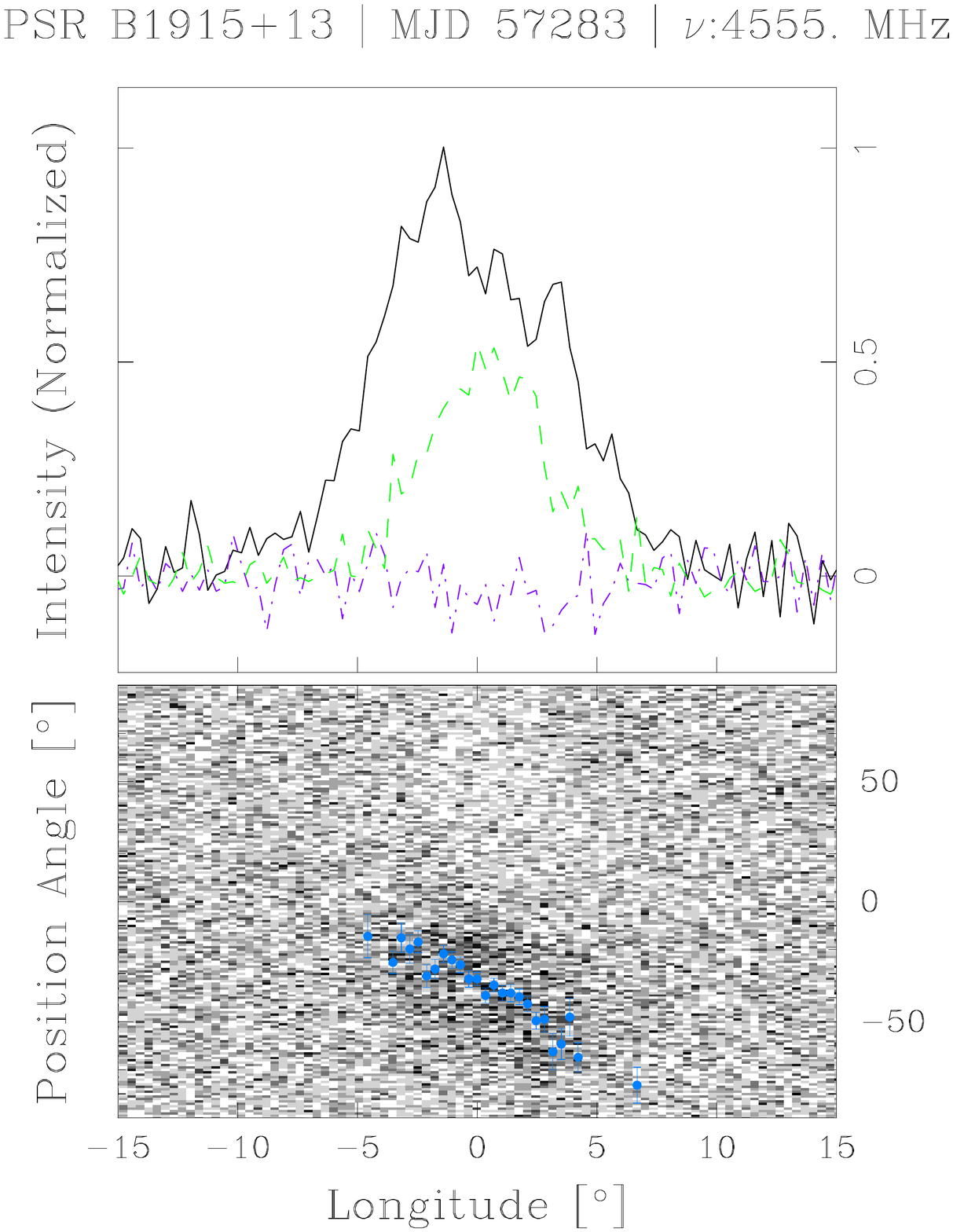} \\ \toprule
\includegraphics[page=1,width=\linewidth]{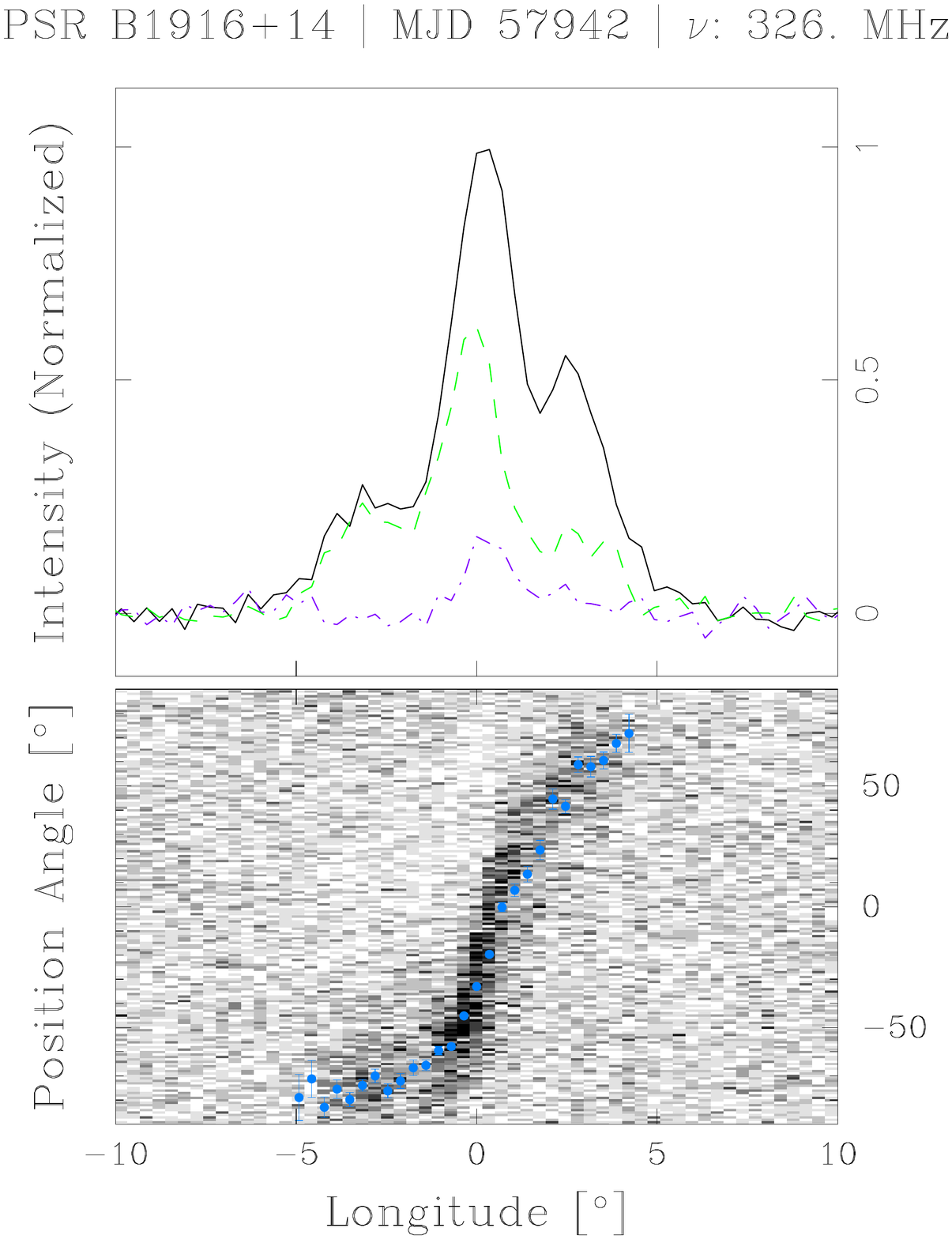} &
\includegraphics[page=1,width=\linewidth]{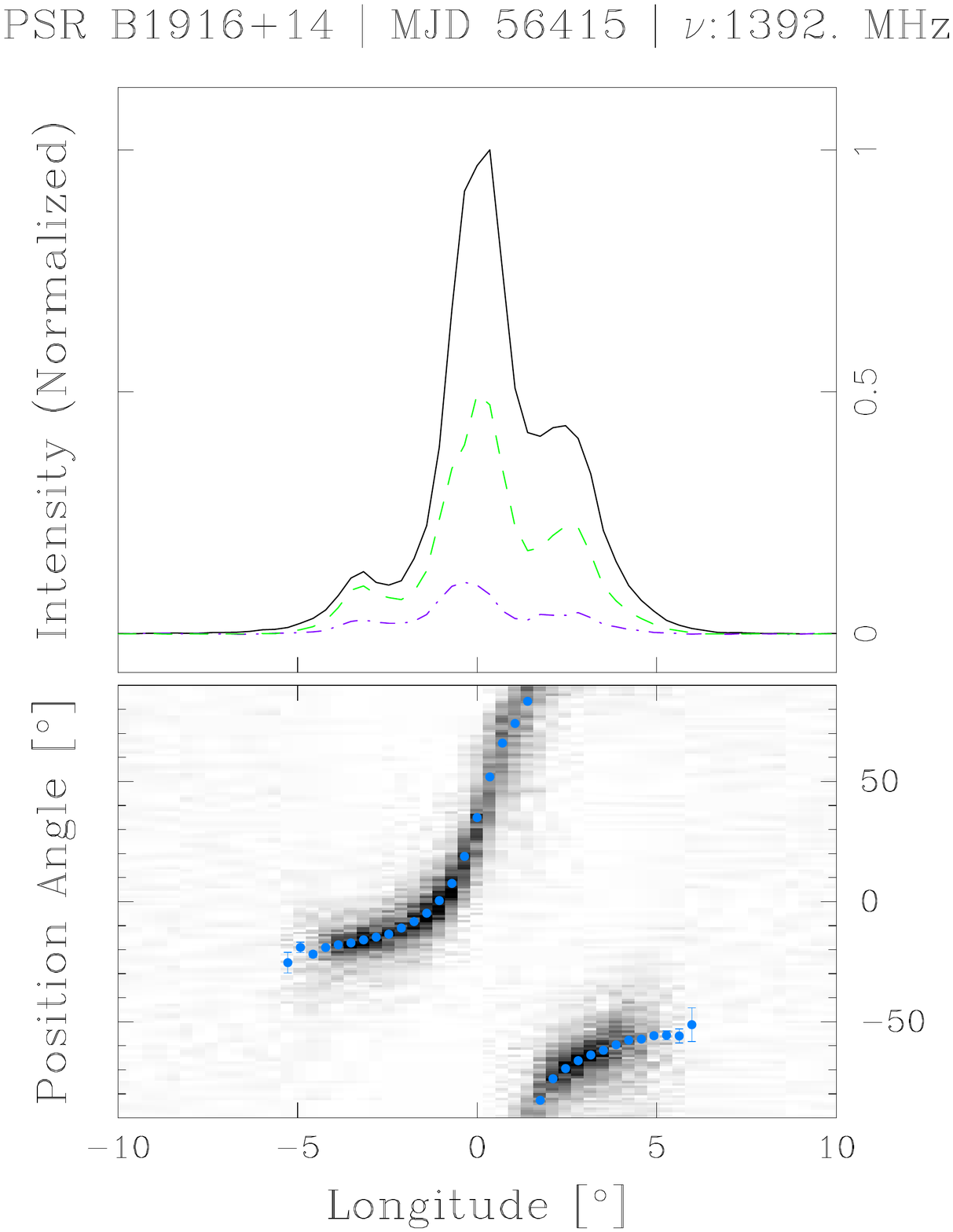} &
\includegraphics[page=1,width=\linewidth]{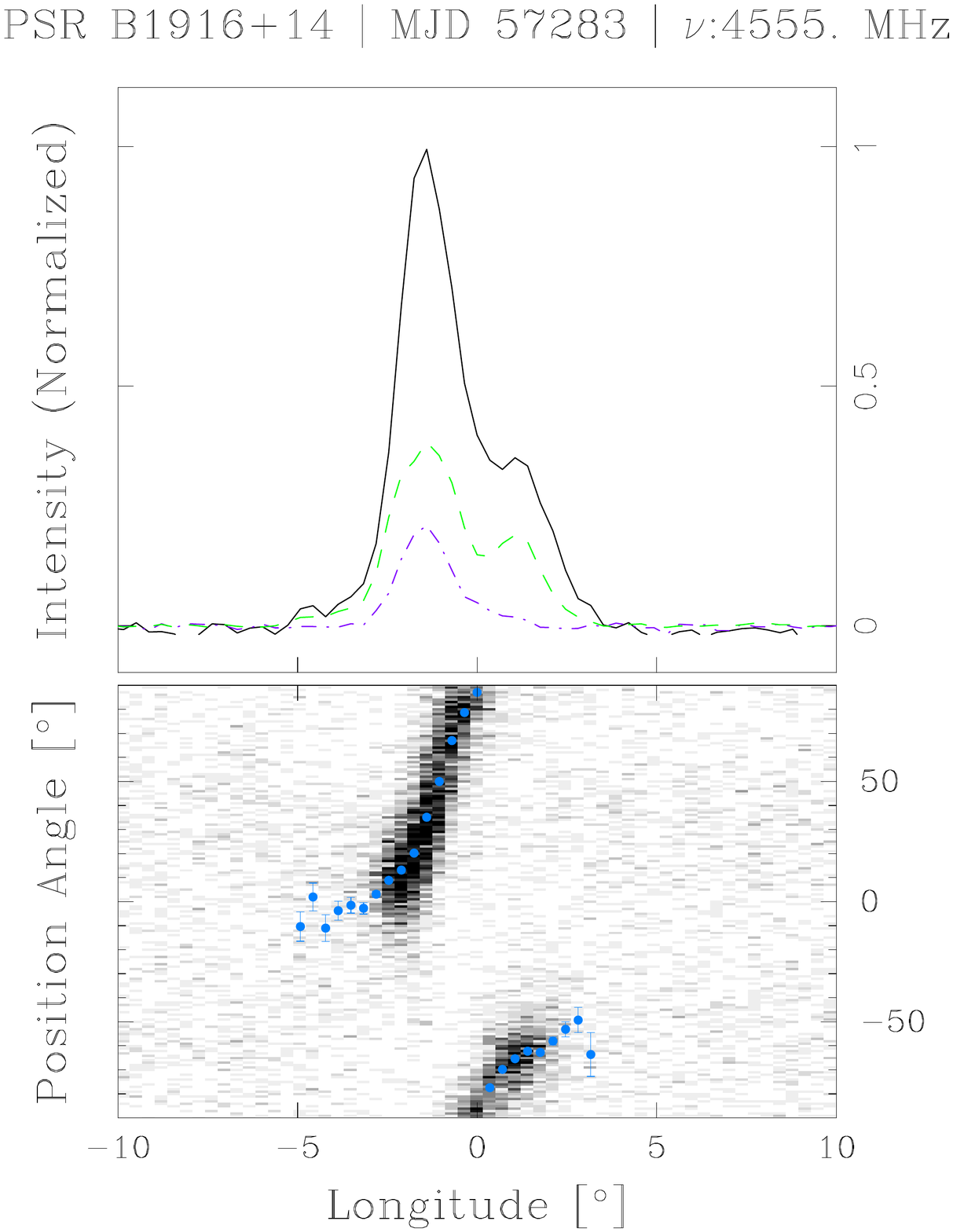} \\ 
     \bottomrule
   \end{tabularx} 
\caption{Average profiles of PSRs B1910+20, B1915+13, and B1916+14.}
 \end{figure*}
\vspace{1cm}
   \begin{figure*} 
 \begin{tabularx}{\textwidth}{YYY}
    \multicolumn{3}{c}{} \\ \toprule
\includegraphics[page=1,width=\linewidth]{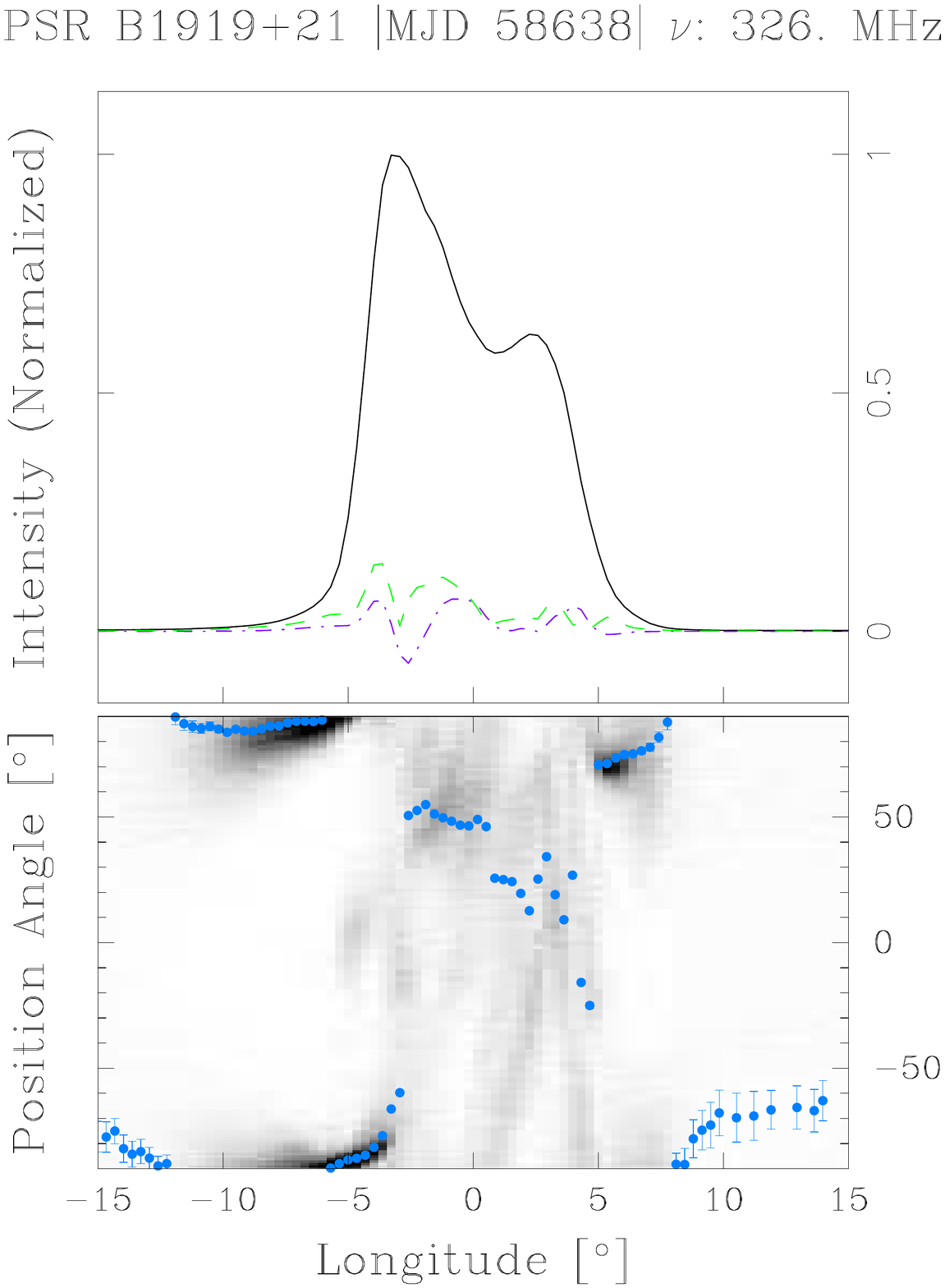} &
\includegraphics[page=1,width=\linewidth]{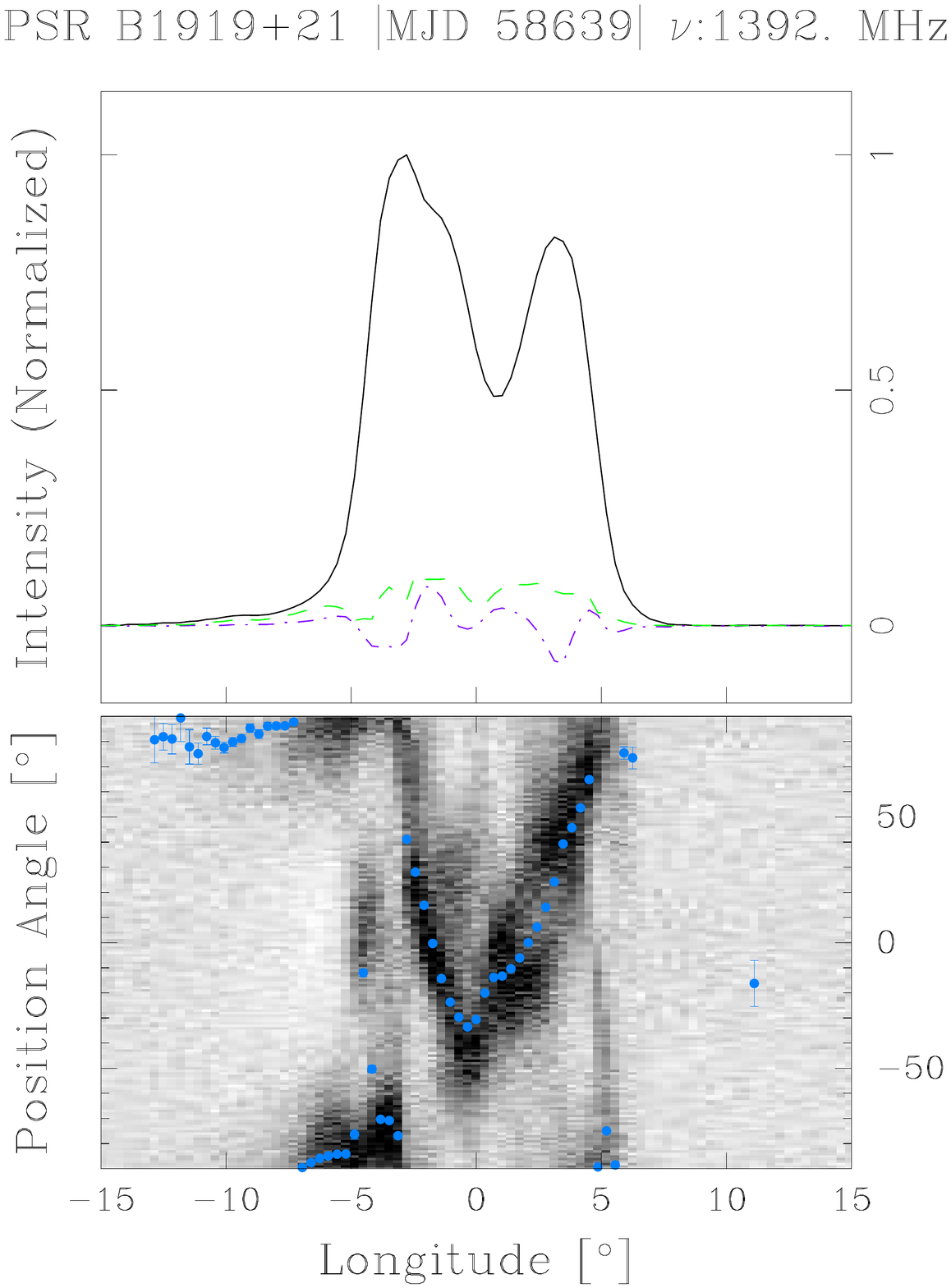} &
\includegraphics[page=1,width=\linewidth]{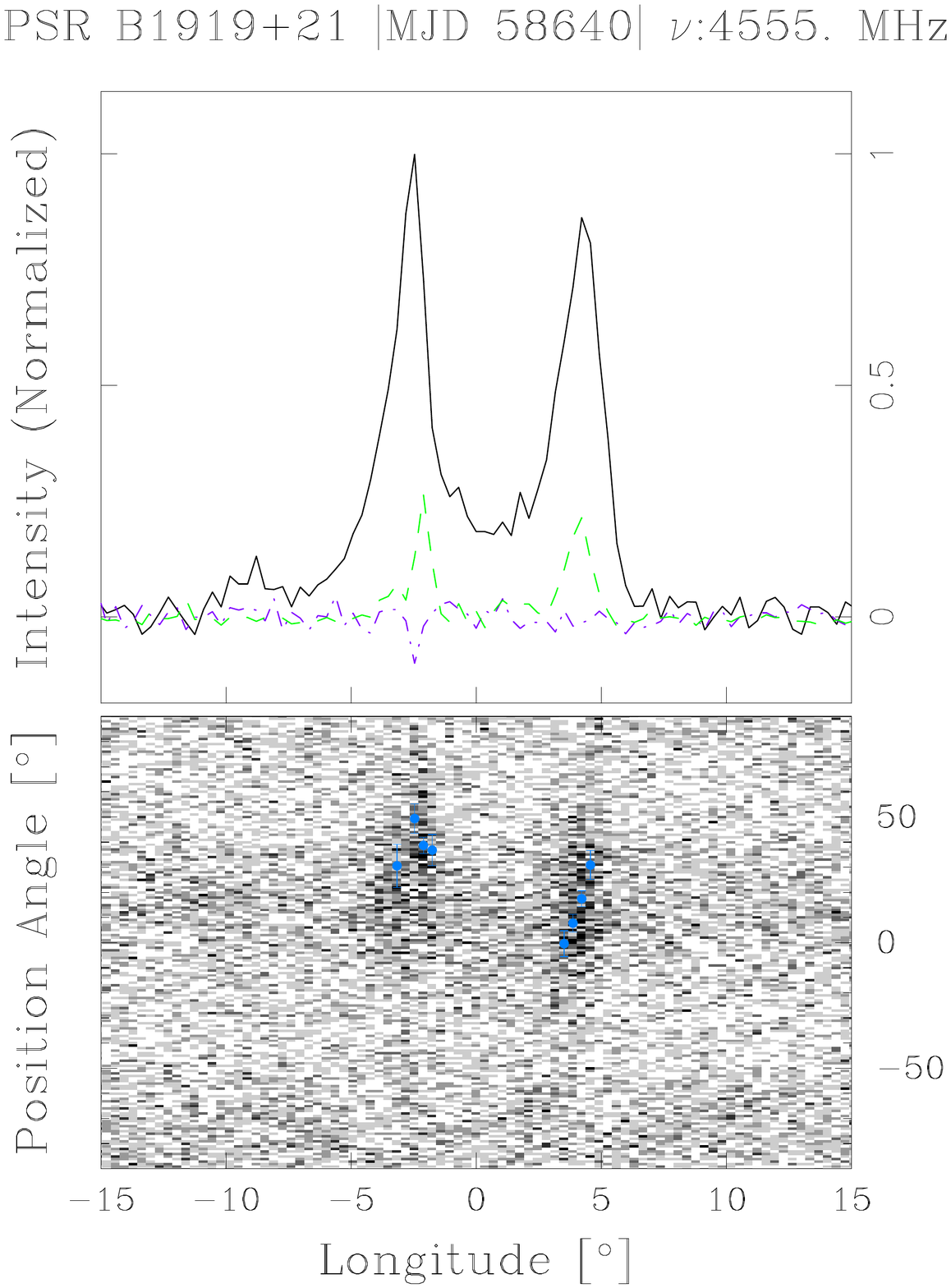} \\ \toprule
\includegraphics[page=1,width=\linewidth]{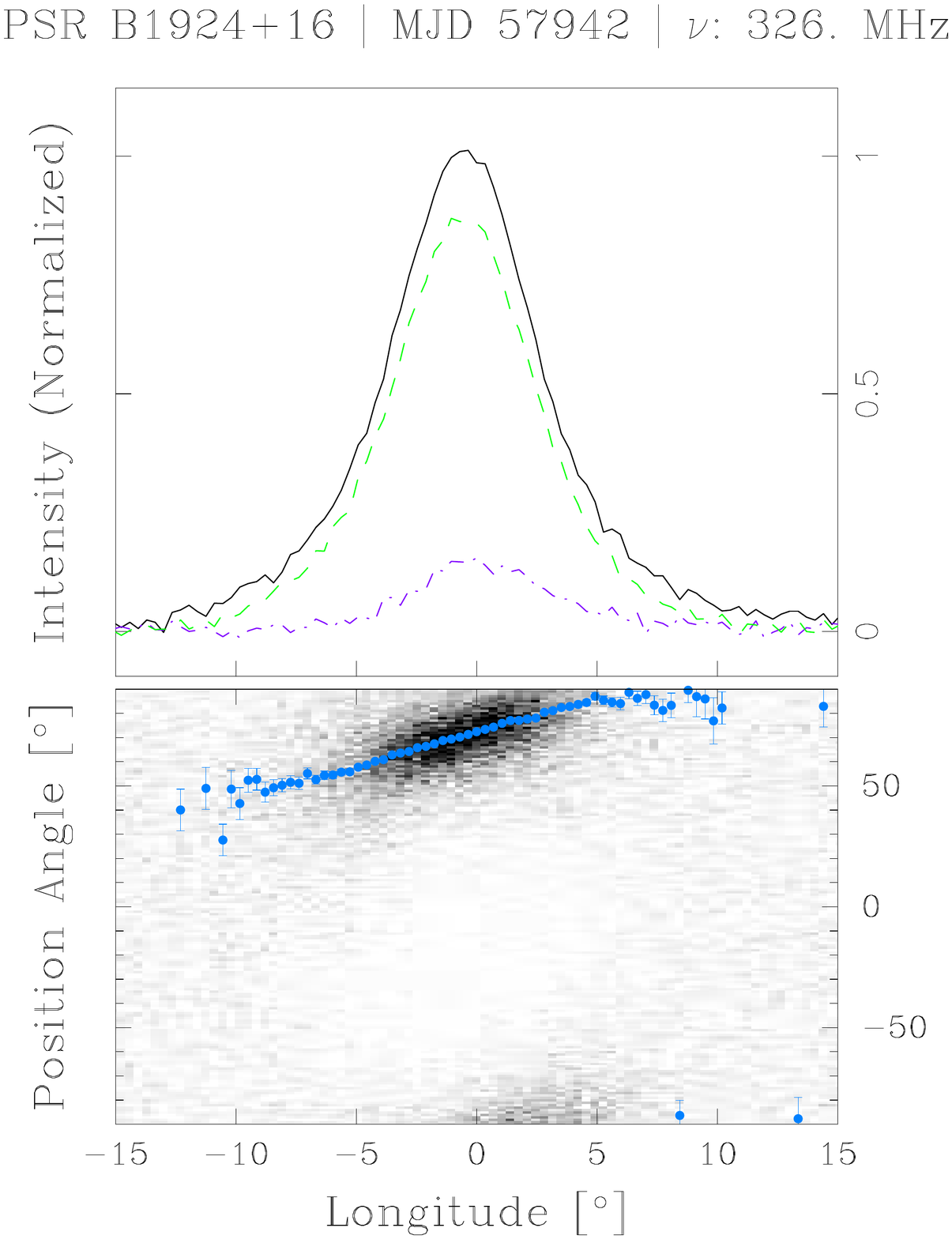} &
\includegraphics[page=1,width=\linewidth]{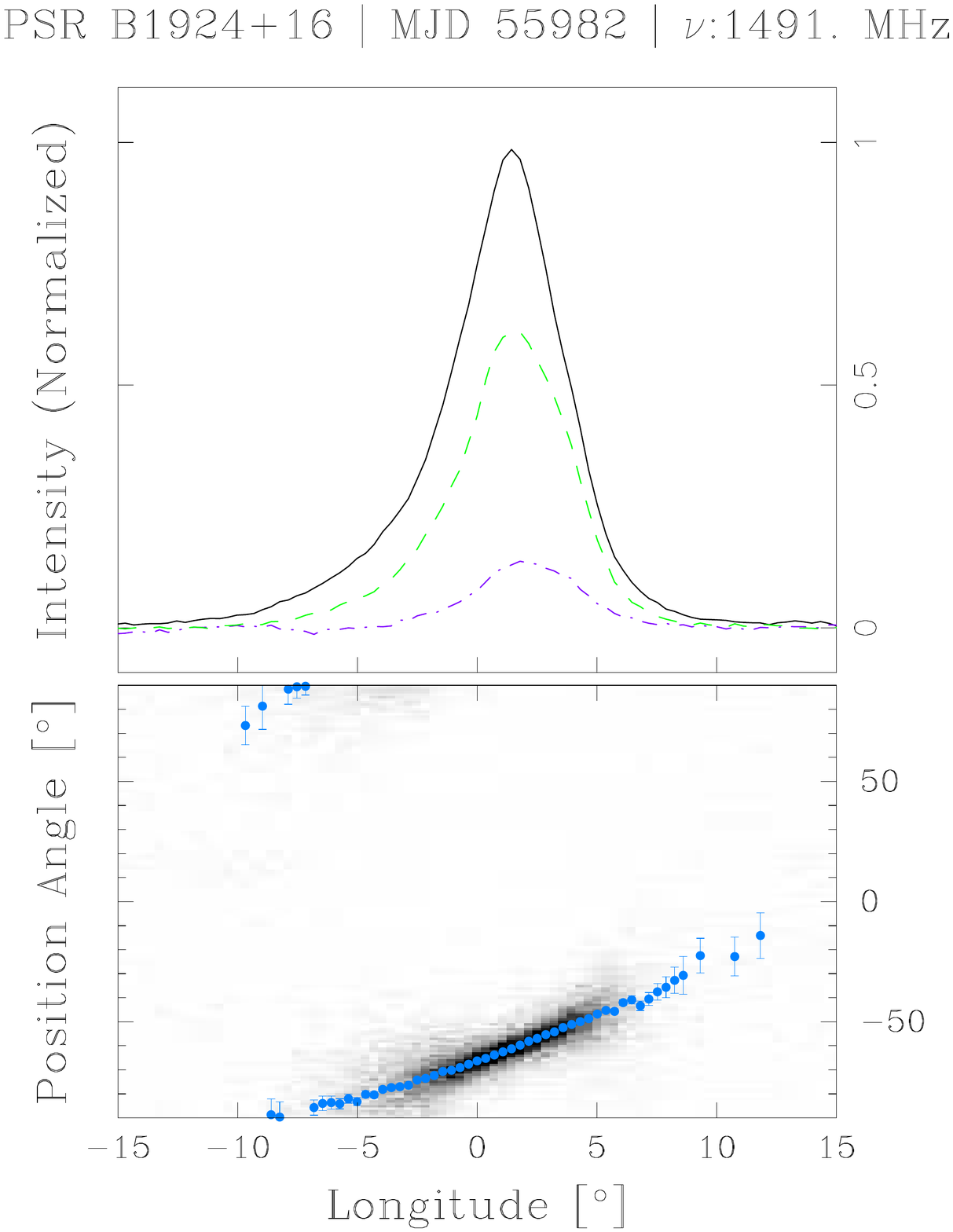} &
\includegraphics[page=1,width=\linewidth]{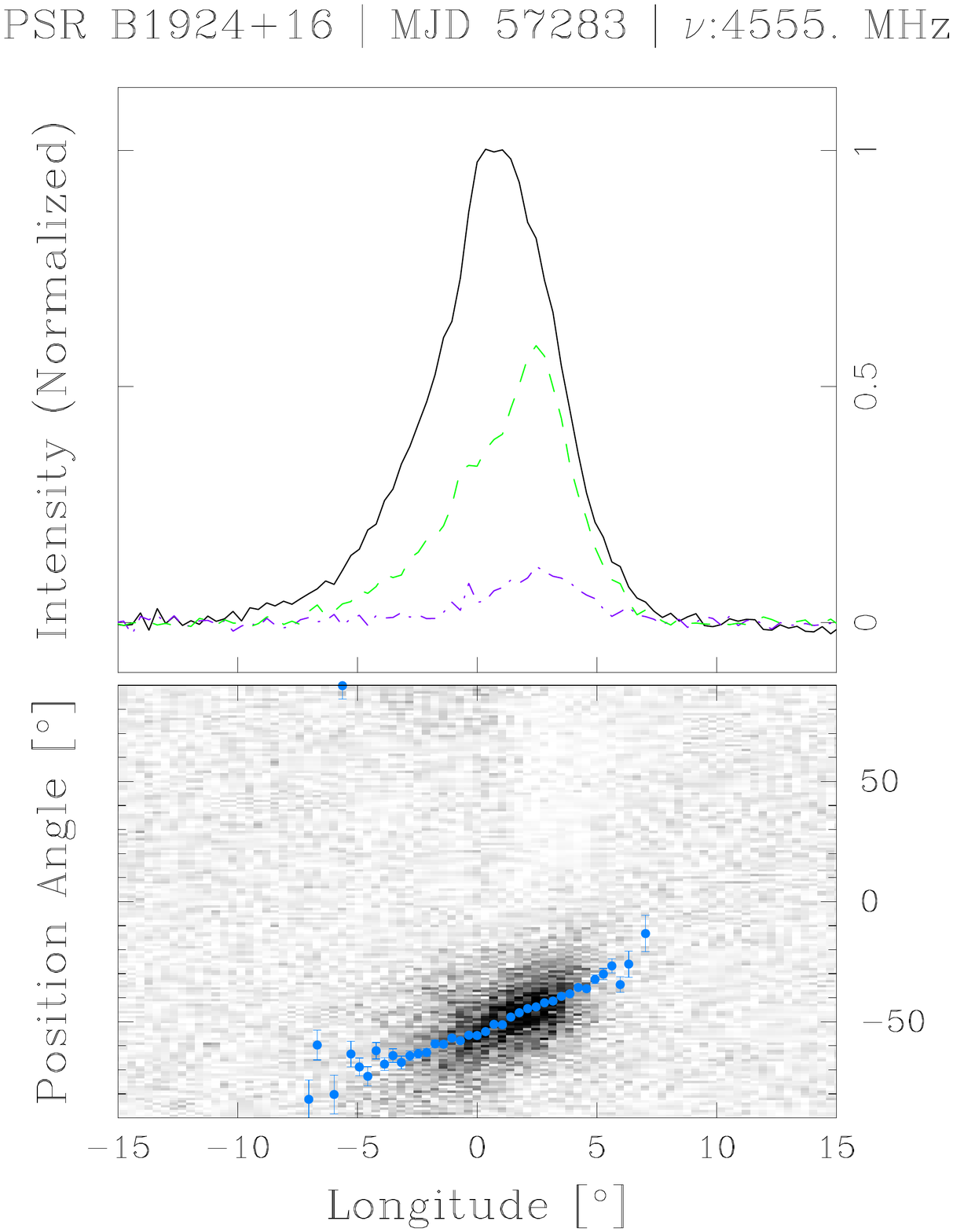} \\ \toprule
\includegraphics[page=1,width=\linewidth]{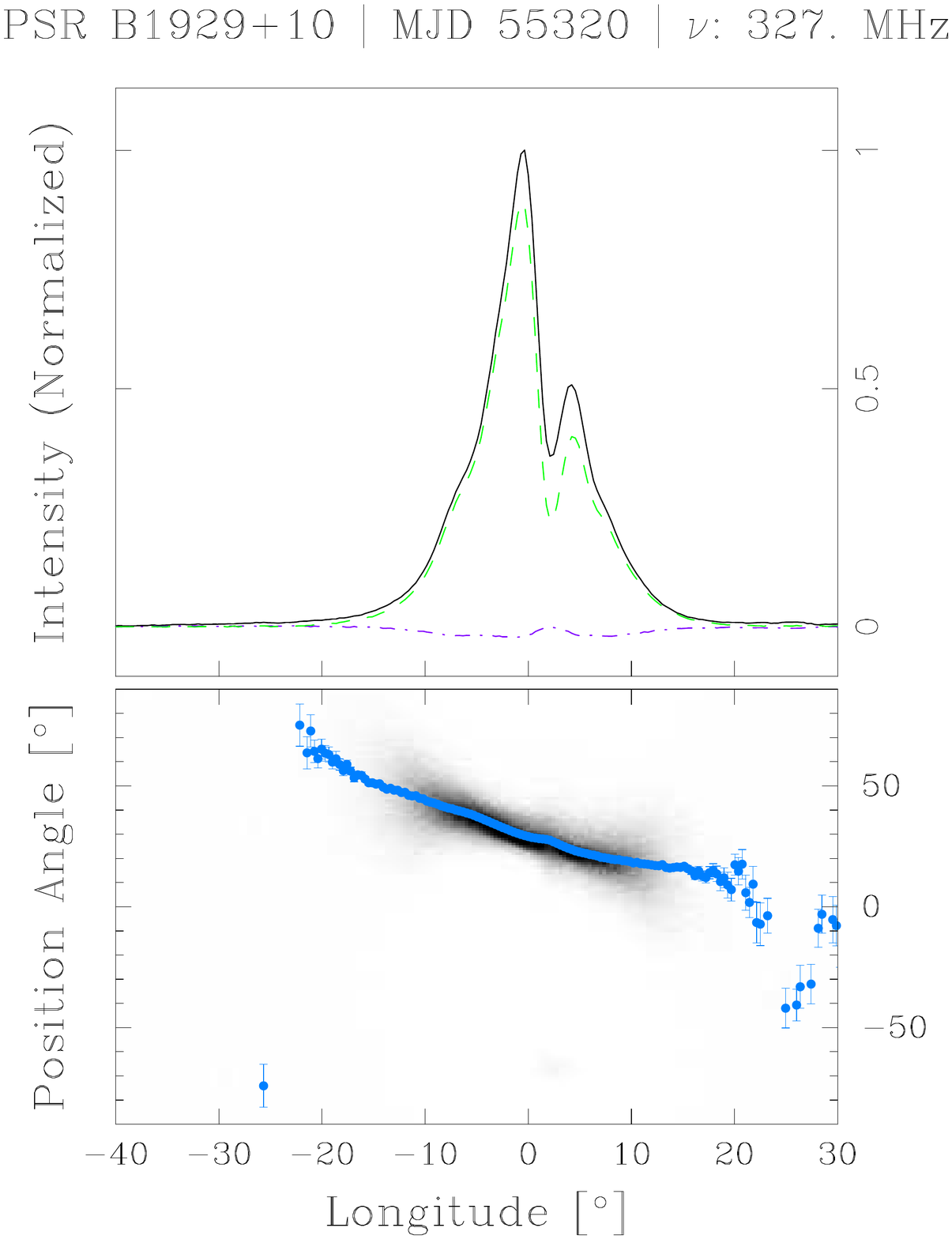} &
\includegraphics[page=1,width=\linewidth]{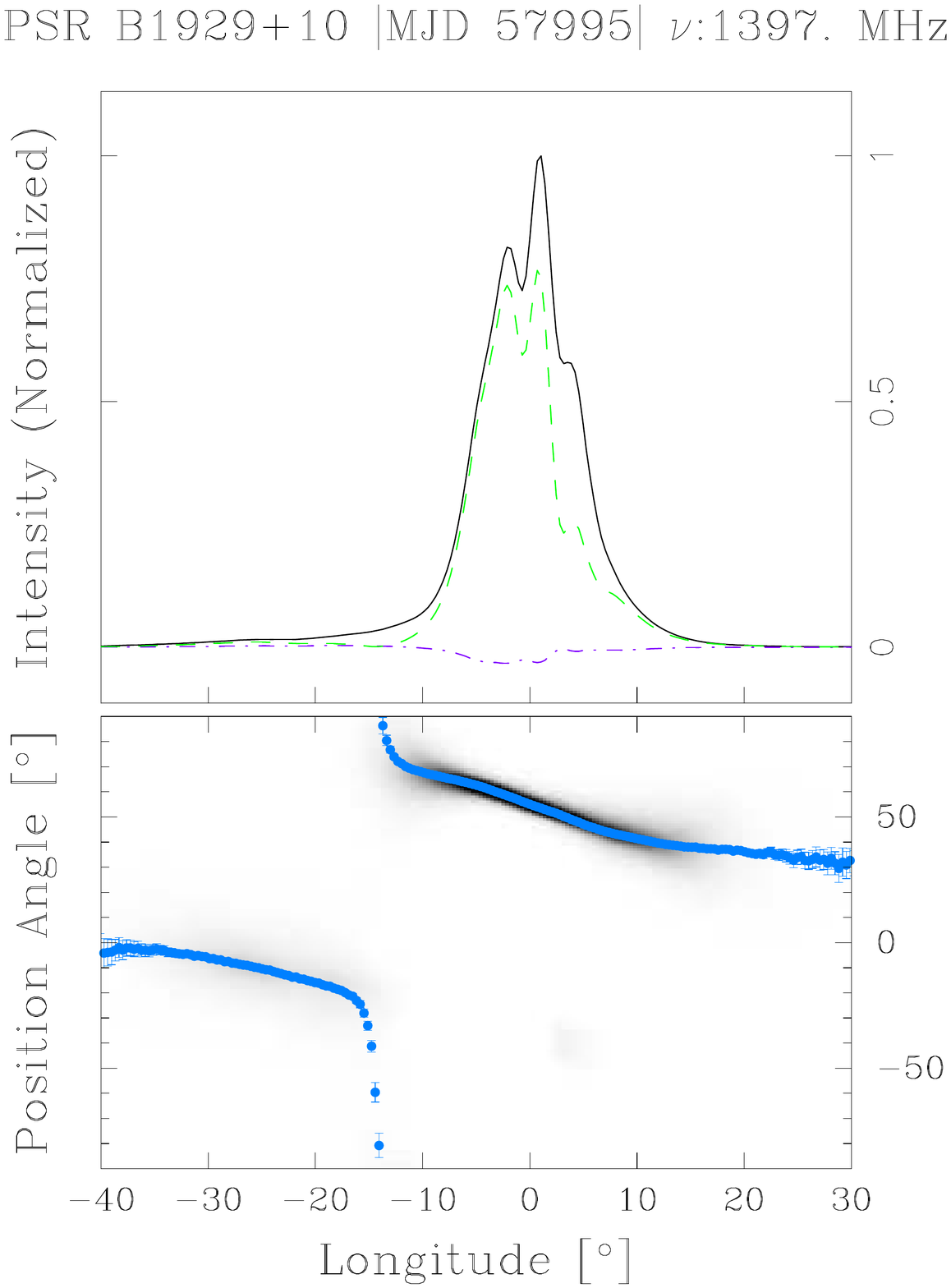} &
\includegraphics[page=1,width=\linewidth]{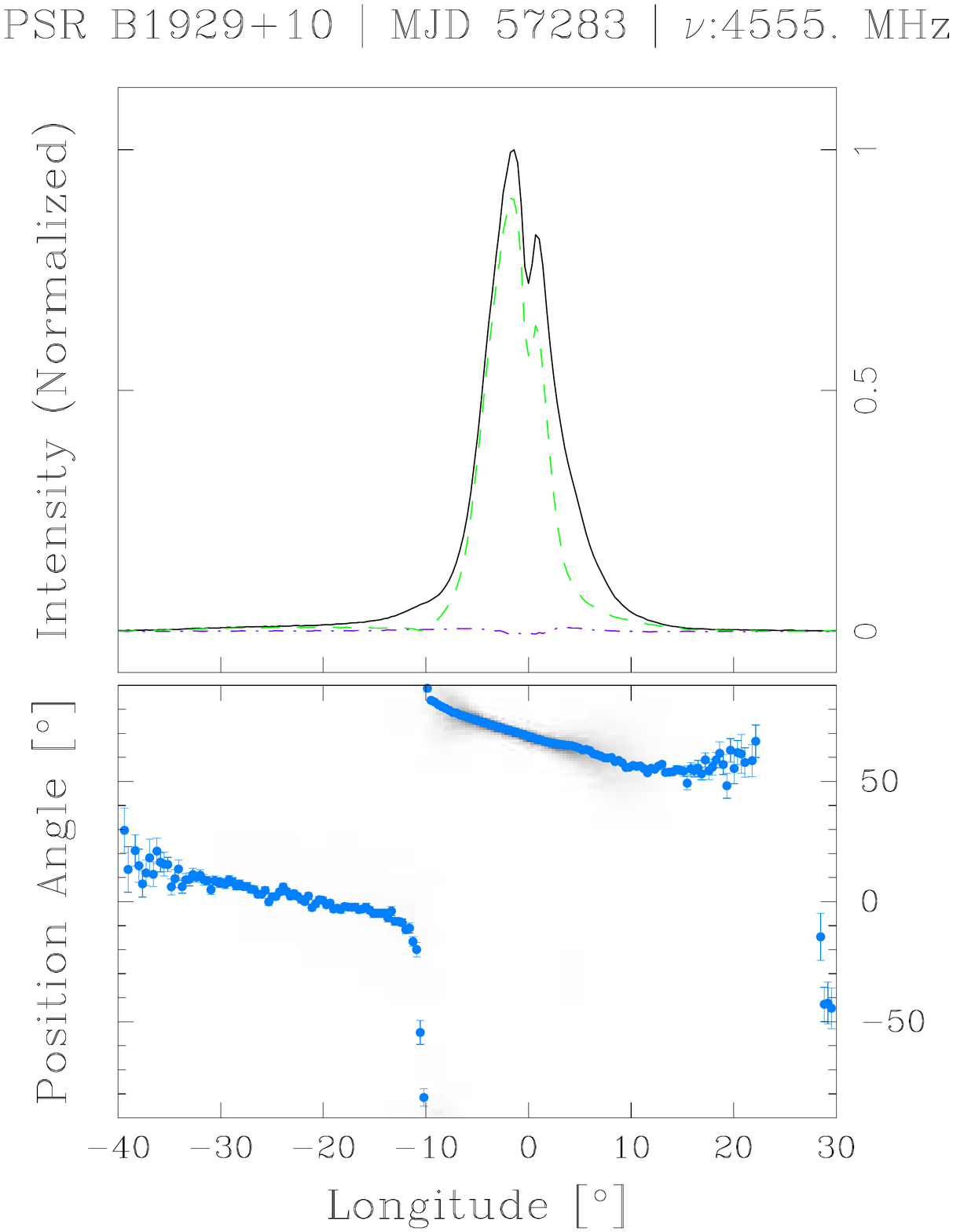} \\ 
     \bottomrule
   \end{tabularx} 
\caption{Average profiles of PSRs B1919+21, B1924+16, and B1929+10.}
 \end{figure*}
\vspace{1cm}
   \begin{figure*} 
 \begin{tabularx}{\textwidth}{YYY}
    \multicolumn{3}{c}{} \\ \toprule
                &
\includegraphics[page=1,width=\linewidth]{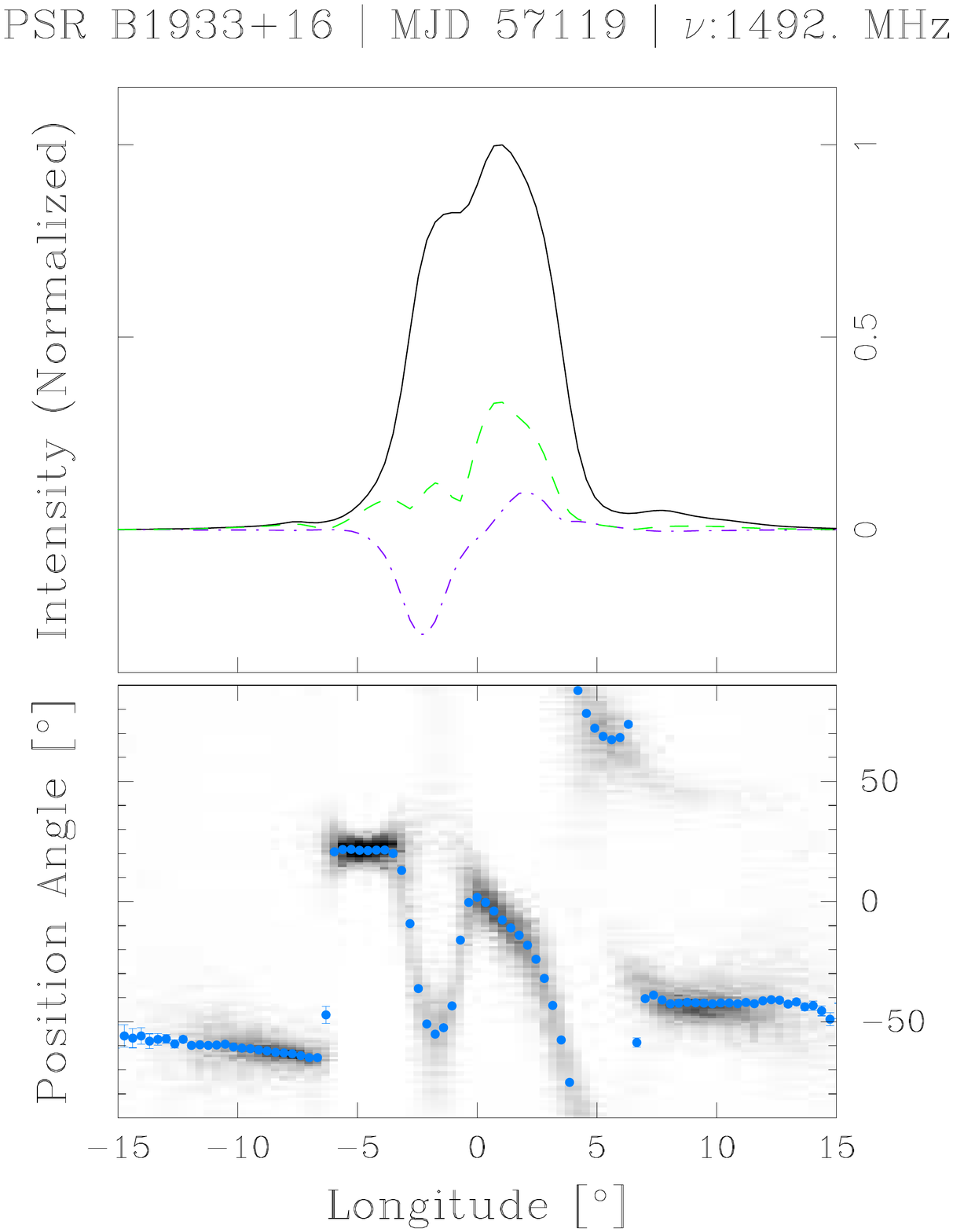} &
\includegraphics[page=1,width=\linewidth]{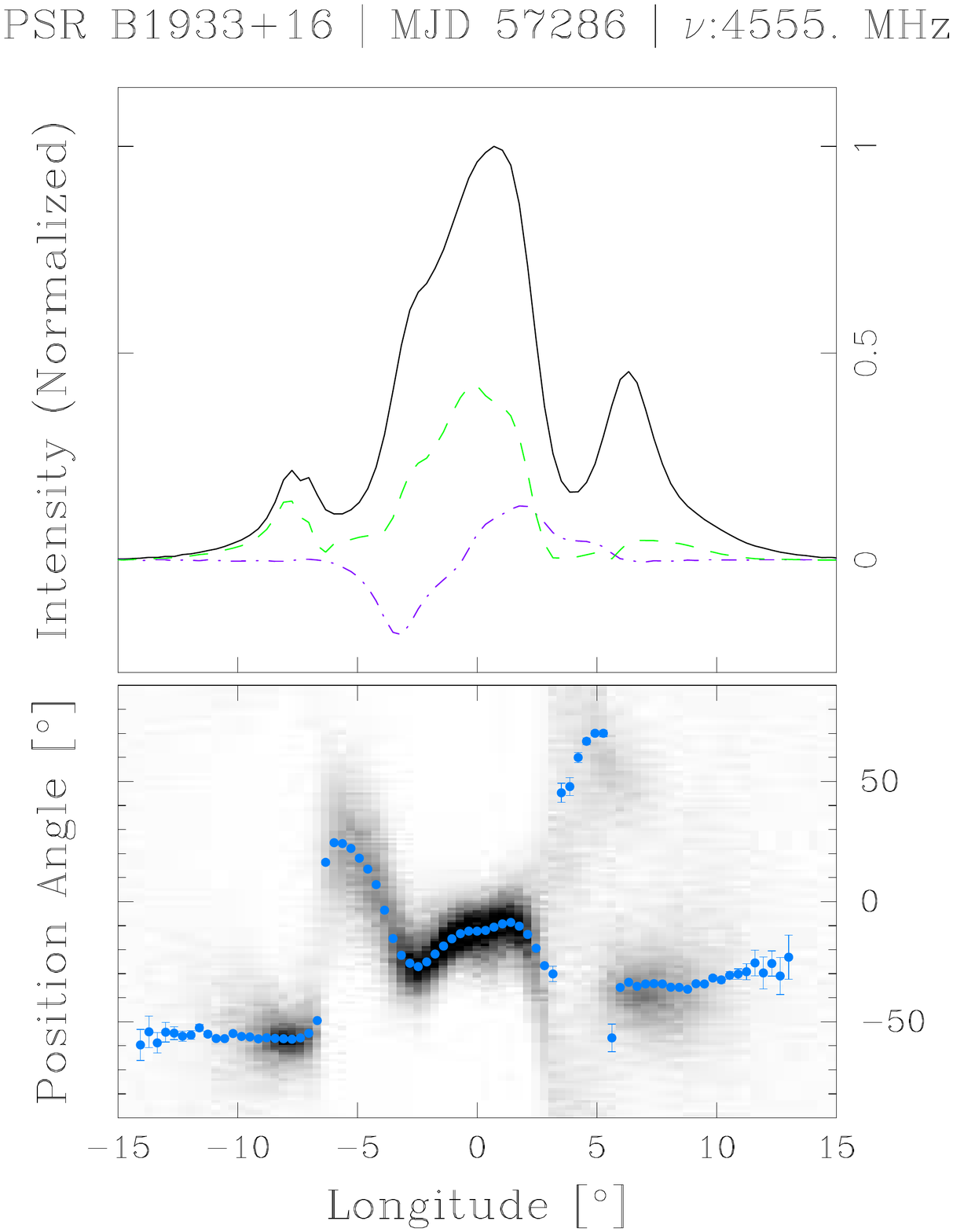} \\ \toprule
\includegraphics[page=1,width=\linewidth]{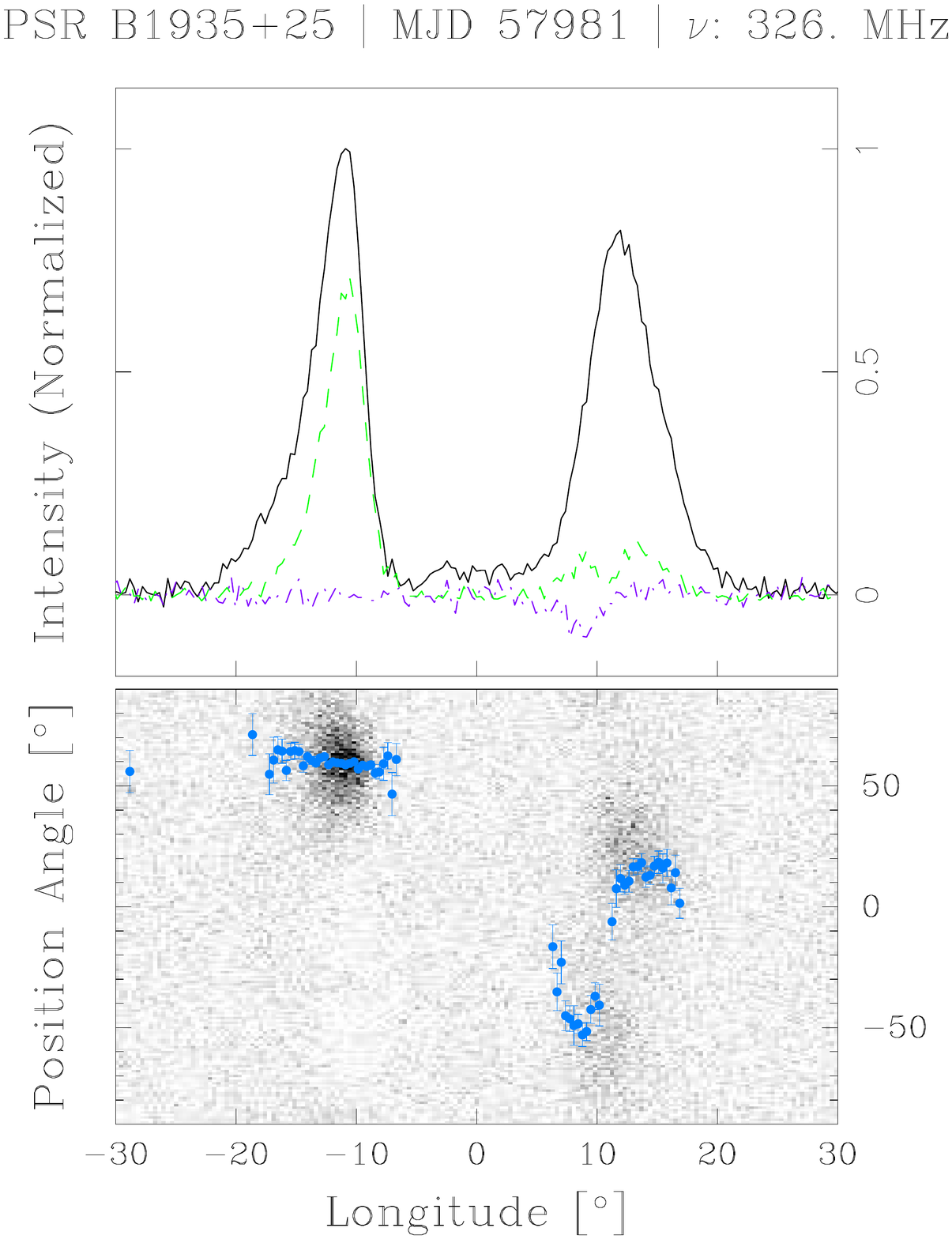} &
\includegraphics[page=1,width=\linewidth]{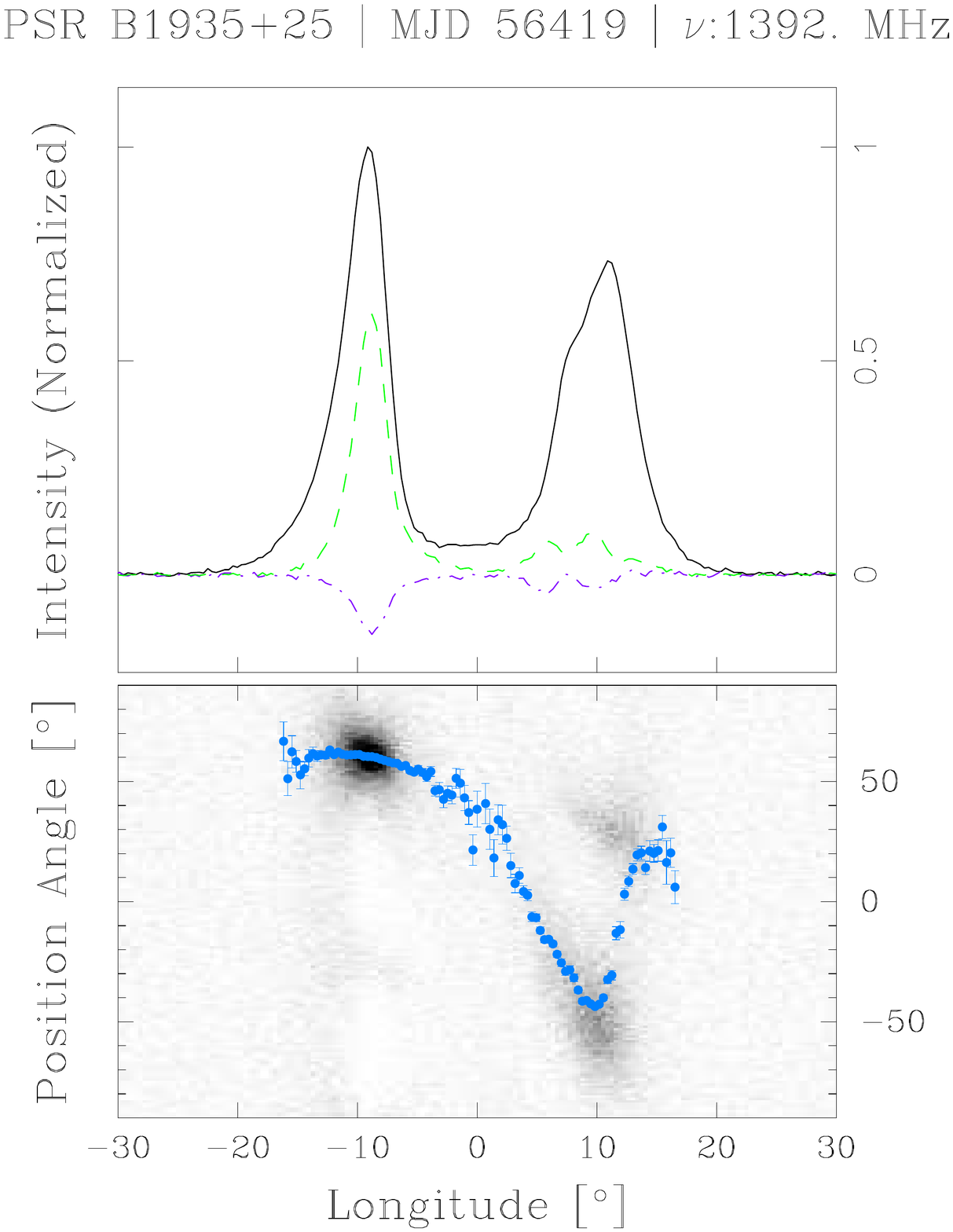} &
\includegraphics[page=1,width=\linewidth]{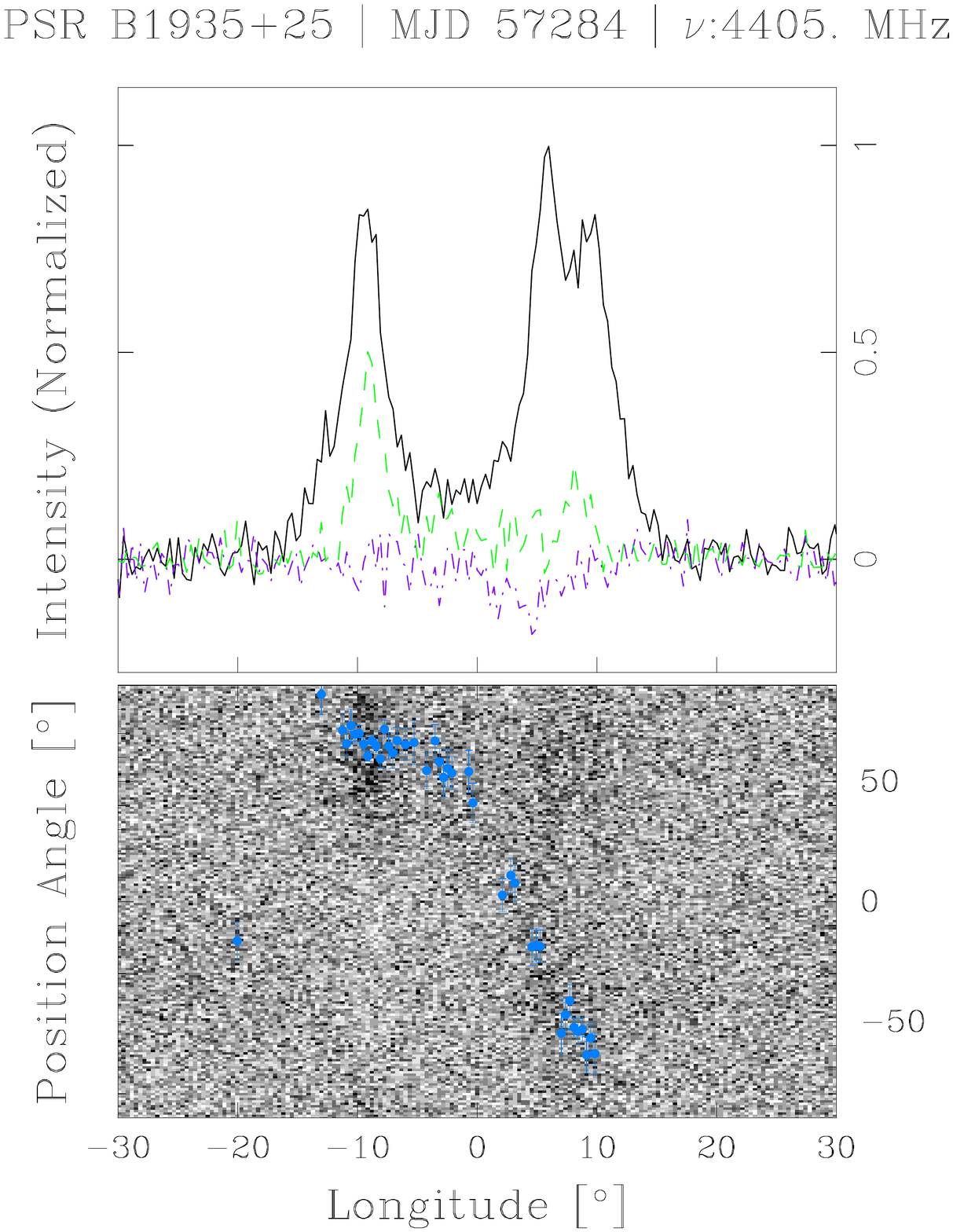} \\ \toprule
                &
\includegraphics[page=1,width=\linewidth]{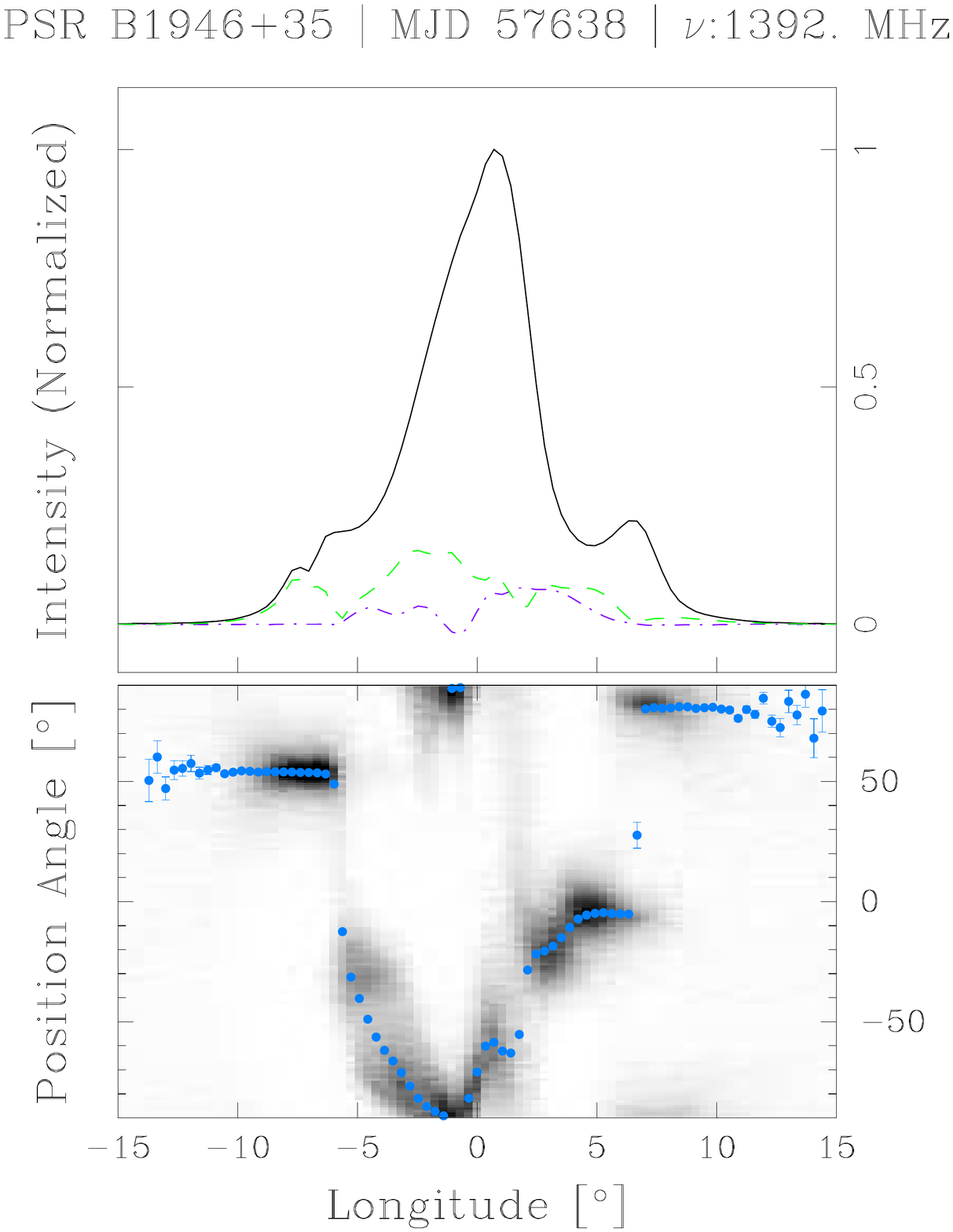} &
\includegraphics[page=1,width=\linewidth]{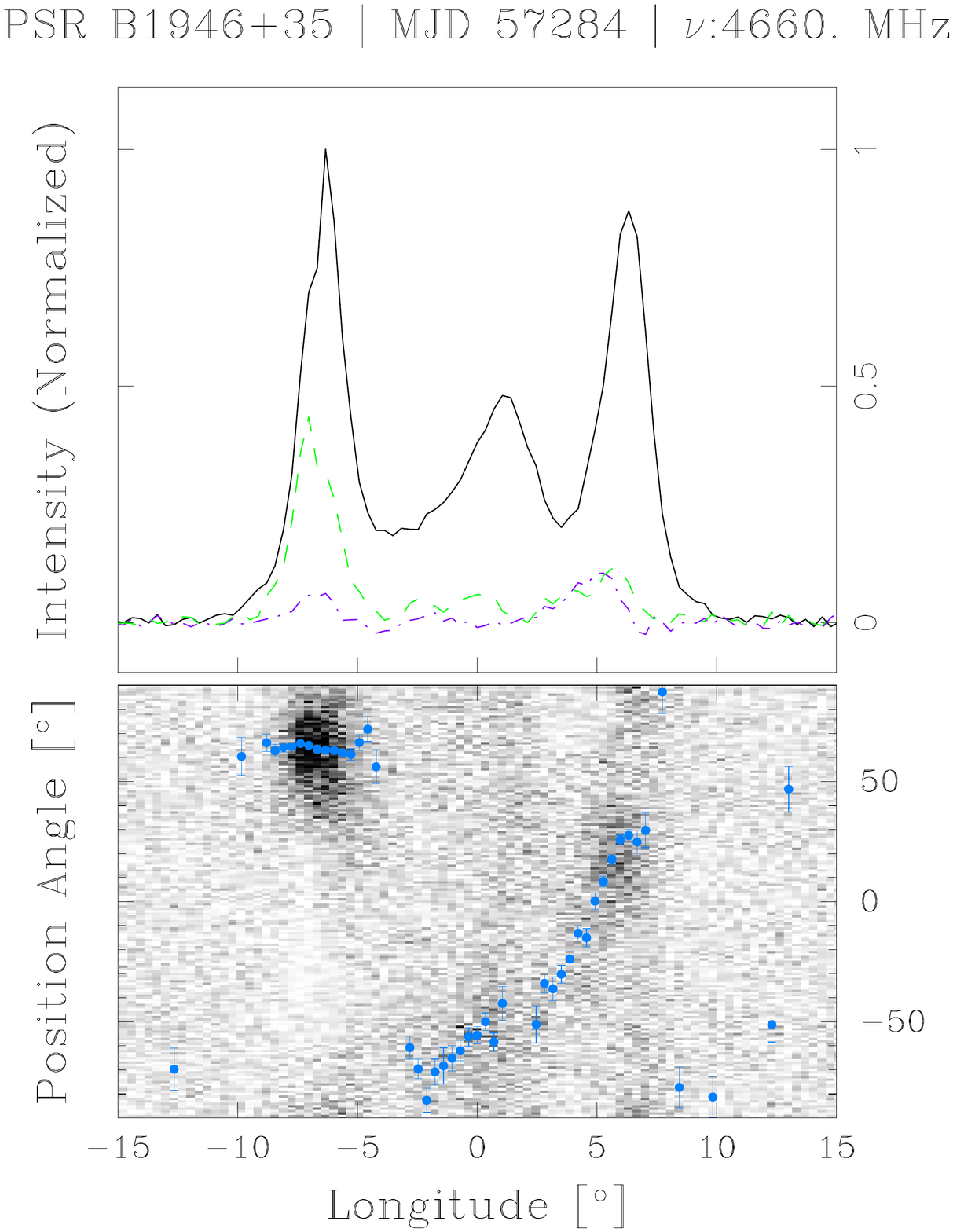} \\ 
     \bottomrule
   \end{tabularx} 
\caption{Average profiles of PSRs B1933+16, B1935+25, and B1946+35.}
 \end{figure*}
\vspace{1cm}

   \begin{figure*} 
 \begin{tabularx}{\textwidth}{YYY}
    \multicolumn{3}{c}{} \\ \toprule
    \includegraphics[page=1,width=\linewidth]{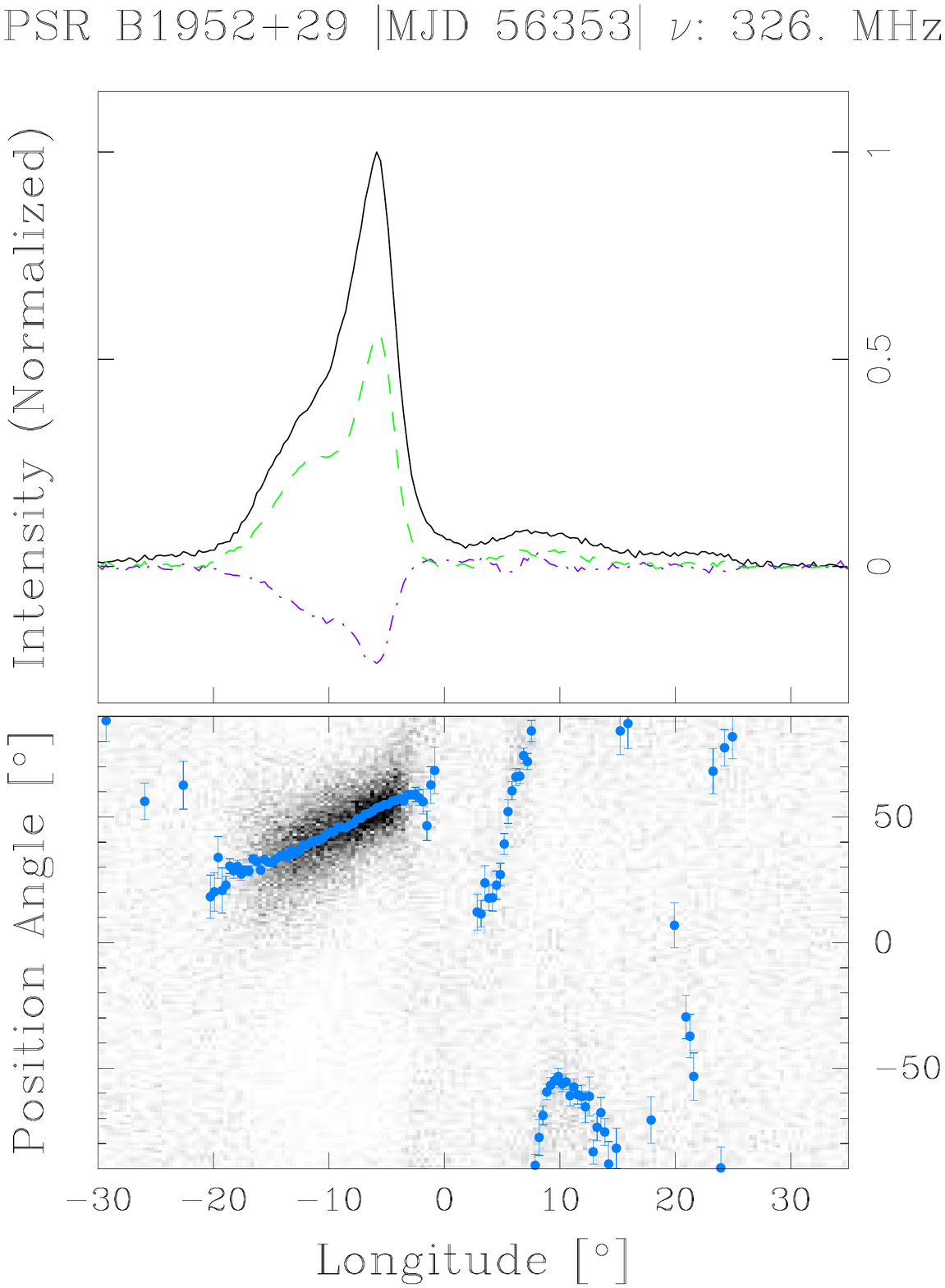} &
\includegraphics[page=1,width=\linewidth]{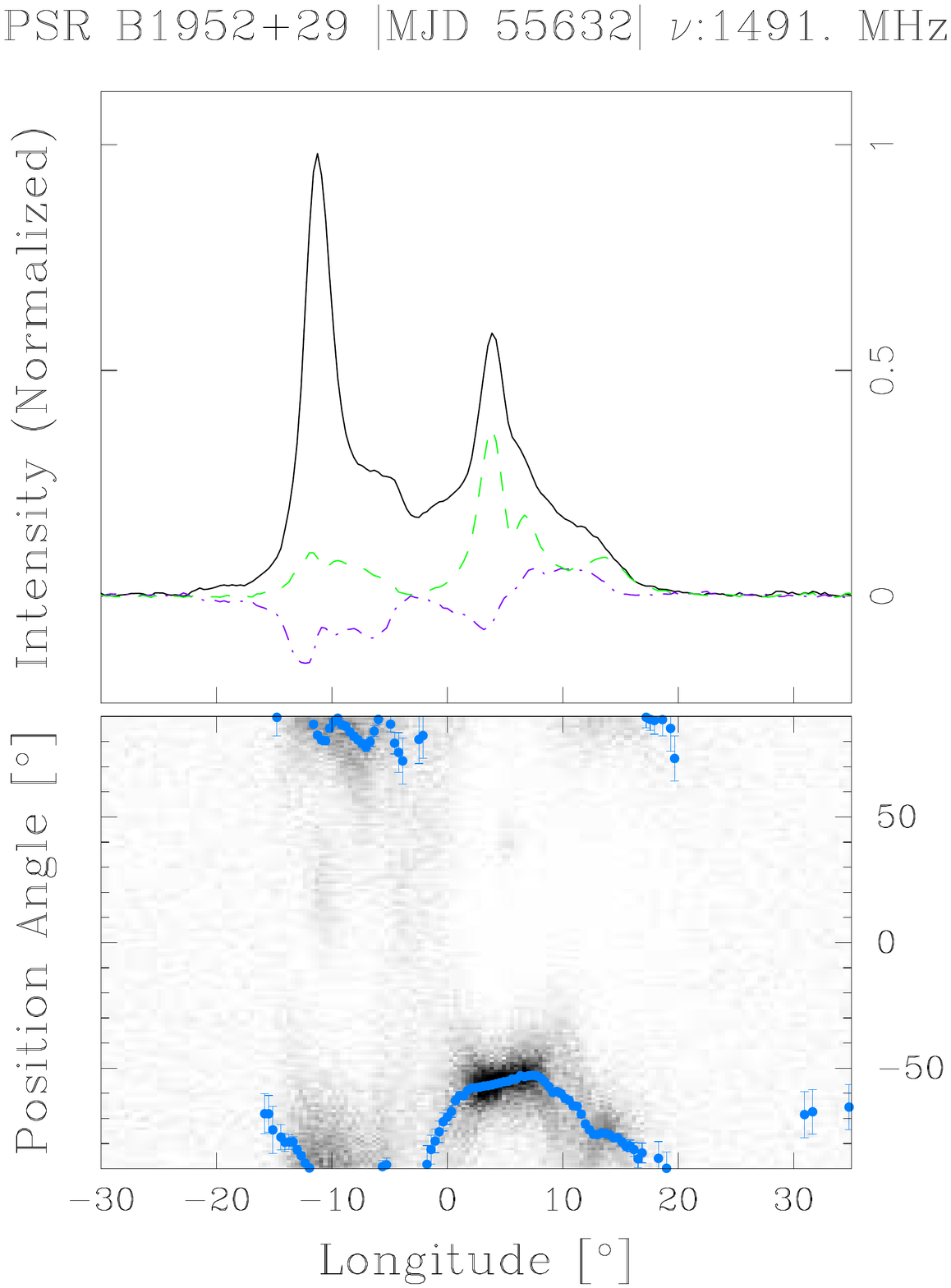} &
\includegraphics[page=1,width=\linewidth]{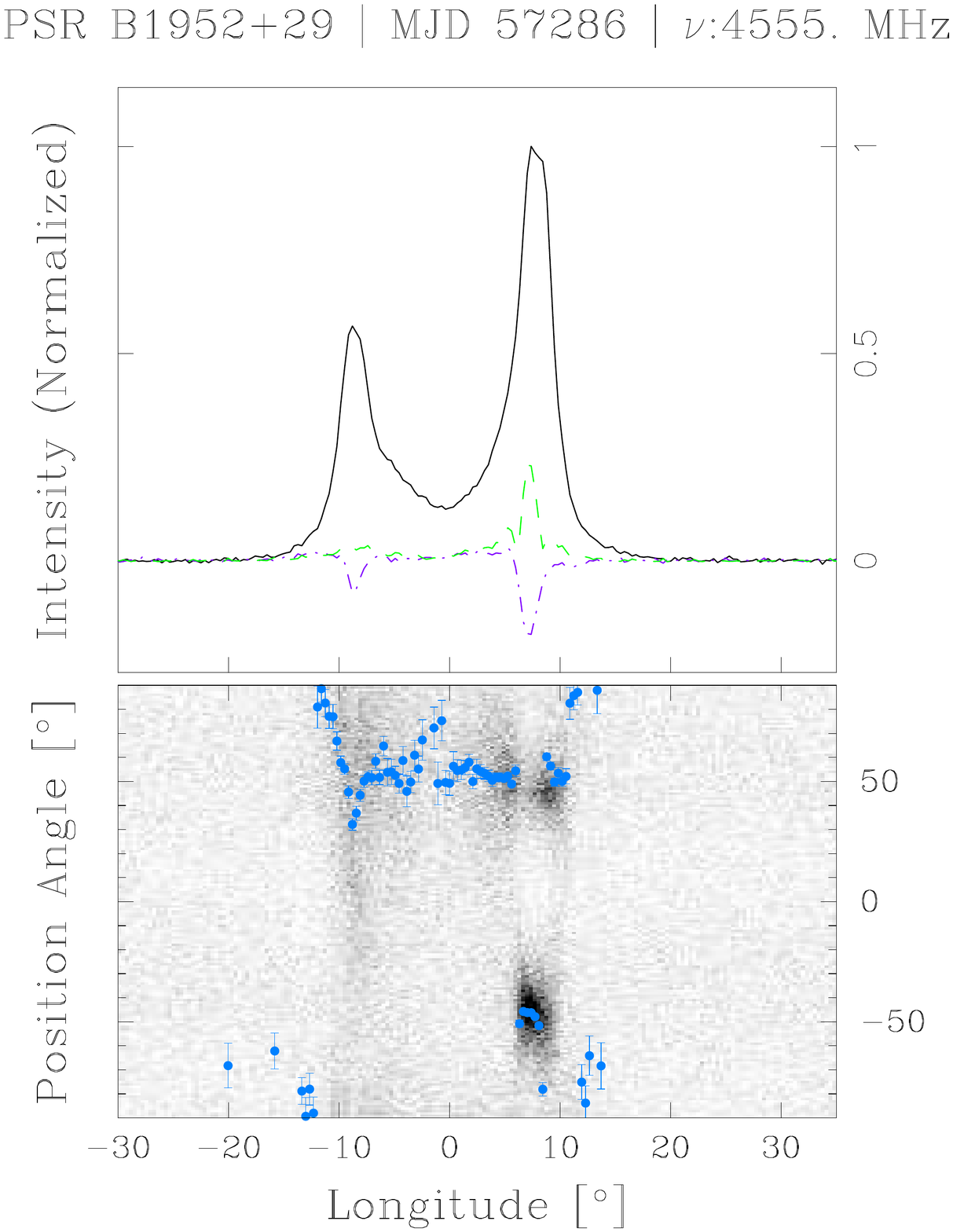} \\ \toprule
                 &
\includegraphics[page=1,width=\linewidth]{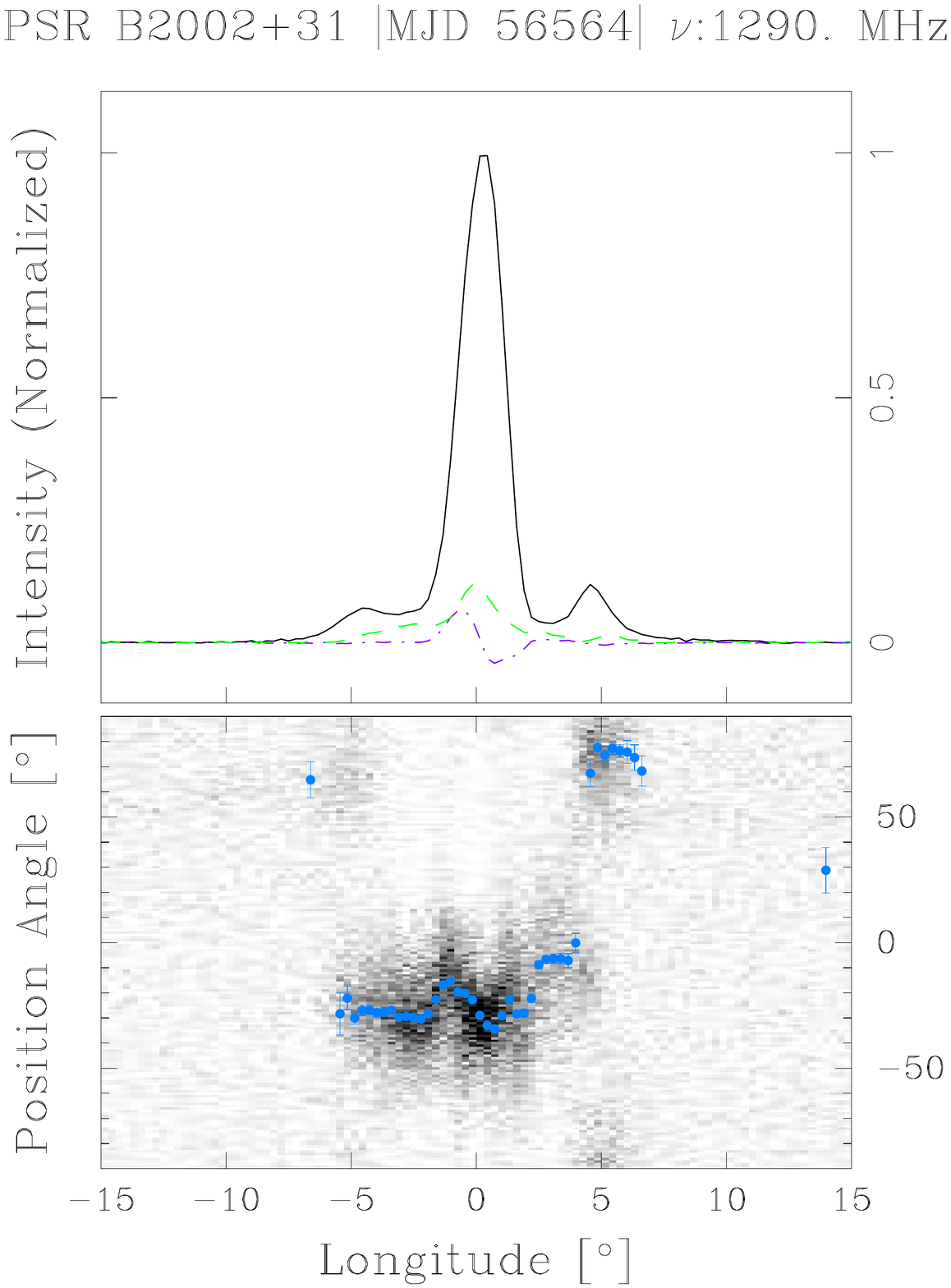} &
\includegraphics[page=1,width=\linewidth]{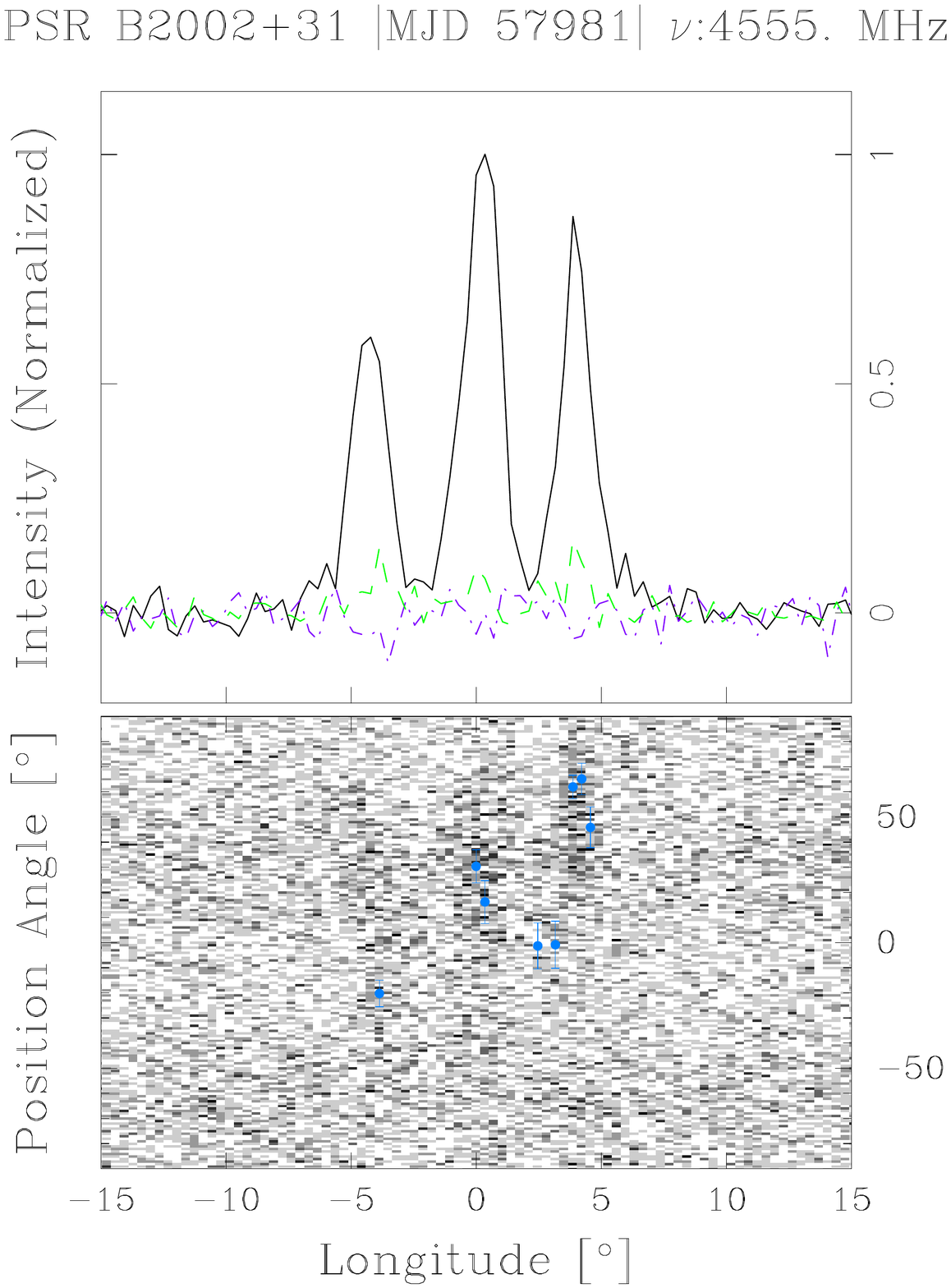} \\ \toprule
\includegraphics[page=1,width=\linewidth]{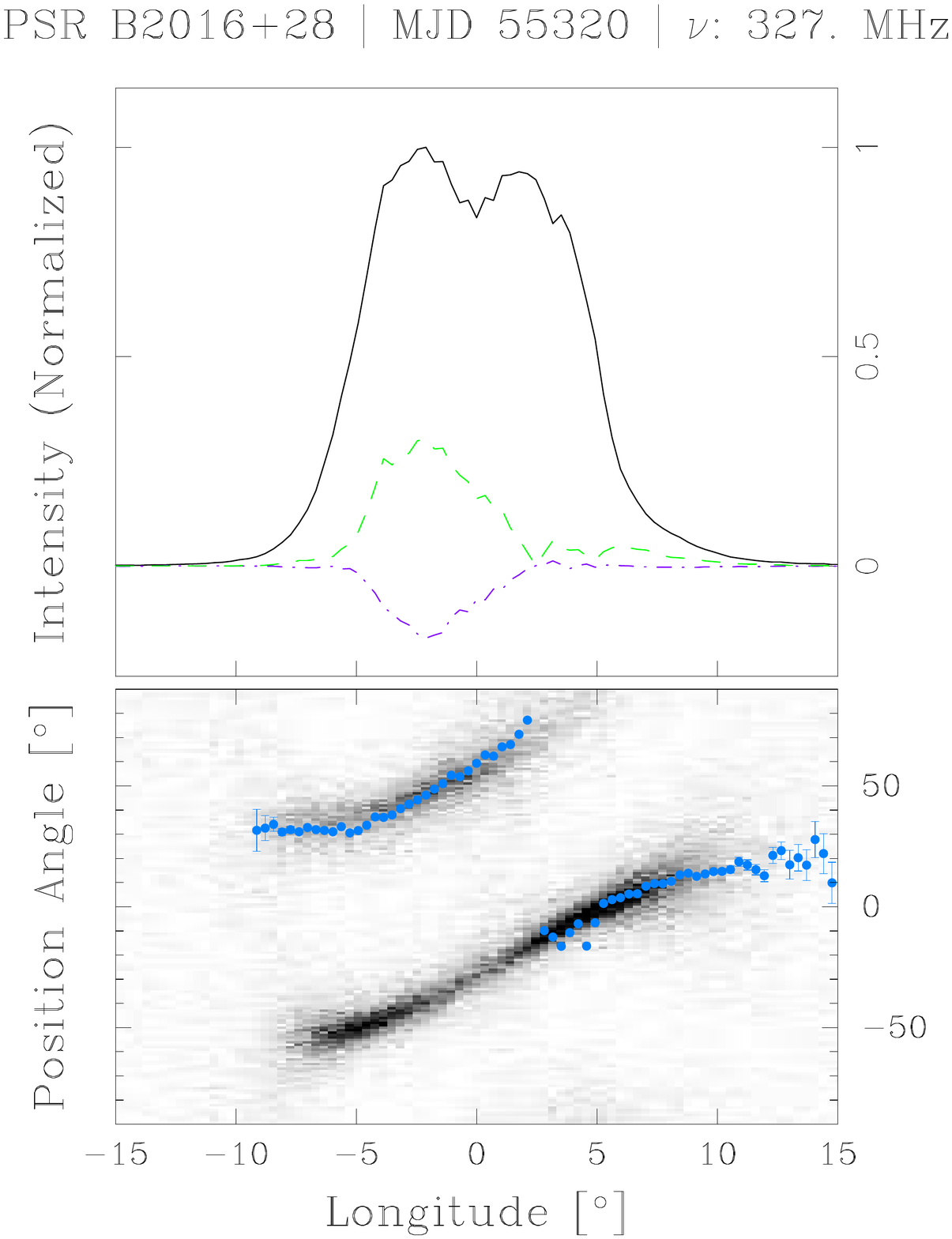} &
\includegraphics[page=1,width=\linewidth]{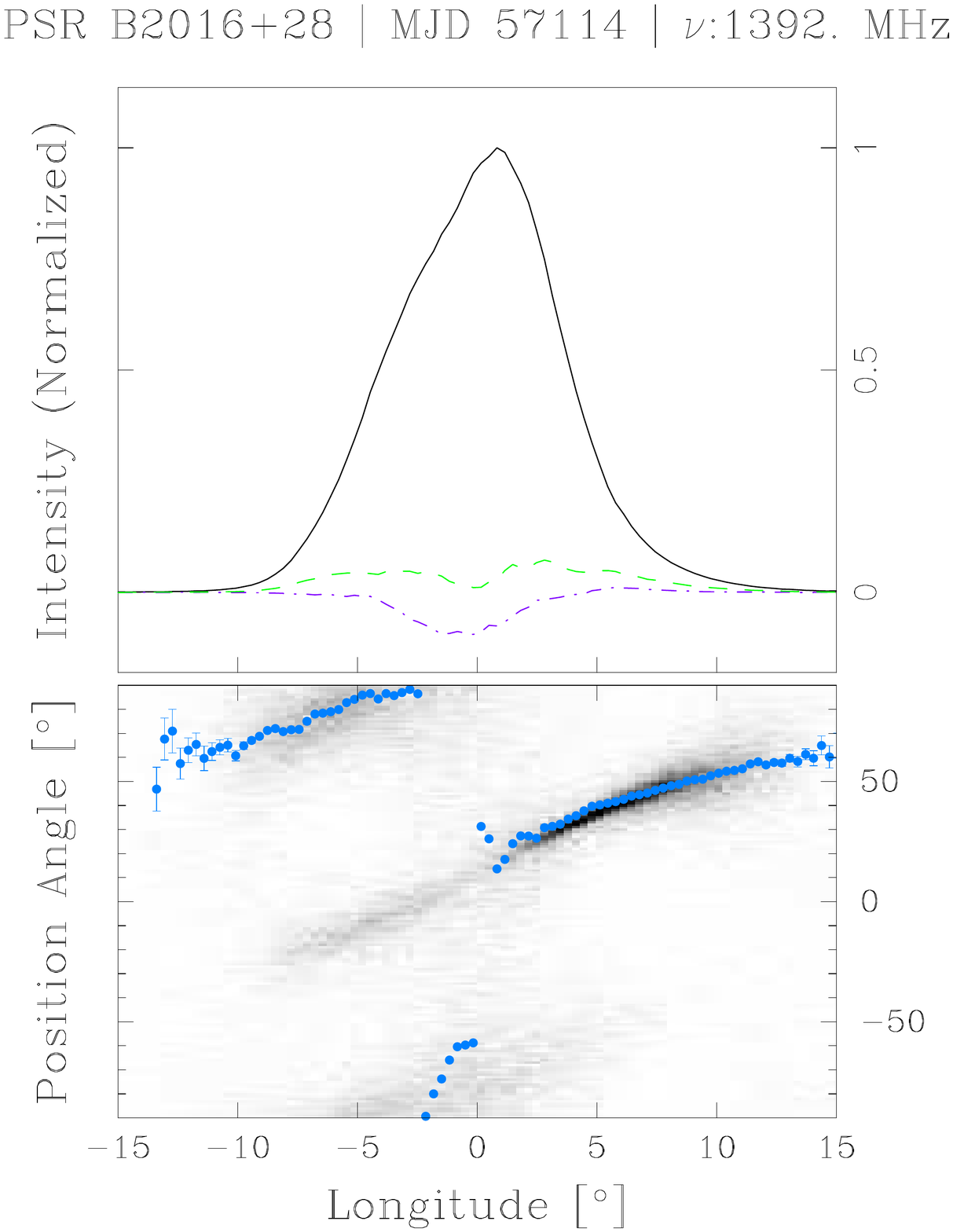} &
\includegraphics[page=1,width=\linewidth]{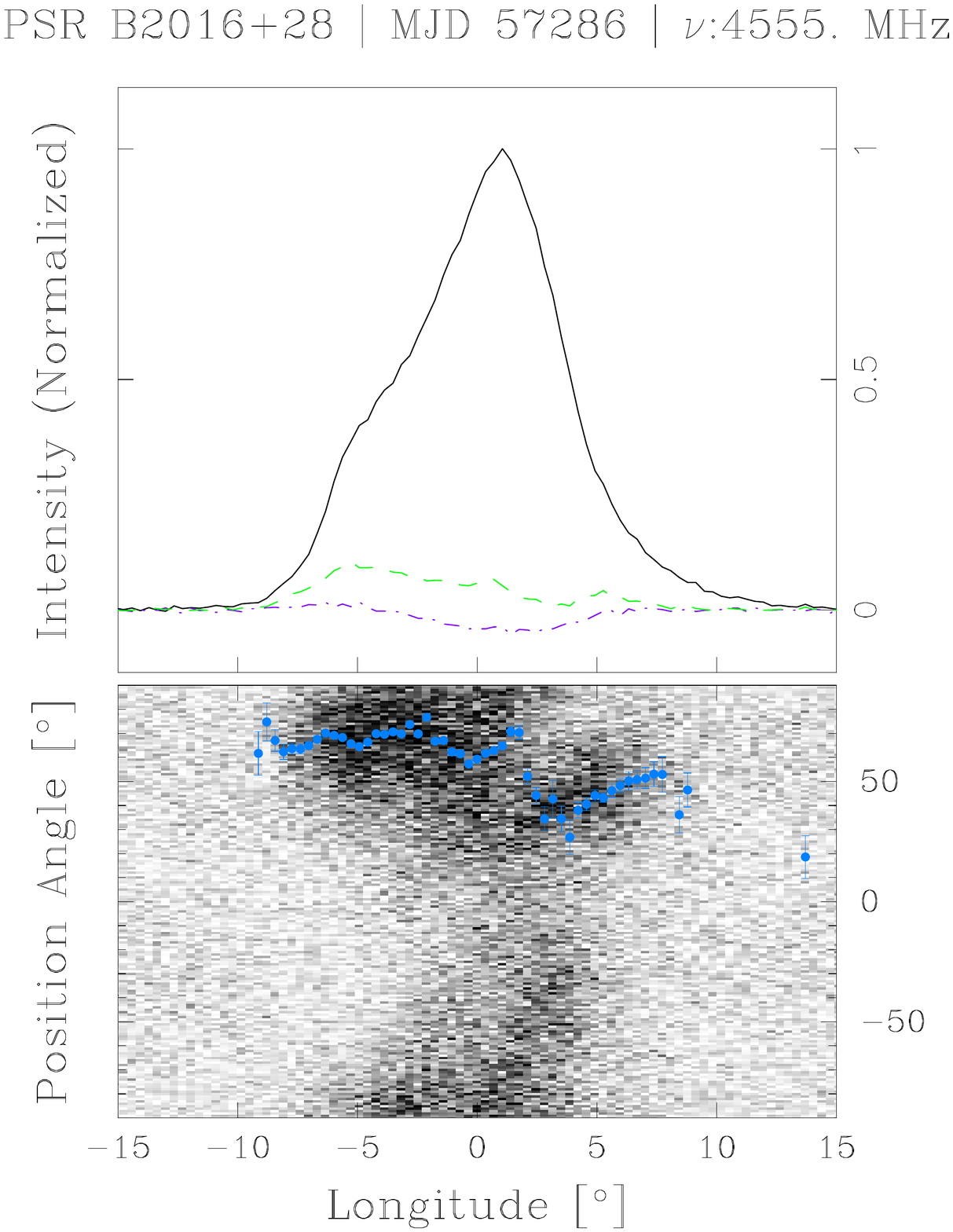} \\ 
     \bottomrule
   \end{tabularx} 
\caption{Average profiles of PSRs B1952+29, B2002+31, and B2016+28.}
 \end{figure*}
\vspace{1cm}

   \begin{figure*} 
 \begin{tabularx}{\textwidth}{YYY}
    \multicolumn{3}{c}{} \\ \toprule
    \includegraphics[page=1,width=\linewidth]{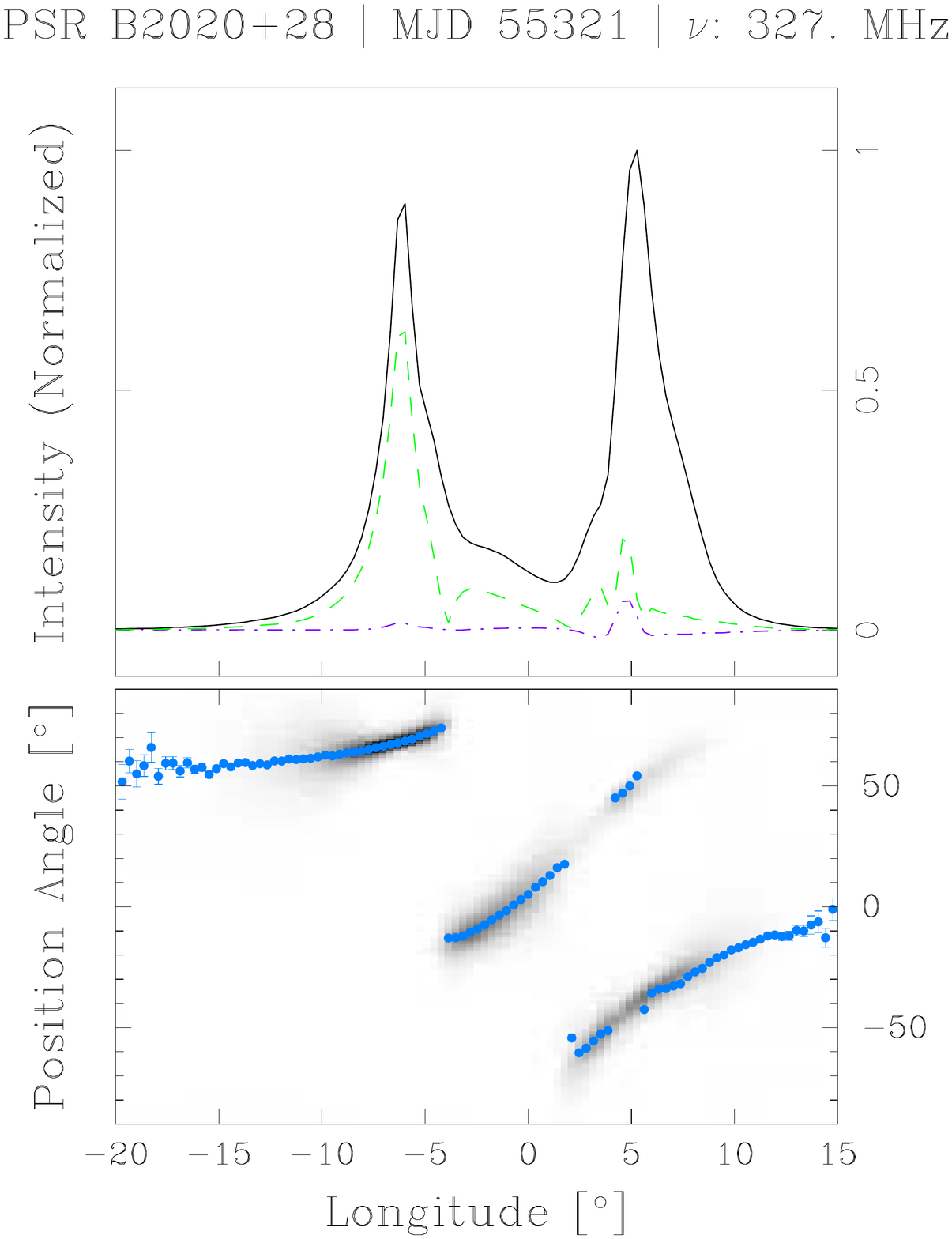} &
\includegraphics[page=1,width=\linewidth]{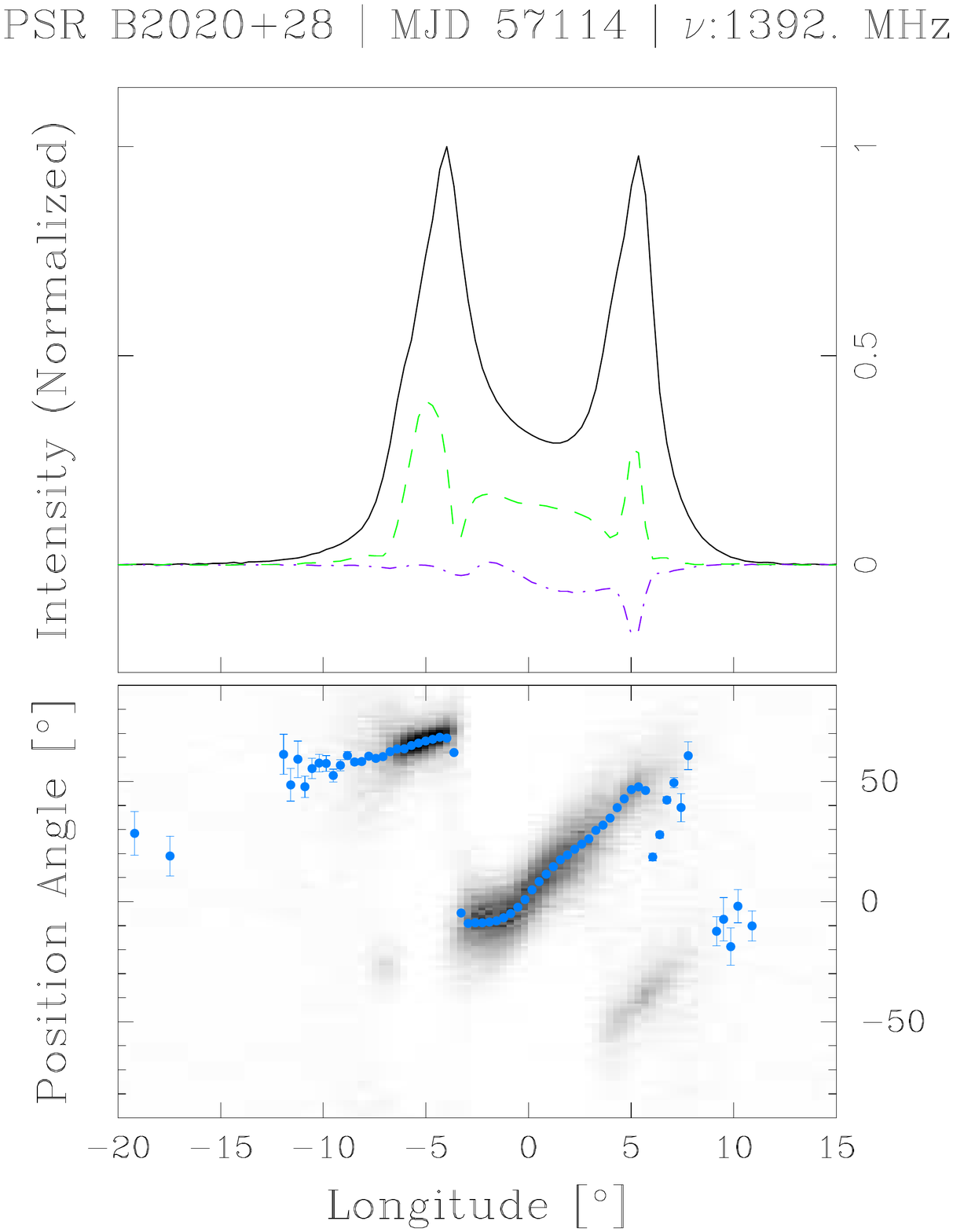} &
\includegraphics[page=1,width=\linewidth]{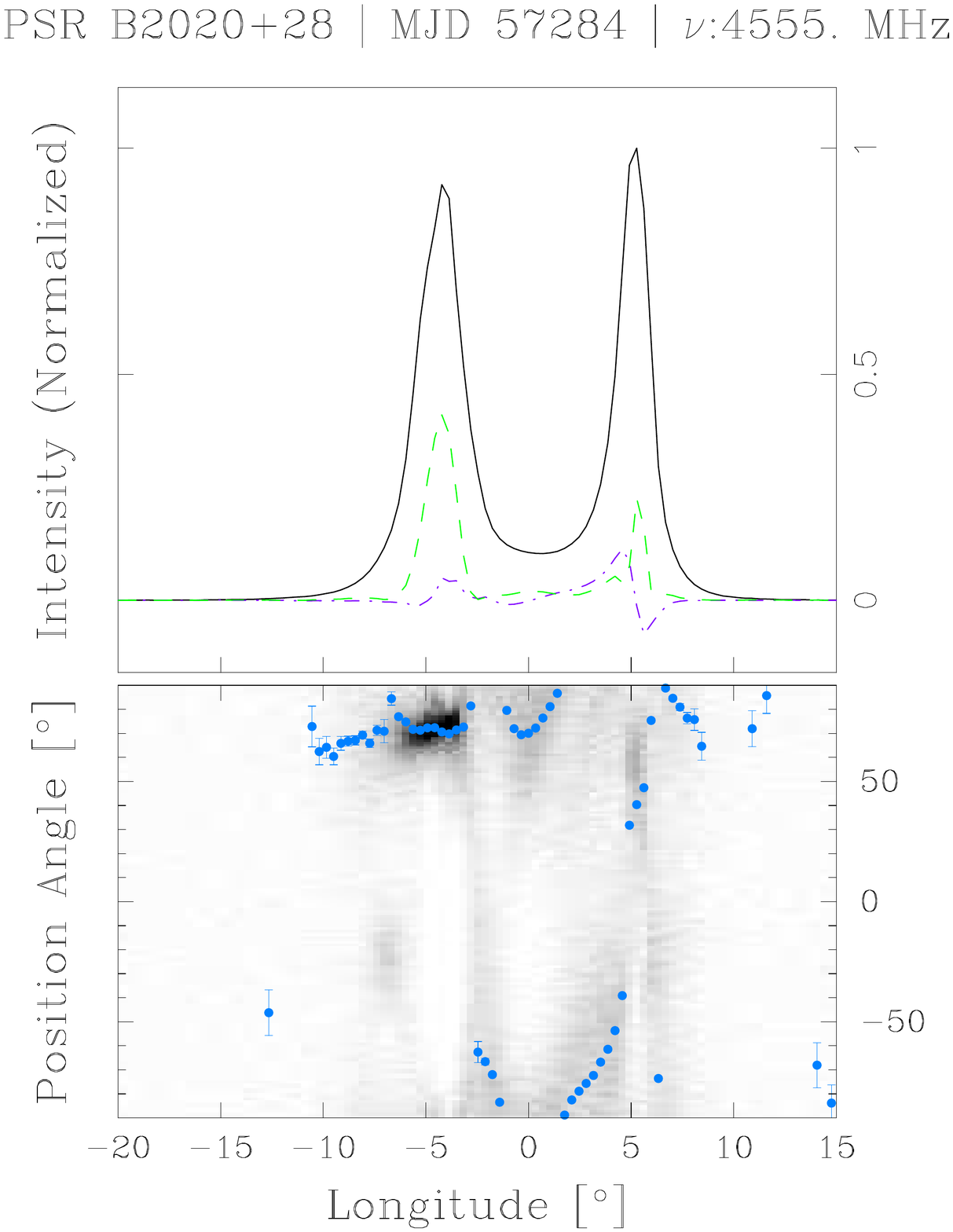} \\ \toprule
\includegraphics[page=1,width=\linewidth]{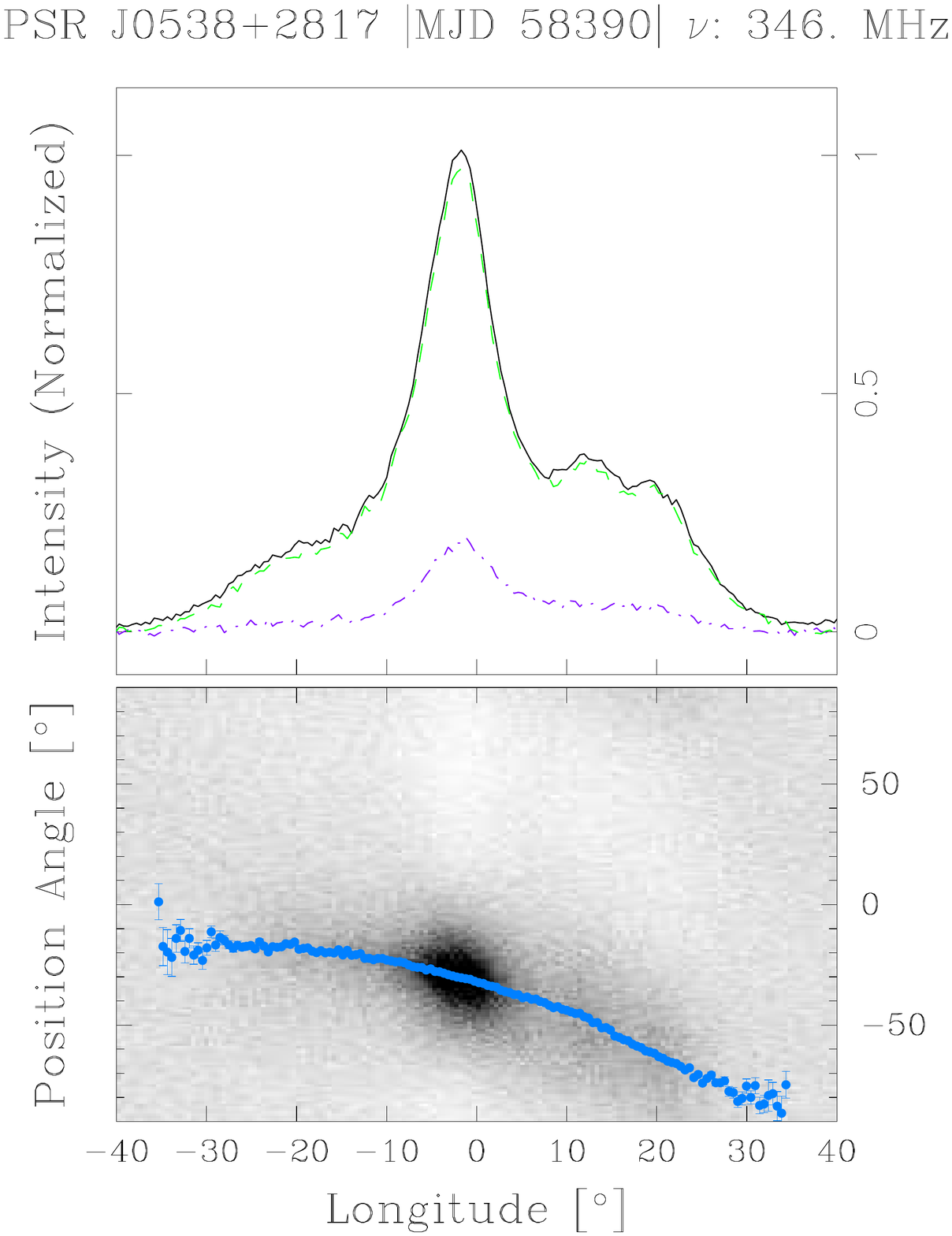} &
\includegraphics[page=1,width=\linewidth]{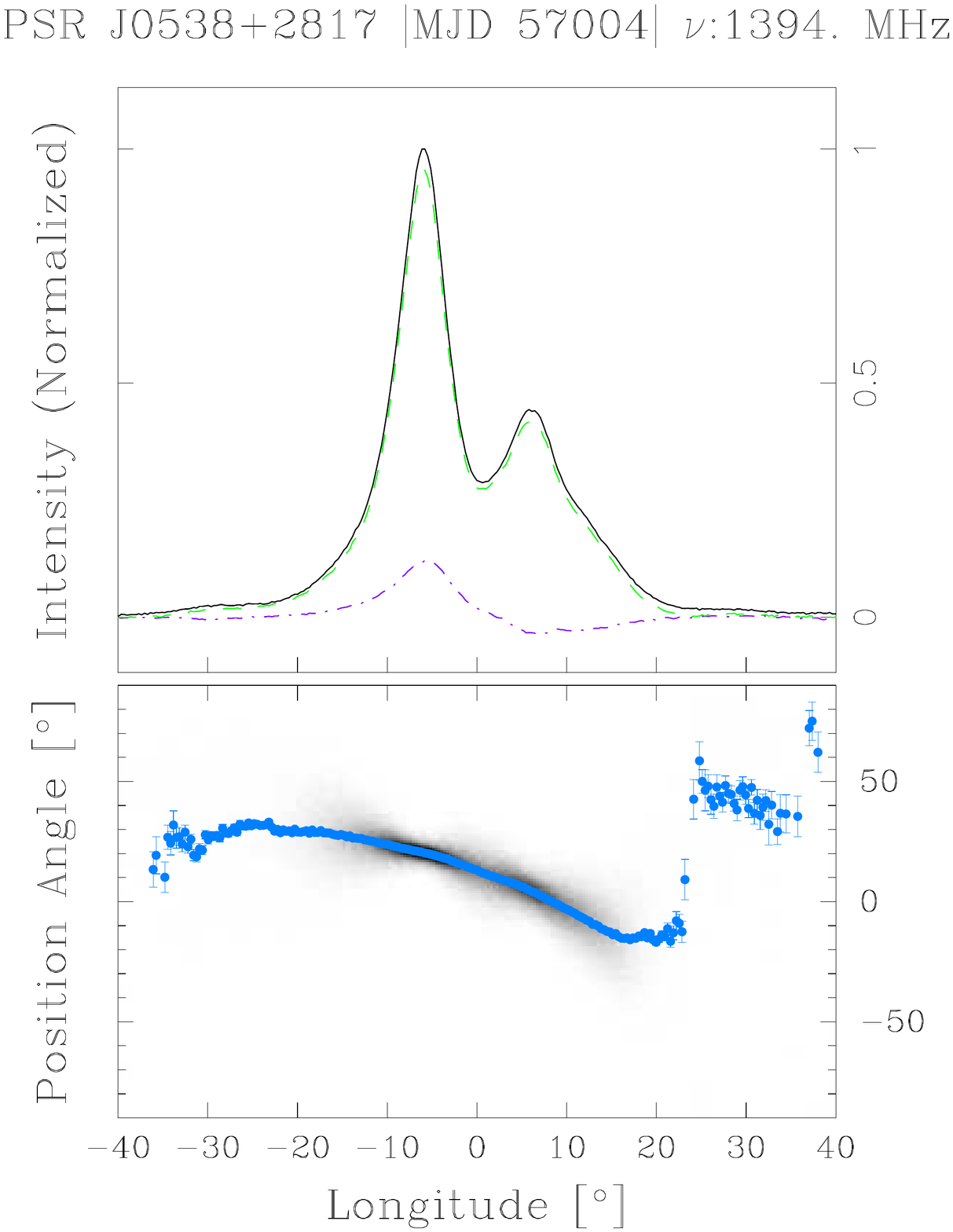} &
\includegraphics[page=1,width=\linewidth]{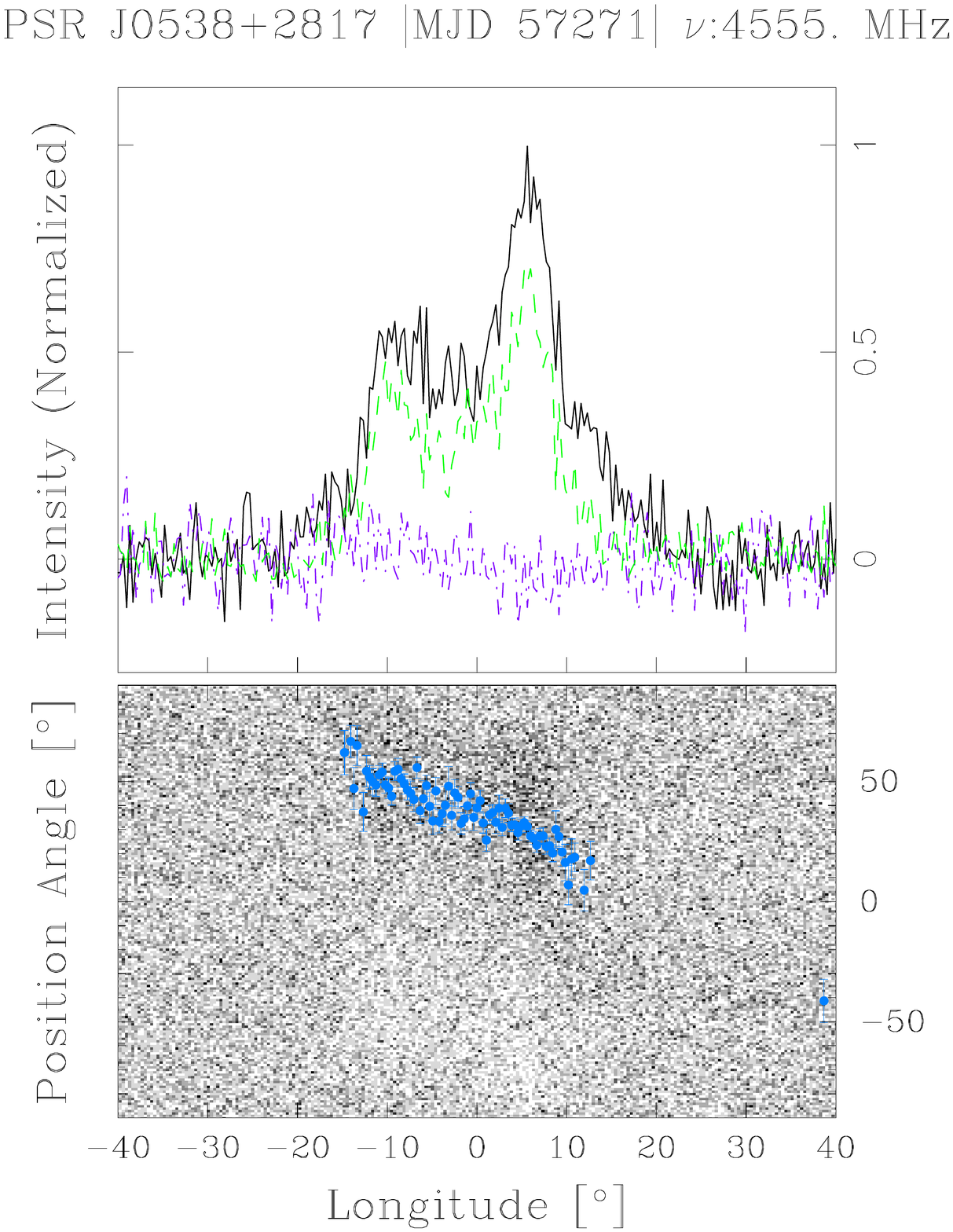} \\ \toprule
\includegraphics[page=1,width=\linewidth]{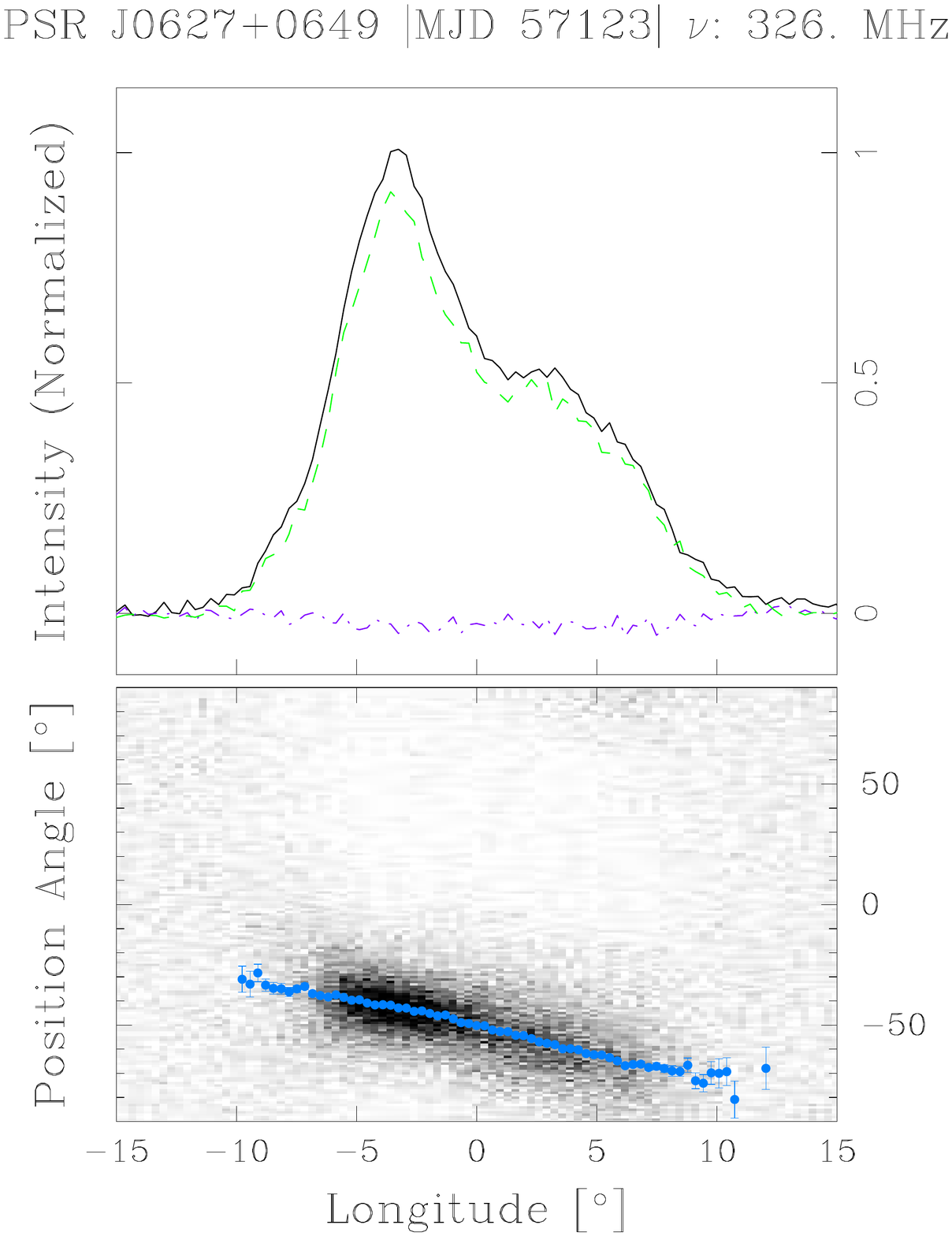} &
\includegraphics[page=1,width=\linewidth]{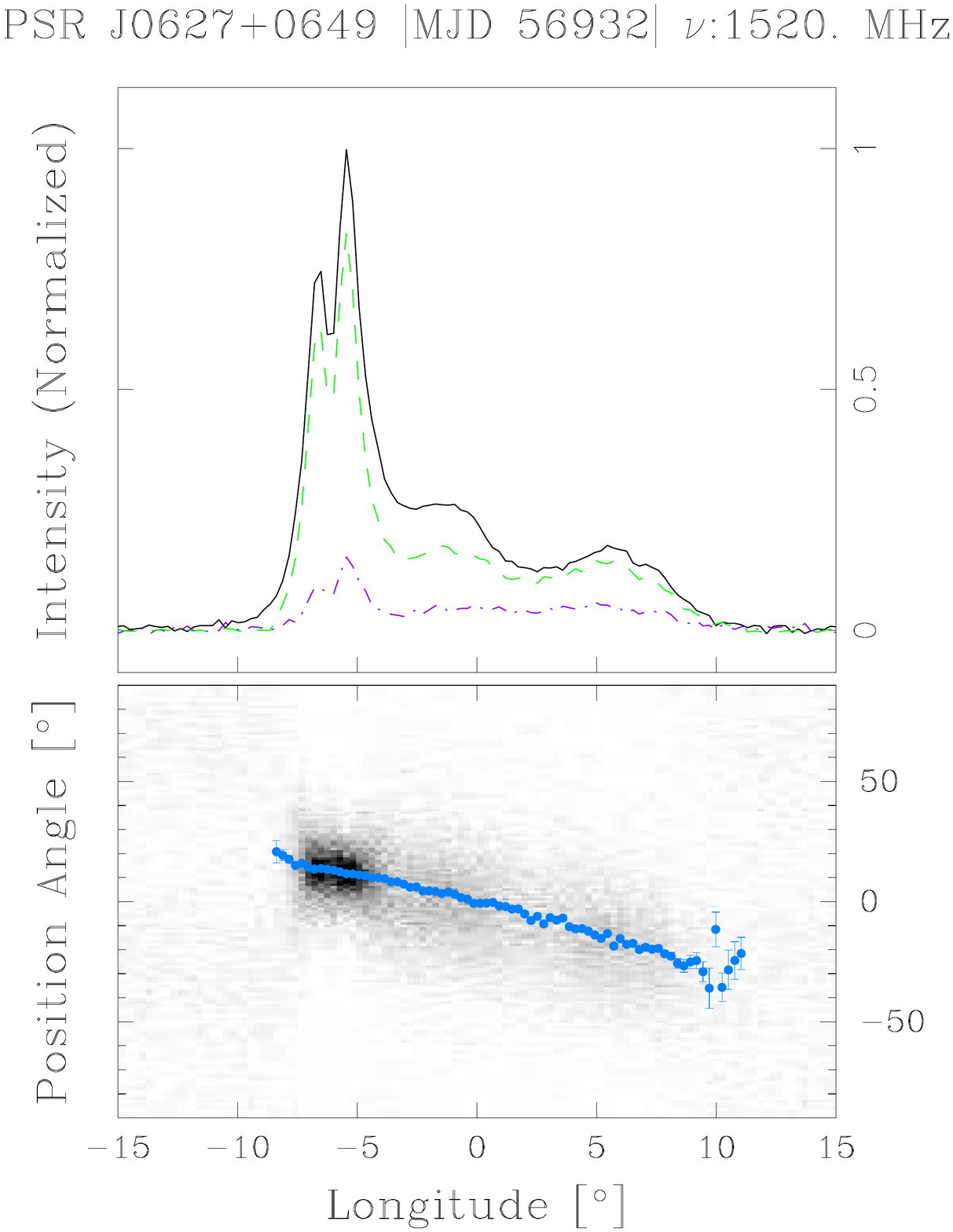} &
\includegraphics[page=1,width=\linewidth]{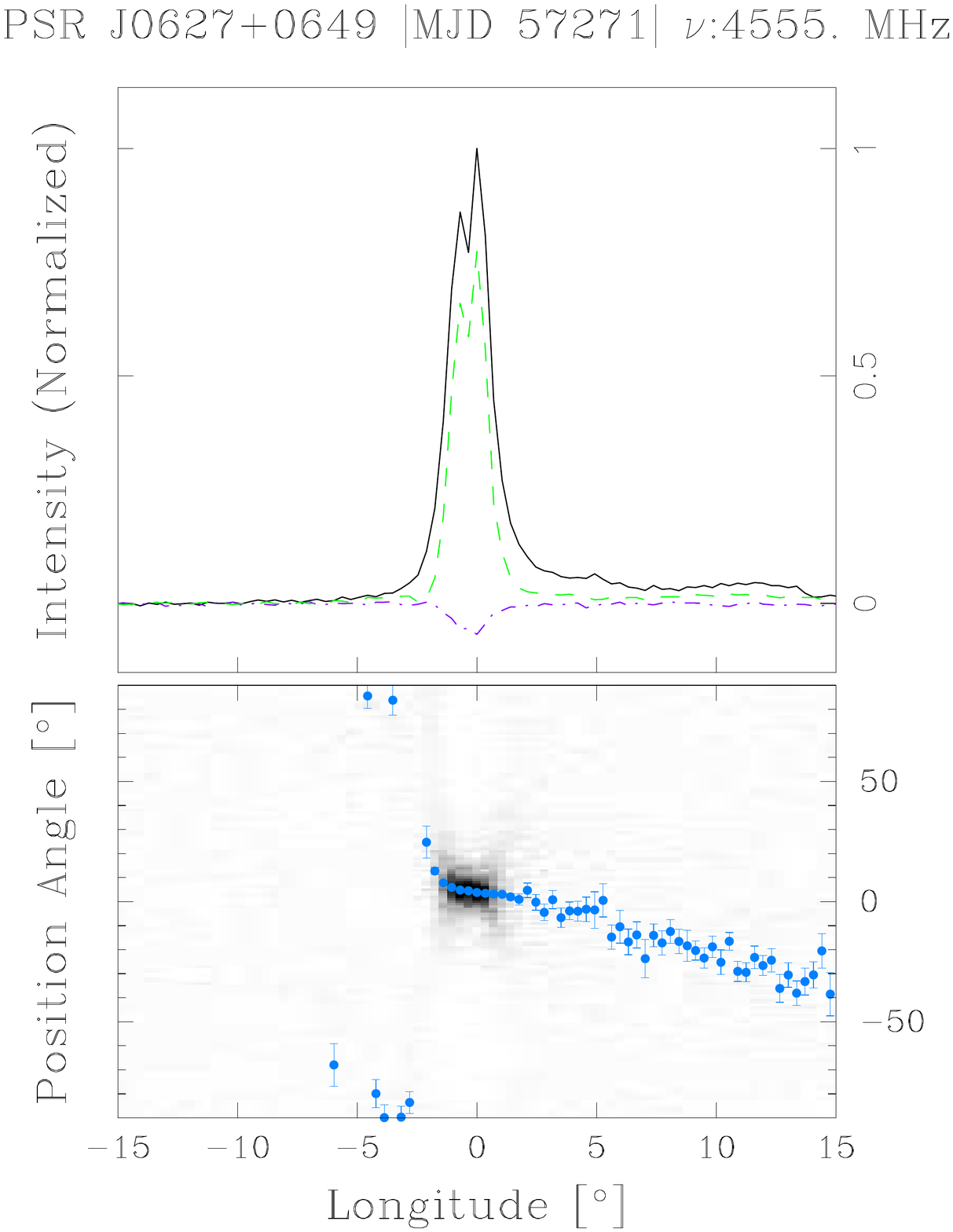} \\ 
     \bottomrule
   \end{tabularx} 
\caption{Average profiles of PSRs B2020+28, J0538+2817, and J0627+0649.}
 \end{figure*}
\vspace{1cm}

   \begin{figure*} 
 \begin{tabularx}{\textwidth}{YYY}
    \multicolumn{3}{c}{} \\ \toprule
\includegraphics[page=1,width=\linewidth]{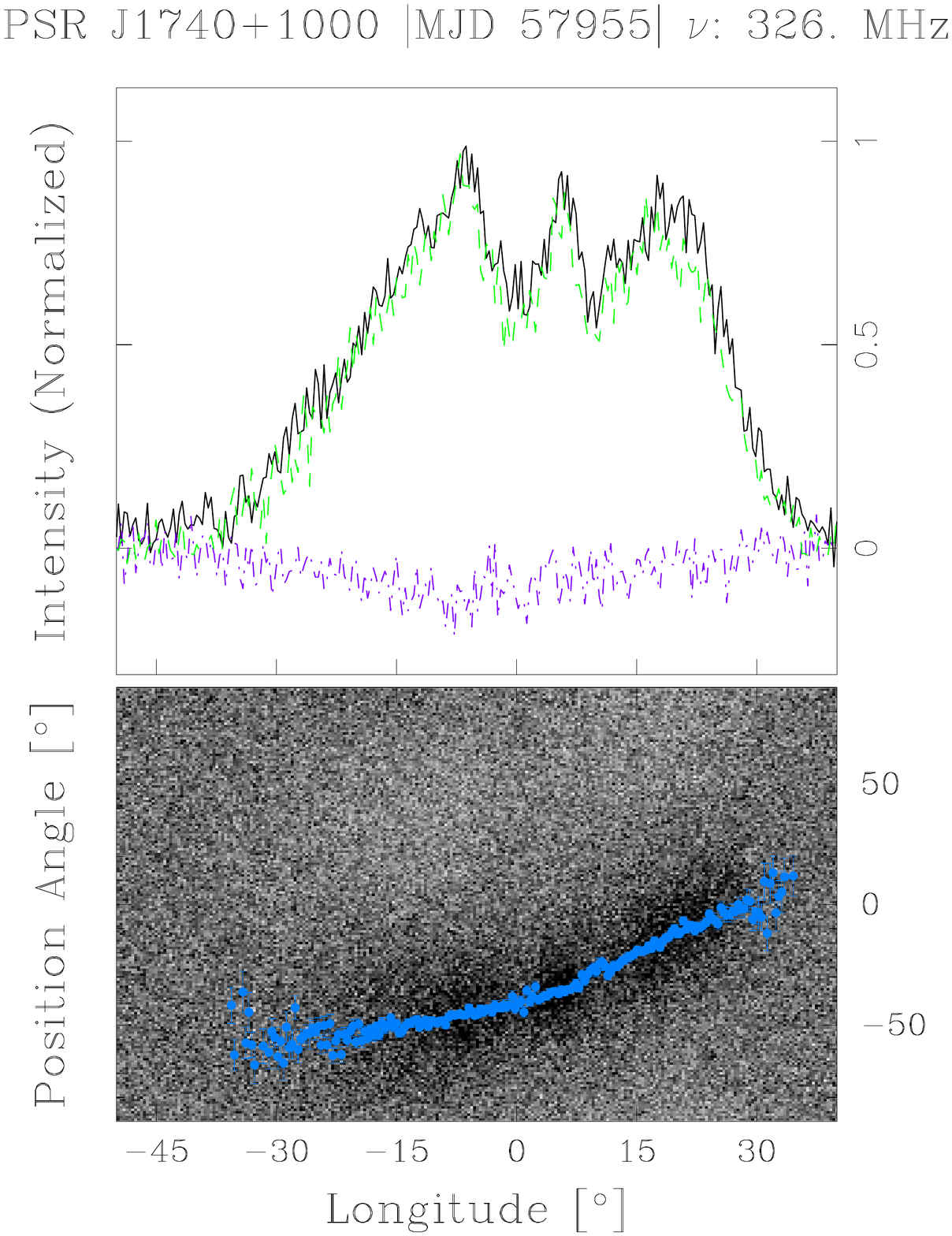} &
\includegraphics[page=1,width=\linewidth]{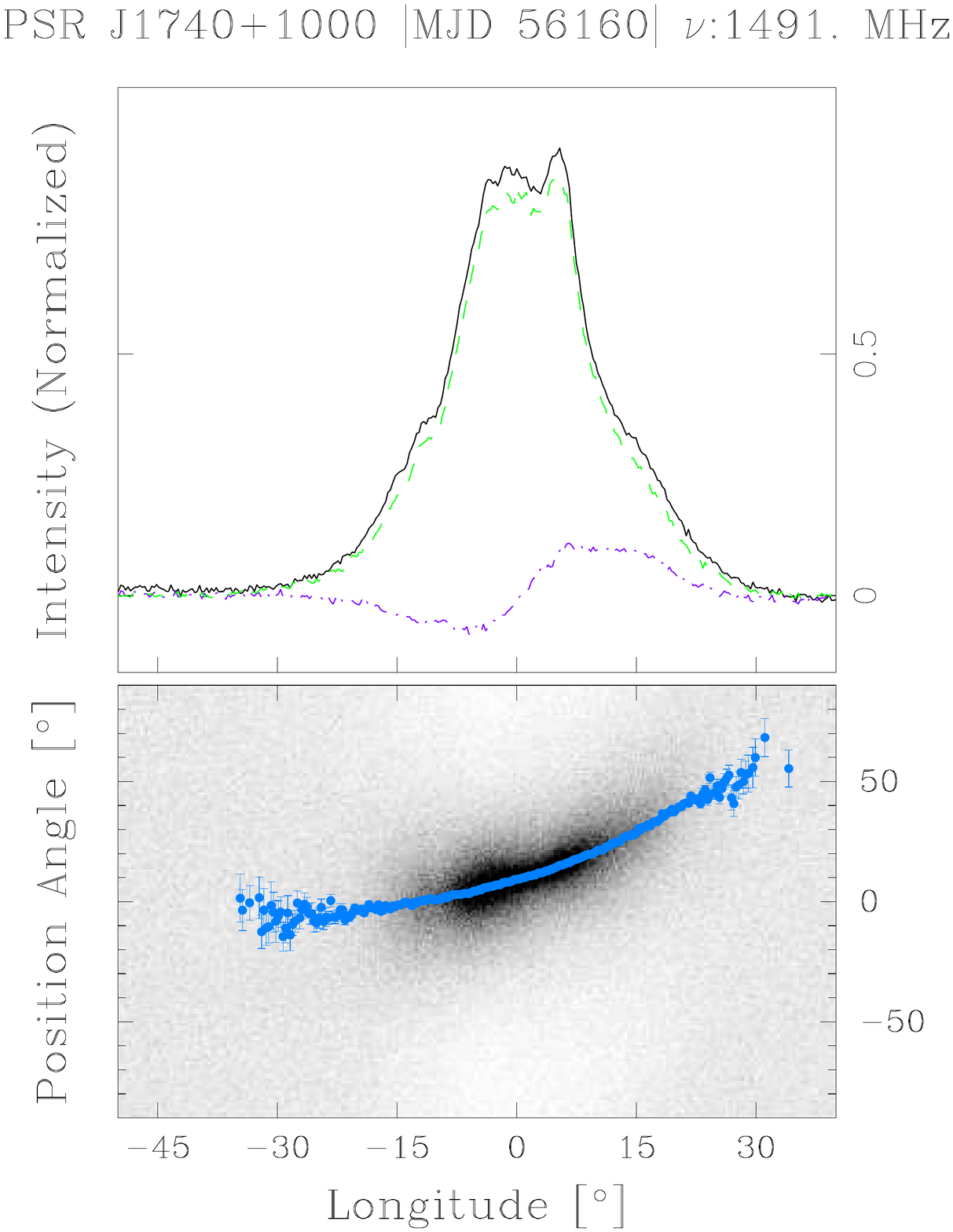} &
\includegraphics[page=1,width=\linewidth]{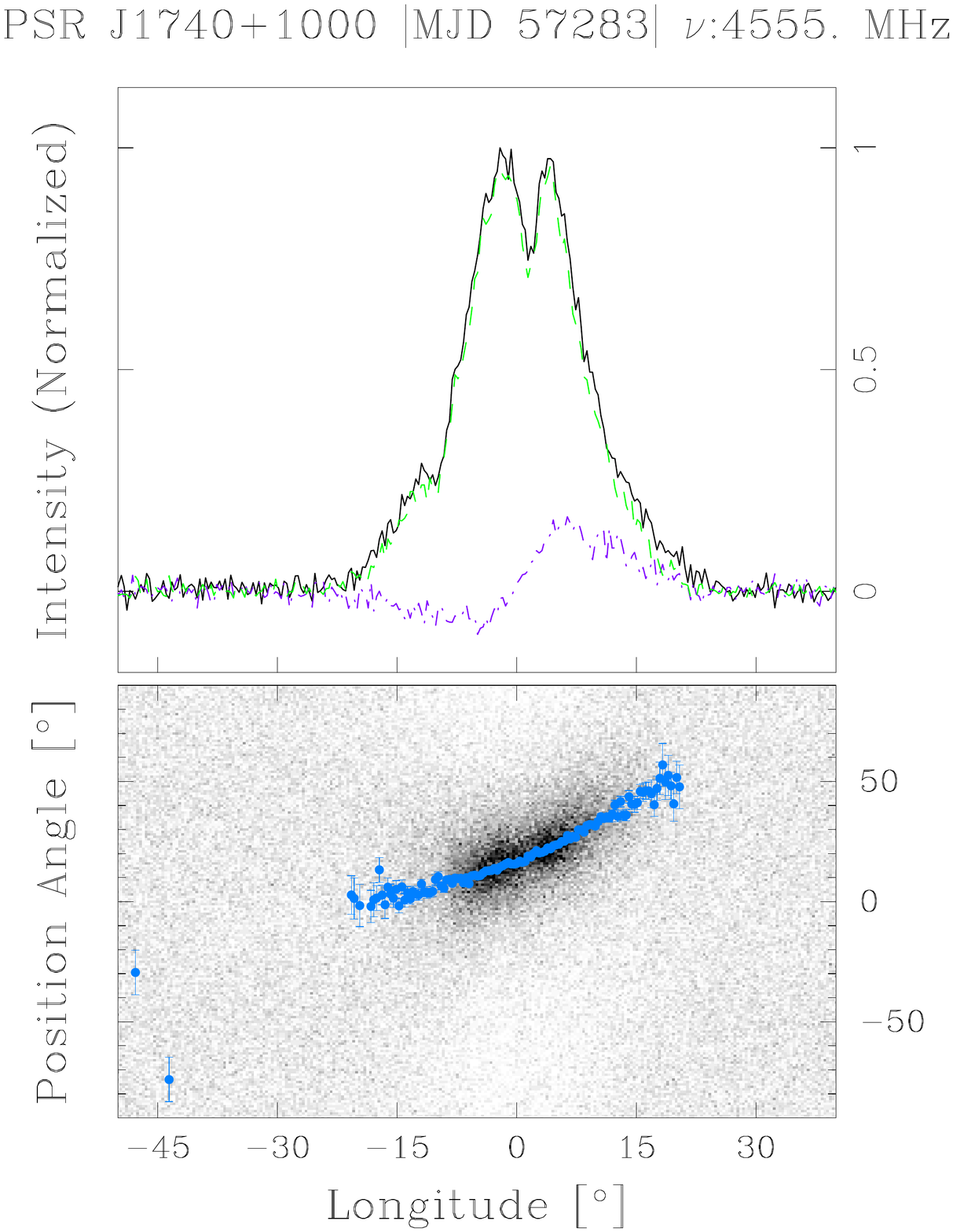} \\ 
     \bottomrule
   \end{tabularx} 
\caption{Average profile of PSR J1740+1000.}
 \end{figure*}
\vspace{1cm}
\vspace{-.4cm}
\twocolumn
\setcounter{figure}{0}
\renewcommand{\thefigure}{B\arabic{figure}}
\setcounter{table}{0}
\renewcommand{\thetable}{B\arabic{table}}
\setcounter{footnote}{0}
\renewcommand{\thefootnote}{B\arabic{footnote}}
\section{Peak Histogram Plots}
\vspace{-.4cm}
\onecolumn
\renewcommand{\thefigure}{B\arabic{figure}}
\begin{figure*}
 \begin{tabularx}{\textwidth}{YYY}
    \multicolumn{3}{c}{} \\ \toprule
\includegraphics[page=1,width=\linewidth]{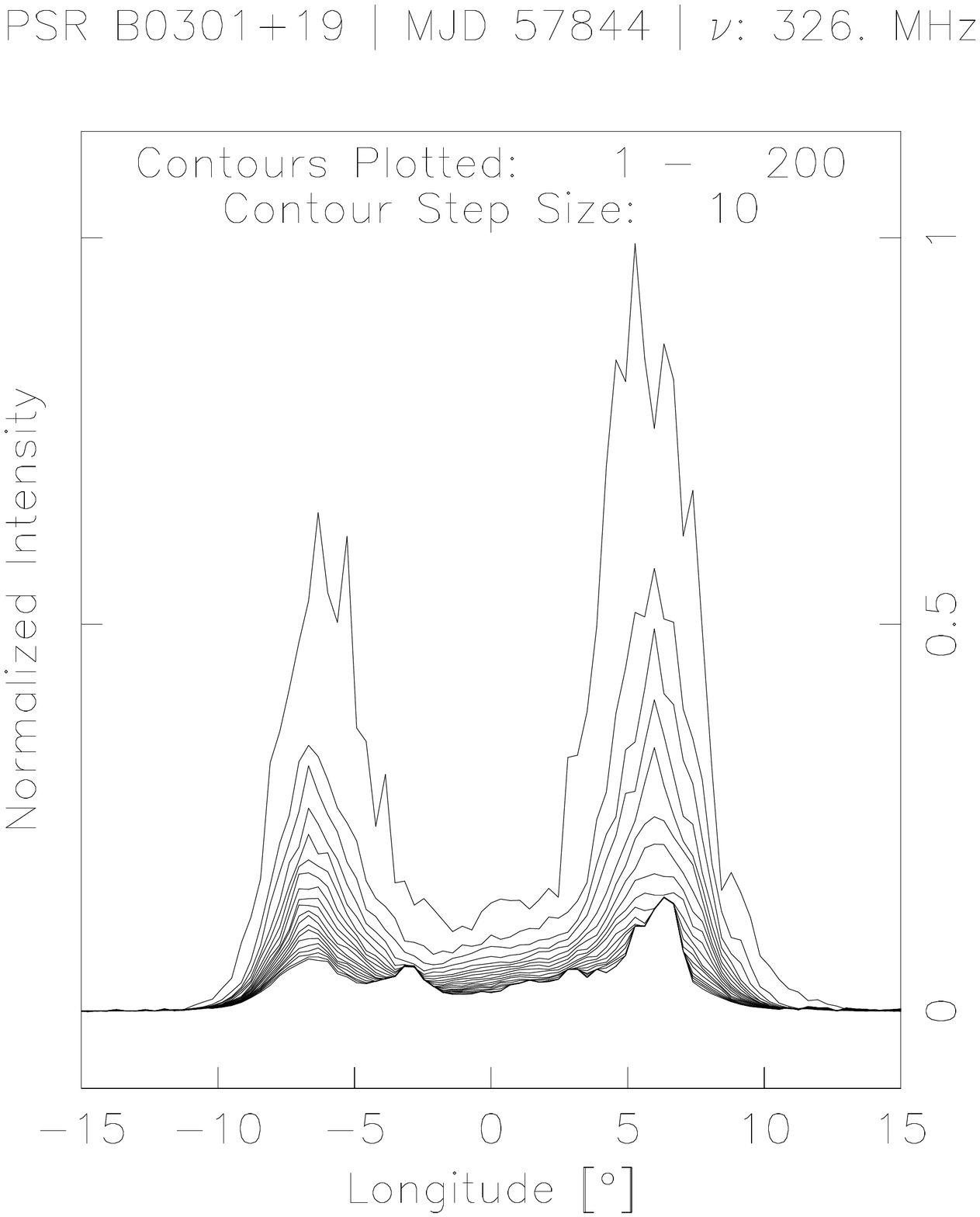} &
\includegraphics[page=1,width=\linewidth]{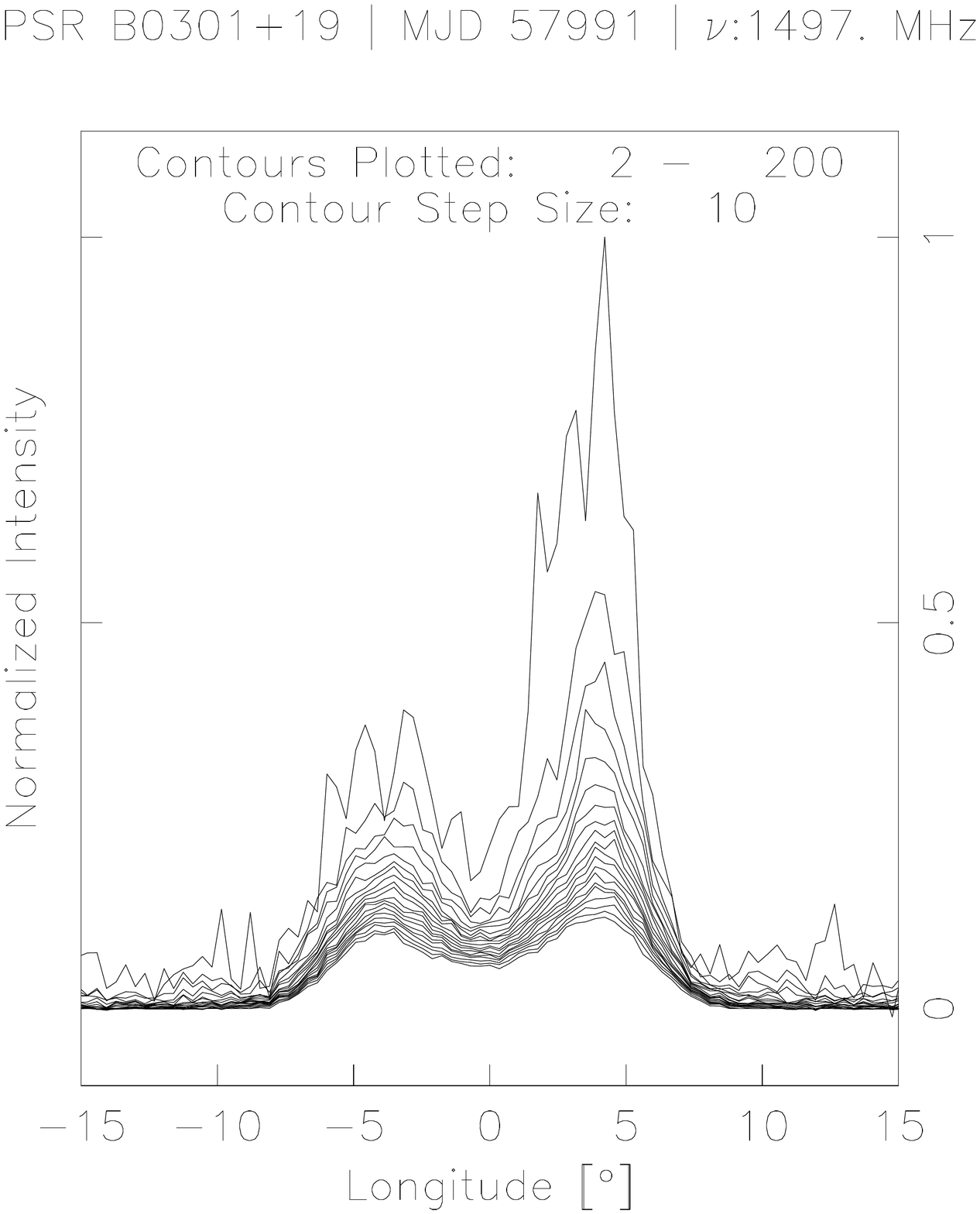} &
\includegraphics[page=1,width=\linewidth]{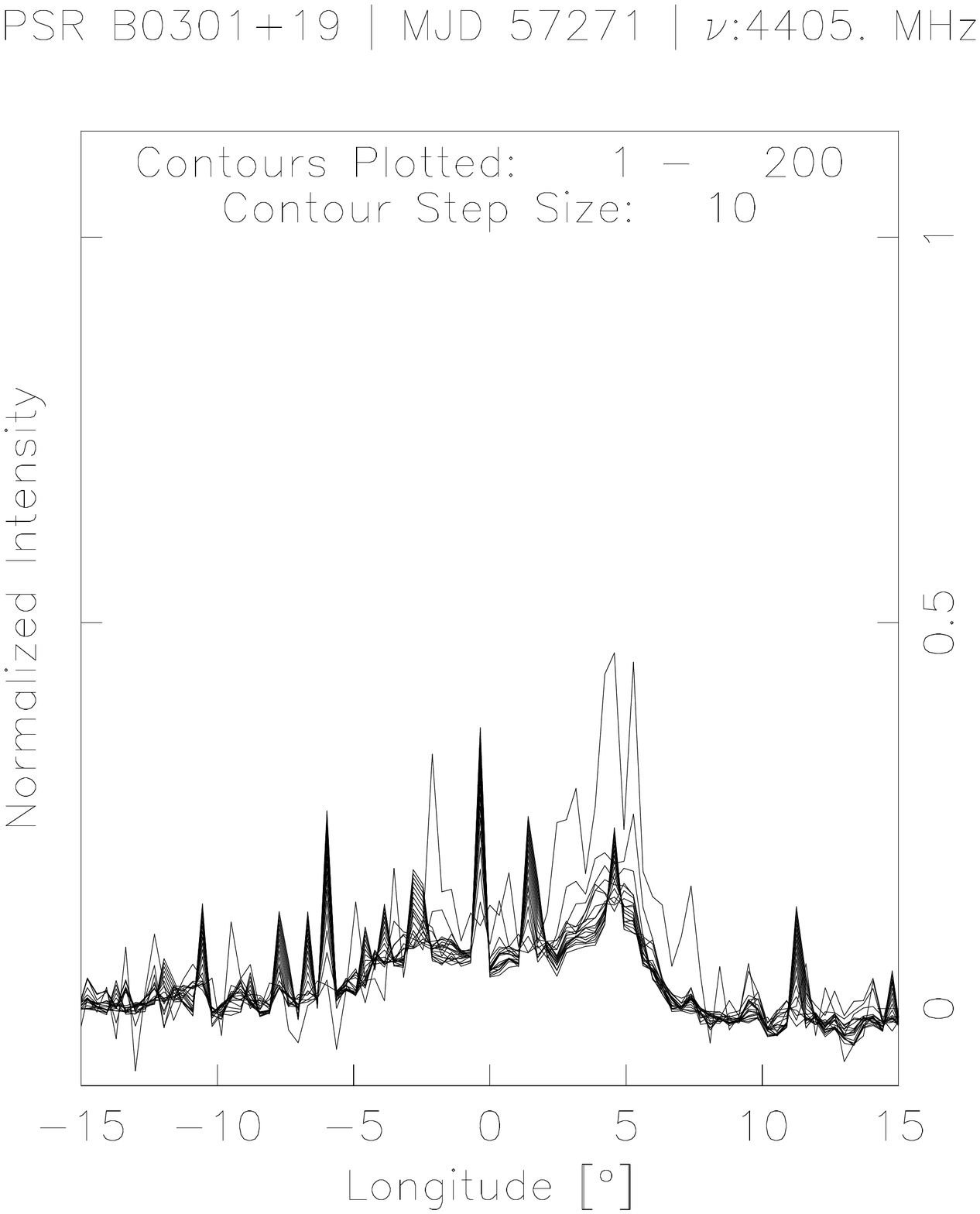} \\ \toprule
\includegraphics[page=1,width=\linewidth]{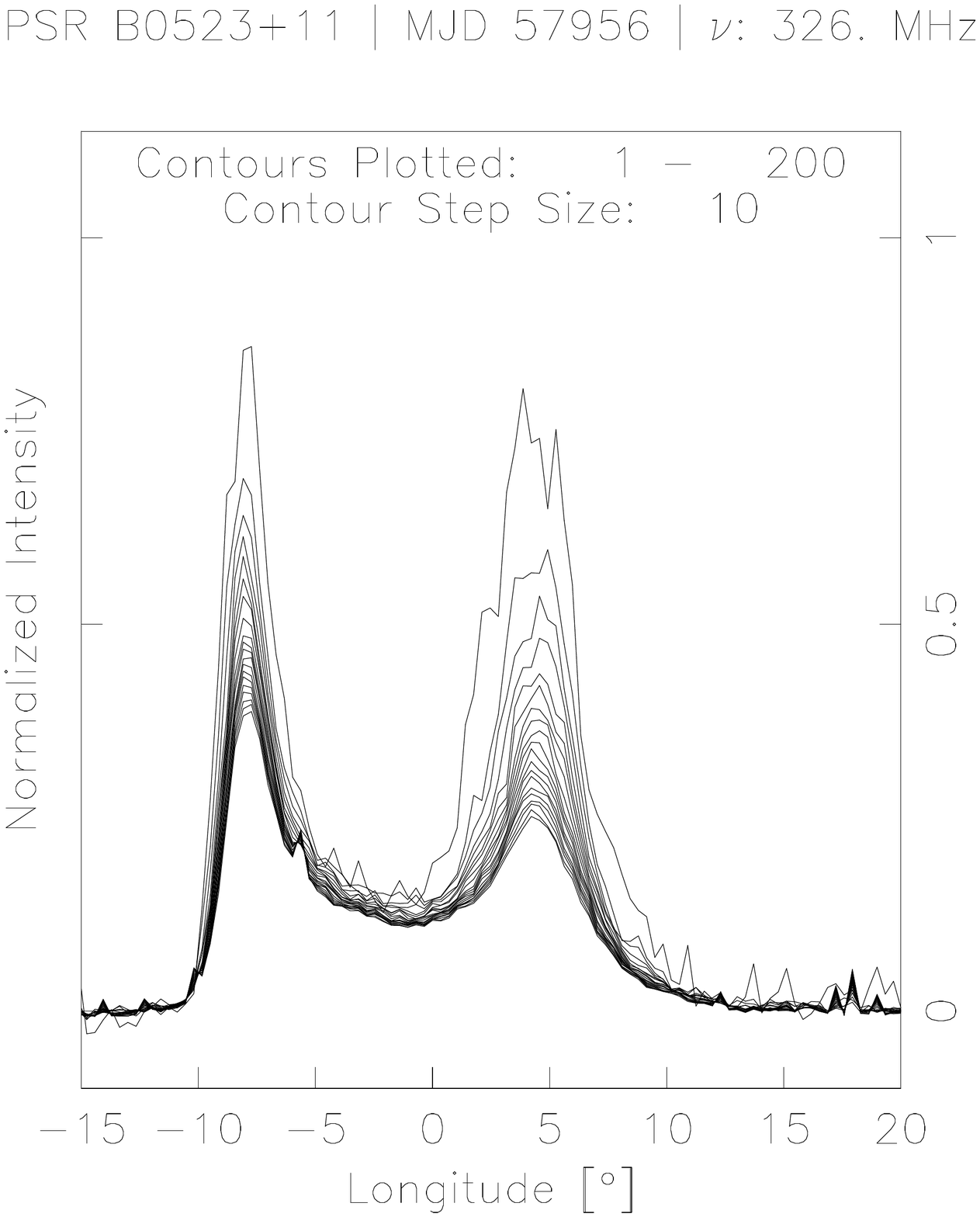} &
\includegraphics[page=1,width=\linewidth]{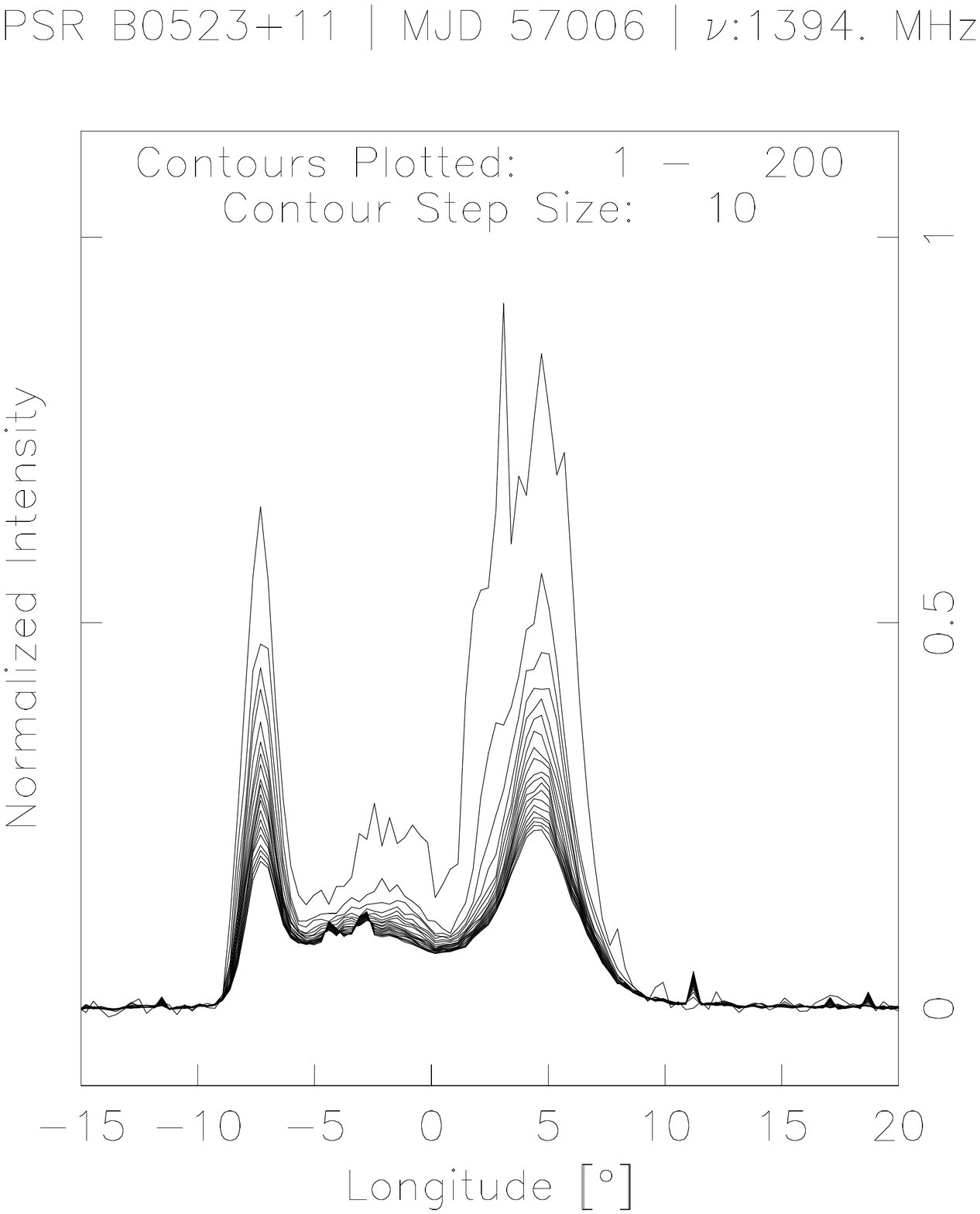} &
\includegraphics[page=1,width=\linewidth]{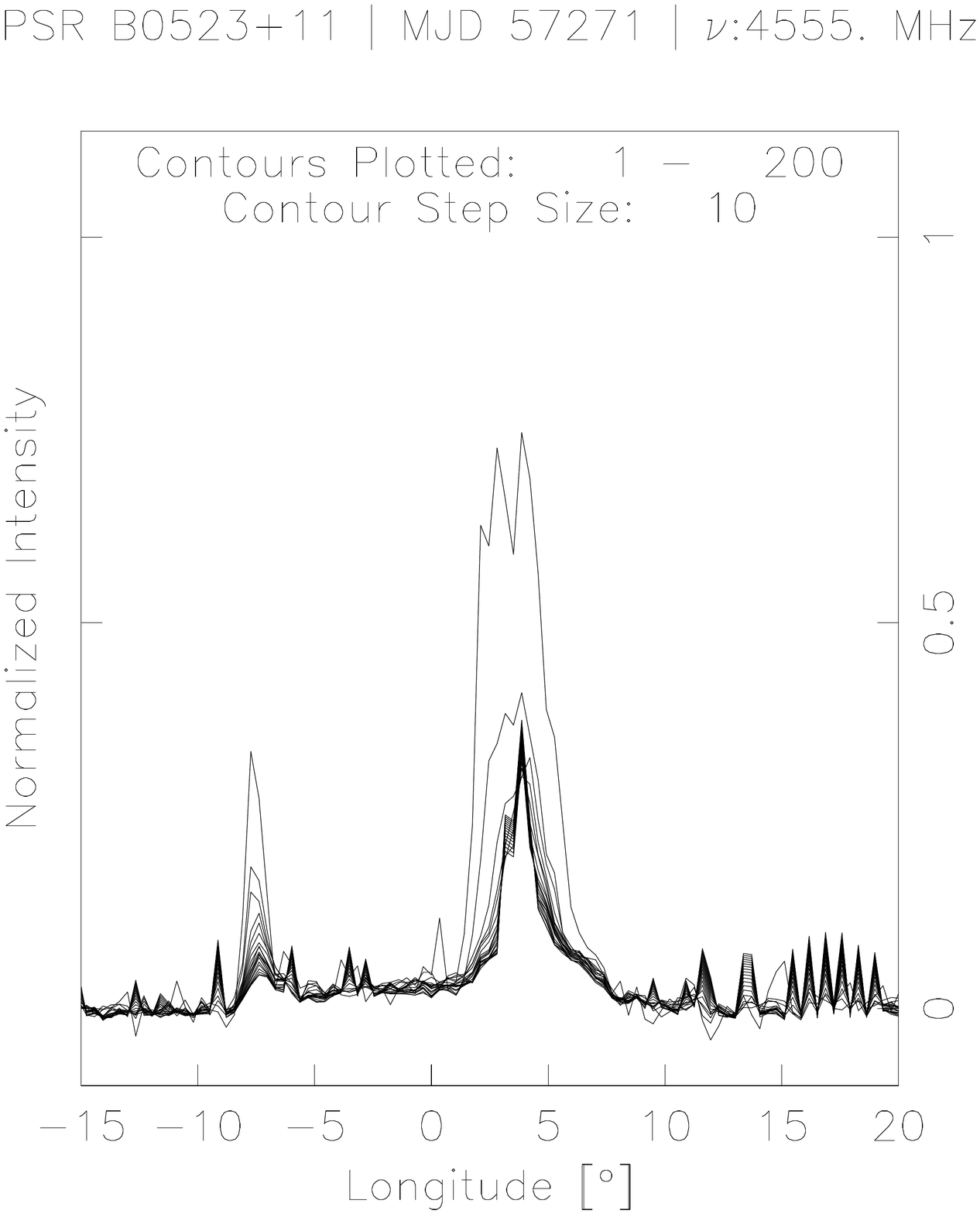} \\ \toprule
\includegraphics[page=1,width=\linewidth]{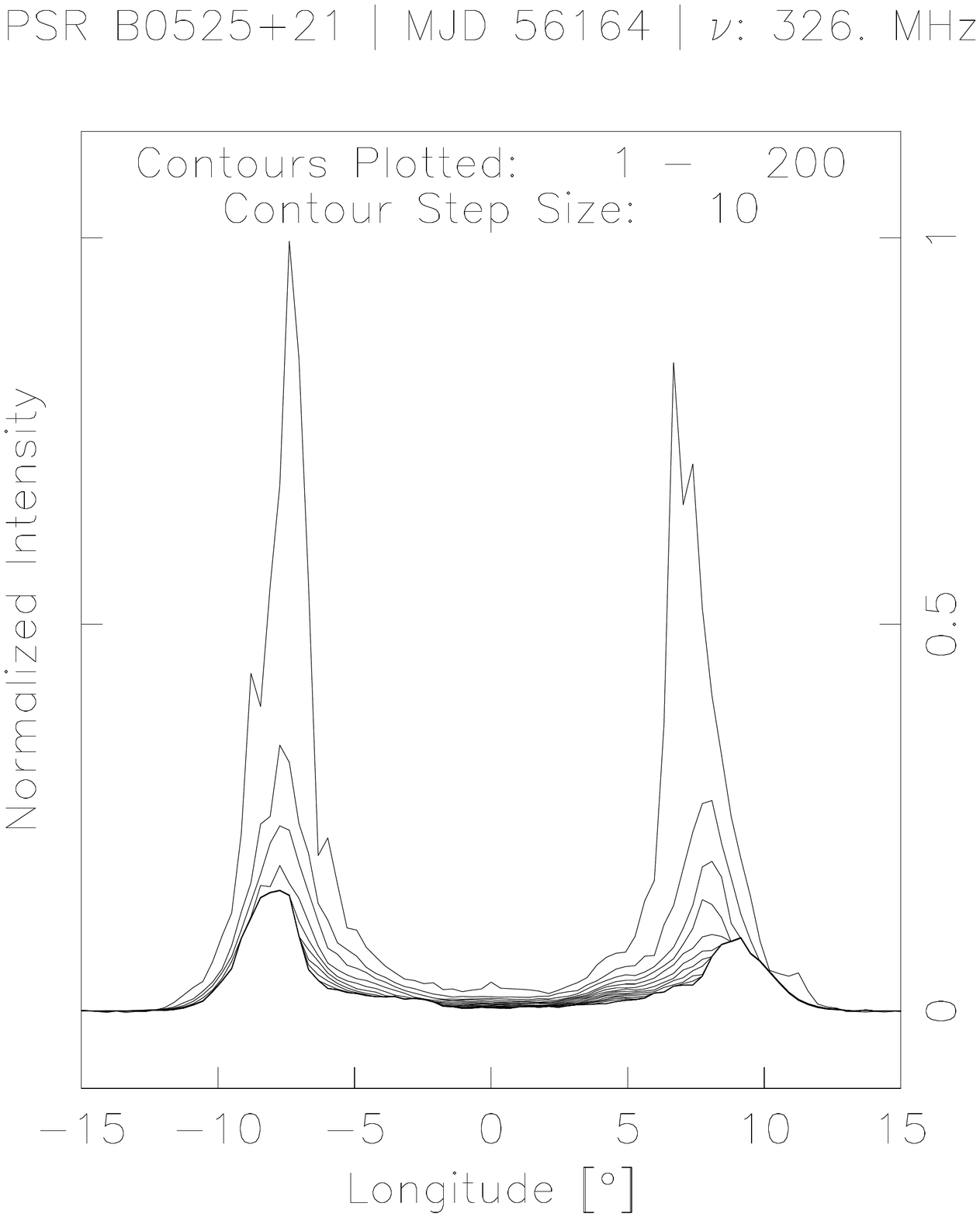} &
\includegraphics[page=1,width=\linewidth]{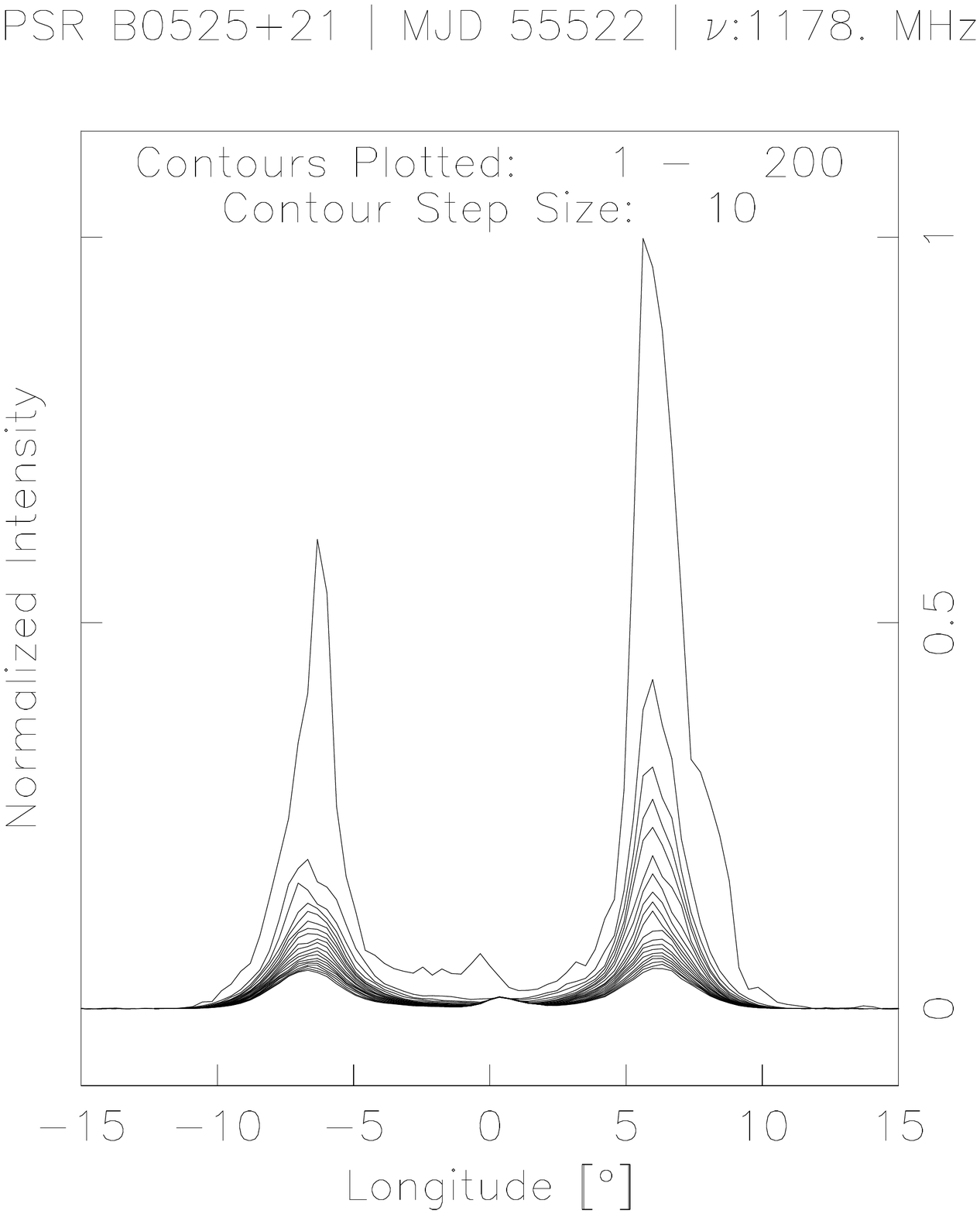} &
\includegraphics[page=1,width=\linewidth]{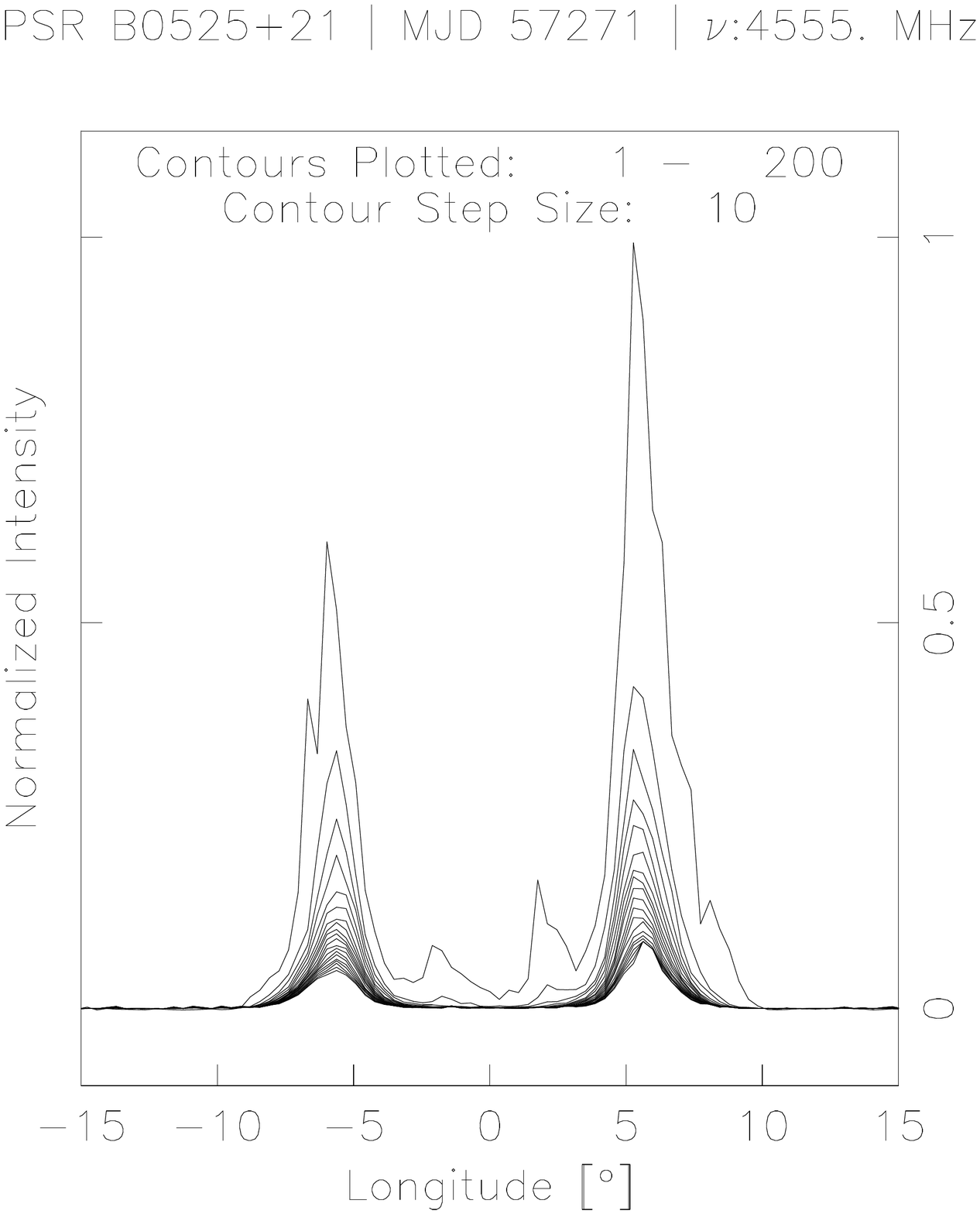} \\ 
     \bottomrule
   \end{tabularx} 
\caption{PHPs of PSR's B0301+19, B0523+11, and B0525+21.}
 \end{figure*}
\vspace{1cm}

\begin{figure*} 
 \begin{tabularx}{\textwidth}{YYY}
 \multicolumn{3}{c}{} \\ \toprule
\includegraphics[page=1,width=\linewidth]{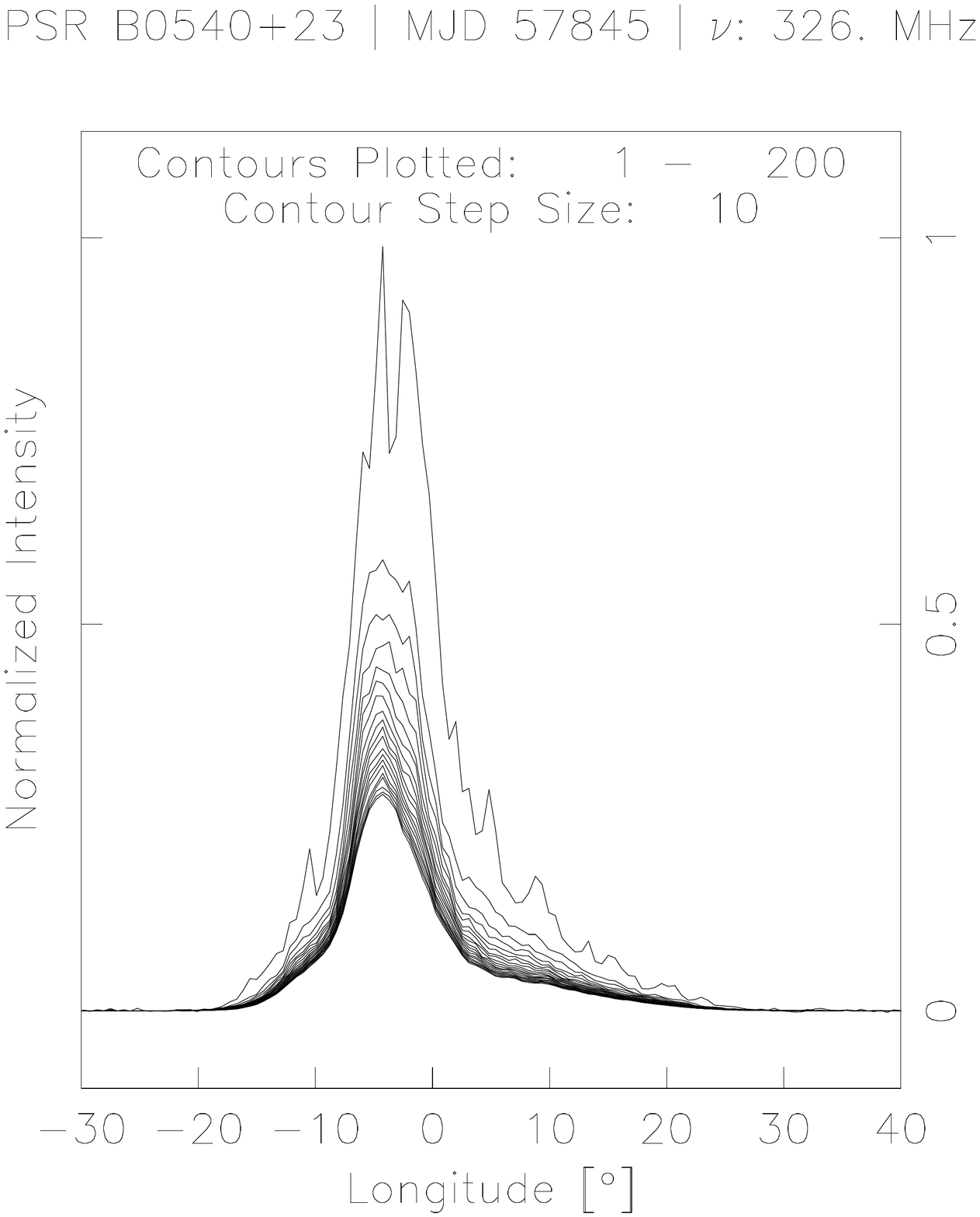} &
\includegraphics[page=1,width=\linewidth]{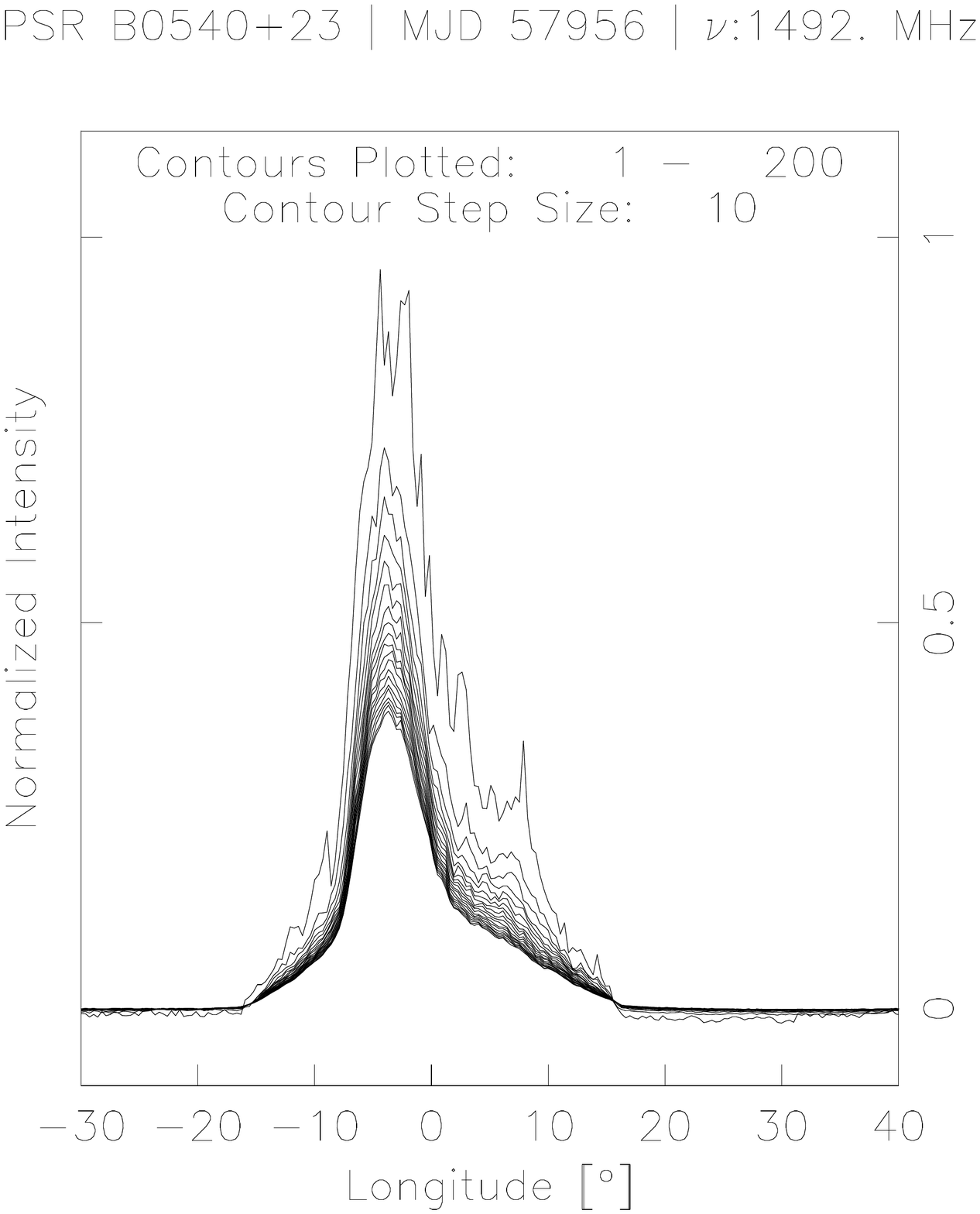} &
\includegraphics[page=1,width=\linewidth]{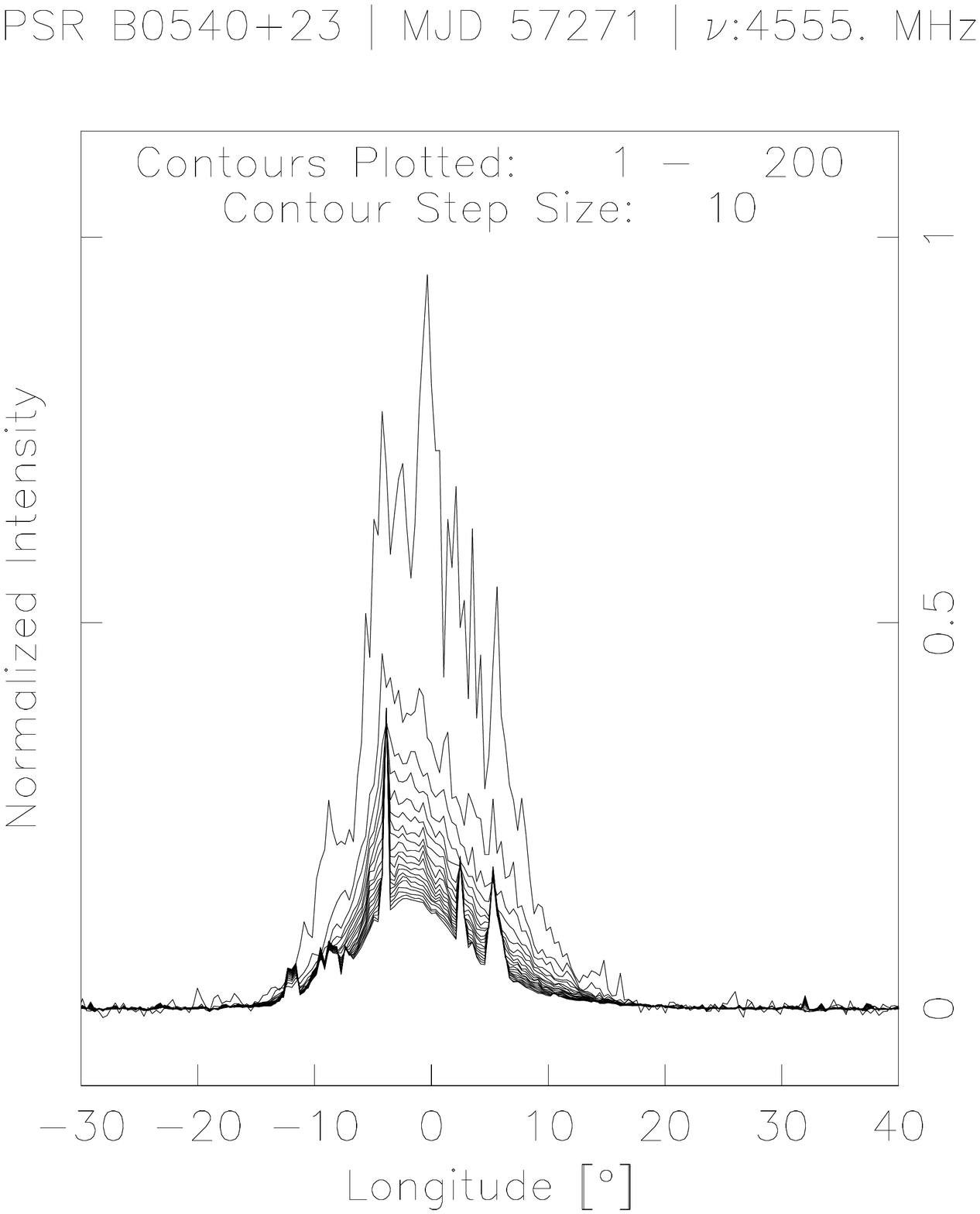} \\ \toprule
\includegraphics[page=1,width=\linewidth]{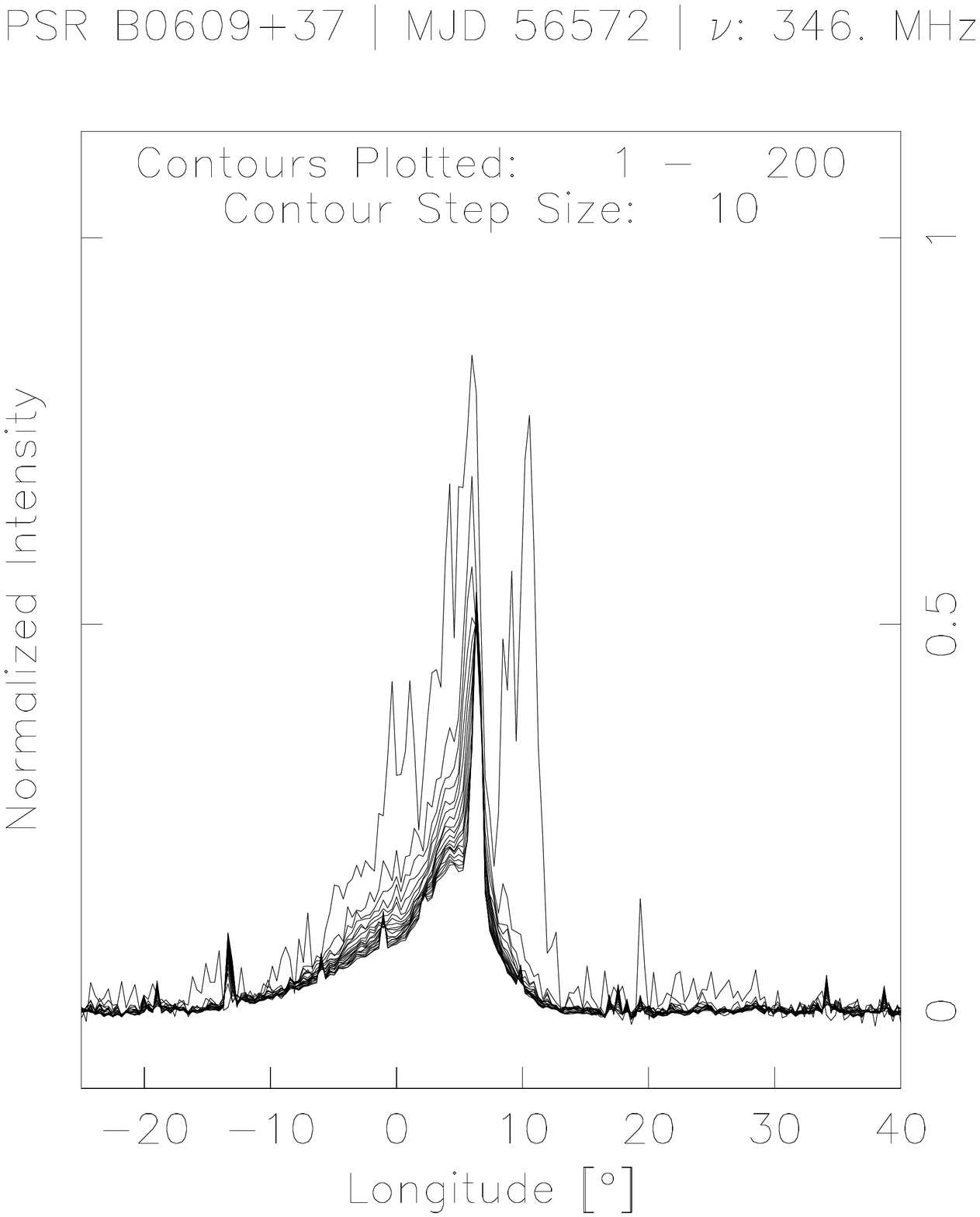} &
\includegraphics[page=1,width=\linewidth]{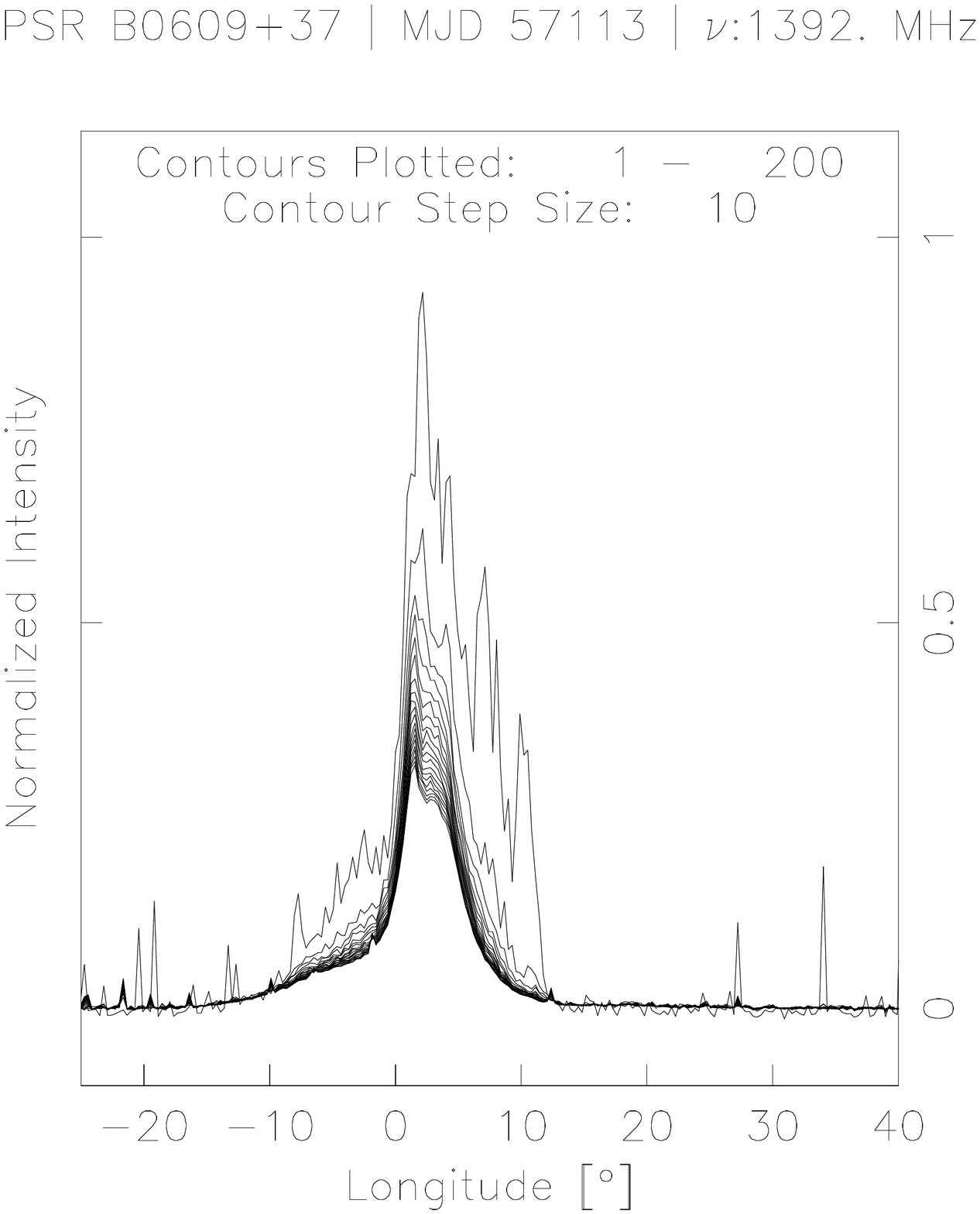} &
\includegraphics[page=1,width=\linewidth]{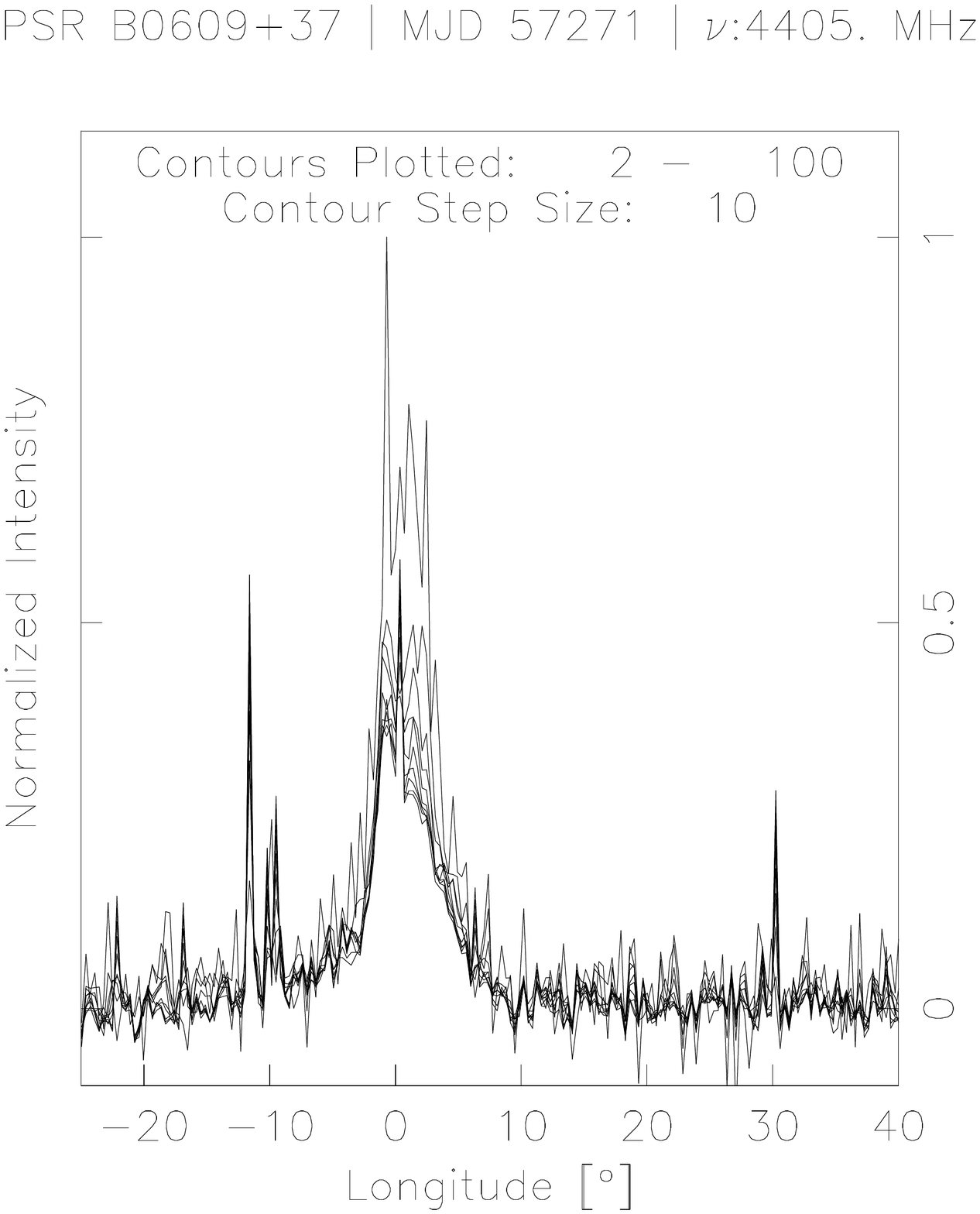} \\ \toprule
\includegraphics[page=1,width=\linewidth]{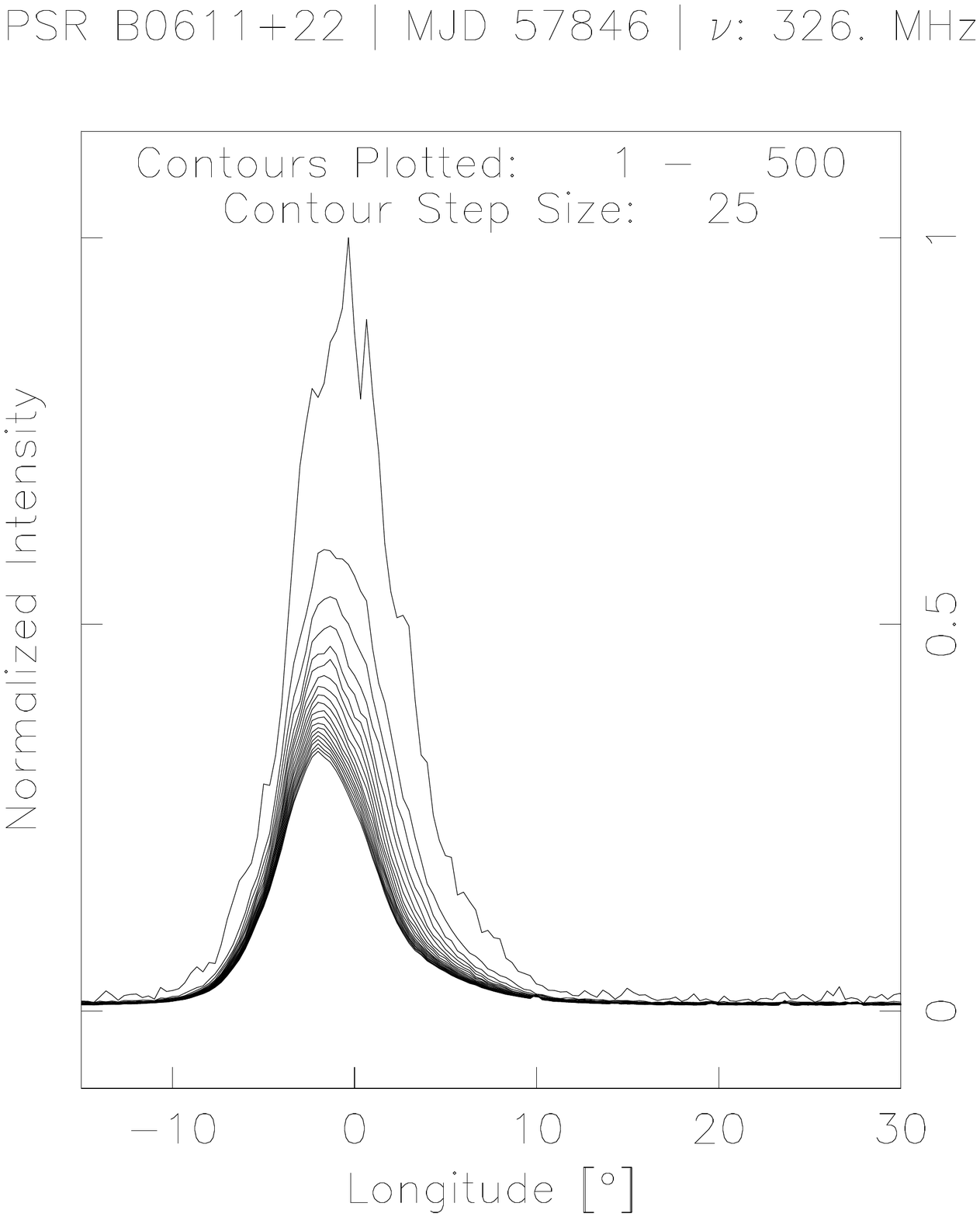} &
\includegraphics[page=1,width=\linewidth]{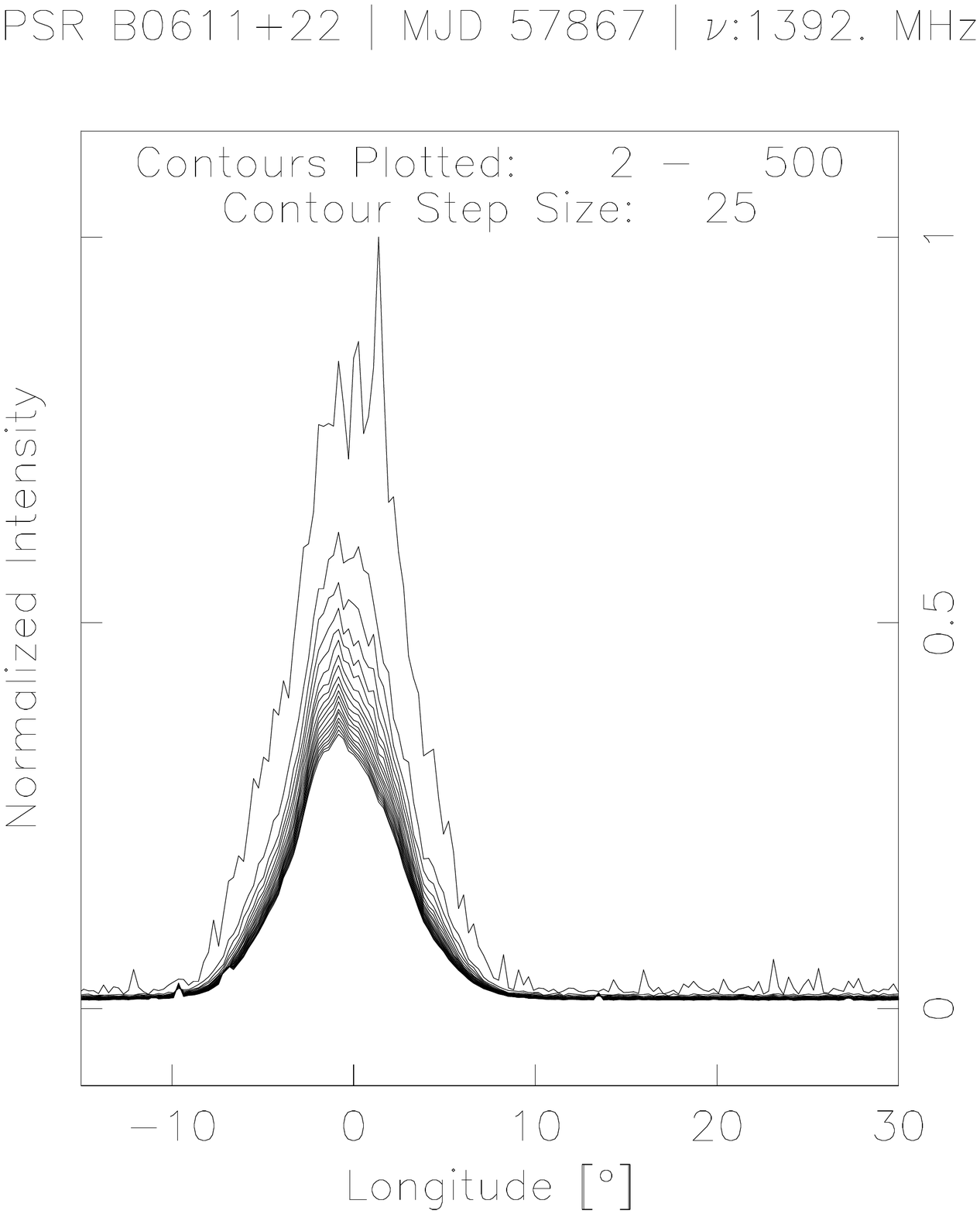} &
\includegraphics[page=1,width=\linewidth]{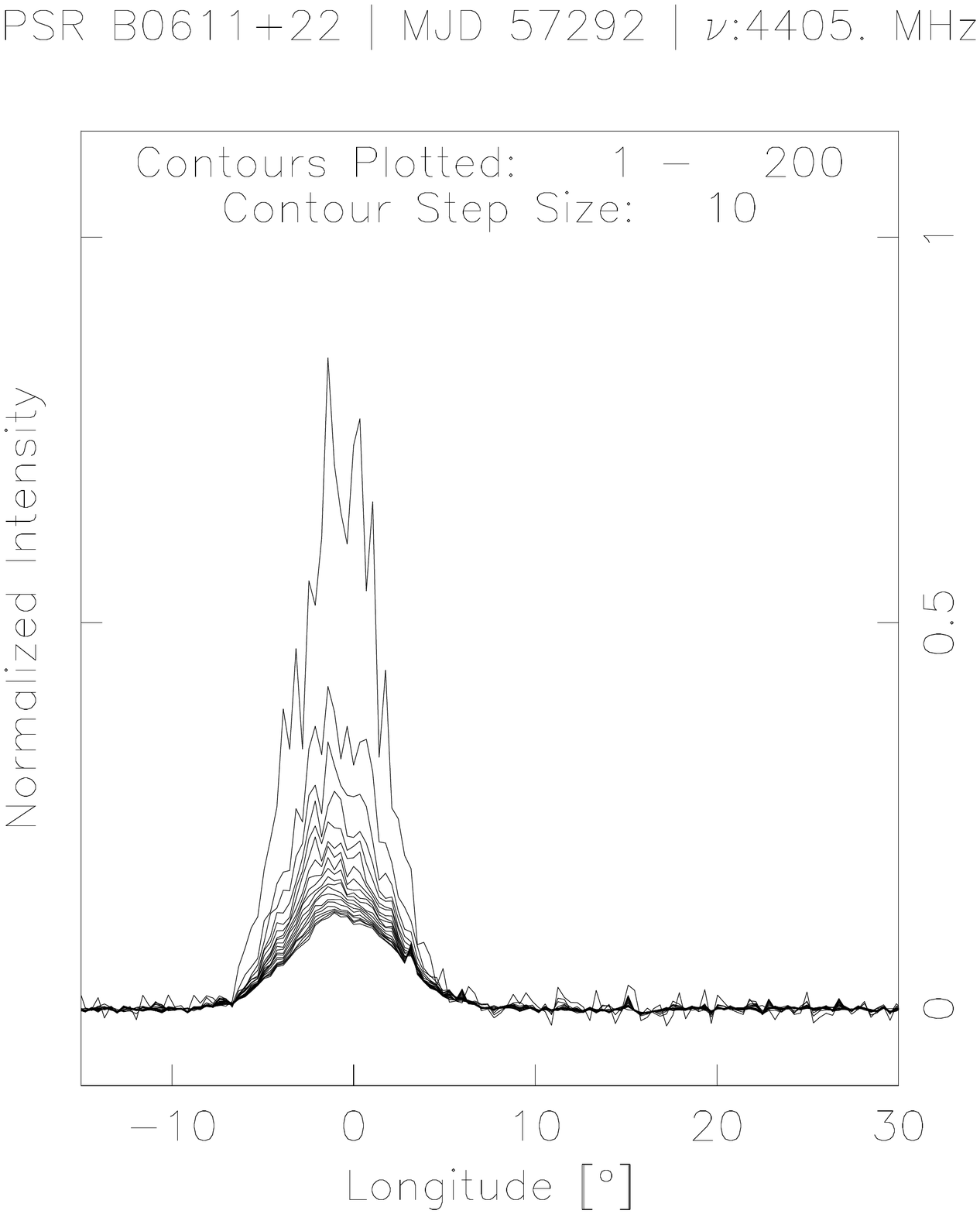} \\ 
     \bottomrule
   \end{tabularx} 
\caption{PHPs of PSR's B0540+23, B0609+37, and B0611+22.}
 \end{figure*}
\vspace{1cm}

\begin{figure*} 
 \begin{tabularx}{\textwidth}{YYY}
 \multicolumn{3}{c}{} \\ \toprule

\includegraphics[page=1,width=\linewidth]{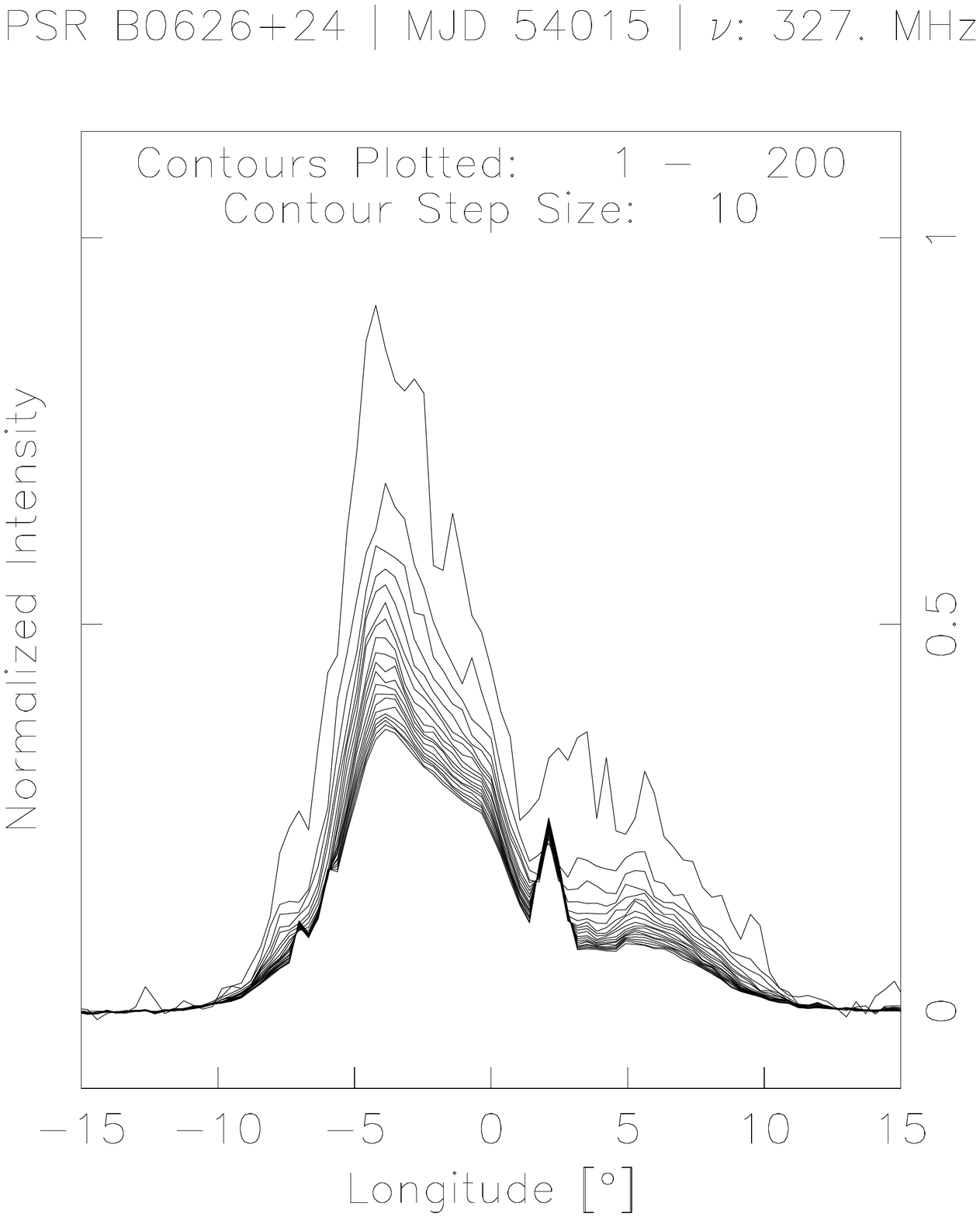} &
\includegraphics[page=1,width=\linewidth]{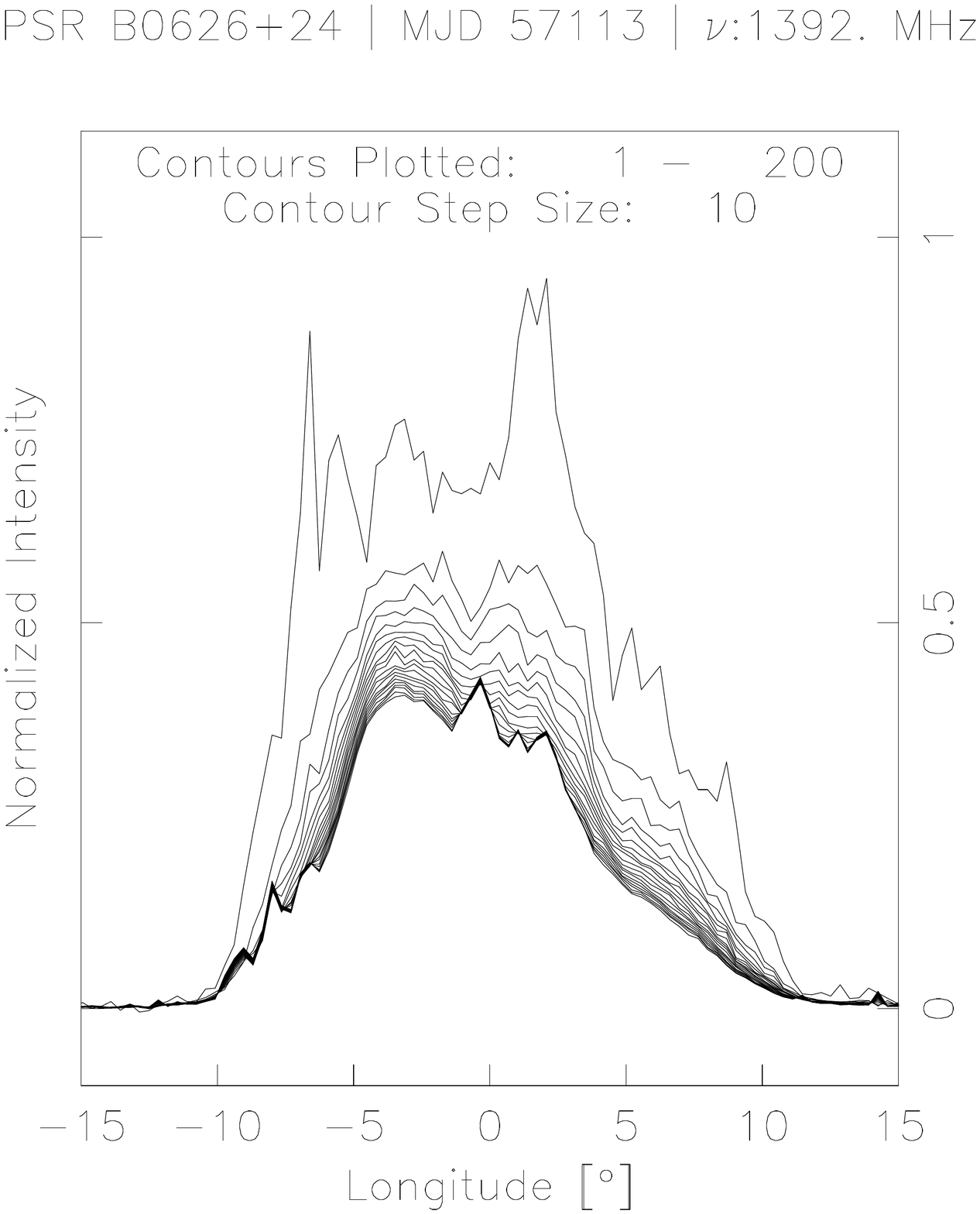} &
\includegraphics[page=1,width=\linewidth]{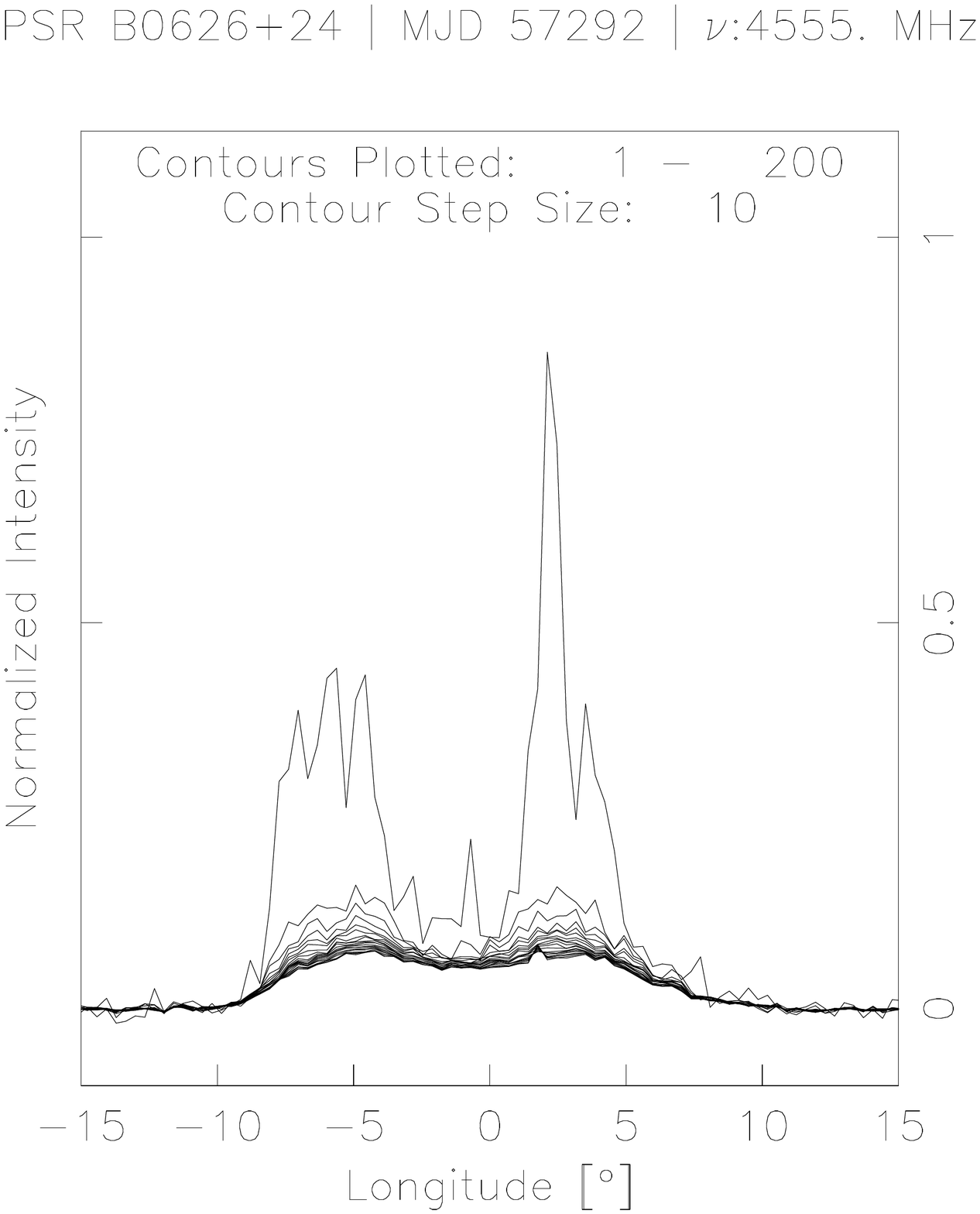} \\ \toprule
\includegraphics[page=1,width=\linewidth]{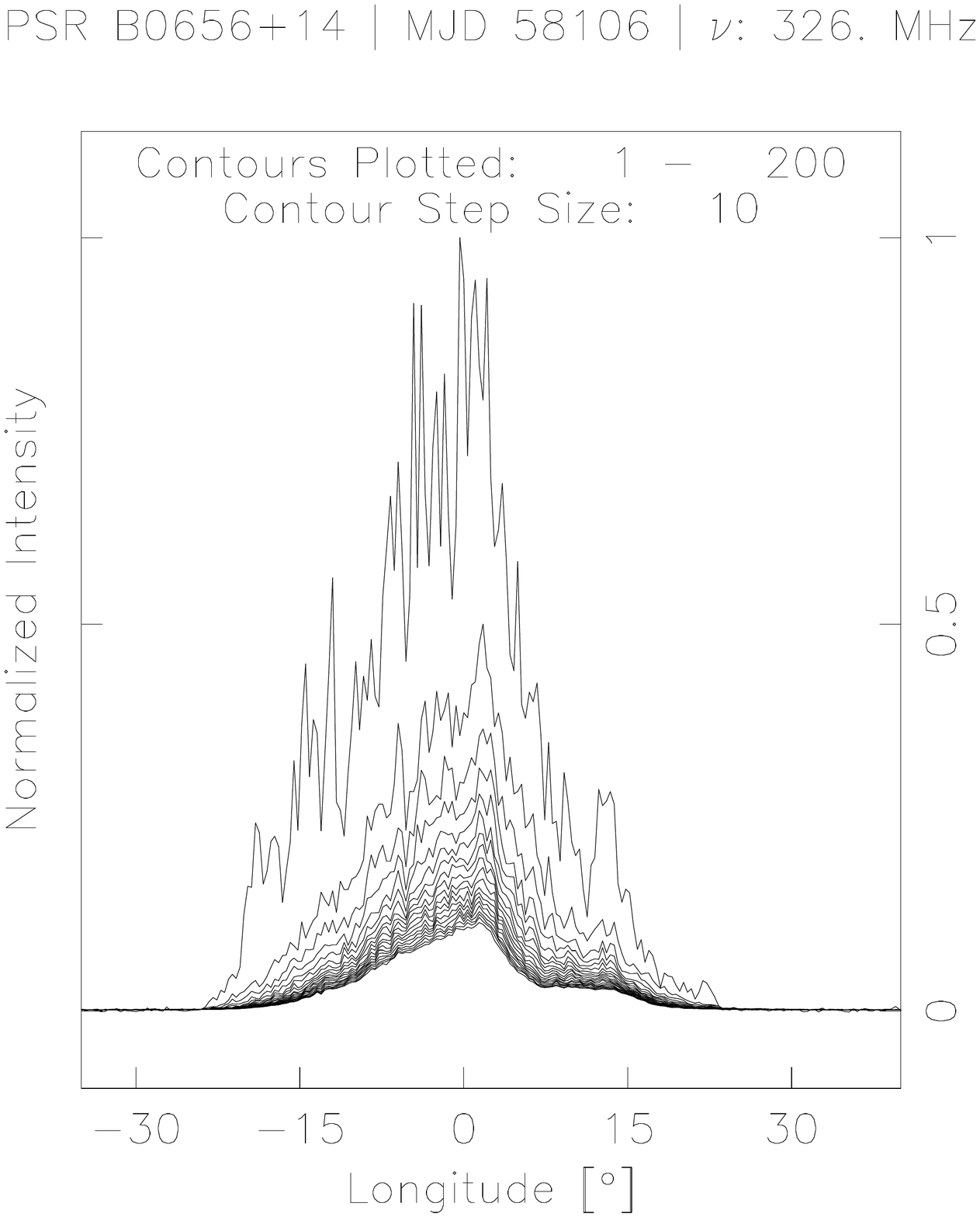} &
\includegraphics[page=1,width=\linewidth]{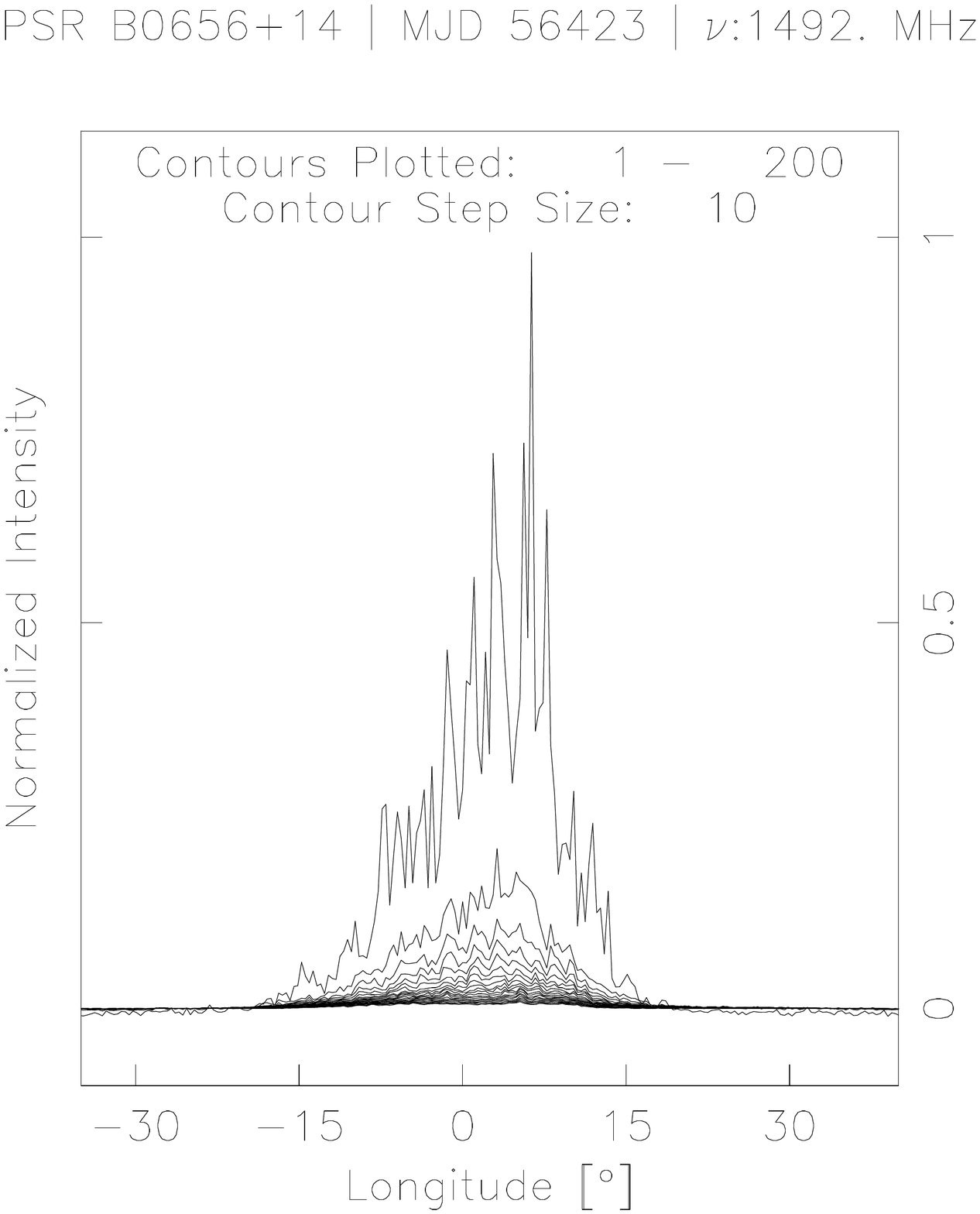} &
\includegraphics[page=1,width=\linewidth]{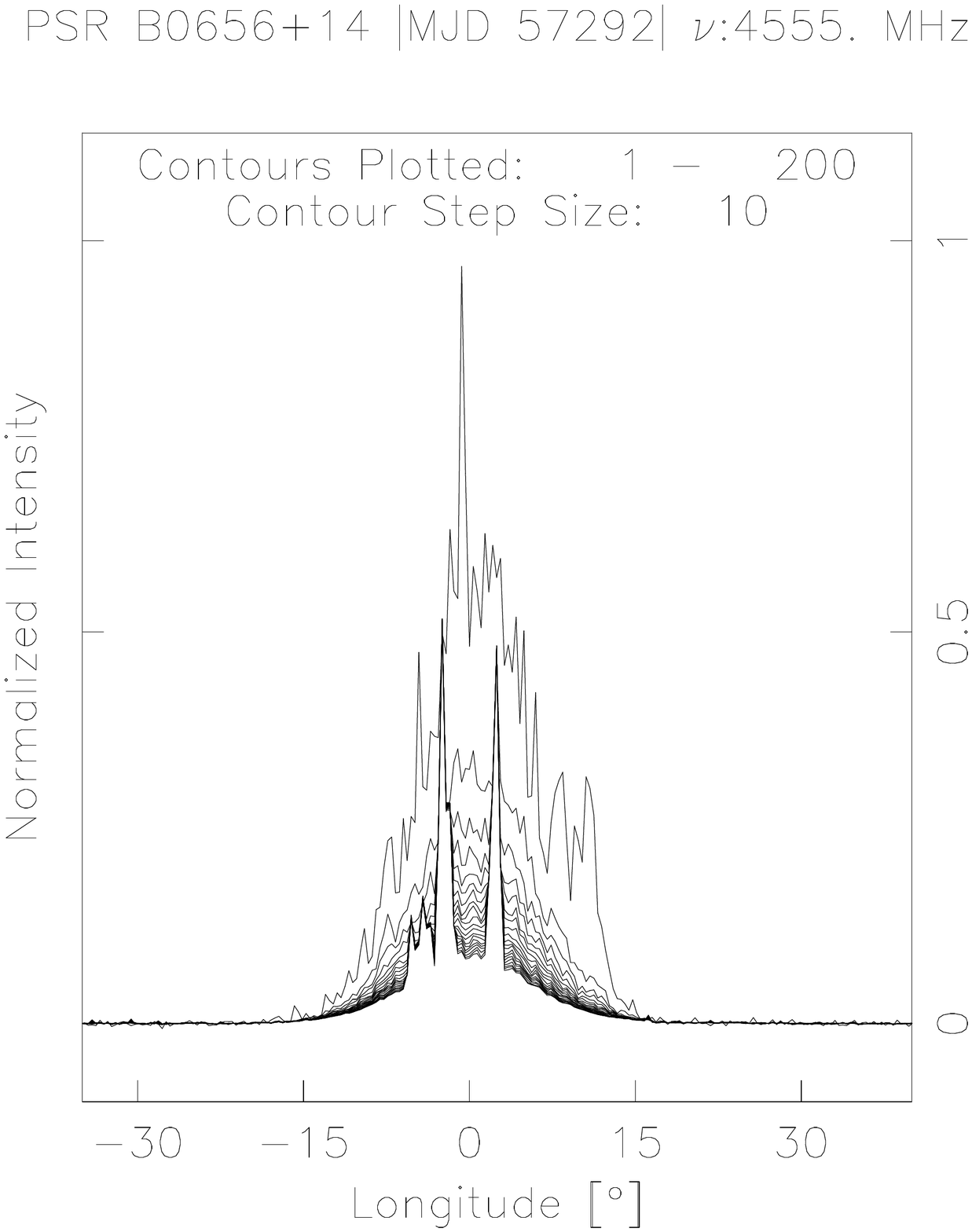} \\ \toprule
\includegraphics[page=1,width=\linewidth]{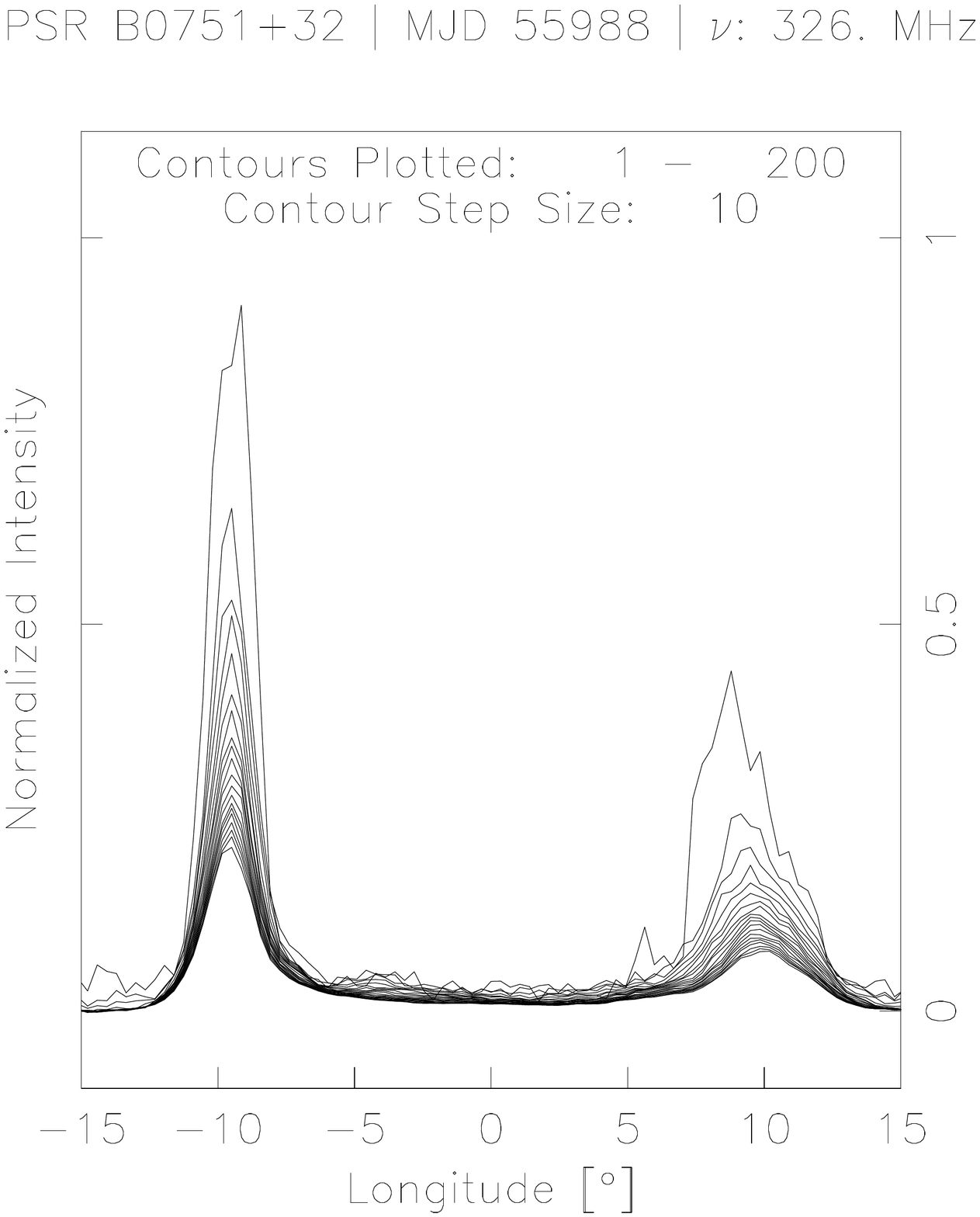} &
\includegraphics[page=1,width=\linewidth]{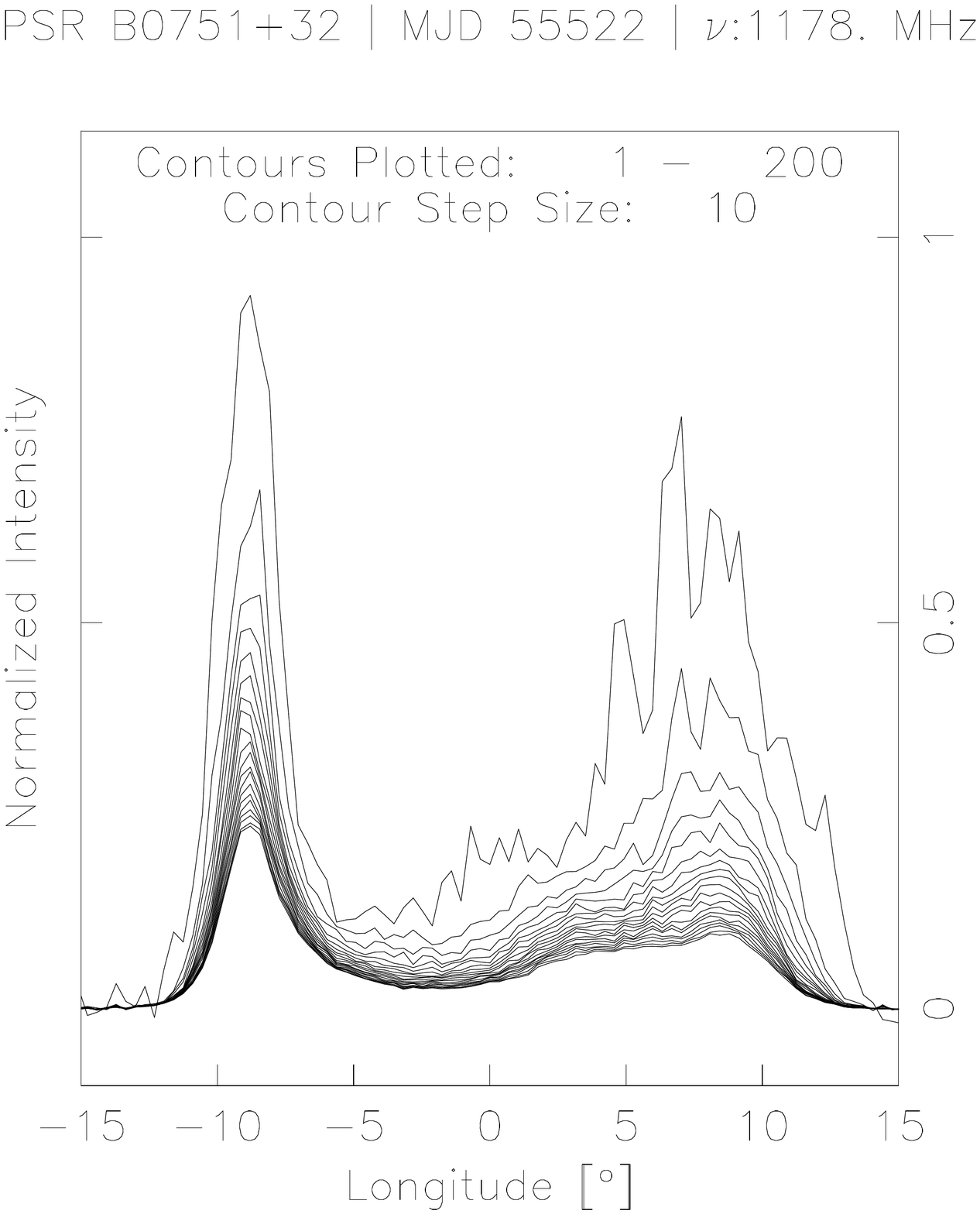} &
\includegraphics[page=1,width=\linewidth]{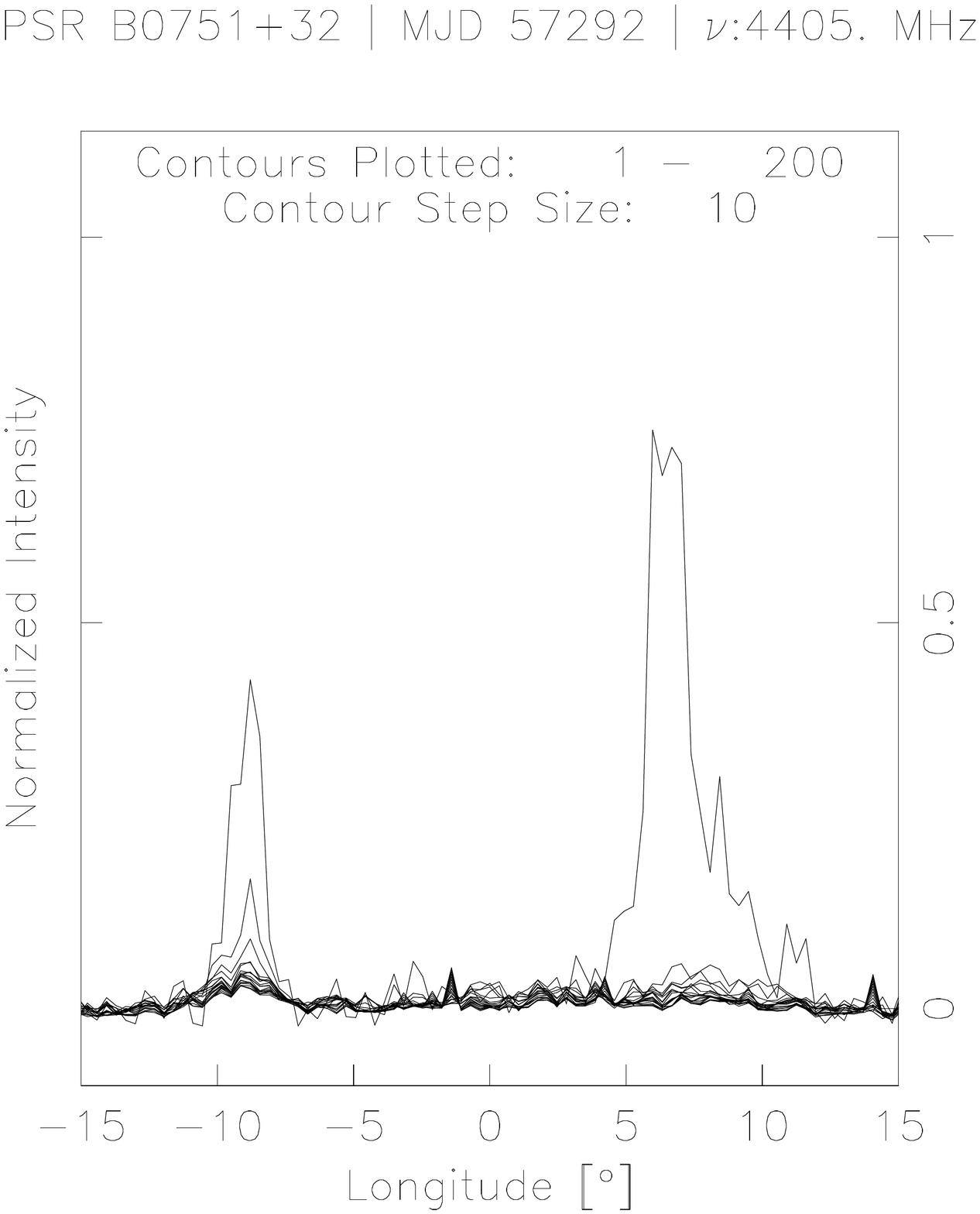} \\ 
     \bottomrule
   \end{tabularx} 
\caption{PHPs of PSR's B0626+24, B0656+14, and B0751+32.}
 \end{figure*}
\vspace{1cm}

\begin{figure*} 
 \begin{tabularx}{\textwidth}{YYY}
 \multicolumn{3}{c}{} \\ \toprule
\includegraphics[page=1,width=\linewidth]{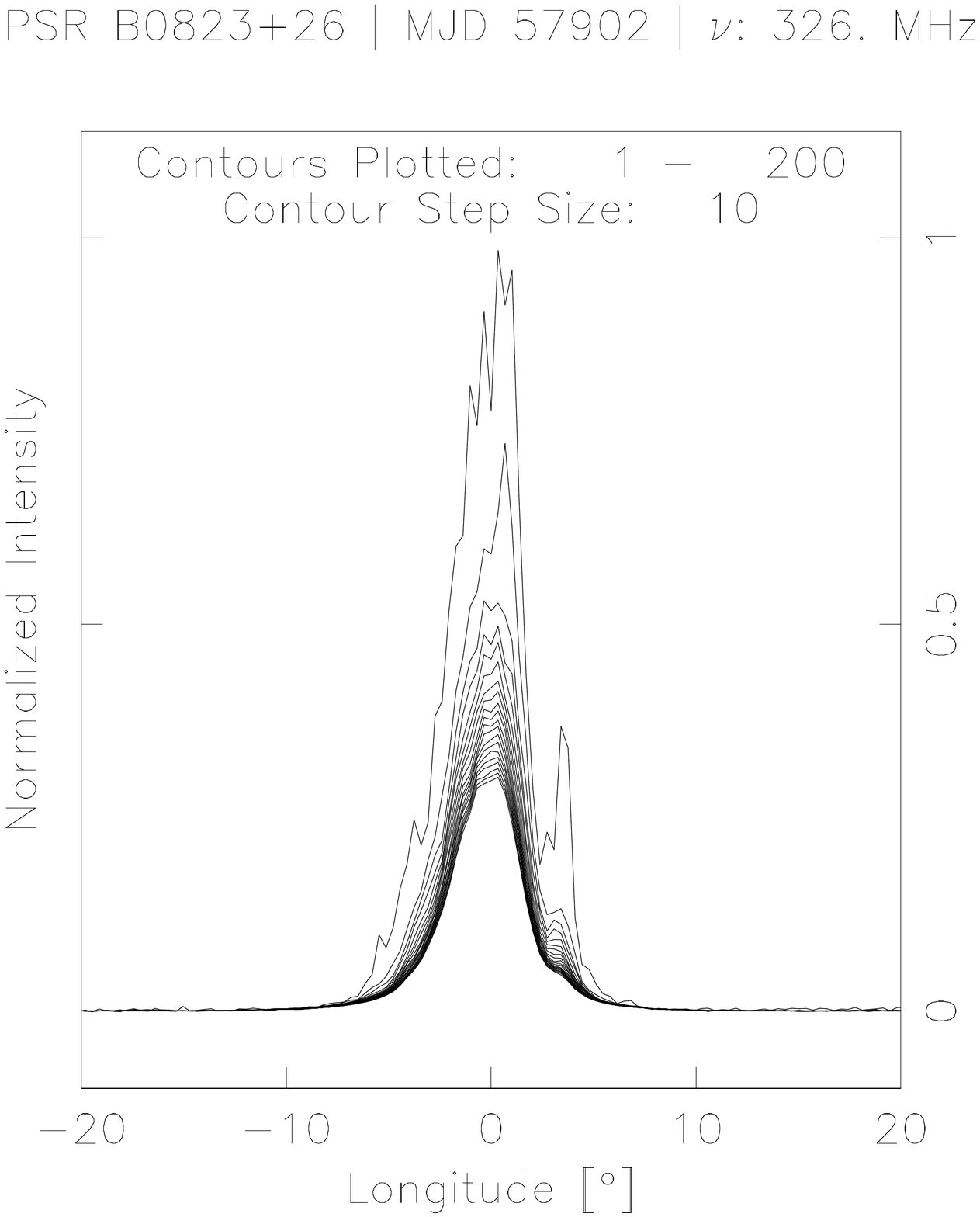} &
\includegraphics[page=1,width=\linewidth]{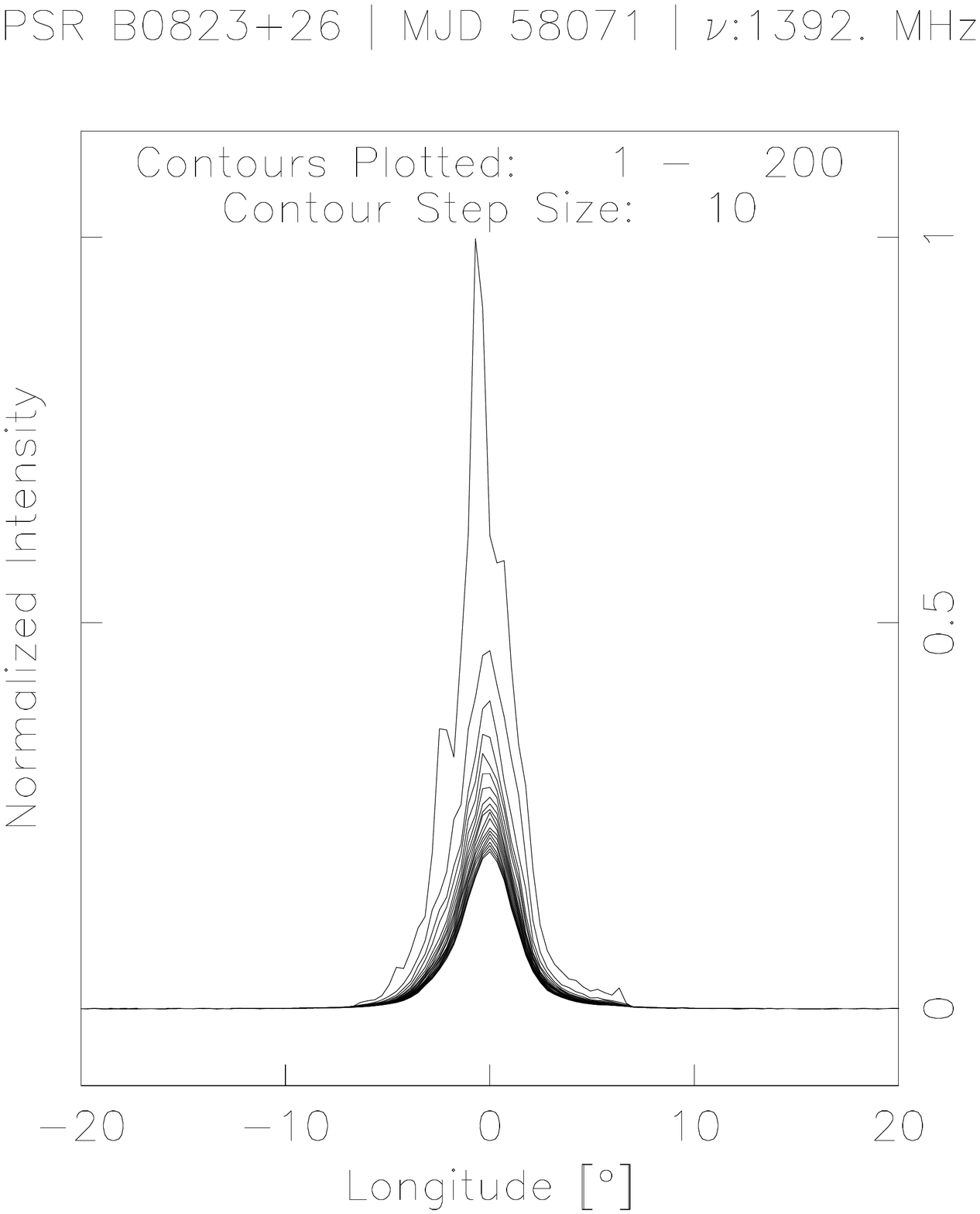} &
\includegraphics[page=1,width=\linewidth]{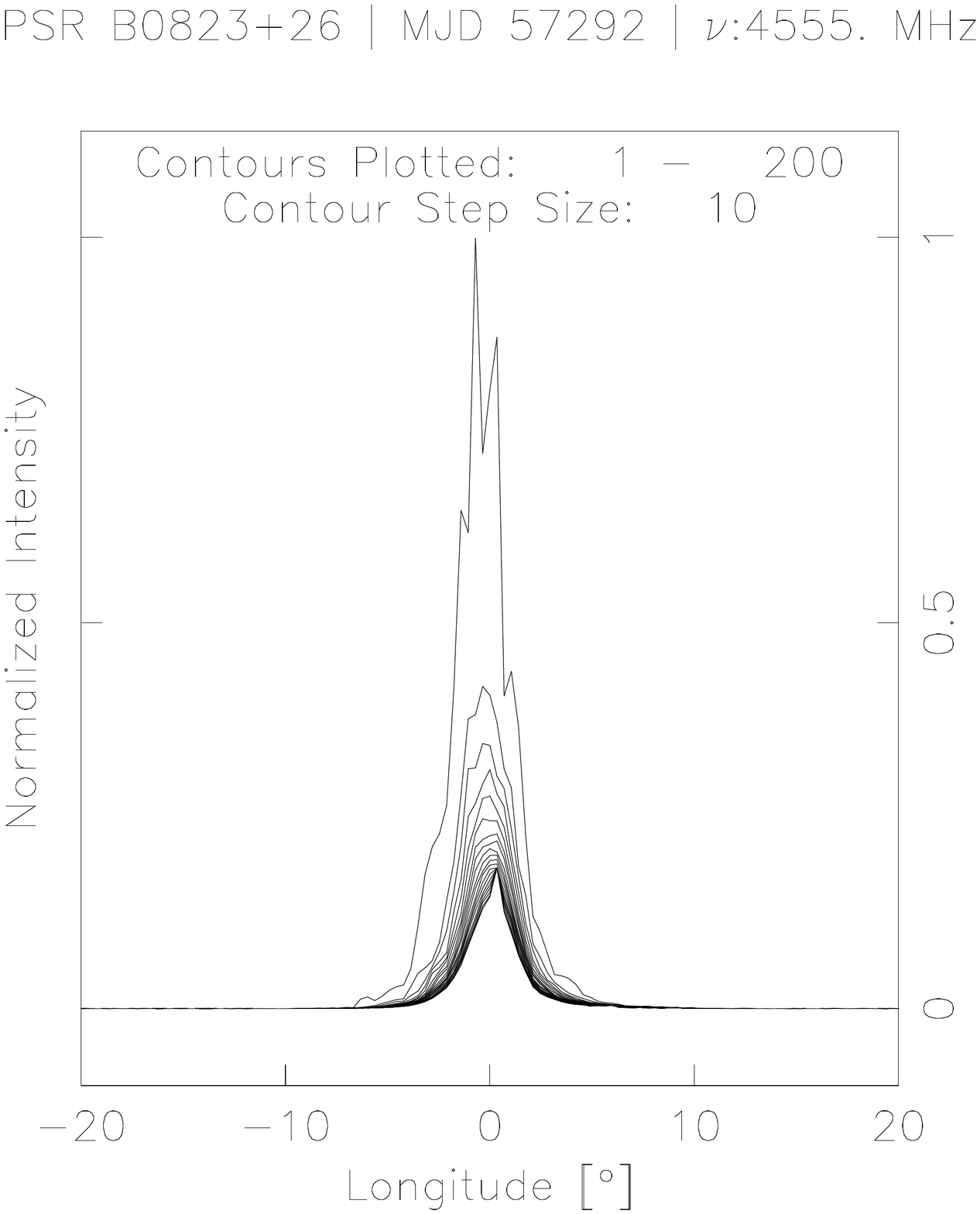} \\ \toprule

\includegraphics[page=1,width=\linewidth]{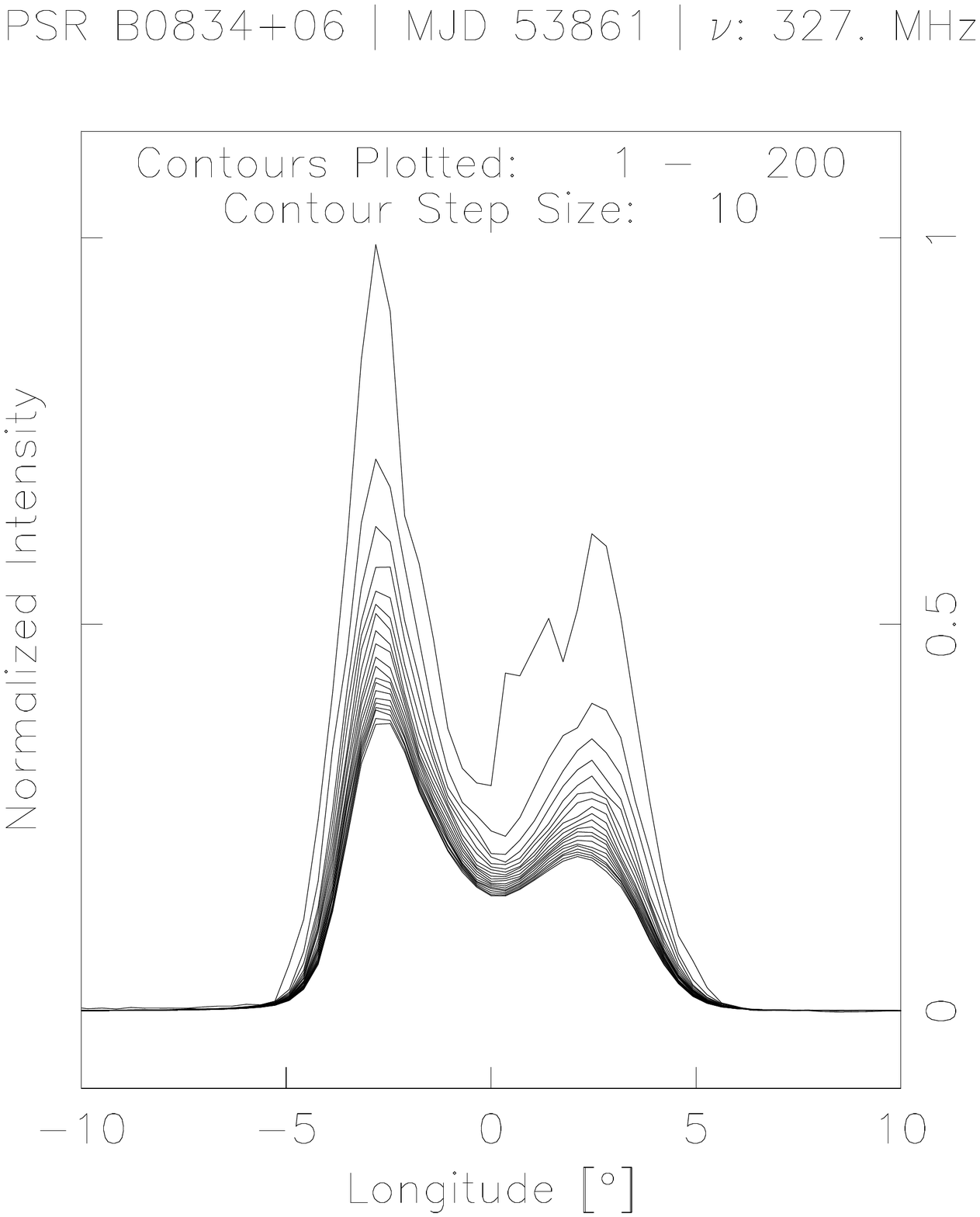} &
\includegraphics[page=1,width=\linewidth]{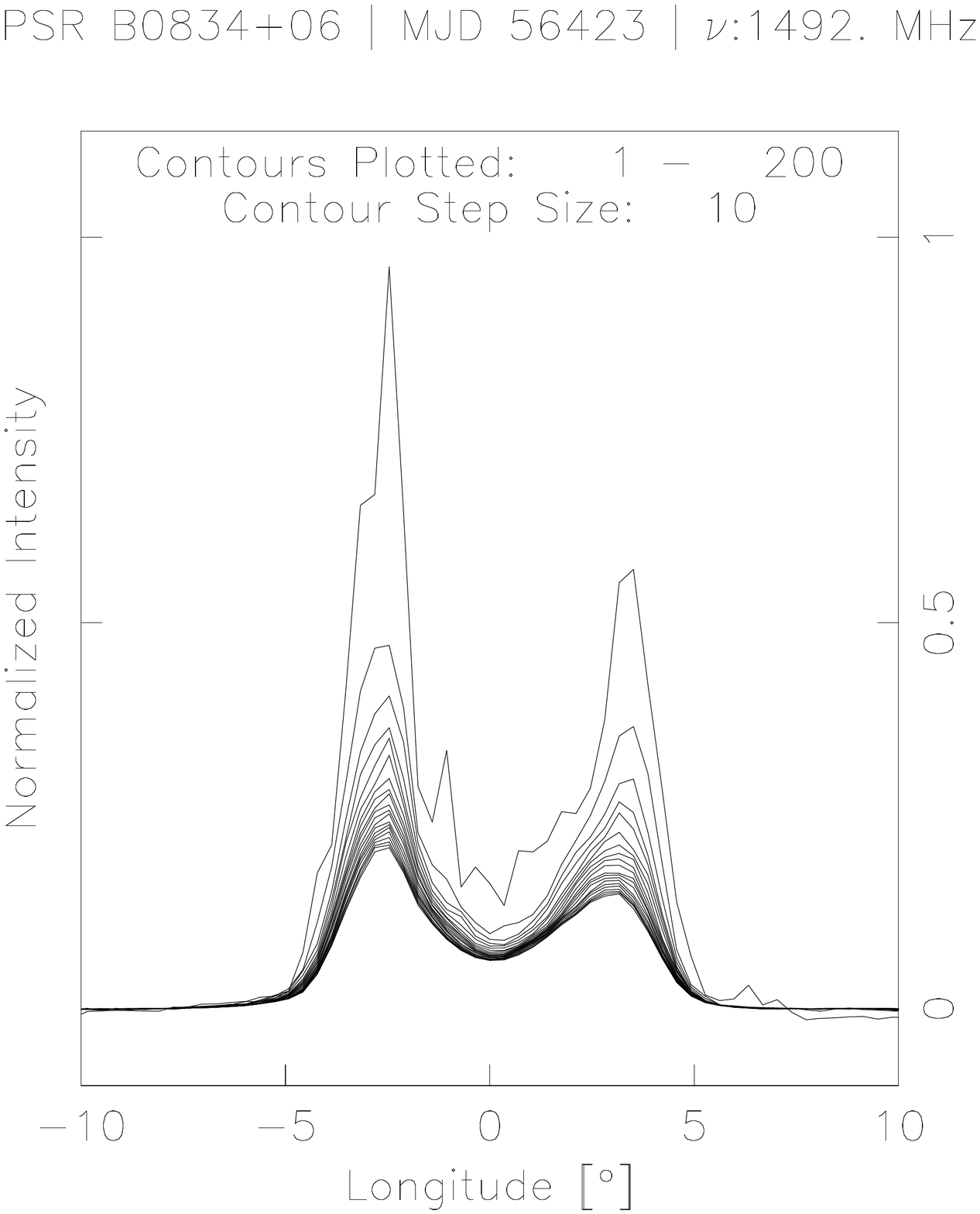} &
\\ \toprule
\includegraphics[page=1,width=\linewidth]{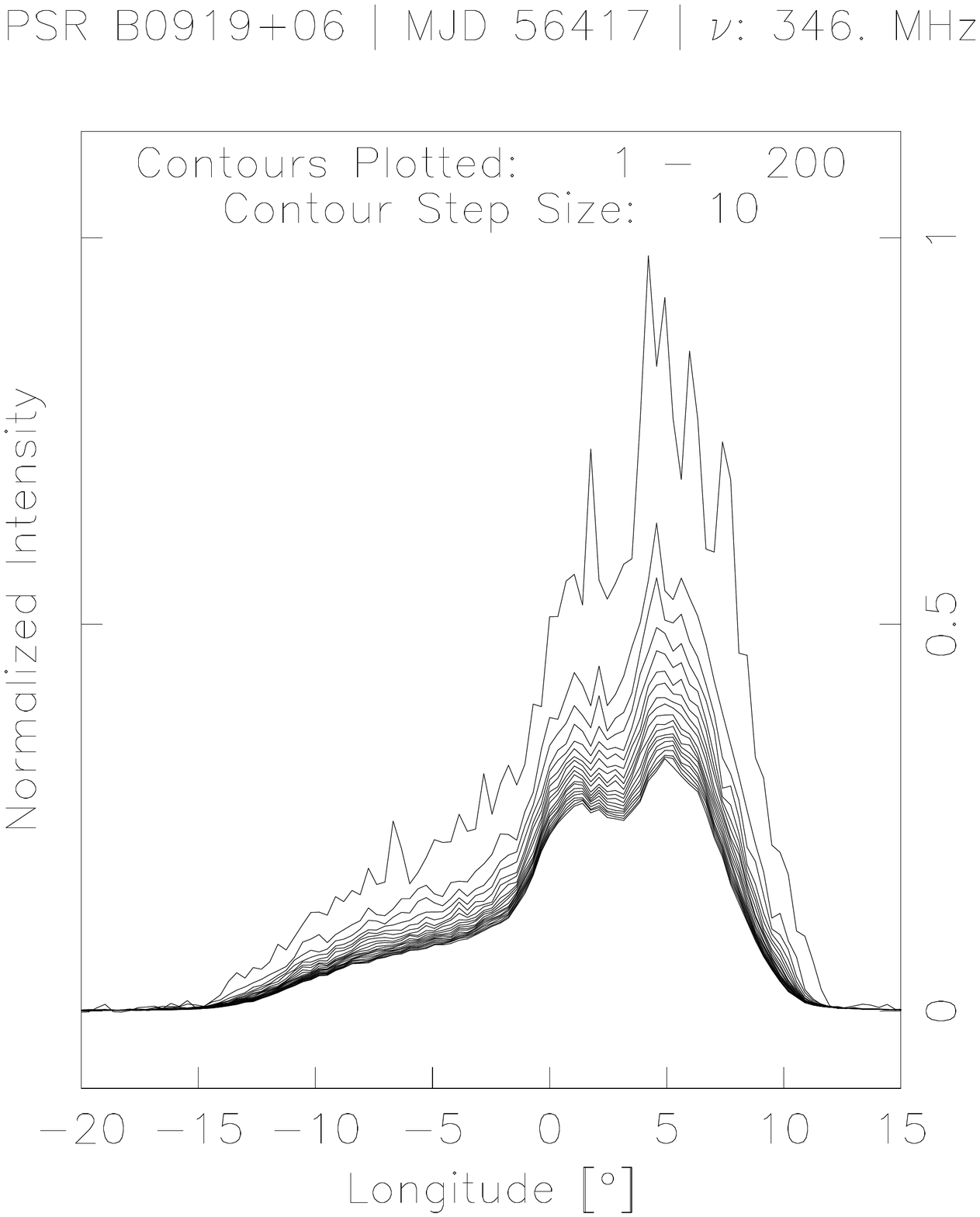} &
\includegraphics[page=1,width=\linewidth]{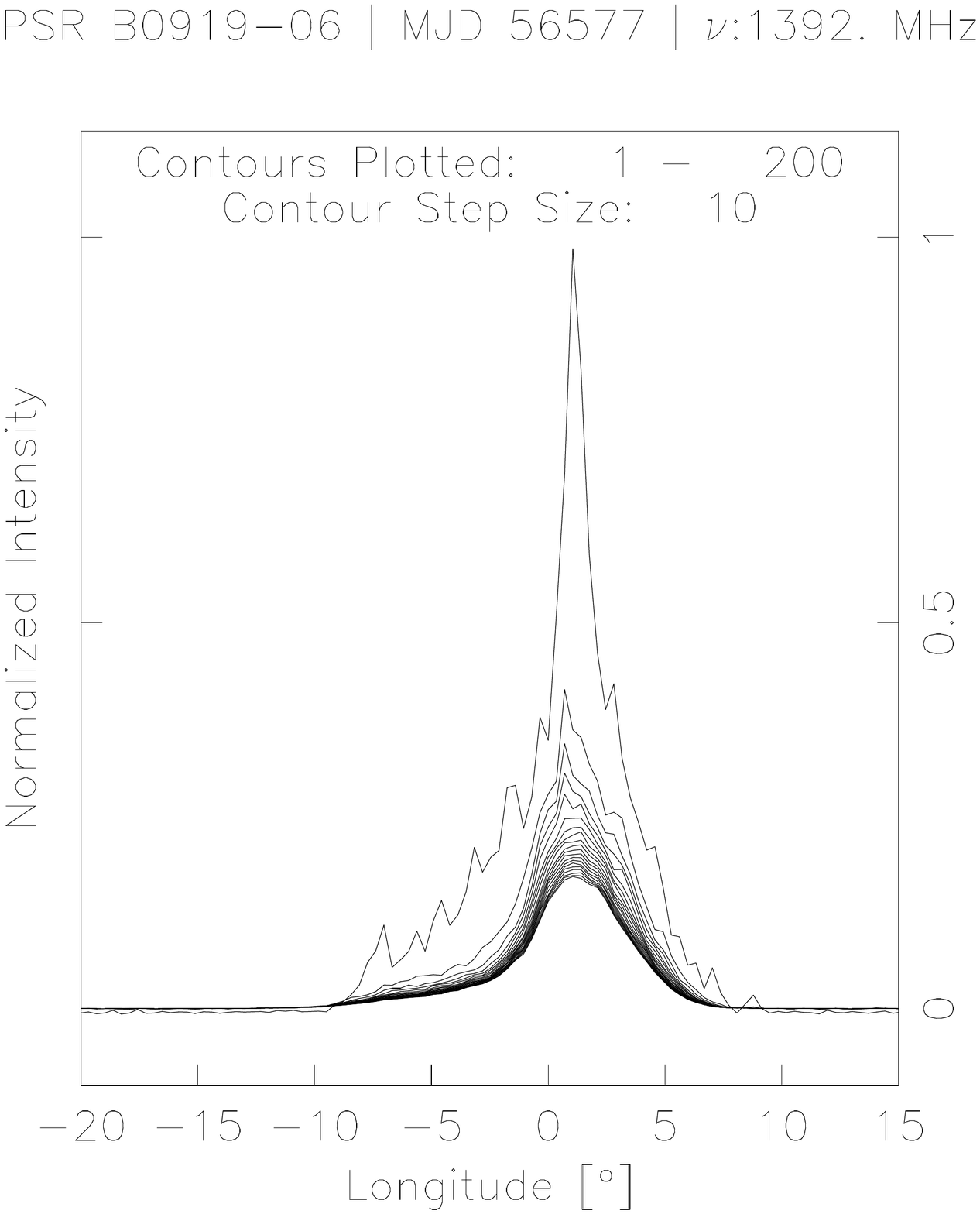} &
\includegraphics[page=1,width=\linewidth]{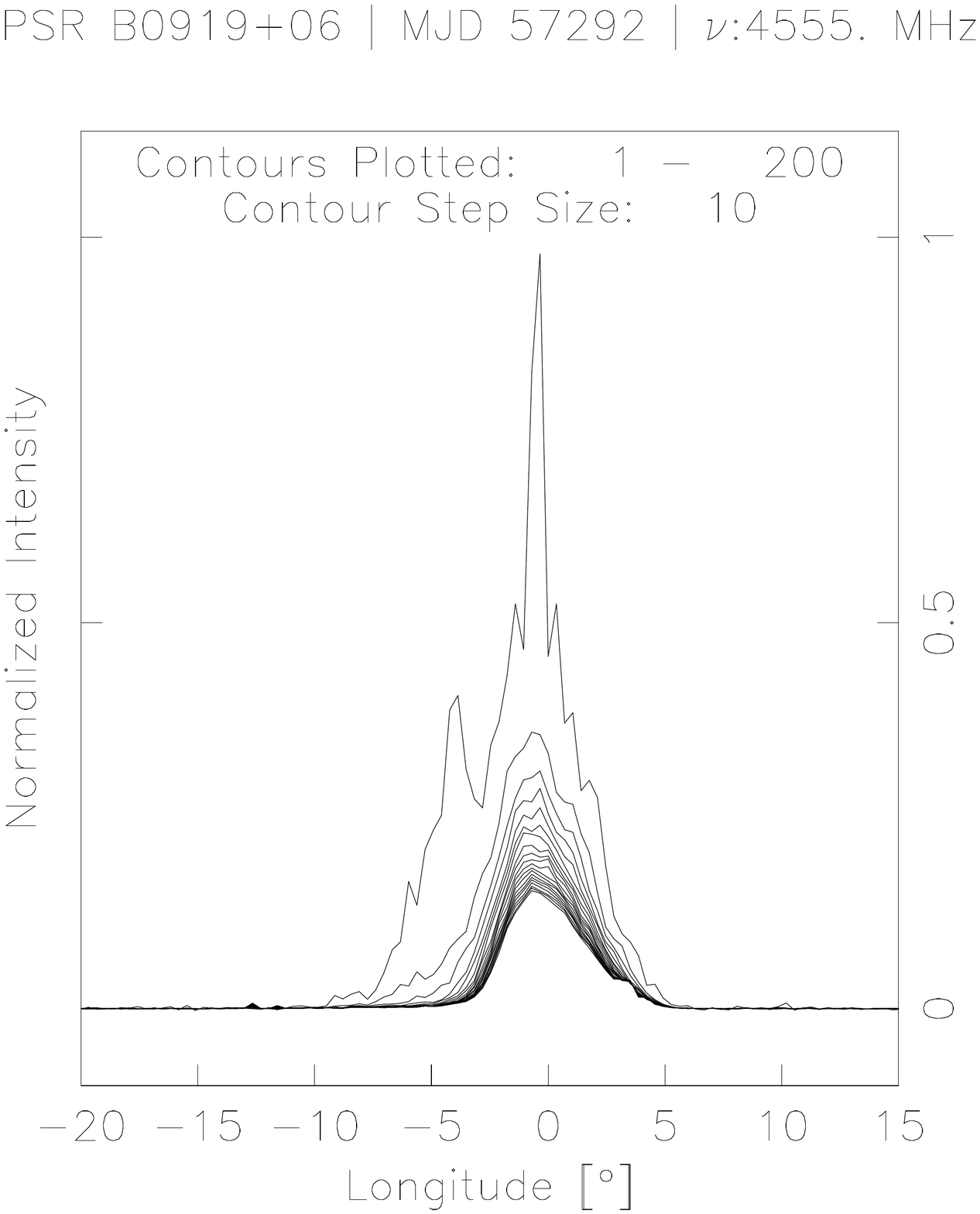} \\ 
     \bottomrule
   \end{tabularx} 
\caption{PHPs of PSR's B0823+26, B0834+06, and B0919+06.}
 \end{figure*}
\vspace{1cm}

   \begin{figure*} 
 \begin{tabularx}{\textwidth}{YYY}
    \multicolumn{3}{c}{} \\ \toprule

\includegraphics[page=1,width=\linewidth]{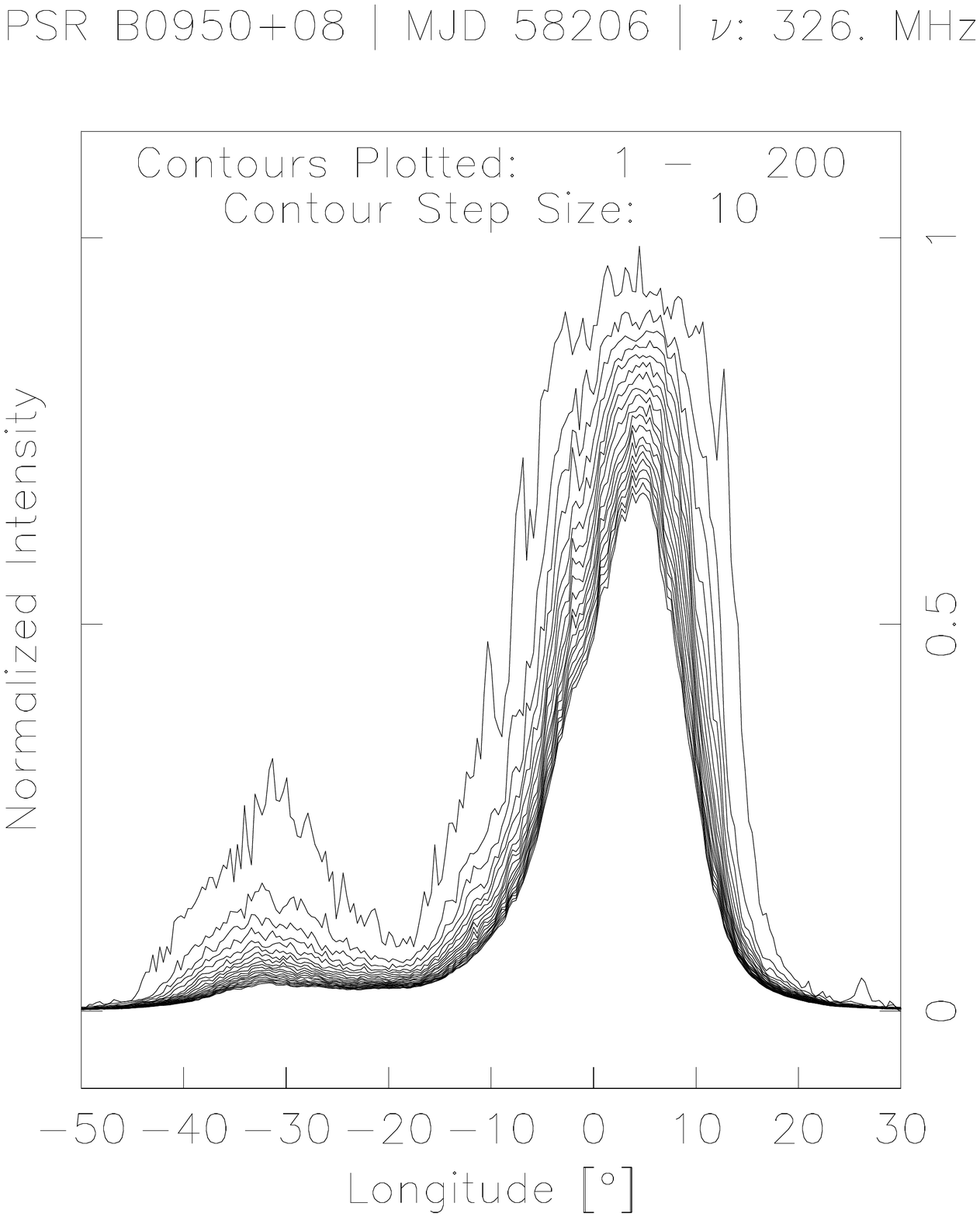} &
\includegraphics[page=1,width=\linewidth]{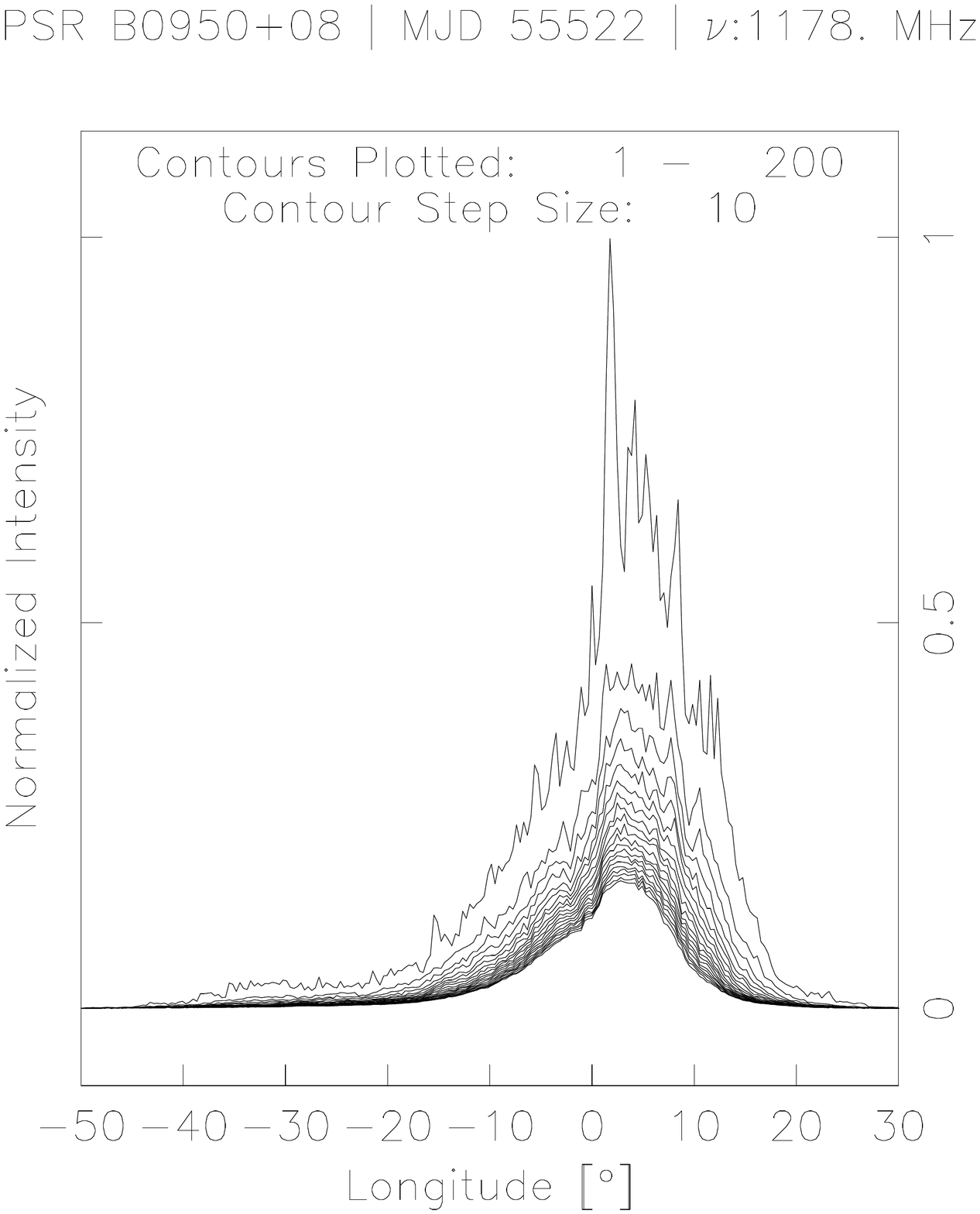} &
\includegraphics[page=1,width=\linewidth]{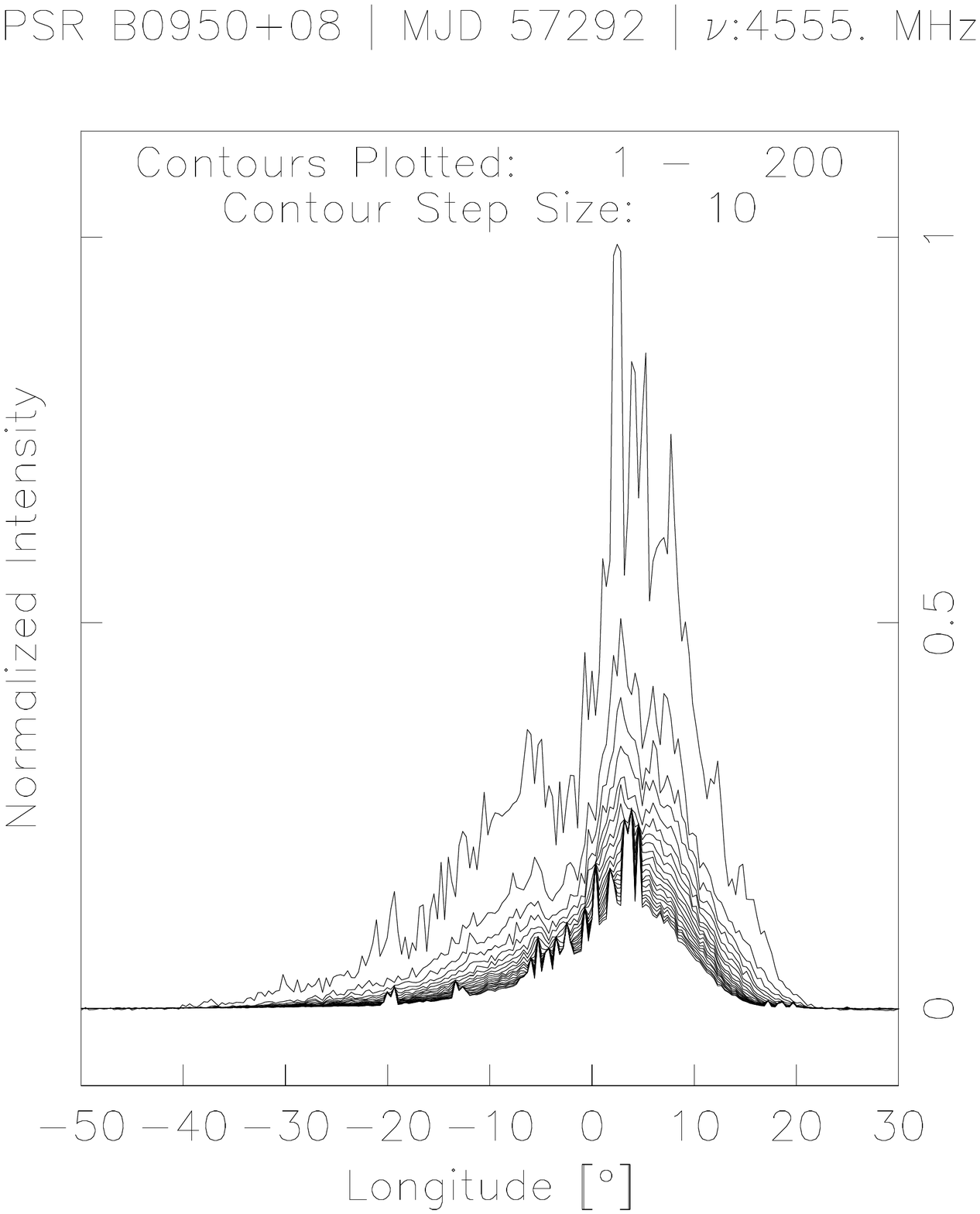} \\ \toprule

\includegraphics[page=1,width=\linewidth]{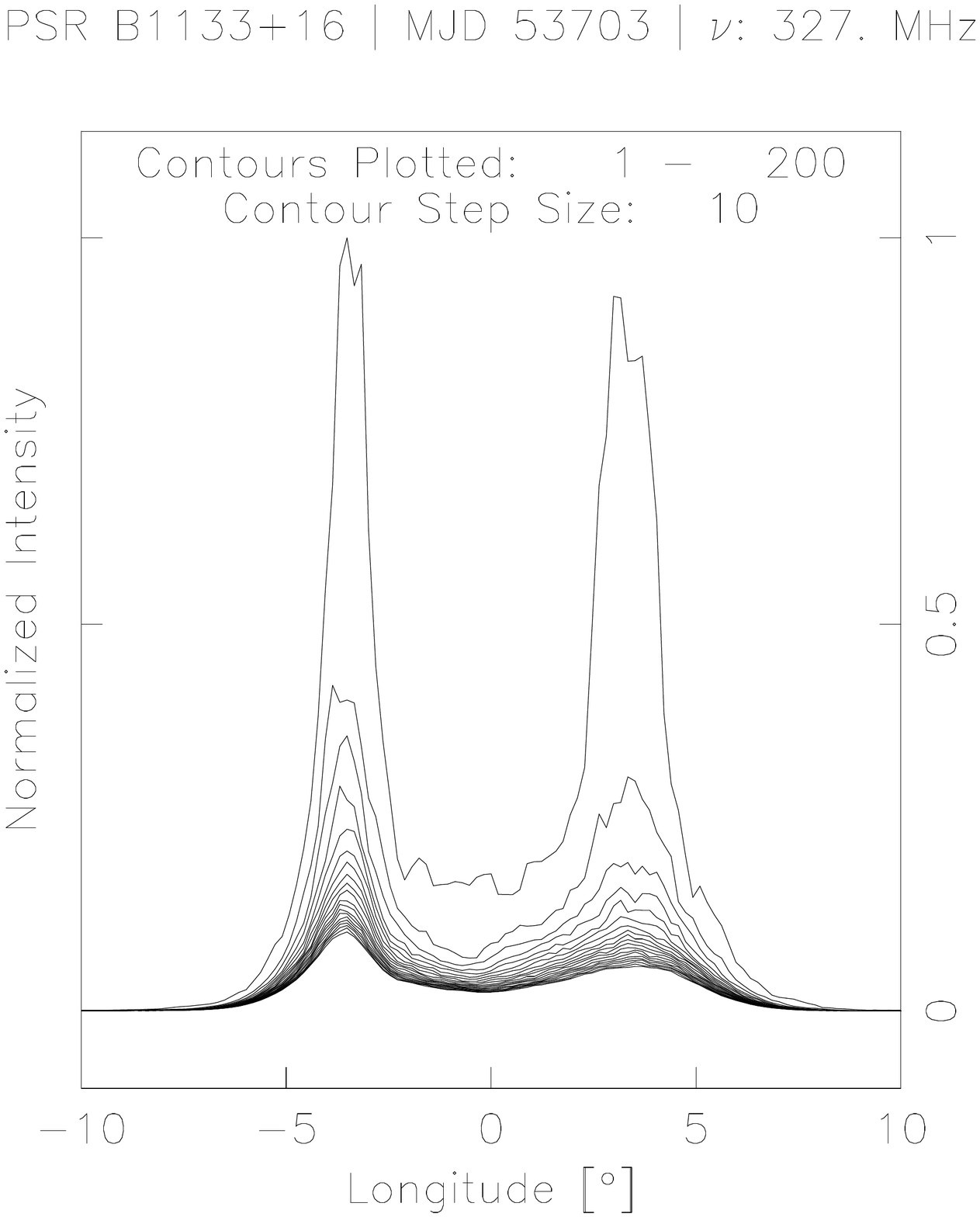} &
\includegraphics[page=1,width=\linewidth]{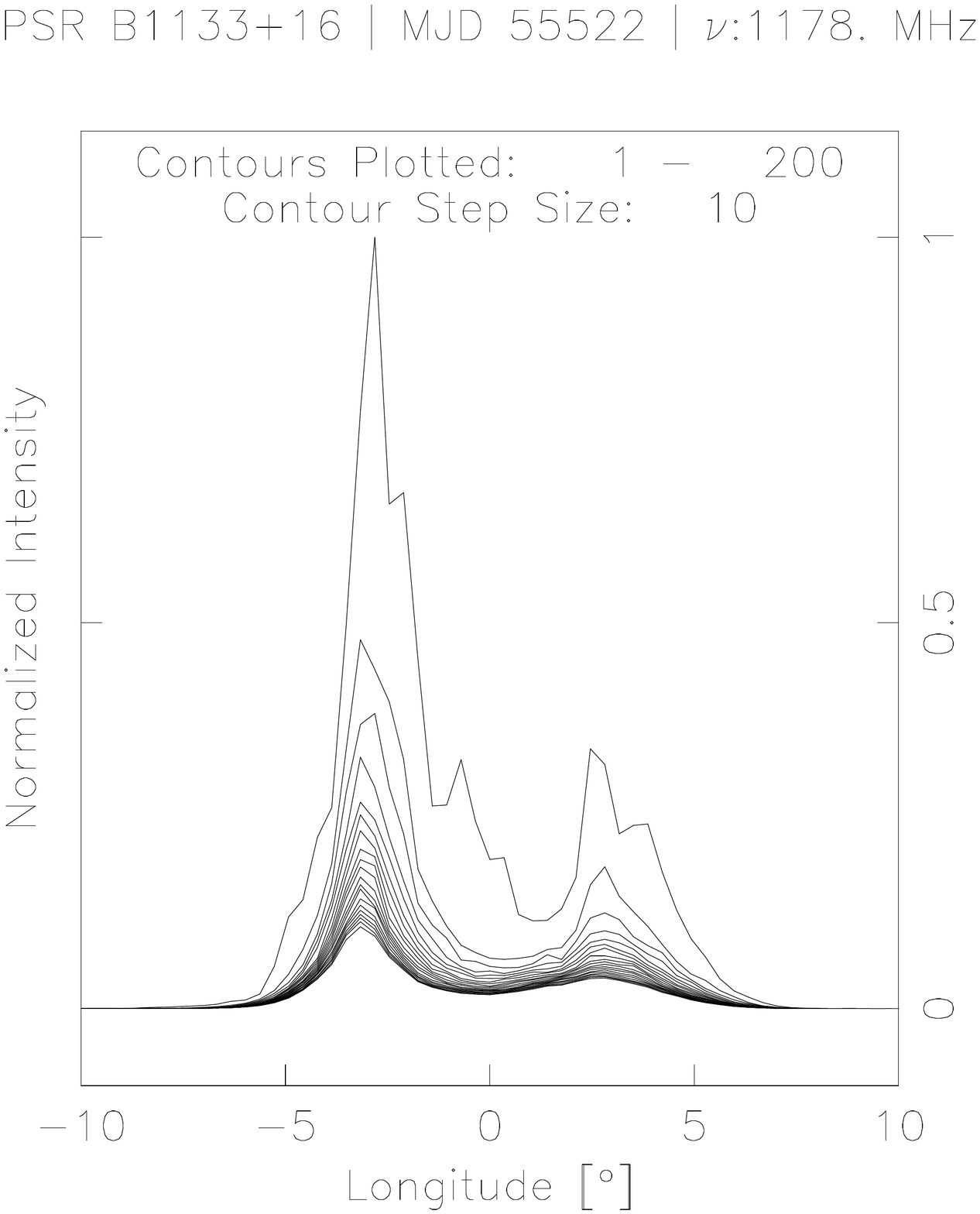} &
\includegraphics[page=1,width=\linewidth]{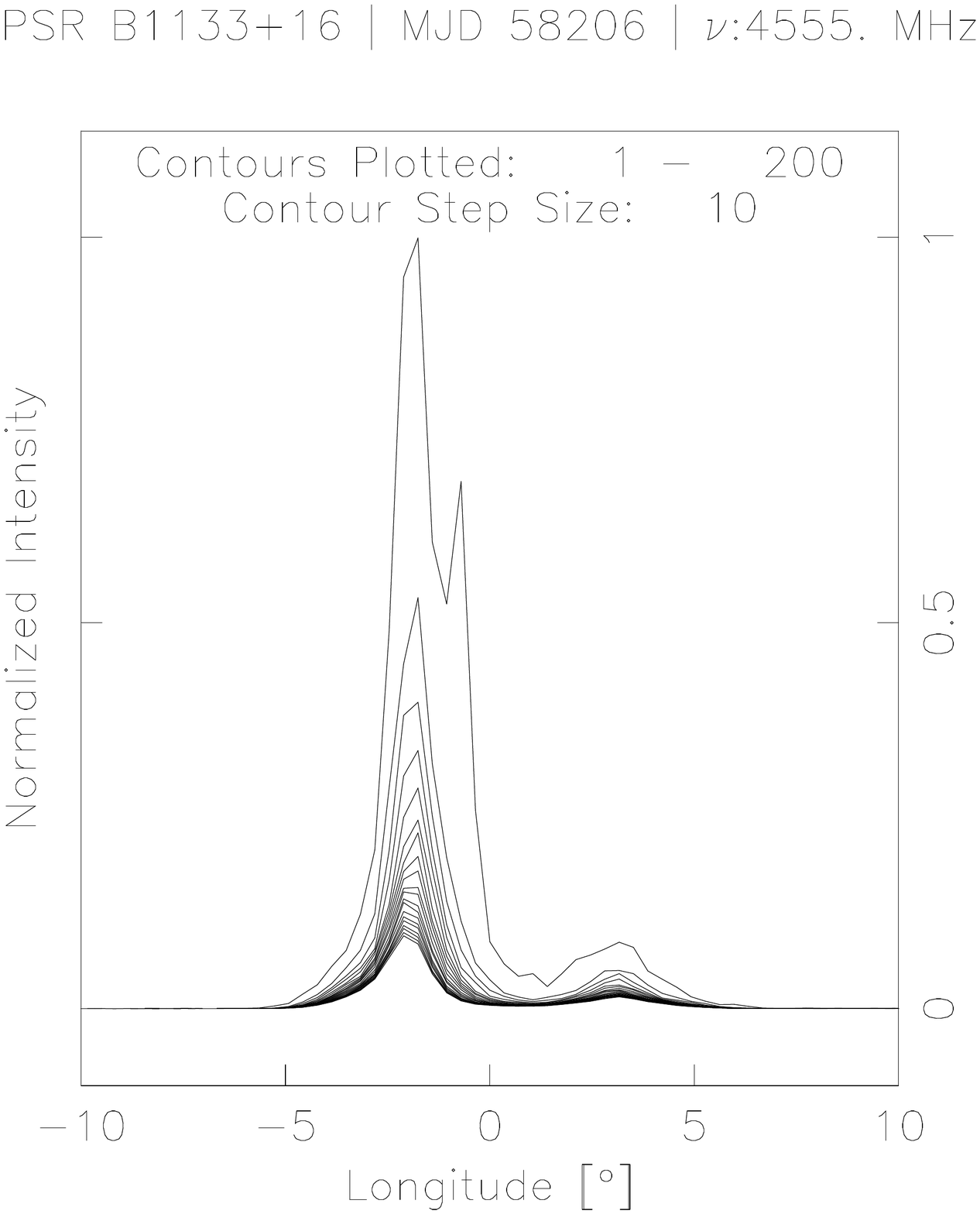} \\ \toprule
\includegraphics[page=1,width=\linewidth]{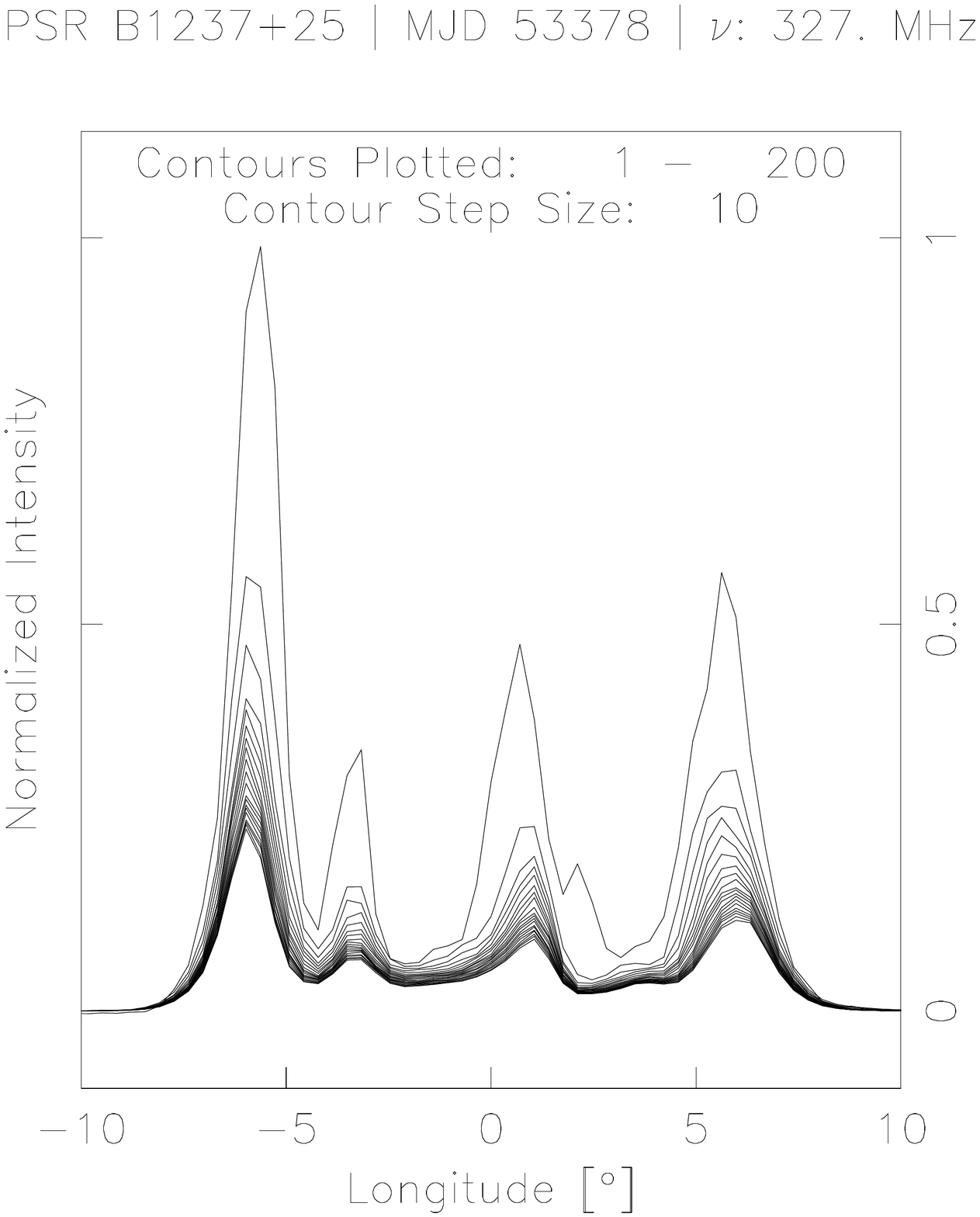} &
\includegraphics[page=1,width=\linewidth]{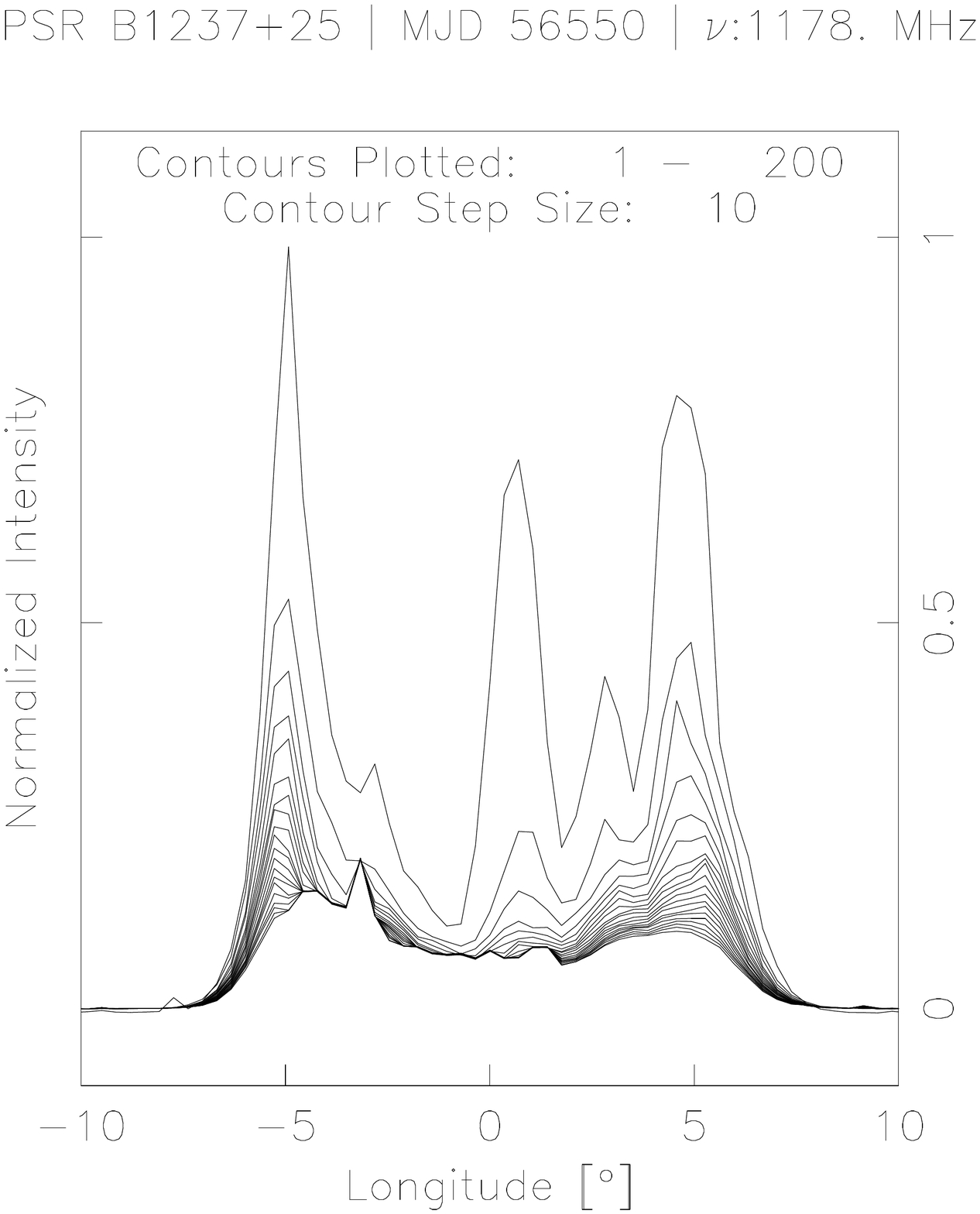} &
\includegraphics[page=1,width=\linewidth]{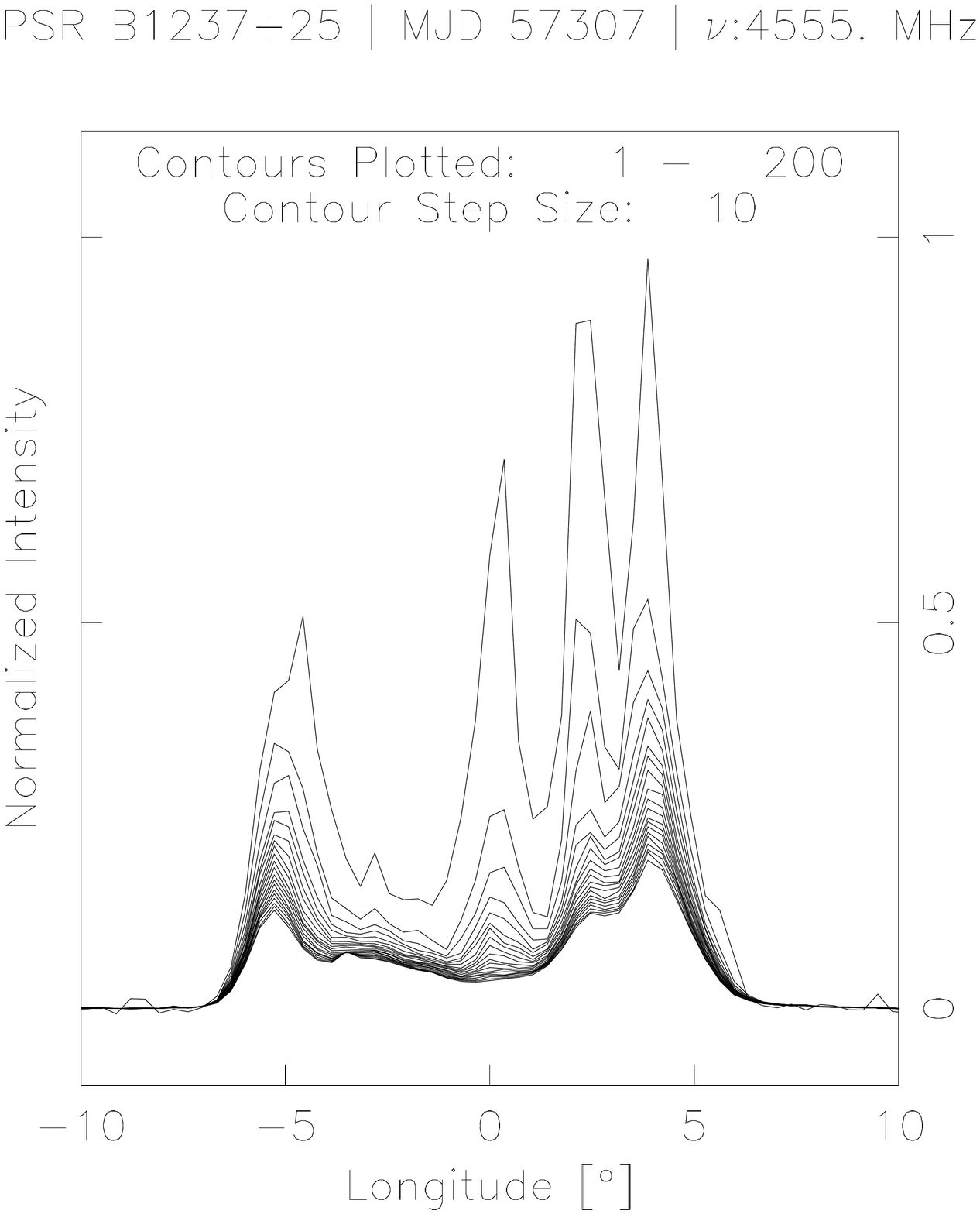} \\ 

     \bottomrule
   \end{tabularx} 
\caption{PHPs of PSR's B0950+08, B1133+16, and B1237+25.}
 \end{figure*}
\vspace{1cm}

   \begin{figure*} 
 \begin{tabularx}{\textwidth}{YYY}
    \multicolumn{3}{c}{} \\ \toprule
\includegraphics[page=1,width=\linewidth]{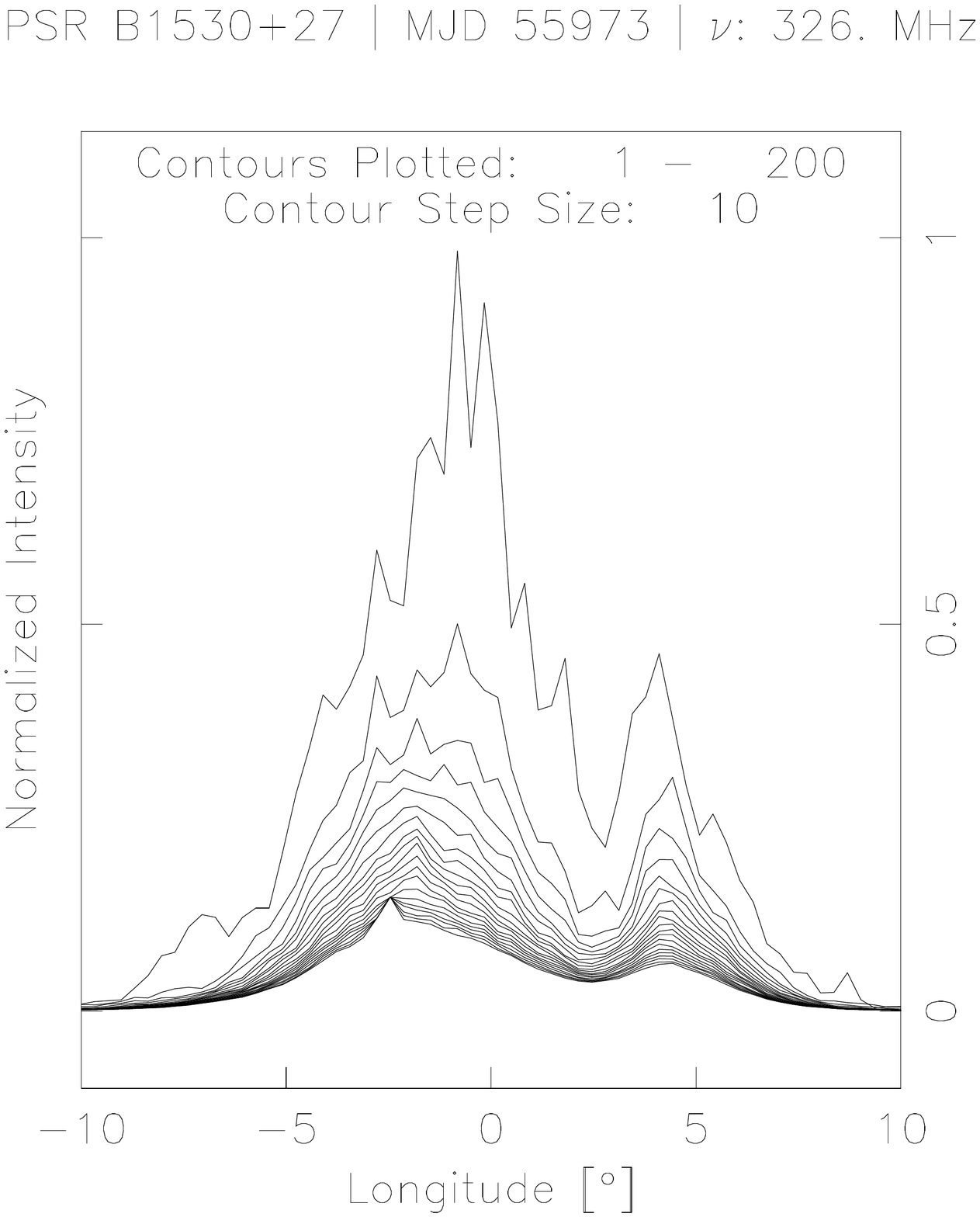} &
\includegraphics[page=1,width=\linewidth]{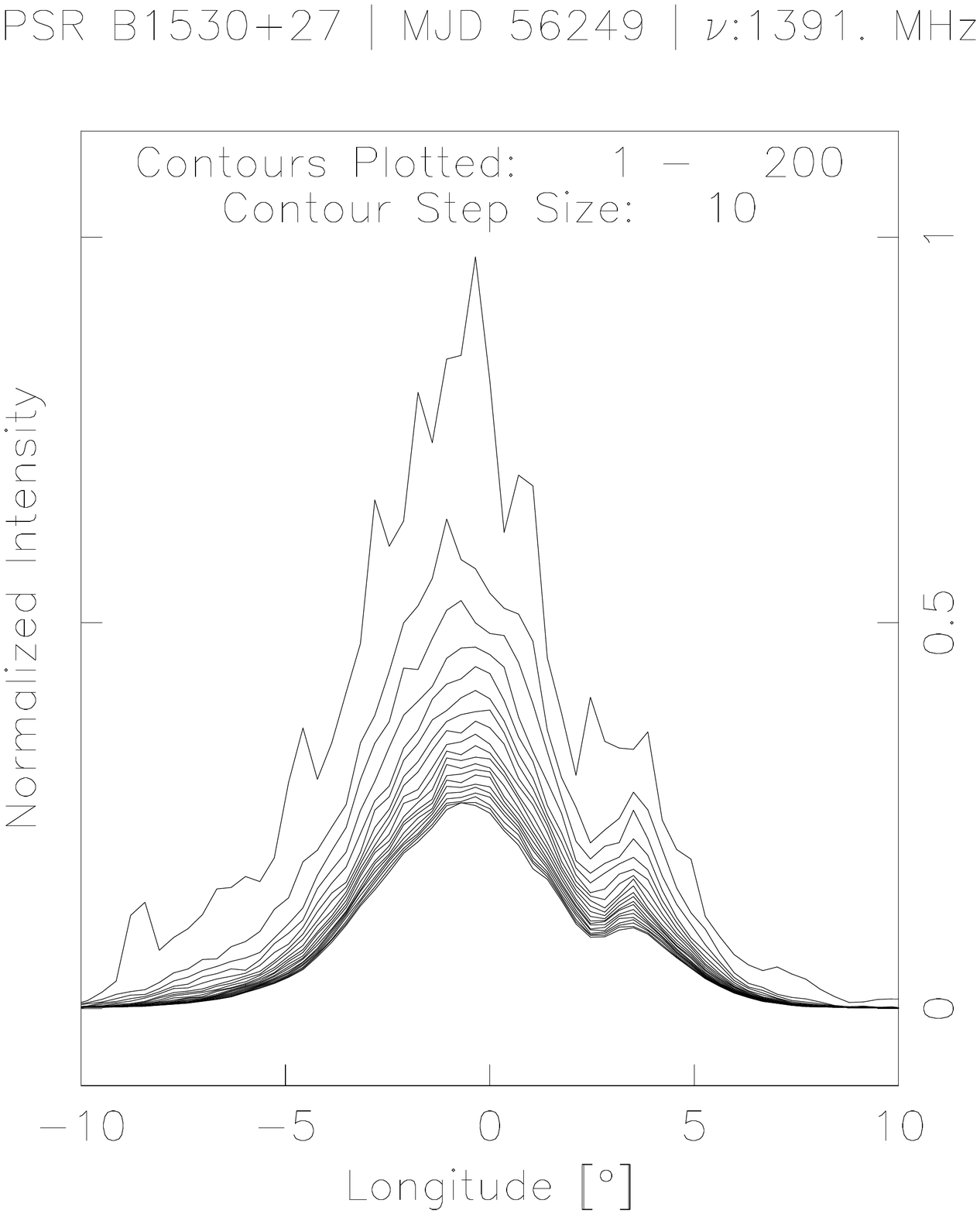} &
\includegraphics[page=1,width=\linewidth]{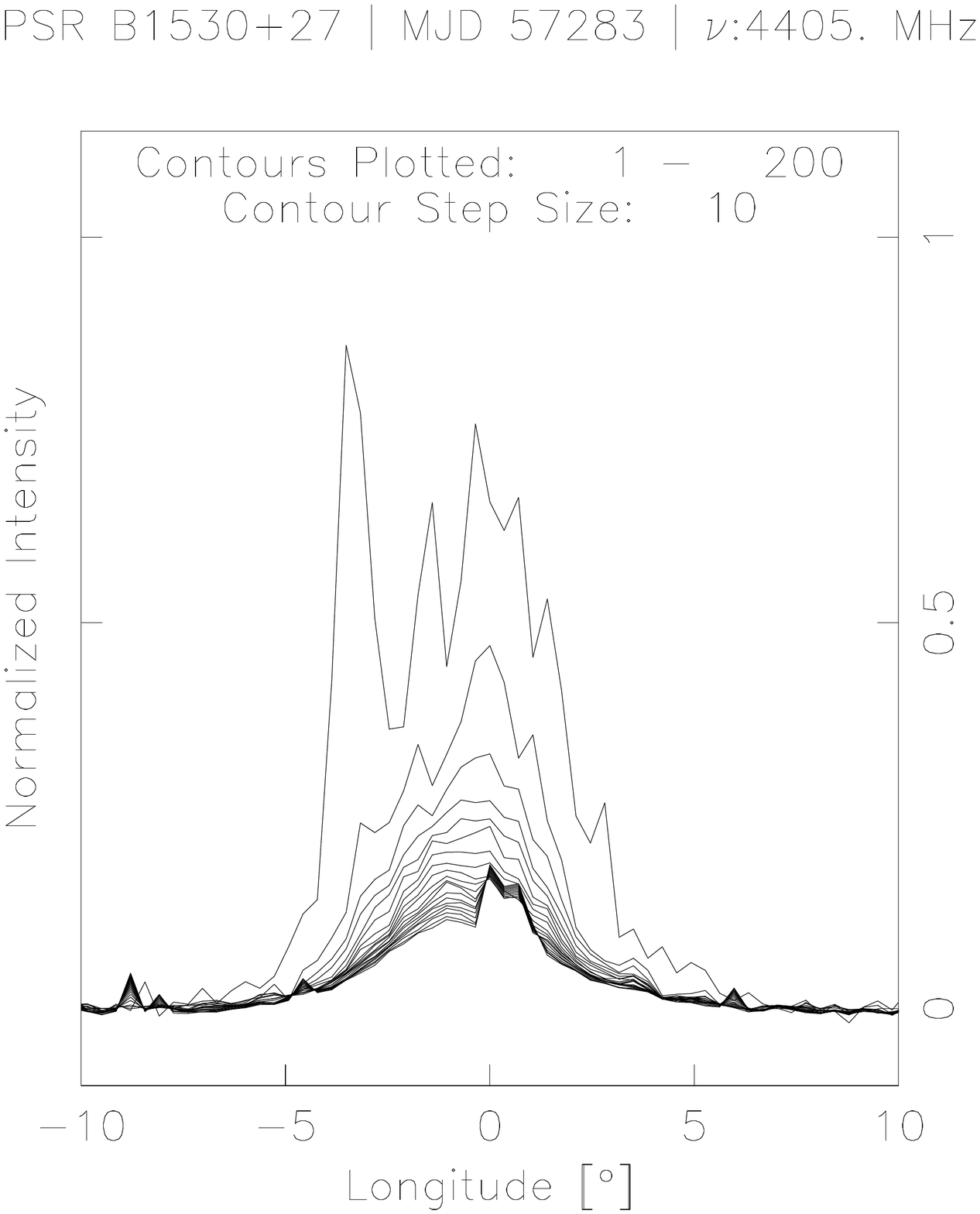} \\ \toprule
\includegraphics[page=1,width=\linewidth]{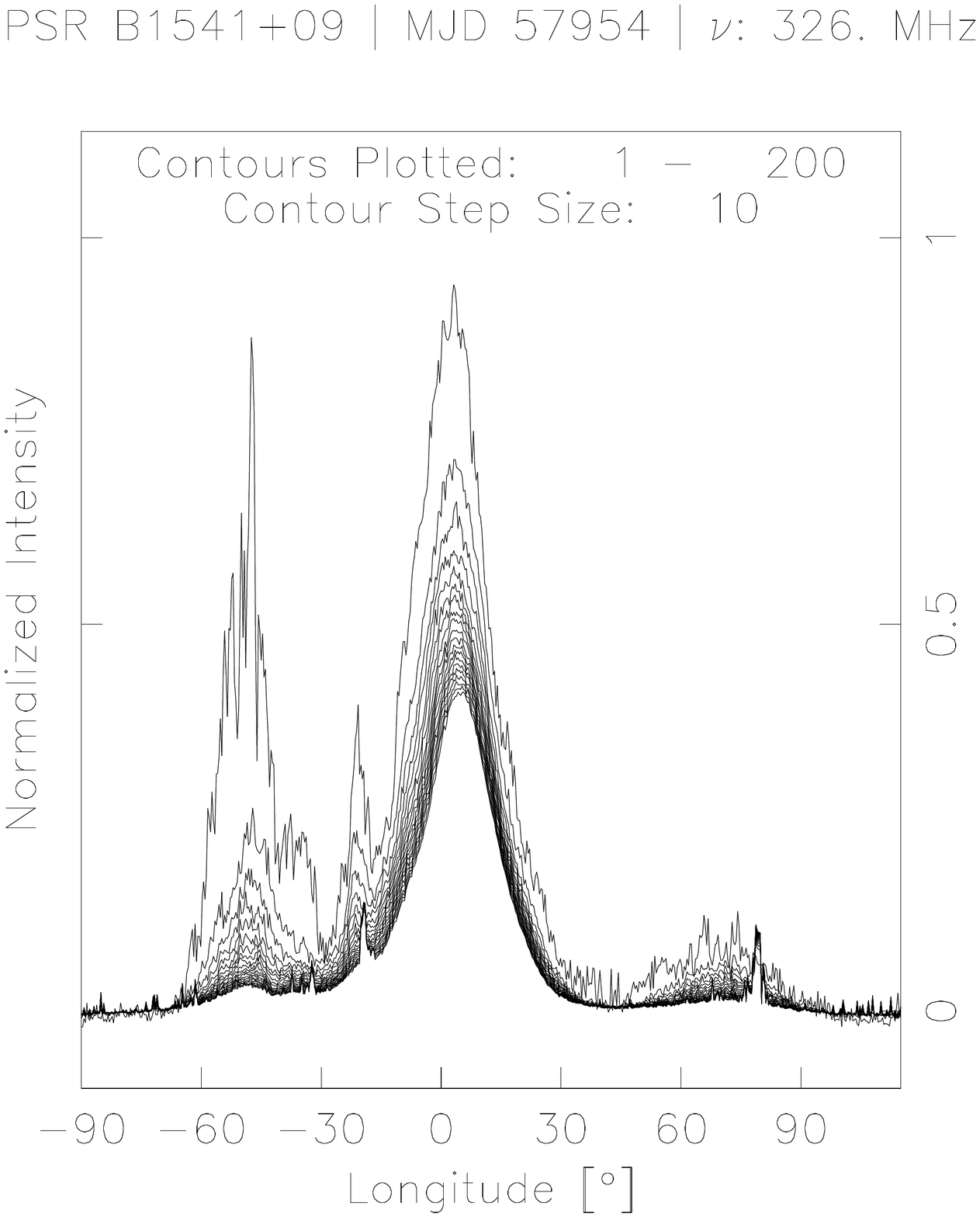} &
\includegraphics[page=1,width=\linewidth]{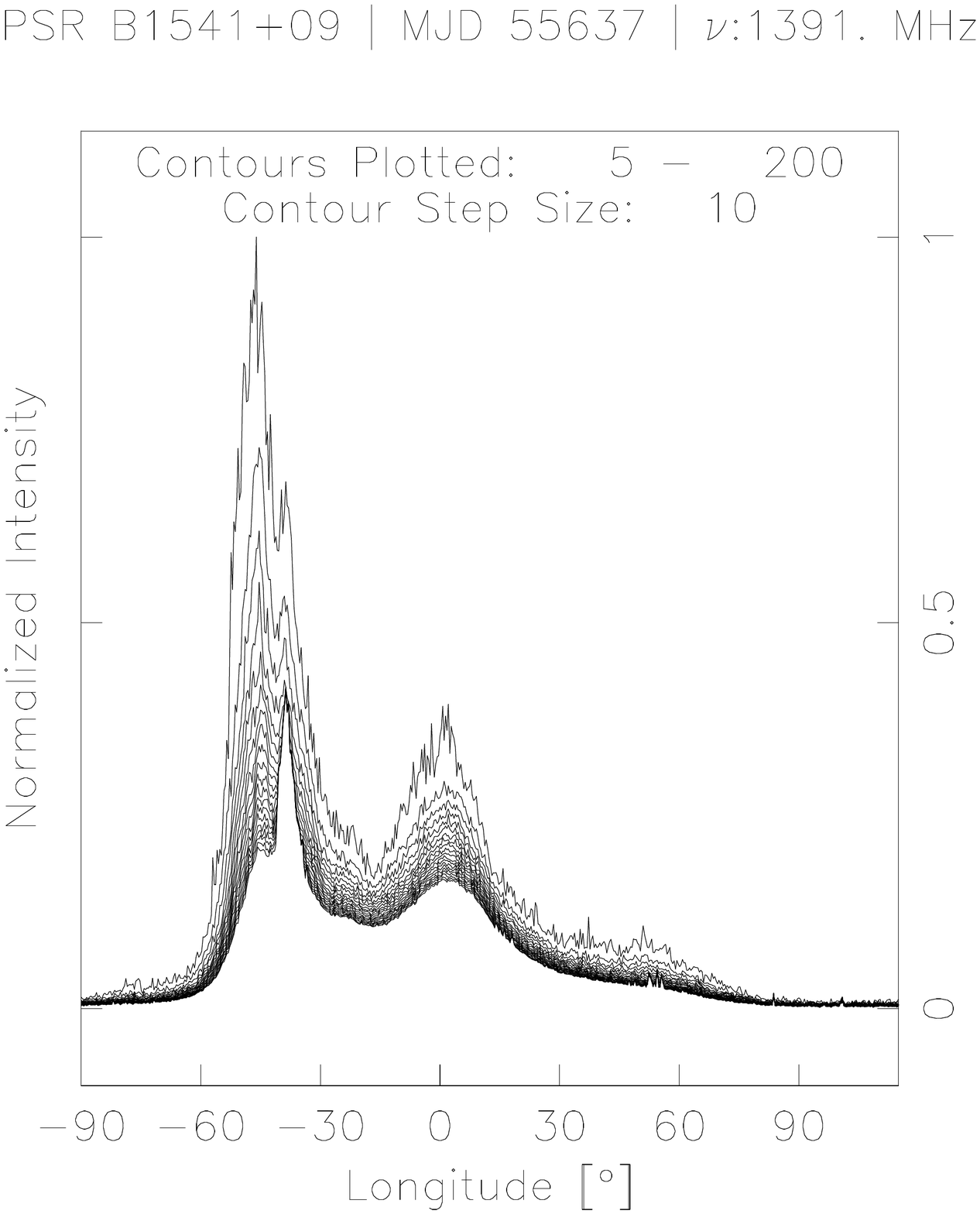} &
\includegraphics[page=1,width=\linewidth]{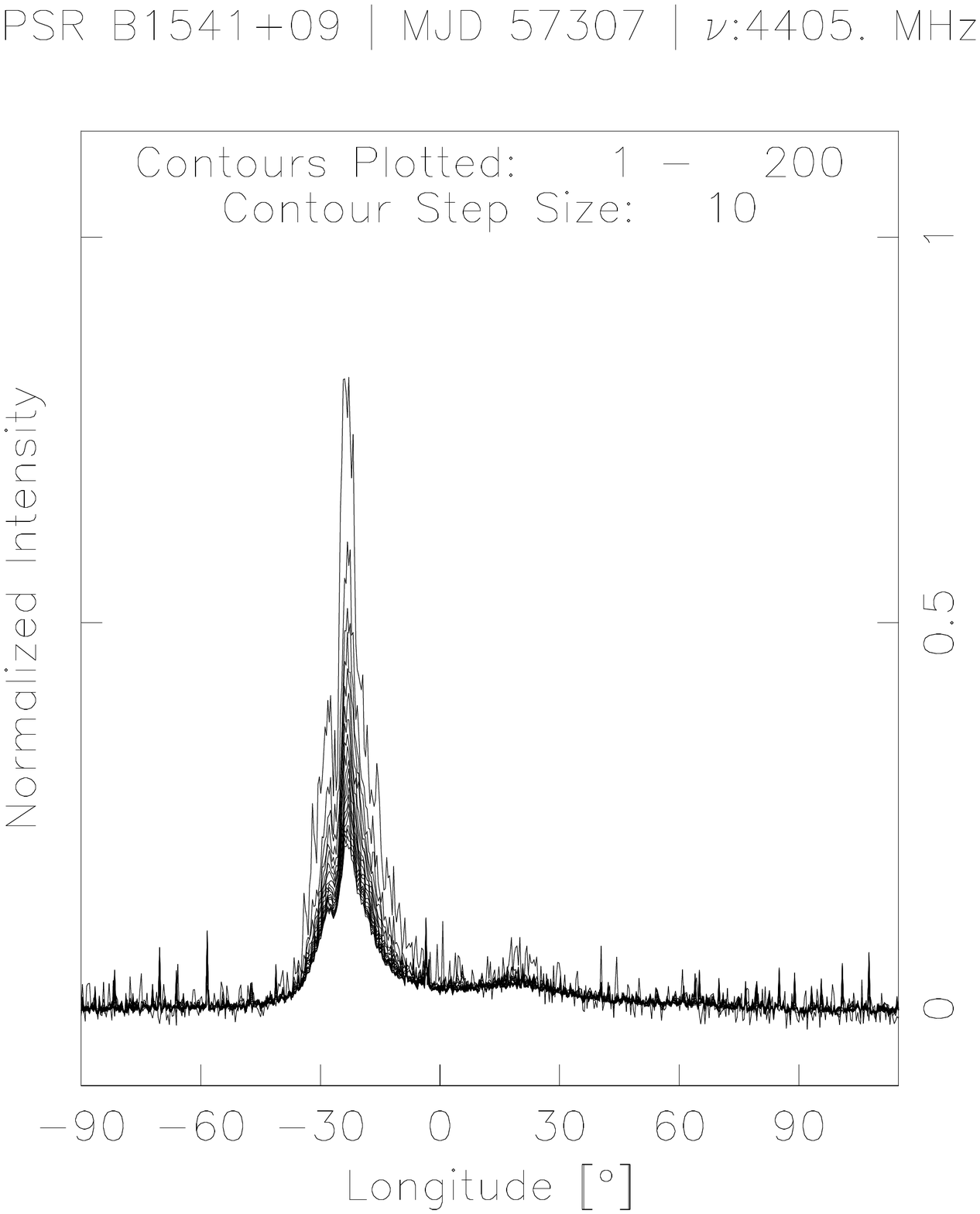} \\ \toprule
\includegraphics[page=1,width=\linewidth]{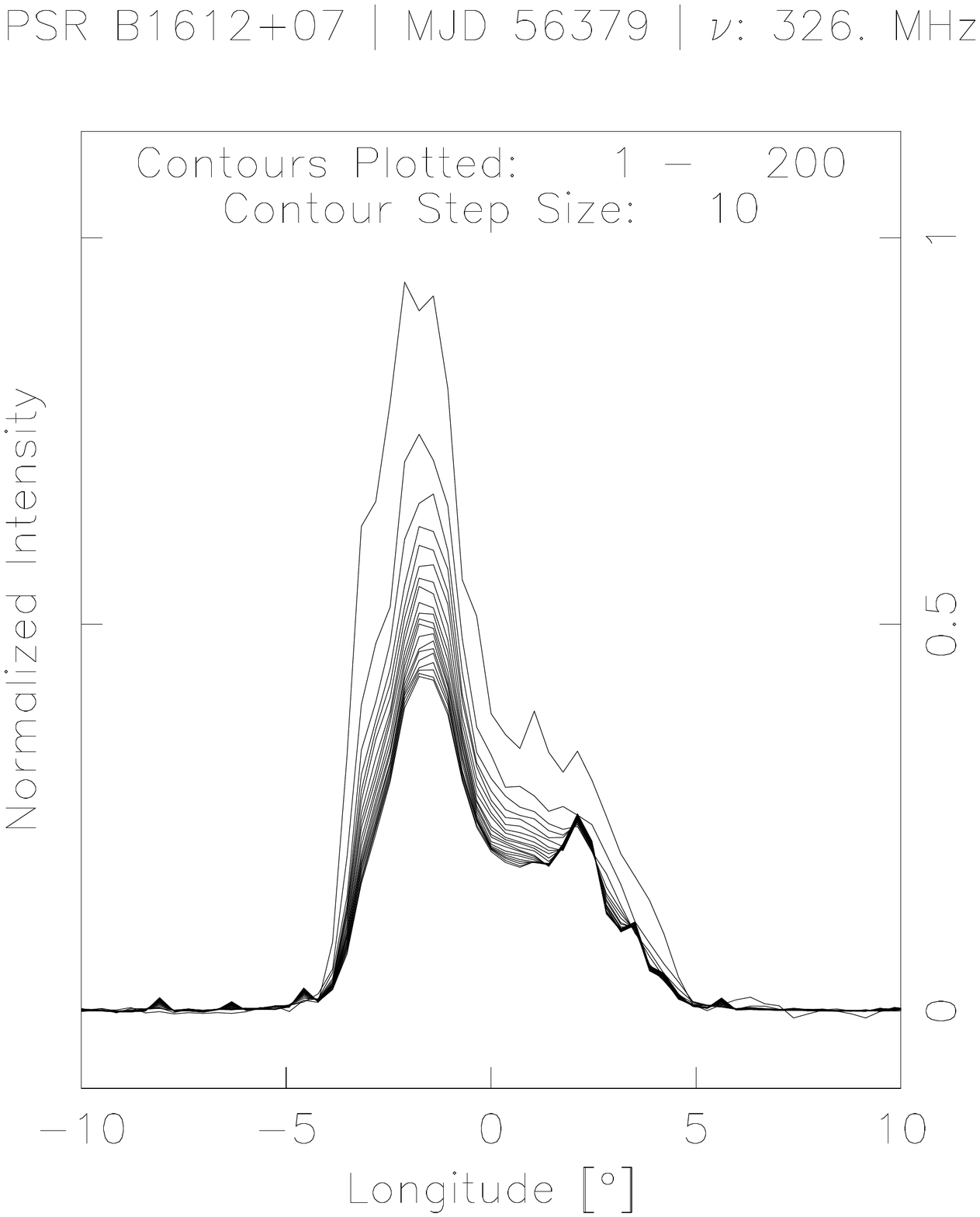} &
\includegraphics[page=1,width=\linewidth]{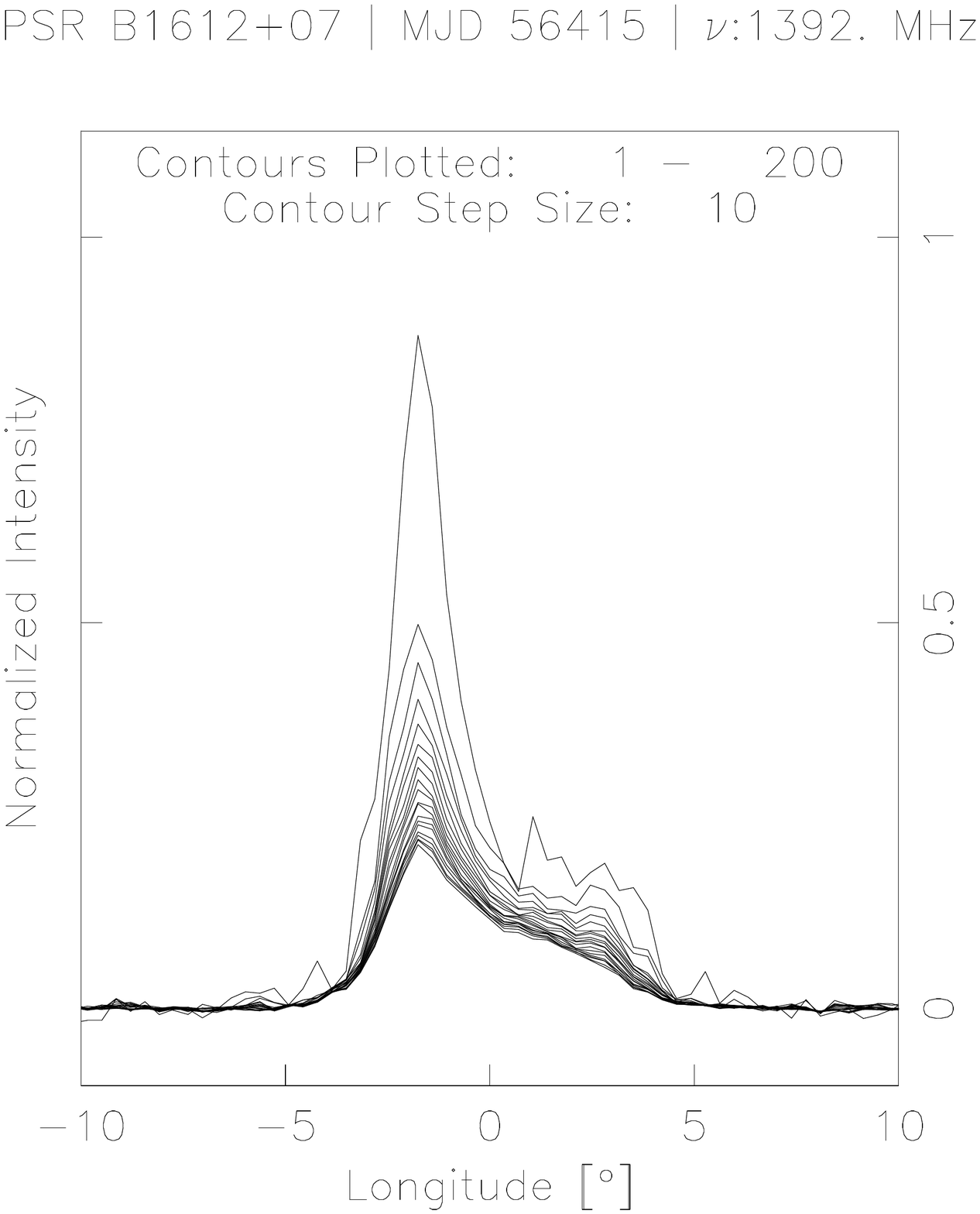} &
\includegraphics[page=1,width=\linewidth]{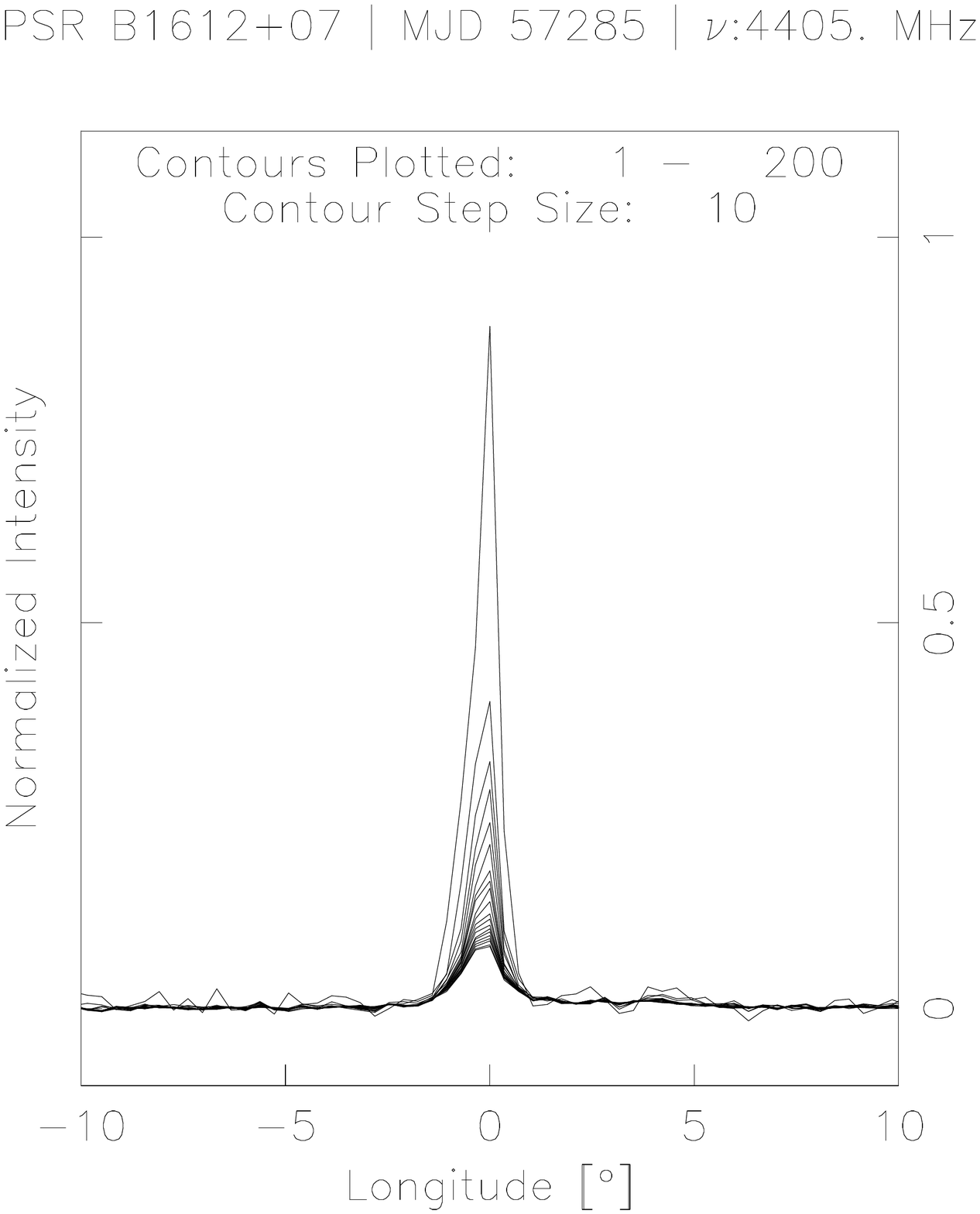} \\ 
     \bottomrule
   \end{tabularx} 
\caption{PHPs of PSR's B1530+27, B1541+09, and B1612+07.}
 \end{figure*}
\vspace{1cm}

   \begin{figure*} 
 \begin{tabularx}{\textwidth}{YYY}
    \multicolumn{3}{c}{} \\ \toprule
\includegraphics[page=1,width=\linewidth]{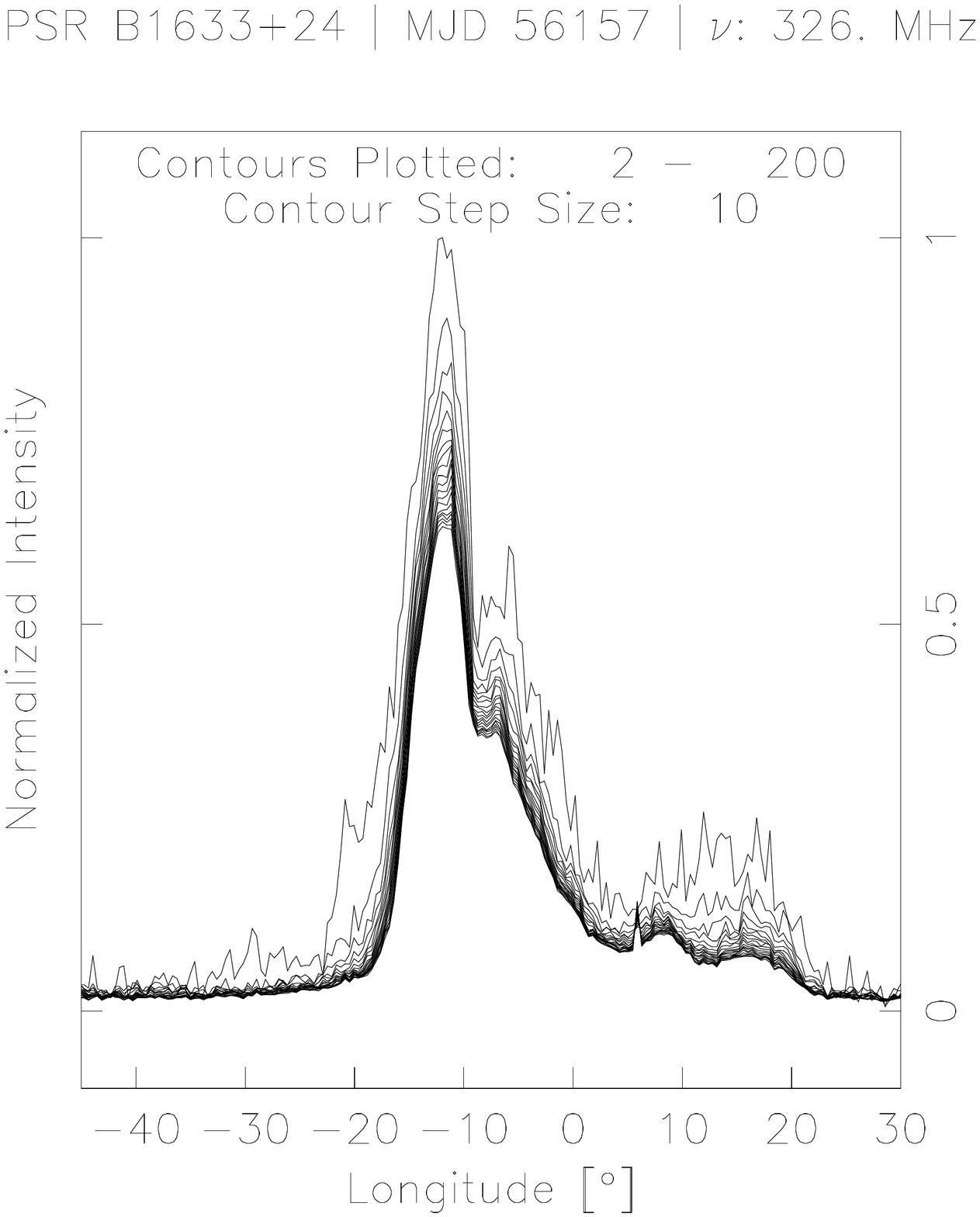} &
\includegraphics[page=1,width=\linewidth]{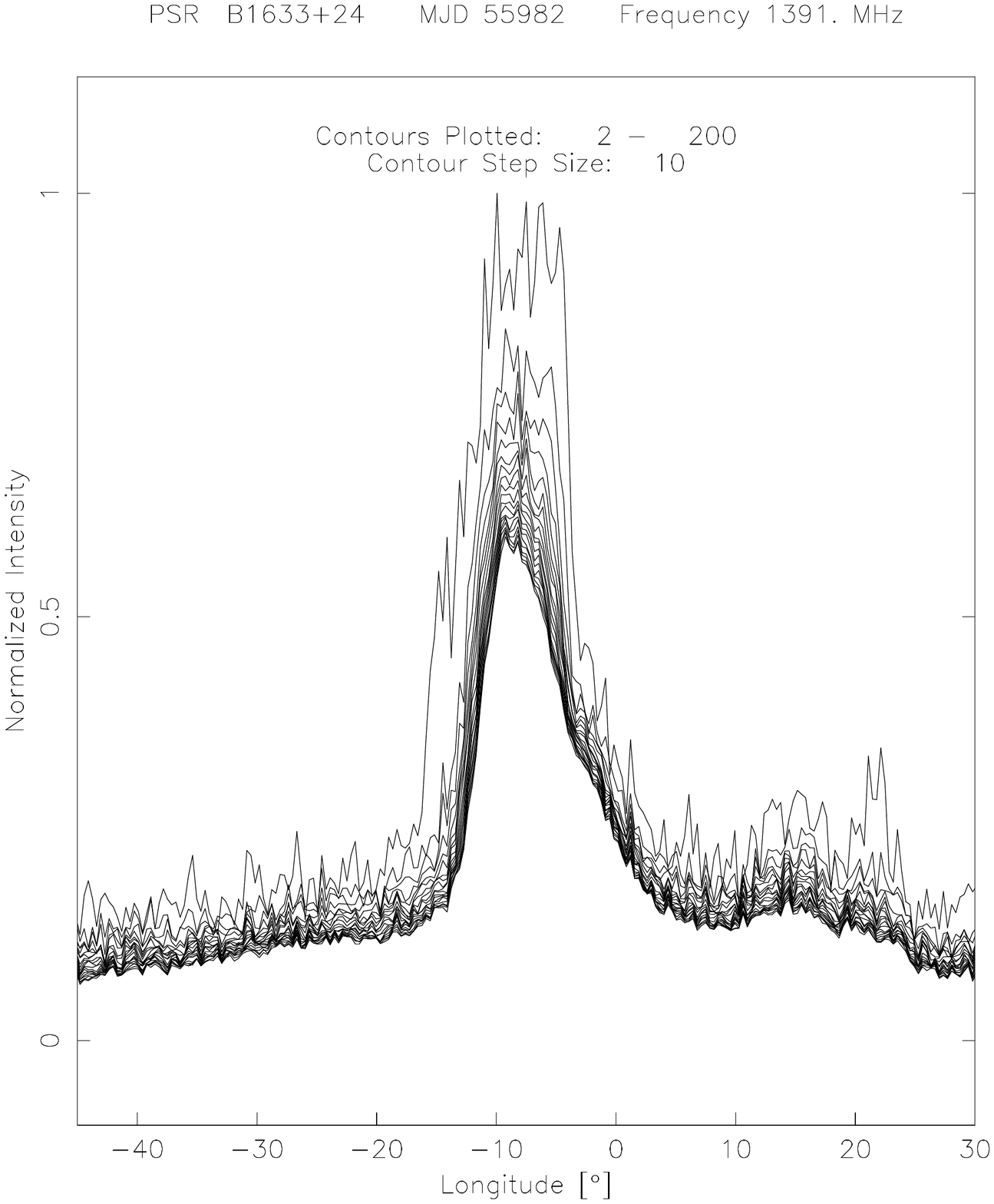} &
\\ \toprule

\includegraphics[page=1,width=\linewidth]{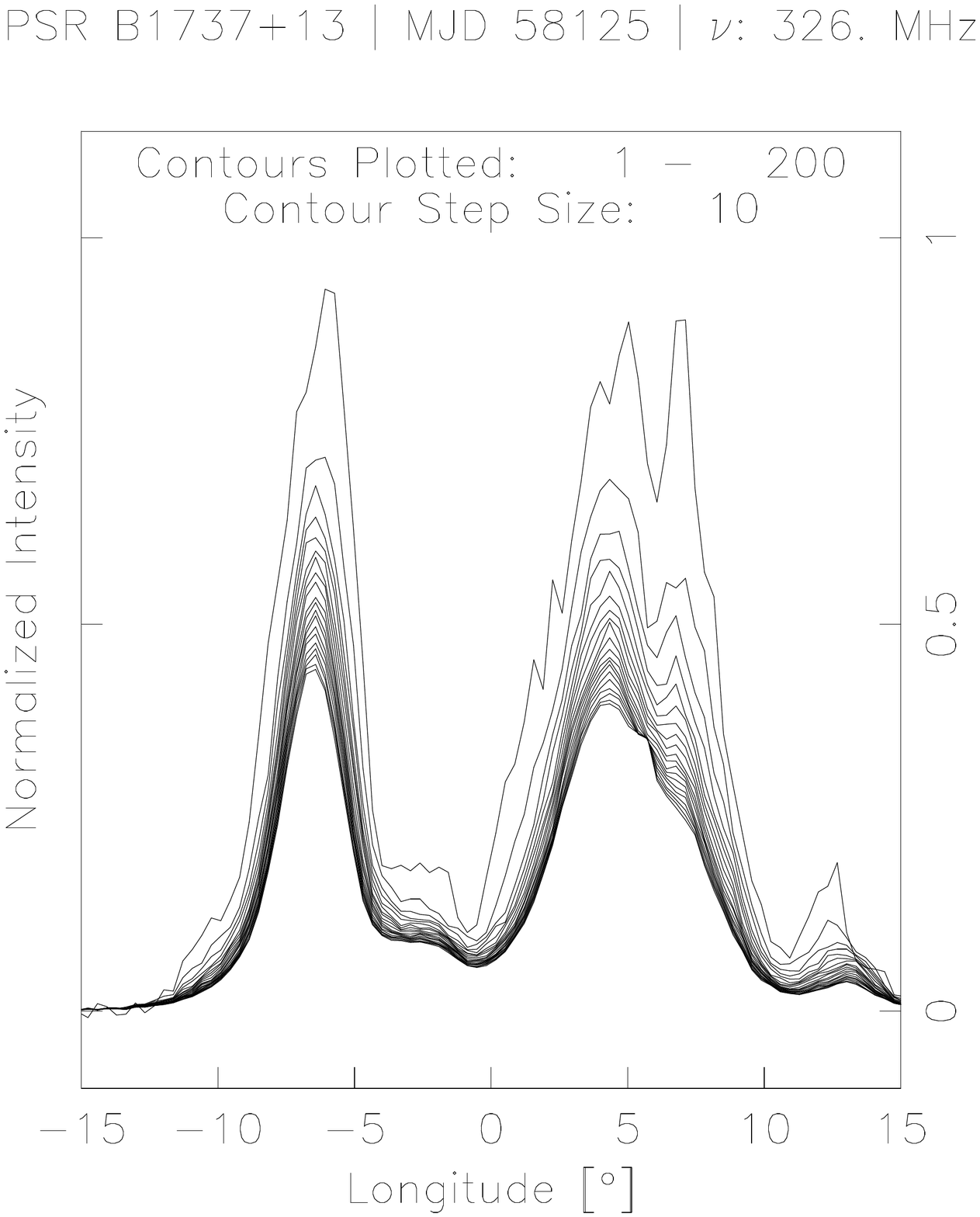} &
\includegraphics[page=1,width=\linewidth]{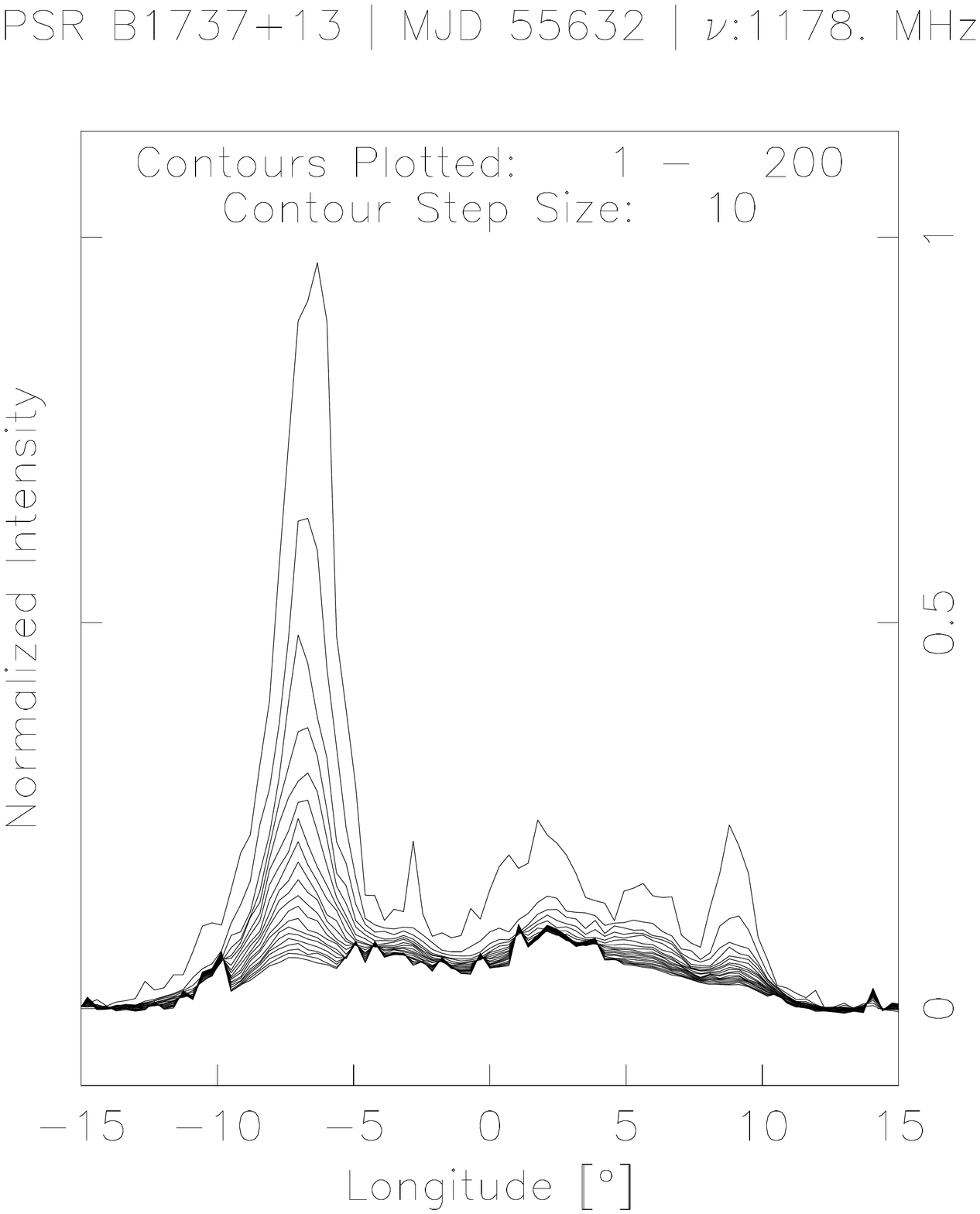} &
\includegraphics[page=1,width=\linewidth]{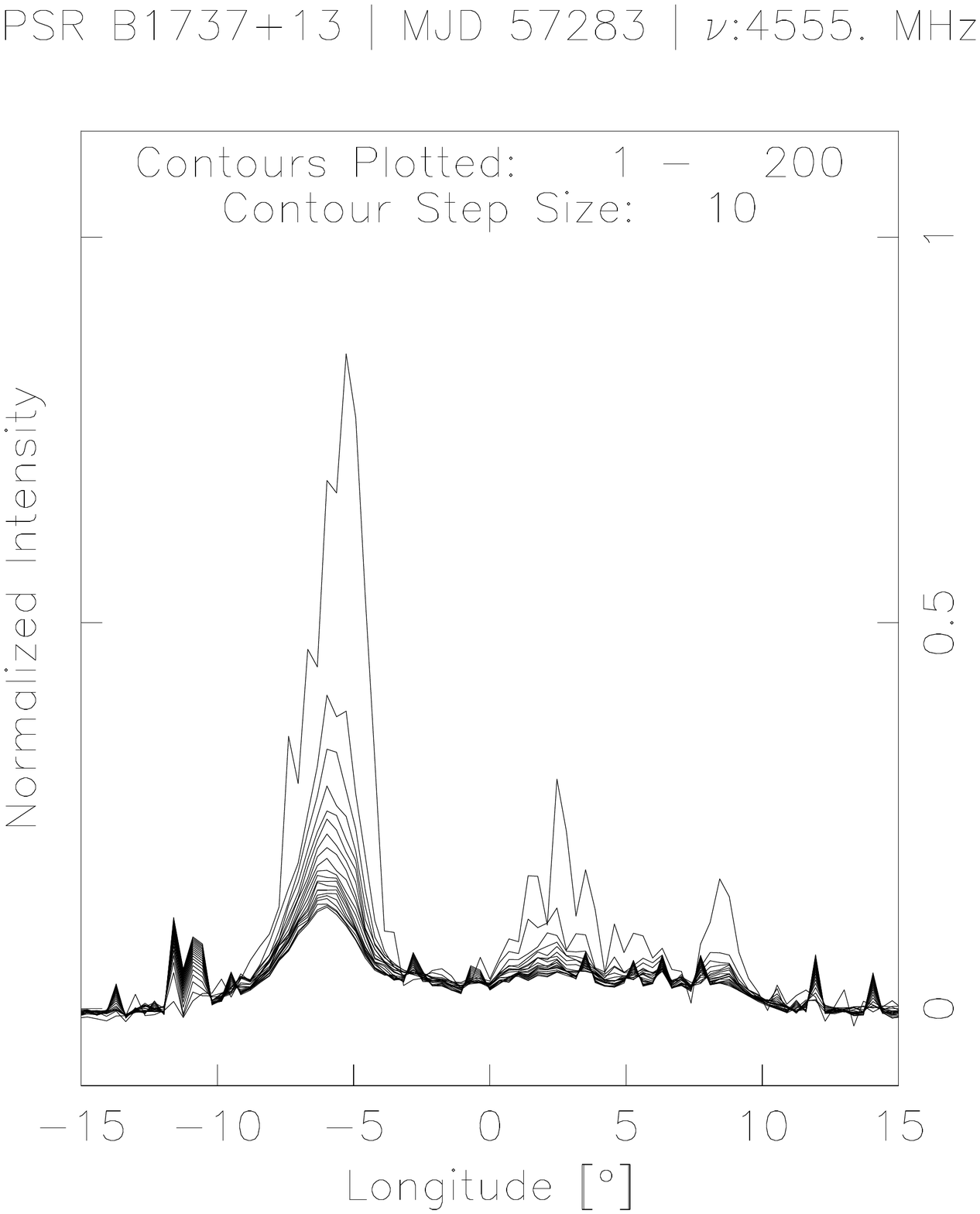} \\ \toprule

\includegraphics[page=1,width=\linewidth]{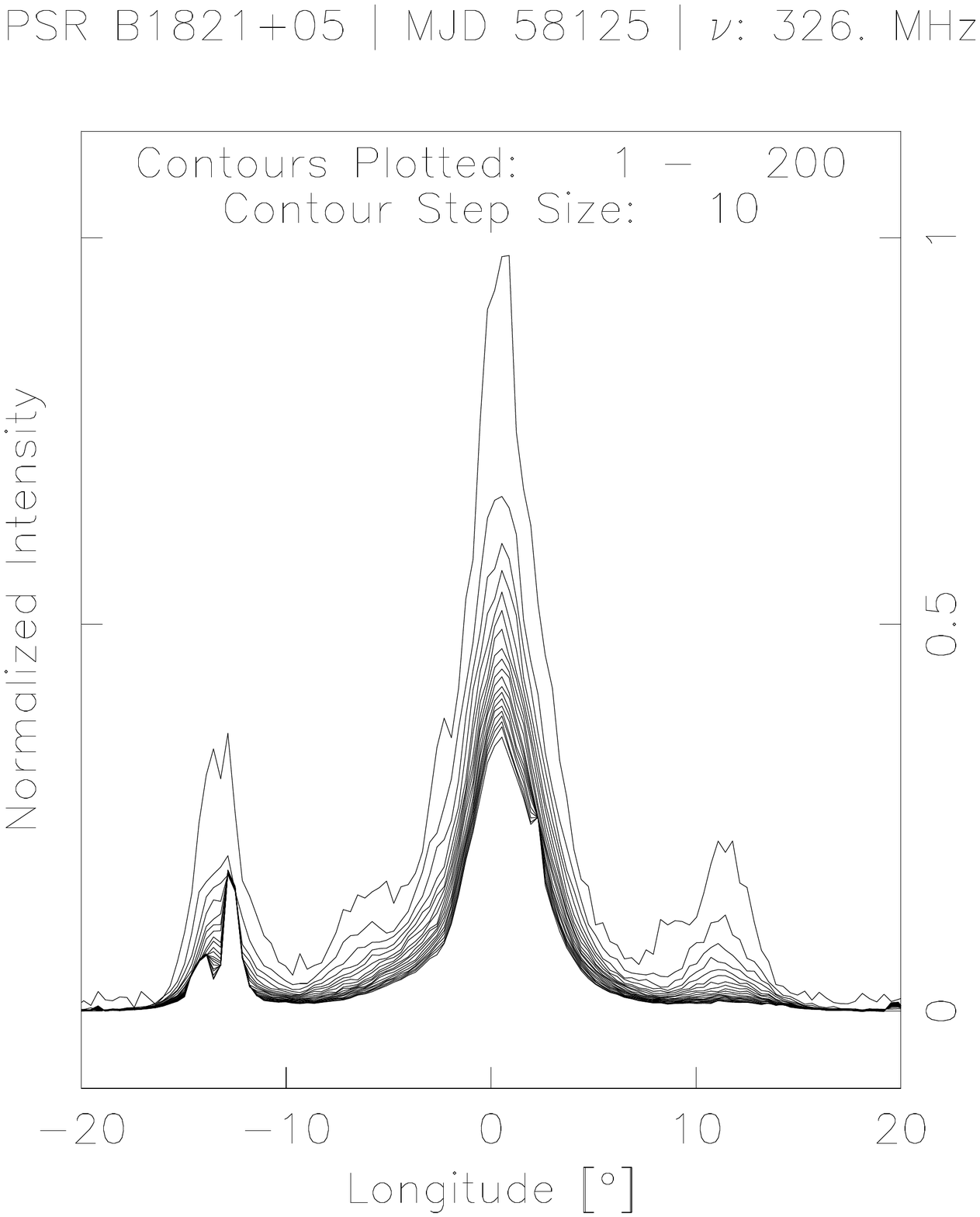} &
\includegraphics[page=1,width=\linewidth]{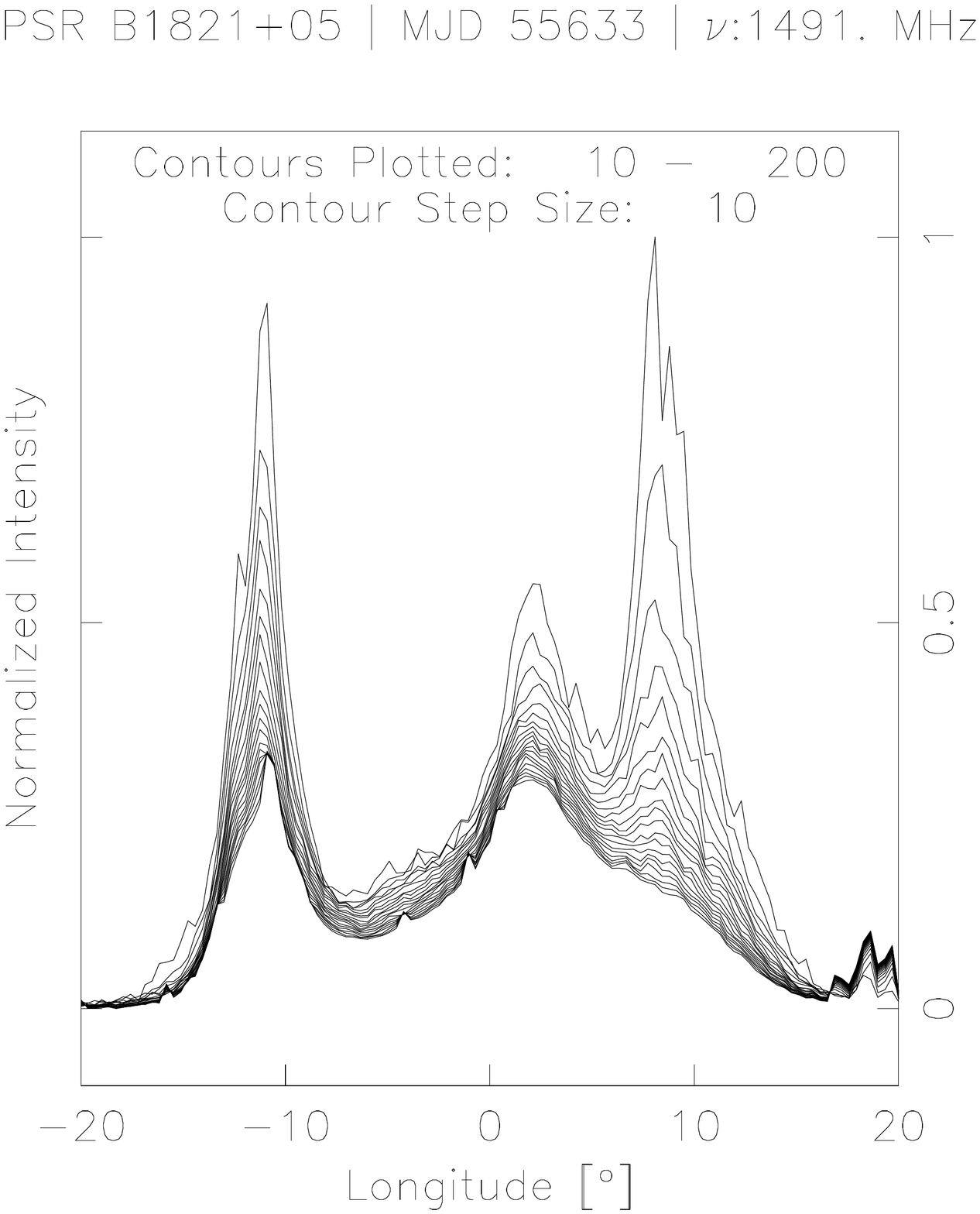} &
\\
     \bottomrule
   \end{tabularx} 
\caption{PHPs of PSR's B1633+24, B1737+13, and B1821+05.}
 \end{figure*}
\vspace{1cm}

   \begin{figure*} 
 \begin{tabularx}{\textwidth}{YYY}
    \multicolumn{3}{c}{} \\ \toprule
\includegraphics[page=1,width=\linewidth]{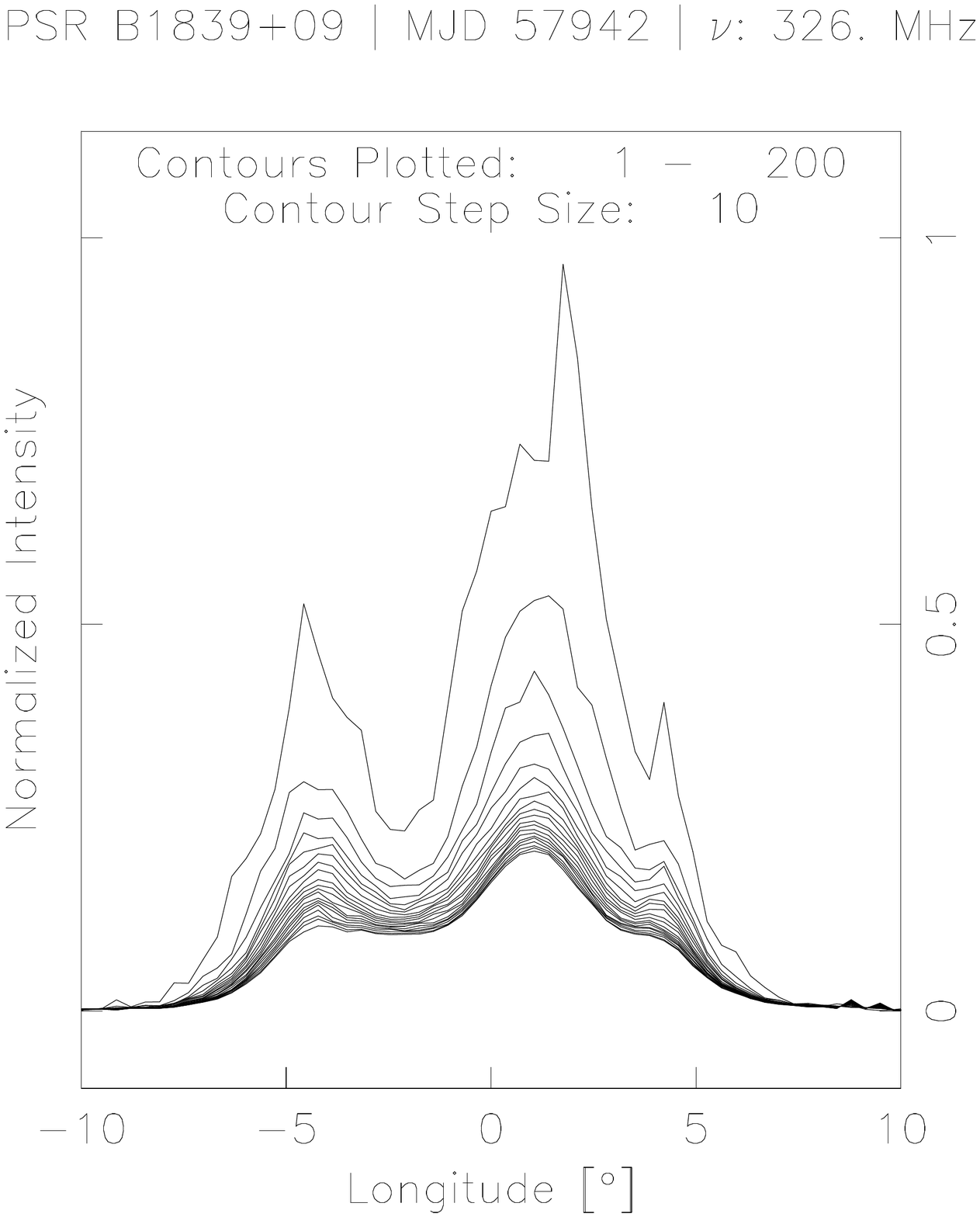} &
\includegraphics[page=1,width=\linewidth]{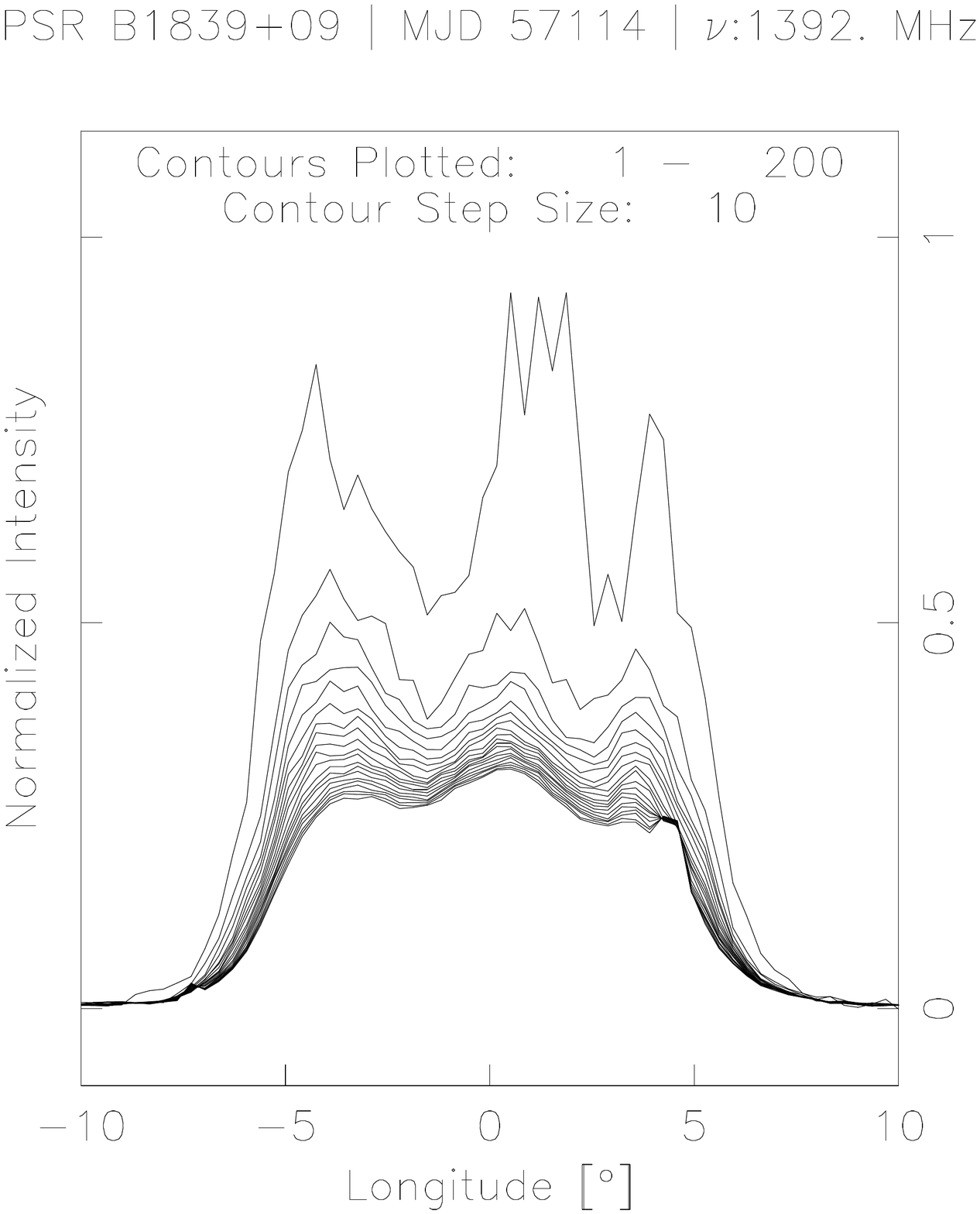} &
\includegraphics[page=1,width=\linewidth]{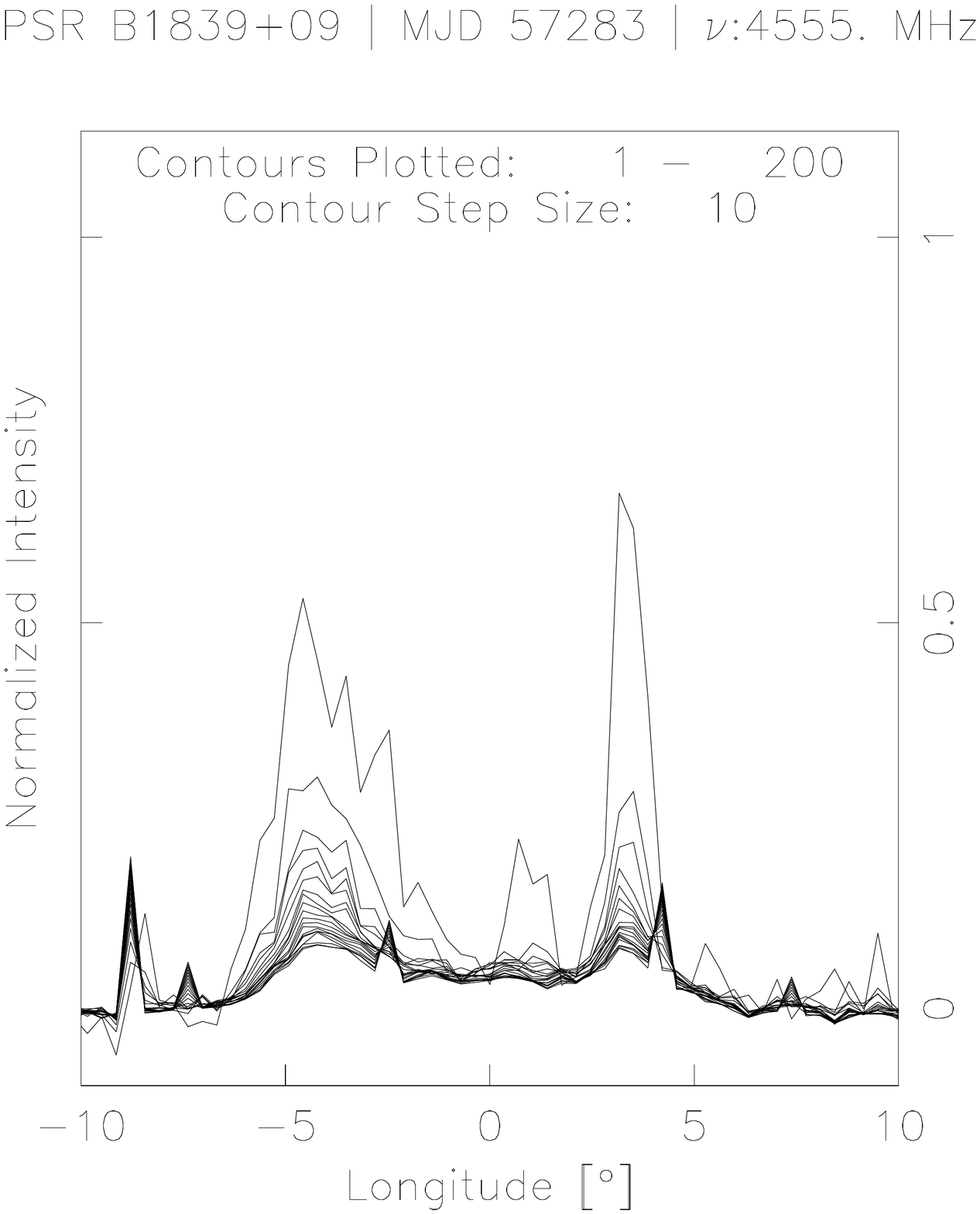} \\ \toprule
\includegraphics[page=1,width=\linewidth]{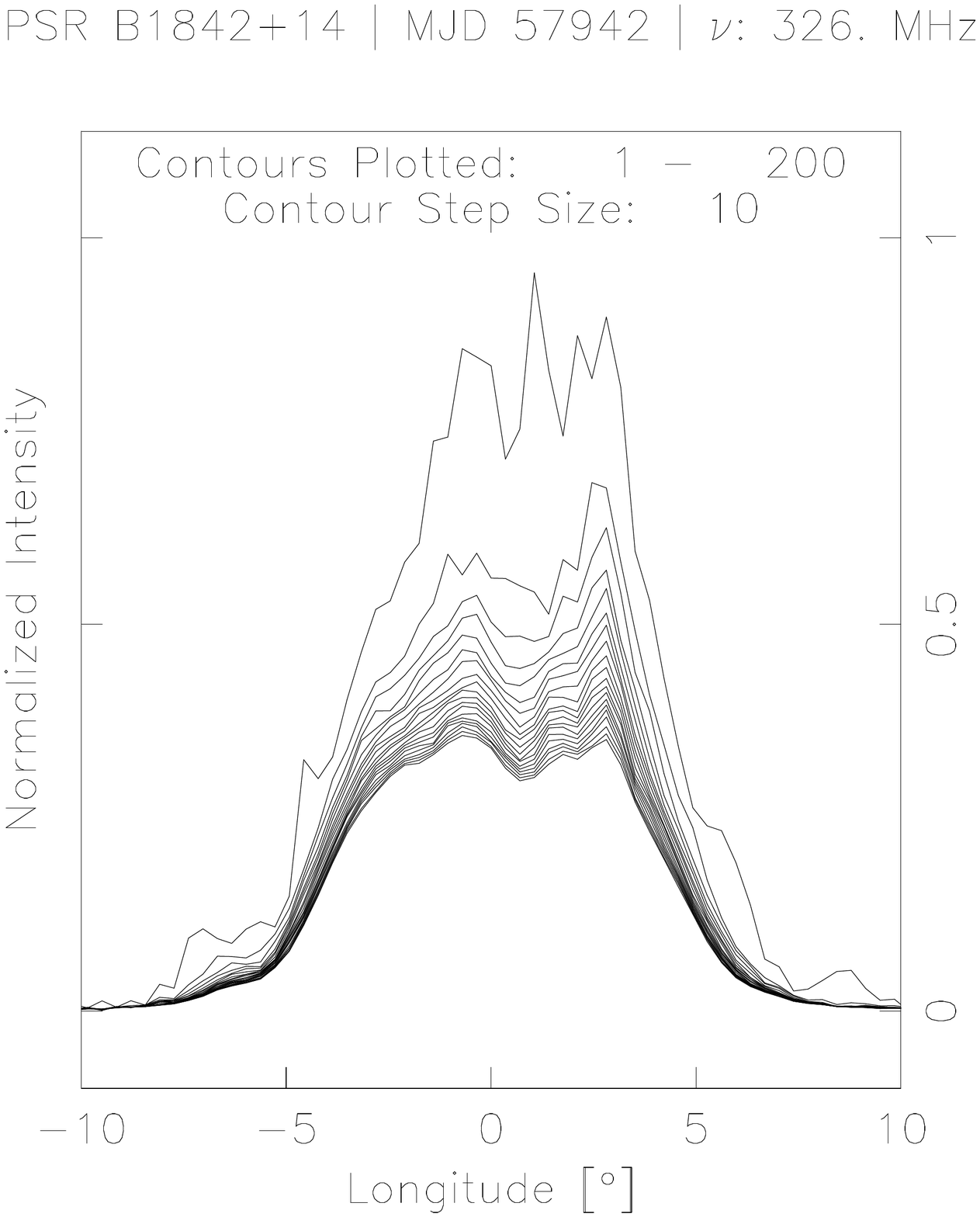} &
\includegraphics[page=1,width=\linewidth]{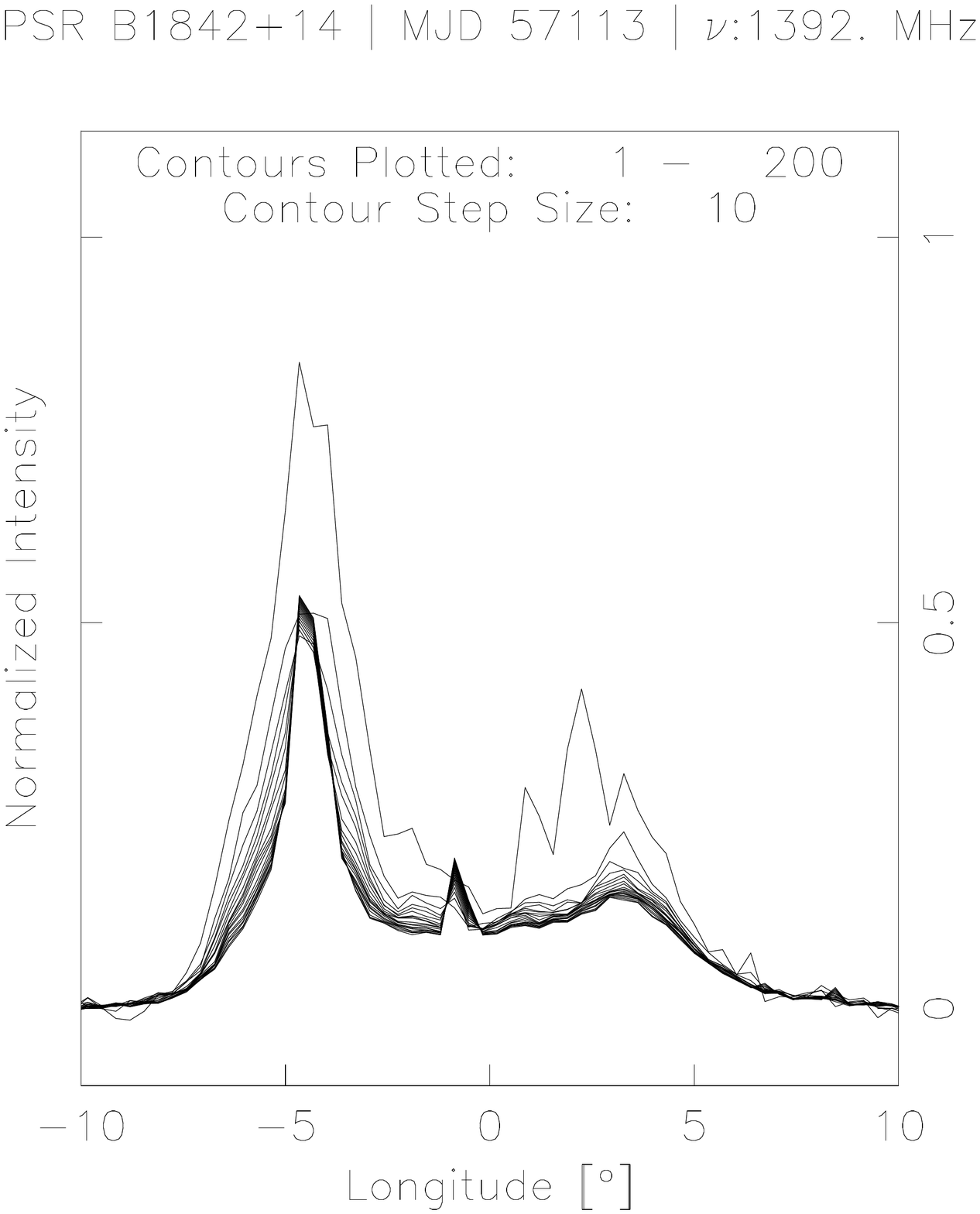} &
\includegraphics[page=1,width=\linewidth]{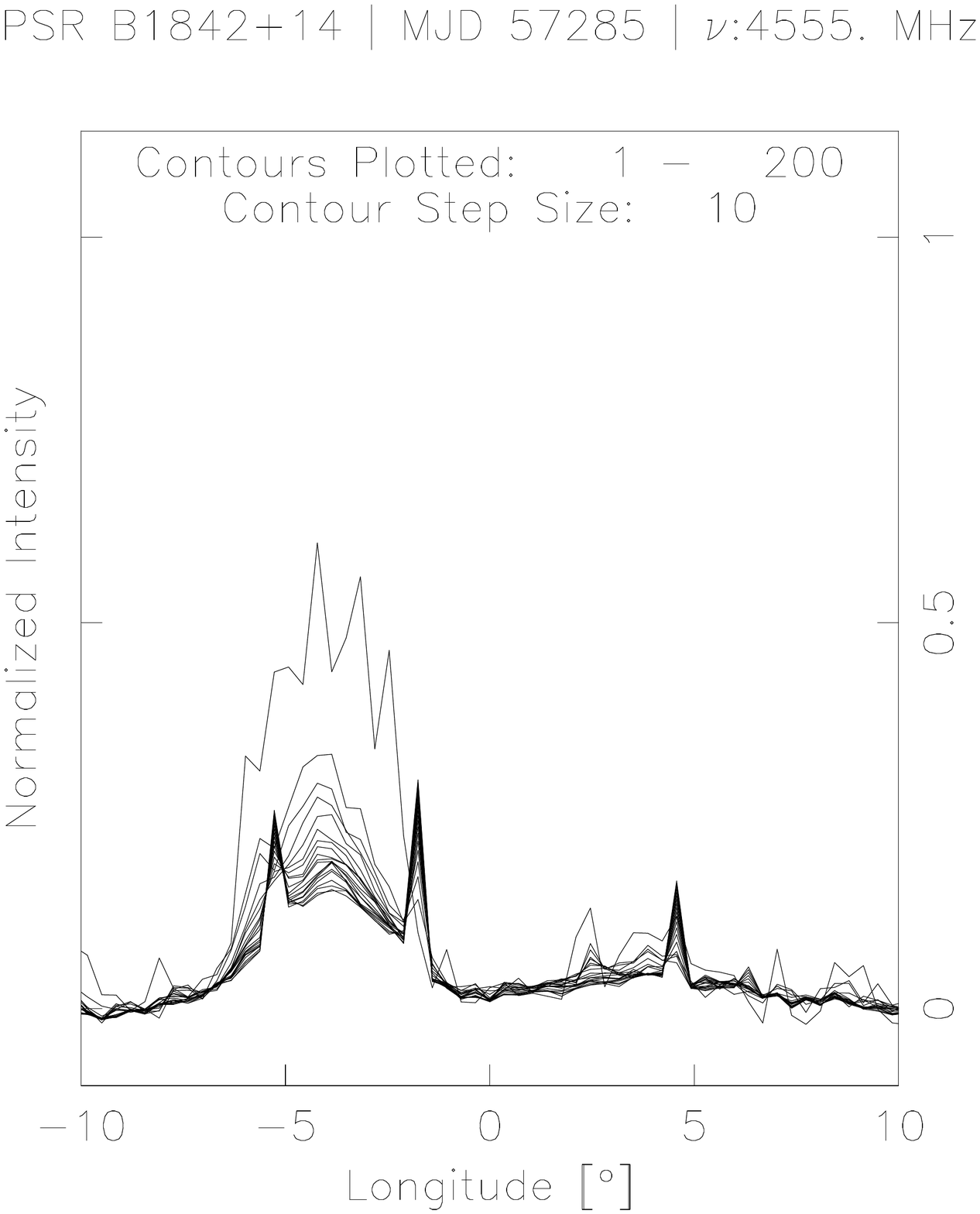} \\ \toprule
\includegraphics[page=1,width=\linewidth]{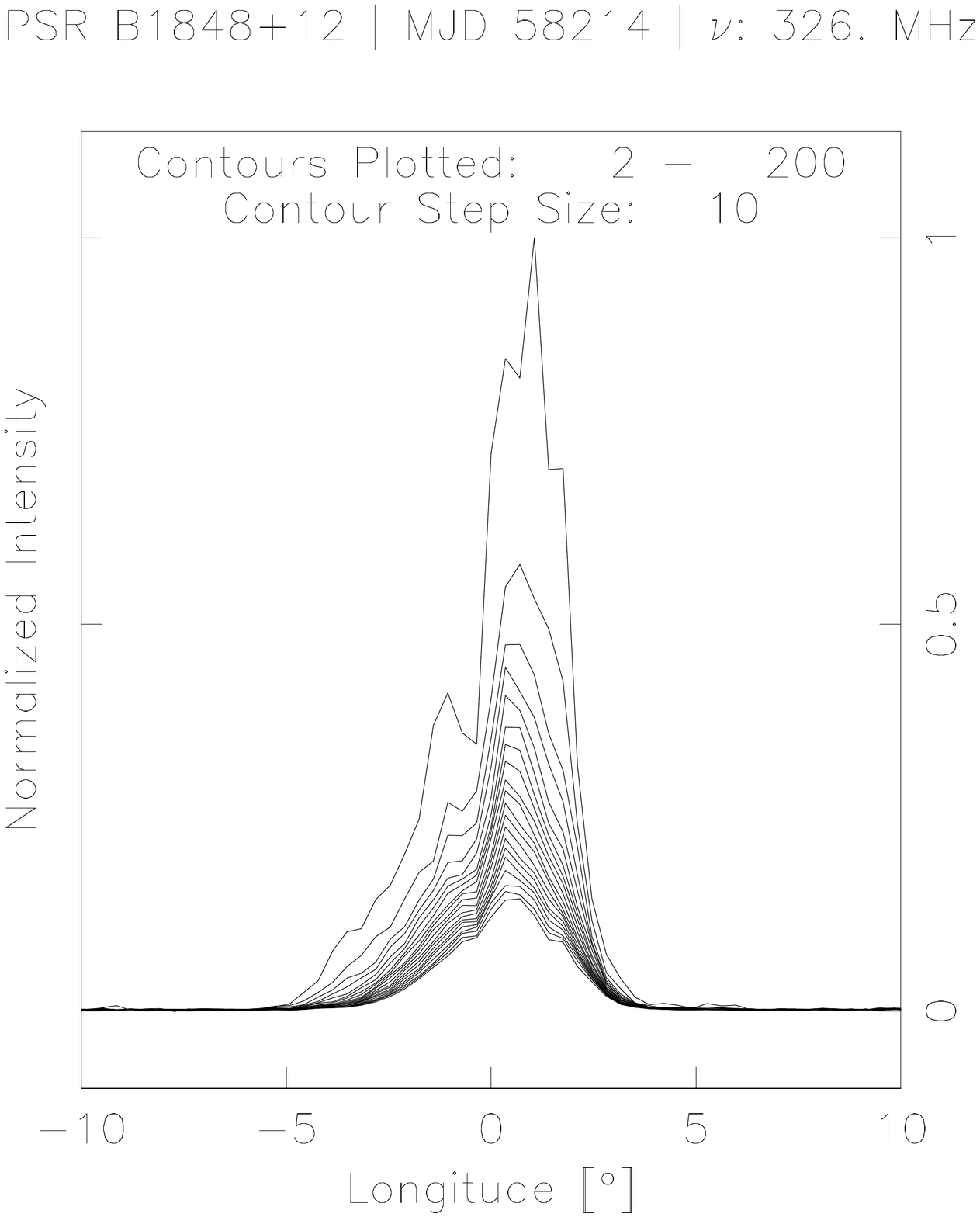} &
\includegraphics[page=1,width=\linewidth]{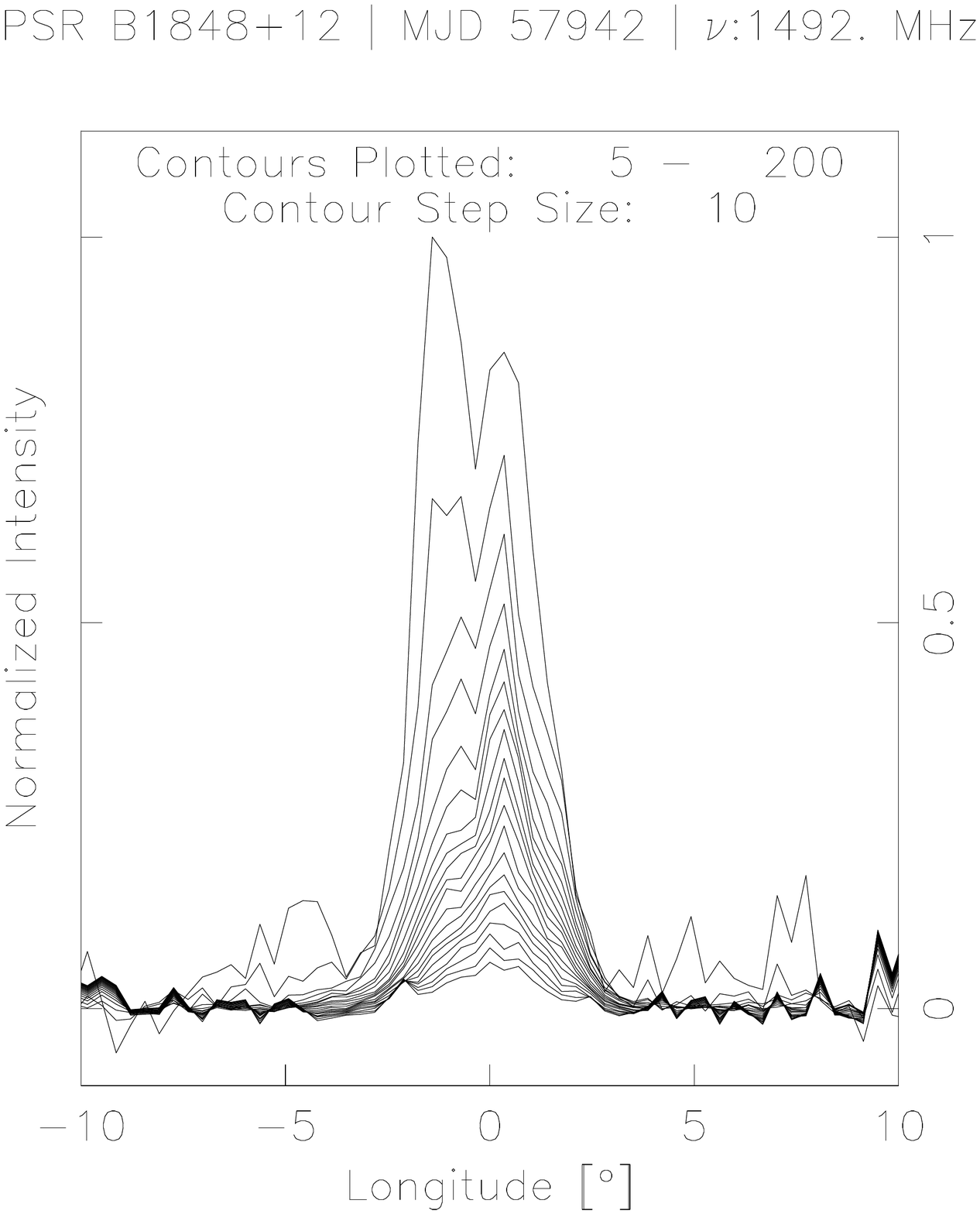} &
\\
     \bottomrule
   \end{tabularx} 
\caption{PHPs of PSR's B1839+09, B1842+14, and B1848+12.}
 \end{figure*}
\vspace{1cm}

   \begin{figure*} 
 \begin{tabularx}{\textwidth}{YYY}
    \multicolumn{3}{c}{} \\ \toprule
\includegraphics[page=1,width=\linewidth]{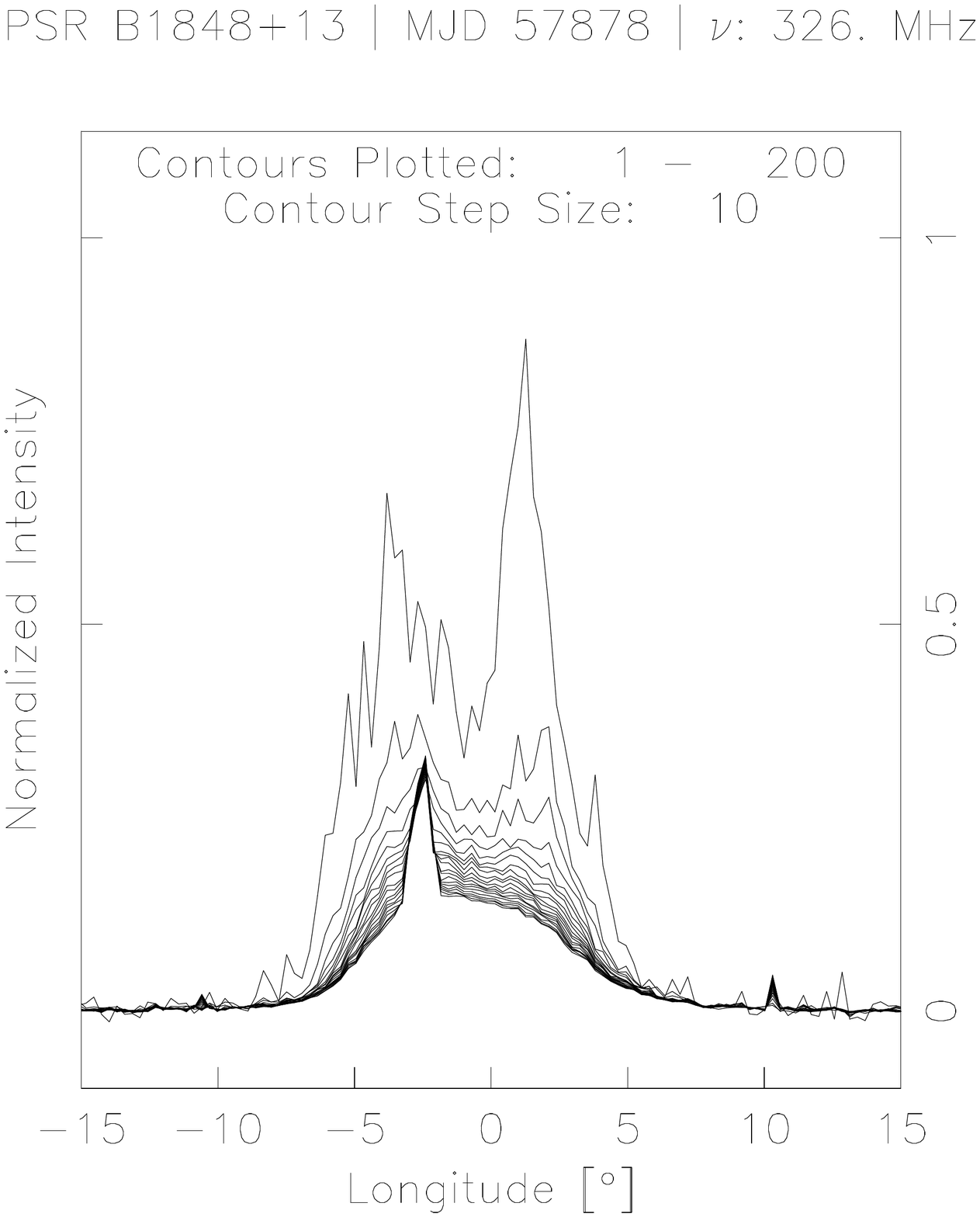} &
\includegraphics[page=1,width=\linewidth]{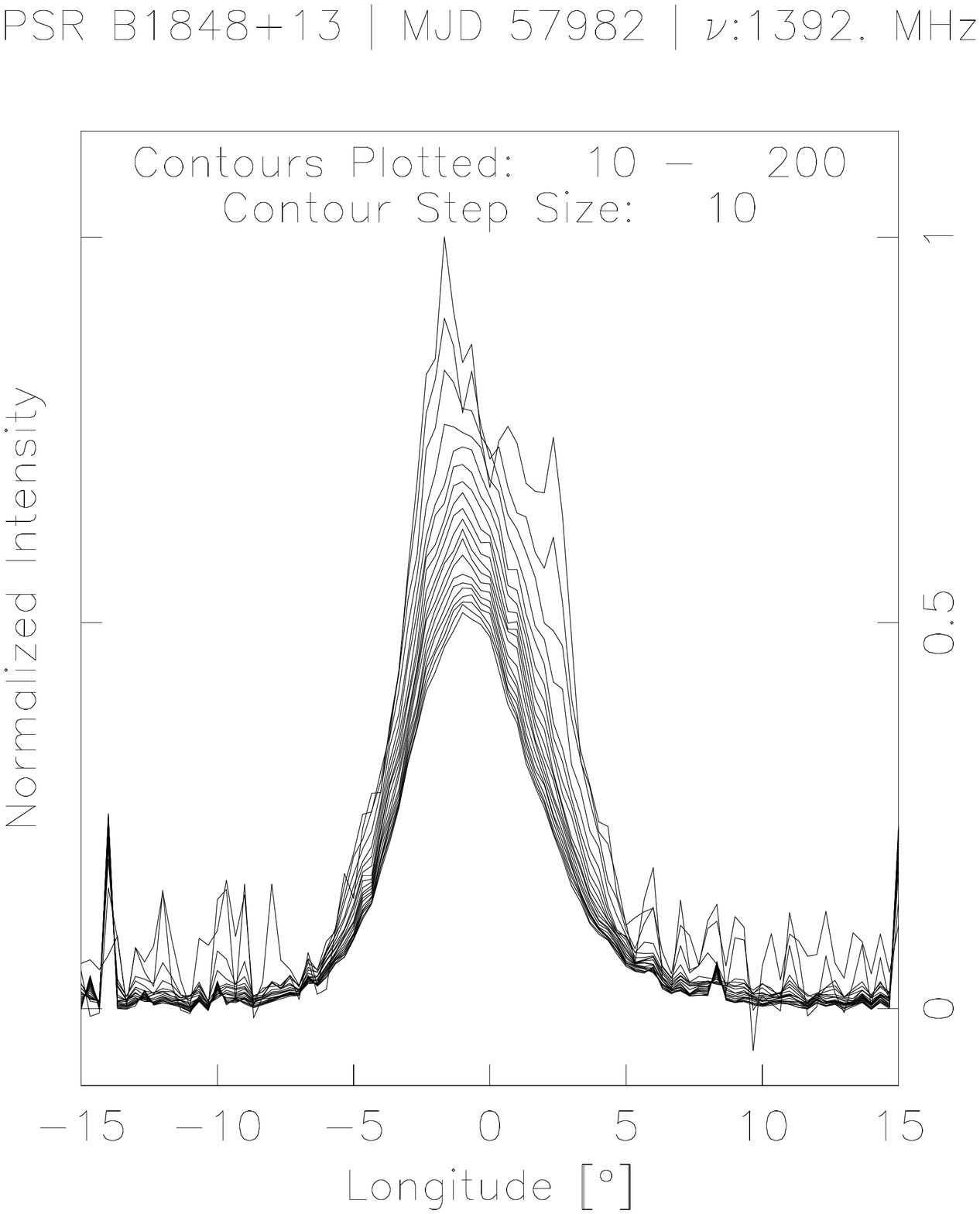} &
\includegraphics[page=1,width=\linewidth]{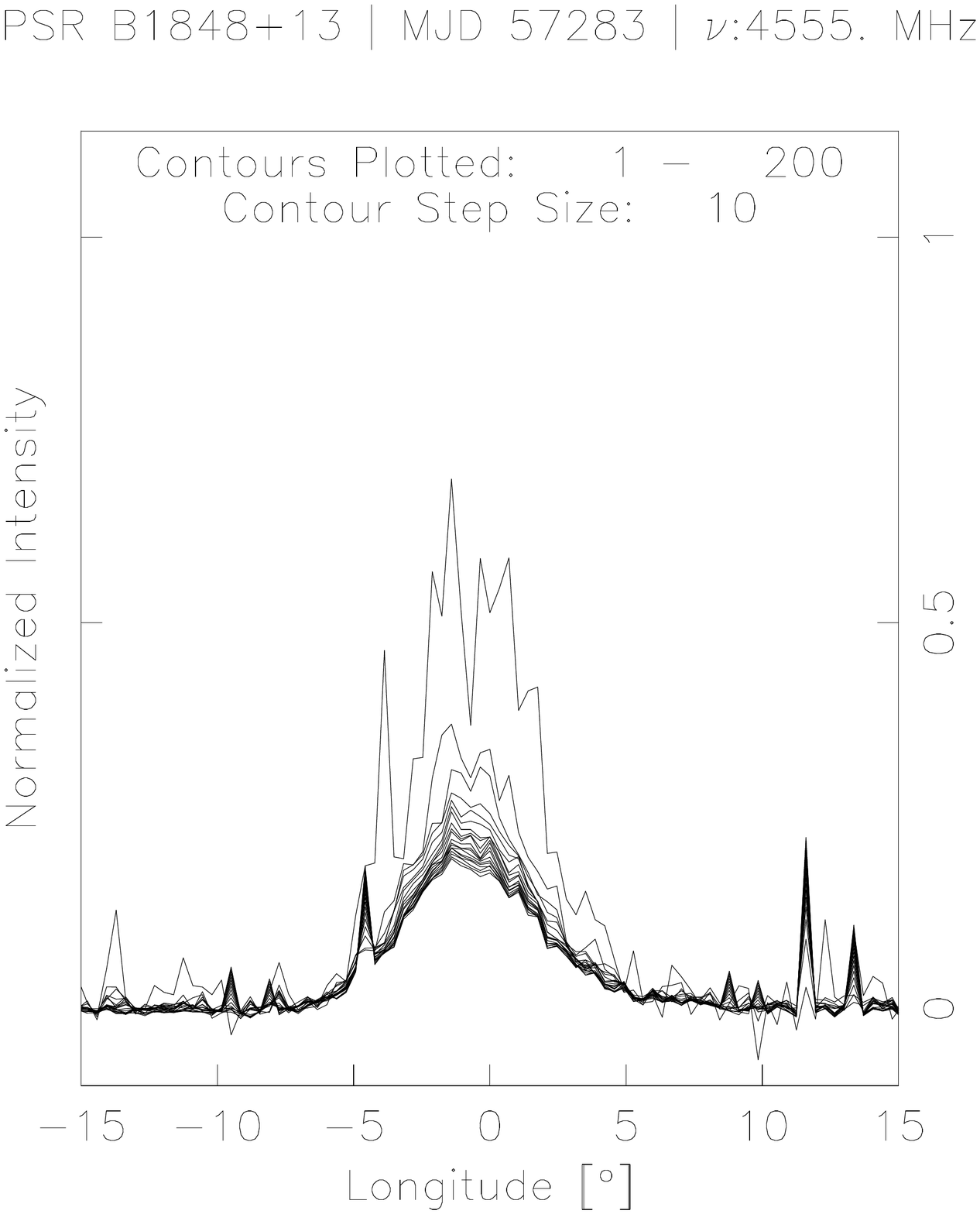} \\ \toprule
                &
\includegraphics[page=1,width=\linewidth]{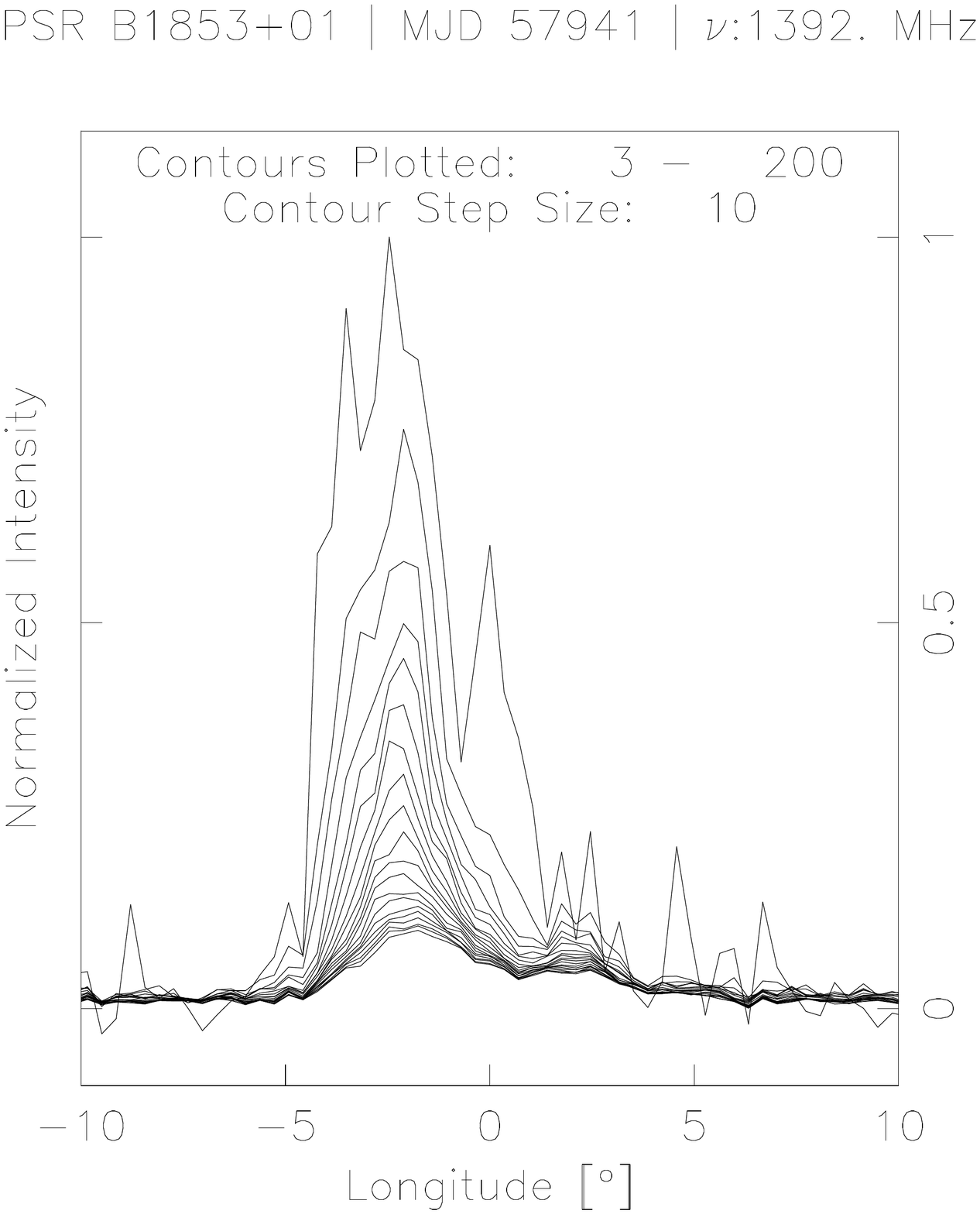} &
\includegraphics[page=1,width=\linewidth]{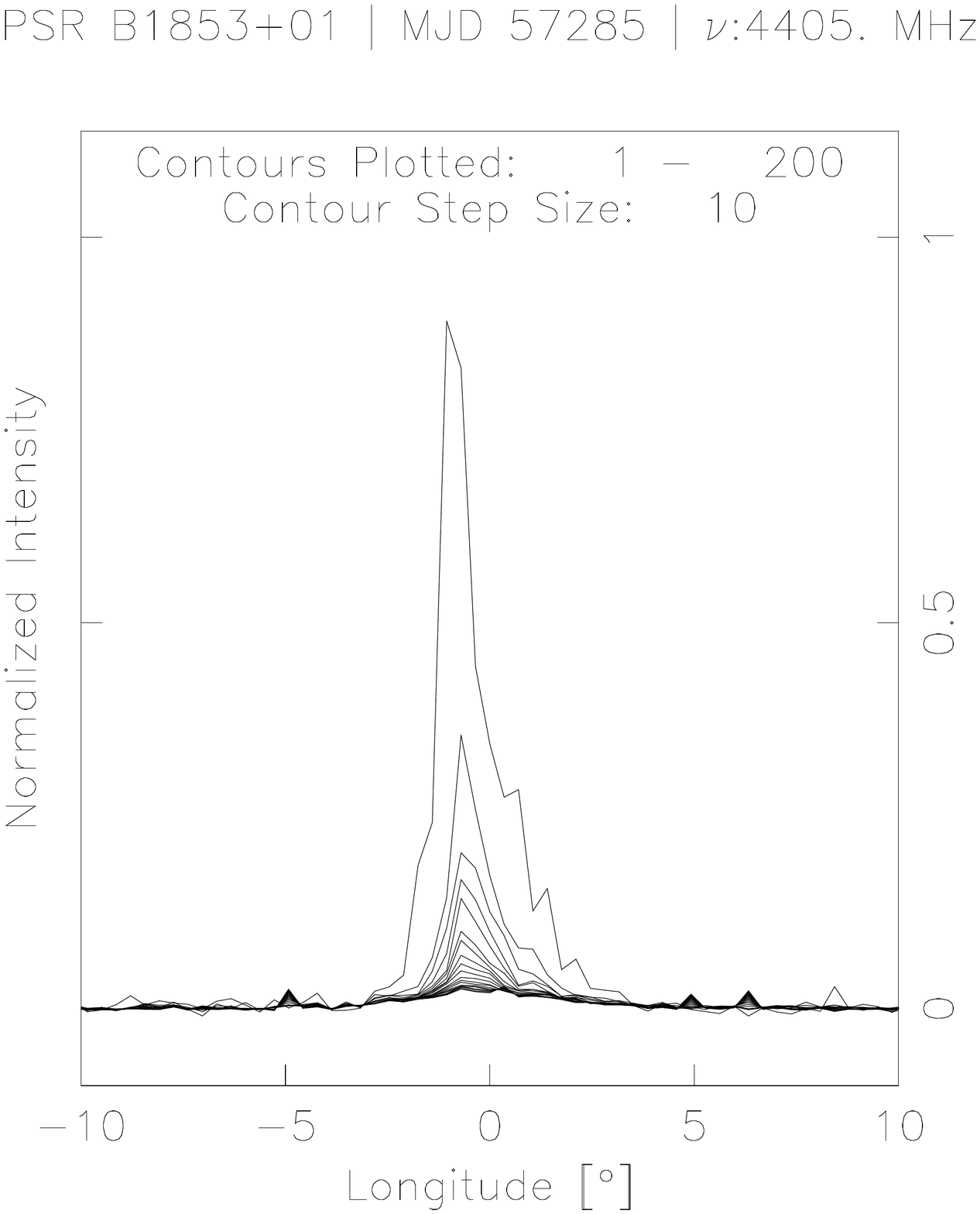} \\

     \bottomrule
   \end{tabularx} 
\caption{PHPs of PSR's B1848+13 and B1853+01.}
 \end{figure*}

\clearpage
\vspace{1cm}
   \begin{figure*} 
 \begin{tabularx}{\textwidth}{YYY}
    \multicolumn{3}{c}{} \\ \toprule
                &
\includegraphics[page=1,width=\linewidth]{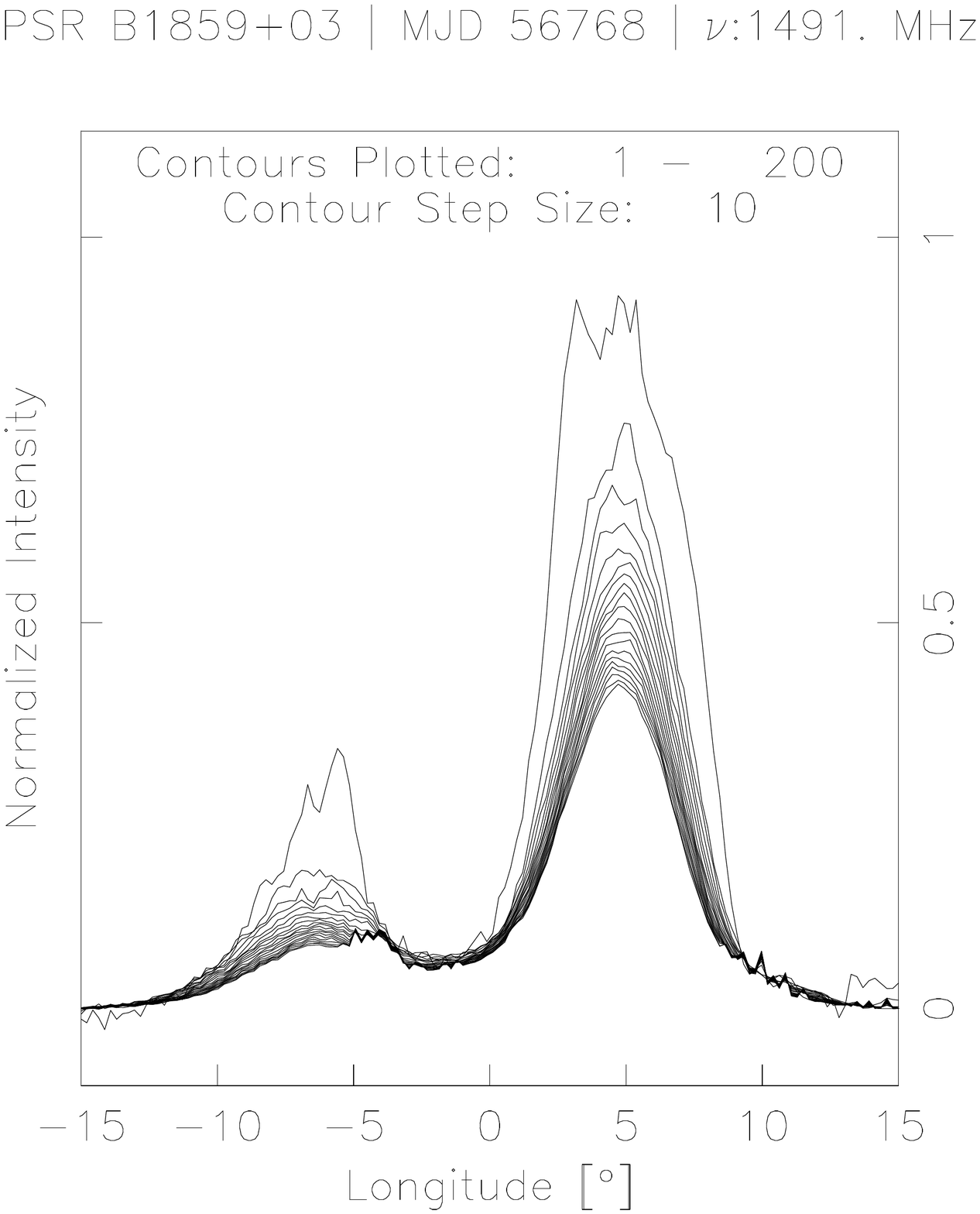} &
\includegraphics[page=1,width=\linewidth]{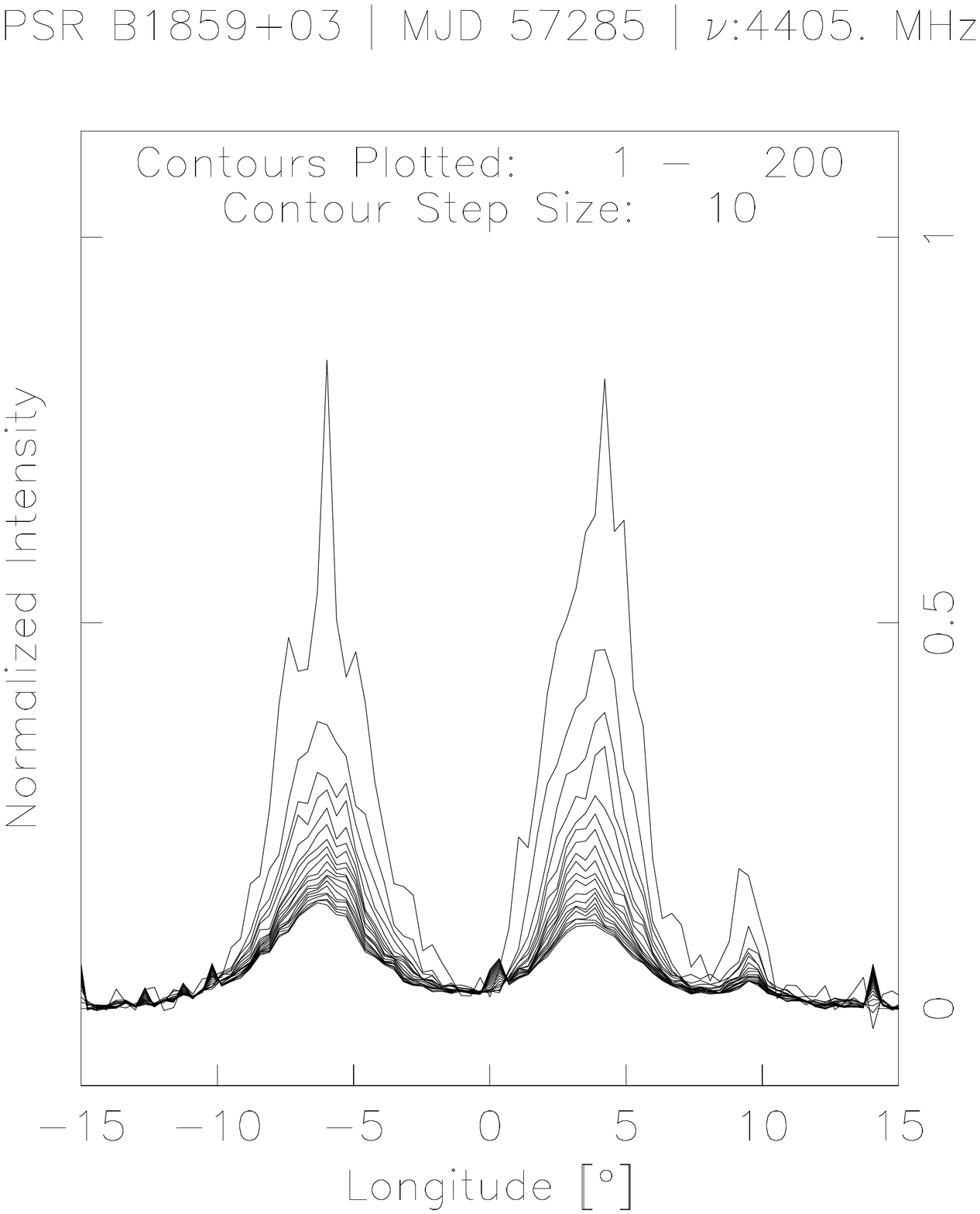} \\ \toprule
&
\includegraphics[page=1,width=\linewidth]{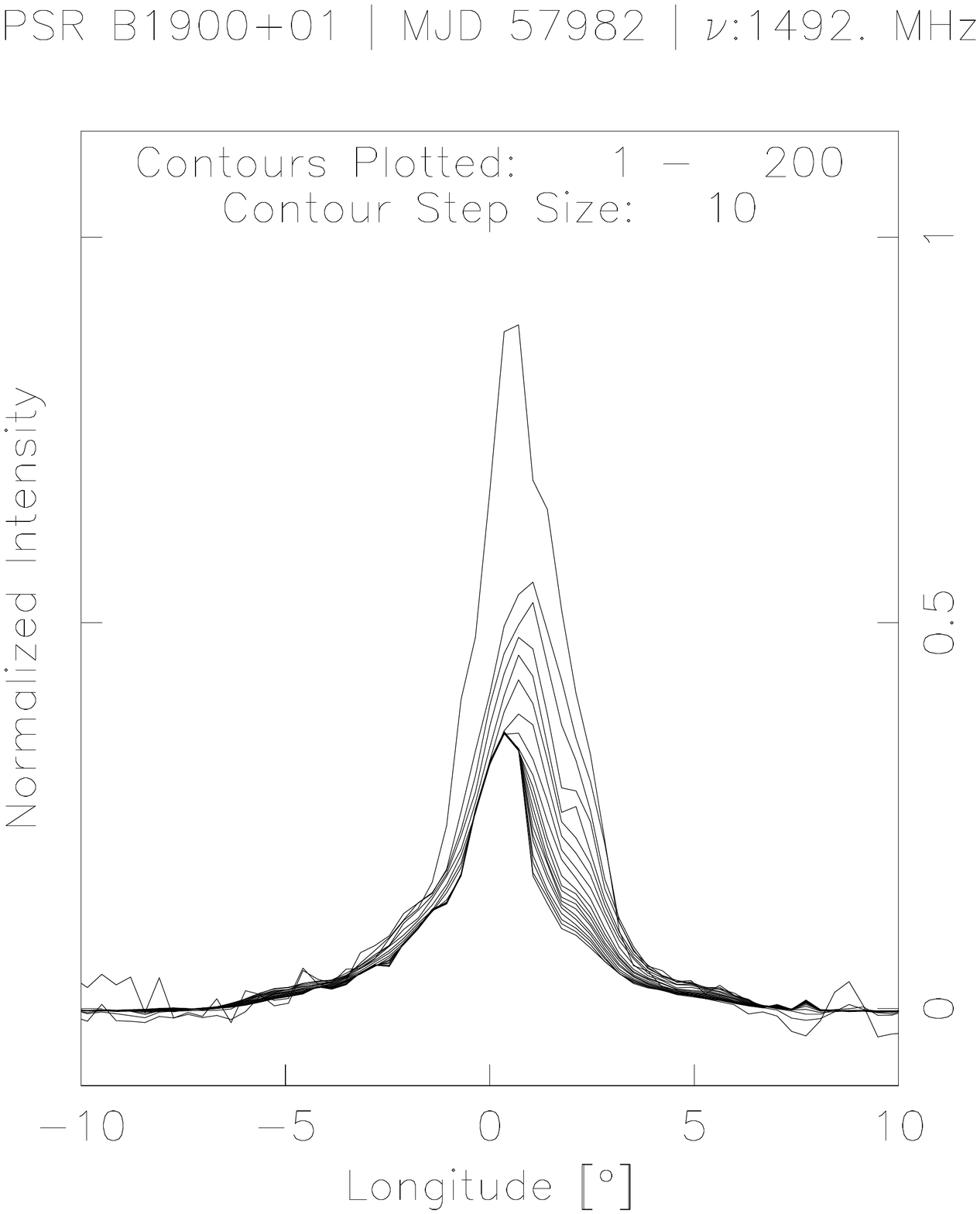}                &
\includegraphics[page=1,width=\linewidth]{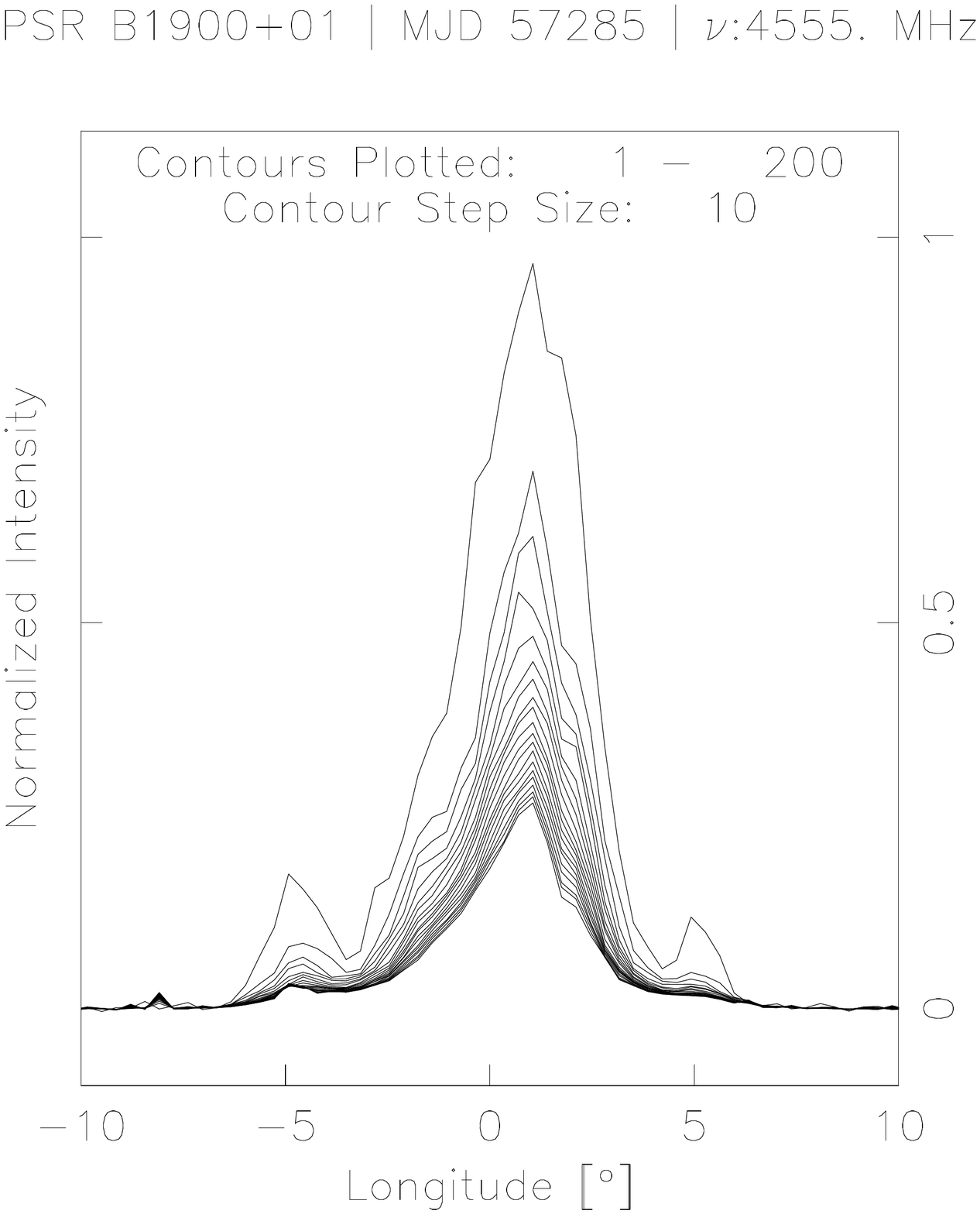} \\ \toprule
                &
                &
                \\
                &
\includegraphics[page=1,width=\linewidth]{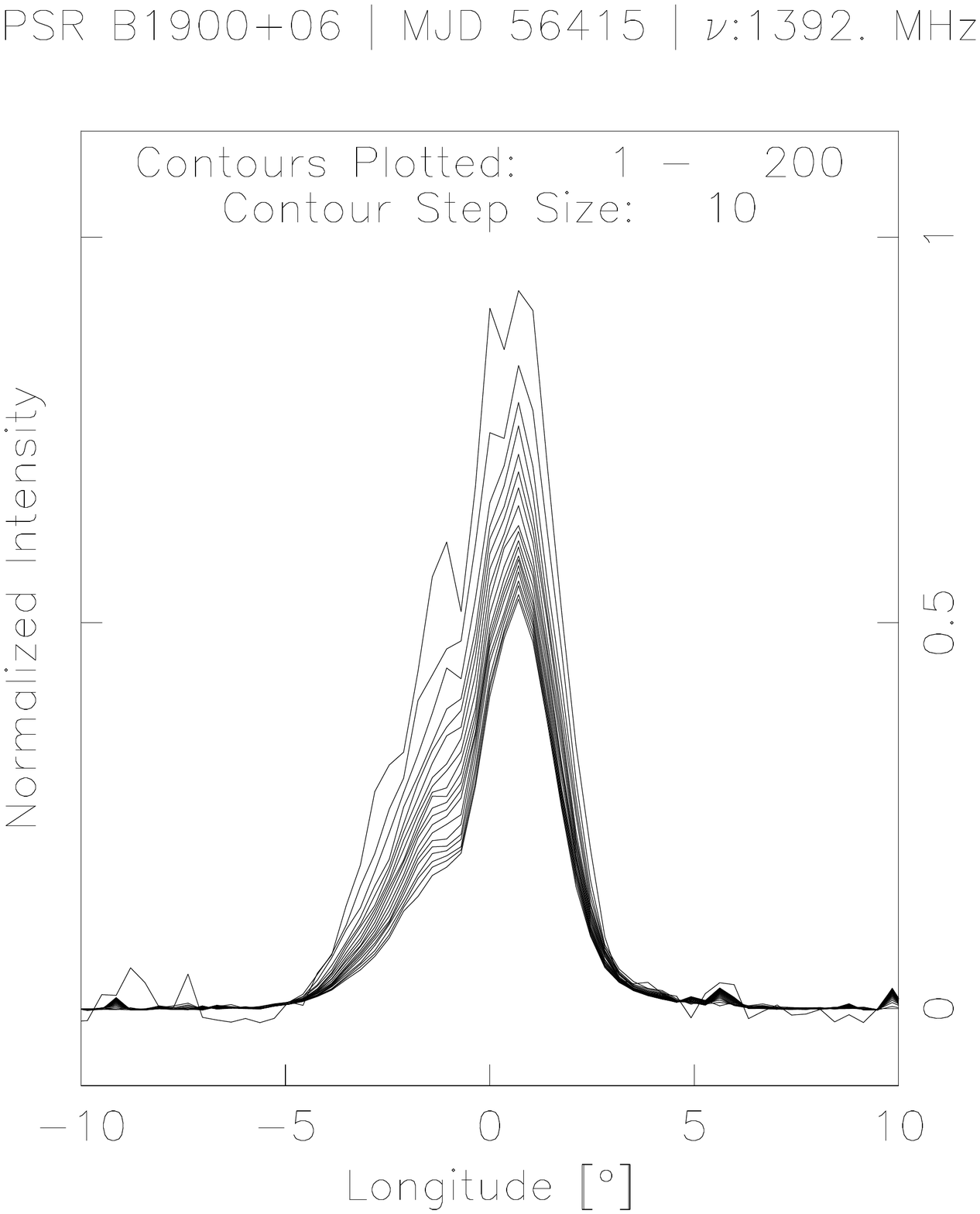} &
\includegraphics[page=1,width=\linewidth]{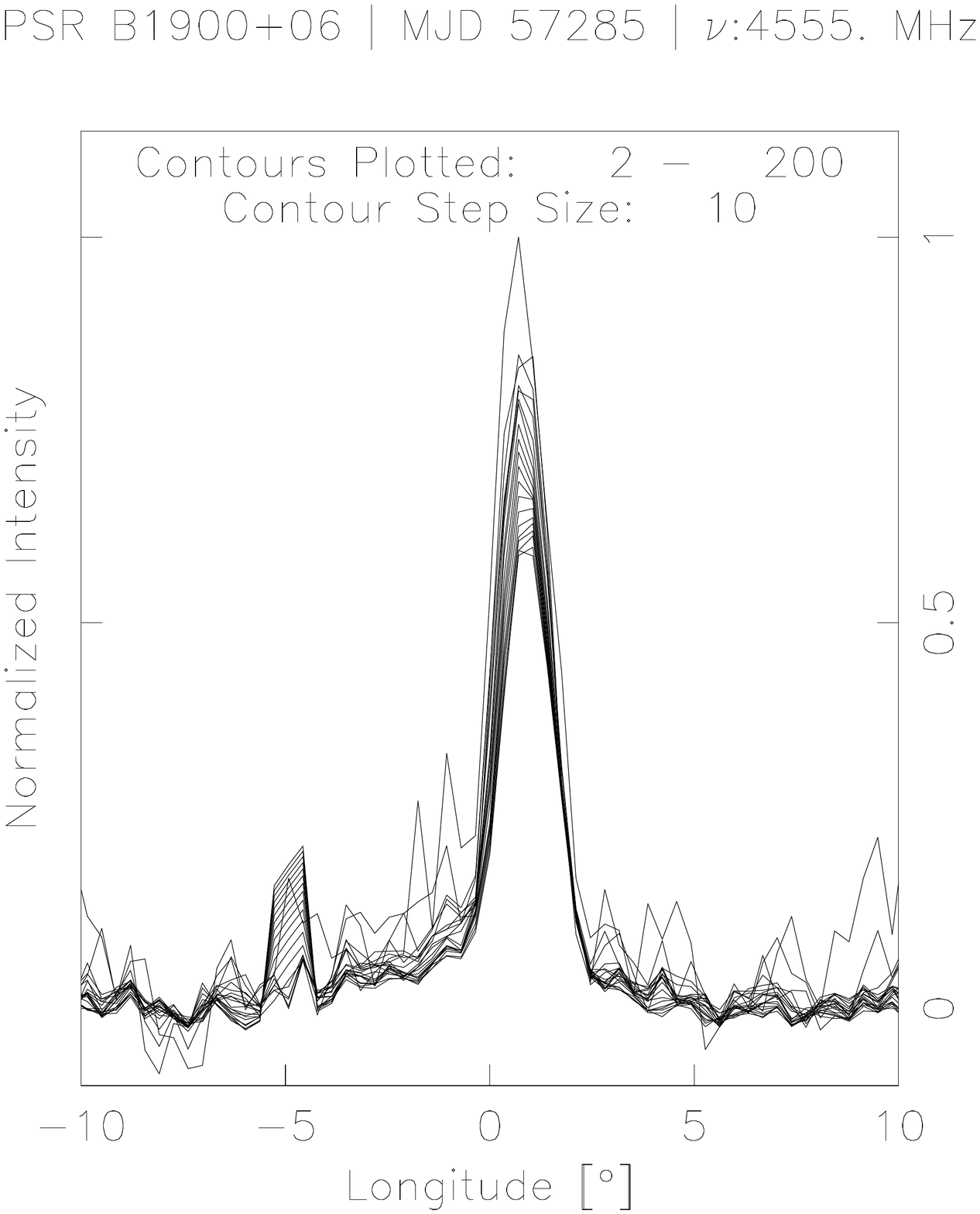} \\ 

     \bottomrule
   \end{tabularx} 
\caption{PHPs of PSR's B1859+03, B1900+01, and B1900+06.}
 \end{figure*}
\clearpage
\vspace{1cm}
   \begin{figure*} 
 \begin{tabularx}{\textwidth}{YYY}
    \multicolumn{3}{c}{} \\ \toprule
\includegraphics[page=1,width=\linewidth]{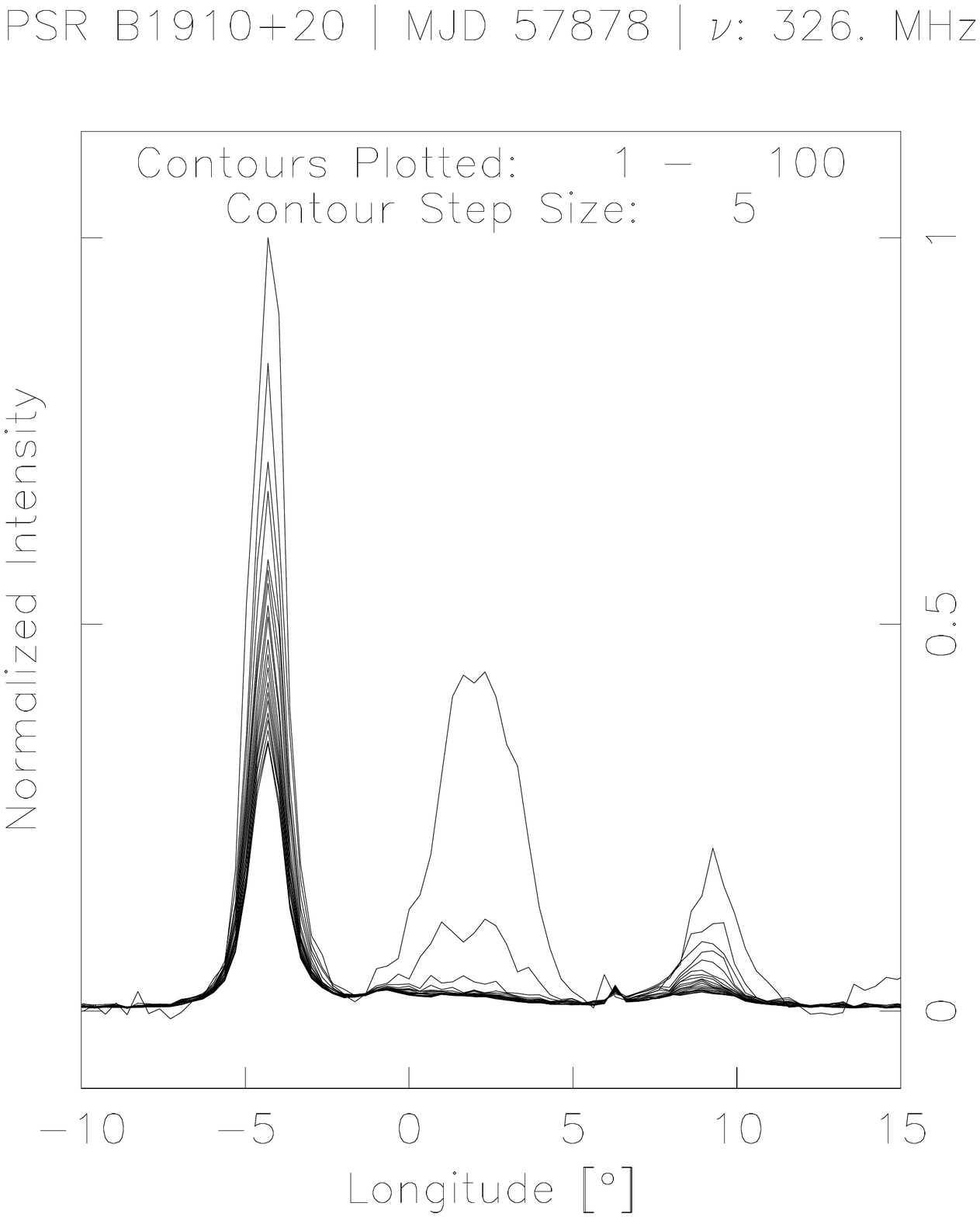} &
\includegraphics[page=1,width=\linewidth]{B1910+20peak_occur_freq2_56563} &
\includegraphics[page=1,width=\linewidth]{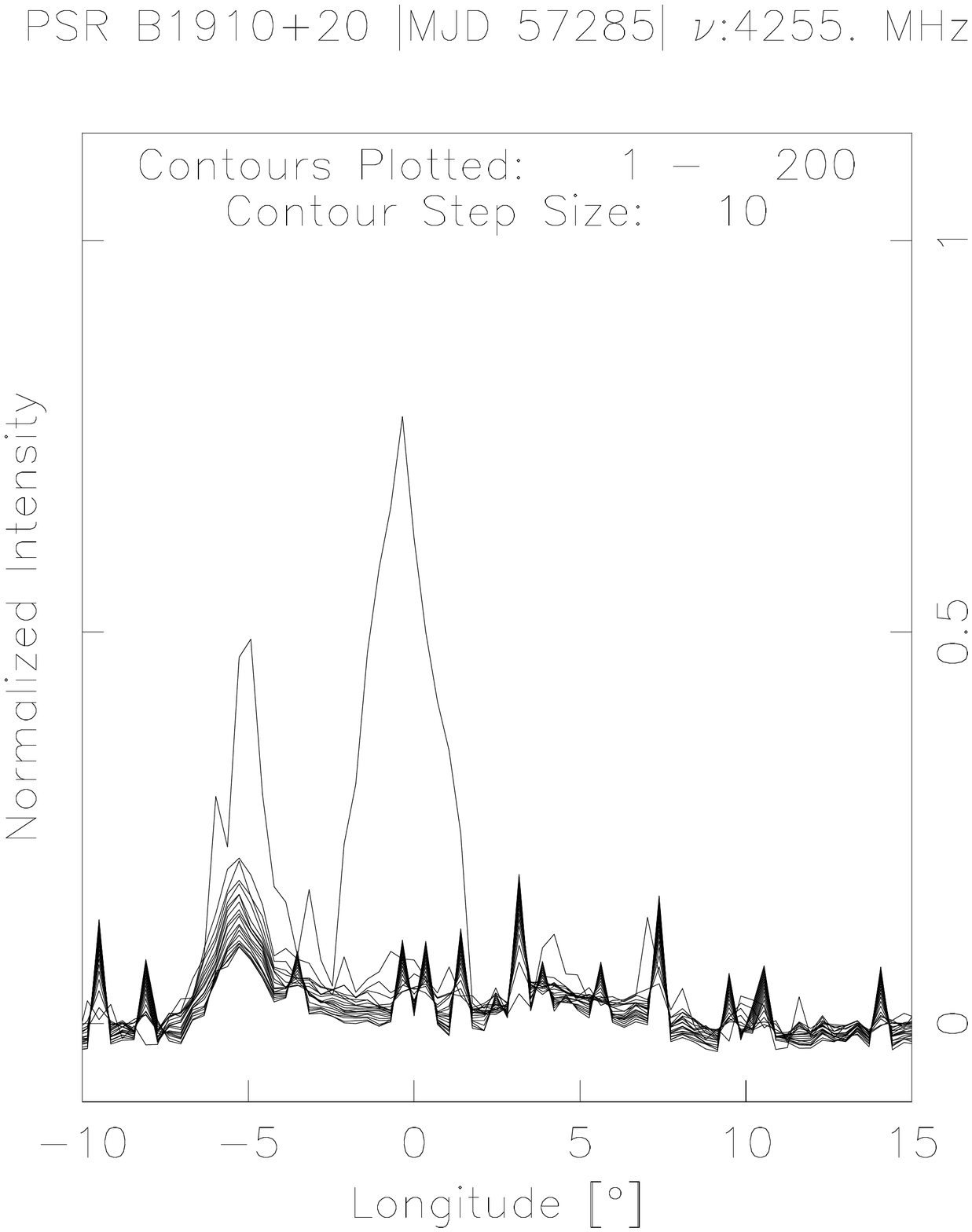} \\ \toprule
\includegraphics[page=1,width=\linewidth]{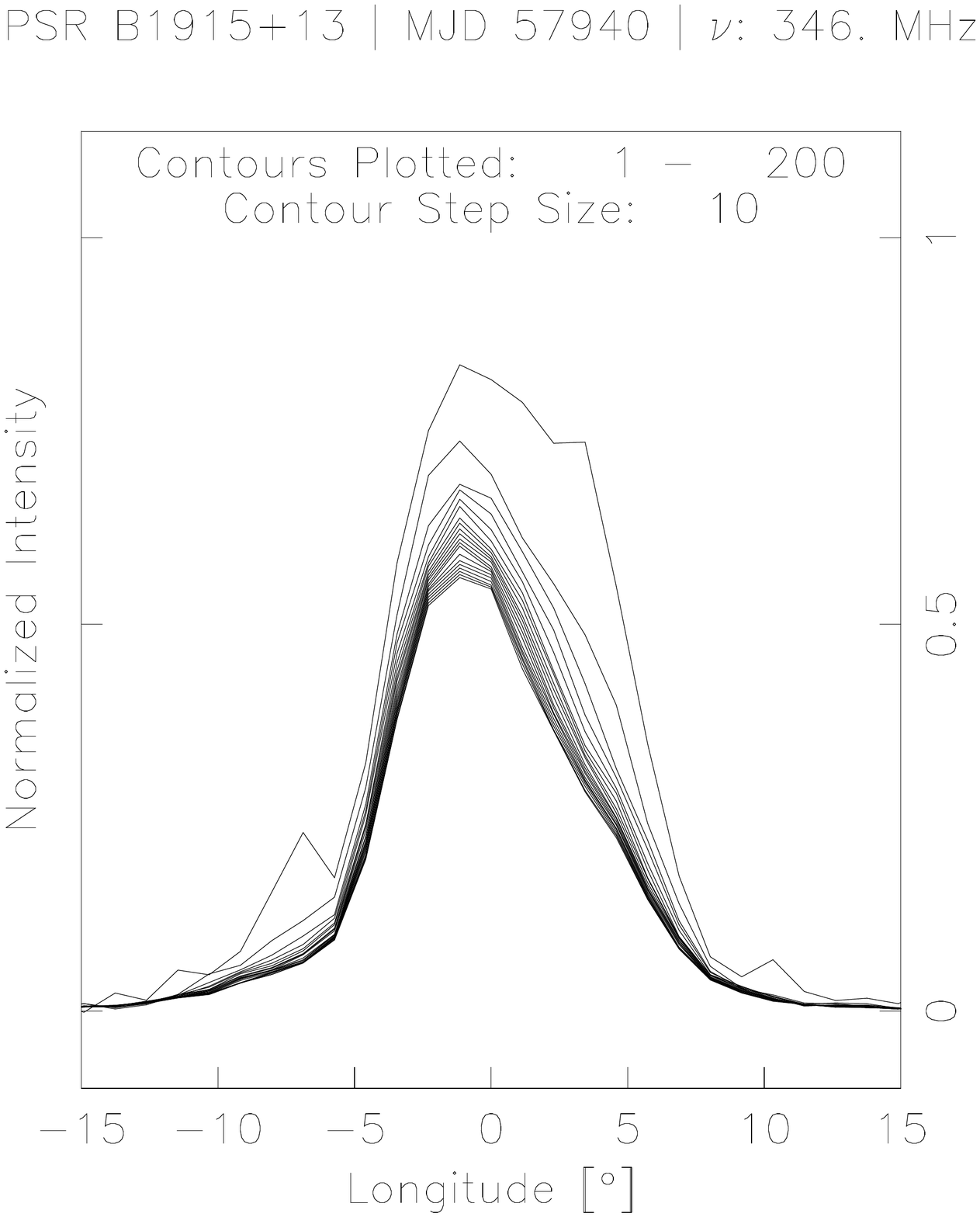} &
\includegraphics[page=1,width=\linewidth]{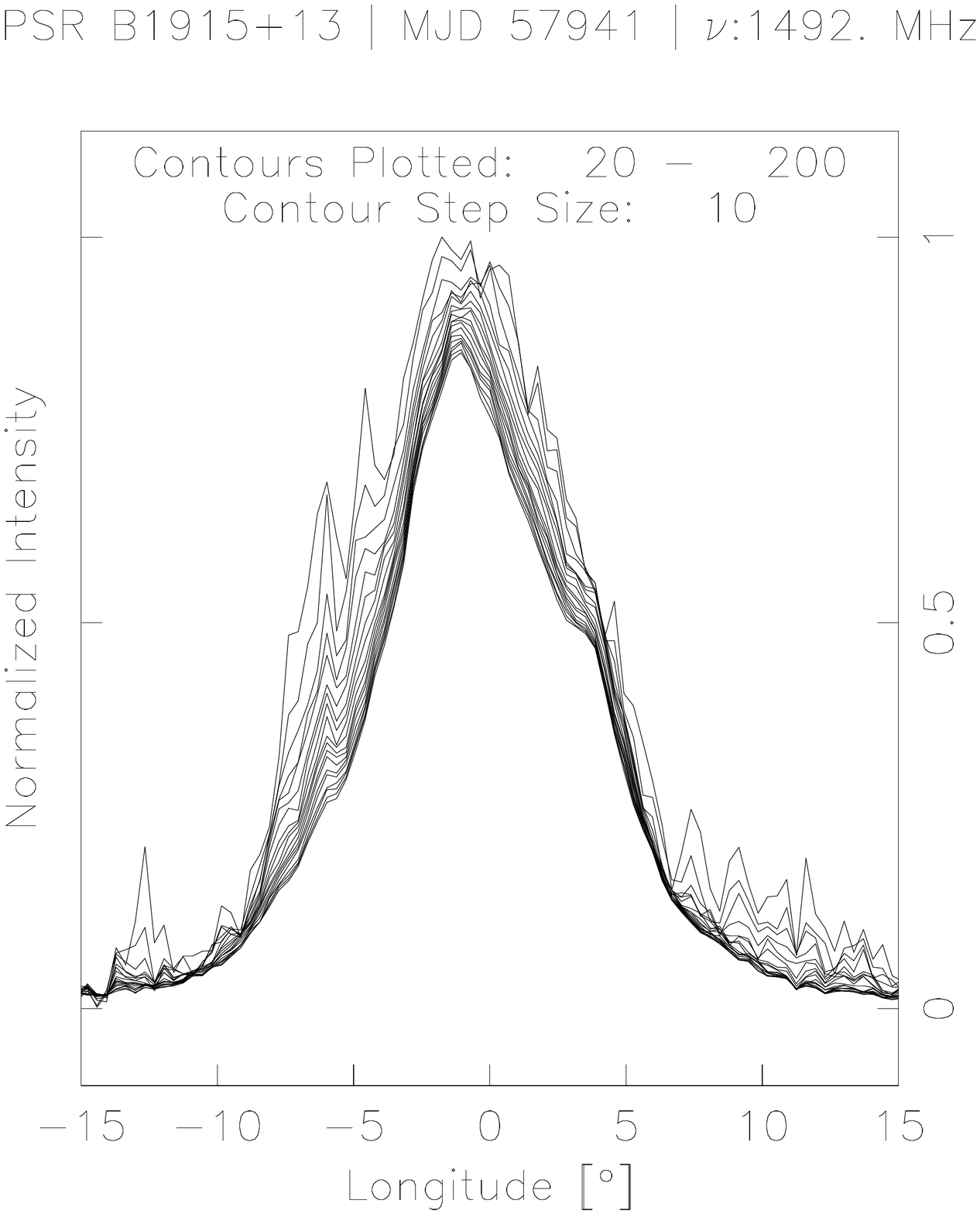} &
\\ \toprule
\includegraphics[page=1,width=\linewidth]{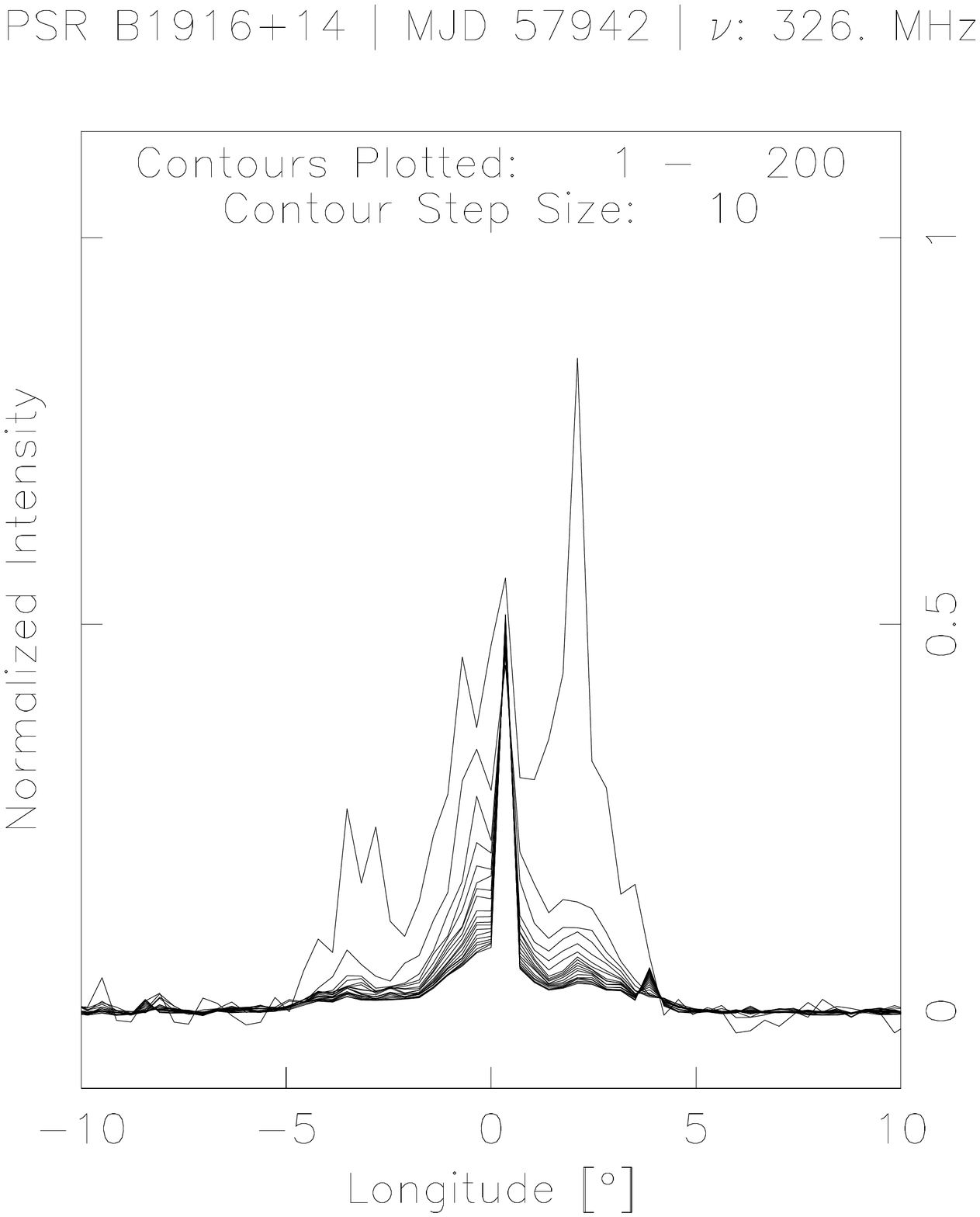} &
\includegraphics[page=1,width=\linewidth]{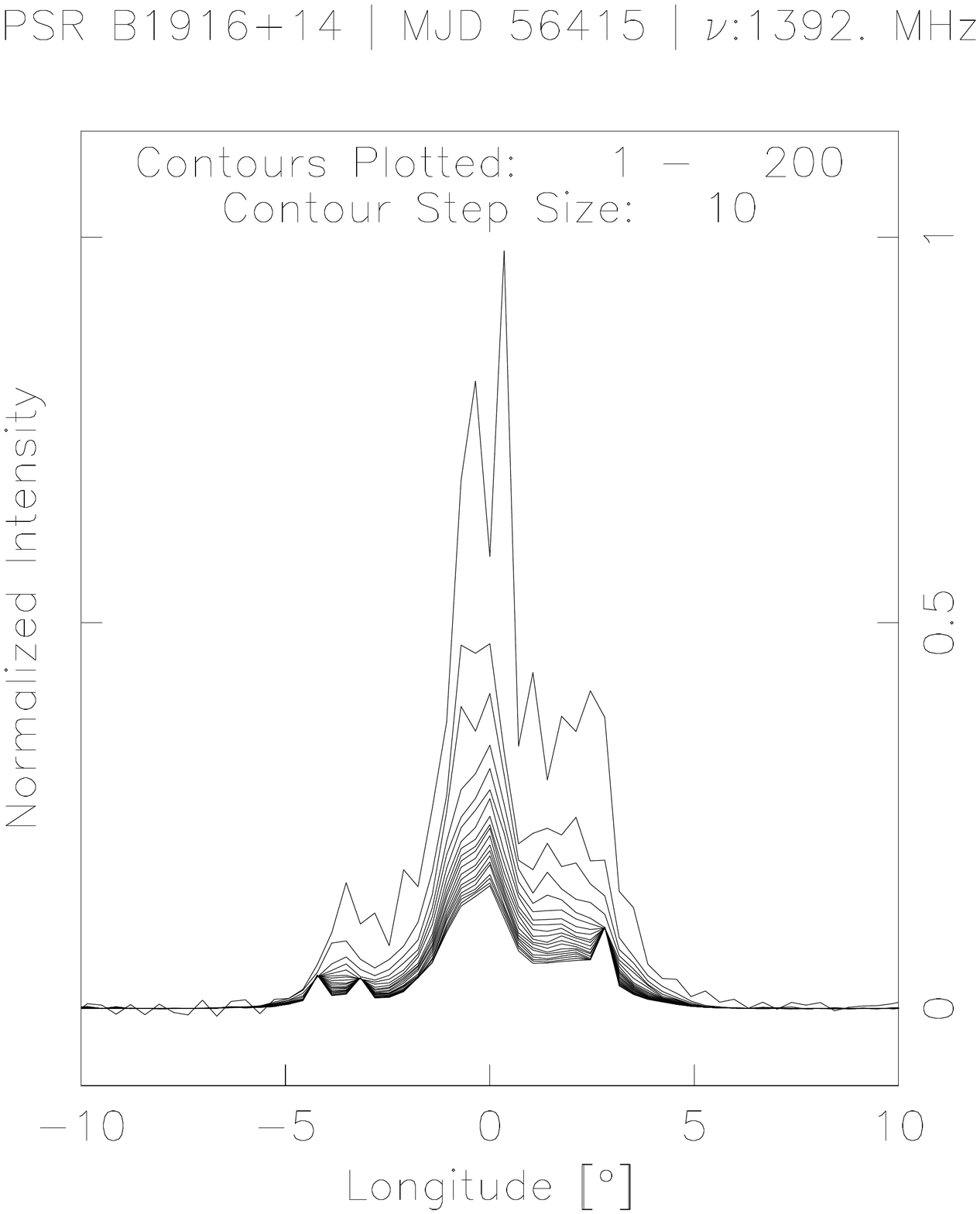} &
\includegraphics[page=1,width=\linewidth]{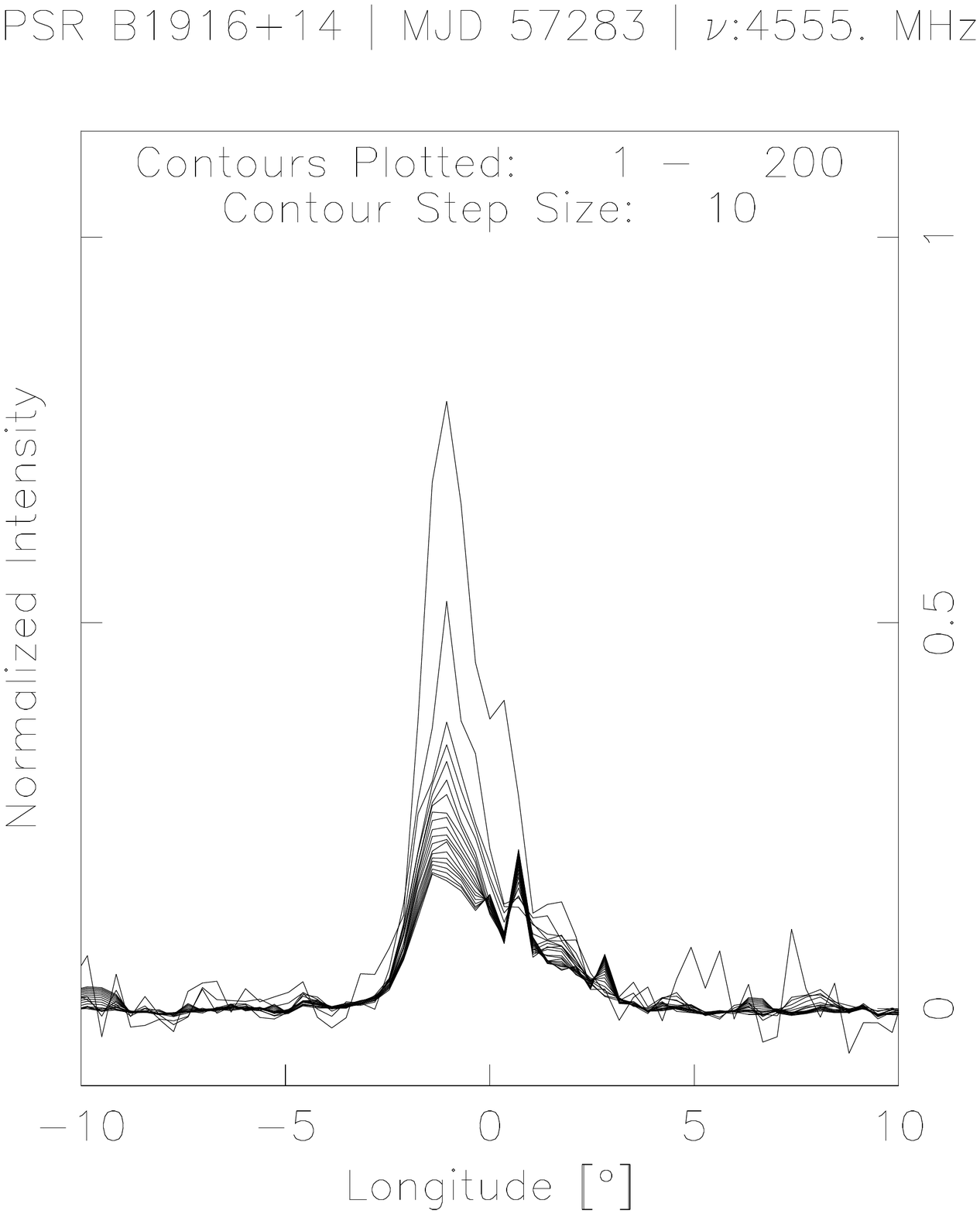} \\ 
     \bottomrule
   \end{tabularx} 
\caption{PHPs of PSR's B1910+20, B1915+13, and B1916+14.}
 \end{figure*}
\vspace{1cm}
   \begin{figure*} 
 \begin{tabularx}{\textwidth}{YYY}
    \multicolumn{3}{c}{} \\ \toprule
\includegraphics[page=1,width=\linewidth]{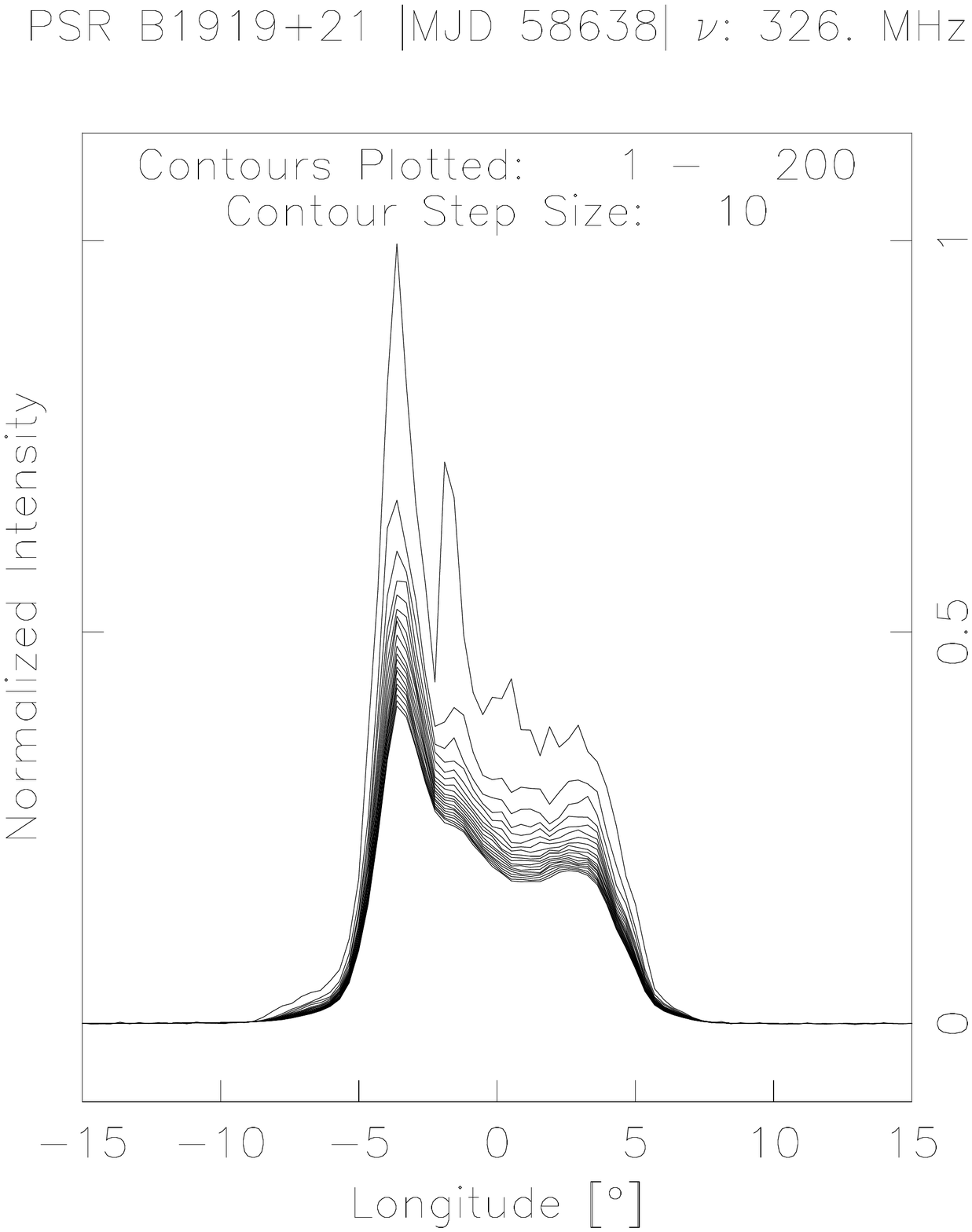} &
\includegraphics[page=1,width=\linewidth]{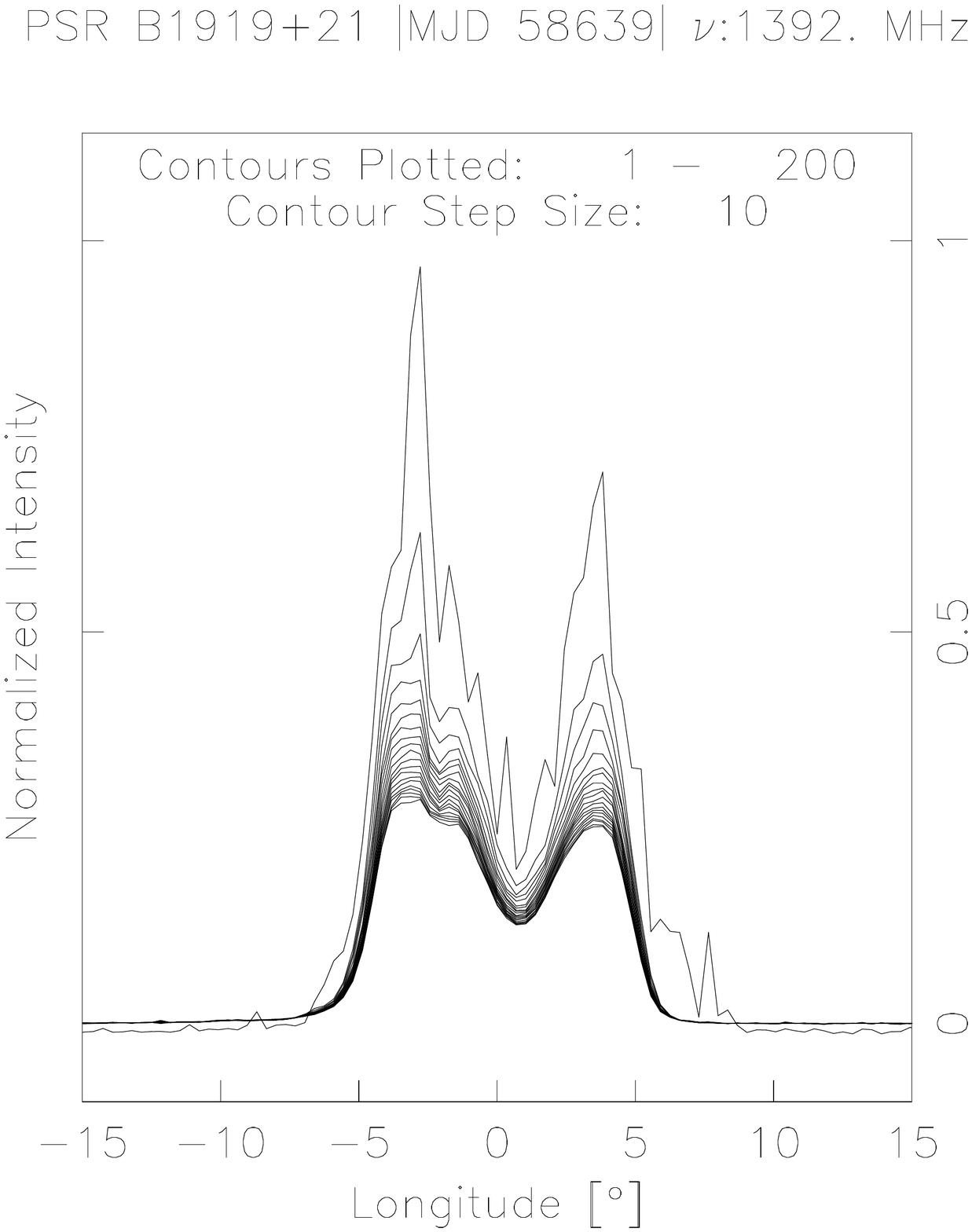} &
\includegraphics[page=1,width=\linewidth]{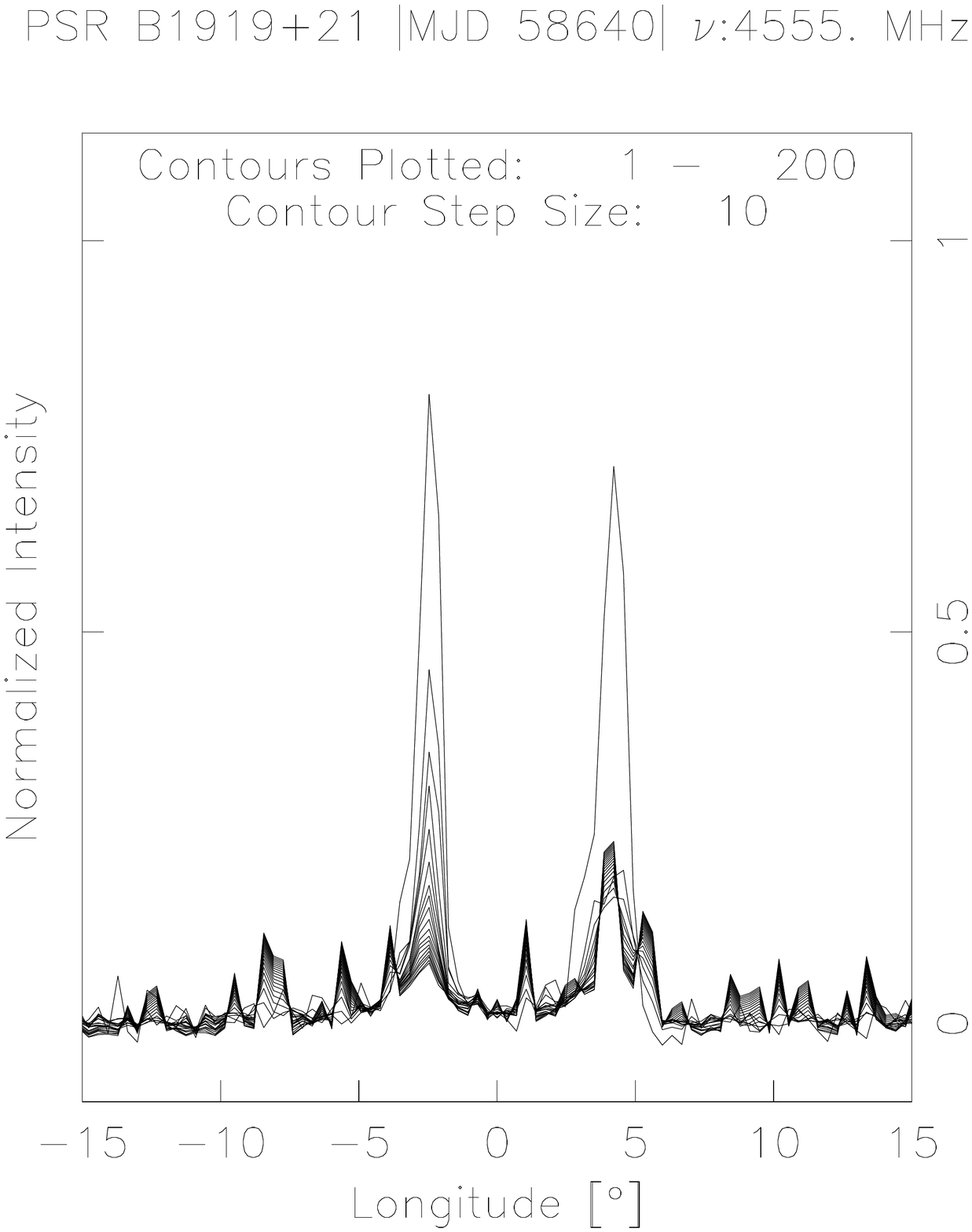} \\ \toprule
\includegraphics[page=1,width=\linewidth]{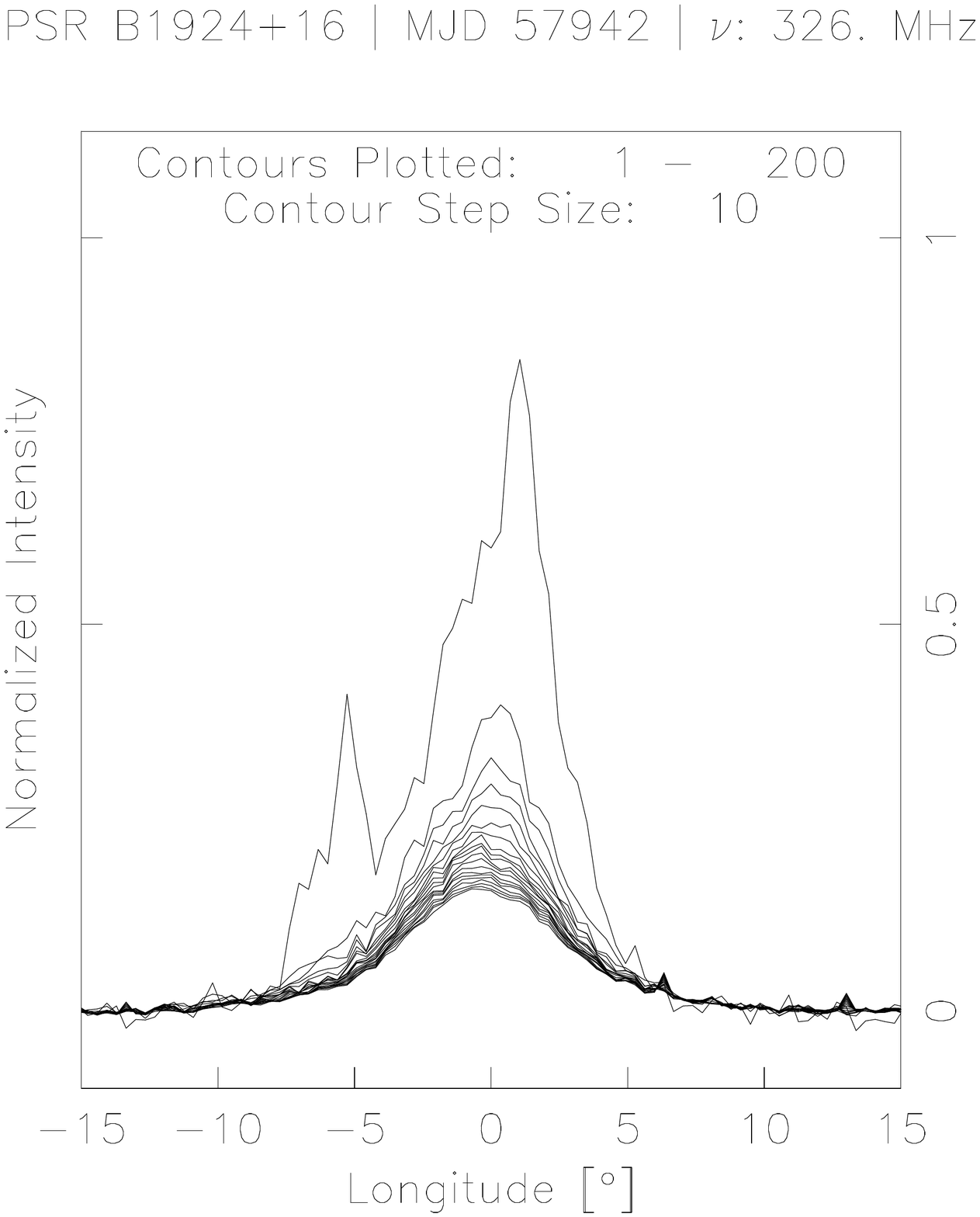} &
\includegraphics[page=1,width=\linewidth]{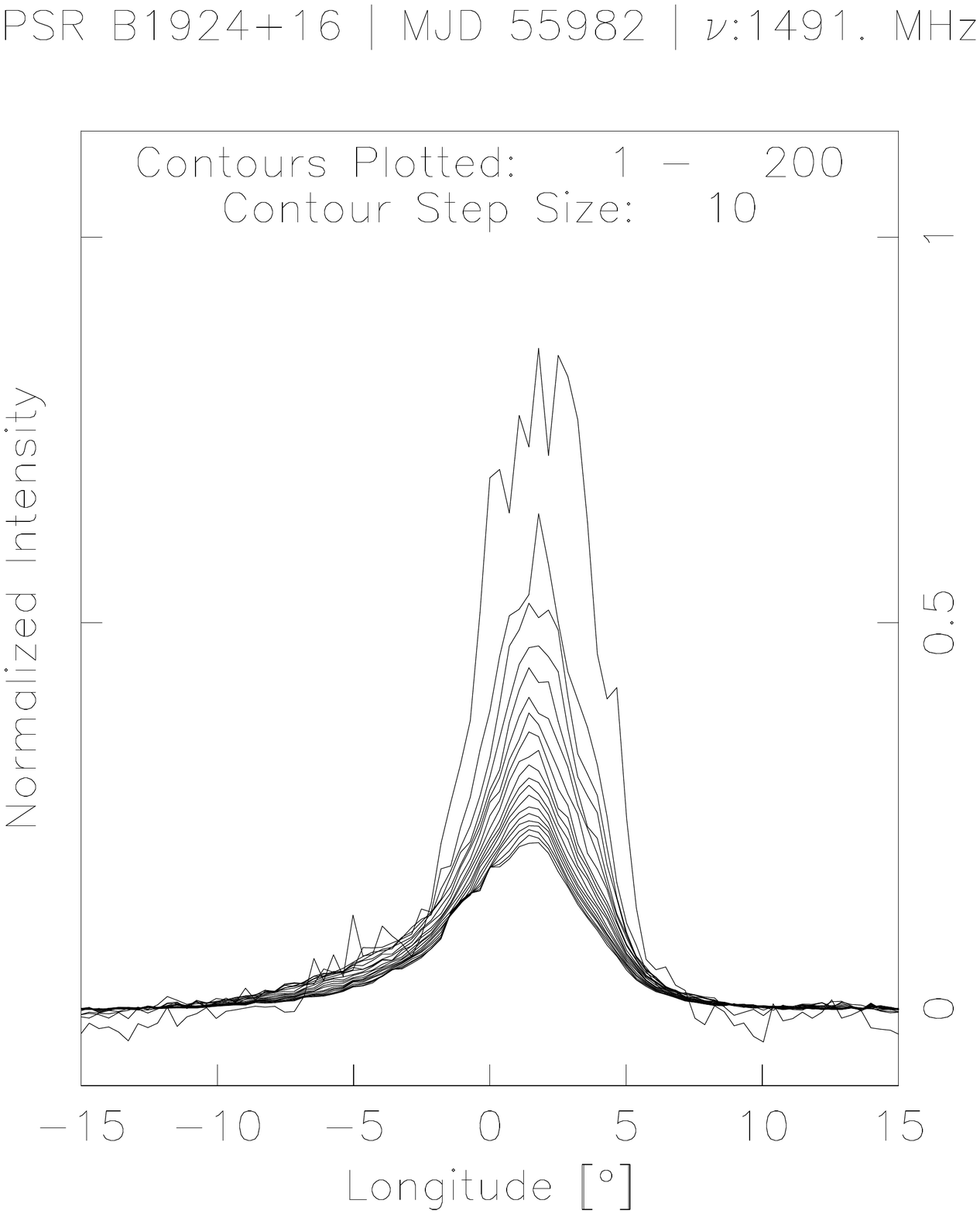} &
\includegraphics[page=1,width=\linewidth]{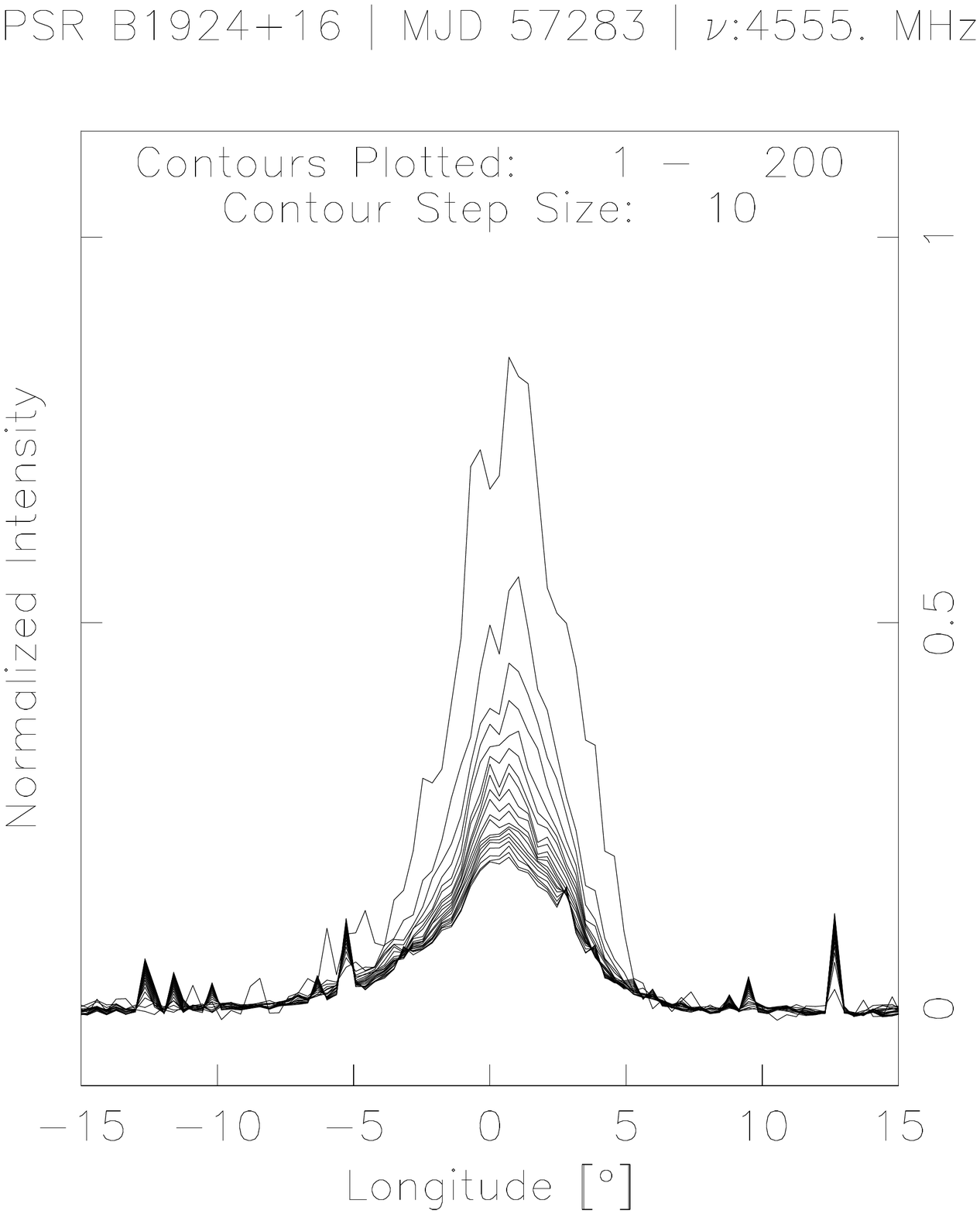} \\ \toprule
\includegraphics[page=1,width=\linewidth]{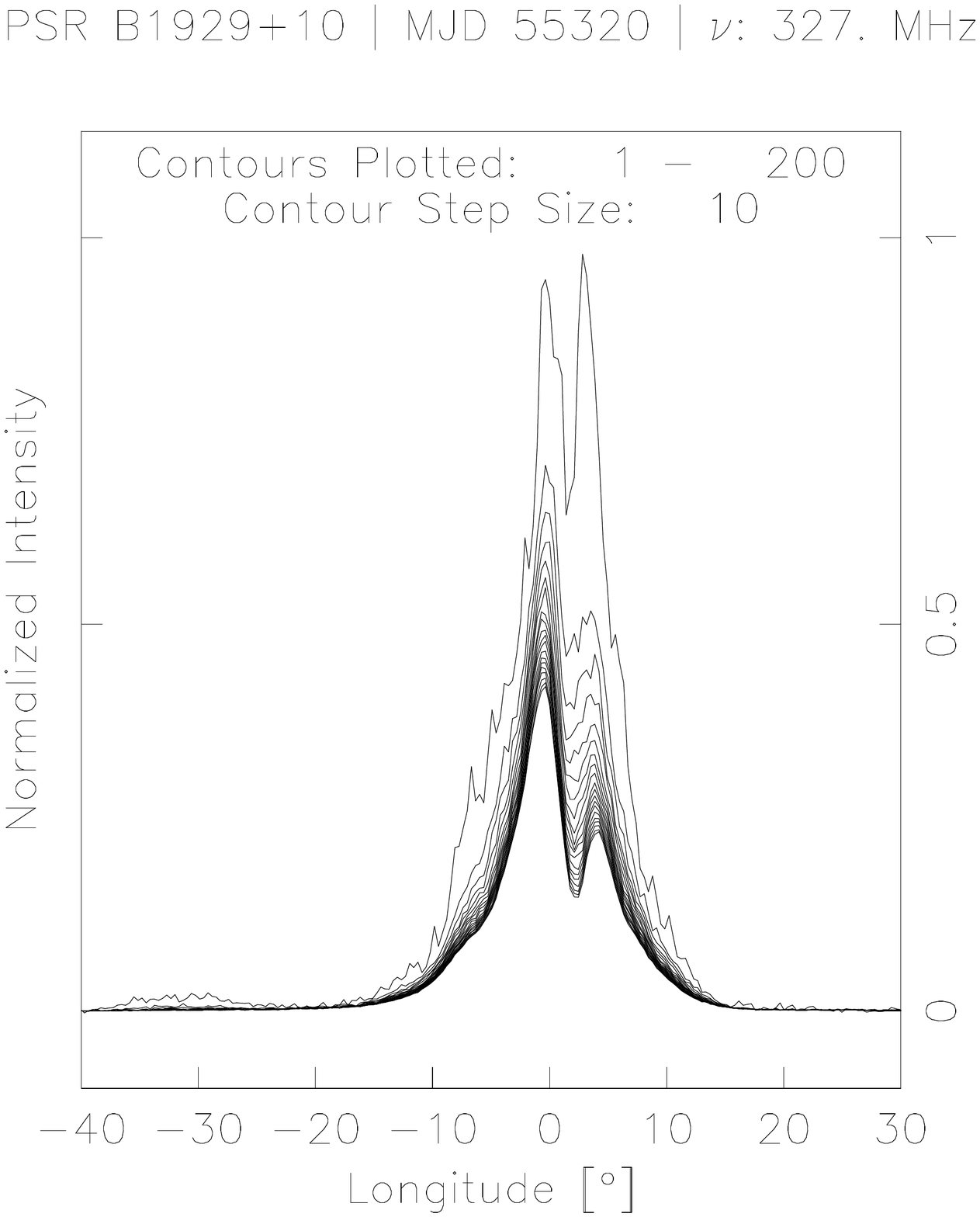} &
\includegraphics[page=1,width=\linewidth]{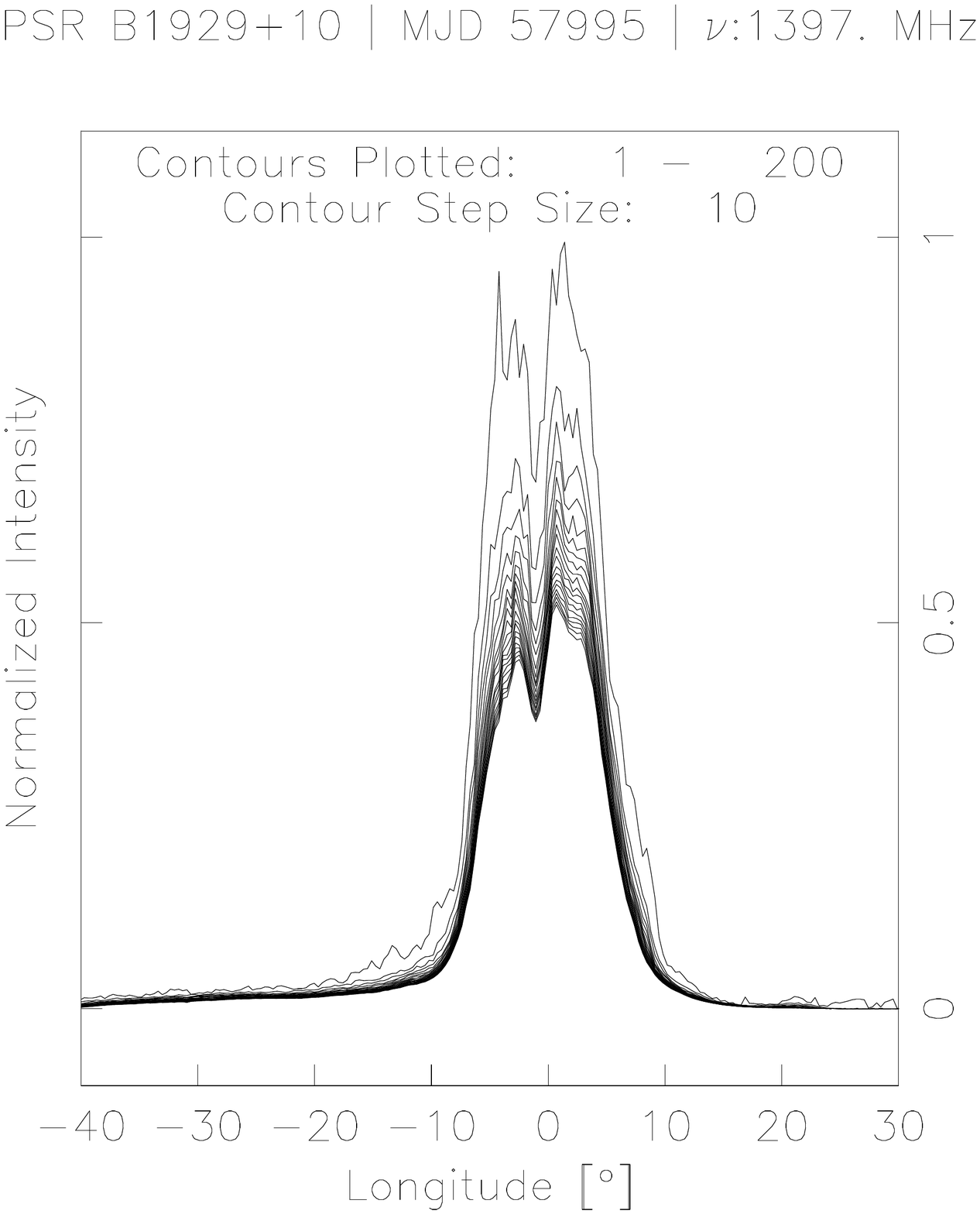} &
\includegraphics[page=1,width=\linewidth]{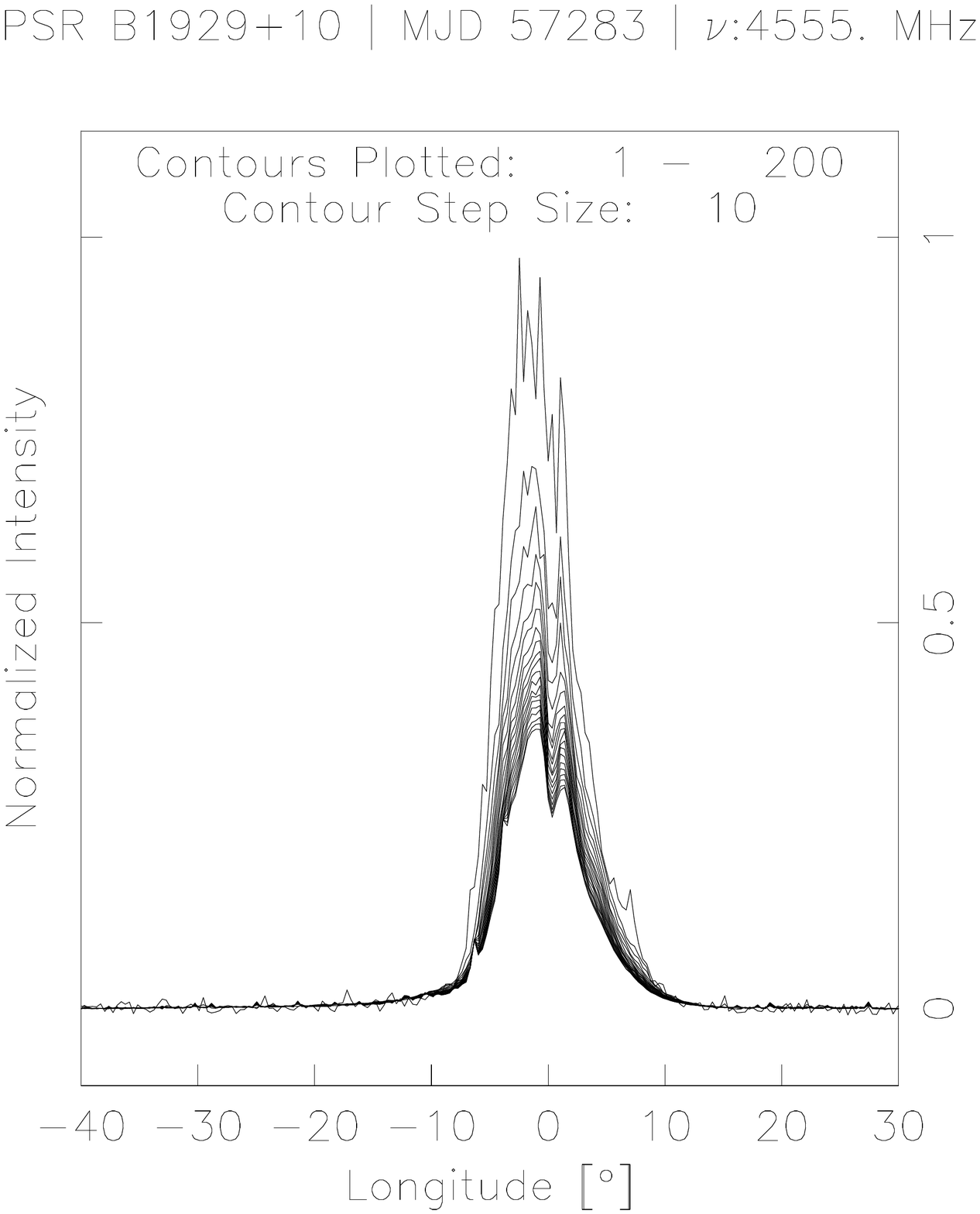} \\ 
     \bottomrule
   \end{tabularx} 
\caption{Average profiles of PSRs B1919+21, B1924+16, and B1929+10.}
 \end{figure*}
\vspace{1cm}
   \begin{figure*} 
 \begin{tabularx}{\textwidth}{YYY}
    \multicolumn{3}{c}{} \\ \toprule
                &
\includegraphics[page=1,width=\linewidth]{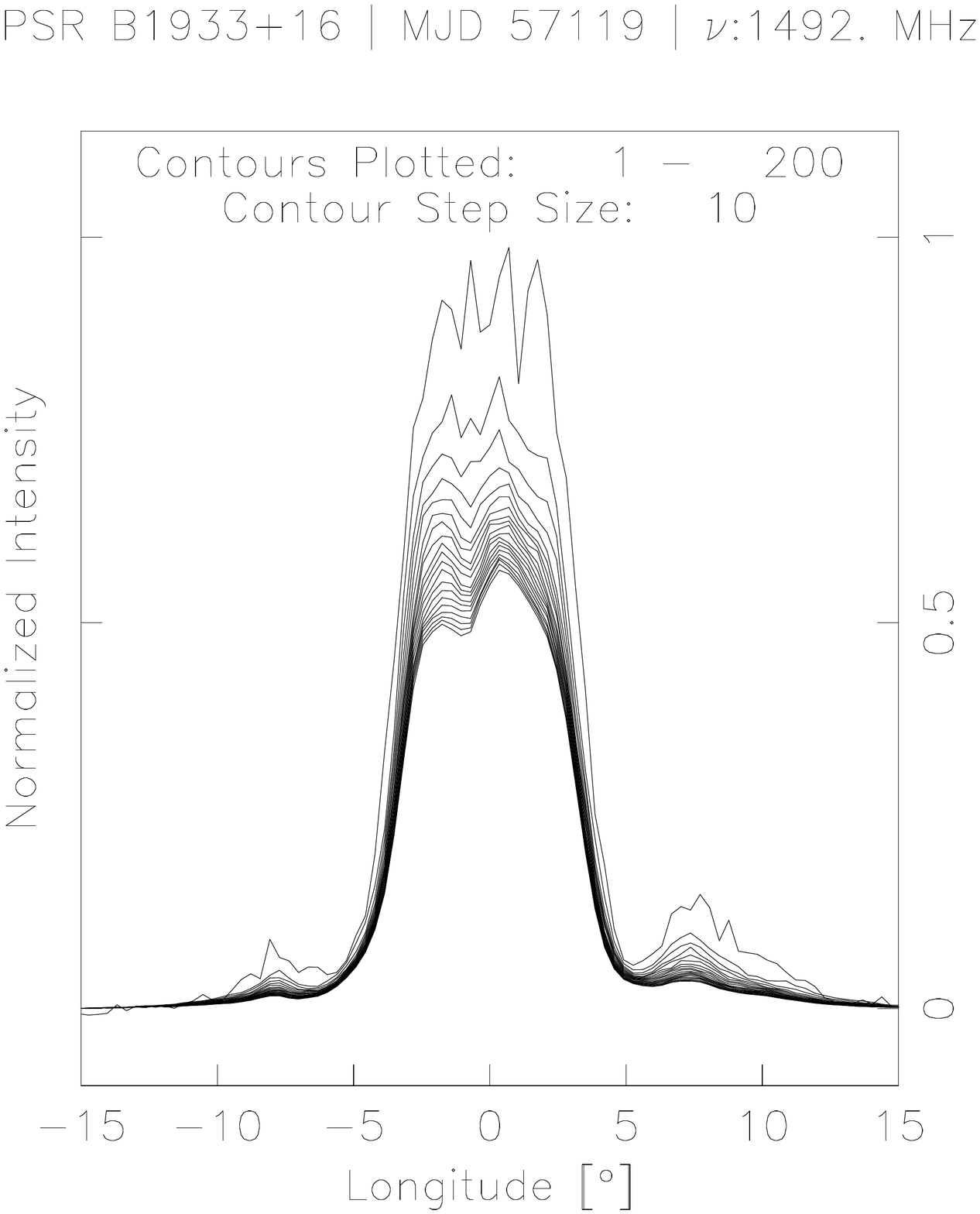} &
\includegraphics[page=1,width=\linewidth]{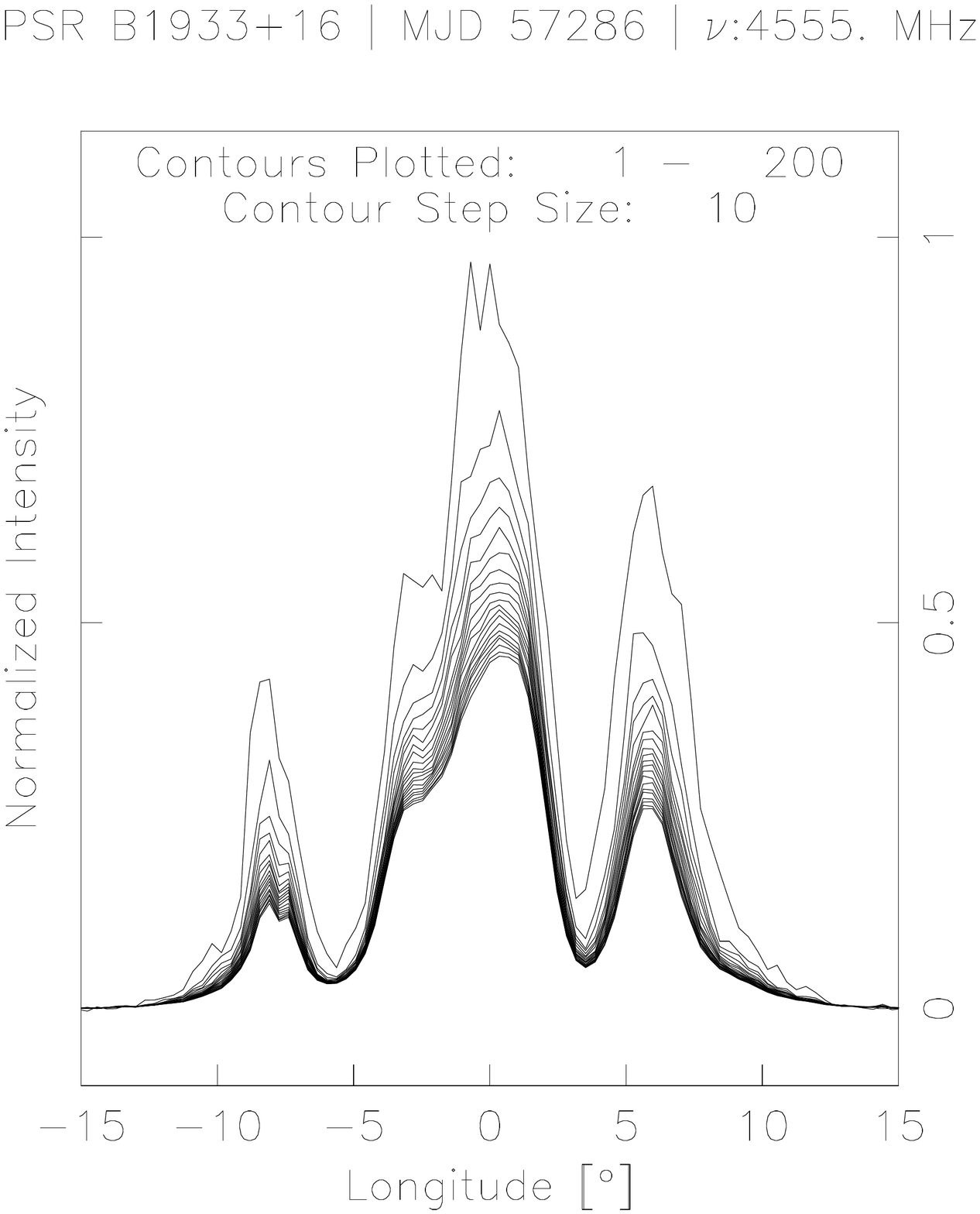} \\ \toprule
\includegraphics[page=1,width=\linewidth]{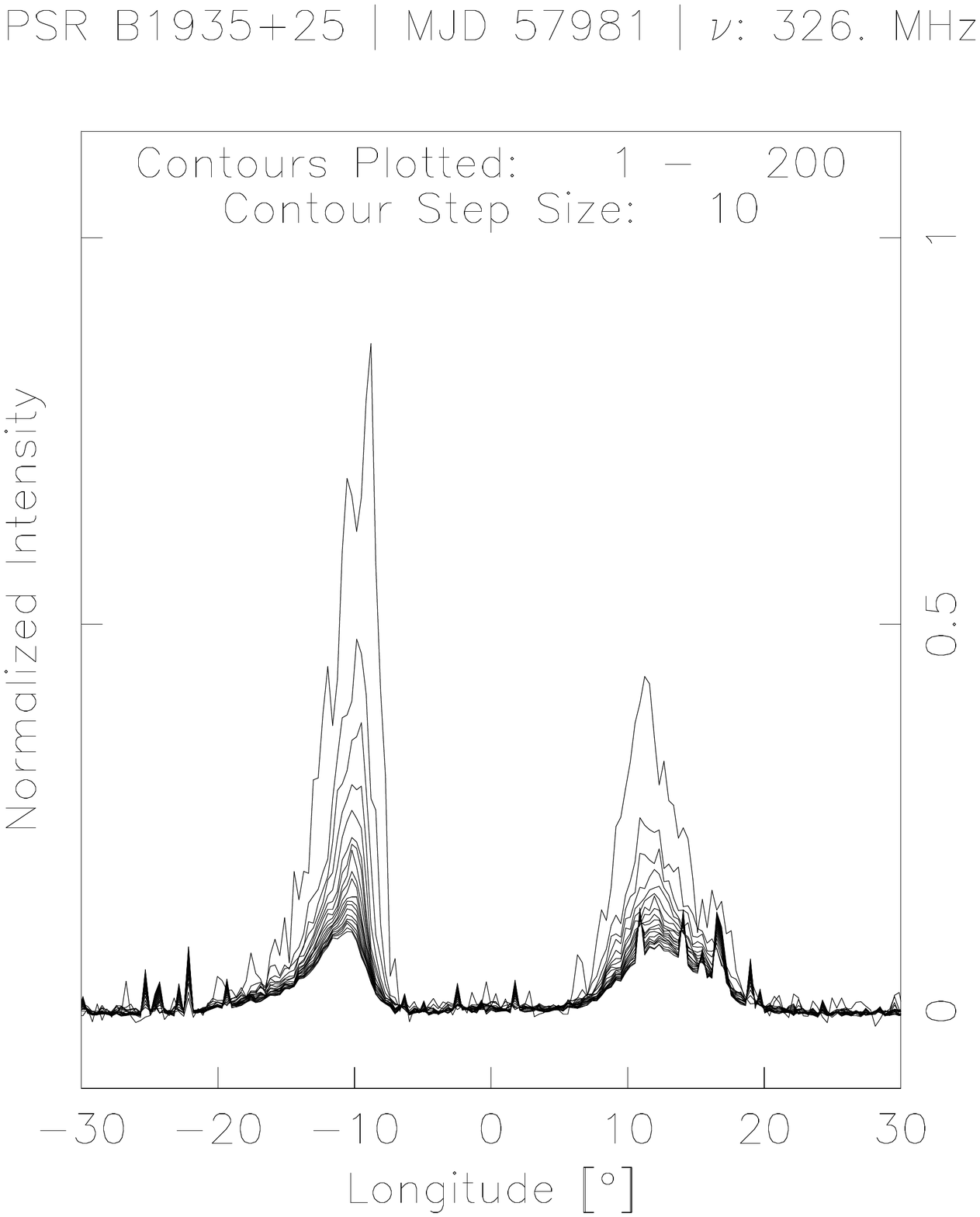} &
\includegraphics[page=1,width=\linewidth]{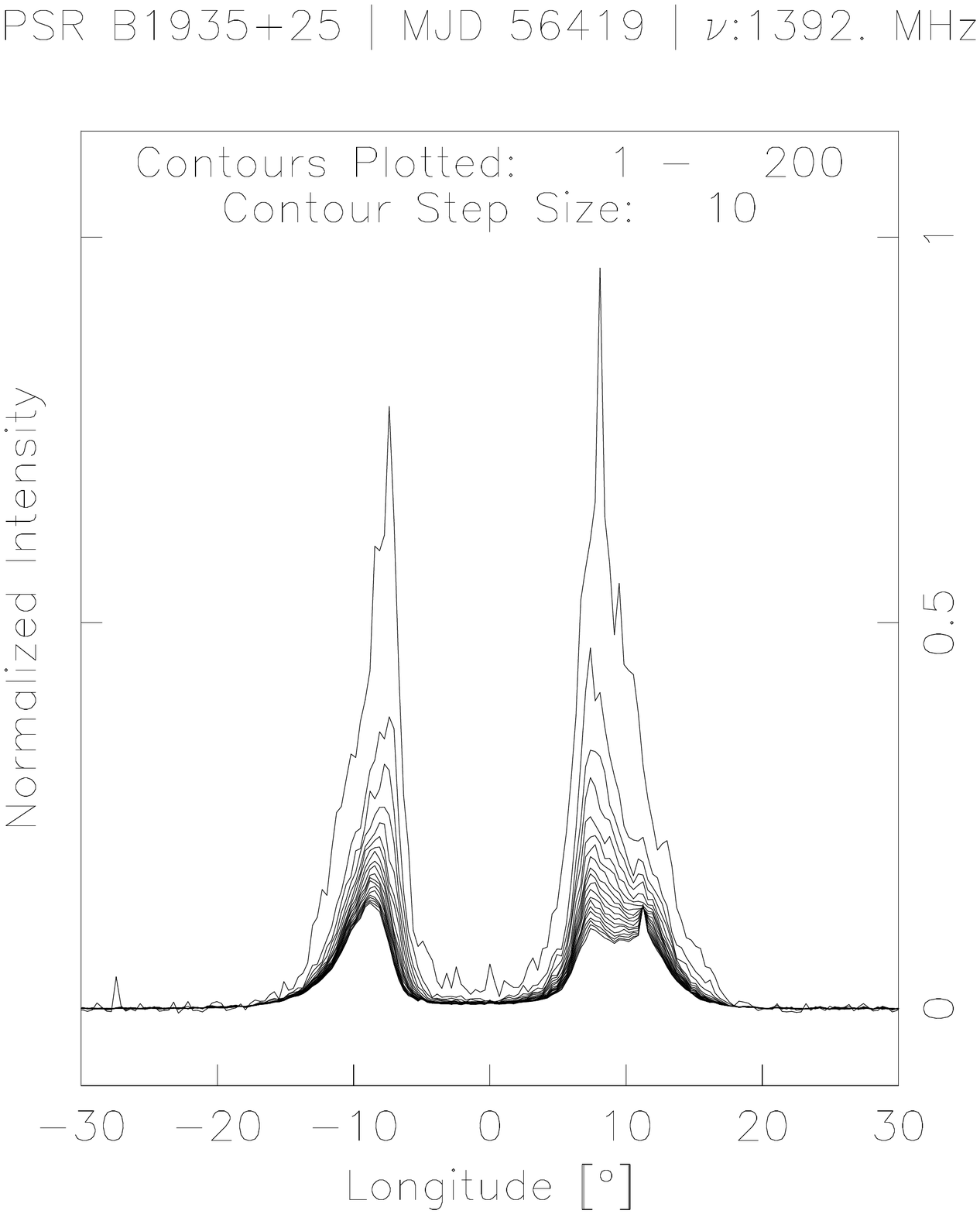} &
\includegraphics[page=1,width=\linewidth]{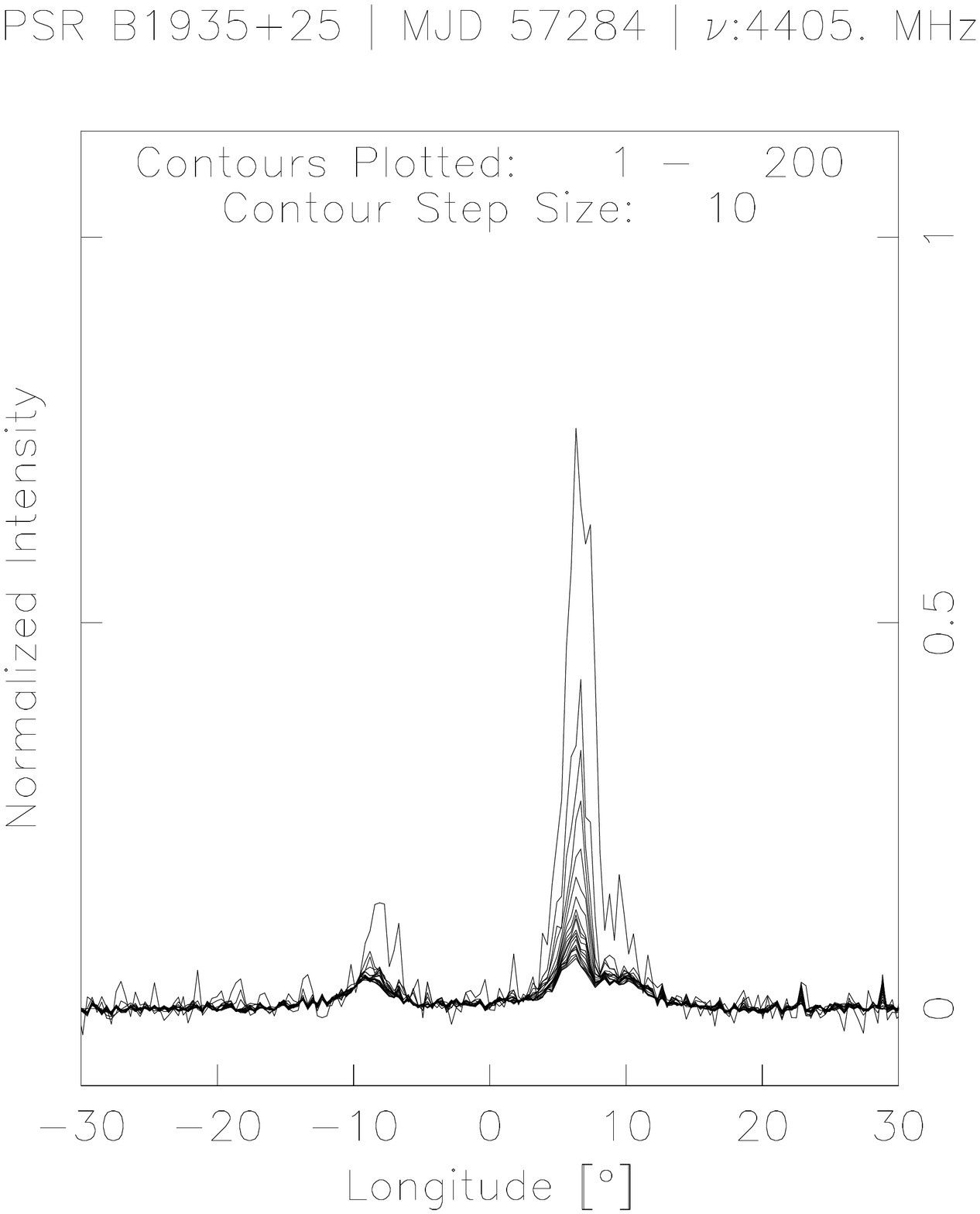} \\ \toprule
                &
\includegraphics[page=1,width=\linewidth]{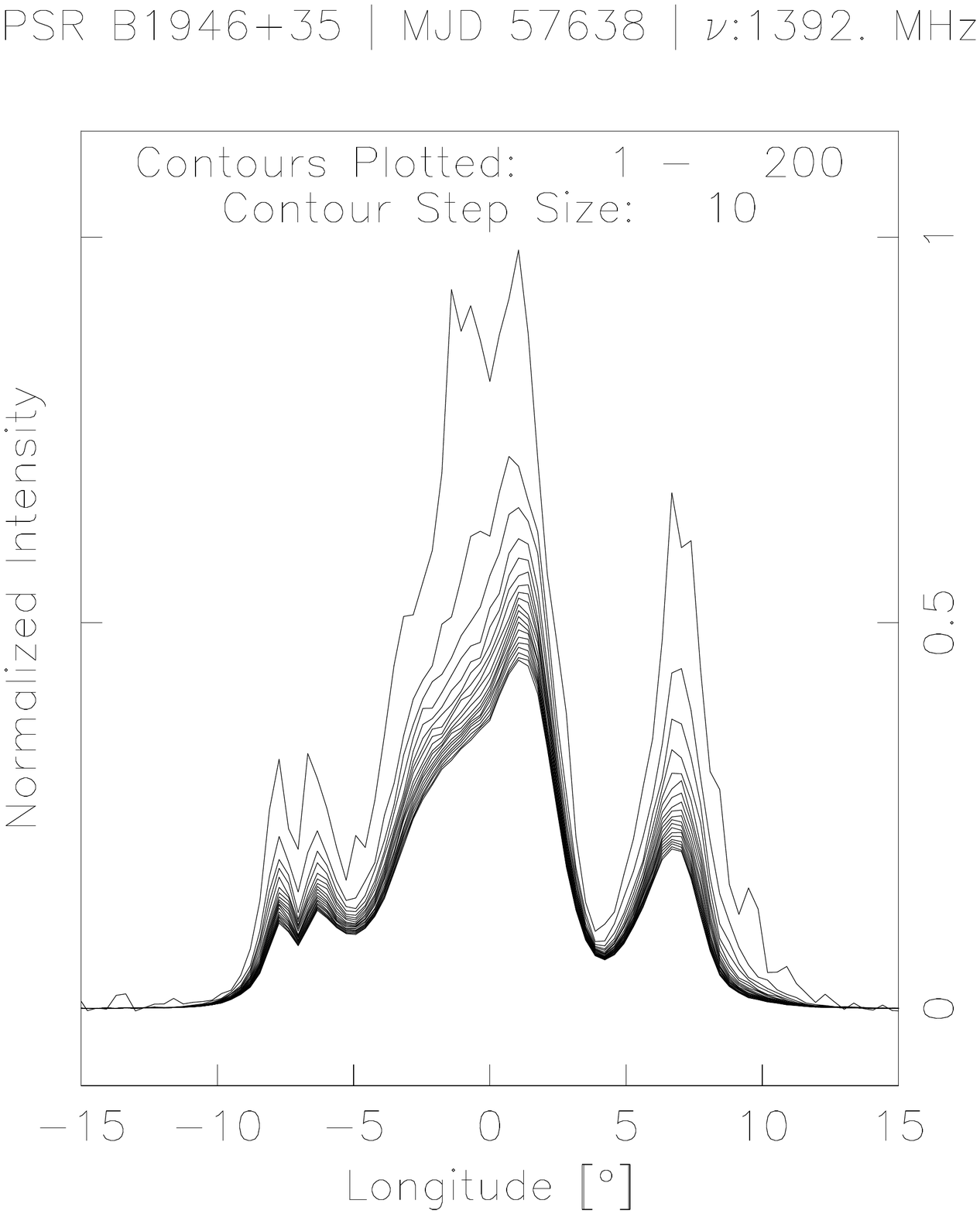} &
\includegraphics[page=1,width=\linewidth]{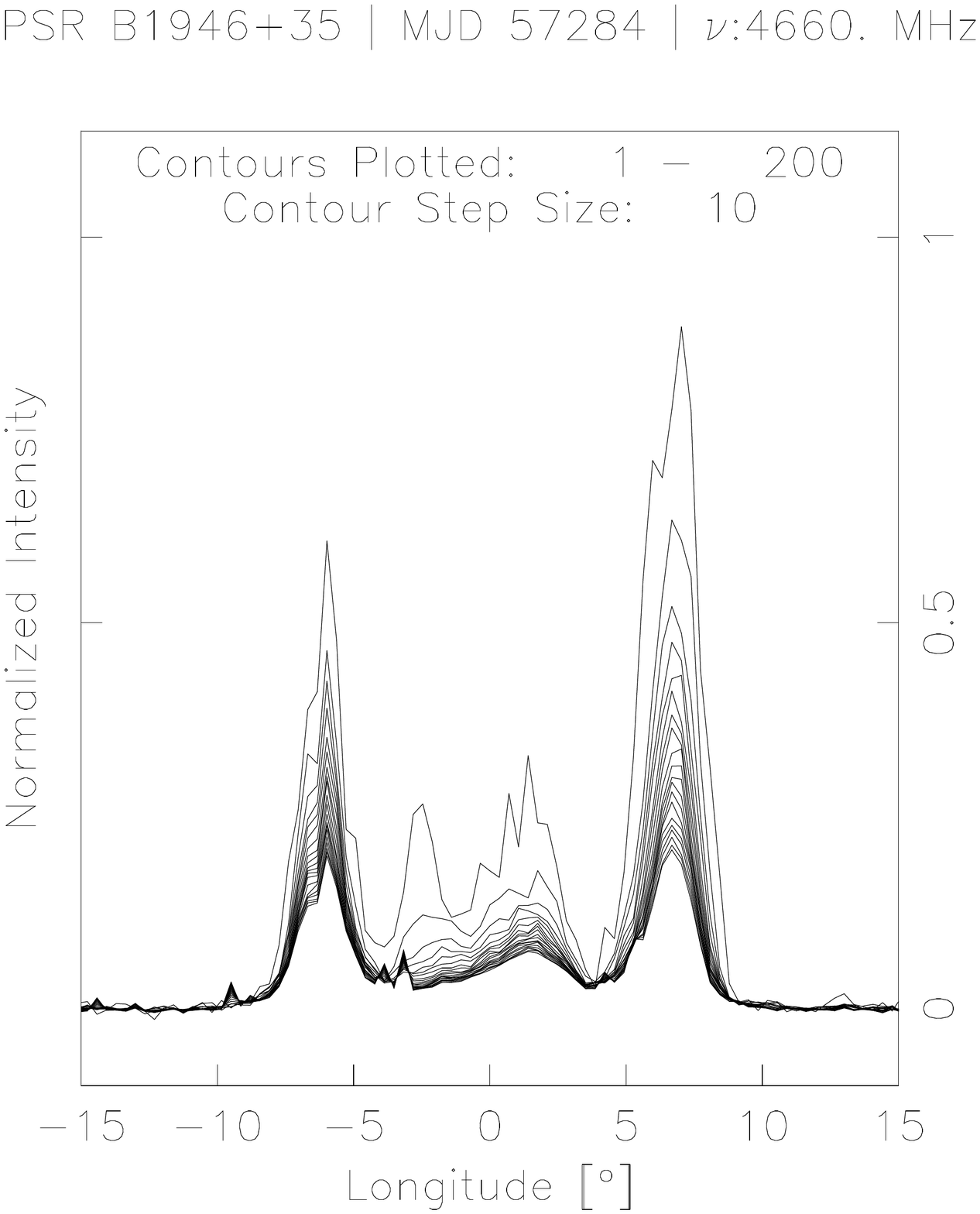} \\ 
     \bottomrule
   \end{tabularx} 
\caption{Average profiles of PSRs B1933+16, B1935+25, and B1946+35.}
 \end{figure*}
\vspace{1cm}

   \begin{figure*} 
 \begin{tabularx}{\textwidth}{YYY}
    \multicolumn{3}{c}{} \\ \toprule
    \includegraphics[page=1,width=\linewidth]{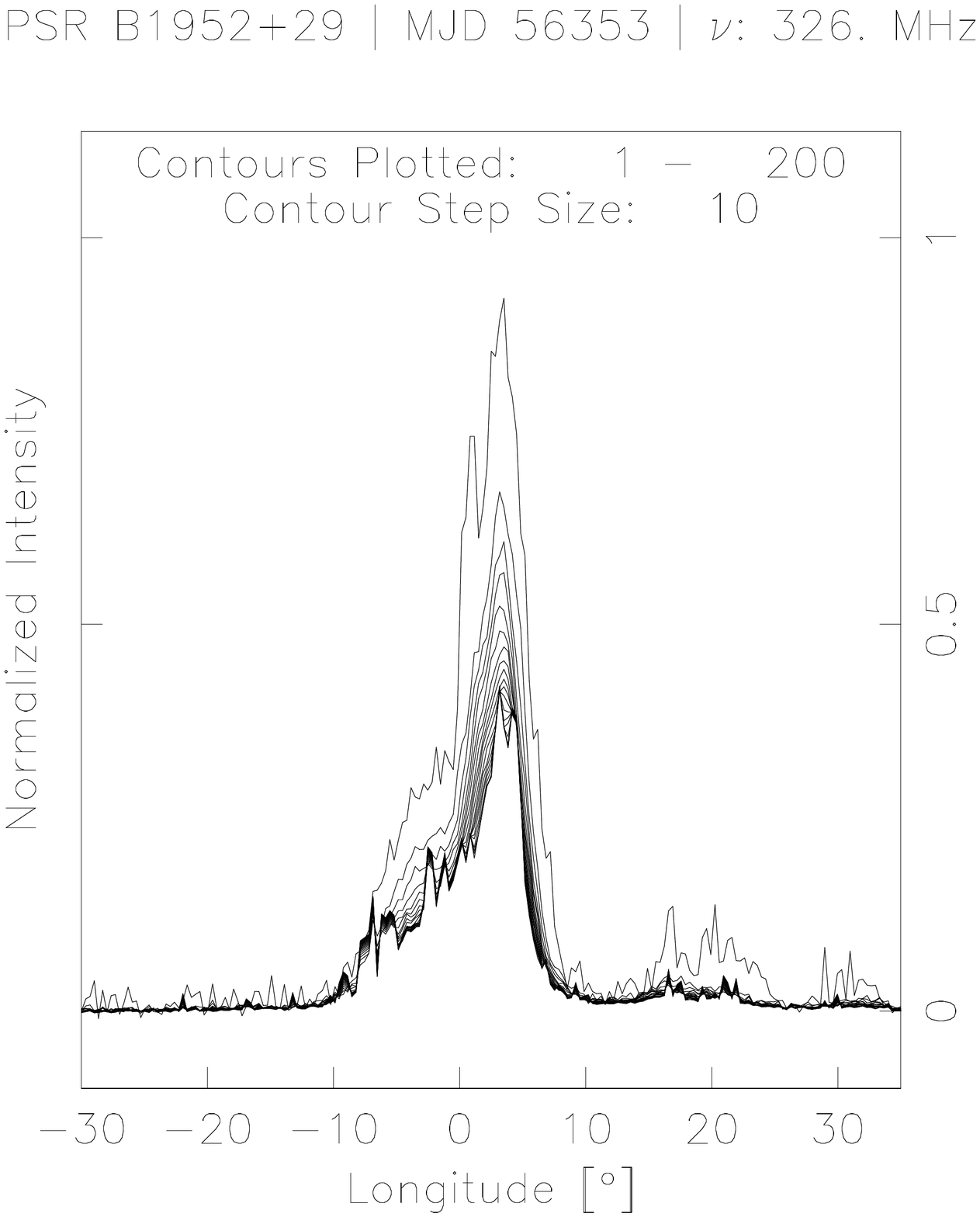} &
\includegraphics[page=1,width=\linewidth]{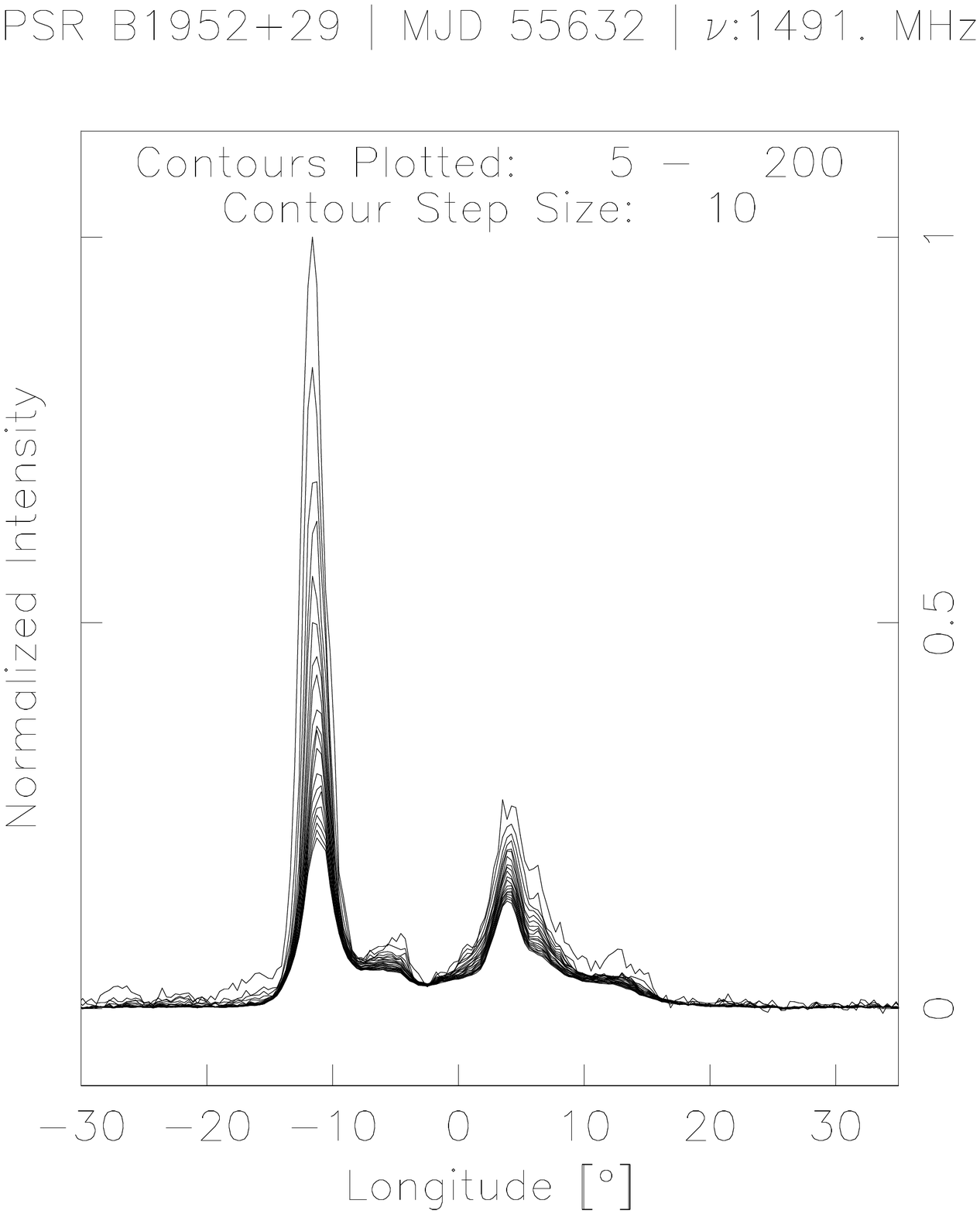} &
\includegraphics[page=1,width=\linewidth]{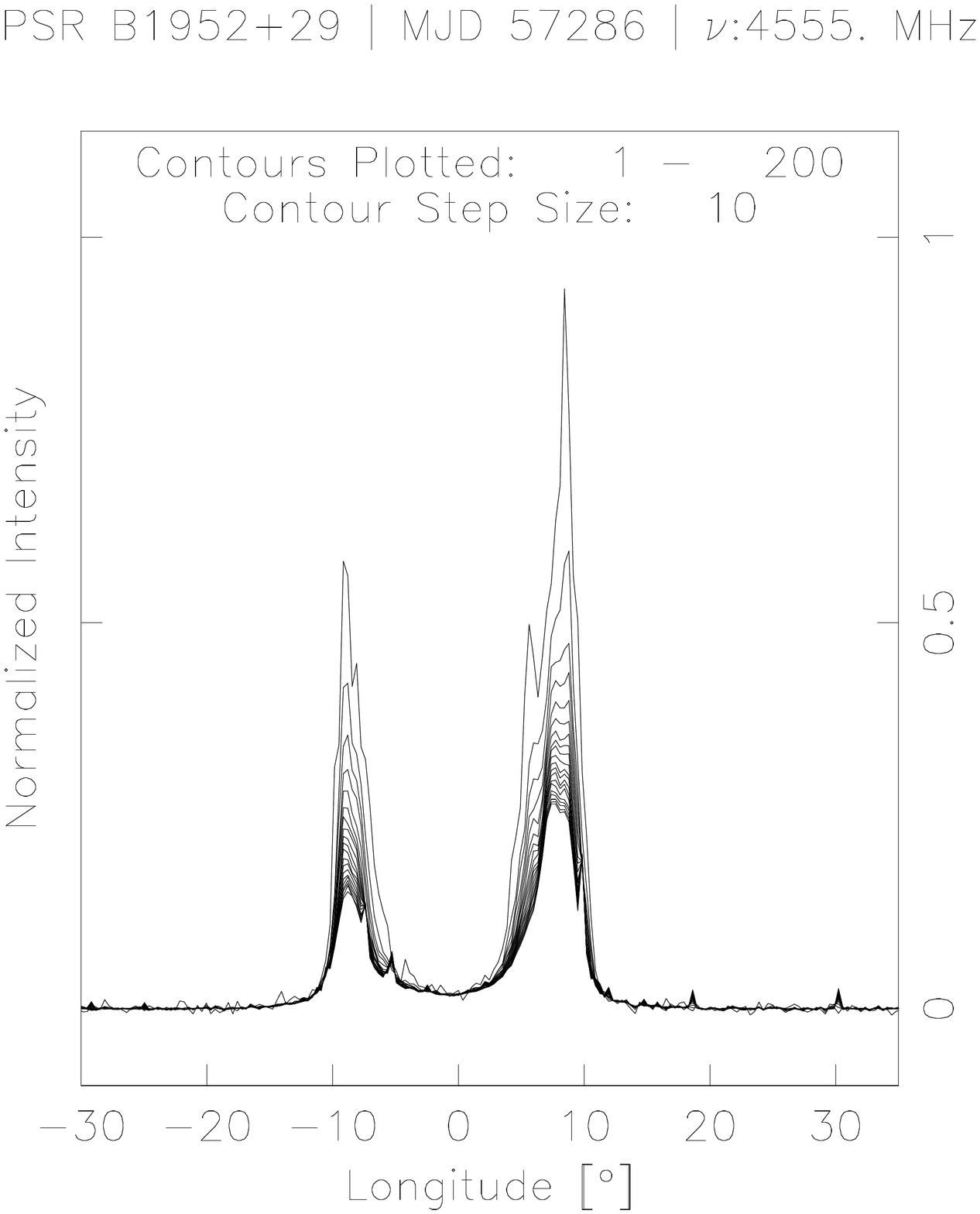} \\ \toprule
                 &
\includegraphics[page=1,width=\linewidth]{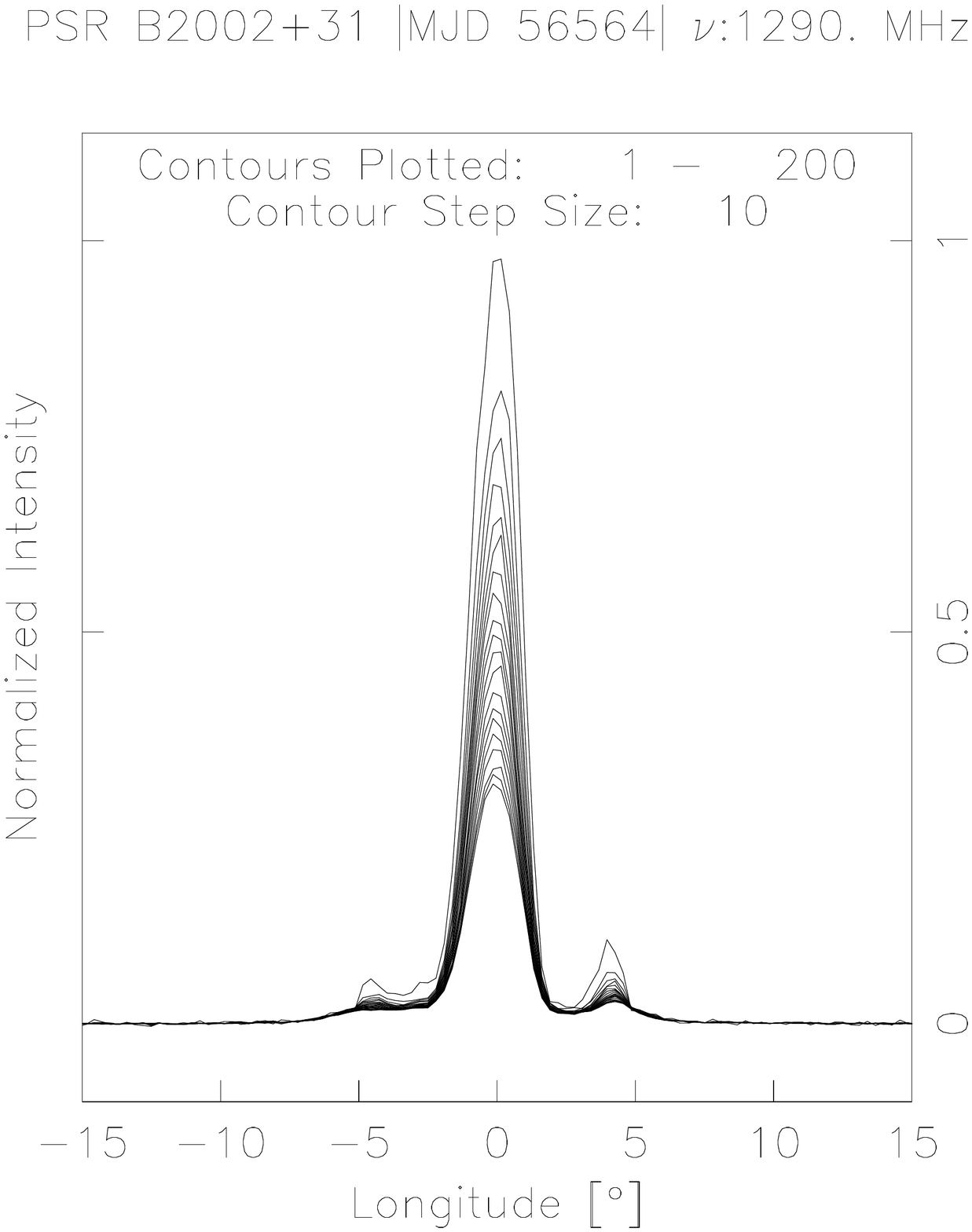} &
\includegraphics[page=1,width=\linewidth]{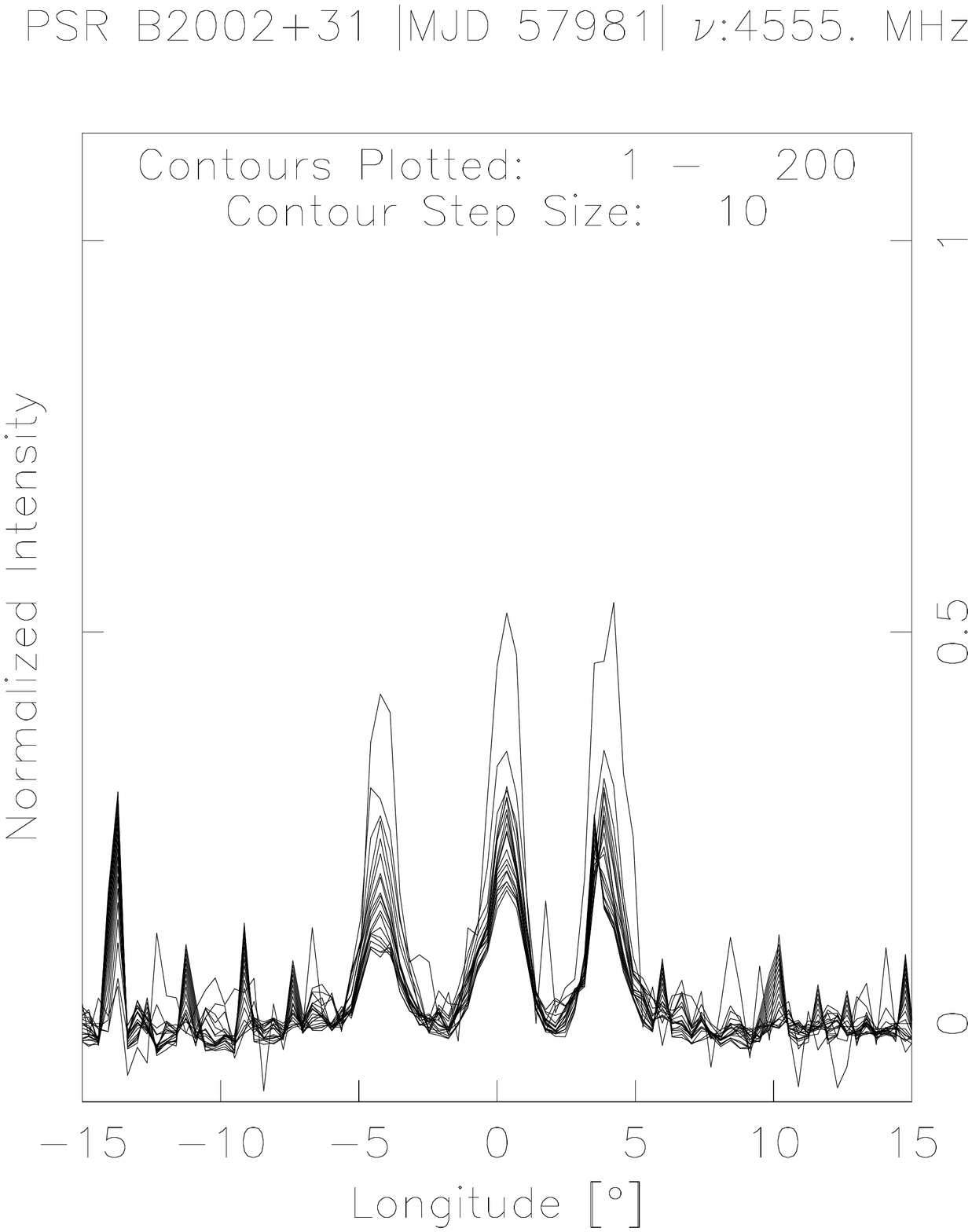} \\ \toprule
\includegraphics[page=1,width=\linewidth]{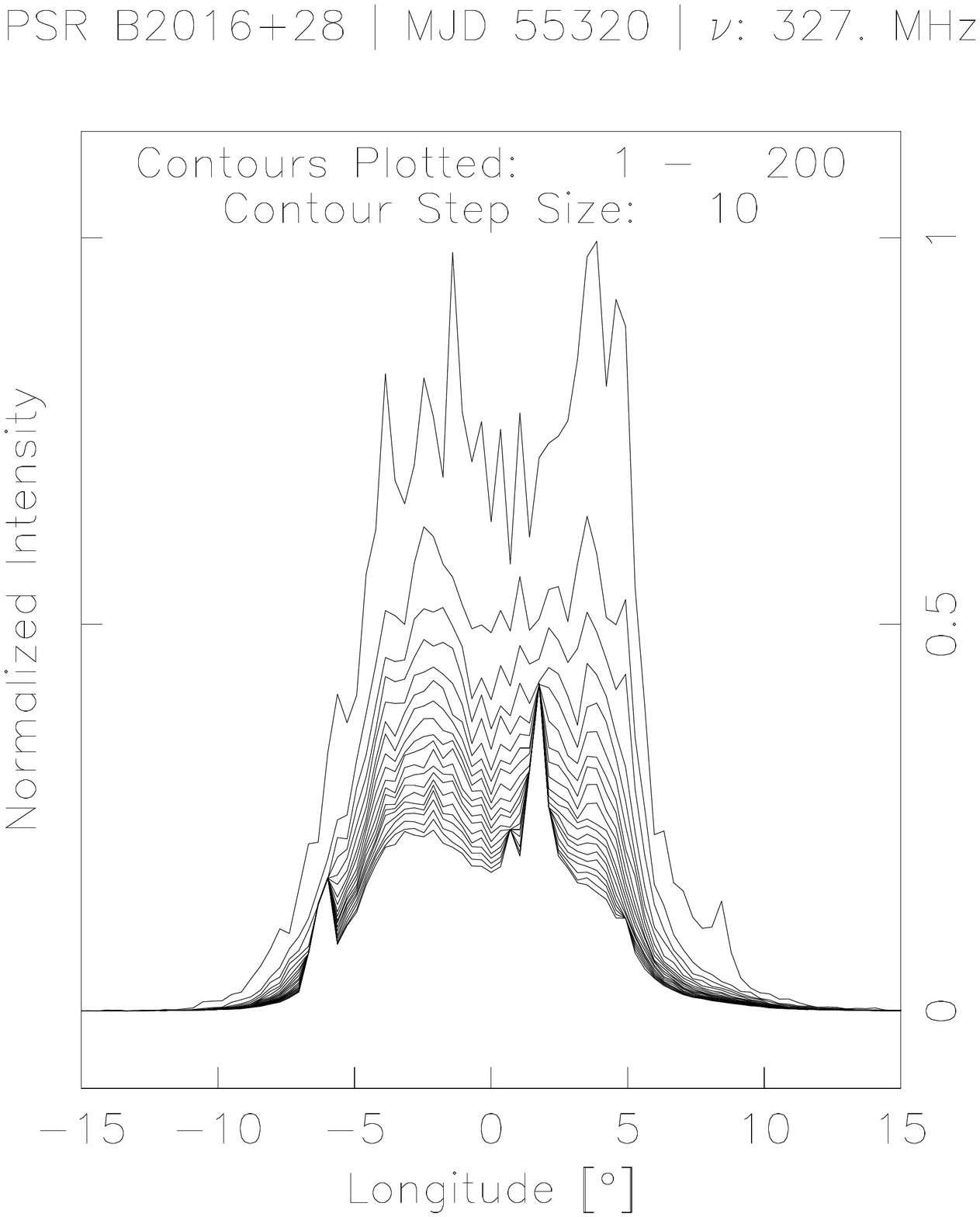} &
\includegraphics[page=1,width=\linewidth]{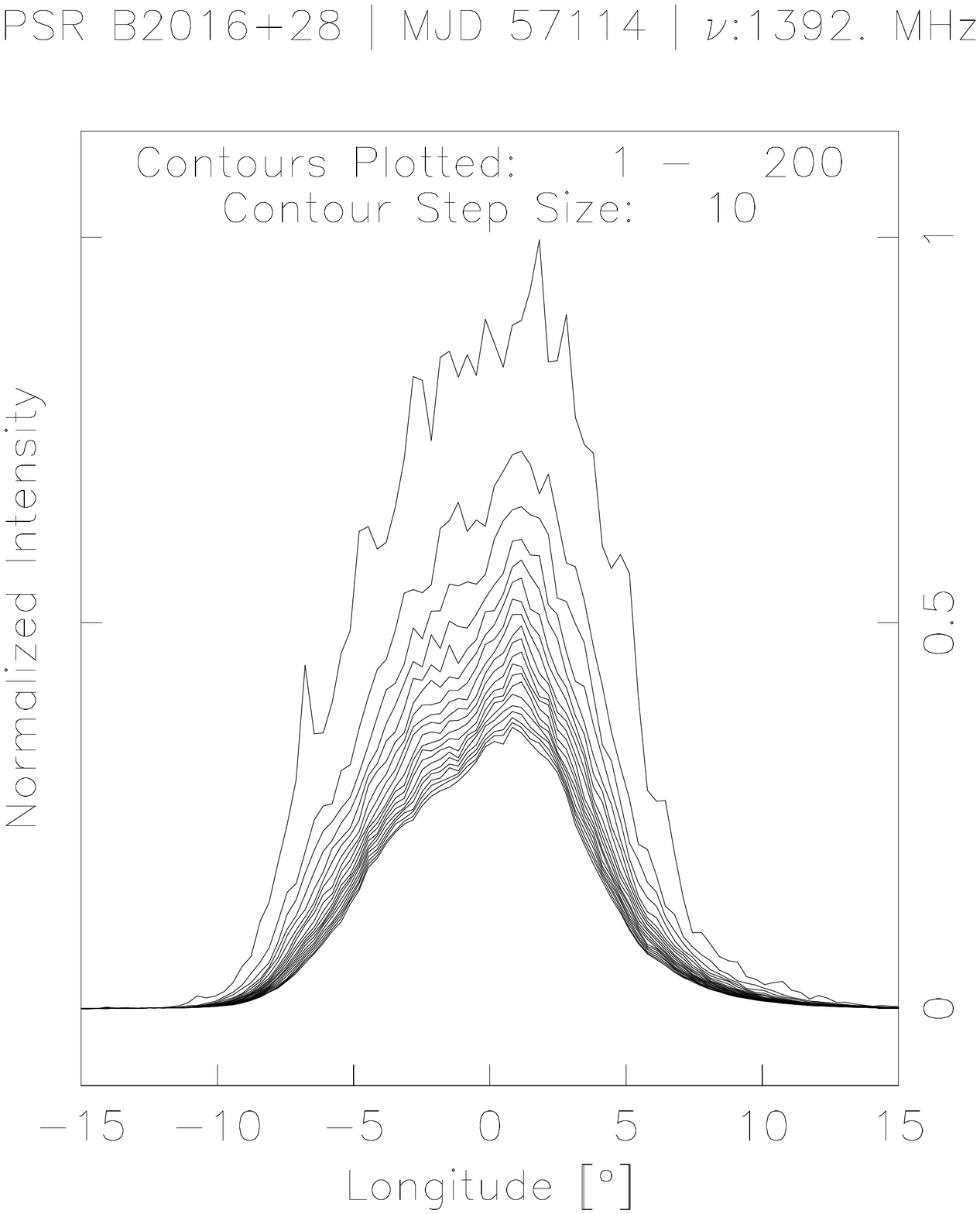} &
\includegraphics[page=1,width=\linewidth]{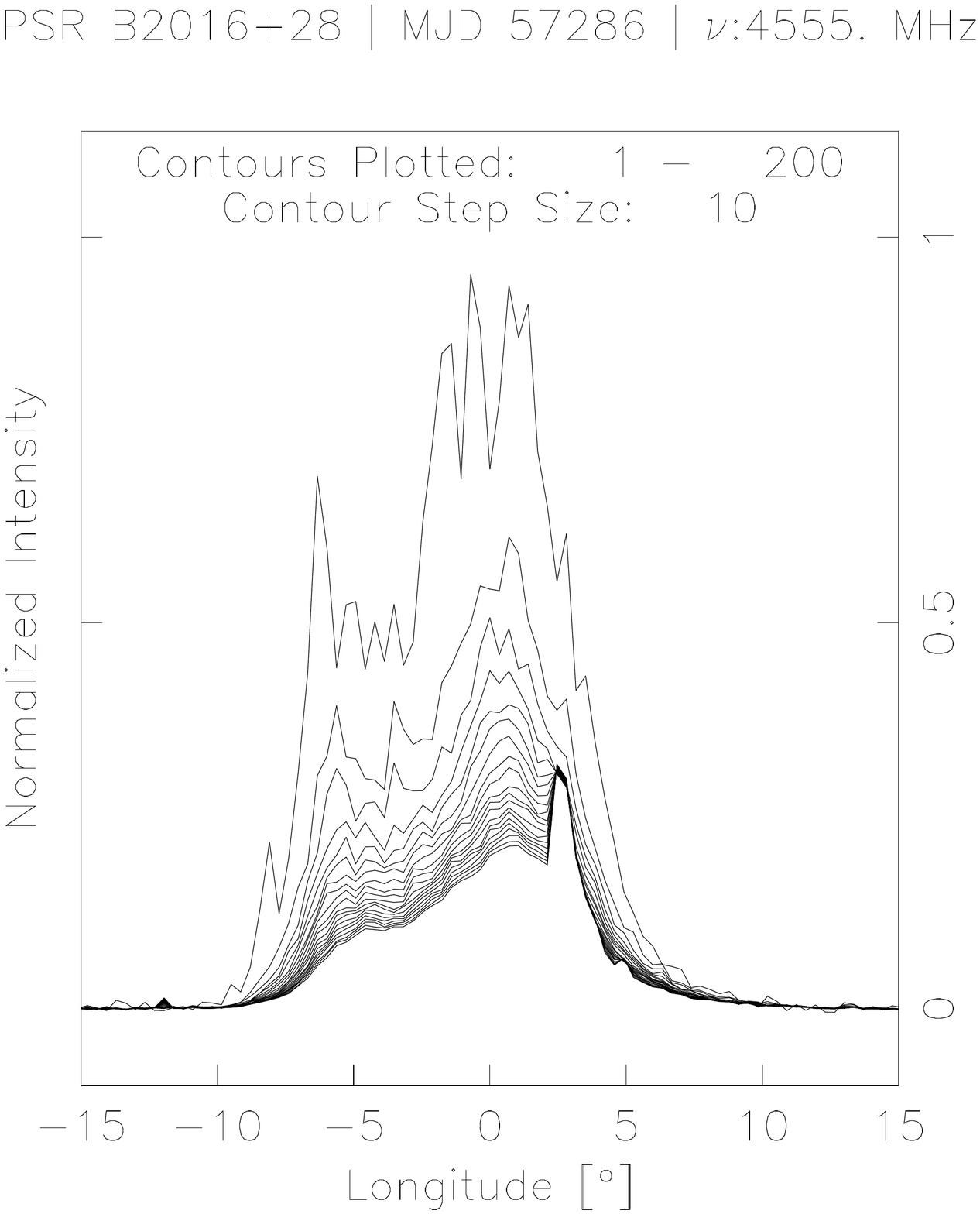} \\ 
     \bottomrule
   \end{tabularx} 
\caption{Average profiles of PSRs B1952+29, B2002+31, and B2016+28.}
 \end{figure*}
\vspace{1cm}

   \begin{figure*} 
 \begin{tabularx}{\textwidth}{YYY}
    \multicolumn{3}{c}{} \\ \toprule
    \includegraphics[page=1,width=\linewidth]{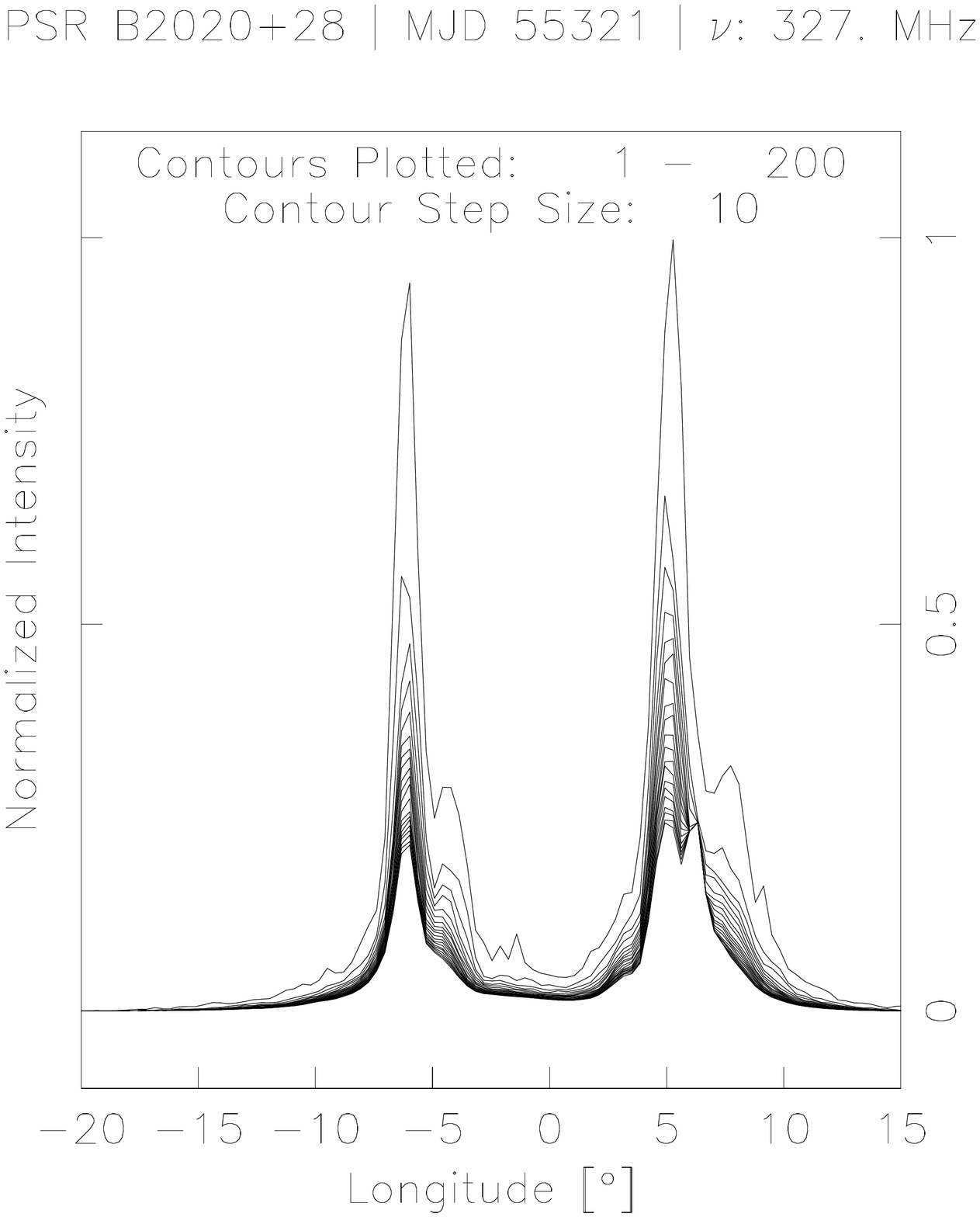} &
\includegraphics[page=1,width=\linewidth]{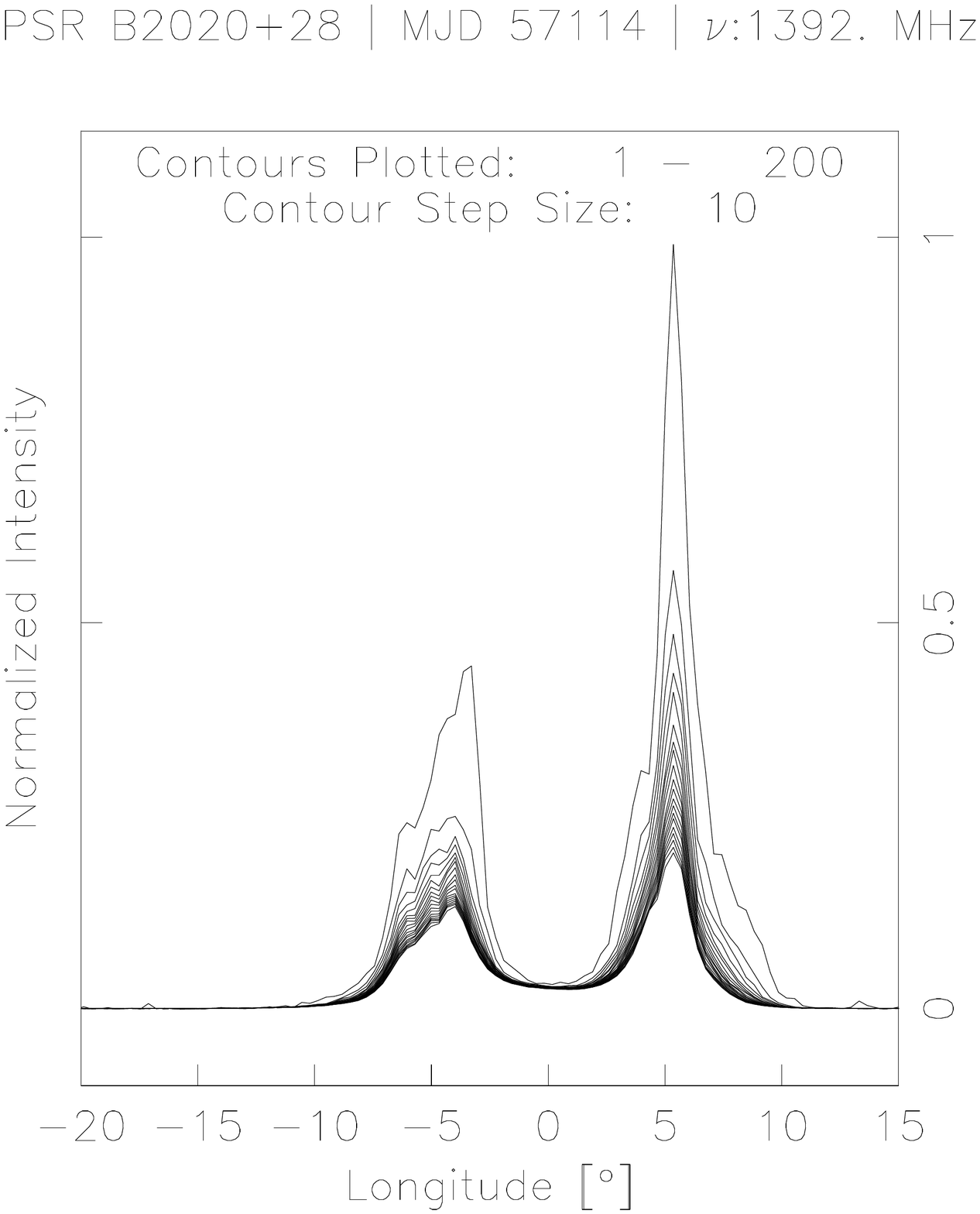} &
\includegraphics[page=1,width=\linewidth]{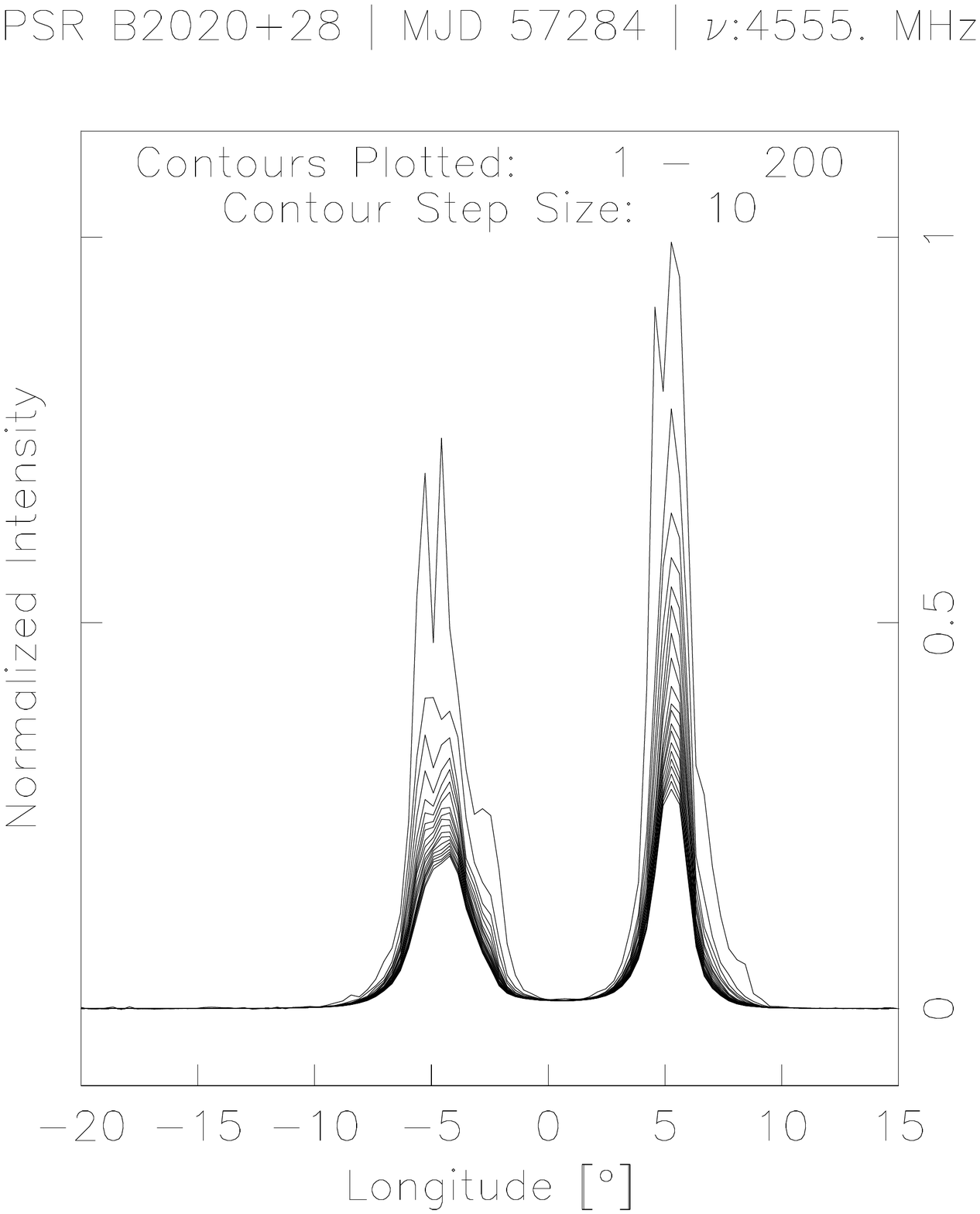} \\ \toprule
\includegraphics[page=1,width=\linewidth]{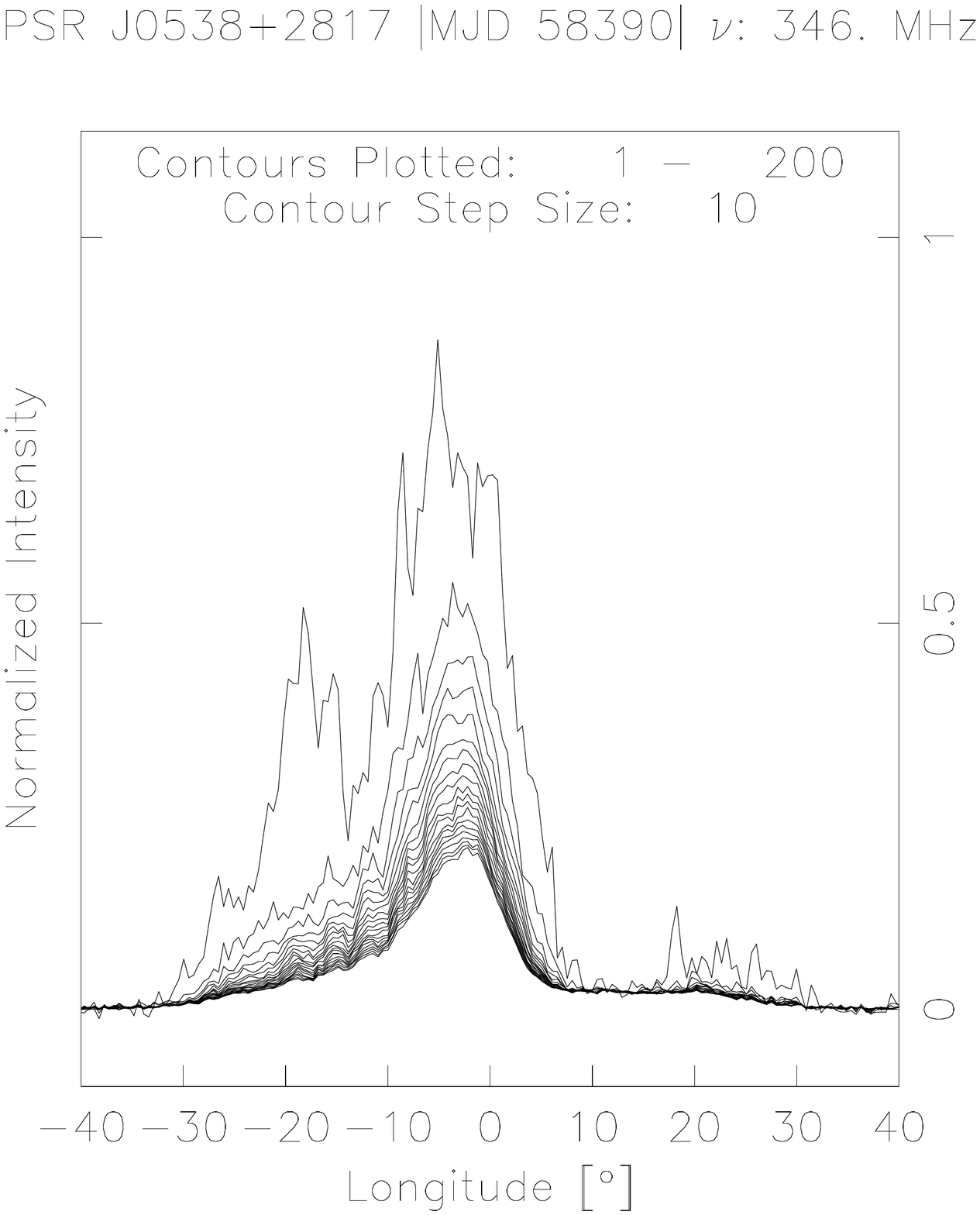} &
 \includegraphics[page=1,width=\linewidth]{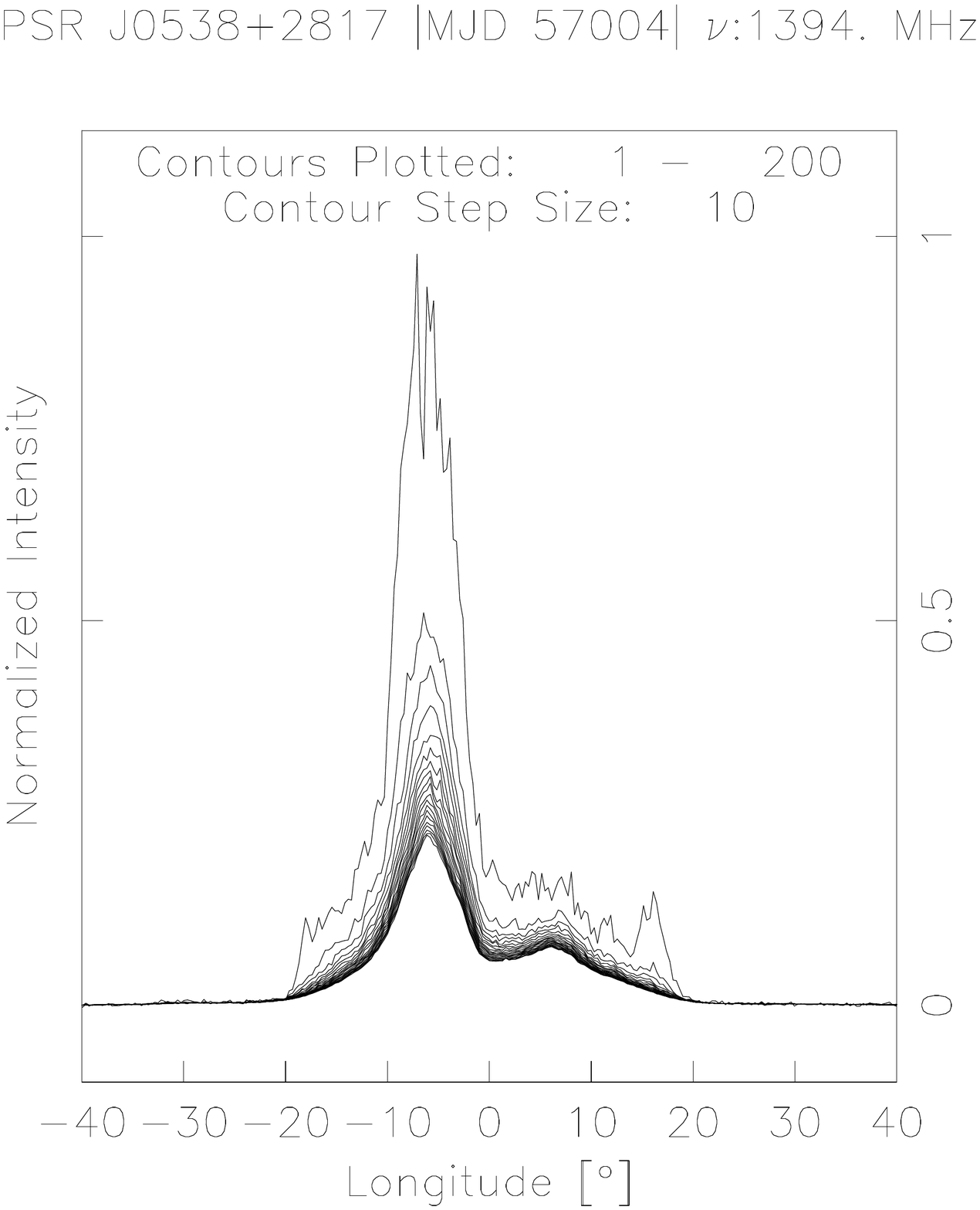} &
 \\
\includegraphics[page=1,width=\linewidth]{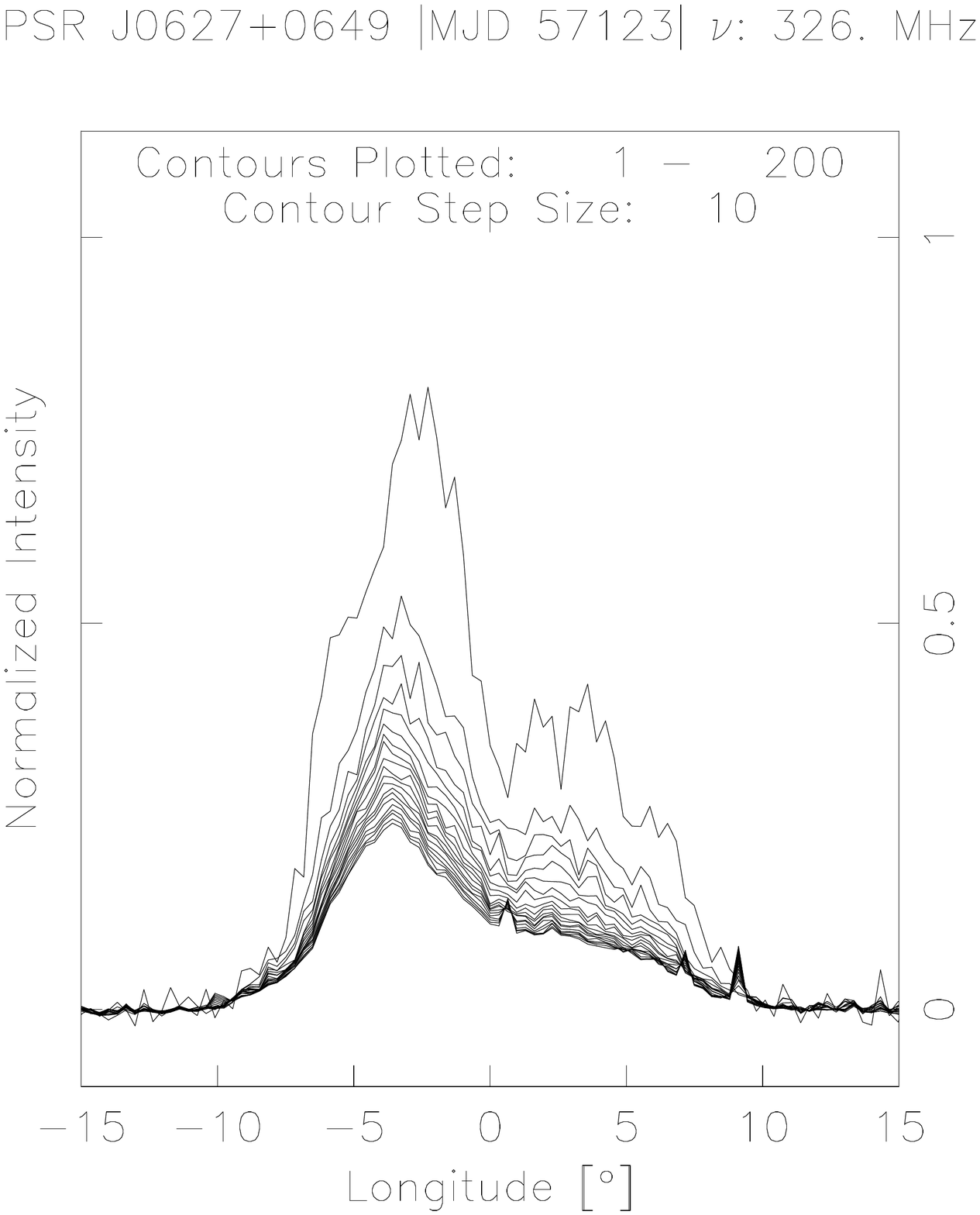} &
\includegraphics[page=1,width=\linewidth]{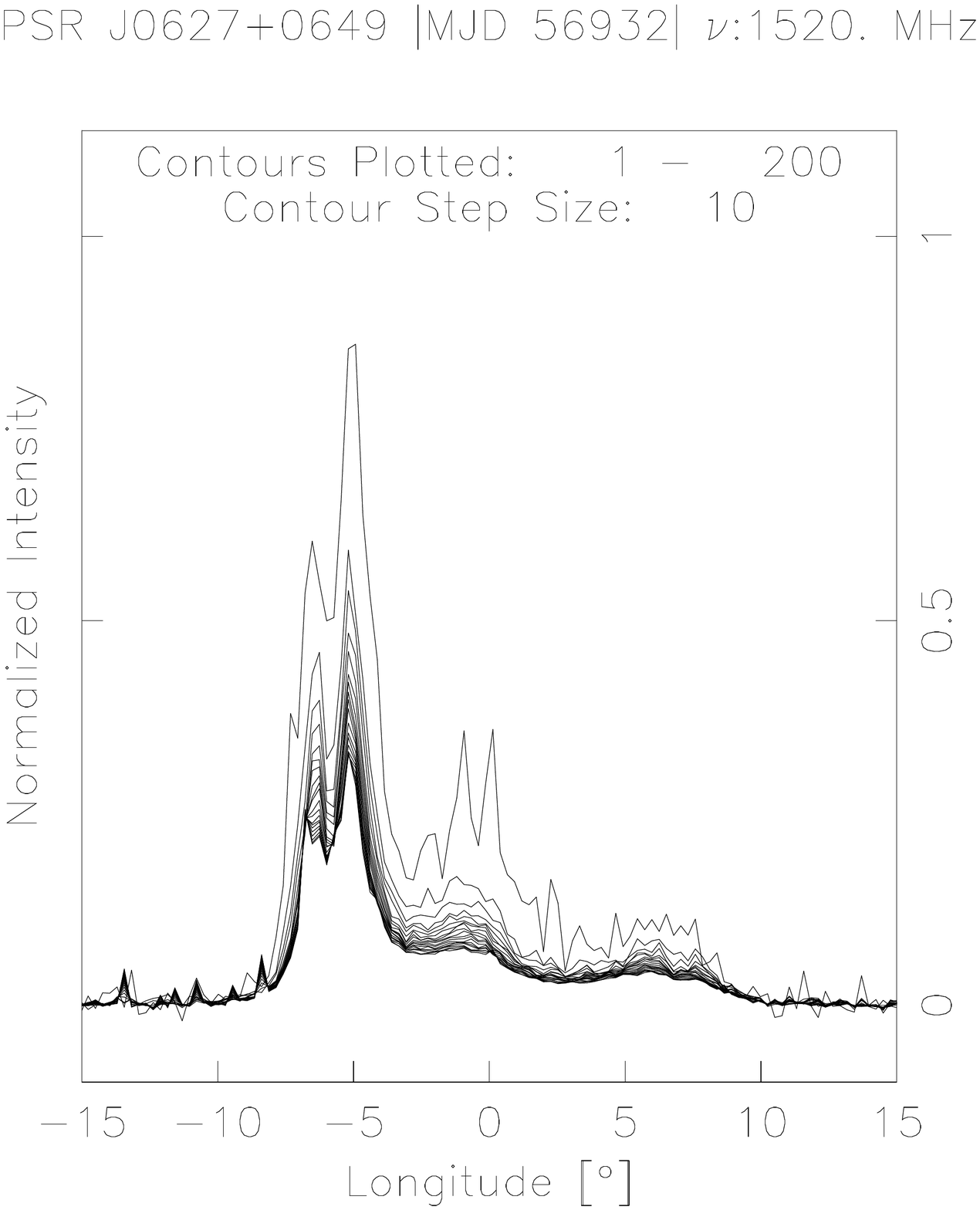} &
\includegraphics[page=1,width=\linewidth]{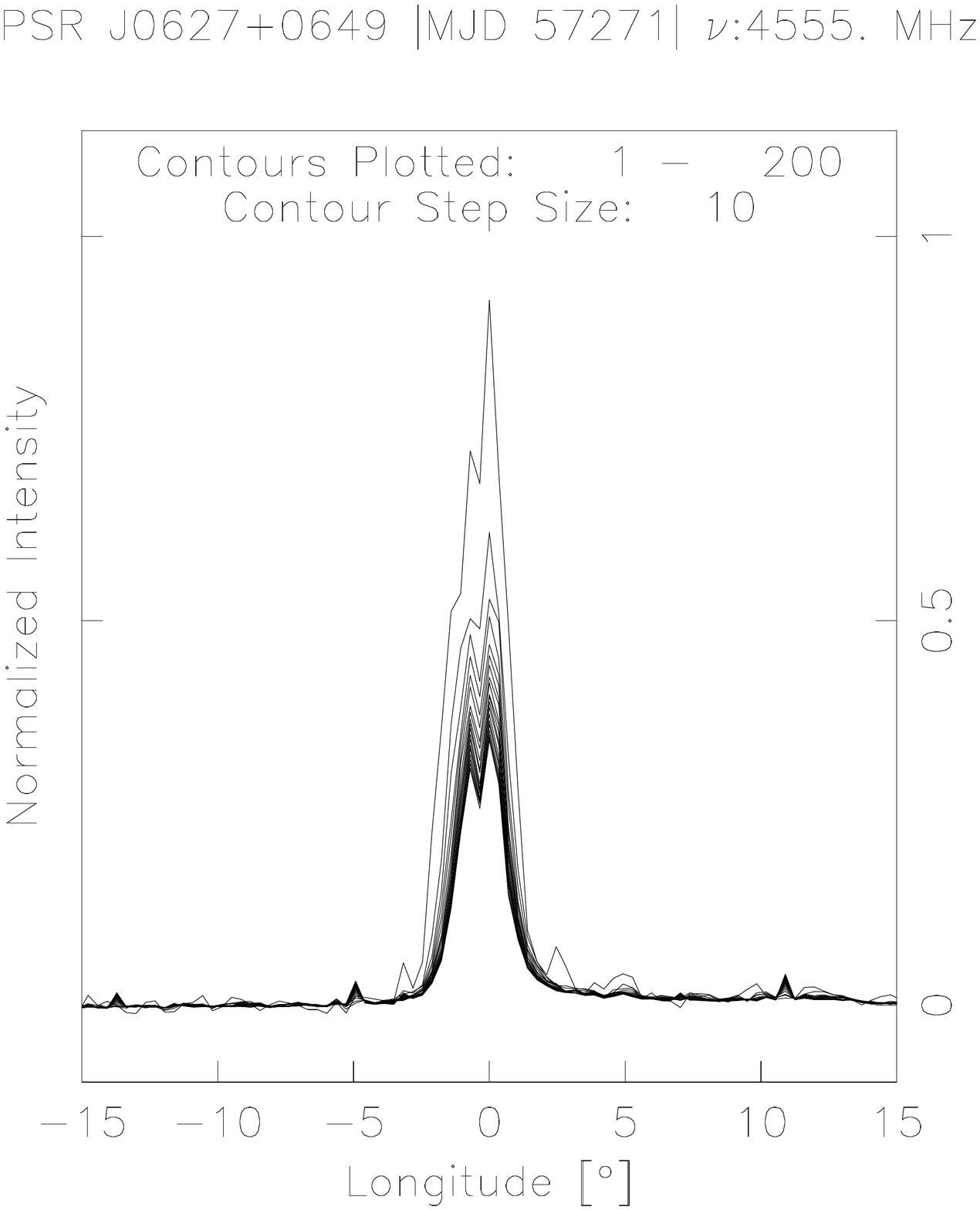} \\ 
     \bottomrule
   \end{tabularx} 
\caption{Average profiles of PSRs B2020+28, J0538+2817, and J0627+0649.}
 \end{figure*}
\vspace{1cm}

   \begin{figure*} 
 \begin{tabularx}{\textwidth}{YYY}
    \multicolumn{3}{c}{} \\ \toprule
\includegraphics[page=1,width=\linewidth]{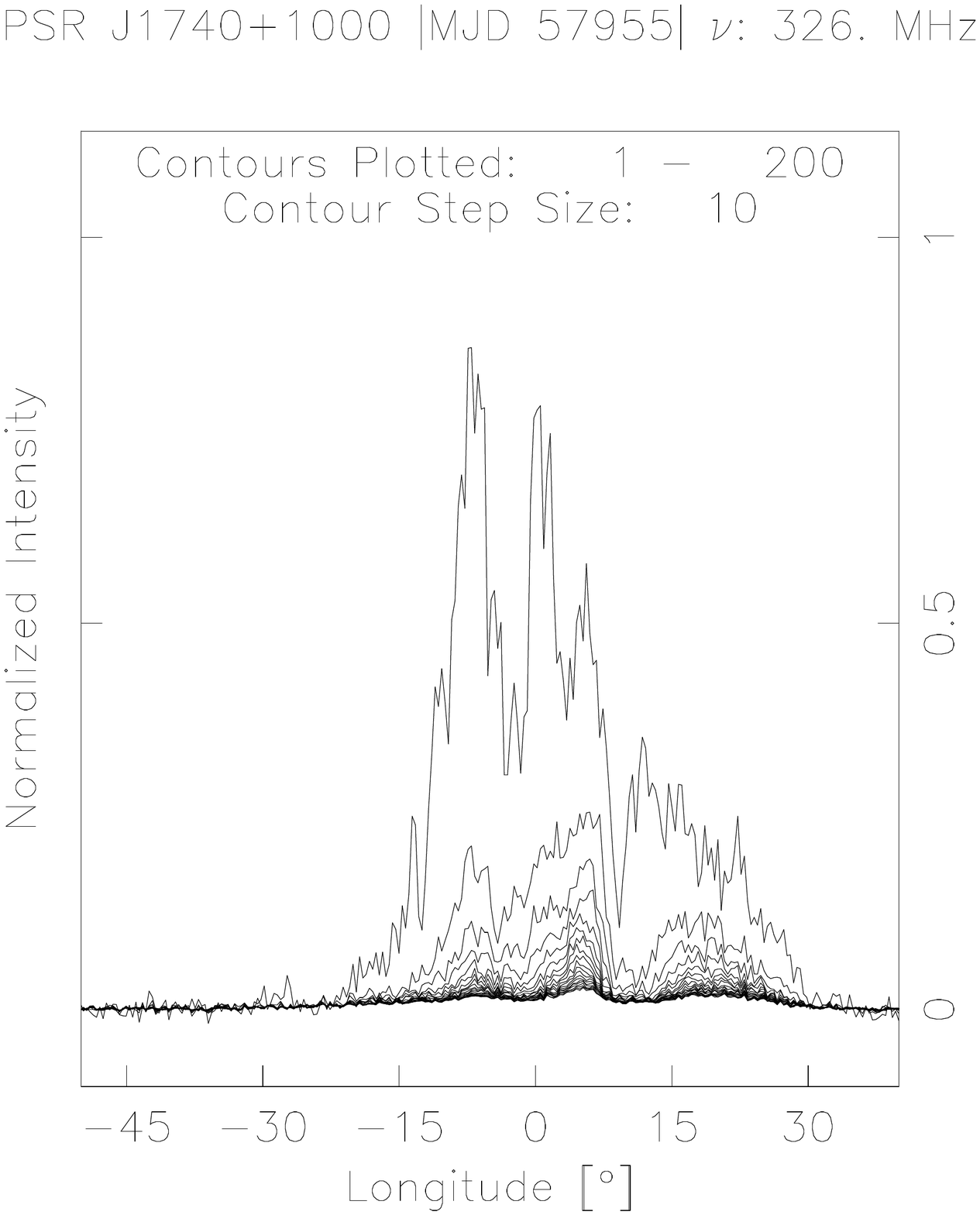} &
\includegraphics[page=1,width=\linewidth]{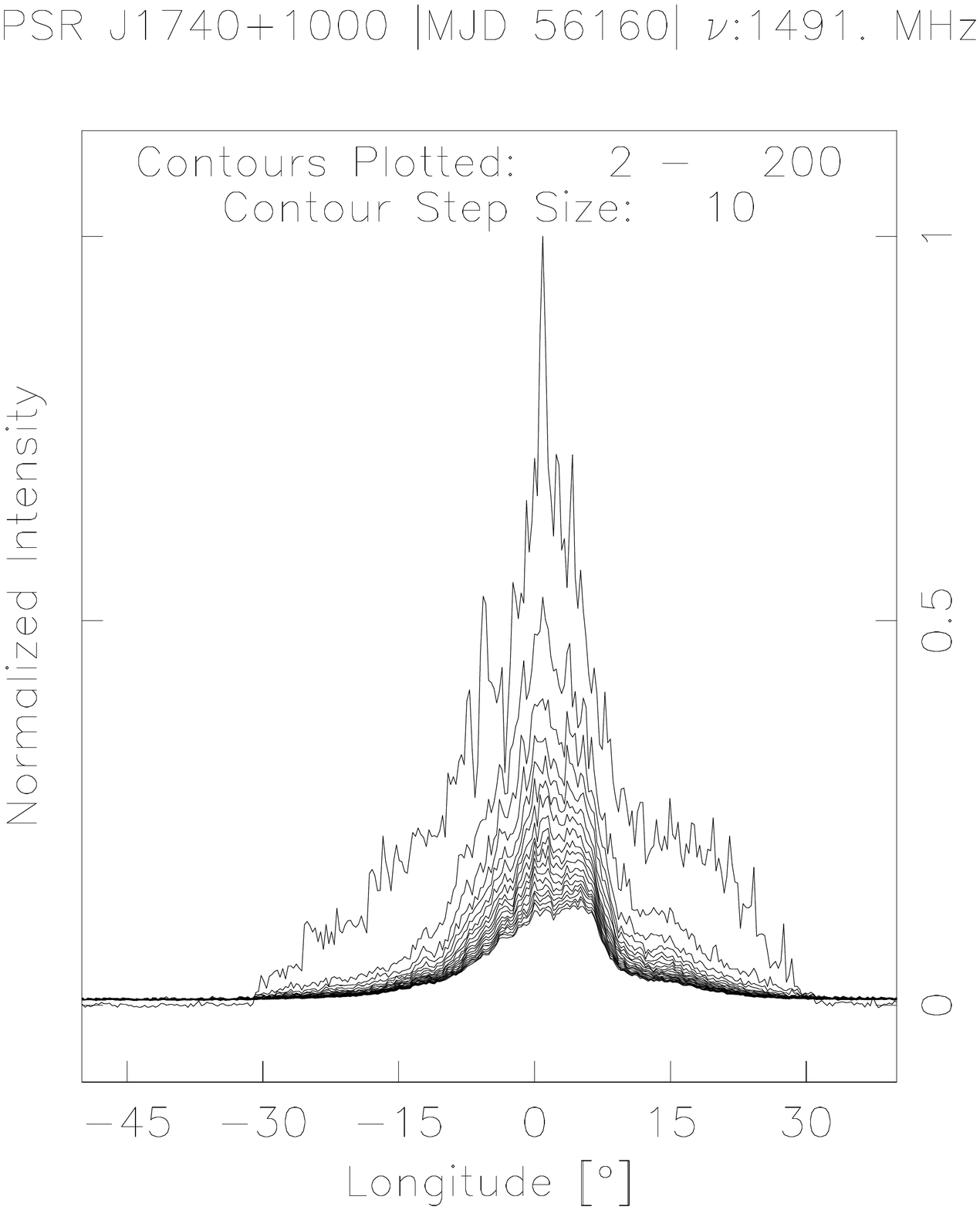} &
\includegraphics[page=1,width=\linewidth]{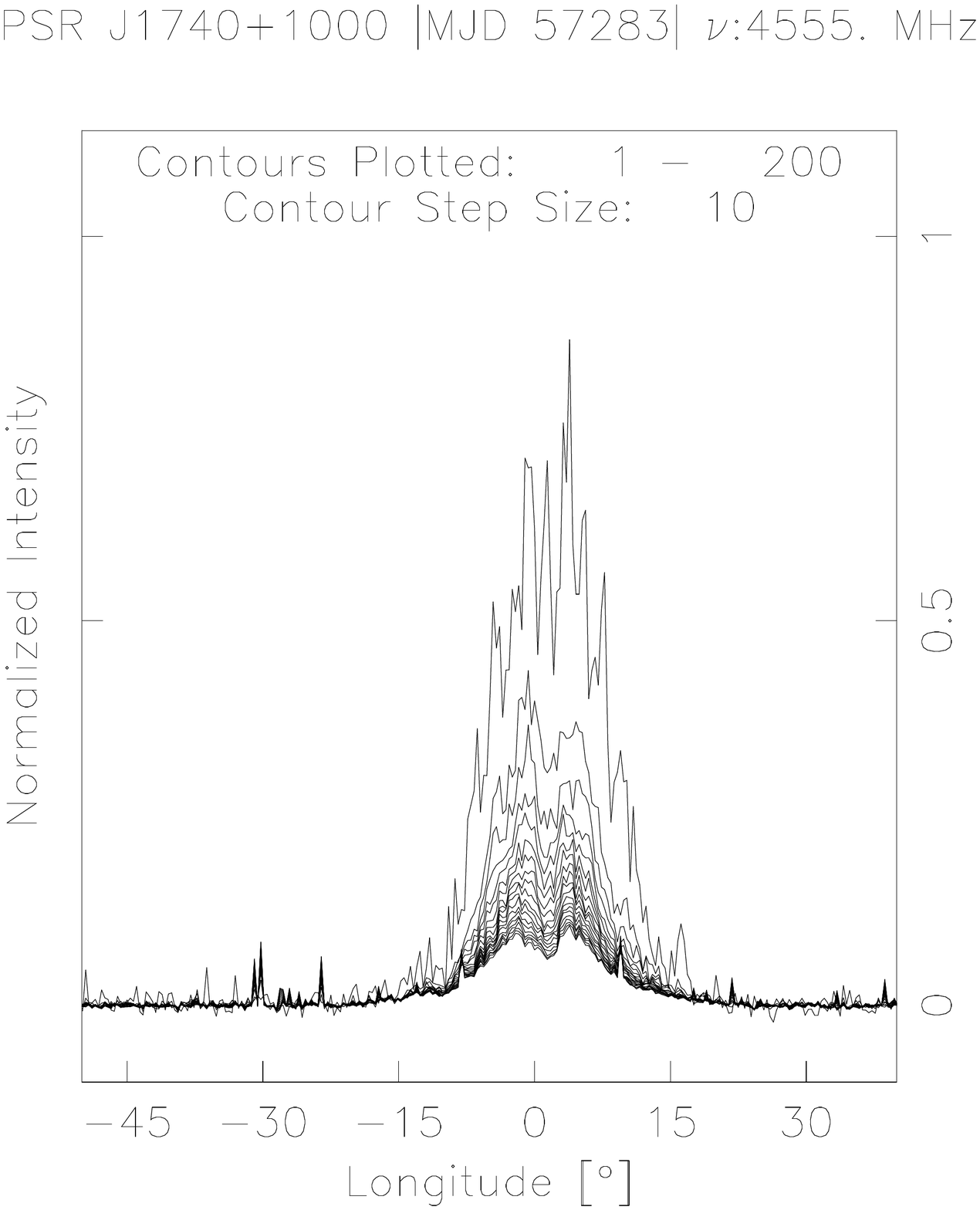} \\ 
     \bottomrule
   \end{tabularx} 
\caption{Average profile of PSR J1740+1000.}
 \end{figure*}
\vspace{1cm}

\twocolumn



\bsp	
\label{lastpage}
\end{document}